%% file: Graph-Homomorphisms-and-Universal-Algebra.tex
\documentclass[a4paper,11pt]{article} 
\usepackage[hidelinks]{hyperref}\usepackage{amssymb}
\usepackage{totcount,xcolor,todonotes}
\usepackage{tikz}
\tikzstyle{bullet}=[circle,fill,inner sep=0pt,minimum size=3pt]
\tikzstyle{bullethole}=[circle,draw,inner sep=0pt,minimum size=3pt]
\usetikzlibrary{arrows}

\input local-defs.tex
\DeclareMathOperator{\Sub}{Sub}

\DeclareMathOperator{\AIP}{AIP}
\DeclareMathOperator{\IP}{IP}
\DeclareMathOperator{\ar}{ar}

\newtheorem{examplex}[theorem]{Example}
\renewenvironment{example}
  {\pushQED{\qed}\examplex}
  {\popQED\endexamplex}
\usepackage{wrapfig}

\begin{document}

\newcounter{probcounter}

\title{Graph Homomorphisms and Universal Algebra \\
Course Notes}
\author{Manuel Bodirsky, \\Institut f\"ur Algebra, TU Dresden, \\
manuel.bodirsky@tu-dresden.de}

\date{September 14, 2026}
\maketitle

% Abstract:
% Constraint satisfaction problems are computational problems that naturally appear in many areas of theoretical computer science. One of the central themes is their computational complexity, and in particular the border between polynomial-time tractability and NP-hardness. In this course we introduce the universal-algebraic approach to study their computational complexity. To keep the presentation accessible, we start the course in the beautiful setting of directed graphs and graph homomorphism problems. 
% Topics: 1) digraphs, homomorphisms, cores, polymorphisms. 
% 2) Arc consistency, power graph, tree duality, totally symmetric polymorphisms. 
% 3) Path consistency, majority polymorphisms, testing for majority polymorphisms. 

{\bf Disclaimer:} these are course notes in draft state, and they might contain mistakes; in case that you find any, please report them to \emph{manuel.bodirsky@tu-dresden.de}. Almost all of the material is not original; I have tried to provide references to the primary sources as much as possible, but might have failed at some places, in which case I am thankful for hints as well. 

\tableofcontents

%PREFACE:
% The "questions" are recognized and well-known questions in 
% the field, and supposedly very difficult; on the 
% other hand, they are probabely not as difficult as the question
% whether P=NP, and they continue to stimulate research activity.
% We might except to hear the answers to some
% of these questions within our lifetimes. 
% It is possible to skip  the section on graph homomorphisms entirely.

%%%%%%%%%%%%%

%\vspace{1cm}

\newpage 
\noindent {\bf Course Goals.} This course has different groups of readers in mind. 
\begin{itemize}
\item For readers interested in theoretical computer science, it provides an introduction to complexity analysis of constraint satisfaction problems.
\item For readers interested in graph theory and graph homomorphisms, it provides an introduction to a universal-algebraic approach to many fundamental questions in the area. 
\item For readers interested in universal algebra, it treats the part of universal algebra that is most relevant for the study of graph homomorphism problems and constraint satisfaction. 
\end{itemize}

\noindent {\bf Prerequisites.} This course is designed for students of mathematics or computer science who  have already had an introduction to discrete structures (so we assume that the reader knows basic set theory, the natural numbers ${\mathbb N}$, the integers ${\mathbb Z}$, the rational numbers ${\mathbb Q}$, and the residue rings ${\mathbb Z}_n$). 
All further notions that we use in this text will be formally introduced, with the following notable exceptions. 
\begin{itemize}
\item Basic concepts from complexity theory. For example, we do not formally introduce the class of polynomial-time computable functions and NP-completeness, even though these concepts are used when we discuss computational aspects of graph homomorphisms. Here we refer to an introduction to the theory of computation as for instance the book of Papadimitriou~\cite{Papa}; we have also added a short appendix with the basics (Appendix~\ref{sect:complexity}). 
\item Basic concepts of first-order logic.  
For example, we assume that the reader knows the difference between \emph{bound} and \emph{free} variables, \emph{quantifier-free formulas}, and \emph{prenex normal form}. We refer to~\cite{BodMathLogic} or~\cite{BodModelTheory}.
\end{itemize}

\medskip 
\noindent
{\bf AI disclaimer.} 
ChatGPT 5.6 Sol and 6 Astra, and Claude Opus 5 were used for proof-reading the final text, solving the exercises and grading their difficulty, finding mistakes and fixing them.

\medskip 
\noindent
{\bf Acknowledgements.} 
Many thanks to Catarina Carvalho, Philipp Grzywaczyk, Lukas Juhrich, Mathison Knight, Sebastian Meyer, Andrew Moorhead, Michael Pinsker, Lukas Schneider, Moritz Sch\"obi, \v{Z}aneta Semani\v{s}inov\'a, Mark Siggers, the participants of the course in the Corona spring semester 2020 and the course in spring 2023 for their suggestions and bug reports. Thanks also to Zarathustra Brady for his lecture notes~\cite{BradyNotes} and to Zarathustra Brady
and Dima Zhuk for answering my questions. 

%\vspace{1cm} 
%\medskip 
\newpage
\noindent 
{\bf Exercises.} 
The text contains \total{exercise} exercises; 
some of them are graded using the old  
Mandala colour scale (no longer in use). \\
%\begin{wrapfigure}{r}{0.6\textwidth}
%    \centering
 \begin{center}
 \includegraphics[scale=.2]{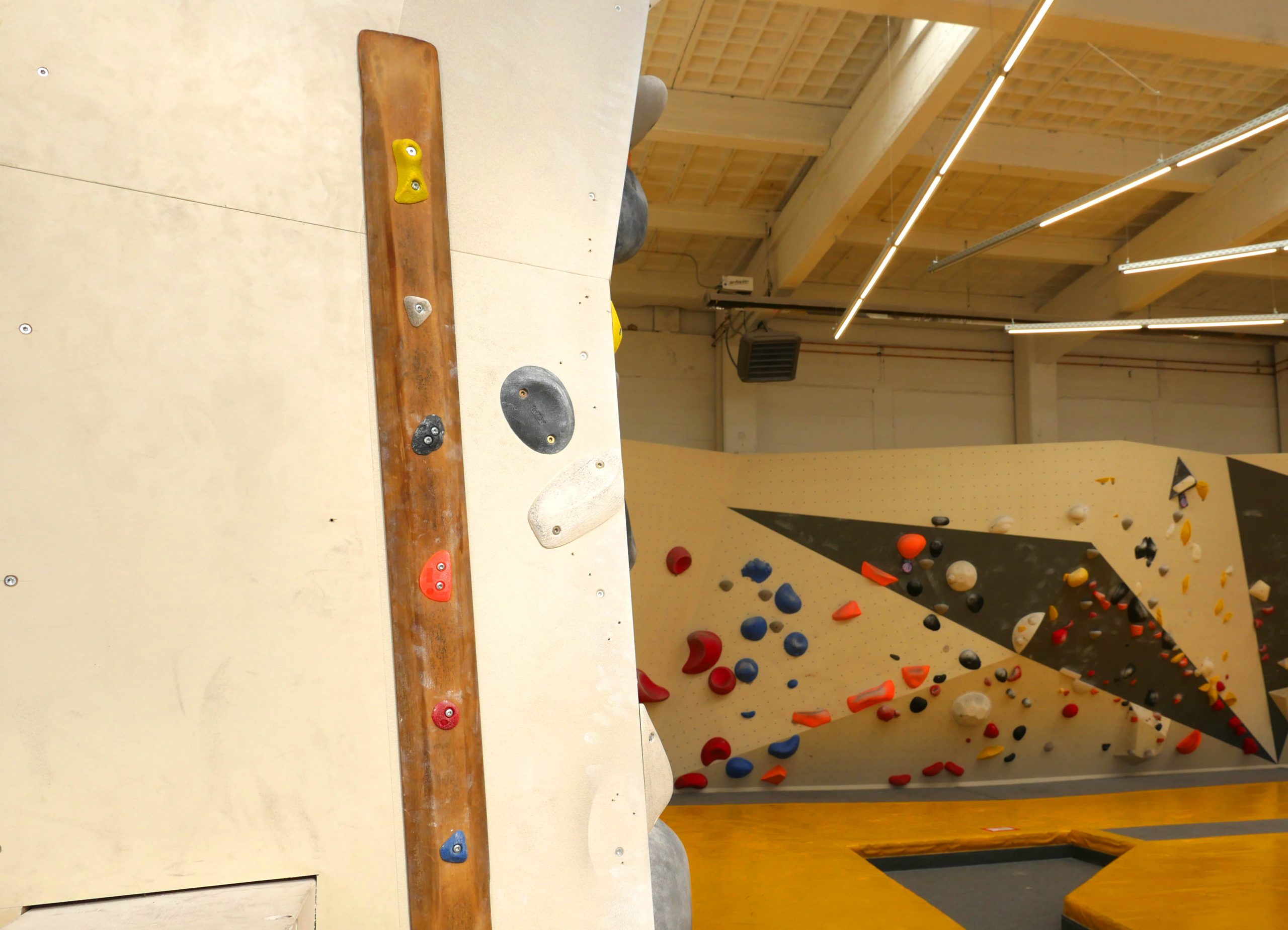}
    \end{center}
    %\vspace{-2cm}
%\end{wrapfigure} 

%\begin{flushright}
%\begin{center}
%\vspace{-1.7cm}
%\includegraphics[scale=.13]{Mandala-scale.jpg}
%\end{center}
%\end{flushright}

\newpage

\medskip 
\noindent
{\bf Possible Routes for Courses.} This course has been taught several times in one semester, but never covering all sections. Possible selections of material can be deduced from the section diagram below;
an arc from section A to section B indicates that section A should typically be treated before section B. 
After treating some basics about digraphs and homomorphisms, I always started with the arc-consistency procedure, because it is a fantastic algorithm (both from a theoretical and practical perspective), and because it is concrete and elementary, but with non-trivial and interesting results quite early in the course. These results will then be important examples for more general and abstract results later. For example, the material about minions strictly speaking does not rely on the arc-consistency procedure, but I prefer to teach it having the concrete example of the minion arising from the
 arc-consistency procedure. 

\begin{center}
%\vspace{-1.7cm}
\includegraphics[scale=.5]{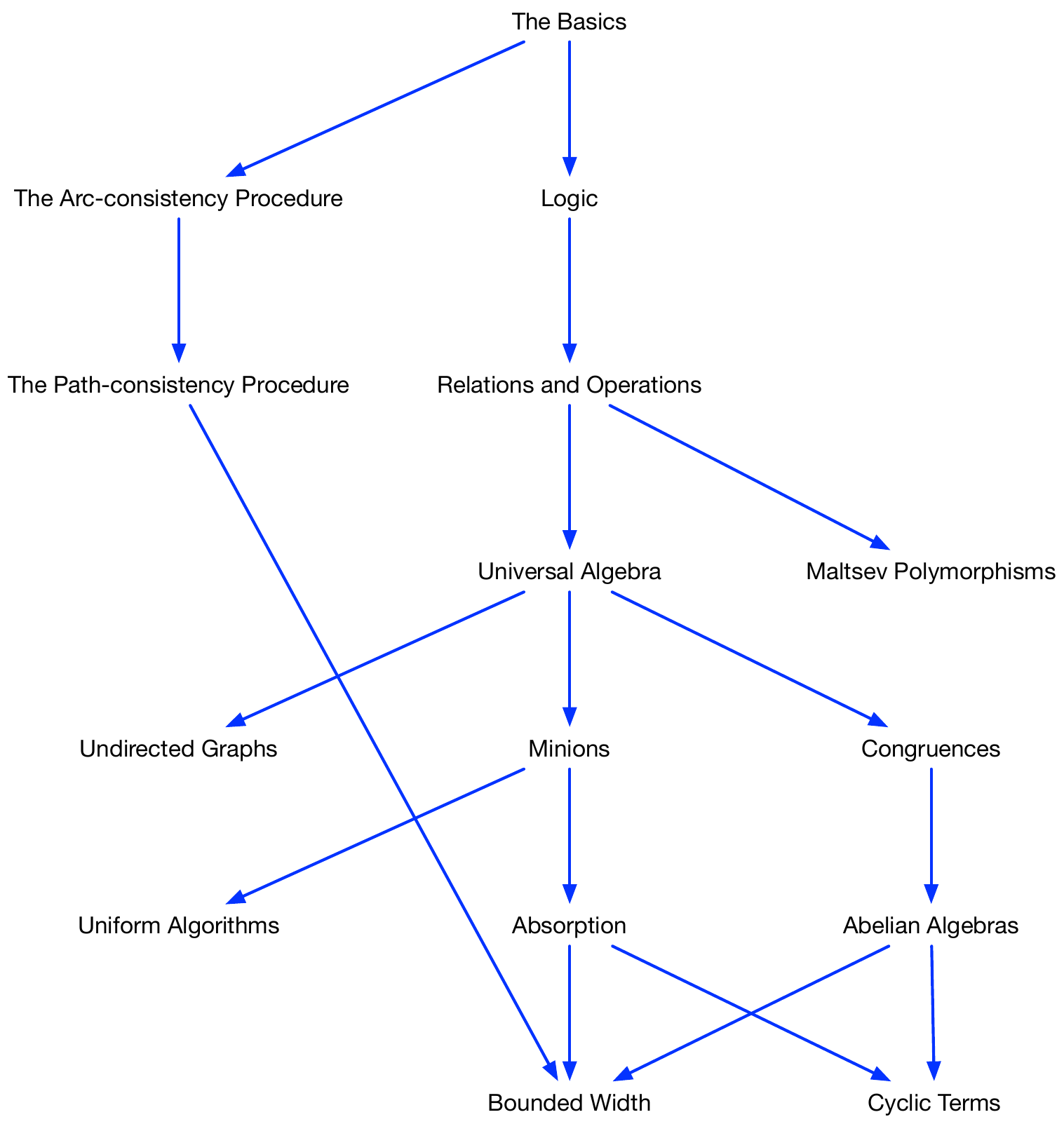}
\end{center}

\newpage

\section{Introduction}
Around the turn of the millennium, several previously separate research communities realised that many of their central questions are essentially the same. 
In the late 80s and 90s, the graph homomorphism community intensively studied the computational complexity of the \emph{$H$-colouring problem}~\cite{HNBook}. 
Independently, the theoretical artificial intelligence community studied \emph{constraint satisfaction problems} and their computational complexity (including the important Boolean CSPs which were  classified by Schaefer~\cite{Schaefer} in 1978). 
In the late 90s, researchers realised that universal algebra provides the right tools for this task~\cite{BulatovJeavons,JeavonsClosure}. 

The paper by Feder and Vardi, whose conference version appeared in 1993, is probably the most influential article in the area and has inspired generations of researchers~\cite{FederVardiSTOC,FederVardi}. It formulates for the first time the \emph{dichotomy conjecture}, which was solved in 2017 by  Bulatov~\cite{BulatovFVConjecture} and by Zhuk~\cite{ZhukFVConjecture}. It also links the topic with finite model theory. For example, it identifies Datalog from database theory as an important framework that captures many of the central consistency algorithms that have been used to solve CSPs. Feder and Vardi prove that a complexity dichotomy for finite-domain CSPs implies a  complexity classification for the fragment of NP called MMSNP. They also prove that every finite-domain CSP is computationally equivalent to the $H$-colouring problem for some finite digraph, thus further substantiating the connection between graph homomorphisms and constraint satisfaction. 

This course starts very concretely, in the setting of digraphs rather than the more general setting of relational structures, because digraphs are notationally simpler than general structures. Digraphs are ideal for black-board teaching, because they are easy to draw and it is easy to come up with interesting examples. 
After having introduced the basics of the theory of graph homomorphisms in Section~\ref{sect:basics}, we present an algorithm of outstanding importance: the so-called \emph{arc-consistency procedure}. This algorithm is theoretically very well understood. Moreover, it is 
practically important, because of its low time and space requirements, because it is easy to implement, and because it is widely applicable. 
Numerous exercises that we formulate at the end of the subsections can be easily solved if the reader properly understands the underlying principles of arc-consistency. 

The arc-consistency procedure can be generalised to the \emph{$k$-consistency procedure}, which is more powerful, and still in P, but more demanding in time and space requirements. 
Theoretical results in later sections of this course show that if $k$-consistency solves the $H$-colouring problem, then so does the $3$-consistency algorithm. This algorithm is sometimes referred to as the \emph{(strong) path-consistency procedure} and it is the topic of Section~\ref{sect:PC}. A full description of when this procedure solves the $H$-colouring problem has to wait until Section~\ref{sect:bounded-width} of the course, but we do see some sufficient conditions that can be used to answer the question for many concrete digraphs $H$. 

At some point, the restriction to digraphs becomes unnatural; we step to general relational structures in Section~\ref{sect:logic}. This will be the appropriate setting for presenting the main tools for complexity classification, which are, in increasing strength, \emph{primitive positive definitions}, \emph{primitive positive interpretations}, and \emph{primitive positive constructions}. Primitive positive constructions are part of the statement of a solution to the Feder-Vardi dichotomy conjecture: if a finite structure admits a primitive positive construction of $K_3$ (the complete graph with three vertices), then its CSP is NP-hard (a statement that we prove in Section~\ref{sect:logic}); otherwise, its CSP is in P (this is the content of the result of  Bulatov~\cite{BulatovFVConjecture} and of Zhuk~\cite{ZhukFVConjecture}). 

All three of these concepts (pp definitions, pp interpretations, and pp constructions) can also be characterised universal-algebraically, in terms of polymorphisms. For 
primitive positive definability, this can be found in Section~\ref{sect:algebra}, where we also apply it to prove Schaefer's complexity dichotomy result for CSPs of two-element structures. 
The universal-algebraic theory that captures primitive positive interpretations is presented in Section~\ref{sect:ua}, and the universal-algebraic theory for primitive positive constructions in Section~\ref{sect:minions}. 

In Section~\ref{sect:HN} we show the Hell-Ne\v{s}et\v{r}il dichotomy for the $H$-colouring problem for finite undirected graphs $H$. From this result we obtain a universal-algebraic formulation of the 
complexity dichotomy for all finite structures in terms of a Siggers polymorphism (of arity six). 
A much more informative formulation of the complexity dichotomy uses \emph{cyclic polymorphisms} in Section~\ref{sect:cyclic}, which  is substantially more difficult to prove. In particular, we need  the fundamental theorem of abelian algebras from Section~\ref{sect:abelian}, and absorption theory, developed by Barto and Kozik~\cite{cyclic} and presented in Section~\ref{sect:absorption}. 

Concerning algorithms for CSPs, we treat the Bulatov-Dalmau algorithm for structures with a Maltsev polymorphism (Section~\ref{sect:Maltsev}), and 
in Section~\ref{sect:bounded-width} the bounded width case (i.e., the CSPs that can be solved by Datalog). 
%Interestingly, the solution to the bounded width, despite the fact that it is proven in the general setting of relational structures, has a graph-theoretic flavour, 
A complete algorithm that solves all tractable finite-domain CSPs is outside of the scope of this course. 

Many other topics are \emph{not} treated by these course notes, most notably 
\begin{itemize}
\item \emph{infinite-domain CSPs}; we refer to~\cite{Book}.
\item \emph{promise CSPs}; we refer to~\cite{BartoBKO21}. 
\item \emph{valued CSPs}; we refer to~\cite{KolmogorovKR17,KozikOchremiak15,KolmogorovThapperZivny15}.
\end{itemize} 

% Classics: \cite{JBK,LLT,LaroseZadori,LaroseTesson}, 

% nice surveys and text-books:
% Brady, polys and how to use them, 
% BEGIN INLINED SOURCE: basics.tex
 % !TEX root = GH-UA.tex

\section{The Basics}
\label{sect:basics}
We mostly work with \emph{finite} graphs; results that also hold for graphs with infinitely many vertices are mentioned only when this requires no extra effort. 

\subsection{Graphs and Digraphs}
The concepts in this section are probably known to most students, and can safely be skipped; the section fixes standard terminology and conventions from graph theory and can be consulted later if needed. 
Almost all definitions in this section
have generalisations to \emph{relational structures}, which will be introduced in Section~\ref{sect:logic};
however, we focus exclusively on graphs in this section 
since they allow us to reach the key ideas of
the underlying theory with a minimum of notation.

A \emph{directed graph} (also \emph{digraph}) $G$ is a pair $(V,E)$ of 
a set $V = V(G)$ of \emph{vertices}
and a binary relation $E=E(G)$ on $V$. Note that
in general we allow that $V$ is an infinite set. For some definitions and results, we require that $V$ is finite, in which case we say
that $G$ is a \emph{finite} digraph.
However, since this course deals mainly with finite digraphs, we will omit this most of the time. 
The elements $(u,v)$ of $E$ are called the \emph{arcs} 
(or \emph{directed edges}) of $G$.
Note that we allow \emph{loops}, i.e., arcs of the form $(u,u)$; a digraph without loops is called \emph{loopless}. 
If $(u,v) \in E(G)$ is an arc, then we say that $v$ is an \emph{out-neighbour} of $u$ 
and that $u$ is an \emph{in-neighbour} of $v$; a \emph{neighbour} is an in-neighbour or an out-neighbour. The \emph{(in-, out-) degree} of a vertex $u$ is the number of (in-, out-) neighbours of $u$. 
We say that the vertices $u$ and $v$ are \emph{incident} to the edge $(u,v)$. 

An \emph{(undirected) 
graph} is a pair $(V,E)$ of a set $V = V(G)$ of \emph{vertices} and 
a set $E=E(G)$ of \emph{edges}, 
each of which is an unordered pair of (not necessarily distinct)
elements of $V$. 
In other words, we explicitly allow \emph{loops},
which are edges that link a vertex with itself. 
Undirected graphs can be viewed as \emph{symmetric} digraphs:
a digraph $G=(V,E)$ is called \emph{symmetric} if $(u,v) \in E$ if and only 
if $(v,u) \in E$. 
For a digraph $G$, we say that $G'$ is the \emph{undirected graph of
$G$} if $G'$ is the undirected graph with $V(G') = V(G)$ and where $\{u,v\} \in E(G')$
if $(u,v) \in E(G)$ or $(v,u) \in E(G)$. 
For an undirected graph $G$, we say that $G'$ is an 
\emph{orientation of $G$} if $G'$ is a directed graph such that
$V(G')=V(G)$ and $E(G')$
contains for each edge $\{u,v\} \in E(G)$ either the arc $(u,v)$
or the arc $(v,u)$, and no other arcs.

For some notions for digraphs $G$ one can just use the corresponding notions
for undirected graphs applied to the undirected graph of $G$; conversely,
most notions for directed graphs, specialised to symmetric graphs, translate
to notions for the respective undirected graphs. 

\subsubsection{Examples of graphs, and corresponding notation}
\begin{itemize}
%\item The digraph without vertices. This graph is unique.
%\item We allow that a digraph may have no vertices at all.
\item The \emph{complete graph} on $n$ vertices $[n] := \{1,\dots,n\}$, denoted by $K_n$. This is an undirected graph on $n$ vertices in which every vertex is joined with any other distinct
vertex (so $K_n$ contains no loops). 
\item The \emph{cyclic graph} on $n$ vertices, denoted by $C_n$; this is the
undirected graph with the vertex set $\{0,\dots,n-1\}$ and
edge set $$\big \{ \{0,1\}, \dots, \{n-2,n-1\}, \{n-1,0\} \big \} = \big \{ \{i,i+1\} : i \in {\mathbb Z}_n \big \} \, .$$ 
Note that $C_1$ has a loop and $C_2 = K_2$. 
\item The \emph{directed cycle} on $n$ vertices, denoted by $\vec C_n$; this is the digraph with the vertex set
$\{0,\dots,n-1\}$ and the arc set 
$$\big \{ (0,1), \dots, (n-2,n-1), (n-1,0) \big \} = \big \{ (i,i+1) : i \in {\mathbb Z}_n \big \}.$$
% = \big \{ (u,v) \; | \; u-v = 1 \mod n \big \}$.
\item The \emph{path} with $n+1$ vertices and $n$ edges, denoted by $P_n$;
this is an undirected graph with the vertex set $\{0,\dots,n\}$ 
and edge set $\big \{\{0,1\},\dots,\{n-1,n\} \big \}$. 
\item The \emph{directed path} with $n+1$ vertices and $n$ edges,
denoted by $\vec P_n$; 
this is a digraph with the vertex set $\{0,\dots,n\}$ 
and edge set $\{(0,1),\dots,(n-1,n)\}$. 
\item A \emph{tournament} is a directed loopless 
graph $G$ with the property that for all distinct vertices $x,y$ either $(x,y)$ or $(y,x)$ is an edge of $G$, but not both. 
\item The \emph{transitive tournament} on $n \geq 2$ vertices, denoted by $T_n$; this is a directed graph with the vertex set $\{1,\dots,n\}$
where $(i,j)$ is an arc if and only if $i < j$.
\end{itemize}

Let $G$ and $H$ be graphs (we define the following notions both for
directed and for undirected graphs). Then $G \uplus H$ denotes
the \emph{disjoint union of $G$ and $H$}, which is the
graph with vertex set $V(G) \cup V(H)$ (we assume that the two vertex sets are disjoint; if they are not, we take a copy of $H$ on a disjoint set of vertices and form the disjoint union of $G$ with the copy of $H$) and edge set $E(G) \cup E(H)$. A graph $G'$ is a \emph{subgraph} of $G$ if $V(G') \subseteq V(G)$ and $E(G') \subseteq E(G)$. 
A graph $G'$ is an \emph{induced subgraph} of $G$ if $V' = V(G') \subseteq V(G)$ and $(u,v) \in E(G')$ if and only if $(u,v) \in E(G)$ for all $u,v \in V'$. We also say that $G'$ is \emph{induced by $V'$ in $G$},
and write $G[V']$ for $G'$. We write $G-u$ for $G[V(G) \setminus \{u\}]$,
i.e., for the subgraph of $G$ where the vertex $u$ and all incident
arcs are removed.

We call $|V(G)|+|E(G)|$ the \emph{size} of a graph $G$.
%; note that the size of a graph $G$ reflects the length of the 
%representation of $G$ as a bit-string (under a reasonable choice of graph representations). 
This quantity will be important when we analyse the efficiency of algorithms on graphs.

%Two graphs $G$ and $H$ are isomorphic
% Define G-e for an edge e?

\subsubsection{Paths and Cycles}
We start with definitions 
for directed paths; the corresponding
terminology is then also used for undirected graphs as explained in the beginning of this section. 

%A sequence $(e_1,\dots,e_k)$ of edges of $G$
%such that for all $i \in \{1,\dots,k-1\}$ 
%if $e_i = (a,b)$ and $e_{i+1} = (c,d)$ then $a=c$. 
%If $e_1 = (a,b)$ and $e_k = (c,d)$ then 
%$a$ is called the \emph{start point} and $d$ is called the end point. ARGH this is directed path. 
%If every vertex of $G$ appears in at most two edges on the path, and the start point and end point of the path only appear in one edge, 
%then the path is called \emph{simple}. 

A \emph{path} $P$ (from $u_1$ to $u_k$ in $G$)
is a sequence $(u_1,\dots,u_k)$ of vertices
of $G$ and a sequence $(e_1,\dots,e_{k-1})$ of edges of $G$ 
such that $e_i = (u_i,u_{i+1})  \in E(G)$ or $e_i = (u_{i+1},u_{i}) \in E(G)$, for every $1 \leq i < k$.
The vertex $u_1$ is called the \emph{start vertex}
and the vertex $u_k$ is called the \emph{terminal vertex} of $P$, and we say that $P$ is a path \emph{from $u_1$ to $u_k$}. 
Edges $(u_i,u_{i+1})$ with $u_i \neq u_{i+1}$ are called \emph{forward edges} and edges $(u_{i+1},u_{i})$ with $u_i \neq u_{i+1}$ are called \emph{backward edges}. A loop edge is
neither forward nor backward.
The path is called \emph{directed} if
\[
 e_i=(u_i,u_{i+1})
\]
for every $1\leq i<k$. Thus, loop edges are permitted in a directed
path.
%If all edges are forward edges then the path is called \emph{directed}.
If $u_1,\dots,u_k$ are pairwise distinct then the path is called \emph{simple}. 
We write $|P| := k-1$ for the \emph{length} of $P$, i.e., we count the number of edges of $P$. 
The \emph{net length} 
of a path $P$ 
without edges of the form $(u,u)$ 
is the 
difference between the number of forward and the number of backward edges. 
Hence, a loop-free path is directed 
if and only if its length equals its net length.

A sequence $(u_0,\dots,u_{k-1})$ of vertices 
and a sequence of edges $(e_0,\dots,e_{k-1})$ 
 is called a 
\emph{cycle} (of $G$) if $(u_0,\dots,u_{k-1},u_0)$
and $(e_0,\dots,e_{k-1})$ form a path. If 
all the vertices of the cycle are pairwise distinct 
then the cycle is called \emph{simple}. 
We write $|C| := k$ for the \emph{length}
of the cycle $C = (u_0,\dots,u_{k-1})$. 
If $C$ has no loop edges, then its \emph{net length} is
the net length of the corresponding path  
$(u_0,\dots,u_{k-1},u_0)$. 
 The cycle $C$  
is called \emph{directed} 
if the corresponding path is a directed path. 

A digraph $G$ is called \emph{(weakly) connected} if there is a path in $G$ from any vertex to any other vertex in $G$. 
Equivalently, $G$ is connected if and only if it cannot be written as $H_1 \uplus H_2$ 
for digraphs $H_1,H_2$ with at least one vertex each. 
A \emph{connected component} of $G$ is a maximal (with respect to inclusion of the vertex sets) connected induced subgraph of $G$. 
A digraph $G$ 
is called \emph{strongly connected} 
if for all vertices $x,y \in V(G)$ there is a directed path from $x$ to $y$ in $G$. 
A \emph{strongly connected component} of $G$ is an inclusion-wise maximal strongly connected induced subgraph of $G$. 
%The \emph{strongly connected compon
Two vertices $u,v \in V(G)$ are \emph{at distance $k$} in $G$ if the shortest 
path from $u$ to $v$ in $G$ has length $k$.

Some particular notions for undirected graphs $G$. 
A \emph{(simple) cycle} of $G$ is a sequence $(v_1,\dots,v_k)$ of $k \geq 3$ pairwise distinct vertices
of $G$ such that $\{v_1,v_k\} \in E(G)$ and $\{v_i,v_{i+1}\} \in E(G)$ for all $1 \leq i \leq k-1$.
 An undirected graph is called \emph{acyclic} if it does not contain 
 a cycle.
 A sequence $u_1,\dots,u_k \in V(G)$ is called a 
 \emph{(simple) path} from $u_1$ to $u_k$ in $G$ 
 if $\{u_i,u_{i+1}\} \in E(G)$
 for all $1 \leq i < k$ and if all vertices $u_1,\dots,u_k$ are pairwise distinct. We allow
 the case that $k=1$, in which case the 
 path consists of a single vertex and no edges. 
%(if we do not assume that all vertices are pairwise distinct, we
 %say that $u_1,\dots,u_k$ is a \emph{walk} from $u_1$ to $u_k$,
 %following Bondy and Murty 1976,
Two vertices $u,v \in V(G)$ are \emph{at distance $k$} in $G$ if
the shortest path in $G$ from $u$ to $v$ has length $k$.
We say that an undirected graph $G$ is \emph{connected} 
if for all vertices $u, v \in V(G)$ there is a path from $u$ to $v$.
The \emph{connected components} of $G$ are the maximal
connected induced subgraphs of $G$. 
A \emph{forest} is an undirected acyclic graph, 
a \emph{tree} is a connected forest.

%\begin{itemize}
%\item 
A \emph{source} in a digraph is a vertex
with no incoming edges, and 
%\item 
a \emph{sink}
is a vertex with no outgoing edges. 
%\end{itemize}

%Most of the definitions for graphs in this text 
%are analogous for directed and for undirected graphs. %We therefore do not state explicitly 
%We therefore sometimes do not explicitely mention whether
%we work with directed or undirected graphs, but just
%state certain concepts for \emph{graphs}, for instance in Subsection~\ref{ssect:cores} or~\ref{ssect:polymorphisms}.

%The following is obvious from the definitions.
%\begin{proposition}
%A digraph $G$ is connected if and only if for all digraphs
%$H$ and $H'$ such that $G=H+H'$ either $H$ or
%$H'$ is the digraph without vertices.
%\end{proposition}
%\paragraph{Exercises.}
%\begin{enumerate}
%\item Show that a digraph $G$ is weakly connected if and only if for all digraphs $H$ and $H'$ such that $G=H \mathbin{\stackrel{.}{\cup}} H'$ the digraph $H$ or
%the digraph $H'$ is the digraph without vertices.
%\setcounter{mycounter}{\value{enumi}}
%\end{enumerate}

\subsection{Graph Homomorphisms}
Let $G$ and $H$ be directed graphs. 
A \emph{homomorphism}\index{Homomorphism} from $G$ to $H$
is a mapping $h \colon V(G) \to V(H)$
which \emph{preserves the edges}, i.e., 
 $(h(u),h(v)) \in E(H)$ 
whenever $(u,v) \in E(G)$. If such a homomorphism exists between $G$ and $H$ we say that $G$ \emph{homomorphically maps} to $H$, and write $G \to H$. Otherwise, we 
write
$G \not\rightarrow H$. 
Two directed graphs $G$ and $H$ are \begin{itemize}
\item \emph{homomorphically equivalent} if $G \to H$ and $H \to G$; in this case, we also write $G \leftrightarrow H$. 
\item \emph{homomorphically comparable} if 
$G \to H$ or $H \to G$; otherwise, we say that $H$ and $G$ are \emph{homomorphically incomparable}.
\end{itemize}

A \emph{strong homomorphism} from a digraph $G$ to a digraph $H$ is a function from $V(G)$ to $V(H)$ such that $(f(u),f(v)) \in E(H)$
if and only if $(u,v) \in E(G)$ for all $u,v \in V(G)$.
An \emph{isomorphism} between two directed graphs $G$ and $H$
is a bijective strong homomorphism from $G$ to $H$. If an isomorphism between $G$ and $H$ exists, we call $G$ and $H$ \emph{isomorphic}, and write $G \simeq H$. 
%, i.e., 
%$f(u) \neq f(v)$ for any two distinct vertices $u,v \in V(G)$. 
Note that a homomorphism
$h \colon G \to H$ is an isomorphism if and only if it is bijective, and $h^{-1}$ is a homomorphism from $H$ to $G$. 
An \emph{automorphism} of a digraph $H$ is an isomorphism from $H$ to $H$.

A homomorphism from $G$ to $H$ is sometimes also called
an \emph{$H$-colouring} of $G$. This terminology originates from the observation
that $H$-colourings generalise classical colourings in the sense
that a graph is $n$-colourable
if and only if it has a $K_n$-colouring. 
Graph $n$-colorability is not the only natural graph property that
can be described in terms of homomorphisms:
\begin{itemize}
\item a digraph is called \emph{balanced}
(in some articles: \emph{layered}) if it homomorphically 
maps to a directed path $\vec P_n$, for some $n$;
\item  
a digraph is called \emph{acyclic} if it homomorphically maps
to a transitive tournament $T_n$, for some $n$.
\end{itemize}

The equivalence classes of finite digraphs
with respect to homomorphic equivalence
  will be denoted by $\cal D$.
  Let $\leq$ be a
binary relation defined on $\cal D$ as follows: we set
$C_1 \leq C_2$ if there exists a digraph $H_1 \in C_1$ and a digraph $H_2 \in C_2$
such that $H_1 \to H_2$ (note that this definition does not depend on the choice of the representatives $H_1$ of $C_1$ and $H_2$ of $C_2$). 
If $f$ is a homomorphism from $H_1$ to $H_2$, and $g$ is a homomorphism
from $H_2$ to $H_3$, then the composition $g \circ f$ of these
functions is a homomorphism from $H_1$ to $H_3$, and therefore
the relation $\leq$ is transitive. Since every graph $H$ homomorphically maps to $H$,
the order $\leq$ is also reflexive.
Finally, $\leq$ is antisymmetric since 
its elements are equivalence classes of directed graphs
with respect to homomorphic equivalence. 
Define $C_1 < C_2$ if $C_1 \leq C_2$ and
$C_1 \neq C_2$. We call $(\cal D,\leq)$ the \emph{homomorphism order} of finite digraphs.

The homomorphism order on digraphs 
turns out to be a \emph{lattice} where every two elements have a supremum (also called \emph{join}) and an infimum (also called \emph{meet}; see Example~\ref{expl:lattice}). 
In the proof of this result, we need the notion of \emph{direct products} 
of graphs. 
This notion of graph product\footnote{Warning: there are several other notions of graph products that have been studied; see e.g.~\cite{HNBook}.} 
%which comes closest
%to the notion of direct products as it is used in universal algebra.
%Indeed it 
can be seen as a special case of the notion
of direct product as it is used in model theory~\cite{Hodges}.
The class of all graphs with respect to homomorphisms forms an interesting category in the sense of category theory~\cite{HNBook}
where the product introduced above is the product in the sense of category theory, 
% TODO: explain more!!!
which is why this product is sometimes also called the
\emph{categorical} graph product.

%Direct products are also called \emph{cross products};
%the reader who wants to 
%one motivation for this definition might be 

\begin{definition}[direct product]\label{def:prod}
Let $H_1$ and $H_2$ be two digraphs. Then the \emph{(direct-, cross-, categorical-) product} $H_1 \times H_2$ 
of $H_1$ and $H_2$ is the graph with vertex set
$V(H_1) \times V(H_2)$; the pair $((u_1,u_2),(v_1,v_2))$ is in 
$E(H_1 \times H_2)$
if $(u_1,v_1) \in E(H_1)$ and $(u_2,v_2) \in E(H_2)$.
\end{definition}
Note that the product is symmetric and associative in the sense that $H_1 \times H_2$ is isomorphic to $H_2 \times H_1$ and $H_1 \times (H_2 \times H_3)$ is isomorphic to $(H_1 \times H_2) \times H_3$, and we
therefore do not specify the order of multiplication 
when multiplying more than two graphs. 
The \emph{$n$-th power} $H^n$ of a graph $H$ is inductively defined as follows.
$H^1$ is by definition $H$. If $H^i$ is already defined, then
$H^{i+1}$ is $H^i \times H$. 
%Again, all definitions in this section are both for directed and for undirected graphs.
%, and for vertex sets of arbitrary cardinality.

\begin{proposition}
The homomorphism order $({\mathcal D},\leq)$ is a \emph{lattice}; i.e., for all $A_1,A_2 \in {\mathcal D}$
\begin{itemize}
\item there exists an element $A_1 \wedge A_2 \in {\mathcal D}$,
the \emph{meet} of $A_1$ and $A_2$, such that $(A_1 \wedge A_2) \leq A_1$ and $(A_1 \wedge A_2) \leq A_2$, and such that for every $U \in {\mathcal D}$ with $U \leq A_1$ and $U \leq A_2$ 
we have $U \leq A_1 \wedge A_2$; 
\item there exists an element $A_1 \vee A_2 \in {\mathcal D}$,
the \emph{join} of $A_1$ and $A_2$, such that $A_1 \leq (A_1 \vee A_2)$ and $A_2 \leq (A_1 \vee A_2)$, and such that for every $U \in {\mathcal D}$ with $A_1 \leq U$ and $A_2 \leq U$
we have $A_1 \vee A_2 \leq U$. 
\end{itemize}
\end{proposition}
\begin{proof}
% TODO: this should be expanded.
Let $H_1 \in A_1$ and $H_2 \in A_2$. For the meet, the
equivalence class of $H_1 \times H_2$ has the desired properties. For the join, the equivalence class of the disjoint union $H_1 \uplus H_2$ has the desired properties.\footnote{For this reason, $H_1 \uplus H_2$ is sometimes called the \emph{co-product} of $H_1$ and $H_2$.} 
\end{proof}

With the seemingly simple definitions of graph homomorphisms and direct products we can already formulate very difficult combinatorial questions.

\begin{conjecture}[Hedetniemi]
Let $G$ and $H$ be finite graphs, and suppose that $G \times H \to K_n$. Then $G \to K_n$ or $H \to K_n$.
\end{conjecture}

For a graph $G$ without loops, 
the smallest $n \in {\mathbb N}$ such that
$G \to K_n$ is also called the \emph{chromatic number} of $G$, and denoted by $\chi(G)$. 
Clearly, $\chi(G \times H) \leq \min(\chi(G),\chi(H))$. Hedetniemi's conjecture can then be rephrased as 
$$\chi(G \times H) = \min(\chi(G),\chi(H)) \, .$$
This conjecture is easy for $n=1$ and $n=2$ (Exercise~\ref{exe:hedetniemi}), and has
been solved for $n=3$ by El Zahar and  Sauer~\cite{ElZaharSauer}. 
The conjecture has been refuted in 2019 by Yaroslav Shitov~\cite{Hedetniemi-false}.

Clearly, $({\mathcal D},\leq)$ has
infinite \emph{ascending chains}, that is,
sequences $E_1,E_2,\dots$ such that 
$E_i < E_{i+1}$ for all $i \in \mathbb N$. 
Take for instance the equivalence class of 
$\vec{P}_i$ for $E_i$. 
More interestingly, $({\mathcal D},\leq)$
also has infinite descending chains (Proposition~\ref{prop:desc}). 
To show this, we use so-called \emph{zig-zags},
which are frequently used in the theory of graph homomorphisms.
We may write an orientation of a path as a word over $\{0,1\}$, where 
$0$ represents a forward arc and $1$ represents a backward arc.

There are two useful invariants of homomorphisms between orientations of 
paths. 
The first one is \emph{height}. 
Let $P$ be an orientation of a path with consecutive vertices
$v_0,\dots,v_m$. Define its \emph{height function}
$\lambda_P$ by $\lambda_P(v_0)=0$ and by adding $1$ for every forward
arc and $-1$ for every backward arc. 
%If $f\colon P\to Q$ is a
%homomorphism of oriented paths, then
%\[
% \lambda_Q(f(v))-\lambda_P(v)
%\]
%is independent of $v$. This follows immediately by considering the
%two endpoints of each arc. Thus,\
Clearly, homomorphisms preserve height
differences.

A second variant is \emph{distance}: since a homomorphism maps the underlying path to a
walk of the same length, homomorphisms cannot increase the distance between vertices. 

\begin{proposition}\label{prop:desc}
The lattice $({\mathcal D},\leq)$ contains 
infinite descending chains $E_1 > E_2 > \cdots$. 
\end{proposition}
\begin{proof}
For $k\geq 1$, let $Z_k$ be the oriented path represented by
\[
 11(01)^{k-1}1.
\]
It has length $2k+1$, height span $3$, and unique vertices
of maximum and minimum height, namely its two endpoints. There is a
homomorphism $Z_{k+1}\to Z_k$: numbering the vertices in their path
order, map the vertices of $Z_{k+1}$ to
\[
 0,1,2,1,2,3,\dots,2k+1
\]
in $Z_k$. 

Conversely, a homomorphism $Z_k\to Z_{k+1}$ would have to map the two
endpoints to the two endpoints, since both paths have height span $3$
and their extreme heights occur only at the endpoints. But the images
of the endpoints are separated by $2k+3$ edges in $Z_{k+1}$, whereas
the image of $Z_k$ gives a walk of length only $2k+1$ between them.
This is impossible. Hence, the classes of $Z_1,Z_2,\dots$ form a
strictly descending chain.
\end{proof}

\begin{proposition}
The lattice $(\cal D,\leq)$ contains infinite \emph{antichains}, that is, sets of pairwise incomparable elements of $\cal D$ with respect to $\leq$. 
\end{proposition}
\begin{proof}
For $k,l\geq 1$, let the \emph{$k,l$ multi zig-zag} $Z_{k,l}$ be the
oriented path represented by
\[
 1\bigl(1(01)^k\bigr)^l1.
\]
The multi zig-zag $Z_{k,k}$ has height span $k+2$ and length $2k^2+k+2$. Suppose that $Z_{k,k} \to Z_{l,l}$. Preservation of height differences implies
that $k\leq l$.

If $k<l$, the endpoints of $Z_{k,k}$ must map to two vertices of $Z_{l,l}$
whose heights differ by $k+2$. 
Let $P$ be the subpath between these vertices.
The higher vertex occurs first in the word, since height can increase
by at most $1$ along any subpath read in the word order.
Decompose $P$ into $m$ complete blocks $1(01)^l$ and at most two
end pieces, each a partial block or an outer single edge.
Each complete block decreases the height by $1$, and each end piece
decreases it by at most $1$.
Hence $k+2\leq m+2$, so $m\geq k$.
Consequently, $P$ contains at least $kl$ forward arcs.
Writing $u$ for the number of forward arcs, its height decrease
implies that it has $u+k+2$ backward arcs, and therefore
\[
 |E(P)|=2u+k+2\geq 2kl+k+2=k(2l+1)+2.
\]
On the other hand, the image of
$Z_{k,k}$ gives a walk between these vertices of length
$2k^2+k+2$. This is impossible, since
\[
 k(2l+1)+2>2k^2+k+2
\]
when $l>k$. Thus, $Z_{k,k} \to Z_{l,l}$ implies $k=l$, and the classes of the
graphs $Z_{k,k}$ form an infinite antichain.
\end{proof}

\begin{exercises}
\exercise[Blau]\label{exe:source-1} How many connected components do we have in $(P_3)^3$?
% Solved by deepseek in 29.1.25: 4. but have to point out that we want the cross product.
\exercise[Blau]\label{exe:source-2} How many weakly and strongly connected components do we have in $(\vec C_3)^3$?
%3:  $(\vec C_3)^2 = \vec C_3 + \vec C_3 + \vec C_3$.
% Solved by deepseek in 29.1.25: but have to point out that we want the cross product.
% But made a mistake with the weakly connected components.
\exercise[Blau]\label{exe:source-3} Let $G$ and $H$ be digraphs. Prove that $G \times H$ has a directed cycle
if and only if both $G$ and $H$ have a directed cycle.
%\usecounter{mycounter}
%\item Show that homomorphic equivalence is an equivalence relation.
%\item If $f$ is a homomorphism from $G$ to $H$, and $a,b \in V(G)$,
%then the distance (i.e., the length of the shortest path) from $a$ to $b$
%in $G$ is not larger than the distance between $f(a)$ and $f(b)$ in $H$.
%\item Show that the number of homomorphisms of a graph $G$ to the graph $H$ shown in Figure~\ref{fig:fig} equals the number of independent sets of $G$.
%\item Which of the three graphs below are homomorphically equivalent? (insert picture)
%\input bild.tex
\exercise[Blau]\label{exe:source-4} Prove the Hedetniemi conjecture for $n=1$ and $n=2$.
\label{exe:hedetniemi}
% Solution sketch: For $n=1$ the assertion says that a product has an edge only if both factors have
% an edge. For $n=2$, if neither factor is bipartite, choose an odd cycle in each factor; their direct
% product contains an odd cycle (of odd length the least common multiple of the two lengths), so the
% product is not bipartite.
\exercise[Blau]\label{exe:source-9} Show that a digraph $G$ homomorphically maps to
$\vec P_1 = T_2$ if and only if $\vec P_2$ does not homomorphically map to $G$. \label{exe:t2}
% Solution sketch: If $G\to T_2$, a directed walk of length two in $G$ would map to one in $T_2$,
% which is impossible. Conversely, if there is no directed walk of length two, no vertex has both an
% incoming and an outgoing edge; map all tails to the first vertex of $T_2$ and all heads to the
% second (isolated vertices arbitrarily).
\exercise[Blau] \label{exe:p3-dual-t3} Show that for all digraphs $G$ we have $G \to T_3$ if and only if $\vec P_3 \not \to G$.
% Solution sketch: This is the standard duality pair $(\vec P_3,T_3)$. If $G\to T_3$, composition
% rules out $\vec P_3\to G$. Conversely, repeatedly assign to a vertex the maximum length of a
% directed path ending there; absence of $\vec P_3$ bounds the value by the three levels of $T_3$ and
% the level map is a homomorphism.
\exercise[Rot] \label{exe:one-common-neighbour} Show that for every $k \in {\mathbb N}$,
every pair of adjacent vertices of $(K_3)^k$
has exactly one common neighbour (that is, every edge lies in a unique
subgraph of $(K_3)^k$ isomorphic to $K_3$).
% Solution sketch: Adjacent tuples differ in every coordinate. In each coordinate there is exactly one
% element of ${1,2,3}$ different from both entries, and the tuple of these third elements is the
% unique common neighbour.
\exercise[Rot]  \label{exe:two-common-neighbours}Show that for every $k \in {\mathbb N}$,
every pair of non-adjacent vertices in $(K_3)^k$ has at least two common neighbours.
% Solution sketch: Non-adjacent distinct tuples agree in at least one coordinate. At a coordinate
% where they differ the common neighbour is forced to use the third value, whereas at an agreeing
% coordinate either of the other two values may be used. Thus at least two common neighbours result.
\exercise[Rot] \label{exe:net-length-1}
Let $G$ be a finite digraph without loops. Show that $G \to \vec P_n$, for some $n \geq 1$, if and only if
any two paths in $G$ that
start and end in the same vertex have the
same net length.
% Solution sketch: In each weak component choose a root and define $h(v)$ as the net length of a path
% from the root to $v$. The hypothesis makes $h$ well-defined; every arc increases it by one. Since
% $G$ is finite, translating the finite range of $h$ gives a homomorphism to a finite directed path.
% The converse follows because a homomorphism to a path preserves net length.
\exercise[Rot]\label{exe:source-14} \label{exe:net-length-2}
Let $G$ be a digraph without loops and 
fix $n \geq 1$.
Show that $G \to \vec C_n$ if and only if
any two paths in $G$ that
start and end in the same vertex have the
same net length modulo $n$.
% Solution sketch: Fix a root in every weak component and assign to $v$ the net length modulo $n$ of
% any path from the root to $v$. The stated condition is exactly what makes this assignment
% well-defined, and every arc increases the residue by one. Conversely, composition with $G\to\vec
% C_n$ shows that two paths with the same endpoints have equal net length modulo $n$.
\exercise[Orange]\label{exe:source-5} Show that the Hedetniemi conjecture
is equivalent to each of the following two statements.
\begin{itemize}
\item Let $n$ be a positive integer. If for two graphs $G$ and $H$ we have $G \not \to K_n$ and $H \not\to K_n$, then $G \times H \not \to K_n$.
% Appears like this in Zhu's Survey, is direct contraposition.
\item Let $G$ and $H$ be graphs
with $\chi(G) = \chi(H) = m$. Then there exists a graph $K$ with $\chi(K) = m$ such that
$K \to G$ and $K \to H$.
% Source: Survey of Hahn and Tardif!
% Forward direction: take K := G \times H
% and use product conjecture.
% Backward direction:
%Let G and H be finite graphs such that G times H -> K_n.
%We have to show that G -> K_n or H -> K_n.
%Let m := min(\chi(G),\chi(H)). We have to show that m \leq n.
%We may remove vertices from G or from H so that \chi(G)=\chi(H) = m.
%By assumption, there exists a graph K wich \chi(K) = m such that K -> G and K -> H.
%Hence, K -> G \times H -> K_n, which means that \chi(K) = m \leq n.
\end{itemize}
\label{exe:hedetniemi-2}
\exercise[Orange]\label{exe:source-10} Construct an orientation of a tree that is not homomorphically equivalent to an orientation of a path. % Hint: the smallest example has 12 vertices.
% Solution sketch: Take two rooted orientations of paths of the same height which are incomparable
% under rooted homomorphisms and identify their roots. The resulting orientation is a tree and maps to
% a path by the height map. If it were homomorphically equivalent to a path, the image of that path
% would dominate both rooted branches, contradicting their incomparability.
\exercise[Orange]\label{exe:source-11} Construct a balanced orientation of a cycle that is not homomorphically equivalent to an orientation of a path. % Hint: glue to incomparable paths of the same length together.
% Solution sketch: Choose two homomorphically incomparable orientations $P,Q$ of paths having the same
% net length and identify their respective initial and terminal vertices. The resulting orientation of
% a cycle is balanced. A path homomorphically equivalent to it would induce a path dominating both $P$
% and $Q$, contrary to the choice of the two paths.
\exercise[Orange]\label{exe:source-15} Let $a$ be an automorphism of
$K_n^k$, for $n \geq 3$. Show that there are permutations
$p_1,\dots,p_k$ of $\{1,\dots,n\}$ and a permutation $q$ of $\{1,\dots,k\}$
 such that
$$a(x_1,\dots,x_k) = (p_1(x_{q(1)}),\dots,p_k(x_{q(k)})).$$
% TODO (have it in the book)
% Solution sketch: The maximal independent sets of $K_n^k$ are precisely the coordinate fibres
% $\{x:x_i=c\}$ when $n\geq3$. An automorphism therefore permutes the $k$ parallel classes of fibres
% and, within every class, permutes its $n$ fibres. Intersections of one fibre from each class are
% singletons, which gives the displayed coordinate formula.
\exercise[Schwarz]\label{exe:source-6} Show that Hedetniemi's conjecture
is false for directed graphs.
\par\noindent {\bf Hint:} there are counterexamples $G$, $H$ with four vertices each.
\label{exe:hedetniemi-3}
% Deepseek fails in 29.1.25
%This is easy to see for $n = 1$, since in this case $G$ or $H$
%must have an empty edge set. This is also easy to see for $n=2$:
%if $G \not\to K_2$, it must contain an odd cycle, and the same for $H$; but when both $G$ and $H$ have an odd cycle, then
%$G \time H$ also has an odd cycle, and hence is not homomorphic to $K_2$.
\end{exercises}

%\section{The Constraint Satisfaction Problem}
\subsection{The $H$-colouring Problem and Variants}
\label{sect:H-col}
When does a given digraph $G$ homomorphically map to a 
digraph $H$?
For every digraph $H$, this question defines a computational problem, called the \emph{$H$-colouring problem}.  
The input of this problem consists of
a finite digraph $G$, and the question is whether there exists a 
homomorphism from $G$ to $H$.

There are many variants of this problem. 
In the \emph{precoloured $H$-colouring problem},
the input consists of a finite digraph $G$, together
with a mapping $f$ from a subset of $V(G)$ to $V(H)$.
The question is whether there exists an extension of $f$ to all of $V(G)$ which is a homomorphism from $G$ to $H$. 
In the \emph{list $H$-colouring problem},
the input consists of a finite digraph $G$, together with a set $S_x \subseteq V(H)$ for every vertex $x \in V(G)$. 
The question is whether there exists 
a homomorphism $h$ from $G$ to $H$
such that $h(x) \in S_x$ for all $x \in V(G)$. 
It is clear that the $H$-colouring problem reduces to the precoloured $H$-colouring problem (it is a special case: the partial map might have an empty domain),
and that the precoloured $H$-colouring
problem reduces to the list $H$-colouring problem (for vertices $x$ in the domain
of $f$, we set $S_x := \{f(x)\}$, and 
for vertices $x$ outside the domain of $f$, we set $S_x := V(H)$).  

% BEISPIELE ANGEBEN WO
% precolored schwerer als normal,
% und wo list schwere ist als precoloored. 
%The graph $ are graphs where the $H$-colouring problem is in P, 

%\begin{proposition}
%Let $H$ be a digraph. Then 
%the $H$-colouring problem 
%\end{proposition}

The \emph{constraint satisfaction problem} 
is a common generalisation of all these
problems, and many more. 
It is defined not only 
for digraphs $H$, 
but more generally for \emph{relational structures}. Relational structures are the generalisation of graphs that can have 
many relations of arbitrary arity instead of just
one binary edge relation. 
%We wanted to mention this, but do not go into the details, since all the results that we present in this course can be presented
%with directed graphs via the $H$-colouring, the precolored $H$-colouring,
%and that list $H$-colouring problem. 
The constraint satisfaction problem will be introduced
formally in Section~\ref{sect:logic}. 
If $H$ is a digraph, then the constraint satisfaction problem for $H$, also denoted $\Csp(H)$, is precisely
the $H$-colouring problem and we use the
terminology interchangeably. 
Note that since graphs can be seen as a special case of digraphs, $H$-colouring 
is also defined for undirected graphs $H$. In this case we obtain essentially the same
computational problem if we only allow undirected graphs
in the input; this is made precise in Exercise~\ref{exe:undirectedHcolouring}.

For every finite graph $H$, the $H$-colouring problem is obviously in NP,
because for every graph $G$ it can be verified in polynomial time whether a given
mapping from $V(G)$ to $V(H)$ is a homomorphism from $G$
to $H$ or not. Clearly, the same holds for the precoloured and the list $H$-colouring problem. 
We have also seen that the $K_n$-colouring problem is the classical $n$-colouring
problem, which is NP-complete~\cite{GareyJohnson} for $n \geq 3$, and therefore,
no polynomial-time algorithm exists for $K_n$-colouring with $n \geq 3$, unless P=NP. 
%. Therefore, we already know graphs $H$ whose $H$-colouring problem is NP-complete. 
However, for many graphs and digraphs $H$ 
(see Exercise~\ref{exe:k2} and~\ref{exe:t2})
%for example for $H=K_2$ (Exercise~\ref{exe:k2}), 
the $H$-colouring problem can be solved in polynomial
time. Since the 1990s, researchers have 
studied the question: 
for which digraphs $H$ can the $H$-colouring problem be solved in polynomial time?
It has been conjectured by Feder and Vardi~\cite{FederVardi} 
that $H$-colouring is for any finite digraph $H$  either NP-complete or can be solved in polynomial time. This
is the so-called \emph{dichotomy conjecture},
and it was proved in 2017,  
independently by Bulatov~\cite{BulatovFVConjecture} and by
Zhuk~\cite{ZhukFVConjecture}.

\begin{theorem}
[Bulatov~\cite{BulatovFVConjecture}, Zhuk~\cite{ZhukFVConjecture}]
\label{thm:dicho}
Let $H$ be a finite digraph. 
Then $\Csp(H)$ is in P or NP-complete. 
\end{theorem}

It was shown by Ladner~\cite{Ladner} that unless P=NP there are infinitely many complexity classes between P and NP; so the conjecture
states that for $H$-colouring these intermediate complexities
do not appear. 
Feder and Vardi also showed that 
if the dichotomy conjecture holds for $H$-colouring problems, then also the more general class of CSPs for finite relational structures exhibits a complexity dichotomy (see Section~\ref{sect:pp}).

%As we will see later, if we solve the
%classification question for the precoloured $H$-colouring problem, then we solve
%the classification question for the $H$-colouring problem, 
%and vice versa (Corollary~\ref{cor:dicho-precol}). Quite surprisingly, the same is true for the complexity
%of constraint satisfaction problems: a complete classification into
%NP-complete and P for the $H$-colouring problem would imply a classification
%for the class of all CSPs, and vice versa~\cite{FederVardi}. 

The list $H$-colouring problem, on the other hand, is quickly NP-hard,
and therefore less difficult to classify. 
And indeed, a complete classification has
been obtained by Bulatov~\cite{Conservative}
already in 2003. 
Alternative proofs can be found in~\cite{Barto-Conservative,Bulatov-conservative-revisited}.
For finite \emph{undirected}
graphs, it is known since 1990 that the dichotomy conjecture holds~\cite{HellNesetril}; this text provides two fundamentally different proofs of the following.
% (Section~\ref{sect:undirected} and Section~\ref{sect:}). 

\begin{theorem}[of \cite{HellNesetril}]\label{thm:HN}
Let $H$ be a finite undirected graph. 
If $H$ homomorphically maps to $K_2$, or contains a loop, then $H$-colouring can be solved in polynomial time. Otherwise, $H$-colouring is NP-complete. 
\end{theorem}
The case that $H$ homomorphically maps to $K_2$ will be the topic of Exercise~\ref{exe:k2}.
%We will see in Section~\ref{sect:tractability}
%how to obtain the other part of the statement of Theorem~\ref{thm:HN} from more general principles. 
The entire proof of Theorem~\ref{thm:HN} 
can be found
in Section~\ref{sect:HN}, and an alternative proof in Section~\ref{sect:undir-revisited}.

\begin{exercises}
\exercise[Blau]\label{exe:source-19} \label{exe:undirectedHcolouring}
Let $G$ and $H$ be directed graphs, and suppose that $H$
is symmetric.
Show that $f \colon V(G) \rightarrow V(H)$
is a homomorphism from $G$ to $H$ if and only if
$f$
is a homomorphism from the undirected
graph of $G$
to the undirected graph of $H$.
% Solution sketch: For a symmetric $H$, an arc $(u,v)$ and its reverse impose the same condition on
% $f(u),f(v)$. Hence the homomorphism condition depends only on the underlying undirected edges of $G$
% and $H$, proving both directions immediately.
\exercise[Blau]\label{exe:source-21} \label{exe:t3}
Prove that $\Csp(T_3)$ can be solved in polynomial time.
% Solution sketch: Use the duality $(\vec P_3,T_3)$ from Exercise~\ref{exe:p3-dual-t3}. Test whether
% the input contains a homomorphic image of the fixed path $\vec P_3$; equivalently, compute the three
% possible levels. This is polynomial (indeed linear) time.
\exercise[Rot]\label{exe:source-16} For graphs $G$ and $H$, order the following sets by inclusion.
\begin{itemize}
\item $\Csp(G) \cup \Csp(H)$.
\item $\Csp(G \times H)$.
\item $\Csp(G \uplus H)$.
\item $\Csp(G) \cap \Csp(H)$.
\end{itemize}
Which of the inclusions are strict?
% Solution: \times = cap < uplus < \cup
% Take G = C2, H = C3 for examples
% that witness strictness.
\exercise[Rot]\label{exe:source-17} \label{exe:strong-homo}
Let $H$ be a finite directed graph. Find an algorithm that decides whether
 there is a strong homomorphism from a given graph $G$ to the fixed
graph $H$.
The running time of the algorithm should be polynomial
in the size of $G$ (note that we consider $|V(H)|$ to be constant).
%One way would be as follows: guess a subgraph S of G that is isomorphic to H -- since H is fixed this can be done in P. If there is no such subgraph, reject. Otherwise, find vertices in G that look like a "double" of one of the vertices of S (can be checked easily), and then contract them with the vertex of S, and keep on doing this until only vertices from S are left. If you get stuck before, reject.
\exercise[Rot]\label{exe:source-18} \label{exe:construct-solution} Let $H$ be a finite digraph such that $\Csp(H)$
can be solved in polynomial time.
Find a polynomial-time algorithm that constructs for a given finite digraph $G$
a homomorphism to $H$, if such a homomorphism exists.
% There is one solution via Exercise~\ref{exe:source-17}: add arcs
% while being satisfiable until you can be sure that there is a
% strong homomorphism.
% the nicer solution is different: contract vertices while being
% satisfiable until you have fewer vertices than the template;
% then you can do a brute-force search in poly-time.
%\newpage
\exercise[Rot]\label{exe:source-20} \label{exe:k2}
Show that for any graph $H$ that homomorphically maps to
$K_2$ the constraint
satisfaction problem for $H$ can be solved in
polynomial time.
% Solution sketch: The assumption $H\to K_2$ makes every nontrivial component of $H$ bipartite. A
% connected input component containing an edge maps to $H$ exactly when it is bipartite and $H$ has an
% edge; isolated vertices only require $H$ to be nonempty. Bipartiteness is testable by breadth-first
% search.
\exercise[Rot]\label{exe:source-22} \label{exe:c3} Prove that $\Csp(\vec C_3)$
can be solved in polynomial time.
% Solution sketch: In each weak component choose a root and propagate residues in $\mathbb Z_3$,
% increasing by one along every arc and decreasing by one against an arc. A conflict occurs exactly
% when a closed walk has nonzero net length modulo three. If there is no conflict, the propagated
% residues define the required homomorphism to $\vec C_3$.
\exercise[Rot]\label{exe:source-23} \label{exe:zigzag-dual-p2}
Let $\cal N$ be the set $\{Z_1,Z_2,Z_3,\dots\}$.
Show that a digraph $G \to \vec P_2$ if and only if
no digraph in $\cal N$ homomorphically maps to $G$.
% Solution sketch: A homomorphism to $\vec P_2$ gives a height function with values $0,1,2$, and every
% zig-zag $Z_i$ violates such a height assignment. Conversely, if the propagation of the three
% possible heights fails, tracing the successive reasons for deletion produces a zig-zag obstruction
% $Z_i\to G$.
\exercise[Rot]\label{exe:source-24}
Suppose that $\Csp(G)$ and $\Csp(H)$,
for two digraphs $G$ and $H$, can be solved
in polynomial time.
Show that $\Csp(G \times H)$ and  $\Csp(G \uplus H)$ can be solved
in polynomial time as well.
% For CSP(G times H), answer YES
% if there is a homo to G and to H.
% For CSP(G oplus H),
% run both algorithms for each weakly connected
% component, and accept if one of the two accepts.
%\vspace{-.4cm}
% Need to check the connected components, because
% 
\exercise[Orange]\label{exe:source-25}
Suppose that $G$ and $H$ are homomorphically incomparable and suppose that
 $\Csp(G) \cup \Csp(H)$ can be solved in polynomial time.
Show that $\Csp(G)$
and $\Csp(H)$ can be solved in polynomial time as well.
% If K to G then K+G to G, and K+G \in CSP(G) \cup CSP(H).
% Conversely, note that K+G can't map to H since G does not map to H. Hence,
% if K+G \in  CSP(G) \cup CSP(H),
% then K + G to G and in particular K \to G.
%\vspace{-.5cm}
%Suppose that $G$ and $H$ are homomorphically incomparable. Give a polynomial-time reduction from $\Csp(G)$ to $\Csp(G) \cup \Csp(H)$.
% If I is the input, then run oracle on I + G.
\exercise[Schwarz]\label{exe:source-26}
Suppose that $G$ and $H$ are homomorphically incomparable and connected,
and suppose that $\Csp(G \uplus H)$ can be solved in polynomial time.
Show that
$\Csp(G)$ and $\Csp(H)$ can be solved in polynomial time as well.
%Can we \\
%get rid of the assumption that $G$ and $H$ are connected?
% This is the next exe
% Solution: the idea is the same as in Exercise~\ref{exe:source-25}. However, at the key step which fails do the following: identify a vertex from each connected component of the input with a vertex from G (alltogether n^|G|, so polynomially many choices). If there is a homo from input to G then one of the choices must work.
\exercise[Schwarz]\label{exe:source-27} Show that the assumption in Exercise~\ref{exe:source-26} that $G$ and $H$ are connected is necessary. Specifically,
find digraphs $G$ and $H$ such that
$\Csp(G \uplus H)$
can be solved in polynomial time, but $\Csp(G)$ is NP-hard.
% Loesung (Leo in Strobl): 
% nehme den harten Baum T, und einen 
% orientierten Pfad P1, der unvergleichbar ist (z.B. drei vor, eins zurueck, drei vor), und 
% einen orientierten (gerichteten!) Pfad P2, so dass T -> P2 aber P1 nicht nach P2 (z.B., 4 vor). 
% Sei G := T \cup P1, und H := P2. 
% Dann geht G nicht nach H, weil P1 nicht nach P2,
% und H geht nicht nach G, weil P2 weder nach T noch nach P1 (beide zu weit). 
% Dann ist CSP(G \cup H) = CSP(P1 \cup P2) einfach,
% aber CSP(G) ist hart wegen T. 
\exercise[Weiss]\label{exe:uplus-drops} Find digraphs $G$ and $H$ such that
$\Csp(G \uplus H)$
can be solved in polynomial time, but $\Csp(G)$ is NP-hard \emph{and} $\Csp(H)$ is NP-hard. 
% Loesung Sebastian Meyer: G ist die disjunkte Vereinigung von [einem orientierten Kreis der L\"ange 3 (orientierter C_3)] und [einem vollst\"angigen gerichteten Graphen mit drei Knoten, wo auf jeder der 6 Kanten ein weiterer Knoten eingef\"ugt wurde]. (G hat 12 Knoten und 15 Kanten.)
%H ist die disjunkte Vereinigung von [einem orientierten Kreis der L\"ange
%2 (orientierter C_2 oder K_2)] und [einem vollst\"angigen gerichteten
%Graphen mit drei Knoten, wo auf jeder der 6 Kanten zwei weitere Knoten
%eingef\"ugt wurden]. (H hat 17 Knoten und 20 Kanten.)
\exercise[Gelb]\label{exe:product-drops} Find finite digraphs $G$ and $H$ such that $\Csp(G)$ is NP-hard, $\Csp(H)$ is NP-hard, but $\Csp(G \times H)$ is in P.
\end{exercises}

\subsection{Cores}\label{ssect:cores}
%The following definitions and Proposition~\ref{prop:core} are again both for directed and
%for undirected graphs.
An \emph{endomorphism} of a digraph $H$ is a homomorphism from $H$ to $H$.
A finite digraph $H$ is called a \emph{core} if 
every endomorphism of $H$ is an automorphism.
A graph $G$ is called \emph{a core of} $H$ if
$H$ is homomorphically equivalent to $G$ and
$G$ is a core. 

\begin{proposition}\label{prop:core}
Every finite digraph $H$ has a core, which is unique up to isomorphism, and which is isomorphic to an induced subgraph of $H$. 
\end{proposition}

\begin{proof} 
Any finite digraph $H$ has a core, since we can select an
endomorphism $e$ of $H$ such that the image of $e$ has smallest cardinality; the subgraph of $H$ induced by $e(V(H))$ 
is a core of $H$.
Let $G_1$ and $G_2$ be cores of $H$, and $f_1 \colon H \rightarrow G_1$, $g_1 \colon G_1 \to H$, $f_2 \colon H \rightarrow G_2$, and $g_2 \colon G_2 \to H$ be homomorphisms. Let $e_1 := f_2 \circ g_1$ and $e_2 := f_1 \circ g_2$. See Figure~\ref{fig:cores}.

\begin{figure}
\begin{center}
\includegraphics[scale=0.5]{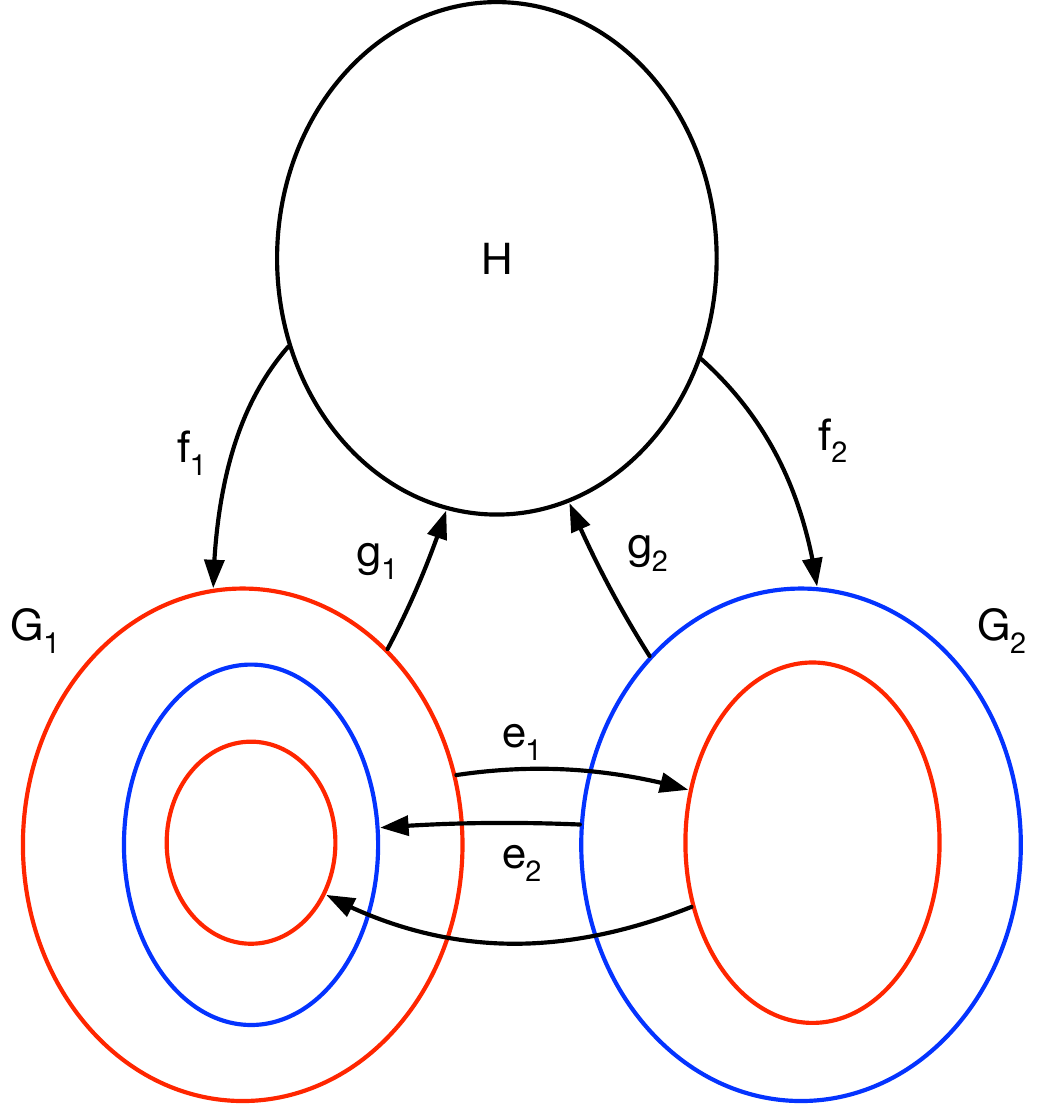} 
\end{center}
\caption{Illustration of the uniqueness proof for cores.}
\label{fig:cores}
\end{figure}

We claim that $e_1$ is the desired
isomorphism. 
Suppose for contradiction that $e_1$ is not injective, i.e., there are distinct $x,y$ in $V(G_1)$ such that $e_1(x)=e_1(y)$. It follows that $e_2 \circ e_1$ cannot be injective either. But $e_2 \circ e_1$ is an endomorphism of $G_1$, contradicting the assumption that
$G_1$ is a core. 
Similarly, $e_2$ is an injective homomorphism 
from $G_2$ to $G_1$, and it follows that 
$|V(G_1)| = |V(G_2)|$ and both $e_1$ and $e_2$ are bijective. 

Now, since $|V(G_1)|$ is finite, there exists an $n \in {\mathbb N}$ such that 
$$e_2 \circ e_1 \circ \cdots \circ e_2 \circ e_1 = (e_2 \circ e_1)^n = \id.$$
Hence, $e_2 \circ e_1 \circ \cdots \circ e_2 = (e_1)^{-1}$, so the inverse of $e_1$ is a homomorphism, and hence an isomorphism
between $G_1$ and $G_2$. 
\end{proof}

Since a core $G$ of a finite digraph $H$ is unique up to isomorphism, we
call $G$ \emph{the} core of $H$. 
We want to mention without proof that it is
NP-complete to decide whether a given digraph $H$ is not a core~\cite{cores}.

 Cores can be characterised in many different ways; for some of them, see Exercise~\ref{exe:cores}. 
 There are examples of infinite digraphs that do not have a core in the sense defined above; see Exercise~\ref{exe:infcore}. 
Since a digraph $H$ and its core have the same CSP,
it suffices to study $\Csp(H)$ for core digraphs $H$ only. Working with cores has advantages; one of them is shown in Proposition~\ref{prop:constants} below. 
In the proof of this proposition, we need a concept that we will use again in later sections.

\begin{definition}\label{def:contract}
Let $H$ be a digraph and a surjective map $q \colon V(H) \to X$ for some set $X$. Then the \emph{homomorphic image} of $H$ under $q$, denoted by $H/q$, is the graph with vertex set $X$ and edge set $\{(q(u),q(v)) \mid (u,v) \in E(H)\}$. 
If $u,v \in V(H)$, then the graph obtained from $H$ by 
\emph{contracting $u,v$}, denoted by $H/_{\{u,v\}}$, is
defined to be the homomorphic image of $H$ with respect to the map $q \colon V(H) \to V(H) \setminus \{u,v\} \cup \big \{ \{u,v\} \big \}$ defined by $q(x) = x$ if $x \in V(H) \setminus \{u,v\}$ and $q(x) = \{u,v\}$ otherwise. 
\end{definition}

%The \emph{precolored CSP} for a digraph $H$
%is the following computational problem.
%Given is a finite digraph $G$ and a partial mapping $c \colon V(G) \rightarrow V(H)$. The question is whether $c$ can be extended
%to a homomorphism from $G$ to $H$.
%In other words, we want to find a homomorphism from
%$G$ to $H$ where some vertices have a pre-described image.

\begin{proposition}\label{prop:constants}
Let $H$ be a core. Then $\Csp(H)$
and 
precoloured $\Csp(H)$ are linear-time 
equivalent.
\end{proposition}

\begin{proof}
The reduction from $\Csp(H)$ to precoloured $\Csp(H)$ is immediate:
an instance $G$ of $\Csp(H)$ is equivalent to the instance $(G,c)$
of precoloured $\Csp(H)$ where $c$ is everywhere undefined.

For the converse reduction, let $(G,c)$ be an instance of precoloured
$\Csp(H)$. Take the disjoint union of $G$ and a copy $H'$ of $H$.
For every $x$ in the domain of $c$, contract $x$ with the vertex of
$H'$ corresponding to $c(x)$. Denote the resulting graph by $G'$.
Clearly, $|V(G')| \leq |V(G)|+|V(H)|$, and $G'$ can be constructed in
linear time.

We claim that $G'\to H$ if and only if $(G,c)$ has a solution.
Suppose first that $f\colon G\to H$ extends $c$. Extend $f$ to the
copy $H'$ by mapping every vertex of $H'$ to the corresponding vertex
of $H$. This is well defined at the identified vertices, since
$f(x)=c(x)$, and it is a homomorphism from $G'$ to $H$.

Conversely, let $f'\colon G'\to H$ be a homomorphism. The restriction
of $f'$ to $H'$ is an endomorphism $\beta$ of $H$. Since $H$ is a
core, $\beta$ is an automorphism. Define
\[
f:=\beta^{-1}\circ f'|_{V(G)}.
\]
Then $f$ is a homomorphism from $G$ to $H$. If $x$ is in the domain
of $c$, then $x$ is identified in $G'$ with the copy of $c(x)$ in
$H'$. Therefore, $f'(x)=\beta(c(x))$,
and hence $f(x)=\beta^{-1}(f'(x))=c(x)$. 
Thus $f$ extends $c$, which proves the claim.
\end{proof}

\begin{corollary}\label{cor:dicho-precol}
If for every finite digraph $H$, the precoloured
$H$-colouring problem is in P or NP-complete,
then CSP$(H)$ is in P or NP-complete for every finite digraph $H$ as well. 
\end{corollary}

The following example shows that the assumption of Proposition~\ref{prop:constants} that $H$ is a core is necessary (unless P = NP). 

\begin{example}
Let $H$ be the disjoint union of $K_3$ and a loop. 
Then $\Csp(H)$ is trivial and in P. The precoloured $H$-colouring problem, however, is NP-complete:
we may prove this by a reduction from the NP-complete $\Csp(K_3)$ as follows. Clearly, this problem is already NP-complete if restricted to input graphs that are connected. 
Let $G$ be a connected finite graph. 
Let $c$ be a partial map sending one vertex of $G$ to some element of $K_3$. 
Then $G$ has a homomorphism to $K_3$ if and only if $c$ can be extended to a homomorphism from $G$ to $H$. 
To see this, let $f \colon G \to K_3$ be a homomorphism. Composing $f$ with a  permutation of $V(K_3)$ is also a homomorphism from $G$ to $K_3$, and hence in particular to $H$. So we may obtain a homomorphism from $G$ to $H$ which extends $c$. Conversely, if $f$ is a homomorphism from $G$ to $H$ which extends $c$ then $f(V(G)) \subseteq V(K_3)$, since $f(x) \in V(K_3)$ and $G$ is connected. 
\end{example} 

We have already seen in Exercise~\ref{exe:construct-solution} 
that the computational
problem to construct a homomorphism from $G$ to $H$, for fixed $H$ and given $G$, can be reduced in polynomial-time to the problem of deciding whether there exists a homomorphism
from $G$ to $H$. The intended solution of Exercise~\ref{exe:construct-solution}  requires in the worst-case $|V(G)|^2$ many executions of the
decision procedure for $\Csp(H)$.
Using the concept of cores and the precoloured CSP (and its equivalence to the CSP) we can give a faster method to construct homomorphisms.

\begin{proposition}
If there is an algorithm that decides $\Csp(H)$ in time $T$, then there is an algorithm that constructs a homomorphism from
a given digraph $G$ to $H$ (if such a homomorphism exists) which runs in time $O(|V(G)|T)$.
\end{proposition}

\begin{proof}
Let $C$ be the core of $H$; we may suppose that
$C$ is a subgraph of $H$. 
By Proposition~\ref{prop:constants}, 
and since $\Csp(C)$ and $\Csp(H)$ are the same problem, there is an algorithm $A$ for 
precoloured $\Csp(C)$ 
with a running time in $O(T)$. 

To construct a homomorphism from a given finite digraph $G$ to $H$, we first apply $A$ to $(G,c)$ for the everywhere
undefined function $c$ 
to decide whether there exists a homomorphism from $G$ to $C$.
If no, then there is also no homomorphism to $H$ and there is nothing to be shown. 
If yes, we select some $x \in V(G)$,
and extend $c$ by defining $c(x)=u$ for some $u \in V(C)$. 
Then we use algorithm $A$ to decide whether there is a
homomorphism from $G$ to $C$ that extends $c$.
If no, we try another vertex $u \in V(C)$. 
Clearly,
for some $u$ the algorithm must give the answer ``yes''. 
We proceed with the extension $c$ where $c(x)=u$, and repeat
the procedure with another vertex $x$ from $V(G)$. 
At the end, $c$ is defined for all vertices $x$ of $G$,
and $c$ is a homomorphism from $G$ to $C$.
Clearly, since $H$ and $C$ are fixed, 
algorithm $A$ is executed $O(|V(G)|)$ times.
\end{proof}

%A digraph $H$ is called \emph{rigid} if the identity mapping is 
%the only endomorphism of $H$.
%The property of being rigid is stronger than the property of being a core, 
%because rigid graphs are \emph{asymmetric}, 
%i.e., the identity is their only automorphism,
%whereas cores need not be asymmetric.

%\newpage 

\begin{exercises}
\exercise[Blau]\label{exe:source-29} Prove that the core of a strongly connected digraph is strongly connected.
% Solution sketch: Realise the core $H$ as a retract of the strongly connected digraph $G$. For
% $u,v\in H$, choose directed paths from $u$ to $v$ and from $v$ to $u$ in $G$ and apply the
% retraction to them. Their images are directed walks in $H$, so $H$ is strongly connected.
\exercise[Blau]\label{exe:source-30} Show that $Z_{k,l}$ is a core for all $k,l \geq 2$.
%\newpage
% Solution sketch: The alternating blocks at the two ends of $Z_{k,l}$ are distinguished by their in-
% and out-distance profiles when $k,l\geq2$. Any endomorphism must consequently fix the two ends and
% then, inductively along the path, every remaining vertex. Hence every endomorphism is the identity
% and $Z_{k,l}$ is a core.
\exercise[Blau]\label{exe:source-34} Prove that cores and products of digraphs without sources and sinks
have no sources and sinks.
% Solution sketch: For a retract $r:G\to H$, an incoming (respectively outgoing) edge at a preimage of
% $h\in H$ maps to an incoming (respectively outgoing) edge at $h$; thus a core retract of a digraph
% without sources or sinks has neither. In a direct product, choose incoming and outgoing neighbours
% coordinatewise.
\exercise[Rot]\label{exe:source-31} \label{exe:cores}
Prove that for every finite digraph $G$ the following is equivalent:
\begin{itemize}
\item Every endomorphism of $G$ is an automorphism.
\item Every endomorphism of $G$ is injective.
\item Every endomorphism of $G$ is surjective.
%\item Every retract is the identity
%\item $G$ does not have a proper retract
\end{itemize}
% Solution sketch: On a finite set, injective and surjective endomorphisms are equivalent, and a
% bijective endomorphism of a finite relational structure has a homomorphic inverse, hence is an
% automorphism. If an endomorphism is not an automorphism, a sufficiently large idempotent power is a
% retraction onto a proper image. Conversely, a proper retraction is a non-automorphic endomorphism.
\exercise[Rot]\label{exe:source-32} Show that the three properties in the previous
exercise are
no longer equivalent
if $G$ is
infinite.
% Example separating two and three: e.g. (Q,<).
% Example of a structure where all endomorphisms
% are bijective:
% (Z,succ,\{(x,x+2) | x \geq 0\}).
% But he map $x \mapsto x+2 is endo, but not auto.
% This separates 1 from {2,3}.
%\item What are the cores of the graphs shown in the exercise~\ref{exe:examples}?
%\item Prove that all cores of a directed graph $G$ are isomorphic.
%\vspace{-1.2cm}
%\newpage
\exercise[Rot]\label{exe:source-33} \label{exe:infcore}
Show that the infinite tournament $(\mathbb Q;<)$ has endomorphisms that are not automorphisms.
Show that every digraph that is
homomorphically equivalent to $(\mathbb Q;<)$ also has endomorphisms
that are not automorphisms.
% Solution sketch: For example, a strictly increasing injection of $\mathbb Q$ into a proper interval
% of $\mathbb Q$ is an endomorphism of $(\mathbb Q;<)$ but not an automorphism. Given homomorphisms in
% both directions between another digraph and $(\mathbb Q;<)$, compose them with a suitably chosen
% proper order embedding; if either original map already collapses vertices it is itself
% non-automorphic, and otherwise the proper embedding makes the composite non-surjective.
\exercise[Rot]\label{exe:source-35} Let $H$ be the core of $G$ which we may assume to be a subgraph of $G$.
Show that there exists a \emph{retraction} from $G$
to $H$, i.e.,
a homomorphism $e$ from
$G$ to $H$ such that
$e(x)=x$ for all $x \in V(H)$.
% Solution sketch: Choose any homomorphism $f:G\to H$. Its restriction to $H$ is an endomorphism of
% the core and therefore an automorphism. Composing $f$ with the inverse of this restriction gives a
% homomorphism $e:G\to H$ whose restriction to $H$ is the identity.
\exercise[Rot]\label{exe:source-36} %The set of automorphisms of a digraph $G$ forms a group; this group
A permutation group on a set $V$
 is called \emph{transitive} if
for all $a,b \in V$ there exists $g \in G$ such that $g(a)=b$.
Show that if $(V,E)$ is a graph with a transitive automorphism group, then the core of $(V,E)$ also has a transitive automorphism group.
% Solution sketch: Let $r:G\to H$ be a retraction onto a core. For an automorphism $a$ of $G$, the map
% $r\circ a|_H$ is an endomorphism of $H$ and hence an automorphism. Given $u,v\in H$, first choose
% $a$ with $a(u)$ in the fibre $r^{-1}(v)$; then $r\circ a|_H$ sends $u$ to $v$.
\exercise[Rot]\label{exe:source-37} Show that the connected components of a core
are cores that form an antichain in $({\mathcal D},\leq)$; conversely, the disjoint union of an antichain of cores
is a core.
% From a script of Cameron
% Solution sketch: An endomorphism of a core cannot collapse a component onto a proper homomorphic
% image, so each component is a core. Homomorphisms between two distinct components would extend by
% the identity elsewhere to a non-automorphic endomorphism, proving the antichain property.
% Conversely, an endomorphism of a disjoint union of pairwise incomparable cores must preserve every
% component and be an automorphism on it.
\exercise[Orange]\label{exe:source-38} Prove that the core of a digraph with a transitive automorphism group
is connected.
% From a script of Cameron;
% use the previous two
% exercises!
% Solution sketch: By Exercises~\ref{exe:source-36} and~\ref{exe:source-37} the components of the core form an antichain, while
% transitivity of the induced automorphism group makes all components isomorphic. Two distinct
% isomorphic components are comparable, contradicting the antichain property. Hence there is only one
% component.
\exercise[Orange]\label{exe:source-40}
Determine the computational complexity of $\Csp(H)$ for
$$H := \big({\mathbb Z}; \{(x,y) \; : \; |x-y| \in \{1,2\} \} \big) \; .$$
% Homomorphically equivalent to K_3!
% Solution sketch: The vertices $0,1,2$ induce a copy of $K_3$. Conversely, reduction modulo $3$ is a
% homomorphism from $H$ to $K_3$, since differences $1$ and $2$ are nonzero modulo $3$. Thus $H$ and
% $K_3$ are homomorphically equivalent and $\Csp(H)$ is NP-complete.
\exercise[Schwarz]\label{exe:source-39} A permutation group $G$ on a set $X$ is called \emph{primitive} if the only
equivalence
relations
on $X$ that are preserved by all permutations
 in $G$ are  the equality relation and the equivalence relation with only
one equivalence class.
Prove that the core of a digraph with a
primitive automorphism group
has a primitive automorphism group.
% Solution:
% Let f be the homo from the digraph to its
% core and let g be the homo from the core
% back to the digraph.
% Let a,b be two distinct points in the core
% and let E be an invariant equivalence relation containing (a,b).
% Pick c,d in the core.
% Let O be the orbit of (g(a),g(b)). Then
% the graph of O must be weakly connected
% by the primitivity of Aut(Digraph).
% Hence, there is in particular a weak O path from h(c) to h(d).
%If (e_i,e_i+1) is an edge,  then (g(e_i),g(e_i+1)) must be in the same orbit as (a,b), so (g(e_i),g(e_i+1)) \in E. By the transitivity of E, this shows that (g(f(c)),g(f(d)) \in E, so (c,d) \in E since g \circ f is an automorphism. Hence, E has only one class.
\end{exercises}

%\section{Graph Endomorphisms}
%{\red what properties does the poset of endomorphic images of 
%a graph/structure have?}

%\section{Graph Products}
\subsection{Polymorphisms}\label{ssect:polymorphisms}
Polymorphisms are a powerful tool for analysing the computational complexity of constraint satisfaction problems; as we will see,
they are useful both for NP-hardness proofs
and for proving the correctness of polynomial-time algorithms for CSPs. 
Polymorphisms can be seen as multi-dimensional variants of endomorphisms.

\begin{definition}
Let $H$ be a digraph and $k \geq 1$.  
Then a \emph{polymorphism of $H$ of arity $k$} is a homomorphism from $H^k$ to $H$. 
\end{definition}

In other words, a mapping $f \colon V(H)^k \to V(H)$ is a polymorphism of $H$ if and only if 
$(f(u_1,\dots,u_k)$,
$f(v_1,\dots,v_k)) \in E(H)$ whenever $(u_1,v_1),\dots,(u_k,v_k)$
are arcs in $E(H)$. 
Note that any digraph $H$ has all \emph{projections} as polymorphisms, i.e., 
all mappings $\pi^k_i \colon V(H)^k \rightarrow V(H)$, for $i \leq k$ given by  $\pi^k_i(x_1,\dots,x_k)=x_i$ for all $x_1,\dots,x_k \in V(H)$. The operation $\pi^k_i$ is called the \emph{$i$-th projection of arity $k$}. 

\begin{example} The operation  
$(x,y) \mapsto \min(x,y)$ 
is a polymorphism of the digraph  
$T_n = (\{1,\dots,n\};<)$.
\end{example}
An operation $f \colon V(H)^k \rightarrow V(H)$ is called 
\begin{itemize}
\item \emph{idempotent}
if $f(x,\dots,x)=x$ for all $x\in V(H)$.
\item \emph{conservative}
if $f(x_1,\dots,x_k) \in \{x_1,\dots,x_k\}$ for all $x_1,\dots,x_k \in V(H)$. 
\end{itemize}

\ignore{
We say that a $k$-ary operation $f$ \emph{depends} on an argument
$i$ iff there is no $k{-}1$-ary operation $f'$ such that $f(x_1,
\dots, x_k) =$ $f'(x_1, \dots, x_{i-1}, x_{i+1},$ $\dots, x_k)$. 
We
can equivalently characterise $k$-ary operations that depend on the
$i$-th argument by requiring that there are elements $x_1, \dots,
x_k$ and $x_i'$ such that $f(x_1, \dots, x_k) \neq f(x_1, \dots,
x_{i-1}, x'_i, x_{i+1},$ $\dots, x_k)$.

\begin{lemma}
An operation $f$ is essentially unary if and only if $f$
depends on at most one argument.
\end{lemma}
\begin{proof}
It is clear that an operation that is essentially unary depends
on at most one argument. Conversely, suppose that $f$ is 
$k$-ary and depends only on the first argument. 
Let $i \leq k$ be the maximal such that there is an operation $g$
with
$f(x_1,\dots,x_k)=g(x_1,\dots,x_i)$. If $i=1$ then $f$ is essentially unary and we are done. Otherwise, observe that since $f$ does
not depend on the $i$-th argument, neither does $g$, and so
there is an $i-1$-ary operation $g'$ such that $f(x_1,\dots,x_n)=g(x_1,\dots,x_i)=g'(x_1,\dots,x_{i-1})$, in contradiction to the choice of $i$.
\end{proof}
}

A digraph $H$ is called \emph{projective}
if every idempotent polymorphism is a projection.
The following will be shown in Section~\ref{ssect:minimal}. 

\begin{proposition}\label{prop:k3}
For all $n \geq 3$, the graph $K_n$ is projective.
\end{proposition}

\begin{exercises}
\exercise[Blau]\label{exe:source-41} \label{exe:hat}
Show that if $f \colon H^k \to H$ is a polymorphism of a digraph $H$, then
$\hat f(x) := f(x,\dots,x)$ is an endomorphism of $H$.
% Solution sketch: Apply $f$ to $k$ copies of an arc $(x,y)$ of $H$. The resulting pair is
% $(f(x,\ldots,x),f(y,\ldots,y))=(\hat f(x),\hat f(y))$, so $\hat f$ preserves all arcs.
% Alternative: compose hat f from f 
% with k times \pi^k_1.
\exercise[Rot]\label{exe:source-42} Show that if $H$ is a finite core digraph with a \emph{symmetric} binary
polymorphism $f$,
that is, $f(x,y) = f(y,x)$ for all $x,y \in V(H)$,
then
$H$ also has an \emph{idempotent} symmetric polymorphism.
% Solution sketch: The diagonal map $\hat f$ is an endomorphism of the finite core and hence an
% automorphism. Put $g=\hat f^{-1}$ and define $f'(x,y)=g(f(x,y))$. Composition preserves
% polymorphisms and symmetry, while $f'(x,x)=g(\hat f(x))=x$.
\end{exercises}

\section{The Arc-consistency Procedure} 
\label{sect:AC}
%\section{Arc-consistency} 
The \emph{arc-consistency procedure}
is one of the most fundamental and well-studied algorithms used to solve CSPs.
This procedure was first discovered for constraint satisfaction problems in artificial intelligence~\cite{Montanari,PathConsistency};
in the graph homomorphism literature, the algorithm is sometimes
called the \emph{consistency check algorithm}.

Let $H$ be a finite digraph, and let $G$ be an instance
of $\Csp(H)$. 
The idea of the procedure is to maintain for each vertex
of $G$ a list of vertices of $H$, and each element in the list
of $x$ represents a candidate for an image of $x$ under a homomorphism from $G$ to $H$.
The algorithm successively
removes vertices from these lists; it only removes 
a vertex $u \in V(H)$
from the list for $x \in V(G)$, if there is no homomorphism from
$G$ to $H$ that maps $x$ to $u$. 
To detect vertices $x,u$ such that $u$ can be removed from
the list for $x$, the algorithm uses two rules (in fact, one rule and
a symmetric version of the same rule): if $(x,y)$ is an edge in $G$, then
\begin{itemize}
\item remove $u$ from $L(x)$ if there is no $v \in L(y)$
with $(u,v) \in E(H)$;
\item remove $v$ from $L(y)$ if there is no $u \in L(x)$
with $(u,v) \in E(H)$. 
\end{itemize}

If eventually we can no longer remove any vertex from any list with these
rules any more, the digraph $G$ together with 
the lists for each vertex is called \emph{arc-consistent}.
The pseudo-code of the entire 
arc-consistency procedure is 
displayed in Figure~\ref{fig:arc-cons}.

\begin{figure}
\begin{center}
\fbox{
\begin{tabular}{l}
$\AC_H(G)$ \\
Input: a finite digraph $G$. \\
Data structure: a list $L(x) \subseteq V(H)$ for each vertex $x \in V(G)$. \\
\\
Set $L(x) := V(H)$ for all $x \in V(G)$. \\
Do \\
\hspace{0.5cm} For each $(x,y) \in E(G)$: \\
\hspace{1cm} Remove $u$ from $L(x)$ if there is no $v \in L(y)$
with $(u,v) \in E(H)$. \\
\hspace{1cm} Remove $v$ from $L(y)$ if there is no $u \in L(x)$
with $(u,v) \in E(H)$. \\
\hspace{1cm} If $L(x)$ is empty for some vertex $x \in V(G)$ then {\bf reject} \\
Loop until no list changes
\end{tabular}}
\end{center}
\caption{The arc-consistency procedure for $\Csp(H)$.}
\label{fig:arc-cons}
\end{figure}

Clearly, if the algorithm removes all vertices 
from one of the lists, then 
there is no homomorphism from $G$ to $H$.
It follows that if $\AC_H$ rejects an instance of $\Csp(H)$, it has no solution. 
The converse implication does not hold in general.
For instance, let $H$ be $K_2$, and let $G$ be $K_3$.
In this case, $\AC_H$ does
not remove any vertex from any list, but obviously there is
no homomorphism from $K_3$ to $K_2$.

However, there are digraphs $H$ where $\AC_H$ is a complete decision procedure for $\Csp(H)$ in the sense
that it rejects an instance $G$ of $\Csp(H)$
if and only if $G$ does not homomorphically map to $H$.
In this case we say that $\AC$ \emph{solves} $\Csp(H)$.

\begin{remark}\label{rem:AC-uniform}
The running time of $\AC_H$ is for
any fixed digraph $H$ polynomial in the size of $G$. 
If $H$ is part of the input, we refer to the procedure as $\AC$. 
Quite remarkably, $\AC$ has polynomial running time as well. 
\end{remark} 

\paragraph{Implementation.}
In a naive implementation of the procedure, 
the inner loop of the algorithm would go over all edges of the digraph,
in which case the running time of the algorithm is quadratic in 
the size of $G$.
In the following we describe an implementation of
the arc-consistency procedure, called AC-3, which is due to Mackworth~\cite{PathConsistency}, and 
has a worst-case running time that is linear in the size of $G$.
Several other implementations of the arc-consistency procedure
have been proposed in the Artificial Intelligence literature, aiming at reducing the costs of the algorithm in terms of the number of vertices of both $G$ and $H$. But here we consider the size of $H$ to be fixed,
and therefore we do not present this line of research. 
With AC-3, we rather present one of the simplest implementations of the arc-consistency procedure
with a linear running time.

\begin{figure}[h]
\begin{center}
\fbox{
\small
\begin{tabular}{l}
AC-3$_H(G)$ \\
Input: a finite digraph $G$. \\
Data structure: a list $L(x)$ of vertices of $H$ for each $x \in V(G)$. \\
\hspace{1.5cm} the \emph{worklist} $W$: a list of arcs of $G$. \\
\\
Subroutine Revise($(x_0,x_1),i$) \\
Input: an arc $(x_0,x_1) \in E(G)$, an index $i \in \{0,1\}$. \\
\hspace{0.5cm}  change = false \\
\hspace{0.5cm}  for each $u_i$ in $L(x_i)$ \\
\hspace{1cm}      If there is no $u_{1-i} \in L(x_{1-i})$ such that $(u_0,u_1) \in E(H)$ then \\
\hspace{1.5cm}        remove $u_i$ from $L(x_i)$ \\
\hspace{1.5cm}        change = true \\
\hspace{1cm}      end if \\
\hspace{0.5cm}  end for \\
\hspace{0.5cm}  If change = true then \\
\hspace{1cm}      For all arcs $(z_0,z_1) \in E(G)$ with $z_0=x_i$ or $z_1=x_i$ add $(z_0,z_1)$ to $W$ \\
\hspace{1cm}      If $L(x_i) = \emptyset$ then {\bf reject} \\
\hspace{0.5cm}  end if \\
\\
Set $L(x) := V(H)$ for all $x \in V(G)$. \\
Set $W := E(G)$ \\
Do \\
\hspace{0.5cm}  remove an arc $(x_0,x_1)$ from $W$ \\
\hspace{0.5cm}  Revise($(x_0,x_1),0$) \\
\hspace{0.5cm}  Revise($(x_0,x_1),1$) \\
while $W \neq \emptyset$
\end{tabular}}
\end{center}
\caption{The AC-3 implementation of the arc-consistency procedure for $\Csp(H)$.}
\label{fig:AC-3}
\end{figure}

The idea of AC-3 is to maintain a \emph{worklist}, which contains a list of
arcs $(x_0,x_1)$ of $G$ that might help to remove a value from $L(x_0)$ or $L(x_1)$.
Whenever we remove a value from a list $L(x)$, we add all arcs that are in $G$ incident to $x$.
Note that then any arc in $G$ might be added at most $2 |V(H)|$ many times to the worklist, which is a constant
in the size of $G$. Hence, the while loop of the implementation is iterated for at most a linear number of times.
Altogether, the running time is linear in the size of $G$ as well.

\paragraph{Arc-consistency for pruning search.}
Suppose that $H$ is such that 
$\AC$ does not solve $\Csp(H)$.
Even in this situation 
the arc-consistency procedure might be useful for \emph{pruning the search space} in exhaustive approaches to solve $\Csp(H)$. In such an approach we might use the arc-consistency procedure as a
subroutine as follows. Initially, we run $\AC_H$
on the input instance $G$. If it computes an empty list, we reject.
Otherwise, we select some vertex $x \in V(G)$, 
and set $L(x)$ to $\{u\}$ for some $u \in L(x)$. 
Then we proceed recursively with the resulting lists.
If $\AC_H$ now detects an empty list, we backtrack,
but remove $u$ from $L(x)$.
Finally, if the algorithm never detects an empty list, we end up with singleton lists for each vertex $x \in V(G)$, which gives rise to a homomorphism from $G$ to $H$.

%TODOS: 
% 1) provide references
% 2) say that this behaves well in practice + provide references
% 3) explain why it is good to have DYNAMIC algorithms and data
% structures for AC.

%\paragraph{Arc-consistency as a decision procedure for H-colouring}
%\paragraph{When is the arc-consistency procedure complete?}
%\paragraph{For which $H$ solves arc-consistency the H-colouring problem?}

\begin{exercises}
\exercise[Blau]\label{exe:source-44} Adapt the arc-consistency procedure to precolored $\Csp(H)$.
%Prove that for core digraphs $H$,
%if $\AC_H$ solves $\Csp(H)$, then the adaptation also solves
%precolored $\Csp(H)$.
% Solution sketch: Run arc consistency with the initial list of every precoloured vertex replaced by
% its prescribed singleton. Perform the usual support deletions thereafter. Soundness is unchanged,
% and for a core template the standard completeness argument applies whenever ordinary arc consistency
% solves the uncoloured problem.
\exercise[Rot]\label{exe:source-45} Let $G$ and $H$ be finite digraphs, let $x \in V(G)$, and let $L(x)$ be the list computed by the arc-consistency procedure. Show that $L(x)$ is preserved by all polymorphisms of $H$.
\label{exe:ac-pol}
% Solution sketch: At the fixed point, take $b_1,\ldots,b_k\in L(x)$ and a polymorphism $f$. Every
% deletion rule is witnessed by a constraint lacking support. Applying $f$ coordinatewise to supports
% for the $b_i$ gives a support for $f(b_1,\ldots,b_k)$, so this value cannot have been deleted. Hence
% $L(x)$ is preserved by every polymorphism.
\end{exercises}

\subsection{The Powerset Graph}
\label{sect:PH}
For which $H$ does the Arc-Consistency procedure  solve $\Csp(H)$? In this section we present an elegant and effective 
characterisation of 
those finite digraphs $H$ where $\AC$ solves $\Csp(H)$, found by Feder and Vardi~\cite{FederVardi}. 

\begin{definition}
For a digraph $H$, the {\em powerset graph} $P(H)$ 
is the digraph whose vertices are non-empty subsets of $V(H)$ and
where two subsets $U$ and $V$ are joined by an arc if the following holds: 
\begin{itemize}
\item for every vertex $u\in U$, there exists a vertex $v\in V$ 
such that $(u,v) \in E(H)$, and
\item for every vertex $v\in V$, there exists a vertex $u\in U$ 
such that $(u,v) \in E(H)$. % (see Figure~\ref{fig-example}).
\end{itemize}
\end{definition}

The definition of the powerset graph resembles the arc-consistency algorithm,
and indeed, we have the following lemma
which describes the correspondence.  

\begin{lemma}\label{lem:ac}
$\AC_H$ accepts $G$ if and only if $G \to P(H)$. 
\end{lemma}
\begin{proof}
Suppose first that $\AC_H$ accepts $G$. For $u \in V(G)$, let $L(u)$ be the list derived at the final stage of the algorithm. Then by definition of $E(P(H))$,
the map $x \mapsto L(x)$
is a homomorphism from $G$ to $P(H)$.

Conversely, suppose that $f \colon G \to P(H)$ is a homomorphism. We prove by induction over the execution of 
$\AC_H$ that for all $x \in V(G)$ the elements of $f(x)$ are never removed from $L(x)$. To see that, let $(a,b) \in E(G)$ be arbitrary. Then $(f(a),f(b)) \in E(P(H))$,
and hence for every $u \in f(a)$ there exists
a $v \in f(b)$ such that $(u,v) \in E(H)$. By inductive assumption, $v \in L(b)$,
and hence $u$ will not be removed from
$L(a)$. The other rule is symmetric. This concludes the inductive step. 
\end{proof}

\begin{theorem}\label{thm:ac}
Let $H$ be a finite digraph. Then $\AC$ solves $\Csp(H)$ if and only if $P(H)$ homomorphically maps to $H$.
\end{theorem}
\begin{proof}
Suppose first that $\AC$ solves $\Csp(H)$. Apply $\AC_H$ to $P(H)$. 
Since $P(H) \rightarrow P(H)$, the previous lemma shows that $\AC_H$ does not reject
$P(H)$. Hence, $P(H) \rightarrow H$ by assumption. 

Conversely, suppose that $P(H) \rightarrow H$. If $\AC_H$ rejects a digraph $G$ then $G \not \rightarrow H$.
If $\AC_H$ does accept $G$, then the lemma asserts that $G \rightarrow P(H)$. 
Composing homomorphisms, we obtain that $G \rightarrow H$. 
\end{proof}

\begin{observation}\label{obs:ac}
Let $H$ be a core digraph. Note that if $P(H)$ homomorphically maps to $H$, then there also exists a homomorphism that maps $\{x\}$ to $x$
for all $x \in V(H)$ (here we use the assumption that $H$ is a core!). 
We claim that in this case 
the precoloured CSP for $H$ can be solved
by the modification of $\AC_H$ which
starts with $L(x) := \{c(x)\}$ for all $x \in V(G)$
in the domain of the precolouring function $c$, instead of $L(x) := V(H)$. 
This is a direct consequence of the proof of Theorem~\ref{thm:ac}. 
If the modified version of $\AC_H$
solves the precoloured $\Csp$ for $H$,
then the classical version of $\AC_H$
solves $\Csp(H)$. Hence, it follows that 
the following are equivalent: 
\begin{itemize}
\item $\AC$ solves $\Csp(H)$;
\item the above modification of $\AC_H$
solves the precoloured $\Csp$ for $H$;
\item $P(H) \to H$. 
\end{itemize}
\end{observation}

Note that the condition given in Theorem~\ref{thm:ac}
can be used to decide algorithmically whether AC solves $\Csp(H)$, because it suffices to 
test whether $P(H)$ homomorphically maps to $H$.
Such problems about deciding properties
of $\Csp(H)$ for given $H$ are often
called algorithmic \emph{meta-problems}. 
A naive algorithm for the above test
would be to first construct $P(H)$, and then 
to search non-deterministically for 
a homomorphism from $P(H)$ to $H$,
which puts the meta-problem for solvability
of $\Csp(H)$ by $\AC$ into the complexity class $\NExpTime$ (\emph{Non-deterministic Exponential Time}). 
This can be improved.

\begin{proposition}\label{prop:power-set-exptime}
There exists a deterministic exponential time algorithm that tests for a given finite core digraph $H$ 
whether $P(H)$ homomorphically maps to $H$. 
\end{proposition}
\begin{proof}
%Since $H$ is a core, there exists a homomorphism from $P(H)$ to $H$
%if and only if there exists such a homomorphism that maps $\{x\}$ to $x$ for all $x \in V(H)$. 
We first explicitly construct $P(H)$, 
and then apply $\AC_H$ to $P(H)$. 
If $\AC_H$ rejects, then there is
certainly no homomorphism from $P(H) \to H$ by the properties of $\AC_H$, and we 
return `false'. 
If $\AC_H$ accepts, then we cannot
argue right away that $P(H)$ homomorphically maps to $H$, since 
we do not know yet whether $\AC_H$
is correct for $\Csp(H)$.

But here is the trick. What we do in this case
is to pick an arbitrary $x \in V(P(H))$, and remove all but one value $u$ from $L(x)$, and continue with the
execution of $\AC_H$. If $\AC_H$
then derives the empty list, we try the same  
with another value $u'$ from $L(x)$.
If we obtain failure for all values 
of $L(x)$, then clearly there is no homomorphism from $P(H)$ to $H$,
and we return `false'. 
Otherwise, if $\AC_H$ does not derive the empty list after removing all values but $u$ from $L(x)$, we continue with another element $y$ of $V(P(H))$, setting $L(y)$ to $\{v\}$ for some $v \in L(y)$. We repeat this procedure until at the end
we have constructed a homomorphism 
from $P(H)$ to $H$. In this case we return `true'.

If $\AC_H$ rejects for some $x \in V(P(H))$ when $L(x) = \{u\}$ 
for all possible $u \in V(H)$,
then the adaptation of $\AC_H$ for the precoloured CSP 
would have given an incorrect
answer for the previously selected variable 
(it said yes while it should have said no).  By Observation~\ref{obs:ac}, this means that $P(H)$ does \emph{not} 
homomorphically map to $H$. 
Again, we return `false'. 
\end{proof}

The computational complexity of deciding for a given digraph $H$ whether
$P(H) \to H$ is not known; 
%the best lower bound is NP-hardness~\cite{MetaChenLarose}.  
% BUT ONLY FOR BINARY STRUCTURES,
% NOT FOR DIGRAPHS. 
%~\cite{bib-dalmau99+,bib-feder93+}. 
%The answer to the following question is not yet known; 
we refer to~\cite{MetaChenLarose} for related questions and results. 

\begin{question}
What is the computational complexity to decide for a
given core digraph $H$ whether $P(H) \to H$? 
Is this problem in P? 
\end{question}

\begin{figure}
\begin{center}
\includegraphics[scale=.4]{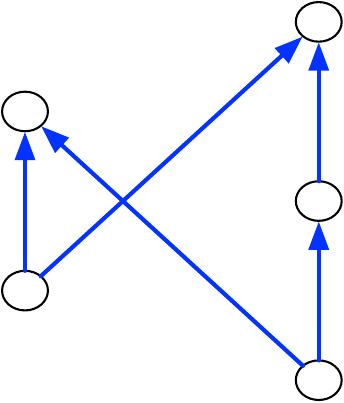}
\end{center}
\caption{The graph from Exercise~\ref{exe:T3T2}.}
\label{fig:T3T2}
\end{figure}

\begin{exercises}
%\item Adapt the arc-consistency procedure to precolored $\Csp(H)$. Prove that for core digraphs $H$,
%if $\AC_H$ solves $\Csp(H)$, then the adaptation also solves
%precolored $\Csp(H)$.
%
\exercise[Blau]\label{exe:source-46} Show that if $G$ and $H$ are homomorphically equivalent, then $P(G)$ and $P(H)$ are also homomorphically equivalent.
% Solution sketch: A homomorphism $h:G\to H$ induces $P(h):P(G)\to P(H)$ by $S\mapsto h(S)$. The two
% support conditions in the definition of the powerset graph are preserved under images. Apply this
% construction to homomorphisms in both directions.
\exercise[Blau]\label{exe:source-48} Solve the problems from Exercise~\ref{exe:source-47} for infinite digraphs $H$.
% Solution sketch: The finite proofs using only a fixed height map remain valid for the first
% assertion. Finiteness is essential for the loop criterion: the two-way infinite directed path is
% acyclic, but its full vertex set has both forward and backward support and therefore is a loop of
% its powerset digraph. This also supplies counterexamples to the remaining finite formulations.
\exercise[Blau]\label{exe:source-51} \label{exe:powergraph-loop}
Let $H$ be a finite digraph. Show that $P(H)$ contains
a loop if and only if $H$ contains a directed cycle.
%Let $H$ be a finite digraph. Show that if the set graph $P(H)$ of $H$ does not contain a loop, then $H$ is acyclic.
% Solution sketch: A directed cycle $C$ gives a loop of $P(H)$ at the set $V(C)$, since every cycle
% vertex has a predecessor and successor in that set. Conversely, if a nonempty finite set $S$ is a
% loop of $P(H)$, repeatedly choose a successor in $S$; finiteness forces a repeated vertex and hence
% a directed cycle in $H$.
\exercise[Blau]\label{exe:source-52} \label{exe:powergraph-inf-digraph}
Show that the previous statement is false for infinite digraphs $H$.
%\item % from my book
%Let $G$ and $H$ be finite digraphs that are homomorphically incomparable
%and suppose that CSP$(G)$ is NP-hard
%or CSP$(H)$ is NP-hard.
%Show that CSP$(G \uplus H)$ is also NP-hard.
% Comes already earlier in contrapositive!
\exercise[Rot]\label{exe:source-47} \label{exe:balanced}
Recall that a digraph is called \emph{balanced} if it homomorphically maps to a directed path.
Let $H$ be a finite digraph. Prove or disprove:
\begin{itemize}
\item if $H$ is balanced, then $P(H)$ is balanced;
% true
\item if $H$ is an orientation of a tree,
then $P(H)$ is an orientation of a forest;
% false
\item $P(H) \to H$ if and only if $H$ is acyclic.
% false
\end{itemize}
% Solution sketch: The first assertion is true: compose the height homomorphism $H\to\vec P_n$ with
% the induced map $P(H)\to P(\vec P_n)$ and use the natural level map. The second is false for a small
% oriented fork, whose powerset graph contains an undirected cycle. The third is false as stated:
% suitable directed cycles satisfy $P(H)\to H$ although $H$ is not acyclic.
\exercise[Rot]\label{exe:source-50} \label{exe:T3T2} Can the CSP for the digraph depicted in Figure~\ref{fig:T3T2} be solved by the arc-consistency procedure?
% Solution: yes, should have finite duality!
%\item Show that if $H$ is a subgraph of $G$, then $P(H)$
%is a subset of $P(G)$.
% too simple, but as an oral exercise
% in course.
%\item % kam schon vorher!
%Show that $(\{\vec P^3\},\vec T_3)$ is a duality pair.
\exercise[Orange]\label{exe:source-49} \label{exe:unbal} Up to isomorphism, there is only one unbalanced cycle $H$ on four vertices that is a core and not the
directed cycle. Show that AC does not solve
$\Csp(H)$.
% Solution sketch: Write down the unique four-vertex unbalanced core cycle and execute arc consistency
% on the standard cycle obstruction obtained by repeating its two possible local orientations. Every
% list retains local support, but the net-length condition around the cycle is inconsistent, so the
% instance is accepted by AC although it has no homomorphism to $H$.
\end{exercises}

\subsection{Tree Duality}
\label{sect:td}
Another mathematical notion that is closely related to the arc-consistency procedure is \emph{tree duality}. 
The idea of this concept is that
for some digraphs $H$, 
we can show that there is no homomorphism from
a digraph $G$ to $H$ by exhibiting
a \emph{tree obstruction}  in $G$:
this is an orientation of a finite tree which homomorphically maps to $G$, but not to $H$. 
This is relevant in the context of the arc-consistency procedure because of the following lemma. 

\begin{lemma}\label{lem:ac-td}
Let $H$ and $G$ be finite digraphs and suppose that $\AC_H$ rejects $G$. Then there exists a tree obstruction in $G$. 
\end{lemma} 

\begin{proof}
By assumption, the arc-consistency procedure applied to $G$
eventually derives the empty list for some vertex of $G$. %$u \in V(G)$. 
We use the computation of the procedure 
to construct an orientation $T$ of a tree, 
following the exposition in~\cite{LLT}. 
When deleting a vertex $u \in V(H)$ 
from the list of a vertex $x \in V(G)$, 
we define an orientation of a rooted tree $T_{x,u}$
with root $r_{x,u}$ such that 
\begin{enumerate}
\item there is a homomorphism from $T_{x,u}$ to $G$ mapping $r_{x,u}$ to $x$;
\item there is no homomorphism from $T_{x,u}$ to $H$ mapping $r_{x,u}$ to $u$.
\end{enumerate}

%The vertex $u$ is deleted from the list of $x$ because we found
%an arc $(x_0,x_1) \in E(G)$ with $x_i = x$ for some $i \in \{0,1\}$ 
%such that there is no arc $(u_0,u_1) \in E(H)$ with $u_i = u$
%and $u_{1-i}$ in the list of $x_{1-i}$. 
%We describe how to construct the tree $T$ for the case that $i=0$, i.e., for the case that there is an arc $(x,y) \in E(G)$
%such that there is no arc $(u,v) \in E(H)$ where $v \in L(y)$;
%the construction for $i=1$ is analogous.

Assume that the vertex $u$ is deleted from the list of $x$ because we found an arc $(x,y) \in E(G)$ such that there is no arc $(u,v) \in E(H)$ with $v \in L(y)$; if it was deleted because of an arc $(y,x) \in E(G)$ the proof follows with the obvious changes. 

%Let $p,q$ be vertices and $(p,q)$ an edge of $T_{x,u}$,
%and select the root $r_{x,u} = p$.
If there is no $v \in V(H)$ such that $(u,v) \in E(H)$, then 
we define $T_{x,u}$ to be the tree that just contains an arc $(p,q)$ with root $r_{x,u} = p$;
clearly, 
% then
%we are done, since 
$T_{x,u}$ satisfies property (1) and (2).
Otherwise, for every arc $(u,v) \in E(H)$ the vertex $v$ 
has already been removed
from the list $L(y)$, and hence by induction 
$T_{y,v}$ having properties (1) and (2) is already defined.
We then add a copy of $T_{y,v}$ to $T_{x,u}$,
contract all the roots of all copies into one vertex $q$, 
 and finally add an arc from the root vertex $r_{x,u}$ to $q$. 
%contract the vertex
%$r_{y,v}$ with $q$. 

We verify that the resulting orientation of a tree $T_{x,u}$
satisfies (1) and (2). 
For every $v \in V(H)$ such that $(u,v) \in E(H)$,   
let $f_v$ be the homomorphism from $T_{y,v}$ mapping
$r_{y,v}$ to $y$, which exists due to (1).
The common extension of all the maps $f_v$ to $V(T_{x,u})$ 
that map $r_{x,u}$ to $x$ is a homomorphism from $T_{x,u}$ to $G$, 
and this shows that (1) holds for $T_{x,u}$.
Suppose for contradiction that there exists a homomorphism $h$ from $T_{x,u}$ to $H$ that maps $r_{x,u}$ to $u$. Let $v = h(q)$;
then $h$ restricts to a homomorphism from $T_{y,v}$ to $H$, a contradiction. 
This shows that (2) holds for $T_{x,u}$.

When the list $L(x)$ of some vertex $x \in V(G)$ becomes empty,
we can construct an orientation of a tree $T$ by contracting the roots of all
$T_{x,u}$ into a vertex $r$. 
We then find a homomorphism from $T$ to $G$
by mapping $r$ to $x$ and extending the homomorphism independently
on each $T_{x,u}$. But any homomorphism from $T$ to $H$ must map $r$ to some element $u \in V(H)$, 
and hence there is a homomorphism from $T_{x,u}$ to $H$
that maps $r_{x,u}$ to $u$, a contradiction.
\end{proof}

Lemma~\ref{lem:ac-td} also has a converse.

\begin{lemma}\label{lem:td-ac} Let $H$ and $G$ be finite digraphs such that there exists an orientation of a tree which maps homomorphically to $G$, but not to $H$. 
Then $\AC_H$ rejects $G$. 
\end{lemma} 
\begin{proof}
We prove the contrapositive, and suppose that $\AC_H$ accepts $G$. 
By Lemma~\ref{lem:ac}, this implies that 
$G \to P(H)$. We have to show that every orientation $T$ of a tree that maps to $G$ also maps to $H$. Let $f$ be a homomorphism from $T$ to $P(H)$. 
We construct a homomorphism $g$ from $T$ to $H$ inductively as follows. 
Let $x$ be 
any vertex of $T$, and define $g(x)$ to be some vertex from $f(x)$. 
Suppose inductively that we have already defined $g$ for a vertex $y$ at distance $i$ from $x$ in $T$ such that $g(y) \in f(y)$, and that we want to define it for a neighbour $y'$ of $y$ at distance
$i+1$ from $x$ in $T$. 
Since $T$ is an orientation of a tree, either $(y,y') \in E(T)$ or $(y',y) \in E(T)$. 
In case that $(y,y') \in E(T)$,
we have $(f(y),f(y')) \in P(H)$. 
Since $u := g(y) \in f(y)$ we get the existence of 
$v \in f(y')$ such that $(u,v) \in E(H)$
by the definition of $P(H)$. 
We then set $g(y') := v$.
In case that $(y',y) \in E(T)$ we proceed analogously. 
By construction, the mapping $g$ is a homomorphism from $T$ to $H$.
%sequence $f_0, \dots, f_n$, for $n=|V(T)|$, where $f_i$ is a homomorphism 
%from the subgraph of $T$ induced by the vertices at distance at most $i$
%to $x$ in $T$, and $f_{i+1}$ 
%is an extension of $f_i$ for all $0 \leq i < n$. 
%The mapping $f_0$ maps $x$ to some vertex from $f(x)$. 
%Suppose inductively that we have already defined $f_i$.
%Let $y$ be a vertex at distance $i+1$ from $x$ in $T$.
%Since $T$ is an orientation of a tree, there is a unique $y' \in V(T)$
%of distance $i$ from $x$ in $T$
%such that $(y,y') \in E(T)$ or $(y',y) \in E(T)$. 
%Note that $u=f_i(y')$ is already defined.
%In case that $(y',y) \in E(T)$,
%there must be a vertex $v$ in $f(y)$
%such that $(u,v) \in E(H)$, since $(f(y'),f(y))$ must be an arc in $P(H)$, and by the definition of $P(H)$.
%We then set $f_{i+1}(y) = v$.
%In case that $(y,y') \in E(T)$ we can proceed analogously. 
%By construction, the mapping $f_n$ is a homomorphism from $T$ to $H$.
\end{proof} 

For easy reference, we combine the previous two lemmas into one.

\begin{corollary}\label{cor:ac-td}
Let $H$ and $G$ be finite digraphs. Then $\AC_H$ accepts $G$ if and only if every orientation of a tree which homomorphically maps to $G$ also maps to $H$. 
\end{corollary}

For some digraphs $H$, to check whether there is a homomorphism from a given digraph $G$ to $H$, it suffices to check for tree obstructions in $G$.  
This is formalised in the following definition. 

\medskip
%The idea is that for digraphs $H$ having tree duality, 
\begin{definition}~\label{def:td}
A digraph $H$ has \emph{tree duality} if
there exists a (not necessarily finite) set $\cal N$ of orientations of finite trees such that for all digraphs $G$ there is a homomorphism from $G$ to $H$ if and only if no digraph in $\cal N$ homomorphically maps to $G$. 
\end{definition}

%Note that the set $\cal N$ is not required to be finite.

We refer to the set $\cal N$ in Definition~\ref{def:td} as an
\emph{obstruction set} for $\Csp(H)$. 
Note that no $T \in {\mathcal N}$ homomorphically 
maps to $H$. 
The pair $({\cal N},H)$ is called a \emph{duality pair}.
We have already encountered such an obstruction set in Exercise~\ref{exe:t2},
where $H=T_2$, and ${\cal N}= \{\vec P_2\}$.
In other words, $(\{\vec P_2\},T_2)$ is a duality pair.
Other duality pairs are $(\{\vec P_3\},T_3)$ (Exercise~\ref{exe:p3-dual-t3}),
and $(\{Z_1,Z_2,\dots\},\vec P_2)$ (Exercise~\ref{exe:zigzag-dual-p2}).

Theorem~\ref{thm:td} is a surprising link between 
the completeness of the arc-consistency
procedure, tree duality, and the powerset graph, 
and was discovered 
by Feder and Vardi~\cite{FederVardiSTOC} 
in the more general context of
constraint satisfaction problems. 
%A formulation and proof of the statement can
%for digraphs can be
%found in~\cite{HNBook}.

\begin{theorem}\label{thm:td}
Let $H$ be a finite digraph. Then the following are equivalent.
\begin{enumerate}
\item $H$ has tree duality;
\item $P(H)$ homomorphically maps to $H$;
\item $\AC$ solves  
$\Csp(H)$; 
\item For every finite digraph $G$, if every orientation of a tree that homomorphically maps to $G$ also homomorphically maps to $H$,
then $G$ homomorphically maps to $H$.
\end{enumerate}
\end{theorem}

% The proof of 4 implies 1 can be intuitively re-done by adapting
% the corresponding proof in the "jelly-fish" paper, as pointed out
% by Victor in our conversation about the final Datalog paper version.

\begin{proof}
The equivalence $2 \Leftrightarrow 3$ has 
been shown in the previous section.
We show $3 \Rightarrow 1$, 
$1 \Rightarrow 4$, and $4 \Rightarrow 2$. 

\medskip 
$3 \Rightarrow 1$: 
Suppose that $\AC$ solves $\Csp(H)$. We have to show that $H$ has tree
duality. Let $\cal N$ be the class of all orientations of finite trees
that do not homomorphically map to $H$. 
If a finite digraph $G$ does not homomorphically map to $H$, by assumption $\AC_H$ rejects $G$, so by Lemma~\ref{lem:ac-td} there exists  $T \in \cal N$ that homomorphically maps to $G$. 
Hence, $({\mathcal N},H)$ is a duality pair. 

\medskip 
$1 \Rightarrow 4$: 
If $H$ has an obstruction
set ${\mathcal N}$ consisting of orientations of trees, and if $G$ does not homomorphically map to $H$, there exists an orientation 
of a tree $T \in {\mathcal N}$ that maps to $G$
but not to $H$. 

%$1 \Rightarrow 2$: Suppose $H$ has tree duality, and let
%$\cal N$ be the tree obstructions from Definition~\ref{def:td}.
%Let $G$ be a digraph, and suppose that every tree that homomorphically maps to $G$ also homomorphically maps to $H$. We
%have to show that $G$ homomorphically maps to $H$.
%No member of $\cal N$ homomorphically maps to $H$: otherwise
%by the definition of tree duality 
%$H$ would not be homomorphic to $H$, a contradiction.
%So, none of the orientations of trees in $\cal N$ homomorphically maps
%to $G$, and by tree duality, $G$ homomorphically maps to $H$.

\medskip 
$4 \Rightarrow 2$: To show that $P(H)$ homomorphically
maps to $H$, it suffices to prove that every orientation $T$ of 
a tree that homomorphically maps to $P(H)$ also homomorphically maps to $H$.
This is the content of Lemma~\ref{lem:td-ac}. 
%$3 \rightarrow 4$: Now, suppose that there is a homomorphism $f$ from $P(H)$ to $H$,
%and let $G$ be an instance of $\Csp(H)$.
%We have to show that if the arc-consistency procedure does not
%derive an empty list for all vertices $u \in V(G)$, then
%there is a homomorphism from $G$ to $H$. We claim that the mapping $g$ 
%that assigns to $x \in V(G)$ the list $L(x)$ of the
%final stage of the arc-consistency algorithm defines a 
%homomorphism from $G$ to $P(H)$. Suppose otherwise that
%there is an arc $(x,y) \in E(G)$ such that $(g(x),g(y)) \notin E(P(H))$.
%Then there is a $u \in g(x)$ such that there is no $v \in g(y)$ 
%with $(u,v) \in E(H)$, or a $v \in g(y)$ such that there is no $u \in g(x)$
%with $(u,v) \in E(H)$. In the first case, the arc-consistency procedure
%would have removed $u$ from $L(x)$, in the second case it would have
%removed $v$ from $L(y)$, a contradiction in both cases.
%Composing the homomorphism $g$ from $G$ to $P(H)$ and
%the homomorphism $f$ from $P(H)$ to $H$, we find a homomorphism
%from $G$ to $H$.
\end{proof}

\begin{exercises}
%\item \label{exe:T3T2} Can the CSP for the digraph depicted in Figure~\ref{fig:T3T2} be solved by the arc-consistency procedure?
% Solution: yes, should have finite duality!
%\item Show that if $H$ is a subgraph of $G$, then $P(H)$
%is a subset of $P(G)$.
% too simple, but as an oral exercise
% in course.
%\item % kam schon vorher!
%Show that $(\{\vec P^3\},\vec T_3)$ is a duality pair.
%\item \label{exe:powergraph-loop}
%Let $H$ be a finite digraph. Show that $P(H)$ contains
%a loop if and only if $H$ contains a directed cycle.
%Let $H$ be a finite digraph. Show that if the set graph $P(H)$ of $H$ does not contain a loop, then $H$ is acyclic.
%\item \label{exe:powergraph-inf-digraph}
%Show that the previous statement is false for infinite digraphs $H$.
%\item % from my book
%Let $G$ and $H$ be finite digraphs that are homomorphically incomparable
%and suppose that CSP$(G)$ is NP-hard
%or CSP$(H)$ is NP-hard.
%Show that CSP$(G \uplus H)$ is also NP-hard.
% Comes already earlier in contrapositive!
\exercise[Blau]\label{exe:source-53} \label{exe:trees} Show that an orientation of a tree homomorphically maps to $H$
if and only if it homomorphically maps to $P(H)$. %(Hint: folgt direkt aus Lemma~\ref{lem:tc-ad}).
% Solution sketch: The forward implication follows from the singleton homomorphism $H\to P(H)$. For
% the converse, root the tree. Starting at the leaves, choose an element from each assigned nonempty
% subset so that it is compatible with the already chosen value of its parent; the two support
% conditions defining arcs of $P(H)$ make the induction possible.
%\item Does Exercise~\ref{exe:k2} remain true for directed graphs?
% No, unless P = NP, look at orientations of
% trees that are hard, but they always map to K_2.
\exercise[Rot]\label{exe:source-54} \label{exe:ac-on-trees} Let $H$ be a finite digraph. Then $\AC_H$ rejects an orientation of a tree $T$ if and only if there is no homomorphism from $T$ to $H$ (in other words, $\AC$ solves $\Csp(H)$ if the input is restricted to orientations of trees).
% Solution sketch: Root $T$ and process it from the leaves. A value survives at a vertex exactly when
% every child subtree has a compatible surviving value. This is precisely the deletion rule of arc
% consistency on a tree. Thus the root list is nonempty exactly when the choices can be extended
% recursively to a homomorphism $T\to H$.
\exercise[Orange]\label{exe:source-55} Show that there is a linear-time algorithm that tests whether a given
orientation of a tree is a core.
% (the author thanks Florian Starke and Lea B\"ander for the idea for this exercise).
% Simply run AC -- see my paper with Bulin Starke Wernthaler.
% Solution sketch: For each vertex $v$ test whether the tree maps to the induced subgraph obtained by
% deleting $v$. Arc consistency is exact on tree inputs, and a proper endomorphic image exists exactly
% when one of these tests succeeds. Leaf-processing data can be shared between the tests, yielding a
% linear-time implementation.
\exercise[Schwarz]\label{exe:source-56} Show that the core of an orientation of a tree can be computed
in polynomial time.
\end{exercises}

\subsection{Totally Symmetric Polymorphisms}
\label{sect:ts}
There is also a characterisation of the power of 
the arc-consistency procedure which is based on
polymorphisms.

\begin{definition}
An operation $f \colon D^k \rightarrow D$ is called 
\emph{totally symmetric} if 
$$f(x_1,\dots,x_k)=f(y_1,\dots,y_k) \text{ whenever }\{x_1,\dots,x_k\} = \{y_1,\dots,y_k\}.$$
\end{definition}

\begin{example}
The operation $(x_1,\dots,x_k) \mapsto \text{minimum}(x_1,\dots,x_k)$ is totally symmetric.
\end{example}
\begin{example}
The majority operation $m \colon \{0,1\}^3 \to \{0,1\}$ given by $$m(x,x,y) = m(x,y,x) = m(y,x,x) = x$$ for all $x \in \{0,1\}$ is
\begin{itemize}
\item not totally symmetric because 
$0 = m(0,0,1) \neq m(0,1,1) = 1$;
\item is \emph{symmetric} in the sense that 
$m(x_1,x_2,x_3) = m(x_{\alpha(1)},x_{\alpha(2)},x_{\alpha(3)})$
for every permutation $\alpha$ of $\{1,2,3\}$. \qedhere
\end{itemize}
\end{example}

\begin{theorem}[from~\cite{DalmauPearson}]\label{thm:totally-symmetric}
Let $H$ be a finite digraph. Then the following are equivalent.
\begin{enumerate}
\item $P(H)$ homomorphically maps to $H$;
\item $H$ has totally symmetric polymorphisms of all arities;
\item $H$ has a totally symmetric polymorphism of 
arity $2 |V(H)|$.
\end{enumerate} 
\end{theorem}
\begin{proof}
$1. \Rightarrow 2.$:
Suppose that $g$ is a homomorphism from $P(H)$ to $H$, and let $k \in {\mathbb N}$. 
Let $f$ be defined by $f(x_1,\dots,x_k) := g(\{x_1,\dots,x_k\})$.
If $(x_1,y_1),\dots,(x_k,y_k) \in E(H)$, then $\{x_1,\dots,x_k\}$
is adjacent to $\{y_1,\dots,y_k\}$ in $P(H)$, and hence 
$(f(x_1,\dots,x_k),f(y_1,\dots,y_k))\in E(H)$. Therefore, $f$ is a polymorphism of $H$, and it is clearly totally symmetric.

The implication $2. \Rightarrow 3.$ is trivial. 
To prove that $3. \Rightarrow 1.$, 
suppose that $f$ is a totally symmetric polymorphism of arity $2 |V(H)|$. 
Let $g \colon V(P(H)) \rightarrow V(H)$ be defined by 
$$g(\{x_1,\dots,x_n\}) := f(x_1,\dots,x_{n-1},x_n,x_n,\dots,x_n)$$ which is well-defined because $f$ is totally symmetric.
Let $(U,W) \in E(P(H))$, and let $x_1,\dots,x_p$ be an enumeration
of the elements of $U$, and $y_1,\dots,y_q$ be an enumeration of the elements of $W$. The properties of $P(H)$ imply that there are $y_1',\dots,y_p' \in W$
and $x_1',\dots,x_q' \in U$
such that $(x_1,y_1'), \dots, (x_p,y_p') \in E(H)$
and $(x_1',y_1), \dots, (x_q',y_q) \in E(H)$.
Since $f$ preserves $E$, 
$$g(U)=g(\{x_1,\dots,x_p\}) = f(x_1,\dots,x_p,x_1',\dots,x_q',x_1,\dots,x_1)$$ is adjacent to 
\begin{align*}
g(W) & =g(\{y_1,\dots,y_q\})= f(y_1',\dots,y_p',y_1,\dots,y_q,y_1',\dots,y_1') \; . \qedhere
\end{align*}
\end{proof}

Given Theorem~\ref{thm:totally-symmetric},
it is natural to ask whether there exists a $k$ so that the existence of a totally symmetric polymorphism of arity $k$ implies totally symmetric polymorphisms of all arities. 
The following example %from~\cite{MetaChenLarose} 
shows that this is not the case. 

\begin{example}\label{expl:ts-p}
For every prime $p \geq 3$, 
the digraph $\vec C_p$ clearly does not have
a totally symmetric polymorphism of arity $p$:
if $f \colon \{0,\dots,p-1\}^p \to \{0,\dots,p-1\}$ is a totally symmetric operation, then
$f(0,1,\dots,p-1) = f(1,\dots,p-1,0)$, and hence $f$ does not preserve the edge relation. 
On the other hand, if $n<p$ then 
$\vec C_p$ has 
the totally symmetric polymorphism 
$$f(x_1,\dots,x_n) := |S|^{-1} \sum_{x \in S} x \mod p$$
where $S = \{x_1,\dots,x_n\}$. 
(Note that $|S| < p$ and hence has a multiplicative inverse.) 
The operation is clearly totally symmetric; 
the verification that it preserves the edge relation of $\vec C_p$ is Exercise~\ref{exe:ts-p}.  
\end{example}

\subsection{Semilattice Polymorphisms}
Some digraphs have a single binary polymorphism that generates operations satisfying the conditions
in the previous theorem. A binary operation $f \colon D^2 \rightarrow D$ 
is called \emph{commutative} if it satisfies $$f(x,y)=f(y,x) \text{ for all } x,y \in D.$$
It is called \emph{associative} if it satisfies $$f(x,f(y,z))=f(f(x,y),z) \text{ for all } x,y,z \in D.$$

\begin{definition}\label{def:semilattice}
A binary operation is called a \emph{semilattice operation} 
if it is associative, commutative, and idempotent.
\end{definition}

Examples of semilattice operations are functions from $D^2 \rightarrow D$ defined as 
$(x,y) \mapsto \min(x,y)$; here the minimum is taken with respect to any fixed linear order of $D$.

\begin{theorem}\label{thm:semilattice}
%Let $H$ be a finite digraph. If $H$ has a semilattice polymorphism, then $P(H) \to H$. 
%If $H$ is a core, then 
%$P(H) \to H$ if and only if $H$ is homomorphically
%equivalent to a directed graph with a semilattice
%polymorphism. 
Let $H$ be a finite digraph. 
Then $P(H) \to H$ if and only if $H$ is 
homomorphically equivalent to a digraph with a semilattice polymorphism. 
\end{theorem}
\begin{proof}
Suppose first that $P(H) \to H$. Thus, $H$ and $P(H)$ are homomorphically
equivalent, and it suffices to show that $P(H)$ has a semilattice
polymorphism. The mapping $(X,Y) \mapsto X \cup Y$ is
clearly a semilattice operation; we claim that 
it preserves the edges of $P(H)$. 
Let $(U,V)$ and $(A,B)$ be edges in $P(H)$. Then for every $u \in U$ there is a $v \in V$ such that $(u,v) \in E(H)$, and for
every $u \in A$ there is a $v \in B$ such that $(u,v) \in E(H)$. 
Hence, for every $u \in U \cup A$ there is a $v \in V \cup B$
such that $(u,v) \in E(H)$. Similarly, we can verify that 
for every $v \in V \cup B$ there is a $u \in U \cup A$ such that
$(u,v) \in E(H)$. This proves the claim.

For the converse, suppose that $H$ is homomorphically equivalent to a digraph 
$G$ with a semilattice polymorphism $f$.
Let $h$ be the homomorphism from $H$ to $G$. 
The operation $(x_1,\dots,x_n) \mapsto f(x_1,f(x_2,f(\dots,f(x_{n-1},x_n) \dots)))$ is a totally symmetric polymorphism of $G$. Then Theorem~\ref{thm:totally-symmetric} implies that $P(G) \to G$.
The map $S \mapsto \{h(u) \mid u \in S\}$
is a homomorphism from $P(H)$ to $P(G)$. 
%Henc, $P(G) \to H$ since $G \to H$. Finally, observe that as $H$ is a subgraph of $G$, the graph
%$P(H)$ is an induced subgraph of $P(G)$. 
Therefore, $P(H) \to P(G) \to G \to H$, as desired. 
\end{proof}

By verifying the existence of semilattice polymorphisms for a concrete class of digraphs,
we obtain the following consequence.

\begin{corollary}
$\AC$ solves $\Csp(H)$ if
$H$ is an orientation of a path.
\end{corollary}
\begin{proof}
Suppose that $1,\dots,n$ are the vertices of $H$ such that
either $(i,i+1)$ or $(i+1,i)$ is an arc in $E(H)$ for
all $i < n$. It is straightforward to verify that the mapping $(x,y) \mapsto \min(x,y)$ 
is a polymorphism of $H$. The statement now follows from Theorem~\ref{thm:semilattice}.
\end{proof}

% JEDESMAL VERTUE ICH MICH, WENN ICH DIESEN
% BEWEIS AN DER TAFEL VORRECHNEN SOLL!!
% REPLACED THIS PROOF BY PROOF BASED ON semilattice min!
%\begin{proof}
%Suppose that $1,\dots,n$ are the vertices of $H$ such that
%either $(i,i+1)$ or $(i+1,i)$ is an arc in $E(H)$ for
%all $i < n$. We claim that 
%the mapping $f \colon V(P(H)) \rightarrow V(H)$ defined
%by $f(S) = \min(S)$ is a homomorphism from $P(H)$ to $H$.

%Let $(U,W)$ be an arc in $P(H)$, and suppose for contradiction
%that $(\min(U),\min(W))$ is not an arc in $H$.
%By the property of $P(H)$
%there must be an element $w$ of $W$ such that $(\min(U),w) \in E(H)$,
%and an element $u$ of $U$ such that $(u,\min(W)) \in E(H)$.
%Note that due to the way in which the vertices are numbered,
%if $(k,l)$ is an edge of $E(H)$ then $k-1 \leq l \leq k+1$.
%Hence,
%$$ \min(U) \geq w+1 \geq \min(W) \geq u+1 \geq \min(U) \; ,$$
%where the second and fourth inequality use the assumption that $(\min(U),\min(W))$ is not an arc in $H$. Therefore, $u=v$ and $\min(U) = \min(W)$. But then
%$(\min(U),w)$ and $(w,\min(U))=(u,\min(W))$ 
%are edges in opposite directions in $H$, which is impossible in an orientation of a path.
%The corollary now follows
%from Theorem~\ref{thm:td}.
%\end{proof}

We want to remark that there are orientations of trees $H$ with an
NP-complete H-colouring problem (the smallest ones have 20 vertices~\cite{otrees}; see Figure~\ref{fig:hard-trees}). It can be shown (using a condition that will be presented in Section~\ref{sect:4ary}) that this digraph does not have tree-duality, without any complexity-theoretic assumptions.
% BEGIN INLINED SOURCE: trees.tex

\def\scale{0.4}
\def\hdist{1cm}
\def\vdist{0.8cm}
\begin{figure}
\centering

\begin{tikzpicture}[scale=1]
% Level 0
\node[bullet] (15) at (3,0) {};
% Level 1
\node[bullet] (16) at (2,1) {};
\node[bullet] (10) at (5,1) {};
\node[bullet] (14) at (3,1) {};
\node[bullet] (5) at (0,1) {};
% Level 2
\node[bullet] (17) at (1,2) {};
\node[bullet] (9) at (5,2) {};
\node[bullet] (7) at (4,2) {};
\node[bullet] (13) at (3,2) {};
\node[bullet] (4) at (0,2) {};
\node[bullet] (0) at (2,2) {};
% Level 3
\node[bullet] (6) at (3,3) {};
\node[bullet] (3) at (0,3) {};
\node[bullet] (1) at (2,3) {};
\node[bullet] (11) at (5,3) {};
\node[bullet] (8) at (4,3) {};
\node[bullet] (18) at (1,3) {};
% Level 4
\node[bullet] (2) at (0,4) {};
\node[bullet] (19) at (1,4) {};
\node[bullet] (12) at (5,4) {};
\path[->,>=stealth']
(16) edge (17)
(16) edge (13)
(17) edge (18)
(10) edge (9)
(9) edge (11)
(9) edge (8)
(7) edge (6)
(7) edge (8)
(13) edge (6)
(4) edge (3)
(3) edge (2)
(1) edge (2)
(15) edge (14)
(14) edge (13)
(11) edge (12)
(0) edge (6)
(0) edge (1)
(5) edge (4)
(18) edge (19)
;
\end{tikzpicture}
\caption{One of the smallest orientations of a tree $H$ such that CSP$(H)$ is NP-complete (assuming P $\neq$ NP; all orientations of trees with fewer vertices can be solved by 
path-consistency~\cite{otrees}).}
%(up to transposition).}
\label{fig:hard-trees}
\end{figure}

% END INLINED SOURCE: trees.tex
%\begin{figure}
%\begin{center}
%\includegraphics[scale=0.5]{Jana.pdf} 
%\end{center}
%\caption{An orientation of a tree $H$ with an NP-complete $H$-colouring problem~\cite{Jana}.}
%\label{fig:jana}
%\end{figure}

%The importance of the concept of the set graph arises
%from the fact that if $2^{\cal H}$ homomorphically maps to $H$, 
%then the CSP for ${\cal H}$ can
%be solved by the arc-consistency algorithm~\cite{FederVardi}.

\begin{exercises}
\exercise[Rot]\label{exe:source-57} Does the following digraph have tree duality?
$$(\{0,1,2,3,4,5\}; \{(0,1),(1,2),(0,2),(3,2),(3,4),(4,5),(3,5),(0,5)\})$$
% Yes, draw it, the core is T3!
% there s subtle pont here about AC_H
% vs AC_core(H)....
%\item Adapt the arc-consistency procedure to precolored $\Csp(H)$. Prove that for core digraphs $H$,
%if $\AC_H$ solves $\Csp(H)$, then the adaptation also solves
%precolored $\Csp(H)$.
\exercise[Rot]\label{exe:source-58} % aus der Nachklausur fuer MPRI 2009/2010
Show that AC solves CSP$(T_n)$, for every $n \geq 1$.
%\item \label{exe:unbal} Up to isomorphism, there is only one unbalanced cycle $H$ on four vertices that is a core and not the
%directed cycle. Show that AC does not solve
%$\Csp(H)$.
%\item \label{exe:T3T2} Can the CSP for the digraph depicted in Figure~\ref{fig:T3T2} be solved by the arc-consistency procedure?
% Solution: yes, should have finite duality!
%\item Show that if $H$ is a subgraph of $G$, then $P(H)$
%is a subset of $P(G)$.
% too simple, but as an oral exercise
% in course.
%\item % kam schon vorher!
%Show that $(\{\vec P^3\},\vec T_3)$ is a duality pair.
%\item % from my book
%Let $G$ and $H$ be finite digraphs that are homomorphically incomparable
%and suppose that CSP$(G)$ is NP-hard
%or CSP$(H)$ is NP-hard.
%Show that CSP$(G \uplus H)$ is also NP-hard.
% Comes already earlier in contrapositive!
%\item \label{exe:trees} Show that an orientation of a tree homomorphically maps to $H$
%if and only if it homomorphically maps to $P(H)$. %(Hint: use parts of the proof of
%Theorem~\ref{thm:td}).
\exercise[Rot]\label{exe:source-60} Does Exercise~\ref{exe:k2} remain true for directed graphs?
% No, unless P = NP, look at orientations of
% trees that are hard, but they always map to K_2.
% Solution sketch: No. There are orientations of trees with NP-complete CSP, while every orientation
% of a tree maps to $K_2$ after forgetting its levels modulo two. Thus the directed analogue would
% imply a polynomial algorithm for an NP-complete problem.
\exercise[Rot]\label{exe:source-59} \label{exe:ts-p} Prove the final statement
in Example~\ref{expl:ts-p}.
% Solution of Albert:
%f(a_1            a_2          ...    a_n) = a
%f(a_1+1        a_2+1     ...    a_n+1) = b.
%Let S_1 be the set of values occurring in the first row, S_2 the set of v. occur. in the second row.
%Note that |S_1| = |S_2|:= t. (If you have two values that are equal in the first line, then  the corresponding ones are equal in the second line).
%Let a = K s  (mod p) , where s is the sum of the elements in S_1, and K is s.t. Kt = 1 (mod p).
%We have b = K (s + t) (mod p) = a + 1 (mod p).
%\item \label{exe:ac-on-trees} Let $H$ be a finite digraph. Then $\AC_H$ rejects an orientation of a tree $T$ if and only if there is no homomorphism from $T$ to $H$ (in other words, $\AC$ solves $\Csp(H)$ if the input is restricted to orientations of trees).
%\item Show that there is a linear-time algorithm that tests whether a given \\
%orientation of a tree is a core.
%\vspace{-1.7cm}
%\begin{flushright}
%\includegraphics[scale=.3]{Schwarz.jpg}
%\end{flushright}
%\vspace{-.8cm}
% (the author thanks Florian Starke and Lea B\"ander for the idea for this exercise).
% Simply run AC -- see my paper with Bulin Starke Wernthaler.
%\item Show that the core of an orientation of a tree can be computed \\
%in polynomial time.
%\item \label{exe:semilattice} % schon im Haupttext.
%Show that CSP$(H)$ can be solved in polynomial time when $H$
%has a polymorphism $f$ that satisfies $f(x,y)=f(y,x)$, $f(x,x)=x$ and $f(x,f(y,z))=f(f(x,y),z)$ for all $x,y,z \in V(H)$.
\end{exercises}

\ignore{
\paragraph{Solutions.}
\begin{itemize}
\item Solution to Exercise~\ref{exe:powergraph-loop}:
If $H$ contains a directed cycle with vertices $v_1,\dots,v_n$,
then the vertex $\{v_1,\dots,v_n\}$ has a loop in $P(H)$.
If $P(H)$ has a loop on a vertex $S$, then for every $u \in S$
there is a $v \in S$ such that $(u,v) \in E(H)$. Hence, we can find a
subset of vertices of $S$ that form a cycle in $H$. (Source of the exercise: mb)
\item Solution to Exercise~\ref{exe:powergraph-inf-digraph}: 
Let $H$ be $(\mathbb Z, \it succ)$ where {\it succ} 
is the binary relation relating $x,y$ iff $x=y+1$.
\item Solution to Exercise~\ref{exe:balanced}: Let $H$ be (picture).
%$\Rnode{A}{\BULLET}\qquad\Rnode{B}{\BULLET}\qquad\Rnode{C}{\BULLET}\qquad\Rnode{D}{\BULLET} \ncline{->}{A}{B} \ncarc[arcangle=-40]{->}{C}{B} \ncarc[arcangle=40]{->}{C}{D}$. 
Then $P(H)$ contains (picture).
%$\Rnode{A}{\BULLET}\qquad\Rnode{B}{\BULLET}\qquad\Rnode{C}{\BULLET}\qquad\Rnode{D}{\BULLET} \ncline{->}{A}{B} \ncarc[arcangle=-40]{->}{C}{B} \ncarc[arcangle=40]{->}{C}{D} \ncarc[arcangle=70]{->}{A}{D}$.
\item Solution to Exercise~\ref{exe:trees}: The difficult direction is
proved verbatim in the implication from 1 to 2 in the proof of Theorem~\ref{thm:td}. For the opposite direction, observe that for any digraph $T$
the singleton subsets of $V(T)$ induce in $P(T)$ a subgraph isomorphic to  $T$.
\end{itemize}
}

\section{The Path-consistency Procedure} % and $k$-consistency}
\label{sect:PC}
The path-consistency procedure is a well-studied generalization of 
the arc-consistency procedure from artificial intelligence. The path-consistency procedure is also known as the pair-consistency check algorithm 
in the graph theory literature.

Many CSPs that cannot be solved by the arc-consistency procedure
can still be solved in polynomial time by the path-consistency procedure.
The simplest examples are $H=K_2$ (see Exercise~\ref{exe:k2}) 
and $H=\vec C_3$ (see Exercise~\ref{exe:c3}).
The idea is to maintain 
a list of pairs from $V(H)^2$ for each
pair of elements from $V(G)$ (similarly to the arc-consistency procedure, where 
we maintained a list of vertices from $V(H)$ for each vertex in $V(G)$).
We successively remove pairs from these lists when the pairs
can be excluded \emph{locally}.
Some authors maintain a list only for each pair of \emph{distinct} vertices of $V(G)$, and they
refer to our (stronger) variant as the 
\emph{strong path-consistency procedure}.  
Our procedure (where vertices need not be distinct) has the advantage that it is at least
as strong as the arc-consistency procedure, because the lists $L(x,x)$ 
and the rules of the path-consistency procedure for $x=y$ 
simulate the rules of the arc-consistency procedure.

\begin{figure}[h]
\begin{center}
\fbox{
\begin{tabular}{l}
PC$_H(G)$ \\
Input: a finite digraph $G$. \\
Data structure: for all $x,y \in V(G)$ a list $L(x,y)$ of elements of $V(H)^2$ \\
\\
Initialise $L(x,y) := V(H)^2$ for all $x,y \in V(G)$. \\
For each $(x,y) \in V(G)^2$ \\
\hspace{0.5cm} If $(x,y) \in E(G)$ then \\
\hspace{1cm} $L(x,y) := L(x,y) \cap E(H)$ \\
\hspace{1cm} $L(y,x) := L(y,x) \cap \{(b,a) \mid (a,b) \in E(H) \}$ \\
\hspace{0.5cm} If $x=y$ then $L(x,y) := L(x,y) \cap \{(u,u) \mid u \in V(H)\}$. \\
Do \\
\hspace{0.5cm} For all vertices $x,y,z \in V(G)$: \\
\hspace{1cm} For each $(u,w) \in L(x,z)$: \\
%\hspace{1.5cm} If there is no homomorphism $f$ from $G[\{u,v,w\}]$ to $H$ \\
%\hspace{1.5cm} such that $f(u)=x,f(v)=y$,$(f(u),f(w)) \in L(u,w)$, and $(f(w),f(v)) \in L(w,v)$ then \\
\hspace{1.5cm} If there is no $v \in V(H)$ such that $(u,v) \in L(x,y)$
and $(v,w) \in L(y,z)$ then \\
\hspace{2cm} Remove $(u,w)$ from $L(x,z)$ \\
\hspace{1cm} If $L(x,z)$ is empty then {\bf reject} \\
Loop until no list changes
\end{tabular}}
\end{center}
\caption{The (strong) path-consistency procedure for $\Csp(H)$.}
\label{fig:pc}
\end{figure}

In Subsection~\ref{ssect:majority} we will see many 
examples of digraphs $H$ where the path-consistency procedure 
solves the $H$-colouring problem, but the arc-consistency procedure does not.
The greater power of the path-consistency procedure comes at the
price of a bigger worst-case running time:
while the arc-consistency procedure has linear-time implementations,  
the best known implementations of the path-consistency procedure
require cubic time in the size of the input (see Exercise~\ref{exe:cubic}).

\begin{remark}
Similarly as for $\AC$, the path-consistency procedure is polynomial even if $H$ is part of the input, in which case we refer to the procedure with $\PC$. 
\end{remark} 

\subsection{The $k$-consistency procedure}
\label{sect:kcons}
The path-consistency procedure can be generalised further 
to the $k$-consistency procedure.
In fact, arc- and path-consistency procedure are just a special case of the $k$-consistency for
$k=2$ and $k=3$, respectively. In other words, for digraphs $H$ the
path-consistency procedure is the $3$-consistency procedure and the
arc-consistency procedure is the $2$-consistency procedure.

The idea of $k$-consistency is to maintain sets of $(k-1)$-tuples from $V(H)^{k-1}$
for each $(k-1)$-tuple from $V(G)^{k-1}$, and to successively remove 
tuples by local inference. If the empty set is derived, the instance is rejected;
otherwise, if eventually no more tuples can be removed from the sets and the empty list has not been derived, the algorithm accepts. 
We say that $\Csp(H)$ is \emph{solved} by the $k$-consistency procedure if it accepts an input graph $G$ if and only if $G \to H$. 
 It is straightforward to generalise also the
details of the path-consistency procedure.
For fixed $H$ and fixed $k$, the running time of the $k$-consistency
procedure is still polynomial in the size of $G$.
But the dependence of the running time on $k$ is clearly
exponential. 

We would like to point out that
path-consistency alias 3-consistency
is of particular theoretical importance, due to the following result.

\begin{theorem}[Barto and Kozik~\cite{BoundedWidth}]\label{thm:width-collaps}
If $\Csp(H)$ can be solved by $k$-consistency for some $k \geq 3$, then $\Csp(H)$ can also be solved 
by 3-consistency.
\end{theorem}

More on the $k$-consistency procedure can be found in Section~\ref{sect:bounded-width}. 

\begin{exercises}
%\item Let $n \geq 3$. For which $k$ is Duplicator winning the existential $k$-pebble game on $\vec P_n$ and $\vec C_n$?
\exercise[Rot]\label{exe:source-61} \label{exe:cubic}
Show that the path-consistency procedure for $\Csp(H)$
can (for fixed $H$) be implemented such that the worst-case running time is cubic in the size of the input digraph. (Hint: use a worklist as in AC-3.)
% Solution sketch: Maintain all currently allowed pairs and a queue of pairs whose support was
% deleted. For fixed $H$ there are $O(|G|^2)$ pair lists of constant size, and processing a deletion
% inspects $O(|G|)$ possible third vertices. Each entry is deleted once, giving $O(|G|^3)$ time.
\exercise[Rot]\label{exe:source-62} Show that if the path-consistency procedure solves $\Csp(H_1)$ and the path-consistency procedure solves $\Csp(H_2)$, then the path-consistency procedure solves $\Csp(H_1 \uplus H_2)$. \label{exe:pc-disj-union}
% Solution sketch: A connected input can map to $H_1\uplus H_2$ only inside one summand. Path
% consistency computes the same pair information as running the two procedures separately and taking
% their disjoint union. Apply this independently to the connected components of the input.
\end{exercises}

%\section{Near-unanimity Operations}
\subsection{Majority Polymorphisms}
\label{ssect:majority}
In this section, we present a powerful criterion that shows
that for certain digraphs $H$ 
the path-consistency procedure solves the $H$-colouring problem.
Again, this condition 
was first discovered in more general form by Feder and Vardi~\cite{FederVardi};
it subsumes many criteria that were studied in artificial intelligence 
and in graph theory before.

%\begin{definition}
%Let $D$ be a set, and $k \geq 3$. A function $f$ from $D^k$ to $D$
%is called a \emph{($k$-ary) near unanimity function} (short, an \emph{nu})
%if $f$ satisfies the following equations, for all $x,y \in D$:
%\begin{align*}
%f(x,\dots,x,y)=f(x,\dots,y,x)=\dots=f(y,x,\dots,x)=x
%\end{align*}
%\end{definition}

\begin{definition}
Let $D$ be a set. An operation $f \colon D^3 \to D$
is called a \emph{majority operation}
if $f$ satisfies the following equations, for all $x,y \in D$:
\begin{align*}
f(x,x,y)=f(x,y,x)=f(y,x,x)=x
\end{align*}
\end{definition}

\begin{example}
As an example, 
%let $T_n$ be the transitive tournament on $n$ vertices.
let $D$ be $\{1,\dots,n\}$, and consider the ternary {\it median} operation, 
which is defined as follows.
For $x,y \in D$, we define $$\median(x,x,y)=\median(x,y,x)=\median(y,x,x) := x.$$ 
If $x,y,z$ are pairwise distinct elements of $D$, 
suppose that $\{x,y,z\}=\{a,b,c\}$, where $a < b < c$.
Then $\median(x,y,z)$ is defined to be $b$.
Note that 
\begin{align*}
\median(x,y,z) & = \min(\max(x,y),\max(x,z),\max(y,z)). \qedhere
\end{align*}
\end{example}

If a digraph $H$ has a polymorphism $f$ that is a majority operation, then $f$ is called a \emph{majority polymorphism of $H$}.

\begin{example}
Let $H$ be the transitive tournament on $n$ vertices, $T_n$.
Suppose the vertices of $T_n$ are the first natural numbers, $\{1,\dots,n\}$,
and $(u,v) \in E(T_n)$ if and only if $u<v$. Then the median operation is a polymorphism
of $T_n$, because if $u_1<v_1$, $u_2<v_2$, and $u_3<v_3$, then clearly
${\it median}(u_1,u_2,u_3) < {\it median}(v_1,v_2,v_3)$.
%This yields a new proof that the $H$-colouring problem for $H=T_n$ is tractable. 
\end{example}

\begin{theorem}[of~\cite{FederVardi}]\label{thm:majority}
Let $H$ be a finite digraph.
If $H$ has a majority polymorphism,
then the $H$-colouring problem can be solved in polynomial time (by the path-consistency procedure).
\end{theorem}

For the proof of Theorem~\ref{thm:majority}
we need the following lemma.
%, which is straightforward to prove
%by induction over the number of times the innermost loop
%is executed in the path-consistency procedure. 
%% (which is of general importance for polymorphisms in general).

\begin{lemma}\label{lem:pres}
Let $G$ and $H$ be finite digraphs. 
Let $f$ be a polymorphism of $H$ of arity $k$ 
and let 
$L := L(x,z)$ be the final list computed by the path-consistency procedure
for $x,z \in V(G)$. Then $f$ \emph{preserves} $L$, i.e., if $(u_1,w_1),\dots,(u_k,w_k) \in L$, then $(f(u_1,\dots,u_k),f(w_1,\dots,w_k)) \in L$.
\end{lemma}

\begin{proof}
Let $(u_1,w_1),\dots,(u_k,w_k) \in L$. 
We prove by induction over the execution of PC$_H$ on $G$ that at all stages of PC$_H$ (i.e., after each synchronous pruning round) 
the pair $(u,w) := (f(u_1,\dots,u_k),f(w_1,\dots,w_k))$ is contained in $L$. Initially, this is true
because $f$ is a polymorphism of $H$.
For the inductive step, let $y \in V(G)$. 
By the definition of the procedure, 
for each $i \in \{1,\dots,k\}$
there exists $v_i$ such that $(u_i,v_i) \in L(x,y)$
and $(v_i,w_i) \in L(y,z)$. By the inductive assumption,
$(f(u_1,\dots,u_k),f(v_1,\dots,v_k)) \in L(x,y)$
and 
$(f(v_1,\dots,v_k),f(w_1,\dots,w_k)) \in L(y,z)$. 
Hence, $(f(u_1,\dots,u_k),f(w_1,\dots,w_k))$ 
will not
be removed in the next step of the algorithm. 
\end{proof}

\begin{proof}[Proof of Theorem~\ref{thm:majority}]
Let $f \colon V(H)^3 \rightarrow V(H)$ be a majority polymorphism of $H$. 
Clearly, if the path-consistency procedure derives the empty list
for some pair $(x,z)$ from $V(G)^2$, then there is no homomorphism
from $G$ to $H$.

Now suppose that after running the path-consistency procedure on $G$ for all pairs $(x,z)$ from $V(G)^2$
the list $L(x,z)$ is non-empty. We have to show that
there exists a homomorphism from $G$ to $H$.
A function $h$ from an induced subgraph $G'$ of $G$ to $H$ is said to \emph{preserve the lists} if
$(h(x),h(z)) \in L(x,z)$ for all $x,z \in V(G')$.
The proof shows by induction on $i$ that every  homomorphism from
a subgraph of $G$ with $i$ vertices that preserves the
lists can be extended to any other vertex in $G$
such that the resulting mapping is a homomorphism to $H$ that again 
preserves the lists.

For the base case of the induction, observe
that for all vertices $x,z \in V(G)$ 
every mapping $h$ from $\{x,z\}$ to $V(H)$ such that
$(h(x),h(z)) \in L(x,z)$ can be extended 
to every $y \in V(G)$ such that $(h(x),h(y)) \in L(x,y)$
and $(h(y),h(z)) \in L(y,z)$ (and hence preserves the lists),
because otherwise the path-consistency procedure
would have removed $(h(x),h(z))$ from $L(x,z)$.

For the inductive step, let $h'$ be any homomorphism 
from a subgraph $G'$ of $G$ on $i \geq 3$ vertices 
to $H$ that preserves the lists, and let $x$ be any
vertex of $G$ not in $G'$. Let $x_1,x_2$, and $x_3$
be some vertices of $G'$, and 
$h_j'$ be the restriction of $h'$ to $V(G') \setminus \{x_j\}$,
for $j \in \{1,2,3\}$. 
By inductive assumption, $h_j'$ can be extended to $x$
such that the resulting mapping $h_j$ is a homomorphism to $H$
that preserves the lists.
We claim that the extension $h$ of $h'$ that maps $x$ to
$f(h_1(x),h_2(x),h_3(x))$ is a homomorphism to $H$ 
that preserves the lists. 

For all $y \in V(G')$, we have to show that $(h(x),h(y)) \in L(x,y)$
(and that $(h(y),h(x)) \in L(y,x)$, which can be shown analogously).
If $y \notin \{x_1,x_2,x_3\}$,
% (in particular, if $y=x$), 
then $h(y)=h'(y)=f(h'(y)$,
$h'(y),h'(y))=f(h_1(y),h_2(y),h_3(y))$, by the properties of $f$.
Since $(h_i(x),h_i(y)) \in L(x,y)$ for all $i \in \{1,2,3\}$,
and since $f$ preserves $L(x,y)$ by Lemma~\ref{lem:pres}, we have $(h(x),h(y)) \in L(x,y)$, and
are done in this case.

Clearly, $y$ can be equal to at most one of $\{x_1,x_2,x_3\}$.
Suppose that $y = x_1$ (the other two cases are analogous).
Since \((h_1(x),h_1(x_2))\in L(x,x_2)\), 
there exists $v \in V(H)$ such that 
\((h_1(x),v)\in L(x,x_1)\) and
\((v,h_1(x_2))\in L(x_1,x_2)\) 
%There must be a vertex $v \in V(H)$ such that $(h_1(x),v) \in L(x,y)$
(otherwise, the path-consistency procedure would have removed
$(h_1(x),h_1(x_2))$ from $L(x,x_2)$).
By the properties of $f$, we have $$h(y)=h'(y)=f(v,h'(y),h'(y))=f(v,h_2(y),h_3(y)).$$
Since $(h_1(x),v),(h_2(x),h_2(y)),(h_3(x),h_3(y)) \in L(x,y)$,
Lemma~\ref{lem:pres} implies that 
$$(h(x),h(y))=(f(h_1(x),h_2(x),h_3(x)),f(v,h_2(y),h_3(y))) \in L(x,y)$$
 and we are done.
We conclude that $G$ has a homomorphism to $H$. 
\end{proof}

\begin{corollary}
The path-consistency procedure solves the $H$-colouring problem
for $H=T_n$.
\end{corollary}

Another class of examples of digraphs having a majority polymorphism are \emph{unbalanced cycles}, i.e., orientations of $C_n$
that do not homomorphically map to a directed path~\cite{FederCycles}.
We only prove a weaker result here.

\begin{proposition}\label{prop:cnmajo}
Directed cycles have a majority polymorphism.
%The directed cylce $\vec C_n$
\end{proposition}

\begin{proof}
%If $u,v,w$ are elements of the directed cycle $\vec C_n$, the majority
Let $\vec C_n$ be a directed cycle. 
Let $f$ be the ternary operation on the vertices of $\vec C_n$ that maps $u,v,w$ to $u$
if $u,v,w$ are pairwise distinct, and otherwise acts as a majority operation.  We claim that $f$ is a polymorphism of $\vec C_n$.
Let $(u,u'), (v,v'), (w,w') \in E(\vec C_n)$ be arcs. If $u,v,w$ are 
all distinct, then $u',v',w'$ are clearly all distinct as well, and
hence $(f(u,v,w),f(u',v',w')) = (u,u') \in E(\vec C_n)$.
Otherwise, if two elements of $u,v,w$ are equal, say $u=v$,
then $u'$ and $v'$ must be equal as well, and hence $(f(u,v,w),f(u',v',w'))=(u,u') \in E(\vec C_n)$.
\end{proof}

\begin{exercises}
\exercise[Blau]\label{exe:source-68} Determine for which $n \geq 1$ the operation $f$ from Proposition~\ref{prop:cnmajo} preserves $T_n$.
% partial solution:
%
\exercise[Blau]\label{exe:source-76} Show that the digraph $({\mathbb Z}; \{(x,y) \; | \; x-y=1\})$
has a majority polymorphism and that its CSP can be solved in polynomial time.
% Solution sketch: The ordinary median on $\mathbb Z$ is a majority polymorphism: translating all
% three entries by one translates their median by one. For the CSP, propagate integer potentials along
% every weak component and reject precisely when two paths impose different differences; otherwise the
% potentials give a homomorphism.
\exercise[Rot]\label{exe:source-63} Show that every orientation of a path has a majority polymorphism.
%\item Find a finite digraph which has a quasi majority polymorphism, but no majority polymorphism.
% Solution sketch: Linearly order the vertices along the underlying path and take the median of three
% vertices. If three arcs are applied coordinatewise, the median of their tails and the median of
% their heads are again consecutive with the same orientation. Thus the median operation is a majority
% polymorphism.
\exercise[Rot]\label{exe:source-64} Show that $C_4$ has a majority polymorphism but $C_6$ does not.
% Flo: 024 kanten zu 133=333, 113=111, 515=555,
% Solution sketch: For $C_4$, identify the graph with $K_2\times K_2$ and take coordinatewise Boolean
% majority. For $C_6$, assume a majority polymorphism and apply it to three suitably rotated edges;
% the majority identities force the image first to be each of two opposite vertices, a contradiction.
\exercise[Rot]\label{exe:source-65} \label{exe:qmaj}
A \emph{quasi majority operation} is an operation from $V^3$ to $V$
satisfying
$$f(x,x,y)=f(x,y,x)=f(y,x,x)=f(x,x,x)$$ for all $x,y \in V$.
\begin{itemize}
\item Show that every digraph with a quasi majority polymorphism is homomorphically equivalent to a digraph with a majority polymorphism.
\item
Use Theorem~\ref{thm:HN} to show that a finite undirected graph $H$
has an $H$-colouring problem that can be solved in polynomial time
if $H$ has a quasi majority polymorphism,
and is NP-complete otherwise.
\end{itemize}
% SOLUTION: Suppose $H$-colouring can be solved in polynomial time. Then
% By Theorem~\ref{thm:HN} and the assumption that N $\neq$ NP
% the graph $H$ must be bipartite. Let $g$ be a homomorphism from $H$
% to $K_2$. It is clear that the graph $K_2$ has a majority operation $f'$.
% Then the operation $f$ defined by $f(x,y,z)=f'(g(x),g(y),g(z))$ is a
% quasi-majority operation.
%
% Conversely, suppose that $H$ has a quasi-majority operation $f$.
% Let $H'$ be the core of $H$, and let $g$
% be the homomorphism from $H$ to $H'$. Since $H'$ is a core, the restriction
% of the endomorphism
% g(f(x,x,x)) to $V(H')$ is an automorphism of $H'$;
% let $g^{-1}$ be its inverse.
% Then the operation $g^{-1}(g(f(x,y,z)))$ is a majority operation of $H'$,
% and hence path-consistency solves the $H'$-colouring problem, which
% equals the $H$-colouring problem.
%
%\item Let $T_n$ be the transitive tournament. Show that there
%is a homomorphism from $(T_n)^3$ to $T_n$ which is a near unanimity operation.
\exercise[Rot]\label{exe:source-66} There is only one unbalanced cycle $H$ on four vertices that is a core and not the
directed cycle (we have seen this digraph already in Exercise~\ref{exe:unbal}).
Show that for this digraph $H$  the $H$-colouring problem can be solved
by the path-consistency procedure.
% Solution sketch: For the unique four-vertex unbalanced core cycle, verify directly the majority
% table obtained by taking the median on each of its two overlapping directed paths. The operation
% preserves all arcs, so the majority correctness theorem for path consistency applies.
\exercise[Rot]\label{exe:source-67} Determine for which $n \geq 1$ there is a linear order on the vertices of $\vec C_n$ such that $\median$ with respect to this linear order is a polymorphism of $\vec C_n$.
% partial solution: works for n = 1,2.
% doesn't work with the standard order for n = 3:
\exercise[Rot]\label{exe:source-70} \label{exe:precol-majority} Modify the path-consistency procedure such that it can
deal with instances
of the precoloured $H$-colouring problem.
Show that if $H$ has a majority polymorphism, then the modified
path-consistency procedure solves the precoloured $H$-colouring problem.
% Solution sketch: Introduce unary singleton constraints for precoloured vertices and initialise the
% pair lists accordingly. The usual minimal-counterexample proof for a majority polymorphism still
% works because singleton relations are preserved by every idempotent majority operation.
\exercise[Rot]\label{exe:source-71} \label{exe:list-majority} Modify the path-consistency procedure such that it can
deal with instances of the list $H$-colouring problem.
Show that if $H$ has a \emph{conservative} majority polymorphism, then the modified
path-consistency procedure solves the list $H$-colouring problem.
% Solution sketch: Treat every input list as a unary constraint. A conservative operation preserves
% every unary subset, so all list relations are preserved by the conservative majority polymorphism.
% The standard path-consistency proof therefore applies without change.
\exercise[Rot]\label{exe:source-72} \label{exe:intervalgraph}
An \emph{interval graph} $H$ is an (undirected)
graph $H=(V;E)$ such that there is an interval $I_x$
of the real numbers for each $x \in V$, and $(x,y) \in E$ if and only if
$I_x$ and $I_y$ have a non-empty intersection.
Note that with this definition interval graphs are necessarily \emph{reflexive}, i.e., $(x,x) \in E$.
Show that the precoloured $H$-colouring problem
for interval graphs $H$ can be solved in polynomial time.
\par\noindent Hint: use the modified path-consistency procedure in Exercise~\ref{exe:precol-majority}.
% Solution sketch: Order the interval endpoints and define a conservative majority by selecting the
% interval whose left endpoint is the median, with a compatible tie rule using right endpoints. The
% Helly property for intervals shows that this operation preserves intersection. Apply
% Exercise~\ref{exe:precol-majority}.
\exercise[Rot]\label{exe:source-77}
Show that $({\mathbb Z}; \neq)$ does not have a majority polymorphism, but a quasi majority polymorphism and that $\Csp({\mathbb Z};\neq)$ can be solved in polynomial time.
% Solution sketch: A majority polymorphism would restrict to a majority operation on every finite
% complete subgraph, contradicting the finite complete-graph argument. On a countably infinite set one
% can choose a quasi-majority operation which returns the repeated value and encodes pairwise-distinct
% triples injectively. A finite graph maps to $(\mathbb Z;\neq)$ exactly when it has no loop, since it
% uses only finitely many colours.
\exercise[Orange]\label{exe:source-73}
% Forward direct: Lemma 4.1 in Feder et all, 2003.
Show that if $H$ is a (reflexive) interval graph,
then $H$ has a conservative majority polymorphism.
%\emph{interval graph}, i.e., $H$ can be represented by intervals of the rational numbers so that two vertices are adjacent if and only if the corresponding intervals intersect.
%[9] T. Feder, P. Hell, and J. Huang, List homomorphisms and circular arc graphs,
%Combinatorica 19 (1999), 487?505.
\exercise[Orange]\label{exe:source-78} Show that the digraph
$H = ({\mathbb Z}; \{(x,y) \; | \; x-y \in \{1,3\} \})$
has a majority polymorphism.
% Solution sketch: Use the median in the natural order on $\mathbb Z$. If $x_i-y_i\in\{1,3\}$ for
% three coordinate pairs, a short case distinction according to the median indices shows that
% $\operatorname{med}(x_1,x_2,x_3)-\operatorname{med}(y_1,y_2,y_3)$ is again $1$ or $3$.
\exercise[Orange]\label{exe:source-81} For any set $V$, the \emph{dual discriminator operation} is the operation $d \colon V^3 \to V$ defined by $d(x,y,z) = y$ if $y=z$, and $x$ otherwise. Note that $d$ is a majority operation. Prove that $E \subseteq V^2$ is preserved by $d$ if and only if there are $A,B \subseteq V$ such that at least one of the following applies:
\begin{itemize}
\item $E = A \times B$,
\item there exists a bijection $\alpha \colon A \to B$ such that $E = \{(a,\alpha(a)) \mid a \in A\}$,
\item there are $a \in A$, and $b \in B$ such that
$E = \{(a,b') \mid b' \in B\} \cup \{(a',b) \mid a' \in A\}$.
\end{itemize}
% Solution sketch: Apply the preservation condition to triples of edges. If two independent edges
% occur, the dual discriminator forces all four corners and iterating gives a rectangle. If no two
% independent edges occur, each component is either the graph of a bijection or all edges are incident
% with one distinguished row or column, yielding the other two forms. Each listed form is checked
% directly to be preserved.
\exercise[Schwarz]\label{exe:source-69} Show that every unbalanced orientation of a cycle
 has a majority polymorphism.
\exercise[Schwarz]\label{exe:source-79} Give a polynomial time algorithm
for the $H$-colouring problem
for the digraph $H$ from Exercise~\ref{exe:source-78}.
% Solution sketch: Associate with every input arc a choice of difference $1$ or $3$. After fixing one
% vertex value per weak component, all other values are path sums. Consistency around a cycle is a
% system of parity and interval constraints on these choices; encode the choices by a 2-SAT instance
% and recover the integer labels from any satisfying assignment.
\exercise[Schwarz]\label{exe:source-80} \label{exe:c2++}
Consider the digraph $C_2^{++}$ depicted in Figure~\ref{fig:c2++} (a so-called \emph{semicomplete digraph}). Show the following statements.\footnote{
The author thanks Florian Starke and Sebastian Meyer for the idea for this exercise.}
\begin{itemize}
\item $\Csp(C_2^{++})$ cannot be solved by the arc-consistency procedure.
\item A finite digraph $G$ homomorphically maps to $C_2^{++}$ if and only if no digraph of the following form maps to $G$: start with any orientation of an odd cycle, and if $(u,v),(w,v)$ are edges on the cycle, append to $v$ an outgoing directed path with two edges.
% one directon s clear snce the graphs clearly
% have no homomorphsm to C_2++.
% Conversely f has no homomorphsm from
%
\item $C_2^{++}$ does not have a majority polymorphism.
\item $\Csp(C_2^{++})$ can be solved by the path-consistency procedure.
% Soluton dea 1: The path consstency procedure may recognse the graphs above (bounded wdth (23)).
%the digraph obtained from $G$ by removing all sinks, and in the resulting graph
\end{itemize}
\begin{figure}
\begin{center}
\includegraphics[scale=.6]{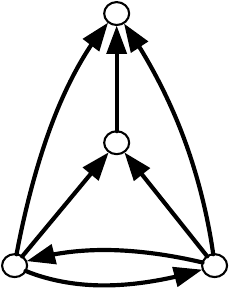}
\end{center}
\caption{The graph $C_2^{++}$ from Exercise~\ref{exe:c2++}.}
\label{fig:c2++}
\end{figure}
\exercise[Weiss]\label{exe:source-74} Let $H$ be a reflexive graph.
%[8] T. Feder and P. Hell, List homomorphisms to reflexive graphs, J Combin
%Theory B 72 (1998), 236?250:
% Note quite, he refers to the paper with
% Huang, but there we have it only
% with complexity theory.
Show that $H$ has a conservative majority polymorphism if and only if $H$ is an
interval graph (see Exercise~\ref{exe:intervalgraph}).
% ChatGPT says:
%"Reflexive graphs with a conservative majority polymorphism are precisely the reflexive bi-arc graphs, which are precisely the interval graphs. A reflexive \(C_4\) already separates the two formulations. See the characterization summarized in "The structure of bi-arc trees."
%Tom\'as Feder, Pavol Hell, Jing Huang 2007.
%\item Show that a graph $G$ (possibly with loops) has a \emph{conservative} majority polymorphism if and only if $G$ is a \emph{bi-arc graph}: these are graphs such that there exists a pair $(N,S)$ of arcs, where $N$ and $S$ are arcs on the unit circle such that all arcs in
%\vspace{-2.1cm}
%\begin{flushright}
%\includegraphics[scale=.3]{Weiss.jpg}
%\hspace{1cm} { }
%\end{flushright}
%\vspace{-.4cm}
%\item Show that a graph $G$ is a bi-arc graph if and only if $G \times K_2$ is a circular arc graph.
%\vspace{-2.1cm}
%\begin{flushright}
%\includegraphics[scale=.3]{Weiss.jpg}
%\hspace{1cm} { }
%\end{flushright}
%\vspace{-.4cm}
%\item Show that a graph $G$ is a bi-arc graph if and only if it is preserved by a near unanmity polymorphism.
%\vspace{-2.1cm}
%\begin{flushright}
%\includegraphics[scale=.3]{Weiss.jpg}
%\hspace{1cm} { }
%\end{flushright}
%\vspace{-.4cm}
\exercise[Weiss]\label{exe:source-75} Let $H$ be an irreflexive graph.
Then $H$ has a conservative majority polymorphism if and only if $H$ is bipartite and its complement is a
\emph{circular arc graph},
 i.e., $H$ can be represented by arcs on a circle so that
 two vertices are adjacent if and only if the corresponding arcs intersect.
% Brewster, Feder, Hell, Huang,
% MacGillivray: Theorem 5.2.
% Forward direct: Lemma 4.1 in Feder et all, 2003.
% Solution sketch: For the forward direction, use conservativity to restrict the majority operation to
% pairs and derive a bipartition; the forbidden configurations obtained from the majority identities
% give a circular-arc model of the complement. Conversely, order the arc endpoints cyclically and
% define the conservative median selector; checking the possible endpoint orders proves that it
% preserves adjacency.
\end{exercises}

\subsection{Testing for Majority Polymorphisms}
In this section we show that the question whether a given digraph has a majority polymorphism can be decided in polynomial time. 
The method that we present is sometimes referred to as a \emph{self-reduction} and can be adapted for several other polymorphism conditions and several other algorithms (see Exercise~\ref{exe:self-red}). 

%\begin{corollary}
%The path-consistency procedure solves the $H$-colouring problem if $H$ is an unbalanced cycle.
%\end{corollary}
%\paragraph{A linear space algorithm}
%The path-consistency procedure requires quadratic space,
%and even though cubic time implementations of the procedure exist,
%applying this procedure

\begin{figure}[h]
\begin{center}
\fbox{
\begin{tabular}{l}
Majority-Test$(H)$ \\
Input: a finite digraph $H$. \\
\\
Let $G := H^3$. \\
For all $u,v \in V(H)$, precolour the vertices $(u,u,v),(u,v,u),(v,u,u),(u,u,u)$ with $u$. \\
If PC$_H(G)$ derives the empty list, then {\bf reject}. \\

For each \(x\in V(G)\): \\
\hspace{0.5cm} For each \((u,u)\in L(x,x)\): \\
\hspace{1cm} Run \(\mathrm{PC}_H(G)\) on a copy \(L_u\) of the current lists with \(L_u(x,x):=\{(u,u)\}\). \\
\hspace{0.5cm} If every such run rejects, then {\bf reject}. \\
\hspace{0.5cm} Otherwise, choose one \(u\) whose run does not reject \\
\hspace{1cm} and replace \(L\) by the resulting lists \(L_u\). \\
%For each $x \in V(G)$ \\
%\hspace{0.5cm} Found-Value := false. \\
%\hspace{0.5cm} For each $(u,u) \in L(x,x)$ \\
%\hspace{1cm} For all $y,z \in V(G)$, let $L'(y,z)$ be a copy of $L(y,z)$.\\
%\hspace{1cm} $L'(x,x) := \{(u,u)\}$. \\
%\hspace{1cm} Run PC$_H(G)$ with the lists $L'$. \\
%\hspace{1cm} If this run does not derive the empty list \\
%\hspace{1.5cm} For all $y,z \in V(G)$, set $L(y,z) := L'(y,z)$.\\
%\hspace{1.5cm} Found-Value := true.\\
%\hspace{0.5cm} End For. \\
%\hspace{0.5cm} If Found-Value = false then {\bf reject}. \\ 
End For. \\
{\bf Accept}.
\end{tabular}}
\end{center}
\caption{A polynomial-time algorithm to find majority polymorphisms.}
\label{fig:majority-test}
\end{figure}

\begin{theorem}
There is a polynomial-time algorithm to decide whether a given digraph $H$ has a majority polymorphism. 
\end{theorem}
\begin{proof}
The pseudo-code of the procedure can be 
found in Figure~\ref{fig:majority-test}. 
Given $H$, we construct a new digraph $G$ as follows. We start from the third power $H^3$,
and precolour all vertices of the form $(u,u,v)$, 
$(u,v,u)$, $(v,u,u)$, and $(u,u,u)$ with $u$. 
Let $G$ be the resulting precoloured digraph. 
Note that there exists a homomorphism from $G$
to $H$ that respects the colours if and only if $H$ has a majority polymorphism.
 
To decide whether $G$ has a homomorphism to $H$, we run the modification of $\PC_H$ for the precoloured $H$-colouring problem 
on $G$ (see Exercise~\ref{exe:precol-majority}). If this algorithm rejects, then we can be sure that there is no homomorphism from $G$ to $H$ that respects the colours, and hence $H$ has no majority polymorphism. Otherwise, we use the same idea as in the proof of Proposition~\ref{prop:power-set-exptime}: 
create a copy $L'$ of the lists $L$. 
Pick $x \in V(G)$ and remove
all but one pair $(u,u)$ from $L'(x,x)$. 
If $\PC_H$ derives the empty list on $L'$ instead of $L$, we
try the same with another pair $(v,v)$ from $L(x,x)$. 

If there exists $x \in V(G)$ such that
$\PC_H$ detects an empty list for all 
$(u,u) \in L(x,x)$ then the adaptation of $\PC_H$
for the precoloured CSP would have given an incorrect answer for the previously selected variable: $\PC_H$ did not detect the empty list even though the input was unsatisfiable.
Hence, $H$ cannot have a majority polymorphism by Theorem~\ref{thm:majority}. 
 
Otherwise, if $\PC_H$ does not derive the empty list after removing all pairs but $(u,u)$ 
from $L(x,x)$, 
we continue with another vertex $y \in V(G)$,
setting $L(y,y)$ to $\{(u,u)\}$ for some $(u,u) \in L(y,y)$. We repeat this procedure; if the algorithm never rejects, then eventually all lists for pairs of the form $(x,x)$ are singleton sets $\{(u,u)\}$;
the map that sends $x$ to $u$ is a homomorphism from $G$ to $H$ that respects the colours. 
In this case we return `\true'.

It is easy to see that the procedure described above has polynomial running time. 
\end{proof}

\begin{exercises}
\exercise[Rot]\label{exe:source-82} Modify the algorithm `Majority-Test' to obtain an algorithm that tests whether a given digraph $H$ has a quasi majority polymorphism.
% Solution sketch: Use the same constraint instance as Majority-Test, but replace idempotence by
% variables for the diagonal endomorphism $x\mapsto f(x,x,x)$ and impose only the quasi-majority
% identities. Solving the resulting fixed-template CSP decides whether the required operation exists.
\end{exercises}

%\paragraph{Near-unanimity Polymorphisms} 
%Majority functions 
%are a special case of so-called \emph{near-unanimity functions}.
%A function $f$ from $D^k$ to $D$ is called 
%a \emph{($k$-ary) near unanimity} (short, an \emph{nu})
%if $f$ satisfies the following equations, for all $x,y \in D$:
%\begin{align*}
%f(x,\dots,x,y)=f(x,\dots,y,x)=\dots=f(y,x,\dots,x)=x
%\end{align*}
%Obviously, the majority operations are precisely the ternary near-unanimity function.
%Similarly as in Theorem~\ref{thm:majority} it can be shown that
%the existence of a $k$-ary nu polymorphism of $H$ 
%implies that the $k$-consistency procedure solves $\Csp(H)$.
%We mention that 
%Theorem~\ref{thm:width-collaps} implies
%that if $H$ be a digraph with a near unanimity polymorphism, then already $\PC$ solves $\Csp(H)$.

% TODO: find examples with k-ary near unanimity
% but no majority, with binary constraints!

%\paragraph{Exercises.}
%\begin{enumerate}
%\setcounter{enumi}{\value{mycounter}}
%\setcounter{mycounter}{\value{enumi}}
%\end{enumerate}

\subsection{Digraphs with a Maltsev Polymorphism}
\label{sect:maltsev}
If a digraph $H$ has a \emph{majority} polymorphism, then the path-consistency procedure solves $\Csp(H)$. How about digraphs $H$ with a \emph{minority} polymorphism of $H$? It turns out that this is an even stronger restriction. 

\begin{definition}
A ternary operation $f \colon D^3 \to D$ is called \begin{itemize}
\item a \emph{minority operation} if it satisfies 
$$ \forall x,y \in D. \, f(y,x,x) = f(x,y,x) =  f(x,x,y) = y$$
\item and a \emph{Maltsev operation} if it satisfies 
$$ \forall x,y \in D. \, f(y,x,x) = f(x,x,y) = y.$$
\end{itemize}
\end{definition}

\begin{example}
%The graph $\vec C_n$ has a Maltsev polymorphism. 
Let $D := \{0,\dots,n-1\}$. Then the operation $f \colon D^3 \to D$ given by
$(x,y,z) \mapsto x-y+z \mod n$ is a Maltsev operation, since $x-x+z=z$ and $x-z+z=x$. For $n \in \{1,2\}$, this is even a minority operation. If $n>2$, this operation is not a minority, since then $1-2+1 = 0 \not\equiv 2 \mod n$. 
Note that $f$ is a polymorphism of $\vec C_n$. To see this, suppose that
$v_1-u_1 \equiv 1 \mod n$,
$v_2-u_2 \equiv 1 \mod n$, and
$v_3-u_3 \equiv 1 \mod n$. Then
\begin{align*}
f(v_1,v_2,v_3) \equiv v_1-v_2+v_3 & \equiv (u_1+1)-(u_2+1)+(u_3+1) \\
& \equiv f(u_1,u_2,u_3)+1 \mod n \, . \qedhere \end{align*}
\end{example}

The following result appeared in 2011. 

\begin{theorem}[Kazda~\cite{Kazda}]
\label{thm:kazda}
If a finite digraph $H$ has a Maltsev polymorphism then
$H$ also has a majority polymorphism. 
\end{theorem}

Hence, for finite digraphs $H$ with a Maltsev polymorphism, the strong path-consistency procedure solves the $H$-colouring problem, and in fact even the 
precoloured $H$-colouring problem. 
Theorem~\ref{thm:kazda} is an immediate
consequence of Theorem~\ref{thm:CEJN} below; to state it, we need the following concepts.  

\begin{definition}
A digraph $G$ is called 
\emph{rectangular} 
if $(x,y), (x',y), (x',y') \in E(G)$ implies that $(x,y') \in E(G)$. %(Draw a picture!) 
\end{definition}

We start with the fundamental observation:
digraphs with a Maltsev polymorphism $m$
are rectangular. This follows immediately from the definition of polymorphisms: we must have
$(m(x,x',x'),m(y,y,y')) \in E(G)$, but $m(x,x',x') = x$
and $m(y,y,y') = y'$, so $(x,y') \in E(G)$. 
The converse does not hold, 
as the following example shows. 

\begin{example}
The digraph $\big (\{a,b,c\};\{(a,a),(a,b),(b,c),(c,c)\}\big )$ is rectangular, %(draw a picture!), 
but has no Maltsev polymorphism $m$.
Indeed, such an $m$ would have to satisfy $m(a,a,c) = c$
and $m(a,c,c) = a$. Note that 
\begin{align*}
(m(a,a,c),m(a,b,c)) & \in E(G) \\
\text{and } (m(a,b,c),m(a,c,c)) & \in E(G), 
\end{align*}
but $G$ has no vertex $x$ such that $(c,x) \in E(G)$ and
$(x,a) \in E(G)$. 
\end{example} 

We are therefore interested in stronger consequences of the existence of a Maltsev polymorphism. 

\begin{definition}
A digraph $G$ is called 
\emph{$k$-rectangular} if whenever $G$ contains directed paths of length $k$ from $x$ to $y$, from $x'$ to $y$, 
and from $x'$ to $y'$, then also from $x$ to $y'$. 
A digraph $G$ is called \emph{totally rectangular} if it is $k$-rectangular for all $k \geq 1$. 
\end{definition}
  
\begin{figure}
\begin{center}
\includegraphics[scale=0.6]{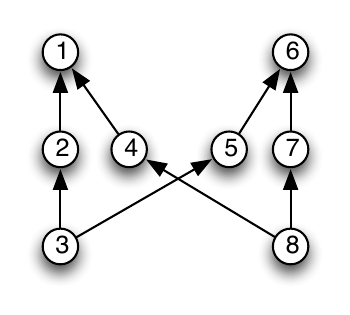} 
\end{center}
\caption{A totally rectangular digraph.}
\label{fig:rect}
\end{figure}

\begin{lemma}\label{lem:rect-easy}
Every digraph with a Maltsev polymorphism $m$ is totally rectangular. 
\end{lemma}
\begin{proof}
%One can easily produce 
%the required directed paths for $k$-rectangularity 
%inductively.  
Let $k \geq 1$, and suppose that $G$ is a digraph with directed paths $(x_0,\dots,x_k)$,
$(y_0,\dots,y_k)$, and $(z_0,\dots,z_k)$
such that $x_k=y_k$ and $y_0=z_0$. 
We have to show that there exists a directed path 
$(u_0,\dots,u_k)$ in $G$ with $u_0=x_0$ and $u_k =z_k$. 
It can be verified that $u_i := m(x_i,y_i,z_i)$ has the desired properties. 
%vertices $x,y,x',y'$ and directed paths of length $k$ from $x$ to $y$, from $x'$ to $y$, and from $x'$ to $y'$. 
%If $k = 1$ then we already observed above that there must be an edge from $x$ to $y'$. 
\end{proof}

An example of a totally rectangular digraph is given in Figure~\ref{fig:rect}. The next lemma points out an important 
consequence of $k$-rectangularity. 

\begin{lemma}%[Cycles and directed paths]
\label{lem:rect-unbalanced-cycle}
Let $G$ be a finite totally rectangular digraph with a cycle of net length $d>0$. 
Then $G$ contains a directed cycle of length $d$. 
%Then there are vertices
%$s,t \in V^G$ and directed paths $Q_1$ and $Q_2$ from $s$ to $t$ in $G$ such that $|Q_1|-|Q_2| = d$. 
\end{lemma}
\begin{proof}
Let \(C\) be a cycle of net length \(d>0\). We argue by induction on
\(k:=|C|\). Decompose \(C\) into maximal directed paths, where each such
path is read in the direction of its arcs. If there is only one such
path, then \(C\) is a directed cycle and \(k=d\).

Otherwise, let \(P\) be a shortest path in this decomposition. Write
\(P\) as a directed path from \(u\) to \(v\), and put
\(\ell:=|P|\). Let \(Q\) be the other directed path in the decomposition
that starts at \(u\), and let \(Q'\) be the other directed path that
ends at \(v\). Thus,
$|Q|,|Q'|\geq \ell$. 
Let \(t\) be the vertex of \(Q\) at distance \(\ell\) from \(u\), and
let \(s\) be the vertex of \(Q'\) at distance \(\ell\) from \(v\).
Consequently, there are directed paths of length \(\ell\) from \(s\)
to \(v\), from \(u\) to \(v\), and from \(u\) to \(t\). By
\(\ell\)-rectangularity, there is a directed path \(R\) of length
\(\ell\) from \(s\) to \(t\).

First suppose that \(Q\neq Q'\). In the cyclic traversal of \(C\), the
path \(P\) is traversed in one direction and the relevant parts of
\(Q\) and \(Q'\) are traversed in the opposite direction. If \(P\) is
traversed backwards, then the portion
\[
 s\longrightarrow v\longrightarrow u\longrightarrow t
\]
consists of the final \(\ell\) edges of \(Q'\), the path \(P\)
traversed backwards, and the first \(\ell\) edges of \(Q\). It has
length \(3\ell\) and net length \(\ell\), and may be replaced by \(R\).
If \(P\) is traversed forwards, the same portion is traversed from
\(t\) to \(s\); it has net length \(-\ell\) and may be replaced by
\(R\) traversed backwards. In either case, we obtain a cycle \(C'\)
with
\[
 |C'|=|C|-2\ell
 \quad\text{and}\quad
 \operatorname{net}(C')=\operatorname{net}(C)=d.
\]
The result follows from the induction hypothesis.

It remains to consider the case \(Q=Q'\). Then \(C\) consists of the
two directed paths \(P\) and \(Q\), both from \(u\) to \(v\). Put
\(q:=|Q|\). Since \(q\geq\ell\) and the net length of \(C\) is positive,
\(Q\) is traversed forwards and \(P\) backwards. Hence
\[
 d=q-\ell>0.
\]

Suppose first that \(q\geq 2\ell\). Then \(t\) occurs on \(Q\) no later
than \(s\), and the portion of \(Q\) from \(t\) to \(s\) is a directed
path of length \(q-2\ell\). Concatenating \(R\) with this portion gives
a directed cycle of length
\[
 \ell+(q-2\ell)=q-\ell=d.
\]

Finally, suppose that \(\ell<q<2\ell\). Then \(s\) occurs before \(t\)
on \(Q\), and the portion \(S\) of \(Q\) from \(s\) to \(t\) has length
\[
 |S|=2\ell-q.
\]
The cycle obtained by traversing \(R\) forwards and \(S\) backwards
has net length
\[
 \ell-(2\ell-q)=q-\ell=d.
\]
Its length is \(3\ell-q\), which is strictly smaller than
\(\ell+q=|C|\), since \(q>\ell\). The induction hypothesis therefore
applies and yields a directed cycle of length \(d\).
\end{proof}

The following is a strengthening 
of Theorem~\ref{thm:kazda};
we only prove that 1 implies 2, and 2 implies 3, which suffices for the already mentioned consequence that for digraphs $H$ with
a Maltsev polymorphism, the path-consistency procedure solves the $H$-colouring problem (cf.~Exercise~\ref{exe:pc-disj-union}). 

\begin{theorem}[Theorem 3.3 and Corollary 4.11 in~\cite{CarvalhoEgriJacksonNiven}]
\label{thm:CEJN}
Let $G$ be a finite digraph. 
Then the following are equivalent. 
\begin{enumerate}
\item $G$ has a Maltsev polymorphism.
\item $G$ is totally rectangular. 
\item If $G$ is acyclic, then the core of $G$ is a directed path. Otherwise, the core of $G$ is a disjoint union of directed cycles. 
%\item $G$ has a Pixley polymorphism. 
\item $G$ has a minority and a majority polymorphism. 
\end{enumerate}
\end{theorem}
\begin{proof}
The implication from 4 to 1 is trivial since every minority operation is in particular a Maltsev operation. The implication from 1 to 2 is Lemma~\ref{lem:rect-easy}. 
For the implication from 2 to 3,
let us assume that $G$ is connected. 
The general case then follows by applying
the following argument to each of its connected components, and the observation that directed paths homomorphically map to longer directed paths and to directed cycles. 

We first consider the case that $G$ is acyclic, and claim that in this case $G$ is balanced, i.e., there exists a surjective homomorphism $h$ from $G $ to $\vec P_n$ for some $n \geq 0$. Otherwise, there exist $u,v \in V(G)$ and two paths $P$ and $Q$  from $u$ to $v$ of different net lengths $\ell_1$ and $\ell_2$
(see Exercise~\ref{exe:net-length-1}). 
Put these two paths together at $u$ and $v$ to form an unbalanced cycle $C$. 
Then Lemma~\ref{lem:rect-unbalanced-cycle} implies that $G$ contains a directed cycle contrary to our assumptions. 

%Now, choose $n$ minimal such that there exists a surjective homomorphism $h$ from $G $ to $\vec P_n$, and fix $u \in h^{-1}(0)$ and $v \in h^{-1}(n)$. 
%Then it is easy to see from total rectangularity that there must exist a path of length $n$ in $G$ from $u$ to $v$, and hence the core of $G$ is $\vec P_n$. 
Fix a surjective homomorphism $h\colon G\to\vec P_n$, and choose
$u\in h^{-1}(0)$ and $v\in h^{-1}(n)$. The case $n = 0$ is trivial. 
Since $G$ is connected, there exists a path $O$ from $u$ to $v$.
We show that every path from $u$ to $v$ which is not directed can be
replaced by a strictly shorter path with the same endpoints. Starting
with $O$ and repeating this replacement, we must eventually obtain a
directed path from $u$ to $v$.

Decompose $O$ into maximal directed subpaths
$P_1,\dots,P_m$, where each $P_i$ is written in the direction of
its arcs. Along $O$, these paths are traversed alternately in the
direction of their arcs and against the direction of their arcs.
Since no edge enters $h^{-1}(0)$ and no edge leaves $h^{-1}(n)$,
the first and the last path in the decomposition are traversed in
the direction of their arcs.

Suppose that $m>1$. We may choose an internal path $P_i$ whose
length $\ell$ is not larger than the lengths of its two neighbours.
Indeed, choose a path of minimum length in the decomposition. If
only $P_1$ has minimum length, then
$|P_2|\leq |P_1|$, since the path cannot descend below level zero.
Similarly, if only $P_m$ has minimum length, then
$|P_{m-1}|\leq |P_m|$, since the path cannot rise above level $n$.
Thus a path of minimum length can always be chosen internal.

First suppose that $P_i$ is traversed against the direction of its
arcs. Write $P_i$ as a directed path of length $\ell$ from $b$ to
$a$. The final $\ell$ edges of $P_{i-1}$ form a directed path from
some vertex $s$ to $a$, and the first $\ell$ edges of $P_{i+1}$
form a directed path from $b$ to some vertex $t$. Hence there are
directed paths of length $\ell$ from $s$ to $a$, from $b$ to $a$,
and from $b$ to $t$. By $\ell$-rectangularity, there is a directed
path of length $\ell$ from $s$ to $t$. Replacing the corresponding
part of $O$, which has length $3\ell$, by this path produces a
shorter path from $u$ to $v$.

Now suppose that $P_i$ is traversed in the direction of its arcs,
and write it as a directed path of length $\ell$ from $a$ to $b$.
The part of $P_{i-1}$ adjacent to $a$ gives a directed path of
length $\ell$ from $a$ to some vertex $s$, and the part of
$P_{i+1}$ adjacent to $b$ gives a directed path of length $\ell$
from some vertex $t$ to $b$. Thus there are directed paths of
length $\ell$ from $t$ to $b$, from $a$ to $b$, and from $a$ to
$s$. By $\ell$-rectangularity, there is a directed path of length
$\ell$ from $t$ to $s$. Traversing this path backwards replaces
the corresponding part of $O$, again reducing its length from
$3\ell$ to $\ell$.

Repeating this shortening procedure eventually yields a path from
$u$ to $v$ whose directed path decomposition has only one member.
It is therefore a directed path. Since $h(u)=0$, $h(v)=n$, and
$h$ increases by one along every arc, this directed path has length
$n$. Write it as
\[
u=w_0,w_1,\dots,w_n=v.
\]
The map from $\vec P_n$ to $G$ that sends $i$ to $w_i$ is a
homomorphism, and its composition with $h$ is the identity on
$\vec P_n$. Consequently, $\vec P_n$ is a retract of $G$. Since
$\vec P_n$ is a core, it is the core of $G$.

Now suppose that $G$ contains a directed cycle. 
If $G$ contains a loop, then $G$ is homomorphically equivalent to $\vec C_1$. We may therefore assume that $G$ is loopless. 
Let $C$ be the shortest directed cycle of $G$. 
We prove that $G$ homomorphically maps to $C$. 
It is easy to see that it suffices
to show that for any two vertices $u,v$ 
of $G$ and for any two paths $P$ and
$Q$ from $u$ to $v$ we have
that their net lengths are congruent modulo $m := |C|$ (see Exercise~\ref{exe:net-length-2}). Suppose for contradiction that there are paths of net length $\ell_1$ and $\ell_2$ from $u$ to $v$ in $G$ such that $d := \ell_1 - \ell_2 \neq 0$ modulo $m$; without loss of generality, $\ell_2 < \ell_1$, so $d > 0$. 
%The paths $P$ and $Q$, appended at $u$ and $v$, form
%a cycle of net length $d$ such that $d$ is not congruent to $0$ modulo $|C|$. 
We can assume that $u$ is an element of $C$, since otherwise we can choose
a path $S$ from a vertex of $C$ to $u$ by connectivity of $G$, and append $S$ to both $P$ and $Q$. We can also assume
that $d < m$ because if not,
we can prepend $C$ to $Q$ to increase
the length of $Q$ by a multiple of $m$,
until $d = \ell_1 - \ell_2 < m$. Lemma~\ref{lem:rect-unbalanced-cycle} then implies that $G$ contains a directed cycle of length $d$, a contradiction to the choice of $C$. 

For the missing implication from 3 to 4, we refer to
%~\cite{Kazda} or~
\cite{CarvalhoEgriJacksonNiven} (Corollary 4.11). 
\end{proof}

%{\bf To be completed here.} We mention the following even more recent than Kazda's result.
%\begin{theorem}[Corollary~4.12 in~\cite{CarvalhoEgriJacksonNiven}]
%A digraph $G$ has a conservative Maltsev polymorphism if and only if $G$ has a conservative
%majority polymorphism. 
%\end{theorem}
%Hence, when $H$ has a conservative Maltsev polymorphism, then strong path consistency solves 
%even the list $H$-colouring problem (see Exercise~\ref{exe:list-majority}). 

% TODO: ideas from the proof.
% In particular, introduce rectangularity!

Rectangularity will be revisited from a universal algebraic perspective in Section~\ref{sect:congr-perm} (Proposition~\ref{prop:rectangularity-variety}).  

\begin{exercises}
\exercise[Blau]\label{exe:source-83} % (MB 2017, from Maltsev Digraphs Paper)
Let $H$ be the digraph $(\{0,1,\dots,6\}; \{(0,1),(1,2),(3,2),(4,3),(4,5),(5,6)\})$. For which $k$ is it $k$-rectangular?
% Solution sketch: Compute the binary path relation $E^k$. The relation $E$ is rectangular. For $k=2$,
% the pairs $(0,2),(4,2),(4,6)$ belong to $E^2$ but $(0,6)$ does not, so rectangularity fails. Since
% the digraph has no directed path of length at least three, $E^k$ is empty and therefore rectangular
% for every $k\geq3$.
\exercise[Blau]\label{exe:source-84} Show that $G = (V,E)$ is rectangular if and only if $E$ is a disjoint union
of sets of the form $A \times B$ where $A,B \subseteq V$.
%\vspace{.2cm}
% Solution sketch: If $E$ is rectangular, two edges lie in the same connected component of the
% associated bipartite graph only if all cross edges are present; hence each component is a biclique
% $A\times B$. The converse follows because three corners in a disjoint union of bicliques necessarily
% lie in one biclique, which contains the fourth corner.
\end{exercises}

\section{Logic}\label{sect:logic}
A \emph{signature} is a  
set of relation and function symbols. 
The relation symbols are
 typically denoted by $R,S,T,\dots$
 and the function symbols are typically denoted by $f,g,h,\dots$;
each relation and function symbol is equipped with an
\emph{arity} from ${\mathbb N}$. Let $\tau$ be a signature.
A \emph{$\tau$-structure} $\bA$ consists of
\begin{itemize}
\item a set $A$ (the {\em domain}, \emph{universe}, or \emph{base set}; we typically use the same letter in the standard font), 
\item a relation $R^\bA \subseteq A^k$ for each relation symbol $R$ of arity $k$ from $\tau$, and 
\item an operation $f^{\bA} \colon A^k \to A$ for each function symbol $f$ of arity $k$ from $\tau$.
\end{itemize} 
Function symbols of arity $0$ are allowed; they are also called \emph{constant symbols} (and the respective operations are called \emph{constants}). 
 In this text it causes no harm to allow structures whose domain is empty (in this case, the signature must not contain constant symbols). A $\tau$-structure $\bA$ is called \emph{finite} if its domain $A$ is finite. 

A \emph{homomorphism} $h$ from a $\tau$-structure $\bA$ 
to a $\tau$-structure $\bB$ is a function from
$A$ to $B$ that \emph{preserves} each relation
and each function: that is, 
 \begin{itemize}
\item if $(a_1,\dots,a_k)$ is in
$R^\bA$, then $(h(a_1),\dots,h(a_k))$ must
be in $R^\bB$;
\item for all $a_1,\dots,a_k \in A$ we have $h(f^{\bA}(a_1,\dots,a_k)) = f^{\bB}(h(a_1),\dots,h(a_k))$. 
\end{itemize}
An \emph{isomorphism} is a bijective homomorphism $h$ such that the
inverse mapping $h^{-1} \colon B \rightarrow A$ that sends $h(x)$ to $x$
is a homomorphism, too. If there exists an isomorphism between two structures $\bA$ and $\bB$, then they are called \emph{isomorphic}, and we write $\bA \simeq \bB$.

A \emph{relational} structure is a $\tau$-structure
where $\tau$ only contains relation symbols,
and an \emph{algebra} (in the sense of universal algebra) is a $\tau$-structure
where $\tau$ only contains function symbols. 
This section is mainly about relational structures; algebras will appear in Section~\ref{sect:ua}. 

Note that in a $\tau$-structure $\bA$, every function symbol of arity $n$ must be defined on all of $A^n$; in some settings, this requirement is not natural and we therefore also define \emph{multi-sorted} structures.

\begin{exercises}
\exercise[Schwarz]\label{exe:source-85} \label{exe:hard-factors-of-easy-product}
Find (binary) structures $\bA$ and $\bB$ such that
$\Csp(\bA \times \bB)$
can be solved
in polynomial time, but $\Csp(\bA)$ and $\Csp(\bB)$ are NP-hard (if you have not already solved the harder Exercise~\ref{exe:product-drops}).
\ignore{ %SOLUTION: 
Let $\tau=\{R,S\}$, where both relation symbols are binary, and let $\bA$ and
$\bB$ have domain $[5]$. Define
\begin{align*}
 R^{\bA}&=\{(x,y)\in\{1,2,3\}^2\mid x\neq y\}, &
 S^{\bA}&=\{(x,y)\in\{4,5\}^2\mid x\neq y\},\\
 R^{\bB}&=S^{\bA}, &
 S^{\bB}&=R^{\bA}.
\end{align*}
The $R$-reduct of $\bA$ is $K_3$ together with two isolated vertices. Hence
$\Csp(\bA)$ is NP-hard. The same argument, using the $S$-reduct, proves that
$\Csp(\bB)$ is NP-hard.

The $R$-relation of $\bA\times\bB$ is $K_3\times K_2$ on
$\{1,2,3\}\times\{4,5\}$, and its $S$-relation is $K_2\times K_3$ on the
disjoint set $\{4,5\}\times\{1,2,3\}$. Both relations are homomorphically
equivalent to $K_2$. Thus, if a variable of an input structure occurs in both
an $R$-constraint and an $S$-constraint, the instance is unsatisfiable.
Otherwise, every connected component of its incidence graph uses only one
relation symbol, and it is satisfiable precisely when its underlying graph is
bipartite. Consequently, $\Csp(\bA\times\bB)$ is solvable in polynomial time.
}
\end{exercises}

\subsection{Multisorted Structures}
\label{sect:multi}
This section is for later reference, and can be skipped at the first reading. Multisorted structures
will be used 
in Section~\ref{sect:abstract-clone} 
%to formalise abstract clones, 
and in Section~\ref{sect:minions}.
% to formalise abstract minions. 

Let $S$ be a set; the elements of $S$ are called \emph{sorts}. 
%We write $S^*$ for the set of words over $S$ (i.e., finite sequences of elements of $S$) and $S^+$ for the set of non-empty words over $S$. 
An \emph{$S$-sorted signature} $\tau$ consists of a set of \emph{function symbols} (typically denoted by $f,g,\dots$) and a set of \emph{relation symbols} (typically denoted by $R,S,\dots$). Each relation symbol $R \in \tau$ of arity $k$ is equipped with a \emph{type} $s \in S^k$, 
and each function symbol $f \in \tau$
of arity $k$ 
 is equipped with a \emph{type} $s \in S^{k+1}$. 
If $\tau$ is an $S$-sorted signature, then an \emph{$S$-sorted $\tau$-structure} $\bM$  consists of 
\begin{itemize}
\item a set $M_s$ for every $s \in S$;  
\item a relation $R^{\bM} \subseteq M_{s_1} \times \cdots \times M_{s_k}$ for each relation symbol $R \in \tau$ of arity $k$ of type  
$(s_1, s_2, \dots,s_k)$;
\item a function $f^{\bM} \colon M_{s_1} \times \cdots \times M_{s_k} \to M_{s_0}$ for each function symbol $f \in \tau$ of arity $k$ and type 
$(s_0,s_1,\dots,s_k)$ (we refer to $s_0$ as the \emph{output sort} of $f$). 
\end{itemize} 

Note that for the one-sorted case, i.e., if $|S|=1$,  we recover the notion of a structure as introduced earlier. Vector spaces or, more generally, modules may be viewed naturally as two-sorted structures; see Example~\ref{expl:module}.
%A natural example of t
%A natural example 
%\begin{example}[Vector spaces]
%TODO. 
%\end{example}

The syntax and semantics of first-order logic over an $S$-sorted signature $\tau$ are defined as follows. Let $V$ be a set; the elements of $V$ are called \emph{variables}. Each variable $x \in V$ is equipped with a sort $s \in S$; we then write $x: s$. 
We require that for every $s \in S$ there are infinitely many variables of sort $s$. 
 
If $x_1,\dots,x_n$ are variables of sort $s_1,\dots,s_n$, respectively, then an
\emph{$S$-sorted $\tau$-term $t$ of type $s = (s_0,s_1,\dots,s_n) \in S^{n+1}$ over the variables $x_1,\dots,x_n$} is defined inductively as follows.
\begin{itemize}
\item $t$ can be of the form $x_i$, for $i \in \{1,\dots,n\}$, in which case it has type $(s_i,s_1,\dots,s_n)$;
\item $t$ can be of the form $f(t_1,\dots,t_\ell)$ (a syntactic object), where $f \in \tau$ is a function symbol of type $(s_0,r_1,\dots,r_\ell)$, for $\ell \in {\mathbb N}$, and for each $i \in \{1,\dots,\ell\}$ the term $t_i$ has type $(r_i,s_1,\dots,s_n)$ over the variables $x_1,\dots,x_n$; in this case, $t$ has type $(s_0,s_1,\dots,s_n)$. The case $\ell=0$ is another base case of the induction and covers terms without variables.
\end{itemize}
As in the case of function symbols of type $s$, we refer to $s_0$ as the \emph{output sort} of $t$.
All $\tau$-terms $t$ over the variables $x_1,\dots,x_n$ are built in this way; we often write $t(x_1,\dots,x_n)$ to indicate that $t$ is a term over these variables.

If $\bM$ is an $S$-sorted $\tau$-structure, $n \geq 1$, $x_1,\dots,x_n \in V$, and $t(x_1,\dots,x_n)$ is a $\tau$-term of type $(s_0,s_1,\dots,s_n)$, then the \emph{term function} $t^{\bM}$ (in the one-sorted case called \emph{term operation}) is the function
$t^{\bM} \colon M_{s_1} \times \cdots \times M_{s_n} \to M_{s_0}$ defined inductively as follows:
\begin{itemize}
\item if $t$ is of the form $x_i$, for $i \in \{1,\dots,n\}$, then $t^{\bM}$ is the function $(a_1,\dots,a_n) \mapsto a_i$;
\item if $t$ is of the form $f(t_1,\dots,t_\ell)$ and $f$ has type $(s_0,r_1,\dots,r_\ell)$, then $t^{\bM}$ is the function
$(a_1,\dots,a_n) \mapsto f^{\bM}(t_1^{\bM}(a_1,\dots,a_n),\dots,t^{\bM}_\ell(a_1,\dots,a_n))$.
\end{itemize}
An \emph{atomic $\tau$-formula over variables $x_1,\dots,x_k$ of sort $(s_1,\dots,s_k)$}
is 
\begin{itemize}
\item an expression of the form $t_1 = t_2$, for $S$-sorted $\tau$-terms $t_1(x_1,\dots,x_k)$ and $t_2(x_1,\dots,x_k)$ 
of type $(s_0,s_1,\dots,s_k)$ and $ (s_0,s_1,\dots,s_k)$, for some $s_0 \in S$;  
\item an expression of the form $R(t_1,\dots,t_l)$,
where $t_i$, for $i \in \{1,\dots,l\}$, is 
a $\tau$-term 
of type $(r_i,s_1,\dots,s_k)$, 
and where $R \in \tau$ has type $(r_1,\dots,r_l)$. 
\end{itemize}
A first-order $\tau$-formula over the variables $x_1,\dots,x_k \in V$ of type 
$(s_1,\dots,s_k)$ 
is defined inductively as one of the following expressions: 
\begin{itemize}
\item an atomic first-order $\tau$-formula over $x_1,\dots,x_k$ of type $(s_1,\dots,s_k)$, 
\item $\phi_1 \wedge \phi_2$ for 
first-order $\tau$-formulas $\phi_1,\phi_2$ over the variables $x_1,\dots,x_k$ of type $(s_1,\dots,s_k)$, 
\item $\neg \phi$ for a first-order $\tau$-formula $\phi$ over the variables $x_1,\dots,x_k$ of type $(s_1,\dots,s_k)$, 
\item $\exists x_0. \phi$ where $x_0 \in V$ and $\phi$ is a first-order $\tau$-formula over the variables $x_0,x_1,\dots,x_k$
of type $(s_0,s_1,\dots,s_k)$. 
\end{itemize} 
If $\bM$ is an $S$-sorted $\tau$-structure, $x_1,\dots,x_k \in V$, and $\phi(x_1,\dots,x_k)$ 
is a $\tau$-formula of type $(s_1,\dots,s_k)$, then $\phi^{\bM}$ is the relation defined as follows. 
\begin{itemize}
\item If $\phi$ is atomic and 
of the form $t_1 = t_2$, then $\phi^{\bM}$ consists of all 
tuples $(a_1,\dots,a_k) \in M_{s_1} \times \cdots \times M_{s_k}$
 such that 
$t_1^{\bM}(a_1,\dots,a_k) = t_2^{\bM}(a_1,\dots,a_k)$. 
\item If $\phi$ is atomic and of the form $R(t_1,\dots,t_l)$ for $S$-sorted $\tau$-terms $t_1,\dots,t_l$ and $R \in \tau$ of type $(r_1,\dots,r_l)$, then
$\phi^{\bM}$ consists of all 
tuples $(a_1,\dots,a_k) \in M_{s_1} \times \cdots \times M_{s_k}$ such that
$$\big (t_1^{\bM}(a_1,\dots,a_k),\dots,t_l^{\bM}(a_1,\dots,a_k) \big ) \in R^{\bM}.$$ 
\item If $\phi$ is of the form $\phi_1 \wedge \phi_2$ for $\tau$-formulas $\phi_1$ and $\phi_2$ over $x_1,\dots,x_k$ of type $(s_1,\dots,s_k)$, 
then $\phi^{\bM} := \phi_1^{\bM} \cap \phi_2^{\bM}$. 
\item If $\phi$ is of the form $\neg \psi$ for some $\tau$-formula $\psi$ over $x_1,\dots,x_k$ of type $(s_1,\dots,s_k)$, then $\phi^{\bM} := (M_{s_1} \times \cdots \times M_{s_k}) \setminus \psi^{\bM}$; 
\item If $\phi$ is of the form $\exists x_0. \psi$ 
 for some $x_0 \in V$ and some 
$\tau$-formula $\psi$  over $x_0,x_1,\dots,x_k$ of type $(s_0,s_1,\dots,s_k)$, then $\phi^{\bM}$ consists of all tuples $(a_1,\dots,a_k) \in M_{s_1} \times \cdots \times M_{s_k}$ such that there exists 
$a_0 \in M_{s_0}$ such that $(a_0,a_1,\dots,a_k) \in \psi^{\bM}$. 
\end{itemize} 
We recover the syntax and semantics of usual first-order logic as the special case of the one-sorted case. 

%It is now straightforward to finish 
%the description of the syntax and semantics of $S$-sorted first-order logic (which I assume is well-known to the reader for the one-sorted case). 

\begin{exercises}
\exercise[Blau]\label{exe:source-86} Show that one can decide in polynomial time whether a given string is an $S$-sorted $\tau$-term over variables $x_1,\dots,x_k$.
% Solution sketch: Parse the string bottom-up. Attach to each variable its prescribed sort and to each
% operation symbol its input and output sorts; accept a node exactly when the sorts of all children
% match the declaration. The parse and all checks take linear, hence polynomial, time.
\exercise[Blau]\label{exe:source-87} Generalise the notion of a homomorphism between $\tau$-structures to $S$-sorted $\tau$-structures.
% Solution sketch: A homomorphism is a family of maps $h_s:A_s\to B_s$, one for every sort $s$, such
% that applying the maps coordinatewise sends every sorted relation tuple of the first structure into
% the corresponding relation of the second structure (and commutes with sorted functions, when
% present).
\end{exercises}

\subsection{Primitive Positive Formulas}
\label{sect:pp}
A first-order $\tau$-formula $\phi(x_1,\dots,x_n)$ is called 
\emph{primitive positive} (in database theory also \emph{conjunctive query}) if it is of the form
$$ \exists x_{n+1},\dots,x_\ell (\psi_1 \wedge \dots \wedge \psi_m)$$
where $\psi_1, \dots, \psi_m$ are \emph{atomic $\tau$-formulas},
i.e., formulas of the form $R(y_1,\dots,y_k)$ with $R \in \tau$ 
and $y_i \in \{x_1,\dots,x_\ell\}$, of the form $y = y'$ for $y,y' \in \{x_1,\dots,x_\ell\}$, or $\top$ and $\bot$ (for \emph{true} and \emph{false}).
As usual, formulas without free variables are called \emph{sentences}. If $\bA$ is a $\tau$-structure and $\phi$ a $\tau$-sentence, then we write $\bA \models \phi$ if $\bA$ satisfies $\phi$
(i.e., $\phi$ holds in $\bA$). 

Note that if we required that all our structures have a non-empty domain, we would not need the symbol $\top$ since we can use the primitive positive sentence $\exists x. \, x=x$ to express it. 
It is possible to rephrase the $H$-colouring problem
and its variants using primitive positive sentences.

\begin{definition}
Let $\bB$ be a structure with a finite relational signature $\tau$. Then \emph{$\Csp(\bB)$} is the 
computational
problem of deciding whether a given primitive positive $\tau$-sentence $\phi$
is true in $\bB$.
\end{definition}

The given primitive positive $\tau$-sentence $\phi$ is also called an
\emph{instance} of $\Csp(\bB)$. 
The conjuncts of an instance $\phi$ are 
called the \emph{constraints} of $\phi$.
A mapping from the variables of $\phi$ to the elements of $B$ that is a satisfying 
assignment for the quantifier-free part of $\phi$ is also called a \emph{solution}
to $\phi$.

\begin{example}[Disequality constraints]\label{expl:basic-ecsp}
Consider the problem $\Csp(\{1,2,\dots,n\};\neq)$.
An instance of this problem can be viewed as an (existentially quantified) set of variables, 
some linked by disequality\footnote{We deliberately use the word \emph{disequality} instead of \emph{inequality}, since we reserve the word \emph{inequality} for the relation $x \leq y$.} constraints.
Such an instance holds in $(\{1,2,\dots,n\};\neq)$ if and only if the graph whose vertices are the variables, and whose edges are the disequality constraints, has a homomorphism to $K_n$. 
\end{example}

The \emph{dichotomy conjecture} of 
Feder and Vardi was that $\Csp(\bB)$ is 
always in P or NP-complete, for every finite structure $\bB$ with finite relational signature; this conjecture was proved by Bulatov~\cite{BulatovFVConjecture} and by
Zhuk~\cite{ZhukFVConjecture}. For a first more informative formulation of their result, see Theorem~\ref{thm:tractability-1}; many more reformulations can be found later in the text. 
Feder and Vardi showed that their conjecture is equivalent to the special case of their conjecture for finite digraphs (see Theorem~\ref{thm:dicho}), 
because for every relational structure $\bB$ there exists a finite digraph $\bH$ such that $\Csp(\bB)$ and $\Csp(\bH)$ are polynomial-time equivalent; this result has later been refined in~\cite{BulinDelicJacksonNiven}.

\begin{exercises}
\exercise[Blau]\label{exe:source-88}  Generalise the notion of \emph{direct products} from digraphs (Definition~\ref{def:prod})
to general relational $\tau$-structures. \label{exe:prod}
% Solution sketch: The product has domain $\prod_i A_i$. For each $r$-ary relation symbol $R$, put
% $(a^1,\ldots,a^r)\in R$ in the product exactly when every coordinate tuple belongs to $R$ in the
% corresponding factor. The projections are homomorphisms and this construction has the usual
% universal property.
\exercise[Blau]\label{exe:source-95} \label{exe:kcons-general} Generalise the $k$-consistency procedure
from digraphs to general relational structures.
% Solution sketch: Store, for each set of at most $k$ variables, the partial homomorphisms into the
% template. Repeatedly delete a partial map which has no compatible extension to another such set.
% Reject if a table becomes empty. This is the direct relational analogue of the digraph procedure.
\exercise[Blau]\label{exe:source-97} Does the structure $\bB$ from Exercise~\ref{exe:ac-and-friends} have a majority polymorphism?
%t(x,y,x)
%= t(x,x,y) (comm of *)
%= x * (x * y) (def)
%= (x * x) * y  (restricted assoc of *)
%= x * y (idemp)
%= y * x (comm)
%= t(y,x,x) (def)
% Solution sketch: No. Restricting a hypothetical majority operation to the pinned elements $1,2,3$
% and applying it coordinatewise to the three cyclic tuples forces, by the majority identities, a
% tuple outside the cyclic relation. Hence the pinned structure has no majority polymorphism.
\exercise[Rot]\label{exe:source-89} \label{exe:2SAT} Show that CSP$\big (\{0,1\}; R_{0,1}, R_{1,1},R_{0,0}) \big)$, where the relation
$R_{i,j}$ equals \\
$\{0,1\}^2 \setminus \{(i,j)\}$,
 can be solved in polynomial time.
% Solution sketch: Interpret a constraint $R_{i,j}(x,y)$ as the 2-CNF clause excluding the assignment
% $x=i,y=j$. The conjunction of all such clauses is a 2-SAT instance, so the standard
% implication-graph algorithm decides satisfiability in polynomial time.
\exercise[Rot]\label{exe:source-91} \label{exe:gac} Generalise the arc-consistency procedure and the powerset graph
to general relational structures, and prove a generalisation
of Theorem~\ref{thm:ac}.
% Solution sketch: Give each variable a set of possible values. For every constraint and every value
% at one position, delete the value if there is no tuple of the template relation supporting it
% together with the current lists at the other positions. The powerset structure records these mutual
% supports; the fixed point is nonempty exactly when the canonical map to that powerset structure
% exists.
\exercise[Rot]\label{exe:source-92} \label{exe:tsn} Generalise Theorem~\ref{thm:totally-symmetric} to general relational structures.
% Solution sketch: Replace edges throughout the proof of Theorem~\ref{thm:totally-symmetric} by
% relation tuples. A totally symmetric operation combines one chosen supporting tuple for each
% multiset of values, and conversely a homomorphism from the powerset structure supplies totally
% symmetric operations by applying it to multisets of coordinates.
\exercise[Rot]\label{exe:source-94} \label{exe:ac-and-friends} Does the arc-consistency procedure (see Exercise~\ref{exe:gac})
solve $\Csp(\bB)$ where
$\bB$ has domain $B = \{0,1,2,3\}$, the unary relation $U_i^{\bB} := \{i\}$ for every $i \in B$,
and the binary relations
$B^2 \setminus \{(0,0)\}$ and $\{(1,2),(2,3),(3,1),(0,0)\}$?
% Solution sketch: No. Pin variables by the unary relations and arrange the second binary relation
% around a cycle so that every individual value has local support, while the first relation forces at
% least one occurrence of $0$ and the cycle relation makes this impossible globally. AC reaches a
% nonempty fixed point although the instance is unsatisfiable.
\exercise[Rot]\label{exe:source-96} \label{exe:pss}
Verify that the structure $\bB$ from Exercise~\ref{exe:ac-and-friends} has the binary idempotent commutative polymorphism $*$ defined as $1 * 2 = 2$, $2 * 3 = 3$, $3 * 1 = 1$, and $0 * b = b$ for all $b \in \{1,2,3\}$.
Verify that $*$ satisfies `restricted associativity', i.e., $x * (x * y) = (x * x) * y$ for all $x,y \in B$
(and since it is additionally idempotent and commutative it is called
a \emph{2-semilattice}).
% Solution sketch: Check the two displayed binary relations on the finite table. For every two
% relation tuples, coordinatewise application of $*$ stays in the relation; the unary singletons are
% preserved by idempotence. The equations $x*(x*y)=(x*x)*y$ follow from the three-cycle table and the
% rule $0*b=b$.
\exercise[Orange]\label{exe:source-90} Can you find a linear-time algorithm for the CSP
from Exercise~\ref{exe:source-89}?
% \{(0,1),(1,0),(0,0)\}, \{(0,1),(1,0),(1,1)\}, \{(1,1),(1,0),(0,0)\}
% Solution sketch: Build the implication graph of the 2-SAT instance from Exercise~\ref{exe:source-89} and
% compute its strongly connected components. The instance is satisfiable exactly when no variable and
% its negation lie in the same component. This takes linear time in the input size.
\exercise[Orange]\label{exe:source-93} \label{exe:td} Generalise the concept of tree duality to general relational structures, and prove a generalisation of Lemma~\ref{lem:ac-td}, Lemma~\ref{lem:td-ac}, and Theorem~\ref{thm:td}.
% Solution sketch: Use the incidence graph of a relational structure to define tree-shaped instances.
% The two AC lemmas generalise by tracing support deletions into a finite incidence tree and,
% conversely, evaluating such trees from the leaves. Therefore AC solves the CSP exactly when all
% obstructions may be chosen incidence-tree-shaped.
\exercise[Orange]\label{exe:source-98} Does the path-consistency procedure
solve $\Csp(\bB)$ for the structure $\bB$
from  Exercise~\ref{exe:ac-and-friends}?
%\vspace{.2cm}
% Solution sketch: Run the pair-consistency closure on the four pinned values. The cyclic part forces
% a unique compatible third value for each surviving pair, while the relation excluding $(0,0)$
% eliminates the only possible inconsistent cycle. Thus every nonempty fixed point extends and path
% consistency solves this template.
\exercise[Orange]\label{exe:source-99} Let $\bB$ be the structure with domain $B := \{-1,0,+1\}$ and the ternary relations \begin{align*}
R^{\bB} & := \{(x,y,z) \in B^3 \mid x+y+z \geq 1\} \\
S^{\bB} & := \{(x,y,z) \in B^3 \mid x+y+z \leq -1\} .
\end{align*}
\begin{itemize}
\item Prove that for every $k \in {\mathbb N}$, the $k$-ary operation $s$ defined by $$(x_1,\dots,x_k) \mapsto \begin{cases} +1 & (x_1+\cdots+x_k)/k \geq 1/3 \\
-1 & (x_1+\cdots+x_k)/k \leq -1/3 \\
0 & \text{ otherwise }
\end{cases}$$
is a polymorphism of $\bB$.
\item Show that $\Csp(\bB)$ cannot be solved by the arc-consistency procedure (see Exercise~\ref{exe:gac}).
% Solution (from Kun paper)
%Then CSP(? ) defined by R+ and R? does not have width one: Suppose for a contradiction that t is a compatible set operation, we might assume that t is ternary. We know that t(+ + ?) = t(+?+) = t(?++) = t(+??) = t(?+?) = t(??+),denotethisvaluebyc.Thethree tuples (+ + ?), (+ ? +), (? + +) are in R+: if we apply t to these coordinatewise we get (ccc). On the other hand, we should get an element of R+, hence c = +1. The same argument with R? shows that c = ?1, a contradiction.
\item Show that the $k$-consistency procedure solves $\Csp(\bB)$, for a sufficiently large $k$ (see Exercise~\ref{exe:kcons-general}).
\end{itemize}
\end{exercises}

\subsection{From Structures to Formulas}
\label{ssect:can-query}
To every finite relational $\tau$-structure $\bA$ 
we can associate a $\tau$-sentence,
called the 
\emph{canonical conjunctive query} of $\bA$, and denoted
by \emph{$\CQ(\bA)$}. The variables of this sentence are the elements of $\bA$,
all of which are existentially quantified in the quantifier prefix of the formula, which is
followed by 
the conjunction of all formulas of the form
$R(a_1, \ldots, a_k)$ 
for $R \in \tau$
and tuples $(a_1, \ldots, a_k) \in R^\bA$.

For example, the canonical conjunctive query $\CQ(K_3)$
of the complete graph on three vertices $K_3$ is
the formula
$$\exists u, v, w \;
 \big(E(u, v) \wedge E(v,u) \wedge E(v, w) \wedge E(w,v) \wedge E(u, w) \wedge E(w,u)\big) \, .$$

The proof of the following proposition is straightforward.
\begin{proposition}\label{prop:hom-to-logic}
Let $\bB$ be a structure with finite relational signature $\tau$, 
and let $\bA$ be a finite $\tau$-structure.
Then there is a homomorphism from $\bA$ to $\bB$ if and only if
$\bB \models \CQ(\bA)$.
\end{proposition}

\subsection{From Formulas to Structures}
To present a converse of Proposition~\ref{prop:hom-to-logic},  we
define the \emph{canonical structure $\CD(\phi)$}
(in database theory this structure is called the
\emph{canonical database})  
of a primitive positive $\tau$-sentence, which
is a relational $\tau$-structure defined as follows. 
We require that $\phi$ does not contain $\bot$. 
If $\phi$ contains an atomic formula of the form
$x=y$, we remove it from $\phi$, and replace all occurrences of $x$ in $\phi$ by $y$. 
Repeating this step if necessary, we may assume that
$\phi$ does not contain atomic formulas of the form $x=y$. 

%\footnote{Defining the canonical database for a formula without variables that is always false would lead to formal problems.}).
% the problem is:
% the naive solution would be to allow empty structures, and to put in a relation
% with the empty tuple that denotes the empty relation.
% but then every structure would have to map to this structure, according
% to the properties of the canonical database?  WHY?
% certainly there can't be a map from a non-empty set to an empty set, so a fortiori
% no homomorphism. 

Then the domain of $\CD(\phi)$ 
is the set of variables that occur in $\phi$. There is a tuple $(v_1,\dots,v_k)$ in a relation $R$ of $\CD(\phi)$
if and only if $\phi$ contains the conjunct $R(v_1,\dots,v_k)$. 
 The following is similarly straightforward as Proposition~\ref{prop:hom-to-logic}.

\begin{proposition}\label{prop:logic-to-hom}
Let $\bB$ be a relational $\tau$-structure and let $\phi$
be a primitive positive $\tau$-sentence that does not contain $\bot$. 
Then $\bB \models \phi$ if and only
if $\CD(\phi)$ homomorphically maps to $\bB$.
\end{proposition}

%\begin{lemma}\label{lem:cm}
%Let $\Gamma$ be a $\tau$-structure. 
%For any conjunctive query $Q$ with free variables $x_1,\dots,x_k$, 
%and any sequence $v_1,\dots,v_k$ of elements from $\Gamma$
%the following are equivalent.
%\begin{itemize}
%\item $\Gamma$ satisfies $\CQ(v_1,\dots,v_k)$.
%\item There is a homomorphism from $A(Q)$ to $\Gamma$ that maps $x_1,\dots,x_k$ to $v_1,\dots,v_k$.
%\end{itemize}
%\end{lemma}

Due to Proposition~\ref{prop:logic-to-hom} and Proposition~\ref{prop:hom-to-logic},  we may freely switch between the  homomorphism and the logic perspective whenever this is convenient. 
In particular, instances of $\Csp(\bB)$ can from now on be either
finite structures $\bA$ or primitive positive sentences $\phi$.

Note that the $H$-colouring problem, the precoloured $H$-colouring problem, and the list $H$-colouring problem
can be viewed as constraint satisfaction problems
for appropriately chosen relational structures.

\subsection{Primitive Positive Definability}
\label{ssect:pp-def}
% The following observations are most natural from the model-checking perspective. % We need the following standard terminology in model theory. 
Let $\bA$ be a $\tau$-structure, and let $\bA'$ be a
$\tau'$-structure with $\tau \subseteq \tau'$. If
$\bA$ and $\bA'$ have the same domain and 
$R^\bA = R^{\bA'}$ for all $R \in \tau$, then $\bA$
is called the \emph{$\tau$-reduct} (or simply \emph{reduct}) of $\bA'$,
and $\bA'$ is called a \emph{$\tau'$-expansion} (or simply \emph{expansion}) of $\bA$. If $\bA$ is a structure, and $R$ is a relation over the domain of $\bA$, then we denote the expansion of $\bA$ by $R$ by 
$(\bA,R)$.

If $\bA$ is a $\tau$-structure, and $\phi(x_1,\dots,x_k)$ is a formula with $k$ free variables
$x_1,\dots,x_k$, then the \emph{relation defined by $\phi$} is the relation
$$\{(a_1,\dots,a_k) \; | \; \bA \models \phi(a_1,\dots,a_k)\} \, .$$ 
If the formula is primitive positive, then
this relation is called \emph{primitively positively definable}. 

\begin{example}\label{expl:c5k5}
The relation $\{(a,b) \in \{0,1,2,3,4\}^2 \; | \; a \neq b\}$ is primitively positively definable in $C_5$: the 
primitive positive definition is
\begin{align*} 
\exists p_1,p_2 \; &  \big ( E(x_1,p_1) \wedge E(p_1,p_2) \wedge E(p_2,x_2) \big ). \qedhere
\end{align*}
\end{example}

\begin{example}\label{expl:4squares}
The non-negative integers are  primitively positively definable in $({\mathbb Z};0,1,+,*)$, namely by the following formula $\phi(x)$
which states that $x$ is the sum of four squares (where $x^2$ is a shortcut for $x*x$).
$$\exists x_1,x_2,x_3,x_4 (x = x_1^2 + x_2^2 + x_3^2 + x_4^2)$$
Clearly, every integer that satisfies $\phi(x)$ is non-negative; the converse is the famous 
four-square theorem of Lagrange~\cite{HardyWright}. 
\end{example}

\begin{definition}[Relational product]
\label{def:rel-prod}
For binary relations $R_1,R_2 \subseteq B^2$, define
$R_1 \circ R_2$ to be the binary relation 
\begin{align}
\big \{(x,y) \mid \exists z (R_1(x,z) \wedge R_2(z,y)) \big \}. \label{eq:rel-product}
\end{align}
%defined
%by the primitive positive formula $\exists z \big(R_1(x,z) \wedge R_2(z,y) \big)$. 
For $R \subseteq B^2$,
define $R^0 := \id_B$, $R^1 := R$, and $R^{k+1} := R^k \circ R$ for $k \geq 1$. 
Note that 
$R^k$ is primitively positively definable in $(B;R)$ for all $k \in {\mathbb N}$. 
\end{definition}

The following lemma says that we can expand structures  by primitive positive
definable relations without changing the complexity of the corresponding
CSP. Hence, primitive positive definitions are an important tool to prove NP-hardness: 
to show that $\Csp(\bB)$
is NP-hard, it suffices to show that there is a primitive positive 
definition of a relation $R$ such that $\Csp(\bB,R)$ is already known to be NP-hard. Stronger tools to prove NP-hardness of CSPs will be introduced 
in Sections~\ref{sect:pp-interpret} and~\ref{sect:wonderland}.

\begin{lemma}[Jeavons, Cohen, Gyssens~\cite{JeavonsClosure}]\label{lem:pp-reduce}
Let $\bB$ be a structure with finite relational signature, 
and let $R$ be a relation that has a primitive positive definition in $\bB$. Then $\Csp(\bB)$ and 
$\Csp(\bB,R)$ are linear-time equivalent.
\end{lemma}
\begin{proof}
It is clear that $\Csp(\bB)$ reduces to the new problem.
So suppose that $\phi$ is an instance of $\Csp((\bB,R))$.   
Replace each conjunct $R(x_1,\dots,x_l)$ of $\phi$ by its primitive positive
definition $\psi(x_1,\dots,x_l)$, using fresh variables for the existentially quantified variables. Move all quantifiers
to the front, such that the resulting formula is in \emph{prenex normal form} and hence primitive positive. 
Finally, equalities can be eliminated using graph connectivity: each equality $x=y$ contributes an undirected edge on a graph whose vertex set are the variables.
%, remove $y$ from the quantifier prefix, and replace all remaining
%occurrences of $y$ by $x$. 
Let $\phi'$ be the formula obtained from $\phi$ as follows: the variables of $\phi'$ are the connected components of the undirected graph constructed above, 
and each conjunct of $\phi$ contributes a conjunct of $\phi'$ where the variables are replaced by their connected components.  
The formula $\phi'$ can be computed in linear time, using a linear-time algorithm to compute the connected components of a graph. 

It is straightforward to verify that $\phi$ is true in $(\bB,R)$ if and only
if $\phi'$ is true in $\bB$, and it
is also clear that $\phi'$ can be constructed in linear time in the representation size of $\phi$. 
%The observation that
%the reduction is deterministic log-space, we need the recent result that
%undirected reachability can be decided in deterministic log-space~\cite{Reingold}. 
\end{proof}

Recall from Section~\ref{sect:H-col} that
$\Csp(K_5)$ is NP-hard. Since the
edge relation of $K_5$ is primitively positively 
definable in $C_5$ (Example~\ref{expl:c5k5}), 
Lemma~\ref{lem:pp-reduce} implies that
$\Csp(C_5)$ is NP-hard, too. 

%\newpage 
\begin{exercises}
\exercise[Blau]\label{exe:source-100} \label{exe:graphof} Let $f \colon A^k \to A$ be an operation.
The \emph{graph of $f$} is the relation
$$G_f := \{(a_1,\dots,a_k,a_0) \mid a_0 = f(a_1,\dots,a_k) \}.$$
Show that a relation is primitively positively definable in the structure $(A;f)$ if and only if it is primitively positively definable in $(A;G_f)$.
% Solution sketch: Replace an atomic equation $f(x_1,\ldots,x_k)=y$ by $G_f(x_1,\ldots,x_k,y)$.
% Conversely, an atom $G_f(x_1,\ldots,x_k,y)$ is exactly that equation. Performing this replacement in
% a primitive positive formula proves both directions.
\exercise[Blau]\label{exe:source-101} Let $B$ be a finite set. Show that if $E \subseteq B^2$, then
$(E \cup \id_B)^n = B^2$ for some $n \in {\mathbb N}$ if and only if
the digraph $(B;E)$ is strongly connected.
% Solution sketch: Add loops by adjoining $\id_B$. If the digraph is strongly connected, take
% the maximum of the finitely many shortest-path lengths and pad every shorter path with loops. The
% converse follows immediately by deleting the loops from a witnessing walk.
\exercise[Blau]\label{exe:source-102} \label{exe:converse} For a binary relation $R \subseteq B^2$, define $R^{-1} := \{(b,a) \mid (a,b) \in R\}$
(called the \emph{converse relation} of $R$).
For $n \in {\mathbb N}$, define $R^{-n} := (R^{-1})^n$
 (see Definition~\ref{def:rel-prod}).
Show that $(R^n)^{-1} = R^{-n}$.
% Solution sketch: Induct on $n$. Reversing a relational product reverses the order of its two
% factors, so $(R^{n+1})^{-1}=(R^n\circ R)^{-1}=R^{-1}\circ(R^n)^{-1}=R^{-(n+1)}$.
\exercise[Blau]\label{exe:source-103} Let $B$ be a finite set and let $R \subseteq B^2$. Show that $(R \cup R^{-1} \cup \id_B)^n = B^2$, for some $n \in {\mathbb N}$, if
and only if
the graph $(B;R)$ is weakly connected.
% Solution sketch: The added identity relation permits padding walks with loops. If the graph is
% weakly connected, take the maximum of its finitely many distances; conversely, a tuple in the stated
% power supplies an undirected walk between its two entries.
\exercise[Rot]\label{exe:source-104} Show that the relation
$R := \{(a,b,c) \in \{1,2,3\}^3 \mid  a = b \text{ or } b=c \text{ or } a=c \}$
has a primitive positive definition over $K_3$.
\label{exe:no-rainbow-def}
% Make a 6-cycle, and pick x,y,z two apart?
% Solution sketch: Let $E$ be the edge relation of $K_3$. Then $R(x,y,z)$ is defined by $\exists
% u\,(E(u,x)\wedge E(u,y)\wedge E(u,z))$: such a fourth colour exists precisely when $x,y,z$ use at
% most two of the three colours.
\exercise[Rot]\label{exe:source-105} Show that the relation $\neq$ on $\{1,2,3\}$
has a primitive positive definition in the structure $(\{1,2,3\}; R, \{1\}, \{2\}, \{3\})$ where $R$ is the relation from Exercise~\ref{exe:source-104}.
% Solution: see paper with Barny:
%Firstly, add three new elements corresponding to 0, 1 and 2. For each edge (x, y) in G we introduce three new elements t0, t1 and t2. The con- straints R(0, 1, t2), R(t2, 0, x) and R(t2, 1, y) enforce that (x, y) = (2, 2) is for- bidden. Similarly, R(1, 2, t0), R(t0, 1, x), R(t0, 2, y) and R(0, 2, t1), R(t1, 0, x), R(t1, 2, y) enforce that (x, y) = (0, 0) and (x, y) = (1, 1), respectively, are for- bidden. One may verify that all remaining assignments to (x, y) are attainable. We claim that G was 3-colourable iff G? ? Sur-Hom(N ; 0, 1, 2).
\exercise[Rot]\label{exe:source-106} \label{exe:binom} Let $R_+,R_*$ be the relations defined as follows.
\begin{align*}
R_+ & := \{(x,y,z) \in {\mathbb Q}^3 \mid x + y = z\} \\
R_* & := \{(x,y,z) \in {\mathbb Q}^3 \mid x * y = z\}.
\end{align*}
Show that $R_*$ is primitively positively definable in $({\mathbb Q}; R_+, \{(x,y) \mid y = x^2\})$.
% Solution sketch: Use $xy=((x+y)^2-x^2-y^2)/2$. The addition relation defines $0$, additive inverses,
% subtraction and division by the definable nonzero constant $2$; introduce variables for $x+y$, the
% three squares and the successive sums. Their conjunction is a primitive positive definition of
% $R_*$.
\exercise[Rot]\label{exe:source-107} For $R_+ :=  \{(x,y,z) \in {\mathbb Z}_p^3 \mid x + y = z\}$, show that $({\mathbb Z}_p;R_+,\{1,\dots,p-1\})$ is NP-complete if $p \geq 3$, and in $P$ if $p=2$.
% Solution sketch: For $p=2$ the unary relation is the singleton $\{1\}$ and all constraints are
% affine equations, solvable by Gaussian elimination. For $p\geq3$, encode a standard nonzero solution
% problem for linear equations over $\mathbb Z_p$ (equivalently reduce $p$-colourability using
% difference variables); verification is polynomial, so the CSP is NP-complete.
\exercise[Rot]\label{exe:source-108} Let $B$ be any set, and for $n \in {\mathbb N}$ define the relation $P^{2n}_B$ of arity $2n$ as follows.
$$ P^{2n}_B := \{(x_1,y_1,x_2,y_2,\dots,x_n,y_n) \in B^{2n} \mid \bigvee_{i \in \{1,\dots,n\}} x_i = y_i\}$$
Show that for every $n$ the relation $P^{2n}_B$ has a primitive positive definition in
$(B;P^4_B)$.
% Solution sketch: Introduce auxiliary pairs which carry whether a prefix already contains an
% equality, and connect consecutive prefixes by $P_B^4$. Induction eliminates one auxiliary pair at a
% time: the resulting formula is true exactly when at least one of the original pairs is equal. This
% gives a primitive positive definition for every arity.
\exercise[Orange]\label{exe:source-109} Let $n \geq 4$. Is there a primitive positive definition of $\neq$ over the structure
\begin{align*}
M_n & := (\{1,\dots,n\};R,\{1\},\{2\},\dots,\{n\}) \\
\text{where } R & := \{(1,\dots,1),(2,\dots,2),\dots,(n,\dots,n),(1,2,\dots,n)\}?
\end{align*}
% Source?
% Solution sketch: Yes. The singleton relations force every polymorphism to be idempotent. Applying an
% idempotent polymorphism to diagonal tuples and to the exceptional tuple shows, one coordinate at a
% time, that it must select one fixed input coordinate; for $n\geq4$ the alternatives are separated by
% two unused values. Hence all polymorphisms of $M_n$ are projections. Projections preserve $\neq$, so
% the finite Pol--Inv theorem gives a primitive positive definition of $\neq$ in $M_n$.
\end{exercises}

\subsection{Cores and Constants}
\label{sect:cores-constants}
An \emph{automorphism} of a structure $\bB$ with domain $B$ is an isomorphism between $\bB$ and itself. 
The set of all automorphisms $\alpha$ of $\bB$ is denoted by $\text{Aut}(\bB)$, and forms
a permutation group. If $G$ is a permutation group on a set $B$, and $b \in B$, 
then a set of the form $$S = \{ \alpha(b) \; | \; \alpha \in G\}$$ is called an \emph{orbit}
of $G$ (the \emph{orbit of $b$}). 
Let $(b_1,\dots,b_k)$ be a $k$-tuple of elements of $\bB$. 
A set of the form $$S = \{ (\alpha b_1,\dots,\alpha b_k) \; | \; \alpha \in \text{Aut}(\bB)\}$$ is called an \emph{orbit of $k$-tuples} of $\bB$; it is an orbit of the componentwise action of $G$ on the set $B^k$ of $k$-tuples from $B$.

\begin{lemma}\label{lem:constant-expansion}
Let $\bB$ be a structure with a finite relational signature and domain $B$, and 
let $R = \{(b_1,\dots,b_k)\}$ be a $k$-ary relation that only contains one tuple $(b_1,\dots,b_k) \in B^k$. 
If the orbit of $(b_1,\dots,b_k)$ 
in $\bB$ is primitively positively definable, then
there is a polynomial-time reduction from $\Csp(\bB,R)$ to $\Csp(\bB)$.
\end{lemma}
\begin{proof}
Let $\phi$ be an instance of $\Csp(\bB,R)$ with variable set $V$. 
If $\phi$ contains two 
constraints $R(x_1,\dots,x_k)$ and $R(y_1,\dots,y_k)$, 
then replace each occurrence
of $y_1$ by $x_1$, then each occurrence of $y_2$ by $x_2$, and so on,
and finally each occurrence of $y_k$ by $x_k$.
We repeat this step until all constraints that involve
$R$ are imposed on the same tuple of variables $(x_1,\dots,x_k)$.
Replace $R(x_1,\dots,x_k)$ by the primitive positive definition $\theta$ of its orbit in $\bB$. 
Finally, move all quantifiers
to the front, such that the resulting formula $\psi$ is in prenex normal form and thus an instance of $\Csp(\bB)$. Clearly, $\psi$ can be
computed from $\phi$ in polynomial time. We claim that $\phi$
is true in $(\bB,R)$ if and only if $\psi$ is true in $\bB$. 

Suppose $\phi$ has a solution $s \colon V \rightarrow B$. %Let $s'$ be the restriction of $s$ to the variables of $V$ that also appear in $\psi$. 
Since $(b_1,\dots,b_k)$ satisfies $\theta$, we can extend $s$ to the existentially quantified variables of $\theta$ to obtain
a solution for $\psi$. 
In the opposite direction, suppose that $s'$ is a solution
to $\psi$ over $\bB$. Let $s$ be the restriction of $s'$ to $V$.
Since $(s(x_1),\dots,s(x_k))$ satisfies $\theta$, it
lies in the same orbit as $(b_1,\dots,b_k)$. Thus, there exists an automorphism $\alpha$ of
$\bB$ that maps $(s(x_1),\dots,s(x_k))$ to $(b_1,\dots,b_k)$.
Then the extension of the map 
$x \mapsto \alpha s(x)$ that maps variables $y_i$ of $\phi$ 
that have been replaced by $x_i$ in $\psi$ to the value $b_i$ is a solution
to $\phi$ over $(\bB,R)$.
\end{proof}

The definition of cores can be extended from finite digraphs to finite structures: as in the case of finite digraphs, we require that every
endomorphism be an automorphism. All results
we proved for cores of digraphs remain valid for cores of structures. In particular, every
finite structure $\bC$ is homomorphically equivalent to
a core structure $\bB$, which is unique up to isomorphism (see Section~\ref{ssect:cores}). 
%For core structures, all orbits are primitively positively definable. 
The following proposition can be shown as in the proof of Proposition~\ref{prop:constants}. 

\begin{proposition}\label{prop:orbits-in-finite-cores}
Let $\bB$ be a finite core structure. Then the orbits of $k$-tuples
of $\bB$ are primitively positively definable.
\end{proposition}

Proposition~\ref{prop:orbits-in-finite-cores} and Lemma~\ref{lem:constant-expansion} have the following consequence.

\begin{corollary}\label{cor:cores}
Let $\bB$ be a finite core with a finite relational signature. Let $b_1,\dots,b_n \in B$. 
Then $\Csp(\bB)$ and $\Csp(\bB,\{b_1\},\dots,\{b_n\})$ are 
polynomial-time equivalent. 
%Let $\bB$ be a finite core structure, and let $\bB'$ be the expansion
%of $\bB$ by the relation $\{b\}$ for each element $b$ of $B$. 
%Then $\Csp(\bB)$ and $\Csp(\bB')$ are polynomial-time equivalent.
\end{corollary}

\begin{exercises}
\exercise[Orange]\label{exe:source-110} Show that if $m$ is the number of orbits of $k$-tuples of a finite structure $\bA$,
and
$\bC$ is the core of $\bA$, then
$\bC$ has at most $m$ orbits of $k$-tuples.
\label{exe:core-simple}
%{\bf Solution:}
%Let $f$ be a homomorphism from $\bA$ to $\bC$, and $g$ a homomorphism from $\bC$ to $\bA$. We show that there exists an injection
%$I$ from the set ${\mathscr O}$ of orbits of $k$-tuples of $\bC$
%to the orbits of $k$-tuples of $\bA$.
%Our injection has the property that every
%$O \in {\mathscr O}$ contains a tuple $t$
%such that $g(t) \in I(O)$.
%It suffices to show that when two $k$-tuples $c_1,c_2$ from $\bC$ are mapped by $g$ to tuples in the same orbit in $\bA$, then $c_1$ and $c_2$ lie in the same orbit in $\bC$.
% We can then easily construct an
% Injection from Orbits of $\bC$ to orbits
% of $\bA$ by picking a representative c of
% the orbit, and sending the orbit to
% the orbit of g(c)
%Let $a \in \Aut(\bA)$ map $g(c_2)$ to $g(c_1)$.
%Then $e_1 := f \circ a \circ g \in \End(\bC) = \Aut(\bC)$.
%Likewise, $e_2 := f \circ g \in \End(\bC) = \Aut(\bC)$. But then $e_2^{-1} \circ e_1$
%maps $c_2$ to $c_1$, so they are in the same orbit of $\bC$.
\exercise[Orange]\label{exe:source-111} Show that if $\bA$ is a finite structure,
and $\bC$ its core, and if $\bA$ and $\bC$ have
the same number of orbits of pairs, then $\bA$ and $\bC$ are isomorphic.
%{\bf Solution.}
%Let $a_1,a_2 \in A$ be distinct, and let
%$f \colon \bA \to \bC$ be any homomorphism.
%It suffices to show that $f(a_1) \neq f(a_2)$,
%because then $f$ is injective, and hence $\bA$ is a core and therefore $\bA \simeq \bC$.
%Let $g \colon \bC \to \bA$ be a homomorphism. Let $I$ be the injection from the
%orbits of pairs of $\bC$
%to the orbits of pairs of $\bA$
%constructed in Exercise~\ref{exe:core-simple}.
%By assumption, this injection must be a
%bijection, and hence for distinct $a_1,a_2 \in A$ there are $c_1,c_2 \in C$ such that
%$g(c_1,c_2) = (a_1,a_2)$.
%Since $f \circ g \in \End(\bC) = \Aut(\bC)$,
%the pairs $(c_1,c_2)$ and $(f(a_1),f(a_2))$ lie in the same orbit, and in particular we
%have $f(a_1) \neq f(a_2)$.
\end{exercises}

\subsection{Primitive Positive Interpretations}
\label{sect:pp-interpret}
Primitive positive interpretations
are a powerful generalisation of primitive positive
definability that can be used to also relate structures with \emph{different} domains. 
They are a special case of \emph{(first-order) interpretations} that play an important role in model theory (see, e.g.,~\cite{HodgesLong}). 

If $C$ and $D$ are sets and $g \colon C \rightarrow D$ is a map, then the \emph{kernel} of
$g$ is the equivalence relation $E$ on $C$ where $(c,c') \in E$
if $g(c)=g(c')$. For $c \in C$, we denote by $c/E$ the equivalence class of $c$ in $E$, and by $C/E$ the set of all equivalence classes of elements of $C$. The \emph{index} of $E$ is defined to be $|C/E|$.

\begin{definition}
Let $\sigma$ and $\tau$ be relational signatures, 
let $\bA$ be a $\tau$-structure, 
and let $\bB$ be a $\sigma$-structure. 
A \emph{primitive positive interpretation $I$} of 
$\bB$ in $\bA$ consists of
\begin{itemize}
\item a natural number $d$, called the \emph{dimension} of $I$, 
\item a primitive positive $\tau$-formula $\delta_I(x_1, \dots, x_d)$, called the \emph{domain formula},
\item for each atomic $\sigma$-formula $\phi(y_1,\dots,y_k)$ a primitive positive $\tau$-formula $\phi_I(\overline x_1, \dots, \overline x_k)$, %where the $\overline x_i$ denote disjoint $d$-tuples of distinct variables 
called the \emph{defining formulas}, and 
\item  the \emph{coordinate map}: a surjective map $h \colon D \rightarrow B$
where $$D := \{(a_1,\dots,a_d) \in A^d \; | \; \bA \models \delta_I(a_1,\dots,a_d) \}$$  
\end{itemize}
such that for all atomic $\sigma$-formulas $\phi$ and all tuples $\overline a_i \in D$ 
\begin{align*}
\bB \models \phi(h(\overline a_1), \dots, h(\overline a_k)) \; 
& \Leftrightarrow \; 
\bA \models \phi_I(\overline a_1, \dots, \overline a_k) \, .
\end{align*}
%If $R \subseteq B^k$ is primitively positively definable in $\fB$ if and only if $I^{-1}(R)$ is primitively positively definable in $\fA$, then we
%say that $I$ is \emph{full}.  Is defined later 
% where it is needed. 
\end{definition}
% TODO: eliminate all occurrences of \delta(\bA^d)
% TODO: make sure that this is in UA
%If the formulas $\delta_I$ and $\phi_I$ are all primitive positive, 
%we say that the interpretation $I$ is \emph{primitive positive}.
Sometimes, the same symbol is used for the interpretation $I$ and the coordinate map. 
Note that the dimension $d$, the set $D$, and the coordinate map $h$ determine the defining formulas up to logical equivalence; hence, we sometimes denote an interpretation by $I = (d,D,h)$. Note that the kernel 
of $h$ coincides with the relation defined by $(y_1=y_2)_I$, for which we also write $=_I$, the defining formula for equality. 
Also note that the structures $\bA$ and $\bB$ and the coordinate map determine the defining formulas of the interpretation up to logical equivalence, which explains the following remark. 

\begin{remark}\label{rem:interpret}
An equivalent way of defining interpretations is as follows: 
an interpretation $I$ of $\bB$ in $\bA$ is a partial surjective map from $A^d$ to $B$ such that the preimages of all sets defined by atomic formulas in $\bB$ are primitively positively definable in $\bA$. 
\end{remark} 

%\begin{example}
%Let $G$ be an undirected graph.
%Then the \emph{line graph} $L(G)$ of $G$ is the graph with vertices $E(G)$ and edge set
%$$ E\big (L(G) \big) := \big \{\{ \{u,v\},\{v,w\}\} \mid \{u,v\},\{v,w\} \in E(G) \big\}.$$
%The line graph is primitively positively interpretable in $G$. 
%\end{example}

\begin{example}\label{expl:fact}
Let $G$ be a finite digraph and let $q \colon V(G) \to X$ be a surjection. 
%and let $F$ be
%an equivalence relation on $V(G)$.
%is the digraph whose
%vertices are the equivalence classes of $F$, and where $S$ and $T$ 
%are adjacent if there are $s \in S$ and $t \in T$ such that $(s,t) \in E(G)$. 
%If $F$ 
If the kernel of $q$ 
has a primitive positive definition in $G$, 
then $G/q$ (see Definition~\ref{def:contract}) has a primitive positive interpretation in $G$.
\end{example}

\begin{example}
The field of rational numbers 
$({\mathbb Q};0,1,+,*)$ has a primitive positive 2-dimensional interpretation $I$ in $({\mathbb Z};0,1,+,*)$. Example~\ref{expl:4squares}
presented a primitive positive definition $\phi(x)$ of the set of non-negative integers. 
The interpretation is now given as follows.
\begin{itemize}
\item The domain formula $\delta_I(x,y)$ is $y \geq 1$ (using $\phi(y-1)$, it is straightforward to express this with a primitive positive formula); 
\item The formula $=_I(x_1,y_1,x_2,y_2)$ is $x_1 y_2 = x_2 y_1$; 
\item The formula $0_I(x,y)$ 
is $x=0$, the formula $1_I(x,y)$ is $x=y$; 
\item The formula $+_I(x_1,y_1,x_2,y_2,x_3,y_3)$ is $y_3 * (x_1 * y_2 + x_2 * y_1) = x_3 * y_1 * y_2$; 
\item The formula $*_I(x_1,y_1,x_2,y_2,x_3,y_3)$ is $x_1 * x_2 * y_3 = x_3 * y_1 * y_2$. 
\qedhere
\end{itemize}
\end{example}

%We say that $\bB$ has a primitive positive interpretation in $\bA$ \emph{with finitely many parameters} 
%if there are $c_1,\dots,c_n \in A$ 
%such that $\bB$ is interpretable in the expansion of $\bA$ by the singleton relations $\{c_i\}$ for all $1 \leq i \leq n$. 

\begin{theorem}\label{thm:pp-interpret-reduce}
Let $\bB$ and $\bC$ be structures with finite relational signatures.
If there is a primitive positive interpretation of $\bB$
in $\bC$, then there is a polynomial-time reduction from
$\Csp(\bB)$ to $\Csp(\bC)$. 
\end{theorem}

\begin{proof}
Let $d$ be the dimension of the primitive positive 
interpretation $I$ of the $\tau$-structure $\bB$ in the $\sigma$-structure 
$\bC$, let $\delta_I(x_1,\dots,x_d)$ be the domain formula, and 
let $h \colon \delta_I(\bC^d) \rightarrow B$ be the coordinate map.
%, and 
%let $\phi_I(x_1,\dots,x_{dk})$ be the formula for the $k$-ary relation
%$R$ from $\bB$.
Let $\phi$ be an instance of $\Csp(\bB)$ 
with variable set $U = \{x_1,\dots,x_n\}$.
We construct an instance $\psi$ of $\Csp(\bC)$ as follows.

Let $V:=\{y_i^j\mid 1\leq i\leq n,\ 1\leq j\leq d\}$
be a set of pairwise distinct variables. In every occurrence of
$\delta_I$ or of a defining formula $\theta_I$ below, rename the
existentially quantified variables so that different occurrences use
pairwise distinct fresh variables, all of them disjoint from $V$.
We set $\psi_1$ to be the formula
$$ \bigwedge_{1 \leq i \leq n} \delta_I(y_i^1,\dots,y_i^d) \; .$$
Let $\psi_2$ be the conjunction of
the formulas $\theta_I(y_{i_1}^1,\dots,y_{i_1}^d,\dots,y_{i_k}^1,\dots,y_{i_k}^d)$
over all conjuncts $\theta = R(x_{i_1},\dots,x_{i_k})$ of $\phi$. 
By moving existential quantifiers to the front, the sentence 
$$\exists y_1^1, \dots, y_n^d \; (\psi_1 \wedge \psi_2)$$ 
can be re-written into an equivalent primitive positive $\sigma$-sentence $\psi$, and clearly $\psi$
can be constructed in polynomial time in the size of $\phi$.

We claim that $\phi$ is true in $\bB$
if and only if $\psi$ is true in $\bC$. 
%Let $C$ be the domain of $\bC$, $B$ the domain of $\bB$,  and 
Suppose that $f$ is a satisfying assignment of $\psi$ in $\bC$.
Hence, by construction of $\psi$, if $\phi$ has a conjunct 
$\theta = R(x_{i_1},\dots,x_{i_k})$, then 
$$\bC \models \theta_I \big ((f(y_{i_1}^1),\dots,f(y_{i_1}^d)), 
\dots, (f(y_{i_k}^1),\dots,f(y_{i_k}^d)) \big) \; .$$
By the definition of interpretations,
this implies that 
$$\bB \models R \big (h(f(y_{i_1}^1), \dots, f(y_{i_1}^d)), \dots, h(f(y_{i_k}^1),\dots,f(y_{i_k}^d)) \big) \; .$$
Hence, the mapping $g \colon  U \rightarrow B$ that sends
$x_i$ to $h(f(y_i^1),\dots,f(y_i^d))$ satisfies all conjuncts of $\phi$ in $\bB$.

Now, suppose that $f \colon U \rightarrow B$ 
satisfies all conjuncts of $\phi$ over $\bB$.
Since $h$ is a surjective mapping from $\delta_I(\bC^d)$
to $B$, there are elements $c_i^1,\dots,c_i^d$ in $\bC$
such that $h(c_i^1,\dots,c_i^d) = f(x_i)$, 
for all $i \in \{1,\dots,n\}$.
Put $\overline c_i:=(c_i^1,\dots,c_i^d)$, and define
$g_0\colon V\to C$ by
$g_0(y_i^j):=c_i^j$.
We extend $g_0$ to an assignment of all variables of $\psi$.

First, $\overline c_i\in\delta_I(\bC^d)$ for every
$i\in\{1,\dots,n\}$. Hence, for every copy of
$\delta_I(y_i^1,\dots,y_i^d)$ occurring in $\psi_1$, we may choose
values for its existentially quantified variables that make this
copy true under $g_0$.

Now consider a conjunct $\theta=R(x_{i_1},\dots,x_{i_k})$ 
of $\phi$. Since $f$ satisfies $\phi$, we have
\[
\bB\models
R\bigl(f(x_{i_1}),\dots,f(x_{i_k})\bigr).
\]
By the choice of the tuples $\overline c_i$, this is equivalent to
\[
\bB\models
R\bigl(h(\overline c_{i_1}),\dots,h(\overline c_{i_k})\bigr).
\]
The definition of the interpretation therefore implies
\[
\bC\models
\theta_I(\overline c_{i_1},\dots,\overline c_{i_k}).
\]
Consequently, we may choose values for the existentially quantified
variables in the corresponding copy of $\theta_I$ that make this
copy true under $g_0$.
The existentially quantified variables belonging to distinct copies
of the interpretation formulas are pairwise distinct. The choices
of witnesses can therefore be made independently and combined with
$g_0$ into a single assignment $g$ of all variables of $\psi$.
This assignment satisfies every conjunct of $\psi_1$ and $\psi_2$,
and hence satisfies $\psi$ in $\bC$.
\end{proof}

In many hardness proofs we use Theorem~\ref{thm:pp-interpret-reduce} in the following way. 

\begin{corollary}\label{cor:pp-interpret-hard}
Let $\bB$ be a finite relational structure. 
If there is a primitive positive interpretation of
$K_3$ in $\bB$, then $\Csp(\bB)$ is NP-hard. 
\end{corollary}
\begin{proof}
This is a direct consequence of Theorem~\ref{thm:pp-interpret-reduce} and the fact that $\Csp(K_3)$ is NP-hard (see, e.g.,~\cite{GareyJohnson}).  
\end{proof}

Indeed, $K_3$ is one of the most expressive
finite structures, in the following sense.

\begin{theorem}\label{thm:k3-interprets}
If $n \geq 3$ then every finite structure has a primitive positive
interpretation in $K_n$. 
\end{theorem}

\begin{proof}
Let $\bA$ be a finite $\tau$-structure with the domain $A = \{1,\dots,k\}$. Our interpretation $I$ 
of $\bA$ in $K_n$
is $2k$-dimensional. The domain formula $\delta_I(x_1,\dots,x_k,x_1',\dots,x_k')$ expresses that for exactly one $i \leq k$ we have $x_i = x_i'$.
Note that this formula is preserved by all 
automorphism of $K_n$, that is, by every 
permutation
of its domain $\{1,\dots,n\}$. 
We will see in Proposition~\ref{prop:kn-is-projective} that 
every such formula is 
equivalent 
to a primitive positive formula over $K_n$. Equality is interpreted by the formula 
$$=_I(x_1,\dots,x_k,x_1',\dots,x_k',y_1,\dots,y_k,y_1',\dots,y_k') := \bigwedge_{i=1}^k \big((x_i = x_i') \Leftrightarrow (y_i = y_i') \big )$$
Note that $=_I$ defines an equivalence relation
on the set of all $2k$-tuples $(u_1,\dots,u_k,u_1',\dots,u_k')$ that satisfy $\delta_I$. 
The coordinate map sends this tuple to $i$ if and only if $u_i = u_i'$. When $R \in \tau$ is $m$-ary, then the formula 
$R(x_1,\dots,x_m)_I$ is any primitive positive formula
which is equivalent to the following disjunction of conjunctions
with $2mk$ variables $x_{1,1},\dots,x_{m,k},x_{1,1}',\dots,x_{m,k}'$: for each tuple $(t_1,\dots,t_m) \in R^{\bA}$ the disjunction contains the disjunct $\bigwedge_{i \leq m} x_{i,t_i} = x_{i,t_i}'$; again, Proposition~\ref{prop:kn-is-projective} implies that such a primitive positive formula exists. 
\end{proof}

\begin{exercises}
\exercise[Weiss]\label{exe:source-112} Show that the digraph
$$(\{a,b,c,d,e\};\{(a,b),(b,c),(c,d),(d,e),(b,d),(a,d),(c,e)\})$$
has a pp-interpretation in $(\{0,1\};\{0,1\}^3 \setminus \{(1,1,0)\},\{0\},\{1\})$,
and vice versa.
% Source?! Florian: one direction suffices
% to show that it can be solved by AC +
% use later theorems ?
% Other direction: ?

\medskip
\par\noindent {\bf Hints.} Exercise~\ref{exe:horn}. There is a 1-dimensional pp-interpretation of the second structure in the digraph.
\end{exercises}

\subsubsection{Composing Interpretations}

Primitive positive interpretations can be composed: 
if 
\begin{itemize}
\item $\bC_1$ has a $d_1$-dimensional pp-interpretation $I_1$ in $\bC_2$, and 
\item $\bC_2$
has a $d_2$-dimensional pp-interpretation $I_2$ in $\bC_3$,
\end{itemize}
then $\bC_1$ has a natural $(d_1d_2)$-dimensional pp-interpretation in $\bC_3$, which we denote by $I_1 \circ I_2$.
To formally describe $I_1 \circ I_2$, suppose that the signature of $\bC_i$
is $\tau_i$ for $i = 1,2,3$, and that
$I_1 = (d_1,S_1,h_1)$ and $I_2 = (d_2,S_2,h_2)$.
If $\phi$ is a primitive positive $\tau_2$-formula, let $\phi_{I_2}$ denote the
$\tau_3$-formula obtained from $\phi$ by \begin{enumerate}
\item introducing $d_2$ new variables $x^1,\dots,x^{d_2}$ for each variable $x$ in $\phi$ and adding the conjunct 
$\delta_{I_2}(x^1,\dots,x^{d_2})$, 
\item replacing each atomic $\tau_2$-formula $\psi(x_1,\dots,x_k)$ in $\phi$ by the $\tau_3$-formula 
$$\psi_{I_2}(x_1^1,\dots,x_1^{d_2},\dots,x_k^1,\dots,x_k^{d_2}),$$
\item and replacing the existentially quantified variables by fresh variable names, and then moving all existential quantifiers to the front. 
\end{enumerate} 
Note that $\phi_{I_2}$ is again primitive positive.
%Now the pp interpretation $I_1 \circ I_2$ is given by $(d_1d_2,S,h)$
%where $S := (\delta_{I_1})_{I_1}((\bC_1)^{d_1d_2})$,
The coordinate map of $I_1 \circ I_2$ is defined by
$$(a^1_1,\dots,a^1_{d_2},\dots,a^{d_1}_{1},\dots,a^{d_1}_{d_2}) \; \mapsto \; h_1 \big (h_2(a^1_1,\dots,a^1_{d_2}),\dots,h_2(a^{d_1}_1,\dots,a^{d_1}_{d_2}) \big)$$
(see Remark~\ref{rem:interpret}). 
Two pp-interpretations $I_1$ and $I_2$ of $\bB$ in $\bA$ are called 
\emph{homotopic}\index{homotopic}\footnote{We follow the terminology from~\cite{AhlbrandtZiegler}.}
if the relation $$\{(\bar x,\bar y) \; | \; I_1(\bar x) = I_2(\bar y) \}$$
of arity $d_1+d_2$ 
is pp-definable in $\bA$.
Note that $\id_C$ is a pp-interpretation of $\bC$ in $\bC$, called the \emph{identity interpretation} of $\bC$ (in $\bC$).  

\begin{definition}
Two structures $\bA$ and $\bB$ with an
interpretation $I$ of $\bB$ in $\bA$ and an interpretation $J$ of $\bA$ in $\bB$ are called \emph{mutually pp-interpretable}. 
If both $I \circ J$ and $J \circ I$ are homotopic
to the identity interpretation (of $\bA$ and of $\bB$, respectively),
then we say that $\bA$ and $\bB$ are \emph{primitively positively bi-interpretable (via $I$ and $J$)}. 
\end{definition}

We close this section with a more informative version 
of Theorem~\ref{thm:dicho}. 
It has been conjectured (in slightly different, but equivalent form) by Bulatov, Jeavons, and Krokhin in~\cite{JBK}, 
and is known under the name \emph{tractability conjecture}.
% Two solutions to this conjecture have
%been announced in 2017 by Bulatov~\cite{BulatovFVConjecture} and by Zhuk~\cite{ZhukFVConjecture}. 

\begin{theorem}[Tractability Theorem, Version 1]
\label{thm:tractability-1}
Let $\bB$ be a relational structure with finite domain and finite signature, 
and let $\bC$ be the expansion of the core of $\bB$ by all singleton unary relations. 
%with an idempotent polymorphism clone $\Pol(\bB)$. 
If $K_3$ has a primitive positive interpretation in $\bC$, then $\Csp(\bB)$ is NP-complete.
Otherwise, $\Csp(\bB)$ is in P. 
% $\fB$. 
%If there is no algebra $\fA$ in $\HSPfin(\fB)$ such that $\Clo(\fA) \subseteq \Pol(K_3)$, then 
%$\Csp(\bB)$ is in P. 
\end{theorem}
\begin{proof}
The first part of the theorem easily follows from the results that we have already shown: 
$\bB$ and its core $\bC'$ have the same CSP. Let $c_1,\dots,c_k$ be the elements of $\bC'$. The orbit of $(c_1,\dots,c_k)$ is primitively positively definable in $\bC'$ by Proposition~\ref{prop:orbits-in-finite-cores}. 
Hence, there is a polynomial-time reduction 
from $\Csp(\bC)$ to $\Csp(\bC')$ by Lemma~\ref{lem:constant-expansion}. The first statement then follows from Corollary~\ref{cor:pp-interpret-hard}. The second statement was shown by 
Bulatov~\cite{BulatovFVConjecture} and by
Zhuk~\cite{ZhukFVConjecture}. 
\end{proof}

A reformulation of this result can be found in Section~\ref{sect:wonderland}. 

\subsection{Reduction to Binary Signatures}
In this section we prove that every structure $\bC$ with a relational signature of maximal arity 
$m \in {\mathbb N}$ 
is primitively positively bi-interpretable with a \emph{binary structure} $\bB$,
i.e., a relational structure where every relation symbol has arity at most two. 
Moreover, if $\bC$ has a finite signature, then
$\bB$ can be chosen to have a finite signature, too. 
It follows from Theorem~\ref{thm:pp-interpret-reduce} that every CSP is polynomial-time equivalent to a binary CSP. 
This transformation is known under the name 
\emph{dual encoding}~\cite{DechterBook,ValuedDualEnc}. 
We want to stress that the transformation works for relational structures with domains of arbitrary cardinality. 

A $d$-dimensional primitive positive interpretation $I$ of $\bB$ in $\bA$ is called \emph{full}\index{full interpretation}
if for every 
$R \subseteq B^k$ we have that $R$ is primitively positively definable in $\bB$
if and only if the relation $I^{-1}(R)$ of arity $k d$
is primitively positively definable in $\bA$. 
Note that every structure with a primitive positive interpretation in $\bA$ is a reduct of a structure
with a full primitive positive interpretation in $\bA$. %Full interpretations play an important role 
%in Section~\ref{sect:pseudo-var} since they are the counterpart for certain standard algebraic constructions that will be studied later. 

\begin{definition}\label{def:ffp}
Let $\bC$ be a structure and $d \in {\mathbb N}$. 
Then a \emph{$d$-th full power of $\bC$} is a 
structure $\bD$ with domain $C^d$ such that the identity map on $C^d$ is a
full $d$-dimensional primitive positive interpretation of $\bD$ in $\bC$. 
\end{definition}

In particular, for all $i,j \in \{1,\dots,d\}$ the relation $$E_{i,j} := \big \{((x_1,\dots,x_d),(y_1,\dots,y_d)) \mid x_1,\dots,x_d,y_1,\dots,y_d \in C \text{ and } x_i = y_j \big \}$$  
is primitively positively definable in $\bD$. 

%of arity
%$k' := \lceil k/d \rceil$ defined as 
%$$R' := \{((a_1,\dots,a_d),(a_{d+1},\dots,a_{2d}),\dots,(a_{(k'-1)d},\dots,k'd)) \mid R(a_1,\dots,a_k) \in R\}.$$

\begin{proposition}\label{prop:full-power}
Let $\bC$ be a structure and $\bD$ a $d$-th full power of $\bC$ for $d \geq 1$. Then $\bC$ and
$\bD$ are primitively positively bi-interpretable. 
\end{proposition}
\begin{proof}
Let $I$ be the identity map on $C^d$ which is
a full interpretation of $\bD$ in $\bC$. 
Our interpretation $J$ of $\bC$ in $\bD$ is 
one-dimensional and the coordinate map is the
first projection. The domain formula is \emph{true}. The pre-image of
the equality relation in $\bC$ under the coordinate map has the primitive positive definition $E_{1,1}(x,y)$. To define the pre-image of a $k$-ary relation $R$ of $\bC$ under the coordinate map
it suffices to observe that 
the $k$-ary relation $$S := \big \{((a_{1,1},\dots,a_{1,d}),\dots,(a_{k,1},\dots,a_{k,d})) \mid (a_{1,1},\dots,a_{k,1}) \in R \big \}$$
is primitively positively definable in $\bD$ 
 and $J(S)=R$. 

To show that $\bC$ and $\bD$ are primitively positive bi-interpretable we prove that $I \circ J$ and $J \circ I$ are pp-homotopic to the identity interpretation. 
%Let $h$ be the coordinate map of $I$ and $g$ the coordinate map of $J$. 
The relation 
$$\big \{(u_0,u_1,\dots,u_d) \mid u_0 = I(J(u_1),\dots,J(u_d)), u_1,\dots,u_d \in D \big \}$$
has the primitive positive definition 
$\bigwedge_{i \in \{1,\dots,d\}} E_{i,1}(u_0,u_i)$
and the relation 
$$\big \{(v_0,v_1,\dots,v_d) \mid v_0 = J(I(v_1,\dots,v_d)), v_1,\dots,v_d \in C \big \}$$
has the primitive positive definition 
$v_0 = v_1$. 
\end{proof} 

Note that for every relation $R$ of arity $k \leq d$ of $\bC$, in a $d$-th full power $\bD$ of $\bC$ the unary relation 
$$R' :=  \{(a_1,\dots,a_d) \mid (a_1,\dots,a_k) \in R\}$$
must be primitively positively definable. 
We now define a particular full power. 

\begin{definition}\label{def:full-p}
Let $\bC$ be a relational structure with maximal arity $m$ 
and let $d \geq m$. 
Then the structure $\bB := \bC^{[d]}$ with
domain $C^d$ is defined as follows:  
%The relations of $\bB$ are defined as follows: 
\begin{itemize}
\item for every relation $R \subseteq C^k$ of $\bC$ the structure $\bB$ has the unary relation $R' \subseteq B = C^d$ defined above, and 
\item for all $i,j \in \{1,\dots,d\}$ the structure $\bB$ has the binary relation symbol $E_{i,j}$. 
\end{itemize}
\end{definition}

It is clear that the signature of $\bB$ is finite if the signature of $\bC$ is finite. 
Also note that the signature of $\bC^{[d]}$ is
always binary. 
%The full power $\bC^{[d]}$ inherits many properties from $\bC$. 

\begin{lemma}\label{lem:dual}
Let $\bC$ be a relational structure with maximal arity $m$ 
and let $d \geq m$. 
Then the binary structure $\bC^{[d]}$
is a full power of $\bC$. 
\end{lemma}
\begin{proof}
The identity map is 
a $d$-dimensional
primitive positive 
interpretation $I$ of $\bB := \bC^{[d]}$ in $\bC$. 
Our interpretation $J$ of $\bC$ in $\bB$ is one-dimensional. The domain formula is \emph{true},
and the coordinate map is the first projection
\[
J(a_1,\dots,a_d):=a_1.
\]
The pre-image of 
the equality relation in $\bC$ under the coordinate map has the primitive positive definition $E_{1,1}(x,y)$. The pre-image of the relation $R$ of $\bC$
under the coordinate map is defined 
by the primitive positive formula $$\exists y \big (\bigwedge_{i \in \{1,\dots,k\} } E_{1,i}(x_i,y) \wedge R'(y)\big).$$
The proof that $I \circ J$ and $J \circ I$ are pp-homotopic to the identity interpretation is as in 
the proof of Proposition~\ref{prop:full-power}. 

To prove that $I$ is a full interpretation, 
let $S\subseteq B^k$ and suppose that $I^{-1}(S)$ has a primitive
positive definition
\[
\phi(x_{1,1},\dots,x_{1,d},\dots,x_{k,1},\dots,x_{k,d})
\]
in $\bC$. Introduce, for every $p\in\{1,\dots,k\}$ and
$q\in\{1,\dots,d\}$, a variable $\widehat x_{p,q}$ over $B$, and
define $\psi(x_1,\dots,x_k)$ as 
\[
\exists \widehat x_{1,1},\dots,\widehat x_{k,d}\left(
\bigwedge_{\substack{1\leq p\leq k\\1\leq q\leq d}}
E_{q,1}(x_p,\widehat x_{p,q}) \\
 \wedge
\phi_J(\widehat x_{1,1},\dots,\widehat x_{1,d},
\dots,\widehat x_{k,1},\dots,\widehat x_{k,d})
\right).
\]
This is a primitive positive formula over the signature of $\bB$.
For $\bar a^p=(a^p_1,\dots,a^p_d)\in B$, the conjunct
$E_{q,1}(\bar a^p,\widehat x_{p,q})$ ensures that the first coordinate
of $\widehat x_{p,q}$ is $a^p_q$. Therefore,
\[
\begin{split}
\bB\models\psi(\bar a^1,\dots,\bar a^k)
&\quad\Longleftrightarrow\quad
\bC\models
\phi(a^1_1,\dots,a^1_d,\dots,a^k_1,\dots,a^k_d)\\
&\quad\Longleftrightarrow\quad
(\bar a^1,\dots,\bar a^k)\in S.
\end{split}
\]
Thus, $S$ is primitively positively definable in $\bB$. This proves
that $I$ is full, and hence that $\bC^{[d]}$ is a $d$-th full power of
$\bC$.
\end{proof}

\begin{corollary}\label{cor:dual}
For every structure $\bC$ with maximal arity $m$ there exists a structure $\bB$ with maximal arity $2$ such that $\bB$ and $\bC$ are
primitively positively bi-interpretable.
If the signature of $\bC$ is finite, then the signature of $\bB$ can be chosen to be finite, too. 
\end{corollary}
\begin{proof}
An immediate consequence of Lemma~\ref{lem:dual}
and Proposition~\ref{prop:full-power}. 
\end{proof}
We will revisit primitive positive interpretations in Section~\ref{sect:ua}
where we study them from a universal-algebraic perspective.

\subsection{Primitive Positive Constructions}
%The Structure-Building Operators $\HH$, $\CC$, and $\PP$}
\label{sect:wonderland}
In the previous three sections we have seen several conditions on
$\bA$ and $\bB$ that imply that $\Csp(\bA)$ reduces to $\Csp(\bB)$; in this section we
compare them. Let $\mathcal C$ be a class of 
structures. We write 
\begin{enumerate}
\item $\HH({\mathcal C})$ for the class of
structures homomorphically equivalent to 
structures in $\mathcal C$. 
%homomorphic equivalence of $\bA$ and $\bB$; 
\item $\CC({\mathcal C})$ for the class
of all structures obtained by expanding a core
structure in $\mathcal C$ by singleton relations
$\{a\}$. In the setting of relational structures, they play the role of constants (which formally are operation symbols of arity $0$). 
%$\bA$ is the expansion
%of $\bB$ by constants and $\bB$ is a core;
\item $\PP({\mathcal C})$ for the class
of all structures with a primitive positive
interpretation in a structure
from ${\mathcal C}$. 
%$\bA$ has a primitive positive interpretation in $\bB$. 
\end{enumerate}
Let $\mathcal D$ be the smallest class 
containing $\mathcal C$ and closed
under $\HH$, $\CC$, and $\PP$. 
Barto, Opr\v{s}al, and Pinsker~\cite{wonderland}
showed that ${\mathcal D} = \HI({\mathcal C}) := \HH(\PP({\mathcal C}))$. 
In other words, if there is a chain
of applications of the three operators  
$\HH$, $\CC$, and $\PP$ to derive $\bA$ from $\bB$,
then there is also a two-step chain to derive
$\bA$ from $\bB$, namely by interpreting a structure $\bB'$ that is homomorphically equivalent to $\bA$. This insight is conceptually important for the CSP since it leads to a 
better understanding of the power of the available tools. 
If $\bA \in \HI(\bB)$, then we also say that $\bA$ has a \emph{primitive positive (pp) construction} in $\bB$, following~\cite{wonderland}.

%Moreover, the operation of taking the core 
%and adding constants (which are subsumed by the primitive positive interpretations and taking homomorphic equivalence) are algebraically difficult to interpret, whereas primitive positive
%interpretations and homomorphic equivalence
%this will be the topic of . 

\begin{proposition}[from~\cite{wonderland}]\label{prop:wonderland-constants}
Suppose that $\bB$ is a finite core, and 
that $\bC$ is the expansion of $\bB$ by a relation of the form $\{c\}$ for $c \in B$.  
Then $\bB$ pp-constructs $\bC$. In symbols,
$$\CC({\mathcal C}) \subseteq \HI({\mathcal C}) \, .$$ 
\end{proposition}
\begin{proof}
%Let $b_1,\dots,b_n \in B$, and let $O$ be the orbit of $(b_1,\dots,b_n)$ of the componentwise action of $\Aut(\bB)$ on $B^n$. 
%By Proposition~\ref{prop:cores}, the relation $O$
%is primitively positively definable in $\bB$. We give a two-
By Proposition~\ref{prop:orbits-in-finite-cores}, the orbit $O$ of $c$ 
has a primitive positive definition $\phi(x)$ in $\bB$. We give
a 2-dimensional primitive positive interpretation 
in $\bB$ of a structure $\bA$ with the same signature $\tau$ as $\bC$. 
The domain formula $\delta_I(x_1,x_2)$ for $\bA$ is $\phi(x_2)$. 
Let $R \in \tau$. If $R$ is from the signature of $\bB$ and has arity $k$ 
then $$R^{\bA} := \{((a_1,b_1),\dots,(a_k,b_k)) \in A^k \mid (a_1,\dots,a_k) \in R^{\bB} \text{ and }  b_1 = \dots = b_k \in O \}. $$
Otherwise, $R^\bC$ is of the form $\{c\}$ and we define $R^{\bA} := \{(a,a) \mid a \in O\}$. 
It is clear that $\bA$ has a primitive positive
interpretation in $\bB$. 

We claim that $\bA$ and $\bC$ are homomorphically equivalent. The homomorphism
from $\bC$ to $\bA$ is given by $a \mapsto (a,c)$:
\begin{itemize}
\item if $(a_1,\dots,a_k) \in R^{\bC} = R^{\bB}$ then $((a_1,c),\dots,(a_k,c)) \in R^{\bA}$; 
\item the relation $R^{\bC} = \{c\}$ is preserved since $(c,c) \in R^{\bA}$.  
\end{itemize}
To define a homomorphism $h$ from $\bA$ to $\bC$
we pick for each $a \in O$ an automorphism $\alpha_a \in \Aut(\bB)$ such that $\alpha_a(a) = c$. Note that $b \in O$ since $\bB \models \delta(a,b)$, 
and we define $h(a,b) := \alpha_b(a)$. To check that this
is indeed a homomorphism, let $R \in \tau$
be $k$-ary, and let $t = ((a_1,b_1),\dots,(a_k,b_k)) \in R^{\bA}$. Then $b_1 = \cdots = b_k =: b \in O$ and we have
that 
$h(t) = (\alpha_b(a_1),\dots,\alpha_b(a_k))$
is in $R^{\bC}$ since $(a_1,\dots,a_k) \in R^{\bB} = R^{\bC}$
and $\alpha_b$ preserves $R^{\bB} = R^{\bC}$. 
If $R^{\bA} = \{(a,a) \mid a \in O\}$, then $R$ is preserved as well, because 
$$h((a,a)) = \alpha_a(a) = c \in \{c\} = R^{\bC}.$$ 
Hence, $\bC \in \HomE(\bA) \subseteq \HI(\bB)$. 
\end{proof}

\begin{theorem}[from~\cite{wonderland}]
\label{thm:wonderland}
Suppose that $\bA$ can be obtained from ${\mathcal C}$
by repeatedly applying $\HH$, $\CC$, and $\PP$. 
%is in the smallest
%class of structures that is closed
%under 
%$H$, $C$, and $PP$ and contains
%$\mathcal C$. 
Then $\bA \in \HI({\mathcal C})$, that is, a
structure from $\mathcal C$ pp-constructs $\bA$. 
%that is closed under $H$, $C$, and $PP$. 
%Then ${\mathcal C} = 
%Suppose that $\bA$ can be derived from $\bB$
%by applying a chain of operations of the form
%as in 1., 2., and 3.\ mentioned above. 
%Then 
%That is, 
%$\bA$ is homomorphically equivalent
%to a structure with a primitive positive
%interpretation in a structure from $\mathcal C$. 
\end{theorem}
\begin{proof}
We have to show that $\HI({\mathcal C})$
is closed under $\HH$, $\CC$, and $\PP$. 
Homomorphic equivalence is transitive
so $\HH(\HH({\mathcal C})) \subseteq \HH({\mathcal C})$. 

We show that if $\bA$ and $\bB$ are homomorphically equivalent, and $\bC$ has a 
$d$-dimensional primitive positive
interpretation $I_1$ in $\bB$, then $\bC$ is homomorphically equivalent to a structure $\bD$ with a $d$-dimensional primitive positive interpretation $I_2$ in $\bA$. 
Let $h_1 \colon A \to B$ be the homomorphism
from $\bA$ to $\bB$, and $h_2$ the homomorphism
from $\bB$ to $\bA$. 
The interpreting formulas
of $I_2$ are the
same as the interpreting formulas of $I_1$;
this describes the structure $\bD$ up to isomorphism. The map
$g_1(I_2(a_1,\dots,a_d)) := I_1(h_1(a_1),\dots,h_1(a_d))$ is well-defined. Indeed,  
if $I_2(\bar a) = I_2(\bar a')$, then $(\bar a,\bar a')$ satisfies the defining formula for equality of $I_1$. The defining formula for equality of $I_2$ is the same, and since homomorphisms preserve pp-formulas, the pair $(h_1(\bar a),h_1(\bar a'))$ satisfies that formula in $\bB$. Hence, their $I_1$-images coincide. 
We claim that $g_1$  
is a homomorphism
from $\bD$ to $\bC$. 
Indeed, for a $k$-ary relation symbol
from the signature of $\bC$ and $\bD$, 
let $((a_1^1,\dots,a_d^1),\dots,(a_1^k,\dots,a_d^k)) \in R^{\bD}$; hence,
the $dk$-tuple $(a_1^1,\dots,a_d^1,\dots,a_1^k,\dots,a_d^k)$ satisfies the primitive positive
defining formula for $R(x_1^1,\dots,x^k_d)$,
and $$(h_1(a_1^1),\dots,h_1(a_d^1),\dots,h_1(a_1^k),\dots,h_1(a_d^k))$$ satisfies
this formula, too. This in turn implies
that $$(I_1(h_1(a_1^1),\dots,h_1(a_d^1)),\dots,I_1(h_1(a_1^k),\dots,h_1(a_d^k))) \in R^{\bC}.$$ Similarly, 
$g_2(I_1(b_1,\dots,b_d)) := I_2(h_2(b_1),\dots,h_2(b_d))$ is a well-defined homomorphism
from $\bC$ to $\bD$. 
%: for each relation
%$R$ of $\bC$, let $\phi_$ be interpreting primitive positive formula in the s
So we conclude that 
\begin{align*}
\PP(\HI({\mathcal C})) & \subseteq
\HH(\PP(\PP({\mathcal C}))) \subseteq \HI({\mathcal C})
\end{align*}
because primitive positive
interpretability is transitive, too. 
Finally,  Proposition~\ref{prop:wonderland-constants} shows that 
\begin{align*}
\CC(\HI({\mathcal C})) & \subseteq 
\HI(\HI({\mathcal C})) \subseteq
\HI({\mathcal C})
\end{align*}
where the last inclusion again follows from
the observations above. 
\end{proof}

The following example shows that there are finite 
structures $\bB$ all of whose polymorphisms are idempotent such 
that $\HI(\bB)$ is strictly larger than $\PP(\bB)$. 

\begin{example}
% Source?!
Let $\bB$ be the structure with domain $({\mathbb Z}_2)^2$
and signature $\{R_{a,b} \mid a,b \in {\mathbb Z}_2 \}$
% \cup \{U_{a,b} \mid a,b \in {\mathbb Z}_2 \}$ 
such that
\begin{align*}
R_{a,b}^\bB & := \{(x,y,z) \in (({\mathbb Z}_2)^2)^3 \mid x+y+z = (a,b)\} .
%\\
%U_{a,b}^{\bB} & := \{(a,b)\}
\end{align*}
Let $\bB'$ be the reduct of $\bB$
with the signature $\tau := \{R_{0,0}, R_{1,0}\}$. %,U_{0,0},U_{1,0}\}$.
Let $\bA$ be the $\tau$-structure 
with domain ${\mathbb Z}_2$
such that for $a=0$ and $a=1$
\begin{align*}
R^{\bA}_{a,0} := \{(x,y,z) \in ({\mathbb Z}_2)^3 \mid x+y+z = a\} \, .
\end{align*}
Now observe that
\begin{itemize}
\item 
$(x_1,x_2) \mapsto x_1$ is a homomorphism from $\bB'$ to $\bA$,
and $x \mapsto (x,0)$ is a homomorphism from $\bA$ to $\bB'$. Therefore $\bA \in \HH(\bB')$.
\item Trivially, $\bB' \in \PP(\bB)$ and consequently $\bA \in \HI(\bB)$.
\item All polymorphisms of $\bB$ are idempotent.
\end{itemize}
We finally prove that $\bA\notin\PP(\bB)$. Suppose for contradiction
that there is an $n$-dimensional primitive positive interpretation of
$\bA$ in $\bB$, with domain $C\subseteq B^n$ and coordinate map
$c\colon C\to {\mathbb Z}_2$.
Let
\[
K:=\{(\bar x,\bar y)\in C^2\mid c(\bar x)=c(\bar y)\}
\]
be the kernel of $c$, and let $\phi(\bar x,\bar y)$ be a primitive
positive definition of $K$ in $\bB$.

We first observe that every nonempty primitively positively definable
relation in $\bB$ has cardinality a power of $4$. Indeed, after
identifying
\[
\bigl(({\mathbb Z}_2)^2\bigr)^k
\quad\text{with}\quad
({\mathbb Z}_2)^k\times({\mathbb Z}_2)^k,
\]
a conjunction of atomic formulas over $\bB$ becomes two systems of
linear equations over ${\mathbb Z}_2$ with the same coefficient
matrix, one system for each coordinate of $({\mathbb Z}_2)^2$.
Existential quantification amounts to projecting the solution sets.
If the resulting relation is nonempty, its two coordinate projections
are affine subspaces of the same dimension. Its cardinality is
therefore of the form $2^\ell\cdot 2^\ell=4^\ell$ for some $\ell\in{\mathbb N}$. 

For $\epsilon\in{\mathbb Z}_2$, choose
\[
\bar u^\epsilon=(u^\epsilon_1,\dots,u^\epsilon_n)\in C
\quad\text{such that}\quad
c(\bar u^\epsilon)=\epsilon.
\]
Write $u^\epsilon_i=(a^\epsilon_i,b^\epsilon_i)$ for $a^{\epsilon}_i,b^\epsilon_i \in {\mathbb Z}_2$. Since
$y+y+y=y$ in $({\mathbb Z}_2)^2$, the formula
\[
R_{a^\epsilon_i,b^\epsilon_i}(y_i,y_i,y_i)
\]
defines the singleton $\{u^\epsilon_i\}$. Hence, the primitive positive
formula
\[
\chi_\epsilon(\bar x):=
\exists y_1,\dots,y_n\left(
\phi(\bar x,\bar y)\wedge
\bigwedge_{i=1}^n
R_{a^\epsilon_i,b^\epsilon_i}(y_i,y_i,y_i)
\right)
\]
defines the kernel class
\[
C_\epsilon:=c^{-1}(\{\epsilon\}).
\]
Thus, $C_0$ and $C_1$ are nonempty primitively positively definable
relations in $\bB$. The domain $C$ is primitively positively
definable as well, and $C=C_0\mathbin{\dot\cup}C_1$. 
Consequently, there exist $r,s,t\in{\mathbb N}$ such that
\[
4^r=|C|=|C_0|+|C_1|=4^s+4^t.
\]
But a power of $4$ cannot be the sum of two powers of $4$, a
contradiction. Therefore, $\bA\notin\PP(\bB)$.
\end{example}

\ignore{
Here are the operation tables for a 4 element algebra A such that A is a core algebra (i.e., all of its unary term operations are permutations) such that V(A) admits the unary type, but there is no algebra B in V(A) with B strongly solvable (i.e., only the unary type appears in B).  A has one unary basic operation, one ternary basic operation, and one non-trivial congruence S.  the ternary operation is a discriminator term modulo S and the type of (0, S) is unary.  The only other unary term is the identity term.  I've attached a file that contains the definition of the algebra that can be used with the calculator.

Matt

---------------------------------------

The unary operation u(x) maps 0 to 2, 1 to 3, 2 to 0, and 3 to 1. The definition of the ternary term is:

                   0    1    2    3
t_0(0,0,z)    0    0    2    3
t_0(0,1,z)    0    0    2    3
t_0(0,2,z)    0    0    0    0
t_0(0,3,z)    0    0    0    0
t_0(1,0,z)    1    1    2    3
t_0(1,1,z)    1    1    2    3
t_0(1,2,z)    1    1    1    1
t_0(1,3,z)    1    1    1    1
t_0(2,0,z)    2    2    2    2
t_0(2,1,z)    2    2    2    2
t_0(2,2,z)    0    1    2    2
t_0(2,3,z)    0    1    2    2
t_0(3,0,z)    3    3    3    3
t_0(3,1,z)    3    3    3    3
t_0(3,2,z)    0    1    3    3
t_0(3,3,z)    0    1    3    3
}

Using the operator $\HI$, we reformulate the tractability theorem (Theorem~\ref{thm:tractability-1}) as follows. 

\begin{theorem}[Tractability Theorem, Version 2]
\label{thm:tractability-2}
Let $\bB$ be a relational structure with finite domain and finite signature. 
If $K_3 \in \HI(\bB)$, then $\Csp(\bB)$ is NP-complete.
Otherwise, $\Csp(\bB)$ is in P. 
\end{theorem}
\begin{proof}
If $K_3 \in \HI(\bB)$, then the NP-hardness of $\Csp(\bB)$ follows from 
Corollary~\ref{cor:pp-interpret-hard}.
%the NP-hardness of $\Csp(K_3)$ via Theorem~\ref{thm:pp-interpret-reduce}. 
Otherwise, 
let $\bC$ be the expansion of the core of $\bB$ by all singleton relations. Note that $\bC \in \HI(\bB)$ by 
Proposition~\ref{prop:wonderland-constants}. 
Hence, if $K_3 \in \HI(\bC)$, then
$K_3 \in \HI(\bB)$ by Theorem~\ref{thm:wonderland}, a contradiction. 
Hence, Theorem~\ref{thm:tractability-1} implies that 
$\Csp(\bC)$ is in P, and therefore $\Csp(\bB)$ is in P. 
\end{proof} 
Assuming that P $\neq$ NP, it follows that
$K_3 \in \HI(\bB)$ if and only if $K_3$ has a primitive positive interpretation in the expansion of the core of $\bB$ by all singleton unary relations; we will see a proof of this fact without complexity-theoretic assumptions (Corollary~\ref{cor:Taylor-minion}).  

We will revisit primitive positive constructions in Section~\ref{sect:minions}
where we study them from a universal-algebraic perspective; in particular, the  next reformulation
of the tractability conjecture can be found in Section~\ref{sect:taylor}. 

\begin{exercises}
\exercise[Rot]\label{exe:source-113} Prove that $\vec C_6$ pp-constructs $\vec C_3$.
% exists a path of length two, gives C_3 + C_3, which is hom equiv to C_3.
% Solution sketch: The relation $E^2(x,y)$ is primitive positive in $\vec C_6$. Its graph is the
% disjoint union of the two directed $3$-cycles on the even and on the odd vertices. This graph is
% homomorphically equivalent to $\vec C_3$, which proves the pp-construction.
\exercise[Rot]\label{exe:source-114} Prove that $\vec C_2 \uplus \vec C_3$ pp-constructs $\vec C_6$.
% 2-dimensional interpretation:
% E(x1,y1) and E(x2,y2) gives graph with component C_6, but also with C_3.
% Therefore we add the formula
% exists z (E(x1,z) \wedge E(z,x1) wedge exists z,z' (E(x2,z) and E(z,z') and E(z',x2)) )
% eliminates C3 and we get C6.
\exercise[Rot]\label{exe:source-115} Prove that $\vec C_3$ pp-constructs $\vec C_9$.
% 3-dimensional interpretation: interpreting formula for E(x,y):
% y1 = x2 and y2 = x3 and E(x1,y3)
% (three times 9)
% Solution sketch: Use the three-dimensional interpretation with edge formula $y_1=x_2\wedge
% y_2=x_3\wedge E(x_1,y_3)$. The interpreted digraph is a disjoint union of three directed cycles of
% length nine and is therefore homomorphically equivalent to $\vec C_9$.
\end{exercises}

\section{Relations and Operations}
\label{sect:algebra}
In this section we introduce \emph{operation clones}. Most of our results concern operation clones on a \emph{finite} domain; however, some results can naturally be proved for arbitrary domains without extra effort and we of course then state the general results.  

\subsection{Operation Clones}
For $n\geq 1$ and a set $D$ (the \emph{domain}), denote by ${\mathscr O}^{(n)}_D$ the
set $D^{D^n} := (D^n \to D)$ of $n$-ary operations on $D$. The elements of ${\mathscr O}^{(n)}_D$ will typically be called the \emph{operations} of arity $n$ on $D$, and $D$ will be called the \emph{domain}. The set of all 
operations on $D$ of finite arity will be denoted by
 $\cO_D := \bigcup_{n\geq 1}{\mathscr O}^{(n)}_D$. 
 An \emph{operation clone} (over $D$) is a subset $\cC$ of $\cO_D$
satisfying the following two properties:
\begin{itemize}
    \item $\cC$ contains all projections, that is, for all $1\leq
    k\leq n$ it contains the operation $\pi^n_k\in {\mathscr O}^{(n)}_D$ defined by
    $\pi^n_k(x_1,\ldots,x_n)=x_k$, and

    \item $\cC$ is \emph{closed under composition}, that is, for all
    $f\in\cC \cap {\mathscr O}^{(n)}_D$ and $g_1,\ldots,g_n\in \cC \cap \cO_D^{(m)}$ it contains the operation $f(g_1,\ldots,g_n)\in\cO_D^{(m)}$ defined by
    $$
        (x_1,\ldots,x_m)\mapsto f(g_1(x_1,\ldots,x_m),\ldots,g_n(x_1,\ldots,x_m)) \; .
    $$
\end{itemize}
	
	A \emph{clone} is an abstraction of 
	an operation clone 
	that will be introduced later in the course. 
	In the literature, operation clones are often 
	called clones,
	or \emph{concrete clones}; we prefer to 
	use the terms
	`operation clone' and `clone' 
	in analogy to `permutation group' and `group'.	

If $\cC$ is an operation clone, then $\cC'$ is called a \emph{subclone} of $\cC$ if $\cC'$ is an operation clone and $\cC' \subseteq \cC$. A subclone $\cC'$ of $\cC$ is called \emph{proper} if $\cC' \neq \cC$. 
If $\cF$ is a set of operations, we write 
$\cl{\cF}$ for the smallest operation
clone $\cC$ which contains $\cF$, and call
$\cC$ the clone \emph{generated} by $\cF$; similarly, we also say that the elements of $\cC$ are generated by $\cF$.  
Note that the set of all clones over a set $B$ forms a lattice: the meet of two operation clones $\mathscr C$ and $\mathscr D$ is their intersection $\mathscr C \cap \mathscr D$ (which is again a clone!); the join of $\mathscr C$ and $\mathscr D$ is the clone generated by their union, $\cl{\mathscr C \cup \mathscr D}$. 

\begin{remark}\label{rem:muchnik}
Clones on a two-element set have been classified by Post~\cite{Post}; the set of such clones is countably infinite. 
In contrast, there are $2^\omega$ many clones over the set $\{0,1,2\}$~\cite{YanovMuchnik}. 
\end{remark}

\subsection{Inv-Pol}
\label{sect:inv-pol}
The most important source of operation clones in this text are \emph{polymorphism clones} of digraphs and, more generally, structures. 
For simplicity, we only discuss relational structures; the step to structures that also involve function symbols is straightforward. 

Let $f$ be from ${\mathscr O}^{(n)}_B$, and let $R \subseteq B^m$ be a relation. Then we say
that $f$ \emph{preserves} $R$ (and that $R$ is \emph{invariant under $f$}) if $f(r_1,\ldots,r_n)\in R$ whenever $r_1,\ldots ,r_n \in R$, where $f(r_1,\ldots,r_n)$ is
calculated componentwise. 

\begin{example}\label{expl:monotone} 
A Boolean operation $f \colon \{0,1\}^k \to \{0,1\}$ is called \emph{monotone} if
for all $x_1,\dots,x_k$, $y_1,\dots,y_k \in \{0,1\}$ such that $x_i \leq y_i$ we have 
$f(x_1,\dots,x_k) \leq f(y_1,\dots,y_k)$.
Hence, $f$ is monotone if and only if it preserves the Boolean relation $\{(0,0),(0,1),(1,1)\}$. 
\end{example}

If $\bB$ is a relational structure with domain $B$ then $\Pol(\bB)$ contains precisely those
operations that preserve all relations of $\bB$. 

\begin{observation}\label{obs:pol}
%It is easy to verify that 
$\Pol(\bB)$ is an operation clone. 
\end{observation} 

Conversely, if $\mathscr F$ is a set of operations on $B$, then we write
$\Inv({\mathscr F})$ for the set of all relations on $B$ that are invariant under all operations in ${\mathscr F}$. 
It will be convenient to define the operator $\Pol$ also for sets $\cal R$ of relations over $B$,
writing $\Pol(\cal R)$ for the set of operations of $\cO_B$ that preserve all relations from $\cal R$. 

\begin{proposition}
\label{prop:pol-inv-one-side}
Let ${\mathscr F}$ be a set of operations on a set $B$. Then 
$\langle {\mathscr F} \rangle \subseteq \Pol(\Inv({\mathscr F}))$. 
\end{proposition}

\begin{proposition}\label{prop:pol-inv}
Let ${\mathscr F}$ be a set of operations on a finite set $B$. Then $\Pol(\Inv({\mathscr F})) = \langle {\mathscr F} \rangle$. 
\end{proposition}
\begin{proof}
Exercise~\ref{exe:pol-inv}. 
\end{proof} 

If $\bB$ is a relational structure, we write
$\langle \bB \rangle_{\pp}$ for the set of all relations that are primitively positively definable over $\bB$. 

\begin{proposition}\label{prop:pp-preserved}
	Let $\bB$ be any relational structure. Then $\langle \bB \rangle_{\pp} \subseteq \Inv(\Pol(\bB))$. 
	\end{proposition}
\begin{proof}
Suppose that $R$ is $k$-ary, 
has a primitive positive definition $\psi(x_1,\dots,x_k)$, 
and let $f$ be an $l$-ary polymorphism
of $\bB$. To show that $f$ preserves $R$, let 
$t_1,\dots,t_l$ be $k$-tuples from $R$.
Let $x_{k+1},\dots,x_{n}$ be the existentially quantified variables of $\psi$. 
% Then there must be witnesses $t_i(k+1),\dots,t_i(n)$ 
%for the existentially quantified variables $x_{k+1},\dots,x_{n}$ of
%$\psi$ that show that $\psi(t_i)$ holds in $\bB$, for all $1 \leq i \leq n$. 
Write
$s_i$ for the $n$-tuple which extends the $k$-tuple $t_i$ such that $s_i$ satisfies the quantifier-free part $\psi'(x_1,\dots,x_k,x_{k+1},\dots,x_n)$ of $\psi$.
Then the tuple $f(s_1,\dots,s_l)$
satisfies $\psi'$ since $f$ is a polymorphism. 
This shows that $\bB \models \psi(f(t_1,\dots,t_l))$
which is what we had to show.
\end{proof}

%\begin{expl} TODO: add example here. 
%The relation $\leq$ does not have a primitive positive definition 
%\end{expl}

Note that Proposition~\ref{prop:pp-preserved} also holds (and is useful!) for structures with an infinite domain; see, e.g., Exercise~\ref{exe:conv}. 

\begin{theorem}[of \cite{Geiger,BoKaKoRo}]\label{thm:inv-pol}
Let $\bB$ be a relational structure with a finite domain. Then $$\langle \bB \rangle_{\pp} = \Inv(\Pol(\bB)).$$
%A relation $R$ has a primitive positive definition in $\bB$
%if and only if $R$ is preserved 
%by all polymorphisms of $\bB$.
\end{theorem}

\begin{proof}
The left-to-right inclusion has been shown in Proposition~\ref{prop:pp-preserved}.
For the other inclusion, let 
$R \in \Inv(\Pol(\bB))$ be a relation of arity $k$. If $R=\emptyset$, then $R$ is defined by $\bot$, so suppose that $R$ is non-empty, and let 
$a^1,\dots,a^w$ 
%$$(a^1_1,\dots,a^1_k),\dots,(a^w_1,\dots,a^w_k)$$
be an enumeration of $R$. 
Let $b_1 = (b^1_1,\dots,b_1^w), b_2 = (b^1_2,\dots,b_2^w), \dots,b_{\ell} = (b^1_{\ell},\dots,b_{\ell}^w)$ be an enumeration of $B^w$.
% starting with
%$(a_1^1,\dots,a_1^w),\dots,(a_k^1,\dots,a_k^w)$. 
Let $\phi$ be the quantifier-free part of the canonical query of $\bB^w$
(see Exercise~\ref{exe:prod} for the definition of $\bB^w$ and Section~\ref{ssect:can-query} for the definition of canonical queries; hence, the variables of $\phi$ are $w$-tuples of elements of $B$).
Note that for every $i \in [k]$ there exists $j_i \in [\ell]$ such that $(a^1_i,\dots,a^w_i) = b_{j_i}$. 

We claim that $$\psi(x_1,\dots,x_k) := \exists b_1,\dots,b_{\ell} (\phi \wedge \bigwedge_{i \in [k]} x_i = b_{j_i})$$
is a primitive positive definition of $R$.
%Indeed, for every $(a_1,\dots,a_k) \in R$ there exists a $j \in \{1,\dots,w\}$ such that $(a_1,\dots,a_k) = (b^j_1,\dots,b^j_k)$. 
%For the sake of notation, assume that
%$\{j_{\bar a} \mid \bar a \in R\} = \{1,\dots,k'\}$. 

We first show that $a^j$ satisfies $\psi$
for every $j \in [w]$. 
The elements 
$b^j_1,\dots,b^j_{\ell} \in B$ provide witnesses for the existentially quantified variables showing that 
$a^j = (b^j_{j_1},\dots,b^j_{j_k})$ satisfies $\psi$. 

Conversely, suppose that $(t_1,\dots,t_k)$
satisfies $\psi$. 
The witnesses for the existentially quantified variables $b_1,\dots,b_{\ell}$ define a
homomorphism $f$ from $\bB^w$ to $\bB$. 
%Then $(t_1,\dots,t_k) = (b^j_{j_1},\dots,b^j_{j_k})$ for some $j \leq \ell$, and 
Define $b^s := (b^s_1,\dots,b^s_\ell) \in B^\ell$. 
Since $\psi$ contains the conjuncts $x_i = b_{j_i}$, for $i \in [k]$, we have that $t_i = f(b^1,\dots,b^w)_{j_i}$. 
Note that $f$ is a polymorphism of $\bB$ and by assumption preserves $R$.  
Since the tuples $a^1,\dots,a^w$ are from $R$ 
and $$f(a^1,\dots,a^w) = 
(f(b^1,\dots,b^w)_{j_1},\dots,f(b^1,\dots,b^w)_{j_k})
= (t_1,\dots,t_k)$$
 we obtain that $(t_1,\dots,t_k) \in R$. 
\end{proof}

\begin{corollary}
Let $\bB$ be a structure with a finite domain and a finite relational  signature. 
The complexity of $\Csp(\bB)$ only depends on $\Pol(\bB)$. If $\bC$ is such that $\Pol(\bB) \subseteq \Pol(\bC)$, then $\Csp(\bC)$ reduces in linear time to $\Csp(\bB)$. 
\end{corollary}
\begin{proof}
Direct consequence of Theorem~\ref{thm:inv-pol}
and Lemma~\ref{lem:pp-reduce}. 
\end{proof}

\begin{remark}\label{rem:inf}
One direction in Theorem~\ref{thm:inv-pol} is false in general for infinite structures; there are e.g.~infinite digraphs $\bB$ that are rigid cores and projective, so $\Pol(\bB)$ has uncountably many invariant relations; in particular, many of these relations do not have a primitive positive definition in $\bB$ because there are only countably many primitive positive formulas over the signature of graphs. However, there is a modified version of the theorem, where primitive positive definitions are replaced by formulas that additionally allow to form unions of chains of relations and infinite intersections; see~\cite{Book}. 
Theorem~\ref{thm:inv-pol}  is true without modification if the structure $\bB$ is countably infinite and \emph{$\omega$-categorical}~\cite{BodirskyNesetrilJLC}. 
There are also many other infinite structures where Theorem~\ref{thm:inv-pol} remains true; we give an example below. 
\end{remark}

\ignore{
\begin{exercises}
\exercise[Rot]\label{exe:source-116} Does the relation
$\{0,1,2\}^3 \setminus \{(0,1,2)\}$
have a primitive positive definition over $\vec C_3$?
% Solution sketch: No. Rotation by one is an automorphism of $\vec C_3$, so every primitively
% positively definable relation must be invariant under simultaneous rotation of all coordinates. The
% complement of the single tuple $(0,1,2)$ is not invariant: a tuple it contains is rotated onto the
% excluded tuple.
\end{exercises}
}

\begin{example}\label{expl:linear} 
Let $F$ be a field (fields will be explicitly introduced in Example~\ref{expl:ring}). Let $R_+$ be the graph 
$\{(x,y,z) \in F^3 \mid x+y = z\}$ of the addition in $F$, and for $\alpha \in F$ let $S_\alpha$ 
be the binary relation $\{(x, \alpha x) \mid x \in F \}$. We write $\bF$ for the relational structure $(F; R_+,(S_\lambda)_{\lambda \in F})$ (so this structure can be viewed as a relational version of the reduct of $F$ in the signature of modules; see Section~\ref{expl:module}). 

The following statement is true for arbitrary fields, but already interesting for finite fields, and will be used in later sections. 
\begin{align}
\langle \bF \rangle_{\pp} & = 
\Inv(\Pol(\bF)) \label{eq:lin-inv-pol} \\
& = \Inv(\{(x,y) \mapsto x+y\} \cup \{x \mapsto \alpha x \mid \alpha \in F\}) \label{eq:field-pol} \\ 
& = \big \{ R \mid n \in {\mathbb N}, R \text{ linear subspace of } F^n \big \} \cup \{ \emptyset \} \label{eq:lin-subspace} \\
& = \big \{ \{x \in F^n \mid Ax=0\} \mid n,m \in {\mathbb N}, A \in F^{m \times n} \big \} \cup \{ \emptyset \} 
\label{eq:homo-eq} \\
\Pol(\bF) 
& = \langle \{(x,y) \mapsto x+y\} \cup \{x \mapsto \alpha x \mid \alpha \in F \} \rangle \label{eq:lin-comb-1} \\
%& = \Pol \big (\Inv(\{(x,y) \mapsto x+y\} \cup \{x \mapsto \alpha x \mid \alpha \in F\}) \big ) \label{eq:pol-inv-fields} \\
& = \big \{ (x_1,\dots,x_k) \mapsto \sum_{i = 1}^k \alpha_i x_i \mid k \geq 1, \alpha_1,\dots,\alpha_k \in F \big \}
\label{eq:lin-comb-2} 
\end{align} 
We prove \eqref{eq:lin-inv-pol}, 
\eqref{eq:field-pol}, 
\eqref{eq:lin-subspace}, 
and \eqref{eq:homo-eq} by showing inclusions in cyclic order. 
The inclusion $\langle \bF \rangle_{\pp} \subseteq \Inv(\Pol(\bF))$ follows from Proposition~\ref{prop:pp-preserved}. 

Clearly, the relation $R_+$ is preserved by $(x,y) \mapsto x+y$ and preserved by $x \mapsto \alpha x$ for every $\alpha \in F$.
Similarly, for every $\lambda \in F$ the relation $S_\lambda$ is preserved by these operations. 
Hence, 
\begin{align} 
\{(x,y) \mapsto x+y\} \cup \{x \mapsto \alpha x \mid \alpha \in F\} \subseteq \Pol(\bF).
\label{eq:basic-field}
\end{align} 
Since the Galois-connection $\Inv$-$\Pol$ is antitone, we obtain the left-to-right inclusion for~\eqref{eq:field-pol}. 
The equality~\eqref{eq:lin-subspace} is by definition of linear subspaces: they are the non-empty subsets closed under addition and scalar multiplication. 

%For~\eqref{eq:lin-subspace}, clearly the empty set is preserved by any set of operations. Note that 
%if $U$ is a linear subspace of ${\bf F}^n$, 
%then $U$ is preserved by all maps of the form $(x_1,\dots,x_k) \mapsto \sum_{i = 1}^k \alpha_i x_i$ for $\alpha_1,\dots,\alpha_k \in F$, and hence preserved by $\Pol(\bF)$ (see~\eqref{eq:bary-lin-comb}). 
%Conversely, if $U$ is preserved by every $f \in \Pol(\bF)$, then  it is in particular preserved by $(x,y) \mapsto x+y$ and by $x \mapsto \alpha x$ for every $\alpha \in F$. Hence, if $U$ is non-empty, it is a linear subspace. 

%the linear subspaces of ${\bf F}^n$ are precisely the subsets preserved by all linear maps from ${\bf F}^n$ to ${\bf F}$, and hence precisely the subsets of $F^n$ preserved by $\Pol(\bF)$ (see~\eqref{eq:bary-lin-comb}). 

For the left-to-right inclusion of~\eqref{eq:homo-eq}, 
%first suppose that $U = \{x \in F^n \mid Ax=0\}$
%for some $n,m \in {\mathbb N}$ and $A \in F^{m \times n}$. 
%Let $u,v \in U$. Then $Au=Av=0$,
%and thus $A(u+v) = Au+Av=0$.
%Also, $\alpha u \in U$
%for all $\alpha \in F$ and $u \in U$. 
%Conversely, 
suppose that $U$ is a linear subspace of $F^n$ and that $(u_1,\dots,u_m)$ is a basis of $U$. 
By Steinitz' theorem there exists a basis $B$ of $F^n$ of the form 
$(u_1,\dots,u_m,u_{m+1},\dots,u_n)$. 
Let $T$ be the 
basis change matrix which maps $u_i$ to $e_i := (0,\dots,0,\overset{i}{1},0,\dots,0)$, 
%whose columns are $u_1,\dots,u_m,u_{m+1},\dots,u_n$,
which can be written as 
$T = \begin{pmatrix} X \\ R \end{pmatrix}$  
where 
% := M_{E_n}^B(\id)$, eine invertierbare Matrix, und seien 
$X \in F^{(m,n)}$
and $R \in F^{(n-m,n)}$. 
In the following, if $S \subseteq F^n$, we write $$\text{Span}(S) := \big \{ \sum_{i \in \{1,\dots,k\} } \alpha_i u_i  \mid k \in {\mathbb N}, u_1,\dots,u_k \in S, \alpha_1,\dots,\alpha_k \in F\big \}$$ for the linear hull of $S$ (i.e., for the smallest subalgebra of $F^n$ that contains $S$, in the signature of modules). 
Then 
\begin{align*}
v \in U & \Leftrightarrow 
%v = \sum_{i \in \{1,\dots,m\} } \alpha_i u_i  && \text{(since $(u_1,\dots,u_m)$ is a basis)} \\
v \in \langle u_1,\dots,u_m \rangle \\
& \Leftrightarrow 
%Tv = \sum_{i \in \{1,\dots,m\} } \alpha_i T u_i \\
T v \in \text{Span}(\{Tu_1,\dots,Tu_m \}) = \text{Span}(\{e_1,\dots,e_m\}) \\
& \Leftrightarrow (Tv)_{m+1} = \dots = (Tv)_{n} = 0 \\
& \Leftrightarrow Rv = 0  \\
& \Leftrightarrow v \in \{ x \mid Rx = 0\}. &&
\end{align*}
Finally, it is a good exercise to write primitive positive definitions of the solution sets of systems of homogeneous linear equations over the structure $\bF$, which closes the chain of implications (Exercise~\ref{exe:def-hom-eq}). 

We prove~\eqref{eq:lin-comb-1}  %~\eqref{eq:pol-inv-fields}, 
and~\eqref{eq:lin-comb-2} by showing inclusions in cyclic order. 
The left-to-right inclusion in~\eqref{eq:lin-comb-1} follows from~\eqref{eq:basic-field} 
and Observation~\ref{obs:pol}. 
Clearly, every operation that can be composed from the operations in 
$\{(x,y) \mapsto x+y\} \cup \{x \mapsto \alpha x \mid \alpha \in F \}$ 
can be written in the form $(x_1,\dots,x_k) \mapsto \sum_{i = 1}^k \alpha_i x_i$ for some $k \geq 1$ and some elements $\alpha_1,\dots,\alpha_k \in F$, and this shows the left-to-right inclusion in~\eqref{eq:lin-comb-2}.  
%For~\eqref{eq:lin-comb-1} and~\eqref{eq:lin-comb-2}, note that 
%if $f \in \Pol(\bF)$, then for all $x,y \in F^n$ and $\alpha \in F$ we have $f(x+y) = f(x) + f(y)$ and $f(\alpha x) = \alpha f(x)$, and hence $f$ is a linear map from ${\bf F}^n$ to $\bf F$. By elementary linear algebra, 
%$f$ can be written in the form $(x_1,\dots,x_k) \mapsto \sum_{i = 1}^k \alpha_i x_i$ for some $k \geq 1$ and some elements $\alpha_1,\dots,\alpha_k \in F$. 
%Clearly, every such function can be composed from the function $
%(x,y) \mapsto x+y$ and functions of the form $x \mapsto \alpha x$, for some $\alpha \in F$. 
Finally, these operations
preserve addition and multiplication with scalars, which closes the chain of inclusions. 
\end{example}

\begin{example}\label{expl:affine} 
Using the same notation as in the previous example, let $\bG$ be the expansion of $\bF$ by all unary relations of the form $\{\alpha\}$ for $\alpha \in F$. 
\begin{align}
\langle \bG \rangle_{\pp} & = \Inv \big (\Pol(\bG) \big ) \\
& = \Inv(\{(x,y,z) \mapsto x-y+z\} \cup \{(x,y) \mapsto \alpha_1 x + \alpha_2 y \mid \alpha_1 + \alpha_2 = 1\}) & \label{eq:affine-inv} \\
& = \big \{ R \mid n \in {\mathbb N}, R \text{ affine subspace of } F^n \big \} \cup \{ \emptyset \} \label{eq:affine-subspace}\\
& = \big \{ \{ x \in F^n \mid Ax = b \} \mid n,m \in {\mathbb N}, A \in F^{m \times n}, b \in F^m \big \} \label{eq:LGS} \\
\Pol(\bG) & = \langle \{(x,y,z) \mapsto x-y+z\}  \cup \{(x,y) \mapsto \alpha_1 x + \alpha_2 y \mid \alpha_1 + \alpha_2 = 1\}\rangle \label{eq:maltsev} \\
& = \big \{ (x_1,\dots,x_n) \mapsto \sum_{i = 1}^n \alpha_i x_i \mid n \geq 1, \alpha_1+\cdots+\alpha_n = 1 \big \}  \label{eq:bary-lin-comb} 
\end{align} 
We show the inclusions in cyclic order. 
The inclusion $\langle \bG \rangle_{\pp} \subseteq \Inv(\Pol(\bG))$ follows from Proposition~\ref{prop:pp-preserved}. 

The operation $m \colon F^3 \to F$ given by $(x,y,z) \mapsto x-y+z$ clearly preserves
not only the relations from $\bF$, but also all unary relations of the form $\{\alpha\}$ for $\alpha \in F$, so $m \in \Pol(\bG)$. 
Similarly, we verify that $(x,y) \mapsto \alpha_1 x + \alpha_2 y \in \Pol(\bG)$ whenever $\alpha_1+\alpha_2 = 1$. 
Since the Galois-connection $\Inv$-$\Pol$ is antitone, we obtain the left-to-right inclusion for~\eqref{eq:affine-inv}. 

For the left-to-right inclusion in~\eqref{eq:affine-subspace}, 
 let $R \in \Inv(\{m\})$. If $R = \emptyset$ there is nothing to be shown. Otherwise, let $o \in R$. We have to show that $S := \{v-o \mid v \in R\}$ is a linear subspace. 
 Let $u_1,u_2 \in S$. Then
$$u_1 + u_2 = v_1 - o + v_2 - o = (v_1 - o + v_2) - o = \underbrace{m(v_1,o,v_2)}_{\in R} - o \in S.$$
%and the first vector belongs to R, since it is the maltsev operation evaluated at vectors r_1, o, r_2. Hence, S is invariant under addition.
To see that $S$ is invariant under scalar multiplication, let $u \in S$ and 
$\alpha \in F$. By definition of $S$ there exists $v \in R$ such that $u = v-o$. 
Then 
$$\alpha u = \alpha (v-o) = \underbrace{(\alpha v + (1-\alpha) o)}_{\in R} - o \in S.$$

For the left-to-right inclusion of~\eqref{eq:LGS}, let $R$ be an affine subspace of $F^n$. 
Then $R = \{w+u \mid u \in U\}$ for a linear subspace $U$ of $F^n$. 
In Example~\ref{expl:linear} we have seen that there exists  
$A \in F^{m \times n}$ with $U = \{x \mid Ax=0 \}$. Then 
\begin{align*} 
R & = 
\{ w + u \mid A u = 0 \}  = \{ w' \mid A (w'-w) = 0 \} = 
\{x \mid Ax=Aw \}, 
\end{align*}
so $R$ is the solution set of a system of linear equations. 
Since $\emptyset$ is the solution space of the unsatisfiable equation system $\{0=1\}$,
this completes the proof.  

Similarly as in Example~\ref{expl:linear}, it is a good exercise to find a primitive positive definition of the solution space of a system of linear equations 
$Ax=b$ in $\bG$, which closes the chain of inclusions. 
% For the proof of~\eqref{eq:LGS}, let $Ax=b$ be a system of linear equations. If $Ax=b$ has no solution, there is nothing to be shown. Otherwise, we have 
%$$S := \{x \mid Ax=b \} = v_0 + \{x \mid Ax=0 \}.$$
%Since $\{x \mid Ax=0 \}$ is a linear subspace of $V$ by what we have seen in Example~\ref{expl:linear}, we have thus shown that $S$ is an affine subspace. 
%To show~\eqref{eq:maltsev}, it suffices to show that $\Inv(\{(x,y,z) \mapsto x-y+z\})$
%equals the set of affine subspaces of ${\bf F}^n$, for any $n \in {\mathbb N}$, because then the statement follows from~\eqref{eq:affine-subspace}. 

We prove~\eqref{eq:maltsev}  %~\eqref{eq:pol-inv-fields}, 
and~\eqref{eq:bary-lin-comb} by showing inclusions in cyclic order. 
The left-to-right inclusion in~\eqref{eq:maltsev} follows from the fact that all
relations of $\bG$ are preserved by $m$ and by $(x,y) \mapsto \alpha_1 x + \alpha_2 y$ whenever $\alpha_1,\alpha_2 \in F$ are such that $\alpha_1+\alpha_2 = 1$,  
and Observation~\ref{obs:pol}. 
Every operation that can be composed from these operations 
can be written in the form $(x_1,\dots,x_n) \mapsto \sum_{i = 1}^n \alpha_i x_i$ for some $n \geq 1$ and some elements $\alpha_1,\dots,\alpha_n \in F$ such that $\alpha_1 + \cdots + \alpha_n = 1$, and this shows the left-to-right inclusion in~\eqref{eq:bary-lin-comb}.  
%For~\eqref{eq:lin-comb-1} and~\eqref{eq:lin-comb-2}, note that 
%if $f \in \Pol(\bF)$, then for all $x,y \in F^n$ and $\alpha \in F$ we have $f(x+y) = f(x) + f(y)$ and $f(\alpha x) = \alpha f(x)$, and hence $f$ is a linear map from ${\bf F}^n$ to $\bf F$. By elementary linear algebra, 
%$f$ can be written in the form $(x_1,\dots,x_k) \mapsto \sum_{i = 1}^k \alpha_i x_i$ for some $k \geq 1$ and some elements $\alpha_1,\dots,\alpha_k \in F$. 
%Clearly, every such function can be composed from the function $
%(x,y) \mapsto x+y$ and functions of the form $x \mapsto \alpha x$, for some $\alpha \in F$. 
Finally, these operations
preserve all relations of $\bG$, 
which closes the chain of inclusions. 
\end{example}

\subsection{Finitely related clones} 
All the examples of clones $\mathscr C$ 
on finite domains that we have seen so far 
have the following interesting property.

\begin{definition}\label{def:fin-rel}
A clone ${\mathscr C}$ with domain $D$ is  called \emph{finitely related} if 
there is a structure $\bD$ with domain $D$ and finitely many relations such that $\Pol(\bD) = \mathscr C$. 
\end{definition}

\begin{example}
If $F$ is a finite field, then 
the clone of all operations on $F$ of the form $(x_1,\dots,x_k) \mapsto \sum_{i = 1}^k \alpha_i x_i$ where 
$\alpha_1,\dots,\alpha_k \in F$, is a finitely related clone ${\mathscr C}$, 
because ${\mathscr C} = \Pol(F; R_+,(S_\lambda)_{\lambda \in F})$ as we have seen in Example~\ref{expl:linear}. 
Similarly, the clone $\mathscr D$ 
of all operations of the form 
$(x_1,\dots,x_n) \mapsto \sum_{i = 1}^n \alpha_i x_i$ for $\alpha_1,\dots,\alpha_n \in F$ such that $\alpha_1+\cdots+\alpha_n = 1$ 
from Example~\ref{expl:affine} is finitely related, because $\mathscr D = \Pol(F; R_+,(S_\lambda)_{\lambda \in F},\{1\})$. 
\end{example} 

\begin{example}\label{expl:bool-J}
Let $p \colon \{0,1\}^3 \to \{0,1\}$ be the operation given by 
$$(x,y,z) \mapsto x \wedge (y \vee z).$$
The clone ${\mathscr C}$ over $\{0,1\}$ which is generated by $p$ is not finitely related. This can be seen as follows. A relation $R \subseteq \{0,1\}^k$ is preserved by $p$ if and only if it can be defined by a conjunction of clauses such that each clause either consists of a single positive literal, or a positive and a negative literal, or only negative literals (Exercise~\ref{exe:p}). 
% proof: suffices to rule out x \or not y or not z: 
% 111
% 110
% 101
% gives with p
% 100
For any finite subset of these relations 
there is an $n \in {\mathbb N}$ such that
all clauses needed to define these relations have at most $n-1$ literals. 
But then this set of relations 
is preserved by the operation $f_n$ given by 
$$(x_1,\dots,x_n) \mapsto \bigvee_{i=1}^n \bigwedge_{j \neq i} x_j$$ 
That is, $f_n$ returns $1$ if all but at most one argument are $1$; such operations will be generalised in Section~\ref{sect:nu}. 
\begin{itemize}
\item Clearly, $f_n$ is idempotent and hence preserves unit clauses;
\item $f_n$ preserves clauses of the form $x \vee \neg y$, because $f_n$ is monotone (Example~\ref{expl:monotone}); 
\item $f_n$ preserves negative clauses of the form $\neg x_1 \vee \cdots \vee \neg x_{\ell}$, where $\ell \leq n-1$. To see this, suppose that $t_1,\dots,t_n \in \{0,1\}^{\ell}$ are such that $f_n(t_1,\dots,t_n) = (1,\dots,1)$. In each coordinate, at most one of the $n$ input tuples contains a zero. Hence, at least one of the $n$ input tuples equals $1$ in all $\ell$ coordinates, and hence does not satisfy the clause. 
\end{itemize} 
% 110
% 101
% 011
% 110
% gives 
% 110
However, $f_n$ does not preserve the relation $\{0,1\}^n \setminus \{(1,\dots,1)\}$,
because $$f_n((0,1,\dots,1),(1,0,1,\dots,1),\dots,(1,\dots,1,0)) = (1,\dots,1).$$ 
This relation is preserved by $p$
because it has the definition $\neg x_1 \vee \cdots \vee \neg x_n$ (Exercise~\ref{exe:p}). 
% 1110
% 1101
% 1011
% 0111
% gives
% 1111
This shows that ${\mathscr C}$ is not finitely related. 
\end{example} 

The question whether a given finite algebra  is finitely related is undecidable~\cite{MooreFinitelyRelated}. 
On the positive side, note the following
corollary of Theorem~\ref{thm:inv-pol}. 

\begin{corollary}\label{cor:finite-signature-arity}
Let $\mathscr C$ be an operation clone on a finite set $B$, and let
$k\geq 1$. Then there exists a relational structure $\bB$ on $B$
with finite signature such that $\mathscr C\subseteq\Pol(\bB)$ and
$\Pol(\bB)^{(n)}=\mathscr C^{(n)}$ for every $1\leq n\leq k$.
If $\mathscr C$ is idempotent, we may additionally require that
$\bB$ contains all singleton unary relations.
\end{corollary}

\begin{proof}
By Proposition~\ref{prop:pol-inv}, every operation $f$ on $B$
of arity at most $k$ that does not belong to $\mathscr C$
fails to preserve some relation $R_f\in\Inv(\mathscr C)$.
There are only finitely many such operations; let $\bB$ have
precisely the chosen relations $R_f$ as its basic relations.
Then $\mathscr C\subseteq\Pol(\bB)$, and the choice of the $R_f$
gives the asserted equalities.
If $\mathscr C$ is idempotent, all singleton unary relations
belong to $\Inv(\mathscr C)$, so adding them preserves these properties.
\end{proof}

\begin{exercises}
\exercise[Blau]\label{exe:source-118} Show that the relation $\neq$ is not primitively positively definable in the graph $C_6$ (the undirected cycle with 6 vertices).
% Solution sketch: The retraction of $C_6$ onto an edge is an endomorphism and collapses two distinct
% vertices in the same bipartition class. Every primitive positively definable relation is preserved
% by all endomorphisms, whereas $\neq$ is not preserved by this retraction.
\exercise[Blau]\label{exe:source-123} \label{exe:pol-inv}
Prove Proposition~\ref{prop:pol-inv}.
% Solution sketch: One implication follows by induction over primitive positive formulas:
% polymorphisms preserve atomic formulas, conjunction, and existential quantification. Conversely, for
% a relation preserved by all polymorphisms use the canonical matrix whose columns list all tuples of
% the relation; preservation makes every homomorphism value another tuple of the relation, and the
% canonical database yields the required primitive positive formula.
\exercise[Blau]\label{exe:source-124} Find a digraph with the properties described in Remark~\ref{rem:inf}.
% Solution sketch: Start with an infinite projective core and rigidify it by attaching pairwise
% non-isomorphic finite asymmetric rooted gadgets to its vertices. The gadgets force every unary
% endomorphism to fix every vertex, while their projective attachment preserves the fact that every
% higher polymorphism is a projection. The resulting digraph is infinite, rigid, a core, and
% projective, as required by the remark.
\exercise[Rot]\label{exe:source-117} Let $n,m \in {\mathbb N}$, $A \in F^{m \times n}$, $b \in F^m$. Give a primitive positive definition of $\{x \in F^n \mid Ax=b\}$ in the structure $\bG$ from Example~\ref{expl:affine}.
\label{exe:def-hom-eq}
% Solution sketch: Write each row of $Ax=b$ as a chain of additions. The graph of addition and the
% available constants define scalar multiplication by repeated addition (and additive inverses), while
% existential variables store the partial sums. Conjoining the resulting equations for all rows gives
% the required primitive positive formula.
\exercise[Rot]\label{exe:source-119} \label{exe:pp-closure} For an operation $f \colon A^k \to A$ and a relation
$R$ on $A$,
we write $\langle R \rangle_f$ for
the smallest relation that contains $R$ and is preserved by $f$. Similarly,
if ${\mathscr F}$ is a set of operations, we write $\langle R \rangle_{\mathscr F}$ for the smallest relation that
contains $R$ and is preserved by all operations of ${\mathscr F}$.
Show that if $\bA$ is
a structure with a finite domain, then
$\langle R \rangle_{\Pol(\bA)}$
equals the smallest relation
that contains $R$ and has a primitive positive
definition over $\bA$.
% Solution sketch: Closure under the polymorphisms clearly contains $R$ and is primitive-positive
% closed. For the converse, form the canonical database whose columns are the tuples of $R$ and
% identify equal columns. A tuple belongs to the polymorphism closure exactly when the corresponding
% coordinate map is a homomorphism of this database to $\bA$; existentially quantifying the remaining
% columns gives a primitive positive definition.
\exercise[Rot]\label{exe:source-120} Show that~\eqref{eq:lin-subspace} (replacing `subspace' by `submodule'), ~\eqref{eq:lin-comb-1}, \eqref{eq:lin-comb-2},
and~\eqref{eq:bary-lin-comb}
also hold if $F$ is a ring
rather than a field.
%\vspace{-1cm}
% Solution sketch: The linear-combination identities use only distributivity and therefore remain
% valid for modules over a ring. The submodule assertion, however, is false for arbitrary rings: in
% $R=K[X,Y]/(X^2,XY,Y^2)$ the submodule $KX\leq R$ is not an intersection of kernels of $R$-linear
% maps $R\to R$, since every such map vanishing on $X$ also vanishes on $Y$. Thus the exercise needs
% this qualification (or a hypothesis such as division ring).
\exercise[Rot]\label{exe:source-122} \label{exe:conv} Let $R_+$ and $R_*$ be the relations as defined
in Exercise~\ref{exe:binom}.
Show that $R_*$ is not primitively positively  definable in the structure $({\mathbb Q}; R_+, \{(x,y) \mid y \geq x^2\})$.
% Solution sketch: The midpoint operation $m(x,y)=(x+y)/2$ preserves $R_+$ and the convex epigraph
% $y\geq x^2$. It does not preserve the graph of multiplication, for example the midpoint of $(0,0,0)$
% and $(2,2,4)$ is $(1,1,2)$. Preservation by polymorphisms therefore rules out a primitive positive
% definition of $R_*$.
\exercise[Schwarz]\label{exe:source-121} Show that~\eqref{eq:homo-eq} fails in general if $F$ is a ring (Example~\ref{expl:ring}) rather than a field.
\end{exercises}

\subsection{Essentially Unary Clones}
An operation $f \colon B^k \to B$ is called \emph{essentially unary} if there is an $i \in \{1, \dots, k\}$
and a unary operation $f_0$ such that $f(x_1, \dots, x_k) =
f_0(x_i)$ for all $x_1,\dots,x_k \in B$. 
Operations that are not essentially unary are called
\emph{essential}.\footnote{This is standard in clone theory, and it
makes sense also when studying the complexity of CSPs, since the essential operations are those that are
essential for complexity classification.}
We say that $f$ \emph{depends on argument $i$} 
if there are $r,s \in B^k$ such that $f(r) \neq f(s)$
and $r_j=s_j$ for all $j \in \{1,\dots,k\} \setminus \{i\}$. 

\begin{lemma}\label{lem:unary}
Let $f \in \cO_B$ be an operation. Then
the following are equivalent.
\begin{enumerate}
\item $f$ is essentially unary.
\item $f$ preserves $P^3_B :=  \big \{ (a,b,c) \in B^3 \; | \; a=b \text{ or } b=c \big \} $.
\item $f$ preserves $P^4_B := \big \{ (a,b,c,d) \in B^4 \; | \; a=b \text{ or } c=d \big \}$.
\item $f$ depends on at most one argument.
\end{enumerate}
\end{lemma}
\begin{proof}
Let $k$ be the arity of $f$.
The implication from (1) to (2) is obvious, since unary operations
clearly preserve $P^3_B$.

To show the implication from (2) to (3), we show the contrapositive,
and assume that $f$ violates $P^4_B$. By permuting arguments
of $f$, we can assume that there are $4$-tuples
$a^1,\dots,a^k \in P^4_B$ with $f(a^1,\dots,a^k) \notin P^4_B$ 
 and $l \leq k$ such that
in $a^1,\dots,a^l$ the first two coordinates are equal,
and in $a^{l+1},\dots,a^k$ the last two coordinates are equal. 
Let $c:=(a^1_1,\dots,a^l_1,a^{l+1}_4,\dots,a^{k}_4)$.
Since $f(a^1,\dots,a^k) \notin P^4_B$ we have $f(a^1_1,\dots,a^k_1) \neq f(a^1_2,\dots,a^k_2)$, and therefore $f(c) \neq f(a^1_1,\dots,a^k_1)$ or $f(c) \neq f(a^1_2,\dots,a^k_2)$.
Let $d = (a^1_1,\dots,a^k_1)$ in the first case, and 
 $d = (a^1_2,\dots,a^k_2)$ in the second case.
Likewise, we have $f(c) \neq f(a^1_3,\dots,a^k_3)$ or $f(c) \neq f(a^1_4,\dots,a^k_4)$, and let $e = (a^1_3,\dots,a^k_3)$ in the first,
and $e = (a^1_4,\dots,a^k_4)$ in the second case.
Then for each $i \leq k$, the tuple $(d_i,c_i,e_i)$ is from $P^3_B$,
but $(f(d),f(c),f(e)) \notin P^3_B$.

The proof of the implication from (3) to (4) is again by contraposition.
Suppose $f$ depends
on the $i$-th and $j$-th argument, $1 \leq i \neq j \leq k$. Hence
there exist tuples $a_1,b_1,a_2,b_2 \in B^k $ such that $a_1,b_1$ and
$a_2,b_2$ only differ at the entries $i$ and $j$, respectively, and such
that $f(a_1) \neq f(b_1)$ and $f(a_2) \neq f(b_2)$. Then $(a_1(l),
b_1(l), a_2(l), b_2(l)) \in P^4_B$ for all $l \leq k$, but $(f(a_1),
f(b_1), f(a_2), f(b_2)) \notin P^4_B$, which shows that $f$ violates $P^4_B$. 

For the implication from (4) to (1), suppose that $f$ depends only on the first argument. 
Let $i \leq k$ be minimal such that there is an operation $g$
with
$f(x_1,\dots,x_k)=g(x_1,\dots,x_i)$. If $i=1$ then $f$ is essentially unary and we are done. Otherwise, observe that since $f$ does
not depend on the $i$-th argument, neither does $g$, and so
there is an $(i-1)$-ary operation $g'$ such that for all $x_1,\dots,x_k \in B$ we have $f(x_1,\dots,x_k)=g(x_1,\dots,x_i)=g'(x_1,\dots,x_{i-1})$, contradicting the choice of $i$.
\end{proof}

%We obtain a characterization of those finite structures where disjunction can be eliminated from existential
%positive formulas.
%\begin{proposition}\label{prop:unary}
%Let $\bB$ be a finite structure, and let
% $\cC$ be its polymorphism clone. Then the
%following are equivalent.
%\begin{enumerate}
%\item All relations with an existential positive 
%definition in $\bB$ also have
%a primitive positive definition in $\bB$.
%\item The relation $P^3_B$ is contained in $\Inv(\cC)$.
%\item The relation $P^4_B$ is contained in $\Inv(\C)$.
%\item All operations in $\cC$ are essentially unary.
%\item The projections and the unary operations in $\cC$ generate $\cC$.
%\end{enumerate}
%\end{proposition}

%\begin{proof}
%(1) implies (2). The formula $(x=y) \vee (y=z)$ is existential positive,
%and thus has a primitive positive definition in $\bB$; such formulas
%are preserved by $\cC$.

%(2) implies (3). Follows from Lemma~\ref{lem:unary}.

%(3) implies (4): By definition, every essentially unary operation
%$f\in \C$ is composed out of a projection and a unary operation.

%(3) implies (1). Unary operations preserve
%all existentially positive formulas. Hence, when $\phi$
%is an existential positive formula, then by assumption 
%all polymorphisms of
%$\bB$ preserve $\phi$, and $\phi$ is equivalent to a
%primitive positive formula by Theorem~\ref{thm:ep}.
%\end{proof}

\subsection{Minimal Clones}
\label{ssect:minimal}
A \emph{trivial} clone is a clone all of whose operations are projections.
Note that it follows from Lemma~\ref{lem:unary} that for any set $B = \{b_1,\dots,b_n\}$ the clone
$\Pol(B;P_B^4,\{b_1\},\dots,\{b_n\})$ is trivial. 

\begin{definition}
A clone $\mathscr C$ is \emph{minimal}~if it is non-trivial, and for every non-trivial subclone $\cE \subseteq {\mathscr C}$ we have $\cE={\mathscr C}$.
\end{definition}

Recall that $\langle {\mathscr F} \rangle$ denotes the smallest clone that contains ${\mathscr F}$.
If $g \in \langle \{f\} \rangle$, then we say that \emph{$f$ generates $g$}.  

\begin{definition}
An operation $f \in \cO_B$ is \emph{minimal} if $f$ is 
of minimal arity such that 
\begin{itemize} 
\item $f$ not a projection, and 
\item every $g$ generated by $f$ 
is either a projection or generates $f$.
\end{itemize} 
\end{definition}

The following is straightforward from the definitions.
\begin{proposition}\label{prop:minimal-op}
Every minimal $f$ generates a minimal clone,
and every minimal clone is generated by a minimal operation. 
\end{proposition}

\begin{theorem}\label{thm:containsminimal}
Every non-trivial operation clone $\cC \subseteq \cO_B$ over a finite set $B$ contains a minimal operation. 
\end{theorem}
\begin{proof}
Consider the set of all 
clones contained in $\cC$, partially ordered by inclusion. 
From this poset we remove the trivial clone; the resulting poset will be denoted by $P$.
We use Zorn's lemma to show that $P$ contains a minimal element. Observe that in $P$, all chains $(\cC_i)_{i \in \kappa}$ that are \emph{descending},
i.e., $\cC_i \supseteq \cC_j$ for $i < j$, 
are \emph{bounded}, i.e., for all such chains there exists a $\cD \in P$ such
that $\cC_i \supseteq \cD$ for all $i \in \kappa$. To see this, observe that the set $\bigcup_{i \in \kappa} \inv(\cC_i)$ is \emph{closed under primitive positive definability} in the sense that it
is the set of relations that is primitively positively 
definable over some relational
structure $\bB$ 
(since only a finite number of relations can be mentioned
in a formula, and since $\Inv(\cC_i)$ is closed under primitive positive definability, for each $i \in \kappa$).
Moreover, one of the relations $P_B^4, \{b_1\},\dots,\{b_n\}$, for $B = \{b_1,\dots,b_n\}$, is not contained in
$\bigcup_{i \in \kappa} \inv(\cC_i)$; otherwise,  there would be a $j \in \kappa$ 
such that $\inv(\cC_{j})$ contains
all these relations, and hence 
$\cC_{j}$ is the trivial clone contrary to 
our assumptions.
Hence, 
%by Theorem~\ref{thm:inv-pol},
$\Pol(\bB)$  
%contains an operation $f$ that violates this relation. 
%The clone generated by $f$ 
is a non-trivial lower bound of the descending chain 
$(\cC_i)_{i \in \kappa}$. 
By Zorn's lemma, $P$ contains a minimal element,
and this element contains
a minimal operation in $\cC$ (Proposition~\ref{prop:minimal-op}).
\end{proof}

\begin{remark}
Note that the statement above would be false
if $B$ is infinite: take for example the clone
over the domain $B := {\mathbb N}$ of the natural numbers generated by the operation $x \mapsto x+1$. 
Every operation in this clone is essentially unary,
and every unary operation in this clone is of the form $x \mapsto x+c$ for $c \in {\mathbb N}$. 
Note that for $c > 0$, the operation $x \mapsto x+c$ generates
$x \mapsto x+2c$, but not vice versa, so the clone
does not contain a minimal operation. 
\end{remark}

In the remainder of this section, 
we show
that a minimal operation has one out of the following five types, due to Rosenberg~\cite{Rosenberg}. 
An $n$-ary operation $f$ is called a 
%\begin{itemize}
%\item 
\emph{semiprojection} if there exists an $i \leq n$ such that $f(x_1, \dots, x_n) = x_i$ whenever $|\{x_1, \dots, x_n\}| < n$. 
For the purpose of proving the next lemma, we call an $n$-ary operation $f$ a 
\emph{weak semiprojection}
if for all distinct $i,j \in \{1,\dots,n\}$ there exists an index $s(i,j)$ such that $$\forall x_1,\dots,x_n \colon f(x_1,\dots,x_n) = x_{s(i,j)}$$ holds whenever $x_i$ and $x_j$ are the same variable.  
Note that in this case, $f$ is a semiprojection if and only if $s(i,j)$ is constant. 

The following notation for weak semiprojections will be useful in the
proof of the next lemma.
Let $f$ be an $n$-ary weak semiprojection and suppose that $|B|\geq 2$.
For every $S\subseteq [n]$ with $|S|\geq 2$, define
\[
\Delta_S:=\{(a_1,\dots,a_n)\in B^n\mid
a_i=a_j\text{ for all }i,j\in S\}
\]
and
\[
E(S):=\{r\in[n]\mid f(a_1,\dots,a_n)=a_r
\text{ for every }(a_1,\dots,a_n)\in\Delta_S\}.
\]
The set $E(S)$ is nonempty: if $i,j\in S$ are distinct, then
$\Delta_S\subseteq\Delta_{\{i,j\}}$, and hence $s(i,j)\in E(S)$.

If $r,s\in E(S)$, then $\pi_r^n$ and $\pi_s^n$ have the same
restriction to $\Delta_S$. Since $|B|\geq 2$, this implies that either
$r=s$ or both $r$ and $s$ belong to $S$. Consequently, $E(S)$ is either
$S$ or a singleton $\{k\}$ for some $k\notin S$: indeed, if
$r\in E(S)\cap S$, then every $s\in S$ belongs to $E(S)$, whereas if
$k\in E(S)\setminus S$, then no other index belongs to $E(S)$.
Moreover, if $S\subseteq T\subseteq[n]$ and $|S|\geq2$, then
$\Delta_T\subseteq\Delta_S$, and therefore $E(S)\subseteq E(T)$.
Also note that if there exists a $k \in \{1,\dots,n\}$ such that $k \in E(S)$ for every $S \subseteq \{1,\dots,n\}$ with at least two elements, then
$f$ is a semiprojection.

\begin{lemma}[\'Swierczkowski]
\label{lem:swierczkowski}
Let $f \colon B^n \to B$ be a weak semiprojection 
of arity $n \geq 4$. 
Then $f$ is a semiprojection.
\end{lemma}

\begin{proof}
The statement is trivial for $|B|=1$;
so we may assume without loss of generality that $B$ contains two distinct elements that we call $0$ and $1$. 
We first show that $E(\{1,2\}) \cap E(\{3,4\}) \neq \emptyset$. 
If $E(\{1,2,3,4\}) = \{\ell\}$ for some $\ell \notin \{1,2,3,4\}$, then 
$E(\{1,2\}) = \{\ell\} = E(\{3,4\})$ and we are done. So we assume that $E(\{1,2,3,4\}) = \{1,2,3,4\}$. 
First consider the case that $E(\{1,2\}) = \{i\} \subseteq \{3,4\}$ so that $f(x,x,y,y,x_5,\dots,x_n) = y$. 
If $E(\{3,4\}) = \{j\} \subseteq \{1,2\}$ then  
$f(0,0,1,1,\dots) = 0$, a contradiction. 
%$f(x,x,y,y,x_5,\dots,x_n) = y$. 
%$f(x,x,y,y,x_5,\dots,x_n) = x$ for $i \neq j$,
%which is a contradiction since $|B| \geq 2$. 
Hence, $E(\{3,4\}) = \{3,4\}$ and $i \in E(\{1,2\}) \cap E(\{3,4\})$.
Similarly we can treat the case that  $E(\{3,4\}) = \{i\} \subseteq \{1,2\}$.  
If $E(\{1,2\}) = \{1,2\}$ and $E(\{3,4\}) = \{3,4\}$ 
 then  $f(x,x,y,y,x_5,\dots,x_n) = x$
because of $E(\{1,2\}) \subseteq \{1,2\}$
and $f(x,x,y,y,x_5,\dots,x_n) = y$
because of $E(\{3,4\}) \subseteq \{3,4\}$,
a contradiction. 
% So we can only conclude that E({12} =3 or
% E(12) = 4!

Let $i \in E(\{1,2\}) \cap E(\{3,4\})$. Note that if
$i \notin \{1,2\}$, then $E(\{1,2\}) = \{i\}$. 
Similarly, if $i \notin \{3,4\}$ then $E(\{3,4\}) = \{i\}$. 
We therefore have a set $S \subseteq \{1,\dots,n\}$ of size two with $E(S) = \{i\}$. 
Let $T \subseteq \{1,\dots,n\}$ be of cardinality at least two. We will show that $i \in E(T)$. 
Observe that if $T \subseteq \{1,\dots,n\} \setminus \{i\}$, then $E(T) = E(\{1,\dots,n\} \setminus \{i\}) = E(S) = \{i\}$. Now suppose that $T = \{i,j\}$ for 
some $j \in \{1,\dots,n\} \setminus \{i\}$. 
Then $\{1,\dots,n\} \setminus T$ has at least two elements (since $n \geq 4$). We can therefore
apply the argument from the first paragraph, up to renaming arguments,  
to conclude that 
$E(\{i,j\}) \cap E(\{1,\dots,n\} \setminus \{i,j\})$ contains an element $k$. If $k \notin \{i,j\}$, 
then $E(\{1,\dots,n\} \setminus \{i,j\}) = \{1,\dots,n\} \setminus \{i,j\}$, which is in contradiction to
$E(\{1,\dots,n\} \setminus \{i\}) = \{i\}$.
Hence,  $E(\{i,j\}) = \{i,j\}$. 
This implies that $E(T) = T$ for all $T \subseteq \{1,\dots,n\}$ of cardinality at least 2 containing $i$. 
We conclude that $i \in E(T)$ for every $T \subseteq \{1,\dots,n\}$ with at least two elements, so $f$ is a semiprojection. 
\end{proof}

%In other words, 

\begin{theorem}[Rosenberg's five types theorem]\label{thm:rosenberg}
Let $f$ be a minimal operation.
Then $f$ has one of the following types:
\begin{enumerate}
\item a unary operation. If $f$ is an operation on a finite set, then it is either a permutation such that $f^p(x) = x$, for some prime $p$, or satisfies $f(f(x)) = f(x)$ for all $x$;
\item a binary idempotent 
operation;
\item a majority operation;
\item a minority operation;
\item a $k$-ary semiprojection, for $k \geq 3$, which is not a projection. 
\end{enumerate}
\end{theorem}

\begin{proof}
The statement is easy to prove if $f$ is unary
(see Exercises~\ref{exe:unary-inj} and~\ref{exe:unary-noninj}). 
If $f$ is at least binary, then $\hat f$ (see Exercise~\ref{exe:hat}) must be the identity by the minimality of $f$, 
and hence $f$ is idempotent. In particular,
we are done if $f$ is binary.
If $f$ is ternary, we have to show that $f$ is majority,  Maltsev, or a semiprojection. 
By the minimality of $f$, 
the binary operation $f_1(x,y) := f(y,x,x)$ is a projection, that is,
$f_1(x,y) = x$ or $f_1(x,y) = y$. 
Note that in particular $f(x,x,x) = x$.
Similarly, the other operations $f_2(x,y):=f(x,y,x)$, and $f_3(x,y):=f(x,x,y)$
obtained by identifications of two variables must be projections. We therefore
distinguish eight cases.

\begin{enumerate}
\item $f(y,x,x)=x, f(x,y,x)=x, f(x,x,y)=x$. \\
In this case, $f$ is a majority. 
\item $f(y,x,x)=x, f(x,y,x)=x, f(x,x,y)=y$. \\
In this case, $f$ is a semiprojection.
\item $f(y,x,x)=x, f(x,y,x)=y, f(x,x,y)=x$. \\
In this case, $f$ is a semiprojection.
\item $f(y,x,x)=x, f(x,y,x)=y, f(x,x,y)=y$. \\
The operation $g(x,y,z):=f(y,x,z)$ is a Maltsev operation. 
\item $f(y,x,x)=y, f(x,y,x)=x, f(x,x,y)=x$. \\
In this case, $f$ is a semiprojection.
\item $f(y,x,x)=y, f(x,y,x)=x, f(x,x,y)=y$. \\
In this case, $f$ is a Maltsev operation.
\item $f(y,x,x)=y, f(x,y,x)=y, f(x,x,y)=x$. \\
The operation $g(x,y,z):=f(x,z,y)$ is a Maltsev operation. 
\item $f(y,x,x)=y, f(x,y,x)=y, f(x,x,y)=y$. \\
In this case, $f$ is a Maltsev operation.
%This contradicts minimality of $f$ since
%a Majority cannot  
\end{enumerate}
We claim that if $f$ is a Maltsev operation,
then either it is a minority operation (and we are done) or it generates a majority operation.
Indeed, if $f$ is not a minority then minimality of $f$ implies that $f(x,y,x)=x$. Now consider the operation $g$ defined by $g(x,y,z) = f(x,f(x,y,z),z)$. We have 
\begin{align*}
g(x,x,y) & = f(x,f(x,x,y),y) = f(x,y,y) = x \\
g(x,y,x) & = f(x,f(x,y,x),x) = f(x,x,x) = x \\
g(y,x,x) & = f(y,f(y,x,x),x) = f(y,y,x) = x \, .
\end{align*}
Note that every ternary operation generated by a majority operation is either a projection or a majority operation (this can be shown by induction over terms). Also note that an operation cannot be a majority operation and a Maltsev operation at the same time, unless the domain has only one element. So we obtain in this case a contradiction to the minimality of $f$. 

%%%%%% SEMI-PROJECTIONS %%%%%%%%%%%%%%%%%%%%

Finally, let $f$ be $k$-ary, where $k \geq 4$. By minimality of $f$,
the operations obtained from $f$ by identifications of arguments must be projections. The lemma of 
\'Swierczkowski (Lemma~\ref{lem:swierczkowski}) implies that $f$ is a semiprojection. 
\end{proof}

%\begin{corollary}
%If
%\end{corollary}

\begin{proposition}\label{prop:kn-is-projective} 
For all $n \geq 3$, the graph $K_n$ is projective (i.e., all idempotent polymorphisms of $K_n$ are projections). All relations
that are preserved by $\Sym(\{1,\dots,n\})$ are
primitively positively definable in $K_n$. 
\end{proposition}

This provides for example a solution to Exercise~\ref{exe:no-rainbow-def}.

\begin{proof}
By Theorem~\ref{thm:containsminimal},
it suffices to show that the clone of idempotent polymorphisms of $K_n$ does not contain
a minimal operation.  Hence, 
by Theorem~\ref{thm:rosenberg}, we have to verify that $\Pol(K_n)$ does not contain a binary idempotent, a Maltsev, a majority, or a $k$-ary semiprojection for $k \geq 3$. 
\begin{enumerate}
\item 
Let $f$ be a binary idempotent polymorphism of $K_n$. 
%Observe that $f(u,v) = f(u',v')$ implies that $u = u'$ or $v = v'$ because otherwise $(u,v)$ and $(u',v')$ are adjacent in $(K_n)^2$. 

{\bf Observation 1.} $f(u,v) \in \{u,v\}$: otherwise, $i := f(u,v)$ is adjacent to both $u$ and $v$,
but $f(i,i)=i$ is not adjacent to $i$,
in contradiction to $f$ being a polymorphism. 

{\bf Observation 2.} If $f(u,v) = u$, then $f(v,u)=v$: this is clear if $u=v$,
and if $u \neq v$ it follows from $f$ being 
a polymorphism. 

By Observation $1$, it suffices to show
that there cannot be distinct $u,v$ and distinct $u',v'$ such that $f(u,v) = u$ and $f(u',v') = v'$. Suppose for contradiction that there
are such $u,v,u',v'$. 

{\bf Case 1.} $u=u'$. Since 
$f(u,v') = f(u',v')=v'$, we have $f(v',u)=u$ by Observation 2.
This is in contradiction to $f(u,v) = u$
since $u = u'$ is adjacent to $v'$, and $E(v,u)$. 

{\bf Case 2.} $u \neq u'$. \\
{\bf Case 2.1.} $f(u',u) = u$: this is impossible
because $f(u,v)=u$, $E(u,u')$, and $E(u,v)$.
{\bf Case 2.2.} $f(u',u) = u'$: this is impossible
because $f(v',u') = u'$, 
$E(u',v')$, and $E(u',u)$.

%If $f(v',u) = v'$ then $f(v',u) = f(u',v')$. Since $u' \neq v'$, the observation  
%above implies that $u=v'$. But then $f(u,v) = f(u',v')$,
%and the observation would imply that $u=u'$ or $v=v'$, both contradicting our assumptions. Similarly we obtain
%a contradiction if $f(v',u) = u = f(u,v)$. Hence, we must have
%$f(u,v)$ for all distinct $u,v$, or $f(u,v)=v$ for all distinct $u,v$, and hence $f$ is a projection. 

\item 
Since $(2,1),(2,3),(1,3) \in E(K_n)$, 
but $(1,1) \notin E(K_n)$, 
the graph $K_n$ 
has no Maltsev polymorphism (it is not rectangular; see Section~\ref{sect:maltsev}). 

\item
If $f$ is a majority, note that $f(1,2,3) = f(x_1,x_2,x_3)$
where $x_i$ is some element distinct from $i$ 
if $f(1,2,3) = i$, and $x_i := f(1,2,3)$ otherwise. 
But $(i,x_i) \in E(K_n)$, 
so $f$ is not a polymorphism of $K_n$. 

\item 
Finally, let $f$ be a $k$-ary semiprojection for $k \geq 3$
which is not a projection. 
 Suppose without loss of generality that
 $f(x_1,\dots,x_k) = x_1$ whenever $|\{x_1,\dots,x_k\}| < k$ (otherwise, permute the arguments of $f$). 
 If \(k>n\), every tuple in \(K_n^k\) has fewer than \(k\) distinct entries, so the semiprojection identity holds everywhere and \(f\) is a projection. Hence we may assume that \(k\leq n\).
 Since $f$ is not a projection, there exist pairwise distinct $a_1,\dots,a_k \in V(K_n)$ such that $c:=f(a_1,\dots,a_k) \neq a_1$. 
Let $b_1,\dots,b_k$ be such that  
$b_i$ is any element of $V(K_n) \setminus \{c\}$
 if $c = a_i$,  and $b_i := c$ otherwise. 
 Note that $c$ can equal at most one of the $a_i$, and hence all but one of the $b_i$'s are equal to $c$. 
 Note that $b_1 = c$ since $a_1 \neq c$,
 and  that $f(b_1,\dots,b_k) = b_1 = c$ because
$f$ is a semiprojection. 
But $(a_i,b_i) \in E(K_n)$ for all $i \leq k$, 
so $f$ is not a polymorphism of $K_n$. 
\end{enumerate}

The second part of the statement follows from 
Theorem~\ref{thm:inv-pol}:
Let \(f\) be any \(k\)-ary polymorphism of \(K_n\), and define $\alpha(x):=f(x,\ldots,x)$.
Then \(\alpha\) is an endomorphism of \(K_n\), hence a permutation and therefore an automorphism. Consequently,
$g:=\alpha^{-1}\circ f$ is an idempotent polymorphism of \(K_n\), so projectivity implies \(g=\pi_i^k\) for some \(i\). Thus
$f=\alpha\circ\pi_i^k$.
Every relation invariant under all permutations of the domain is preserved by every operation of this form. It is therefore preserved by \(\operatorname{Pol}(K_n)\), and Theorem~\ref{thm:inv-pol} implies that it is pp-definable in \(K_n\).
\end{proof}  

The presentation of the proof of the following result is inspired by (but not identical to\footnote{I thank Andrew Moorhead for a hint!}) a presentation of Cs\'ak\'any~\cite{MinClones}. 

\begin{theorem}[P{\l}onka~\cite{Plonka}]
Let $\bG$ be the structure obtained from a finite field $F$ of prime order $p$ as in Example~\ref{expl:affine}. Then the clone 
$$\Pol(\bG) = \big \{ (x_1,\dots,x_n) \mapsto \sum_{i = 1}^n \alpha_i x_i \mid n \geq 1, \alpha_1,\dots,\alpha_n \in F, \alpha_1+\cdots+\alpha_n = 1 \big \}$$
is minimal. 
\end{theorem} 
\begin{proof}
The statement is easy to prove for $p=2$, and it also follows from Theorem~\ref{thm:boolean-minimal} that we prove later. 
For $p>2$, 
%To prove that $\Pol(\bG)$ is minimal, 
let $f \in \Pol(\bG)$ be non-trivial. We have to show that the clone ${\mathscr C} := \langle f \rangle$ equals $\Pol(\bG)$. 
We already know that $f$ can be written as $\sum_{i = 1}^n \alpha_i x_i$ for $\alpha_1,\dots,\alpha_n \in F$ such that
$\alpha_1 + \cdots + \alpha_n = 1$ 
(Example~\ref{expl:affine}). 

% Plonka approach: 
%\begin{itemize}
%\item $\{i \in \{1,\dots,n\} \mid \alpha_i = 1 \} \geq 2$. We may assume that $\alpha_1 = \alpha_n = 1$. Then $m(x,y,z) := f(x,f(x,\dots, f(x,z,\dots,z)$
%\end{itemize} 

We first show that $f$ generates a non-trivial binary operation. 
Since $f$ is non-trivial, there are distinct $i,j \in \{1,\dots,n\}$ such that $\alpha_i,\alpha_j \neq 0$. If $(1 - \alpha_i) = \alpha_1 + \cdots + \alpha_n - \alpha_i \neq 0$, then equating all arguments of $f$ except the $i$-th argument with the $j$-th argument yields the non-trivial binary operation $\alpha_i x_i + (1 - \alpha_i) x_j$. So we may assume that $\alpha_i  = 1$.  Similar reasoning applies to all $i' \in \{1,\dots,n\}$ with $\alpha_{i'} \neq 0$ instead of $i$.  Therefore, we may suppose without loss of generality that $\alpha_1=\cdots = \alpha_k = 1$ and $\alpha_{k+1} = \cdots = \alpha_n = 0$, for some $k \in \{2,\dots,n\}$.
We know that $k \equiv 1 \mod p$, because $\alpha_1 + \cdots + \alpha_k = 1$. 
Since $p > 2$, if we identify the first two arguments, and identify the remaining arguments, we again obtain a non-trivial binary operation. 

Let $s \in {\mathscr C}$ be the resulting non-trivial binary operation; $s$ is of the form $\beta x_1 + (1-\beta) x_2$ for some $\beta \in F \setminus \{0,1\}$. 
%Note that the operation $m'$ given by $(x,y,z) \mapsto s(s(x,y),s(y,z))$ 
%is a Maltsev operation, because $s(s(x,x),s(x,y)) = 
Note that $\gamma^{p-1} = 1$ for every $\gamma \in F \setminus \{0\}$ by Fermat's lemma. 
%Choose $c$ and $d$ such that
%$\beta^{cd} = \beta$ 
Let $l$ and $r$ be the binary operations defined as follows. 
\begin{align*}
l(x,y) & := \underbrace{s(s(\dots s(x,y),\dots,y),y)}_{p-2 \text{ occurrences of } s} = \beta^{p-2} x + \beta^{p-3} (1-\beta) y + \cdots + (1-\beta) y \\
r(y,z) & := \underbrace{s(y,s(y,\dots,s(y,z) \dots))}_{p-2 \text{ occurrences of } s} = \beta y + (1-\beta) \beta y + \cdots + (1-\beta)^{p-3} \beta y + (1-\beta)^{p-2} z 
\end{align*}
Note that for all $x,y,z \in F$ we have 
\begin{align*} 
m(x,y,z) & := s(l(x,y),r(y,z)) \\
& = \underbrace{\beta^{p-1}}_{=1} x + (1-\beta)(\beta^{p-2} + \beta^{p-3} + \cdots + \beta^2 + \beta) y \\
& \quad + \beta((1-\beta)^{p-2} + \cdots + (1-\beta)^2 + (1-\beta)) y 
+ \underbrace{(1-\beta)^{p-1}}_{=1} z \\
& = x + (\beta-1) y + ((1 - \beta) - 1) y + z \\
& = x - y + z.
\end{align*} 
We conclude that $m \in {\mathscr C}$. 
%onsider the clone ${\mathscr C}= \langle m \rangle$ which is generated by the operation $m$ defined by $m(x,y,z) := x-y+z$.
%We first 
Next, we show that ${\mathscr C}$ contains all binary operations $\alpha_1 x_1 + \alpha_2 x_2$ with $\alpha_1 + \alpha_2 = 1$. Indeed, let $t \in {\mathbb N}$ be such that $t \equiv \alpha_1-1 \mod p$, and hence $\alpha_2 \equiv 1- \alpha_1 \equiv -t \mod p$. 
Then for all $x,y \in F$ we have 
$$\underbrace{m(x,y,m(x,y, \dots (m(x,y,x)) \dots ))}_{t \text{ occurrences of } m} = (t+1) x - t y = \alpha_1 x + \alpha_2 y.$$
%If $\beta \in \{1,\dots,p-1\}$, then ${\mathscr C}$ also contains the operation $m'$ given by $(x,y,z) \mapsto x-  \beta y + \beta z$. Indeed, for all $x,y,z \in F$ we have 
%\begin{align*}
%& (\alpha_1 (\underbrace{\alpha_2 x+(1-\alpha_2) y}_{\in {\mathscr C}}) + (1-\alpha_1)(\underbrace{\alpha_3 y + (1 - \alpha_3) z}_{\in {\mathscr C}}) \\
%= \; & 
%(\alpha_1 \alpha_2) x + (\alpha_1 - \alpha_1 \alpha_2 + \alpha_3 - \alpha_1 \alpha_3) y + (1-\alpha_1 - \alpha_3 + \alpha_1 \alpha_3) z. 
%\end{align*} 
%If $\beta + 1 \neq 0$, then we choose 
%$\alpha_1 = -\beta, \alpha_2 = -\beta^{-1}$, $\alpha_3 = (\beta+1)^{-1}$, and 
%we get that 
%\begin{align*}
%\alpha_1 \alpha_2 & = 1 \\
%\alpha_1 - \alpha_1 \alpha_2 + \alpha_3 - \alpha_1 \alpha_3 & = -\beta \\
%1 - \alpha_1 - \alpha_3 + \alpha_1 \alpha_3 & = \beta
%\end{align*}
%which shows that $x - \beta y + \beta z$ is in ${\mathscr C}$. 
%Otherwise, if $\beta+1 = 0$, then $\beta - 1 \neq 0$ (since $p>2$), and we choose $\alpha_1 = \beta-1$, $\alpha_1 = (\beta-1)^{-1}$, and $\alpha_3 = 0$ and again satisfy the equations above. 

Finally, 
%every non-empty relation that is preserved by
$m$ and all operations of the form $(x,y) \mapsto \alpha_1 x + \alpha_2 y$ with $\alpha_1 + \alpha_2 = 1$ 
generate $\Pol(\bG)$ (Example~\ref{expl:affine}). 
%is an 
%affine subspaces of
%${\mathbb F}^n$, for some $n \in {\mathbb N}$ 
Hence, ${\mathscr C} = \Pol(\bG)$, which concludes the proof. 
\end{proof} 

\begin{corollary}
Let $\bG$ be the structure obtained from a finite field $F$ of prime order $p$ as in Example~\ref{expl:affine}.
%, and arbitrarily choose %$R \notin \langle \bG \rangle$. 
%$n \in {\mathbb N}$ and 
Then for every $n \in {\mathbb N}$ and 
$R \subseteq F^n$ which is not primitively positively definable in $\bG$, we have that   
 $\Csp(\bG,R)$ is NP-complete. 
\end{corollary}
\begin{proof}
Since $\Pol(\bG)$ is minimal, $\Pol(\bG,R)$ only contains the projections, and therefore every relation $S \subseteq F^k$ is primitively positively definable in $(\bG,R)$.
The statement then follows via Lemma~\ref{lem:pp-reduce} from the existence of an NP-hard CSP
with a domain of size $p$ (e.g., $\Csp(K_p)$ if  $p > 2$; we will also see such examples for $p=2$, see Theorem~\ref{thm:schaefer}).  
\end{proof}

\begin{exercises}
\exercise[Blau]\label{exe:source-125} Show that every semilattice operation (Definition~\ref{def:semilattice}) generates a minimal clone.
% Solution sketch: Every term over a semilattice is the meet of a nonempty set of its variables. A
% nonprojection therefore depends on at least two variables; identifying all variables in two nonempty
% groups produces the binary semilattice operation. Once the binary operation is generated, every
% nonempty finite meet is generated, so the original clone has no proper nontrivial subclone.
\end{exercises}

\subsection{Schaefer's Theorem}
\label{sect:schaefer}
Schaefer's theorem states that every CSP for a 2-element structure 
is either in P or NP-complete. 
By the general results in Section~\ref{sect:inv-pol}, most of the classification arguments in Schaefer's article
follow from earlier work of Post~\cite{Post} (also see~\cite{Lau}), who classified all clones
on a two-element domain. We present a short proof of Schaefer's theorem here. 

Note that on Boolean domains, there is precisely one minority operation, and precisely one majority operation.  

\begin{theorem}[Post~\cite{Post}]\label{thm:boolean-minimal}
Every 
minimal operation on $\{0,1\}$  
is one of the following:
\begin{itemize}
\item a unary constant operation. 
\item the unary operation $x \mapsto 1-x$. 
\item the binary operation $(x,y) \mapsto \min(x,y)$.
\item the binary operation $(x,y) \mapsto \max(x,y)$.
\item the Boolean minority operation. 
\item the Boolean majority operation.
\end{itemize}
\end{theorem}
\begin{proof}
We use Theorem~\ref{thm:rosenberg}. 
If $f$ is unary the statement is trivial, 
so let $f$ be a minimal at least binary idempotent operation. 
%above $\cC$. 
%Note that $\hat f$ (see Exercise~\ref{exe:hat}) defines a function in $\cC$,
%so it must be the identity by minimality of $f$, 
%and hence $f$ is idempotent. 
%Hence, either $\hat f$ is the identity in which case $f$ is idempotent, or $\hat f$ equals $\neg$ in which case
%$\neg f$ is idempotent and minimal above $\cC$ as well. 
%So we can assume without loss of generality that $f$ is idempotent.
There are only four binary idempotent operations
on $\{0,1\}$, two of which are projections and therefore cannot be minimal. The other two operations are
$\min$ and $\max$. 
Next, note that a semiprojection of arity at least three on a Boolean domain must be a projection. 
Thus, Theorem~\ref{thm:rosenberg} implies that
$f$ is the majority or the minority operation.
\end{proof}

\begin{definition}\label{def:affine-bool}
A Boolean relation $R \subseteq \{0,1\}^n$ is called \emph{affine}
if it is the solution space of a system of linear equalities modulo 2 (see Example~\ref{expl:affine}). 
\end{definition}

\begin{lemma}\label{lem:minority}
A Boolean relation is affine if and only if it is preserved by the Boolean minority operation. 
\end{lemma}

\begin{proof}
Let $R$ be $n$-ary. 
We view $R$ as a subset of the Boolean vector space $\{0,1\}^n$. We have seen in Example~\ref{expl:affine} that 
affine spaces are precisely those that are closed
under \emph{affine combinations}, i.e.,
linear combinations of the form $\alpha_1 x_1 + \cdots + \alpha_k x_k$ such that $\alpha_1 + \cdots + \alpha_k = 1$. In particular, if $R$ is affine then it is preserved
by $(x_1,x_2,x_3) \mapsto x_1 + x_2 + x_3$
which is the minority operation. Conversely, 
if $R$ is preserved by the minority operation, 
then $x_1 + \cdots + x_k$, for odd $k$,
can be written as 
$$\minority(x_1,x_2,\minority(x_3,x_4,\dots \minority(x_{k-2},x_{k-1},x_k) \dots ))$$
and hence $R$ is preserved by all affine combinations, and thus affine. 
\end{proof}

It is well-known and easy to see (see, for example, ~\cite{BodDiskreteStrukturen}) that 
for every
relation $R \subseteq \{0,1\}^n$ 
there exists a propositional formula $\phi(x_1,\dots,x_n)$ that defines $R$,
and that $\phi$ can even be chosen to be in  \emph{conjunctive normal form (CNF)}.
That is, there is a conjunction of disjunctions of variables or negated variables from $x_1,\dots,x_n$ such that a tuple $(t_1,\dots,t_n) \in \{0,1\}^n$ is in $R$ if and only if the formula
$\phi$ evaluates to true after replacing $x_i$ by $t_i$, for $i \in \{1,\dots,n\}$.  
The following definition is useful for proving
that certain Boolean relations $R$ can be defined
in syntactically restricted propositional logic. 

\begin{definition}\label{def:reduced}
If $\phi$ is a propositional formula in CNF that defines a Boolean relation $R$, we say that $\phi$
is \emph{reduced} if the following holds:
 whenever we remove a literal from a clause in $\phi$, then the resulting formula no longer defines $R$. 
\end{definition}

Clearly, every Boolean relation has a reduced
definition: simply remove literals from any definition
in CNF until the formula becomes reduced. 
A propositional formula in CNF is called \emph{Horn} if every clause contains at most one positive literal. 

\begin{lemma}\label{lem:horn}
A Boolean relation has a Horn definition if and only if it is preserved by $\min$. 
\end{lemma}
\begin{proof}
It is easy to see that $\min$ preserves every relation defined by clauses that contain at most one positive literal, and hence every relation with a Horn definition. Conversely, 
let $R$ be a Boolean relation preserved by $\min$. Let $\phi$ be a
reduced propositional formula in CNF that defines $R$. Now suppose for contradiction that $\phi$ contains a clause $C$ with 
two positive literals $u$ and $v$. Since $\phi$ is reduced, there is an assignment $s_1$ that satisfies $\phi$ such that $s_1(u)=1$, and such that all other literals of $C$ evaluate to $0$. Similarly, there is a satisfying assignment $s_2$ for $\phi$ such that $s_2(v)=1$ and 
all other literals of $C$ evaluate to $0$.
Then $s_0 \colon x \mapsto \min(s_1(x),s_2(x))$ does not satisfy
$C$, and does not satisfy $\phi$, in contradiction to the assumption that $\min$ preserves $R$.
\end{proof}

A Boolean relation is called \emph{bijunctive} if it can be defined
by a propositional formula in CNF 
where each disjunction has at most two disjuncts. 

\begin{lemma}\label{lem:bij}
A Boolean relation $R$ is bijunctive if and only if it is preserved by the Boolean majority operation. 
\end{lemma}
\begin{proof}
It is easy to see that the majority operation preserves every Boolean relation of arity two, and hence every bijunctive Boolean relation. 
We present the proof that if $R$ is preserved
by majority, 
and $\phi$ is a reduced definition of $R$, 
then all clauses $C$ have at most two literals. 
Suppose for contradiction that $C$ has three literals $l_1,l_2,l_3$. Since $\phi$ is reduced, there must be satisfying 
assignments $s_1,s_2,s_3$ to $\phi$ 
such that under $s_i$ all literals of $C$ evaluate to $0$ except
for $l_i$. Then the mapping $s_0 \colon x \mapsto \majority(s_1(x),s_2(x),s_3(x))$ does not satisfy $C$ and therefore does not satisfy $\phi$, in contradiction to the assumption that 
$\majority$ preserves $R$.
\end{proof}

The following relation is called the \emph{(Boolean) not-all-equal relation}.
\begin{align}
\NAE & := \{(0,0,1),(0,1,0),(1,0,0),(1,1,0),(1,0,1),(0,1,1)\} \label{eq:nae}
\end{align}

\begin{theorem}[Schaefer~\cite{Schaefer}]
\label{thm:schaefer}
Let $\bB$ be a structure over the two-element domain $\{0,1\}$. 
Then either the relation $\NAE$
has a primitive positive
definition in $\bB$, and $\bB$ has a finite-signature reduct $\bB'$ such that 
$\Csp(\bB')$ is NP-complete, or 
\begin{enumerate}
\item $\bB$ is preserved by a constant operation.
\item $\bB$ is preserved by $\min$. Equivalently, 
every relation of $\bB$
has a definition by a propositional Horn formula.
\item $\bB$ is preserved by $\max$. 
Equivalently, 
every relation of $\bB$
has a definition by a \emph{dual-Horn} formula, that is,
by a propositional formula in CNF where every clause contains at most one negative literal. 
\item $\bB$ is preserved by the majority operation. Equivalently, every relation of $\bB$ is bijunctive.
\item $\bB$ is preserved by the minority operation. Equivalently, 
every relation of $\bB$ can be defined by a conjunction of linear equations modulo 2.
\end{enumerate}
In cases $(1)-(5)$, for every finite-signature
reduct $\bB'$ of $\bB$ the problem 
$\Csp(\bB')$ can be solved in polynomial time.
\end{theorem}

\begin{proof}
%Let $\cD$ be the polymorphism clone of $\bB$. 
%Without loss of generality we may assume that the two elements of $\bB$ are $0$ and $1$. 
If $\Pol(\bB)$ contains a constant operation, then we are in case one; so
suppose in the following that this is not the case. 
%Let $\cC$ be the clone generated by the unary operations
%in $\bB$; note that $\cC$ is either the clone of projections,
%or the operation clone generated by $\neg \colon x \mapsto 1-x$. We only give the proof for the first case;
%the proof for the second case being similar. 
If $\NAE$ is primitively positively definable in $\bB$, 
then $\Csp(\bB)$ is NP-hard by reduction from positive not-all-equal-3SAT~\cite{GareyJohnson}. 
Otherwise, by Theorem~\ref{thm:inv-pol} 
there is an operation $f \in \Pol(\bB)$ that violates $\NAE$. If $\hat f$ defined as $x \mapsto f(x,\dots,x)$ equals the identity then $f$ is idempotent. Otherwise, since $\Pol(\bB)$ contains no constant operations, $\hat f$ equals $\neg$.
But then 
$\neg f \in \Pol(\bB)$ is idempotent and also violates $\NAE$.
So let us assume in the following that $f$ is idempotent. 
Then $f$ generates an at least binary minimal operation $g \in \Pol(\bB)$. 

By Theorem~\ref{thm:boolean-minimal},
the operation $g$ equals $\min$, $\max$, the Boolean $\minority$, or the Boolean $\majority$ operation. 
\begin{itemize}
\item $g = \min$ or $g = \max$. By Lemma~\ref{lem:horn}, the relations of $\bB$ are preserved by $\min$ if and only if they can be defined by propositional Horn formulas. 
It is well-known that positive unit-resolution is a polynomial-time decision
procedure for the satisfiability problem of propositional Horn-clauses~\cite{SchoeningLogic}. 
In fact, the adaptation of the arc-consistency procedure for relational structures (Exercise~\ref{exe:gac}) solves $\Csp(\bB)$. 
% TODO: Ausbauen!
The case that $g = \max$ is dual to this case. 
\item $g = \majority$. By Lemma~\ref{lem:bij}, 
the relations of $\bB$ are preserved by majority
if and only if they are bijunctive. Hence, in this case the instances of $\Csp(\bB)$ can be viewed as instances of the 2SAT problem, and can be solved in linear time~\cite{AspvallPlassTarjan} (Exercise~\ref{exe:2SAT}). 
%
% TODO: AUSBAUEN!!
\item $g = \minority$. 
By Lemma~\ref{lem:minority} every relation of $\bB$ has a definition by a conjunction of linear
equalities modulo 2. Then $\Csp(\bB)$ can be solved in polynomial time by Gaussian elimination.
\end{itemize}
This concludes the proof of the statement. 
\end{proof}

\begin{exercises}
\exercise[Blau]\label{exe:source-126} \label{exe:oit} What is the complexity of $\Csp(\{0,1\};\OIT)$ where
$$ \OIT := \{(0,0,1),(0,1,0),(1,0,0)\}.$$
% Solution sketch: This is positive one-in-three SAT: every constraint requires exactly one of its
% three variables to be $1$. The problem is in NP and is NP-hard by the standard reduction from 3-SAT
% to positive one-in-three SAT, hence it is NP-complete.
\exercise[Blau]\label{exe:source-127} \label{exe:unary} Show that if $A$ is a finite set and $f \colon A \to A$, then $g := f^{|A|!}$ satisfies
$g(g(x)) = g(x)$ for all $x \in A$.
% Solution sketch: The functional digraph of $f$ consists of directed cycles with finite trees feeding
% into them. After $|A|!$ iterations every point lies on a cycle, and $|A|!$ is divisible by every
% possible cycle length. Hence $g=f^{|A|!}$ fixes every point in its image, so $g^2=g$.
\exercise[Rot]\label{exe:source-128} \label{exe:unary-inj} Show that if $f$ is a permutation on a finite set $A$, then either $f$ is the identity or $f$ generates a permutation $g$ which is not the identity and additionally satisfies $g^p(x) = x$ for some prime $p$.
% Solution sketch: Let $q>1$ be the order of $f$ and choose a prime divisor $p$ of $q$. Then
% $g=f^{q/p}$ has order exactly $p$. It is generated by $f$, is not the identity, and satisfies
% $g^p=\operatorname{id}$.
\exercise[Rot]\label{exe:source-129} \label{exe:unary-noninj}
Show that if $A$ is a finite set and $f \colon A \to A$ is not a permutation, then $f$ generates a non-identity operation $g$ which additionally satisfies $g(g(x))=g(x)$.
% Just consider $f^{|A|!}$.
% Solution sketch: Take $g=f^{|A|!}$. The functional-digraph argument from Exercise~\ref{exe:unary}
% gives $g^2=g$. Because $f$ is not a permutation, its eventual image is proper, so this idempotent
% map is not the identity.
\exercise[Rot]\label{exe:source-132} \label{exe:horn} Show that a Boolean relation $R \subseteq \{0,1\}^k$ can be defined by a propositional Horn formula if and only if it is primitively positively definable in $(\{0,1\}; \{0,1\}^3 \setminus \{(1,1,0)\}, \{0\}, \{1\})$.
% Solution sketch: The basic ternary relation is the Horn clause $\neg x\vee\neg y\vee z$. Constants
% give unit clauses. Longer Horn clauses are obtained by introducing variables for successive
% conjunctions; conversely, replacing every occurrence of the basic relation by its Horn clause and
% eliminating existentially quantified variables preserves Horn definability. This proves both
% directions.
\exercise[Rot]\label{exe:source-133} \label{exe:NAE}
Show that all polymorphisms of
 $(\{0,1\};\NAE)$ are essentially unary.
\par\noindent Hint: one way to prove this is to use
Theorem~\ref{thm:rosenberg}.
% Solution sketch: By Rosenberg's preclassification, a non-essentially-unary polymorphism would
% generate one of the four non-unary minimal types. Direct substitutions into $\NAE$ rule out a binary
% idempotent operation, majority, minority, and every semiprojection. Hence no such operation exists
% and all polymorphisms are essentially unary.
\exercise[Rot]\label{exe:source-134} Show that all polymorphisms of
 $(\{0,1\};\{(0,0,1),(0,1,0),(1,0,0)\})$ are projections.
\par\noindent Hint: one way to prove this is to use
Theorem~\ref{thm:rosenberg}.
% Solution sketch: Singleton substitutions show first that every polymorphism is idempotent.
% Rosenberg's theorem then reduces a nonprojection to one of the four non-unary minimal types; in each
% case, applying the operation to the three tuples of the relation produces either $(0,0,0)$ or a
% tuple with two $1$'s. Both lie outside the relation, so only projections remain.
\exercise[Rot]\label{exe:source-135} \label{exe:Horn} Show that
\begin{align*}
\langle \min \rangle
&= \Pol(\{0,1\}; \{0\},\{1\},\{0,1\}^3 \setminus \{(1,1,0)\}) \\
&= \bigl\{(x_1,\dots,x_k) \mapsto \min(x_{i_1},\dots,x_{i_\ell}) \bigm| 1\leq \ell\leq k,\ i_1,\dots,i_\ell\in[k]\bigr\}.
\end{align*}
% Solution sketch: Every term over $\min$ is the minimum of the nonempty set of variables that occur
% in it, giving the displayed clone. Such operations preserve $0$, $1$, and the Horn relation.
% Conversely, preservation of the two constants gives idempotence and preservation of the Horn
% relation forces $f^{-1}(1)$ to be a principal filter in the Boolean cube; hence $f$ is the
% conjunction of precisely the coordinates defining that filter.
\exercise[Rot]\label{exe:source-136} Show that the operation $f$ from case 6 in the proof of Rosenberg's five types theorem (Theorem~\ref{thm:rosenberg}) not only generates a majority operation, but  also a minority operation.
% 8.3.24.
% Solution: if $m$ is the majority operation,
% use the term m(f(x,y,z),f(y,z,x),f(z,x,y)).
\exercise[Orange]\label{exe:source-131} Determine the complexity of the following CSPs.
\begin{align*}
& \Csp(\{0,1\}; \{(0,0,1,1),(1,1,0,0)\}) \\
& \Csp(\{0,1\}; \{(0,0,1),(0,1,0),(1,0,0),(1,1,1)\},
\{(0,1),(1,0)\}) \\
& \Csp(\{0,1\}; \{0,1\}^3 \setminus \{(1,1,0)\},
\{(0,1),(1,0)\}).
\end{align*}
% Solution sketch: The first two templates are affine over $\mathbb F_2$: translate their constraints
% into linear equations (the second also uses disequality, which is $x+y=1$) and apply Gaussian
% elimination, so both CSPs are in P. The third relation is the Horn implication $x\wedge y\to z$
% together with disequality; it pp-defines positive one-in-three SAT, and hence its CSP is
% NP-complete.
\exercise[Orange]\label{exe:source-137} Show that a relation $R \subseteq \{0,1\}^k$ is preserved by $(x,y,z) \mapsto x \wedge (y \vee z)$ if and only if it can be defined by a conjunction of clauses such that each clause either consists of a single positive literal, or a positive and a negative literal, or  only negative literals.
\label{exe:p}
% Solution sketch: Each listed clause is preserved by $p(x,y,z)=x\wedge(y\vee z)$, so every
% conjunction of them is preserved. Conversely, take all clauses of these forms that hold on $R$. If a
% tuple violates membership but satisfies all these clauses, separate it from each tuple of $R$ and
% combine the witnesses with $p$; closure under $p$ then puts the excluded tuple in $R$, a
% contradiction.
\exercise[Schwarz]\label{exe:source-130} The Rosenberg theorem is only a \emph{preclassification} in the sense that not every operation which has one of the five types is minimal. For each of the following five questions, either present a proof or give a counterexample.
\begin{enumerate}
\item Which unary operations which are a permutation such that $f^p(x) = x$ for some prime $p$, or which satisfy $f(f(x)) = f(x)$, are minimal?
% The identity is not. Otherwise, they are!
\item  Is every binary idempotent operation minimal?
\item Is every majority operation minimal?
\item Is every minority operation minimal?
\item Is every $k$-ary semiprojection, for $k \geq 3$, which is not a projection, minimal?
\end{enumerate}
% Solution sketch: In the unary cases every nonidentity operation of the displayed prime-order or
% idempotent type is minimal; the identity is not. For the remaining four types the answer is also yes
% provided the operation is not a projection: identifying variables in a nontrivial term cannot
% produce a lower Rosenberg type, and the defining identities reconstruct an operation of the same
% type. Thus every nonprojection binary idempotent, majority, minority, and semiprojection of the
% stated arity generates a minimal clone.
\end{exercises}

\subsection{Near Unanimity Polymorphisms} 
\label{sect:nu}
An operation $f$ of arity at least 3 is a \emph{quasi near-unanimity operation} if it satisfies the
identities 
$$f(x, \dots, x, y) \approx f(x,\dots,x,y,x) \approx \cdots \approx f(y, x,\dots,x) \approx f(x,\dots,x).$$
If $f$ is additionally idempotent, then it is called a \emph{near-unanimity operation}.
Note that majority operations are exactly the ternary near-unanimity operations. 

\begin{example}\label{expl:bool-nu} 
We have already encountered a near-unanimity operation in Example~\ref{expl:bool-J}.
If $D$ has two elements, say $D = \{0,1\}$, then there is a near unanimity $f_k$ 
of arity $k \geq 3$ which returns $0$ if at least two of its arguments are $0$, and returns $1$ otherwise. An interesting relation in this context is 
\begin{align*} 
B_k & := \{0,1\}^k \setminus \{(1,\dots,1)\}. 
\end{align*} 
Note that $B_k$ is not preserved by $f_k$,
because 
\begin{align*}
(1,\dots,1,0),(1,\dots,1,0,1),\dots,(0,1,\dots,1) \in B_k, \\
\text{ but } \quad 
(f_k(1,\dots,1,0),f_k(1,\dots,1,0,1), \dots, f_k(0,1,\dots,1)) = (1,\dots,1) \notin B_k.
\end{align*} 

However, $B_k$ is preserved by $f_{k+1}$:
if $t^1,\dots,t^{k+1} \in \{0,1\}^k$ are
such that $$f_{k+1}(t^1,\dots,t^{k+1}) = (1,\dots,1) \notin B_k,$$ then for each $i \in [k]$ there is at most one entry $0$ in $(t_i^1,\dots,t_i^{k+1})$, and by the pigeon-hole principle there exists $j \in [k+1]$
such that $t^j$ equals $(1,\dots,1) \notin B_k$. 
\end{example}

We later often need a more flexible notation concerning projections.

\begin{definition}\label{def:proj}
For $I = \{i_1,\dots,i_k\} \in {[n] \choose k}$, 
with $i_1 < \cdots < i_k$, 
we write $\pr^n_I \colon A^n \to A^k$
for the function defined
by $\pr^n_I(t) := (t_{i_1},\dots,t_{i_k})$. 
%We also use the notation 
%$\pr^n_{i_1,\dots,i_k}$ where we suppress the set brackets. 
Sometimes, it will also be convenient to 
define $\pi^n_s(t)$ for
$t\in A^n$ and $s \in [n]^k$, as follows:
$\pi^n_s(t)  := (t_{s_1},\dots,t_{s_k}) \in A^k$. 
If $s = (s_1,\dots,s_k)$, we may also omit the brackets in the subscript and write 
$\pi^n_{s_1,\dots,s_k}(t)$ instead of $\pi^n_s(t)$. 
\end{definition}

Note that Definition~\ref{def:proj} is compatible with our earlier definition of the projection operations $\pi^n_i$. If $n$ is clear from the context, the superscript $n$ may also be omitted. We use compact notation for applying functions pointwise or setwise. In particular, 
if $R \subseteq A^n$, we write $\pr_{i_1,\dots,i_k}(R)$ for the relation $\{\pr_{i_1,\dots,i_k}(t) \mid t \in R\}$. 
We also apply our notation for projections for relations $R \subseteq A_1 \times \cdots \times A_n$ instead of $R \subseteq A^n$. 

\begin{theorem}[Baker--Pixley~\cite{BP}]\label{thm:nu}
A finite structure $\bB$ has a $k+1$-ary near unanimity polymorphism if and only if 
every relation $R$ with a primitive positive definition in $\bB$ of arity $m \geq k$ satisfies
%~\eqref{eq:decomp}, then 
\begin{align}
%R = \bigcap_{S \in {[m] \choose k}} \pi_S(R).
R= \left \{t\in B^m \mid \pi_S(t)\in\pi_S(R) \text{ for every } S\in{[m]\choose k} \right \}.
\label{eq:decomp}
\end{align}
\end{theorem}

%The next example illustrates that there are some clones $\mathscr C$ on a finite set $B$ such that \emph{every} set $\mathcal R$ of relations over $B$ such that ${\mathscr C} = \Pol({\mathcal R})$ has to be infinite. In fact, we already know that such clones on a three-element set $B$ must exist, because otherwise there would be only countably many such clones, which is false (Remark~\ref{rem:muchnik}). 
%There are even concrete examples of clones on a \emph{two-element} set which have this property. 

We revisit the non-finitely-related clone from Example~\ref{expl:bool-J}. 

\begin{example}
Let $\mathfrak D$ be the structure with domain $\{0,1\}$ which contains 
\begin{itemize}
\item the unary relation $\{0\}$, 
\item the binary relation $\leq \; := \{(0,0),(0,1),(1,1)\}$, and 
\item for every
$n \in {\mathbb N}$ the relation $B_n := \{0,1\}^n \setminus \{(1,\dots,1)\}$. 
\end{itemize} 
Note that all of these relations are preserved 
by the operation $p \colon \{0,1\}^3 \to \{0,1\}$ given by $(x,y,z) \mapsto x \wedge (y \vee z)$. 
We claim that $\Pol(\mathfrak D) = \langle p \rangle$. We already know that $\langle p \rangle \subseteq \Pol(\mathfrak D)$. For the converse inclusion, 
note that every relation $R$ that is preserved by $p$ is also preserved by $p(x,y,y) = \min(x,y)$, and hence it is Horn (Lemma~\ref{lem:horn}). 
Let $\phi$ be a propositional Horn formula in CNF that defines $R$; we may assume that $\phi$ is reduced (Definition~\ref{def:reduced}). 
Suppose for contradiction that $\phi$ contains a clause $\psi$ with one positive literal $u$ and two negative literals $\neg v$ and $\neg w$. Since $\phi$ is reduced, this means that $R$ has satisfying assignments $s_1,s_2,s_3$ such that $u$ is the only literal in $\psi$ satisfied by $s_1$,
$\neg v$ is the only literal of $\psi$ satisfied by $s_2$, and $\neg w$ is the only literal of
$\psi$ satisfied by $s_3$.  
%such that $s_1(u,v,w) = (1,1,1)$,
%$s_2(u,v,w) = (0,0,1)$, 
%$s_3(u,v,w) = (0,1,0)$. 
Define $s := p(s_1,s_2,s_3)$.
Then $s(u) = p(1,0,0) = 0$, 
$s(v) = p(1,0,1) = 1$, 
and $s(w) = p(1,1,0) = 1$. Hence, $s$ satisfies none of the literals $u, \neg v, \neg w$; moreover, the other literals of $\psi$ are not satisfied as well, a contradiction to the assumption that $R$ is preserved by $p$. 
Therefore, all clauses of $\phi$ either consist of a single positive literal, or of one positive and one negative literal, or only of negative literals. %In each case, ${\mathfrak D}$ contains a relation 
This shows that $R$ can even be defined by a conjunction of relations of ${\mathfrak D}$ (no existential quantification is needed), and hence 
\begin{align*} 
\langle p \rangle & = \Pol(\Inv(\{p\})) && 
\text{(Proposition~\ref{prop:pol-inv})} \\
& \supseteq \Pol(\bD_{\pp}) && \text{(by the above)} \\
& \supseteq \Pol(\Inv(\Pol(\bD))) && \text{(Proposition~\ref{prop:pp-preserved})} \\
& = \Pol({\mathfrak D}) && \text{(Proposition~\ref{prop:pol-inv}).}
\end{align*}
  
Suppose that $\mathfrak E$ is a reduct of 
$\mathfrak D$ with finite signature; let $k$ be the maximal arity of the relations in $\mathfrak E$. Then $\mathfrak E$ is preserved by the $k+1$-ary near unanimity polymorphism $f_{k+1}$ from Example~\ref{expl:bool-nu}. 
We have already mentioned that this operation does not preserve the relation $B_{k+1} = \{0,1\}^{k+1} \setminus \{(1,\dots,1)\}$. It follows that $\Pol({\mathfrak D})$ is a proper subclone of $\Pol({\mathfrak E})$. 
Note that this shows that there is no structure $\bD'$ over the domain $\{0,1\}$ with finitely many relations such that $\Pol(\bD') = \Pol(\bD)$, because the relations of $\bD'$ would have a primitive positive definition in a reduct $\bE$ of $\bD$ with finite relational signature, as we have seen above. We have also seen that $\Pol(\bD)$ is a proper subclone of $\Pol(\bE)$, and hence $\Pol(\bD') \subseteq \Pol(\bE)$ is a proper subclone as well. 
\end{example}

\begin{exercises}
\exercise[Blau]\label{exe:source-140} Show that the digraph $C_2^{++}$ from Exercise~\ref{exe:c2++} does not have near unanimity polymorphisms.
% Solution sketch: Use the standard inconsistent instance associated with $C_2^{++}$ from
% Exercise~\ref{exe:c2++}$. It survives $k$-consistency for every fixed $k$ although it has no
% homomorphism to $C_2^{++}$. By Exercise~\ref{exe:nu-kcons}, a $k+1$-ary near-unanimity polymorphism
% would make that procedure complete, a contradiction.
\exercise[Rot]\label{exe:source-139} %Let $k \geq 2$.
Show that if $H$ is a digraph with a $k+1$-ary near unanimity polymorphism, then the $k$-consistency procedure (see Section~\ref{sect:kcons}) solves $\Csp(H)$.
\label{exe:nu-kcons}
% Solution sketch: At the fixed point, every partial homomorphism on at most $k$ variables has
% compatible restrictions and extensions. Choose extensions one variable at a time. Whenever the
% choices disagree, combine the $k+1$ available partial maps coordinatewise with the near-unanimity
% polymorphism; it preserves all constraints and fixes their common coordinates. Induction produces a
% total homomorphism.
\exercise[Rot]\label{exe:source-141} % Brewster, Feder, Hell, Huang,
% MacGillivray: Theorem 5.2.
Let $H$ be an irreflexive graph.
Then $H$ has a conservative near unanimity polymorphism if and only if it has a
conservative majority polymorphism.
% Solution sketch: The reverse implication is immediate. For the forward implication, restrict a
% conservative near-unanimity operation to every two-element subset. Irreflexivity and preservation of
% edges make these pairwise restrictions coherent; the standard pair-selection construction then
% defines a conservative majority operation. Equivalently, this is the pairwise form of the
% conservative near-unanimity theorem for irreflexive graphs.
\exercise[Schwarz]\label{exe:source-138} Prove Theorem~\ref{thm:nu}.
% Solution sketch: For a $k+1$-ary near-unanimity polymorphism, combine the $k+1$ tuples that agree
% with a target tuple on all but one coordinate; the identities show that every relation is determined
% by its projections to at most $k$ coordinates. Conversely, apply this $k$-decomposability property
% to the canonical relation whose columns enumerate all $k$-tuples. The resulting homomorphism
% supplies a $k+1$-ary near-unanimity polymorphism.
\end{exercises}

\section{Universal Algebra}
\label{sect:ua}
We have seen in Section~\ref{sect:algebra}
that for finite relational structures $\bB$ with finite relational signature, the computational complexity of $\Csp(\bB)$ only depends on the polymorphisms of $\bB$. For more advanced results that use this perspective, it will be useful to view the set of all polymorphisms of $\bB$ as an algebra, since we may then use ideas and results from universal algebra. 

\subsection{Algebras and Clones}
\label{sect:algebras-clones}
In universal algebra, an \emph{algebra} is simply a structure with a purely functional signature.
We will typically use bold font letters, like $\fA$, to denote algebras, and the corresponding capital roman letters, like $A$, to denote their domain. 
%A \emph{subalgebra} of an algebra $\fA$ is a 
%subset $B$ of $A$ which is closed under applications of operations of $\fA$, that is,
%if $b_1,\dots,b_k \in B$, and $f$ is an operation of $\fA$, then $f(b_1,\dots,b_k)$ must also be in $B$. Note that in particular, if $\fA$ contains constant symbols, then the elements denoted by these constants must belong to all subalgebras of $\fA$. If $B$ is a subalgebra of $\fA$, then we may also view $B$ as an algebra $\fB$ of the same signature $\tau$ as $\fA$, so that if $f \in \tau$ then $f^{\fB}$ denotes the restriction of $f^{\fA}$ to $B$. We may view subalgebras of $\fA$  sometimes as sets, sometimes as algebras, and this should never cause confusion. 

\begin{example}[Group] \label{expl:groups}
A \emph{group} is an algebra with a binary function symbol $\circ$ for composition, 
a unary function symbol 
$^{-1}$ for taking the inverse, and a constant denoted by $e$,
satisfying 
\begin{itemize}
\item $\forall x,y,z. \, x \circ (y \circ z) = (x \circ y) \circ z$,
\item $\forall x. \, x \circ x^{-1} = e$, 
\item $\forall x. \, e \circ x = x$, and $\forall x. \, x \circ e = x$.
\end{itemize}
Note that all axioms are \emph{universal}
in the sense that all the variables are universally quantified (more on that comes later). 
A group is called \emph{abelian} if it additionally satisfies 
\begin{align*}
& \forall x,y. \, x \circ y = y \circ x.  
\end{align*} 
For abelian groups we sometimes use the signature $\{+,-,0\}$ instead of $\{\circ,^{-1},e\}$. 
\end{example}

\begin{example}[Ring]\label{expl:ring}
A \emph{(unital) ring} 
is an algebra $\fR$ 
with the signature 
$\{\cdot,+,-,0,1\}$ where $\cdot,+$ are binary\footnote{as usual, we use the convention that $\cdot$ binds stronger than $+$ to reduce the number of brackets in terms.}, $-$ is unary, and $0,1$ are constants, such that
$(R;+,-,0)$ is an abelian group and additionally 
\begin{align*}
\forall x,y,z. \; (x \cdot y) \cdot z & = x\cdot(y\cdot z) && \text{(associativity)} \\
\forall x. \; 1 \cdot x  & = x \cdot 1 = x && \text{(multiplicative unit)} \\
\forall x,y,z. \; x \cdot (y+z) & = x \cdot y + x \cdot z && \text{(left distributivity)} \\
\forall x,y,z. \; (y+z) \cdot x& = y \cdot x + z \cdot x && \text{(right distributivity)} 
\end{align*}
A ring is called \emph{commutative} if it additionally satisfies 
\begin{align*}
\forall x,y. \; x \cdot y & = y \cdot x && \text{(commutativity).}
\end{align*}
\emph{Fields} are commutative rings where $0 \neq 1$ 
and where every non-zero element $x$ has a multiplicative inverse $y$, i.e., $x \cdot y = 1$. We often omit the symbol $\cdot$ for readability. 
\end{example}

The next example generalises vector spaces.  

\begin{example}[Module]
\label{expl:module}
Let $\fR$ be a ring. 
An \emph{$\fR$-module} is an algebra $\fM$ with the signature $\{+,-,0\} \cup \{f_r \mid r \in R\}$ 
such that $(M;+,-,0)$ is an abelian group and
for all $r,s \in R$ it holds that 
%\begin{itemize}
%\end{itemize}
\begin{align}
\forall x,y. \; f_r(x+y) & = f_r(x) + f_r(y) \label{eq:d1} \\
\forall x. \; f_{r+s}(x) & = f_r(x) + f_s(x) \label{eq:d2} \\
\forall x. \; f_r(f_s(x)) & = f_{rs}(x). \label{eq:d3}
\end{align}
An $\fR$-module is called \emph{unitary} if it additionally satisfies $\forall x. f_1(x) = x$. 
We usually write $rx$ instead of $f_r(x)$. 

An alternative formalisation of modules is to view them as structures with two sorts, one sort for $R$ and one for $M$ above (see Section~\ref{sect:multi}). The details of this perspective are the content of Exercise~\ref{exe:modules} and omitted because we do not need it in this text. 
\end{example} 

\begin{example}[Semilattice] \label{expl:semilattice}
A \emph{meet-semilattice}\index{meet-semilattice} $\bS$ is a $\{\leq\}$-structure 
with domain $S$ such that $\leq^{\bS}$ denotes a partial order where any two $u,v \in S$ 
have a (unique) \emph{greatest lower bound}\index{greatest lower bound} $u \wedge v$,
i.e., an element $w$ such that $w \leq u$,
$w \leq v$, and for all $w'$ with $w' \leq u$ and $w' \leq v$ we have $w' \leq w$. 
Dually, a \emph{join-semilattice}\index{join-semilattice} is a partial order 
with least upper bounds\index{least upper bound}, denoted by $u \vee v$. 
A \emph{semilattice}\index{semilattice} is a
meet-semilattice or a join-semilattice where the distinction between meet and join is either not essential or clear from the context. 

Semilattices can also be characterised as 
$\{\wedge\}$-algebras where $\wedge$ is a binary operation\index{semilattice operation} that must satisfy the following axioms
\begin{align*}
\forall x,y,z \colon \, x \wedge (y \wedge z) & = (x \wedge y) \wedge z  &&
\emph{(associativity)}\index{associativity} \\
\forall x,y \colon \, x \wedge y & = y \wedge x && 
\emph{(commutativity)}\index{commutativity} \\
\forall x \colon x \wedge x & = x && \emph{(idempotency, }\index{idempotency} or  \emph{ idempotence)}\index{idempotence}. 
\end{align*}
Clearly, the operation $\wedge^{\bS}$, defined as above in a semilattice $\bS$ viewed as a poset, satisfies these axioms. Conversely, if $(S;\wedge)$ is a semilattice, then the formula $x \wedge y = x$ defines a partial order on $S$ which is a meet-semilattice (and $x \wedge y = y$ defines a partial order on $S$ which is a join-semilattice). 

Note that the two ways of formalising semilattices differ when it comes to the notion of a substructure; a \emph{subsemilattice} refers to the substructure of a semilattice when formalised as an algebraic structure.
\end{example}

\begin{example}[Lattice] \label{expl:lattice}
A \emph{lattice}\index{lattice} $\bL$ is a $\{\leq\}$-structure 
with domain $L$ such that $\leq^{\bL}$ denotes a partial order such that any two $u,v \in L$ 
have a greatest lower bound $u \wedge v$ (\emph{meet}) and
a least upper bound, denoted by $u \vee v$ 
(\emph{join}). 
A lattice $\bL$ is called \emph{complete}\index{complete lattice} if every
family $(u_i)_{i\in I}$ of elements of $L$ has a greatest lower bound,
denoted by $\bigwedge_{i\in I}u_i$, and a least upper bound, denoted by
$\bigvee_{i\in I}u_i$. This includes the empty family: its meet is the
greatest element of $L$, and its join is the least element of $L$.
Every non-empty finite lattice is complete.

Lattices can also be viewed as 
$\{\wedge,\vee\}$-algebras $\fL$ where $\wedge^\fL$ 
and $\vee^\fL$ are semilattice operations (Example~\ref{expl:semilattice}) that additionally satisfy
\begin{align*}
\forall x,y \colon \, x \wedge (x \vee y) & = x \text{ and } x \vee (x \wedge y) = x  &&
\text{(absorption)}\index{absorption}.
\end{align*}
If $\fL$ is a lattice and the operations $\wedge$ and $\vee$ are defined as above for semilattices, then these two operations also satisfy the absorption axiom. Conversely, if we are given an algebra $(S;\wedge,\vee)$ satisfying the mentioned axioms, then the formula $x \wedge y = x$ (equivalently, the formula $x \vee y = y$) defines a partial order on $S$ which is a lattice. 
Of course, there is potential danger of confusion of the symbols for lattice operations $\wedge$ and $\vee$ with the propositional connectives $\wedge$ for conjunction and $\vee$ for disjunction (which can be seen as lattice operations on the set $\{0,1\}$) which luckily should not cause trouble here.  
A lattice $\fL = (L;\wedge,\vee)$ is called \emph{distributive}\index{distributive}
if it satisfies 
\begin{align*}
\forall x,y,z \colon \, x \wedge (y \vee z) & = (x \wedge y) \vee (x \wedge z) && \text{(distributivity).}  \qedhere 
\end{align*}
\end{example}

% Draw a picture: 
% Vertices: 
% Bold - Algebras,
% Frak: Structures,
% Mathscr: Clones
% Arcs:
% structures - Pol -> clones,
% algebras - Clo -> clones,
% clones - Inv -> structures, 

\begin{exercises}
\exercise[Rot]\label{exe:source-143} \label{exe:modules}
Formalise modules (Example~\ref{expl:module}) as two-sorted structures as introduced in Section~\ref{sect:multi}.
% Solution sketch: Use two sorts, a ring sort $R$ and a module sort $M$. Put the ring operations on
% $R$, the abelian-group operations on $M$, and represent scalar multiplication by the sorted function
% $R\times M\to M$. The module axioms are precisely the usual two-sorted identities for these
% operations.
\exercise[Orange]\label{exe:source-142} Let $\fA = (A;+,-,0)$ be an abelian group. Let $R$ be the set of all endomorphisms of $\fA$. Prove the following operations define a ring $\fR$ on $R$:
addition is defined pointwise, and multiplication is defined as function composition. The constant $0^{\fR}$ denotes the endomorphism which is constant $0$,
and the constant $1^{\fR}$ denotes the identity. %What are the units of $\fR$?
%\item Show for every $r \in R$ the map $e(x) \mapsto f_r(x)$ is an endomorphism
% of an $R$-module $M$. REALLY?
% e(x) \mapsto f_r(x) preserves + by the first axiom,
% it preserves 0 since 0 = 0 u + -(0 u)
% = (0+0) u + (- 0 u)
% = 0 u + 0 u - 0 u
% = 0 u
% it preserves - since 0 = f_r(0) = f_r(x-x) = f_r(x) + f_r(-x) by the first axiom, so -f_r(x) = f_r(-x).
% it preserves f_r by third axiom and commutativity of ?? HMH don't have this.
\end{exercises}

\paragraph{The clone of an algebra.}
If $\fA$ is an algebra with the signature $\tau$ and $n \geq 1$,
then a $\tau$-term $t(x_1,\dots,x_n)$ gives rise
to a \emph{term operation} $t^{\fA} \colon A^n \to A$;
the value of $t^{\fA}$ at $a_1,\dots,a_n \in A$
can be obtained by replacing the variables
$x_1,\dots,x_n$ by $a_1,\dots,a_n$ and evaluating in $\fA$.

\begin{example}
If $\fA$ is a group, then the term operation for the term 
($x \circ y^{-1}) \circ z$ is a Maltsev operation on $A$. 
\end{example}

\begin{example}
If $t(x_1,x_2)$ is the term that just consists of the variable $x_1$, then $t^{\fA}$ equals the projection $\pi^2_1$. 
\end{example}

Algebras give rise to clones in the following way. 
We denote by $\Clo(\fA)$
the set of all term operations  of $\fA$ 
of arity at least one.
%, that is, operations with domain $A$ of the form
%$(x_1,\dots,x_n) \mapsto t^{\fA}(x_1,\dots,x_n)$ where $t$ is any
%term over the signature $\tau$ whose set of variables is contained in $\{x_1,\dots,x_n\}$.
Clearly, $\Clo(\fA)$ is an operation clone since it is 
closed under compositions, and contains the projections.

\paragraph{Polymorphism algebras.}
In the context of complexity classification of CSPs, algebras arise as follows. 

\begin{definition}\label{def:poly-algebra}
Let $\bB$ be a relational structure with domain $B$. 
An algebra $\fB$ with domain $B$ such that $\Clo(\fB) = \Pol(\bB)$
 is called a \emph{polymorphism algebra of $\bB$}.
\end{definition}

Note that a structure $\bB$ has many different polymorphism algebras, since Definition~\ref{def:poly-algebra} does not prescribe how to assign function symbols to the polymorphisms of $\bB$. 

%But the precise choice of the signature never plays a role, and therefore we sometimes refer to \emph{the} polymorphism algebra of $H$, and denote it by $\Alg(H)$. Formally, 

Any clone
$\mathcal C$ on a set $D$ can be viewed as an algebra $\fA$ with domain $D$ whose signature consists 
of the operations of $\mathcal C$ themselves;
that is, if $f \in \mathcal C$, then
$f^{\fA} := f$. 
We will therefore use concepts defined for algebras also for clones.  %for example, we may write $\HSP(\mathcal C)$ for the variety generated by the clone ${\mathcal C}$ viewed as an algebra. 
In particular, the polymorphism clone $\Pol(\bB)$ of a structure $\bB$ might be viewed as an algebra, which we refer to as \emph{the} polymorphism algebra of $\bB$. 
Note that the signature of the polymorphism algebra is always infinite, since we have polymorphisms of arbitrary finite arity.

\subsection{Subalgebras, Products, Homomorphic Images}
\label{sect:hsp}
In this section we recall some basic universal-algebraic facts that 
will be used in the following subsections.

\paragraph{Subalgebras.}
Let $\fA$ be a $\tau$-algebra with domain $A$.
A $\tau$-algebra
$\fB$ with domain $B \subseteq A$ is called
a \emph{subalgebra} of $\fA$ if 
for each $f \in \tau$ of arity $k$ we have $f^\fB(b_1,\dots,b_k) = f^\fA(b_1,\dots,b_k)$ for all $b_1,\dots,b_k \in B$; in this case, we write $\fB \leq \fA$. 
We write $\fB < \fA$ if $\fB \leq \fA$ and additionally $A \neq B$. 
A \emph{subuniverse} of $\fA$
is the domain of some subalgebra of $\fA$. Note that as for structures, we do not exclude algebras whose domain is empty (which is of course only possible if the signature does not contain any constant symbols). 
A subalgebra $\fB$ of $\fA$ is called \emph{proper} if $\emptyset \neq B \neq A$ (we use the same terminology also for subuniverses; this applies to all the terminology that we introduce for subalgebras and subuniverses). 
The smallest subuniverse of $\fA$ that contains
a given set $S \subseteq A$ is called the \emph{subuniverse of $\fA$ generated by $S$},
and the corresponding subalgebra is called the
\emph{subalgebra of $\fA$ generated by $S$}, and denoted by $\langle S \rangle_{\fA}$.

\paragraph{Products.}
Let $\fA,\fB$ be $\tau$-algebras with domain $A$ and $B$, respectively. 
Then the \emph{product} $\fA \times \fB$ is the $\tau$-algebra with domain $A \times B$ such that
for each $f \in \tau$ of arity $k$ we have $f^{\fA \times \fB}\big((a_1,b_1),\dots,(a_k,b_k)\big) = \big(f^\fA(a_1,\dots,a_k),f^\fB(b_1,\dots,b_k)\big)$ for all $a_1,\dots,a_k \in A$ and $b_1,\dots,b_k \in B$. 
%As usual, we write $\fA^2$ for $\fA \times \fA$,
%and $\fA^{n}$ for $n>2$ for $\fA \times \fA^{n-1}$. 
More generally, when $(\fA_i)_{i \in I}$ is a sequence
of $\tau$-algebras, indexed by some set $I$,
then $\prod_{i \in I} \fA_i$ is the $\tau$-algebra $\fA$
with domain $\prod_{i \in I} A_i$ such that for $a^1_i,\dots,a^k_i \in A_i$
$$f^{\fA}\big((a^1_i)_{i \in I},\dots,(a^k_i)_{i \in I}\big) := \big(f^{\fA_i}(a^1_i,\dots,a^k_i)\big)_{i \in I}. $$

\begin{lemma}
Let $\fA$ be the polymorphism algebra of a finite structure $\bA$. Then the (domains of the) subalgebras of $\fA^k$ are precisely the relations that have a primitive positive definition in $\bA$. 
\end{lemma}
\begin{proof}
A relation $R \subseteq A^k$ is a subalgebra of
$\fA^k$ if and only if for all $m$-ary $f$ in the signature of
$\fA$ and $t^1,\dots,t^m \in R$, we have
$\big(f(t^1_1,\dots,t^m_1),\dots,f(t^1_k,\dots,t^m_k)\big) \in R$, which is the case if and only if $R$ is preserved by all polymorphisms of $\bA$, which is the case if and only if $R$ is primitively positively definable in $\bA$ by Theorem~\ref{thm:inv-pol}. 
\end{proof}

%The direct product of two algebras $\fA$ and $\fB$ is the 

\paragraph{Homomorphic Images.}
Let $\fA$ and $\fB$ be $\tau$-algebras.
Then a \emph{homomorphism} from $\fA$ to $\fB$
is a mapping $h \colon A \to B$ such that
for all $k$-ary $f \in \tau$ and $a_1,\dots,a_k \in A$
we have
$$h\big(f^{\fA}(a_1,\dots,a_k)\big) = f^{\fB}\big(h(a_1),\dots,h(a_k)\big) \, .$$
Note that if $h$ is a homomorphism from $\fA$ to $\fB$ then the image of $h$ is the domain of a subalgebra of $\fB$, which is called a  \emph{homomorphic image} of $\fA$. 

\begin{definition}\label{def:congruence}
A \emph{congruence} of an algebra $\fA$ is 
an equivalence relation $C$ that is preserved by all operations in $\fA$.\footnote{This is one of the rare cases where we deviate from the literature, which often uses small greek letters, in particular $\theta$, to denote congruences. Since we use small greek letters already for formulas and for maps, we will use capital Roman letters for congruences, following the convention to denote relations by capital Roman letters.}
\end{definition}

\begin{lemma}\label{lem:congruence-pp}
Let $\bB$ be a finite structure,
and let $\fB$ be a polymorphism algebra of $\bB$. Then the congruences of $\bf B$ are exactly the primitively positively definable
equivalence relations over $\bB$. 
\end{lemma}
\begin{proof}
A direct consequence of Theorem~\ref{thm:inv-pol}.
\end{proof}

\begin{example}
Let $G = (V,E)$ be the undirected graph 
with $V = \{a_1,\dots,a_4,b_1,\dots,b_4\}$ such that $a_1,\dots,a_4$ and $b_1,\dots,b_4$ induce a clique,  for each $i \in \{1,\dots,4\}$ there is an edge between $a_i$ and $b_i$, and otherwise there are no edges in $G$. Let $\fA$ be a polymorphism
algebra of $G$. Then $\fA$ 
has a congruence with two classes. 
%homomorphically maps to a two-element algebra $\bf B$. By Proposition~\ref{prop:congruence-homo}, 
%it suffices to show that $\fA$ has a congruence with two equivalence classes. 
By Lemma~\ref{lem:congruence-pp}, it suffices to show that
an equivalence relation of index two is 
primitively positively definable. 
The following primitive positive formula 
$$\exists u,v \, \big (E(x,u) \wedge E(y,u) \wedge E(x,v) \wedge E(y,v) \wedge E(u,v) \big)$$
defines an equivalence relation whose classes are precisely $\{a_1,\dots,a_4\}$ and $\{b_1,\dots,b_4\}$. 
\end{example}

\begin{example}
Let $\fA$ be the algebra with domain $$A := S_3 = \big\{\id,(123),(132),(12),(23),(13)\big\}$$
(the symmetric group on three elements),
and a single binary operation, the composition
function of permutations. 
Note that $\bf A$ has the subalgebra induced by
$\big \{\id,(123),(132)\}$.
Also note that $\bf A$ homomorphically maps to $(\{0,1\},+)$
where $+$ is addition modulo 2:
the preimage of $0$ is $\{\id,(123),(132)\}$
and the preimage of $1$ is $\{(12),(23),(13)\}$. 
\end{example}

\begin{definition}
If $\fA$ is a $\tau$-algebra with a 
congruence $K$, then we write 
$\fA/_K$ for the \emph{quotient algebra} whose domain 
is the set $A/_K$
of congruence classes of $K$, 
and where 
$$ f^{{\fA}/K}(a_1/_K, \dots,a_k/_K)=f^{\bf A}(a_1,\dots,a_k)/_K$$ for 
all $a_1,\dots,a_k \in A$ and $k$-ary $f \in \tau$. This is well-defined since $K$ is preserved by all operations of ${\bf A}$. 
\end{definition}

Note that $a \mapsto a/_K$ is a surjective homomorphism from ${\fA}$ to ${\fA}/K$, called the \emph{natural homomorphism} from $\fA$ to ${\fA}/K$. 
This essentially proves the following (see~\cite{BS}). 

\begin{proposition} 
\label{prop:congruence-homo}
Let $\bf A$ be an algebra. Then $K$ is a congruence of $\bf A$ if and only if
$K$ is the kernel of a homomorphism
from $\bf A$ to some other algebra $\bf B$. 
\end{proposition}

%If $\fA$ is a $\tau$-algebra, and $h \colon A \rightarrow B$ is a mapping
%such that the kernel of $h$ is a congruence of $\fA$, we define the \emph{quotient algebra
%${\bf A}/h$ of $\bf A$ under $h$} to be the algebra with domain $h(A)$ where 
%$$ f^{{\fA}/h}(h(a_1), \dots,h(a_k))=h(f^{\bf A}(a_1,\dots,a_k))$$ where $a_1,\dots,a_k \in A$ and $f \in \tau$
%is $k$-ary. This is well-defined since the kernel of $h$ is preserved by all operations of ${\bf A}$. 
%The following is well-known.
%\begin{lemma}[The Homomorphism Lemma]\label{lem:hom}
%Let $\bf A$ be a $\tau$-algebra, let $K$ be a congruence of ${\bf A}$, 
%and let $h_1 \colon A \rightarrow B_1$ and $h_2 \colon A \rightarrow B_2$ 
%be two mappings with kernel $K$. Then ${\bf A}/h_1$ is isomorphic to ${\bf A}/h_2$.
%\end{lemma}
The following is well known (see e.g.~Theorem 6.3 in~\cite{BS}).
%The isomorphism and homomorphism theorems. 
%If $B \subset A$, then $K|_B$ denotes $B^2 \cap K$.
%Let $B^{K}$ be the set $\{ a \in A \; | \; B \cap a/K \neq \emptyset\}$. 
%\begin{theorem}[The third isomorphism theorem; see Theorem 6.18 in~\cite{BS}] If ${\bf B}$ is a subalgebra of ${\bf A}$, and
%$\theta$ a congruence of ${\bf A}$, then ${\bf B}/(\theta |_{\bf B})$ is isomorphic to ${\bf B}^{\theta}/(\theta|_{{\bf B}^\theta}$.
%\end{theorem}
%\begin{proof}
%The mapping that sends $b/(\theta|_B)$ to $b/(\theta|_{B^\theta})$
%is the desired isomorphism.
%\end{proof}

\begin{lemma}%[Theorem 6.3 in~\cite{BS}]
\label{lem:congruences}
Let $\fA$ and $\fB$ be algebras with the same signature, and let $h \colon \fA \rightarrow \fB$ be a homomorphism.
Then
\begin{itemize}
\item  the image of any subalgebra ${\bf A}'$ of ${\bf A}$ under $h$ is a subalgebra of ${\bf B}$, and
\item the preimage of any subalgebra ${\bf B}'$ of ${\bf B}$ under $h$ is a subalgebra of ${\bf A}$.
\end{itemize} 
%In both cases, the restriction of $h$ to $A'$ is a homomorphism from ${\bf A}'$ to ${\bf B}'$.
% THIS IS TRIVIAL!
\end{lemma}
\begin{proof}
Let $f \in \tau$ be 
$k$-ary. Then for all $a_1,\dots,a_k \in A'$, 
$$f^{\bf B}(h(a_1),\dots,h(a_k)) = h(f^{\bf A}(a_1,\dots,a_k)) \in h(A') \; ,$$
so $h(A')$ is a subalgebra of $\fB$. 

Now suppose that $h(a_1),\dots,h(a_k)$ are elements of $B'$;
then $f^{\bf B}(h(a_1),\dots,h(a_k)) \in B'$ and hence $h(f^{\bf A}(a_1,\dots,a_k)) \in B'$.
So, $f^{\bf A}(a_1,\dots,a_k) \in h^{-1}(B')$ which shows that $h^{-1}(B')$ induces a subalgebra
of $\bf A$.
\end{proof}

\begin{remark}\label{rem:congnot}
If $\fA' \leq \fA$ and $K$ is a congruence of $\fA$, then %the argument in  
Lemma~\ref{lem:congruences} shows that 
the restriction $K' := K \cap (A')^2$ is a congruence of $\fA'$; 
we then sometimes abuse notation and write $\fA'/_K$ instead of $\fA'/_{K'}$. 
\end{remark}

\begin{exercises}
\exercise[Blau]\label{exe:source-144} Show that for all $\tau$-algebras $\fA$ and $\fB$ with $\Clo(\fA) = \Clo(\fB)$ we have $\Clo(\fA^2) = \Clo(\fB^2)$.
% Solution sketch: Every term operation of a direct square is obtained by applying the corresponding
% term operation coordinatewise. Thus $\Clo(\fA)=\Clo(\fB)$ implies equality of the coordinatewise
% operations on $A^2$ and $B^2$ after the given identification of the term clones, and the converse
% inclusion is symmetric.
\exercise[Blau]\label{exe:source-145} Find a relational signature $\tau$ and
$\tau$-structures $\bA$, $\bB$ such that $\Pol(\bA) = \Pol(\bB)$ but $\Pol(\bA^2) \neq \Pol(\bB^2)$.
% Loesung: Gleichheit versus volles Produkt.
\exercise[Blau]\label{exe:source-146} Prove Proposition~\ref{prop:congruence-homo}.
%\newpage
%\vspace{-2.3cm}
%\begin{flushright}
%\includegraphics[scale=.3]{Blau.jpg}
%\end{flushright}
%\vspace{.2cm}
% Solution sketch: The kernel of any homomorphism is an equivalence relation and is preserved by every
% basic operation, hence is a congruence. Conversely, if $K$ is a congruence, the natural map
% $a\mapsto a/_K$ from $\fA$ onto the quotient algebra $\fA/K$ is a homomorphism and has kernel
% exactly $K$.
\exercise[Rot]\label{exe:source-148} Let $p$ be a positive prime. Show that
the only proper subalgebras of $\fA_p$ from Exercise~\ref{exe:source-147} are of the form $\{a\}$ for some $a \in \{0,\dots,p-1\}$.
\label{exe:subalgebras-Ap}

\par\noindent {\bf Hint.} Use Exercise~\ref{exe:affine-maltsev-terms} and Example~\ref{expl:linear}.
%Indeed, the operations of $\fA$ are of the form $(x_0,\dots,x_{k-1}) \mapsto \sum_i a_i x_i$
%where $a_0,\dots,a_{k-1} \in {\mathbb Z}$ are such that $\sum_i a_i = 1$ (Exercise~\ref{exe:affine-maltsev-terms}),
%and hence every invariant relation is the solution space to a linear equation system (see Example~\ref{expl:affine}). This holds in particular for every subuniverse, which therefore must have size $p^1$ or $p^0$.
\exercise[Orange]\label{exe:source-147} \label{exe:affine-maltsev-terms}
Consider the algebra $\fA_n := ({\mathbb Z}_n;m)$ where $m(x,y,z) := x - y + z$. Then for every $k \geq 1$ the clone $\Clo(\fA_n)^{(k)} $ consists of precisely the operations defined as
$$g(x_0,\dots,x_{k-1}) := \sum_i a_i x_i$$
where $a_0,\dots,a_{k-1} \in {\mathbb Z}$ with $\sum_i a_i = 1$.
%\vspace{-.5cm}
% Clearly, $m$ satisfies the given condition,
% and every term over $m$ satisfied the given condition, by induction over the term structure.
% Conversely, alpha1 x + alpha2 y can be produced as follows.
% may assume that alpha1 < 0 < 1 < alpha2, otherwise have a projection and statement is trivial (or flip arguments).
% Then repeat alpha2-1 many times:
% m(y,x,m(y,x, ... m(y,x,y))).
% The general case is similar.
\end{exercises}

\subsection{Algebras and CSPs}
\label{sect:alg-csp}
\begin{figure}
\begin{center}
\includegraphics[scale=.5]{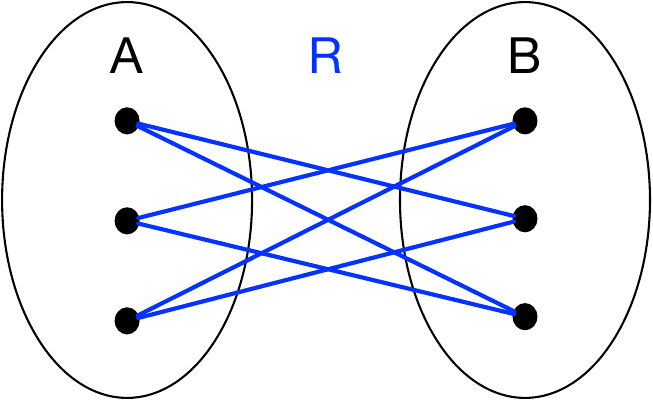}
\end{center}
\caption{Illustration of $E(K_3) \leq \fA^2$, where $\Clo(\fA) = \Pol(K_3)$, as a bipartite graph.}
\label{fig:potato-K3}
\end{figure}

Let $\fA$ and $\fB$ be algebras with the same signature, and let $R \leq \fA \times \fB$ be a subalgebra. 
The relation $R$ can be viewed as the edge relation of a bipartite graph with colour classes $A$ and $B$. Note that if $\fA = \fB$ and $\Clo(\fA) = \Pol(\bA)$, then the relations $R$ that arise in this way are precisely the binary relations on $A$ that are primitively positively definable in $\bA$. For example, if $\Clo(\fA) = \Pol(K_3)$,
then $E(K_3) \leq \fA^2$ and the corresponding bipartite graph is drawn in Figure~\ref{fig:potato-K3}. 

 The importance of the set-up of $R \leq \fA \times \fB$ for CSPs is that we may imagine $A$ as the possible values for a variable $x$ in an instance of the CSP, and $B$ as the possible values for a variable $y$ in the CSP, and $R$ represents a binary constraint between $x$ and $y$. The advantage of this perspective is that many important definitions are very intuitively phrased in the language of bipartite graphs.

%$R \leq \fA^2$ if and 

We start with the following fundamental definition from universal algebra which is highly relevant for the universal-algebraic approach to CSPs, in particular in Section~\ref{sect:absorption}. 

\begin{definition}\label{def:subdirect}
Let $k \geq 1$ and $\fA_1,\dots,\fA_k$ be $\tau$-algebras.
Then $R \leq \fA_1 \times \cdots \times \fA_k$ is called \emph{subdirect}
if $\pi_i(R) = A_i$ for every $i \in \{1,\dots,k\}$. 
%\begin{align*}
%\{a \in A \mid \text{there exists } b \in B \text{ such that }  (a,b) \in R\} & = A \\
%\text{ and } \quad \{b \in B \mid  \text{there exists } a \in A \text{ such that } (a,b) \in R\} & = B. 
%\end{align*}
\end{definition}

\begin{remark} 
If $\fA$ is the polymorphism algebra of a finite digraph $H$, then 
there is a link between the notion of subdirect subalgebras and arc-consistency. Let $G$ be a finite digraph and let $L(x) \subseteq V(H)$ be the list for $x \in V(G)$ at the final stage of the evaluation of $\AC_H(G)$. Recall that
for every $x \in V(G)$, the set $L(x)$ is a subuniverse of ${\fA}$ (Exercise~\ref{exe:ac-pol}). 
Note that for every $(x,y) \in E(G)$ we have that
$E(H) \cap (L(x) \times L(y))$ is subdirect in $L(x) \times L(y) \leq \fA^2$. 
\end{remark} 

\begin{remark} 
$E(H)$ is subdirect in $\fA^2$ if and only if $H$ has no sources and no sinks. 
Digraphs without sources and sinks are also called \emph{smooth}. 
\end{remark}

\begin{proposition}\label{prop:multisorted} 
Let $\bB$ be a finite structure and let $\fB$ be such that $\Pol(\bB) = \Clo(\fB)$. Then 
for any $R \subseteq B^k$, the following are equivalent.
\begin{itemize}
\item $R$ is primitively positively definable in $\bB$; 
\item $R \leq \fB^k$; 
\item $R$ is a subdirect subalgebra of 
$\fB_1 \times \cdots \times \fB_k$ for some uniquely determined subalgebras $\fB_1,\dots,\fB_k$ of $\fB$. 
%\item there are $\fA_1, \dots, \fA_k \leq \fB$ such that  
%$A_1,\dots,A_k \subseteq B$ such that 
%$R$ is a unary relation with a primitive positive definition 
%in the substructure of $\bB$ with domain $A_1,\dots,A_k$. 
%in the relational structure with the domain 
%$A_1 \times \cdots \times A_k$ 
%and %relations that are 
%the relations $\Inv(\Clo(\fA_1 \times \cdots \times \fA_k))$. 
\end{itemize}
\end{proposition} 

The perspective of viewing relations of $\bB$ 
%(and constraints in instances of $\Csp(\bB)$) 
as
subdirect subalgebras of finite products of subalgebras of $\fB$ (as in Proposition~\ref{prop:multisorted}) is sometimes also referred to as \emph{multisorted}.  

%Let $x,y \in V(G)$ and let $L(x)$ and $L(y)$ be the lists computed by the arc-consistency procedure. 
%Recall from  that $L(x)$ and $L(y)$ are subuniverses of the polymorphism algebra $\fA$ of $H$. Note that $E(G) \cap (L(x) \times L(y)) \leq \fA^2$ is subdirect! 

\begin{exercises}
\exercise[Blau]\label{exe:source-149} Prove Proposition~\ref{prop:multisorted}.
% Solution sketch: A multisorted relation is primitive positively definable exactly when it is
% preserved by all sorted polymorphisms; the proof is the canonical-database proof of
% Proposition~\ref{prop:pol-inv} with every variable kept in its sort. Applying this to subuniverses
% of a product gives precisely the stated multisorted subalgebras and proves the proposition.
\exercise[Blau]\label{exe:source-150} Show that a digraph $G = (V,E)$ is rectangular if and only if $E$,
when regarded as a bipartite graph with color classes $A$ and $B$
as described in this section, is a disjoint union of \emph{bicliques}, i.e.,
if $a \in A$ has a path to $b \in B$, then $(a,b) \in E$.
% Solution sketch: If three corners of a rectangle lie in $E$, the two corresponding edges lie in one
% connected component of the bipartite representation, and rectangularity supplies the fourth corner.
% Repeating this along a path shows that every component is complete bipartite. Conversely, three
% corners in a disjoint union of bicliques must belong to one biclique, which contains the fourth.
\end{exercises}

\subsection{Pseudovarieties and Varieties}
\label{sect:pseudo-var}
Varieties are a fascinatingly powerful concept to study classes of algebras. 
%They are also central for the study of the complexity of CSPs: 
%We will see that 
%the complexity of $\Csp(\bB)$ for finite $\bB$
%only depends on the variety generated by a polymorphism algebra $\fB$ of $\bB$. 
%This comes from the fact that a finite algebra is in the variety generated
%by a finite algebra $\fB$ if and only if it is in the pseudovariety generated by $\fB$;  
%and the link between the pseudovariety generated by $\fB$ and  CSP$(\bB)$ has already been
% ined in Section~\ref{sect:tractability}. 
The fundamental result about  varieties is 
Birkhoff's theorem, which links varieties with equational theories
(Section~\ref{sect:birkhoff}). 
By Birkhoff's theorem, 
there is also a close relationship between varieties and the concept of an \emph{abstract clone} (Section~\ref{sect:abstract-clone}). 

\medskip 
Throughout this section, we study classes $\cal K$ of algebras with the same fixed signature. 
% $\cal K$ is a class of algebras of the same signature, then
\begin{itemize}
\item $\PPP({\cal K})$ denotes the class of all products of algebras from $\cal K$.
\item $\PPPfin({\cal K})$ denotes the class of all finite products of algebras from $\cal K$.
\item $\SSS({\cal K})$ denotes the class of all subalgebras of algebras from $\cal K$.
\item $\HHH({\cal K})$ denotes the class of all homomorphic images of algebras from $\cal K$.
\end{itemize}
Note that closure under homomorphic images implies in particular closure under
isomorphism. For the operators $\PPP$, $\PPPfin$, $\SSS$ and $\HHH$ we often omit the brackets when applying them
to single singleton classes that just contain one algebra, i.e., we write $\HHH(\fA)$ instead of $\HHH(\{\bf A\})$.
The elements of $\HS(\bf A)$ are also called the \emph{factors}
of $\bf A$.

A class $\cal V$ of algebras with the same signature $\tau$ is called
a \emph{pseudovariety} if $\cal V$ contains all homomorphic
images, subalgebras, and finite direct products
of algebras in $\cal V$, i.e., $\HHH({\cal V})=\SSS({\cal V})=\PPPfin({\cal V}) = \cal V$.
The class $\cal V$ is called a 
\emph{variety} if $\cal V$ also contains all (finite and infinite) 
products of algebras in $\cal V$.
So the only difference between pseudovarieties
and varieties is that pseudovarieties need not be closed under direct products of infinite cardinality.
The smallest pseudovariety (variety) that contains an algebra $\bf A$
is called the pseudovariety (variety) \emph{generated} by $\bf A$.

\begin{lemma}[HSP lemma]\label{lem:hsp}
Let $\fA$ be an algebra. 
\begin{itemize}
\item The pseudovariety generated by $\fA$ equals
 $\HSPfin(\fA)$. 
\item The variety generated by $\fA$ equals $\HSP(\fA)$. 
\end{itemize}
\end{lemma}
\begin{proof}
Clearly, $\HSPfin(\fA)$ is contained in the pseudovariety generated by $\fA$, and $\HSP(\fA)$ is contained in the variety generated by $\fA$.
For the converse inclusion, it suffices to verify that 
$\HSPfin(\fA)$ is closed under $\HHH$, $\SSS$, and $\PPPfin$.
It is clear that $\HHH(\HSPfin(\fA))=\HSPfin(\fA)$. 
The second part of Lemma~\ref{lem:congruences} implies that $\SSS(\HSPfin(\fA)) \subseteq \HS(\SP^{\fin}(\fA)) = \HSPfin(\fA)$. Finally, 
$$\PPPfin(\HSPfin(\fA)) \subseteq \HHH\PPPfin\SSS\PPPfin(\fA) \subseteq \HSPfin\PPPfin(\fA) = \HSPfin(\fA) \; .$$
The proof that $\HSP(\fA)$ is closed under $\HHH$, $\SSS$, and
$\PPP$ is analogous. 
\end{proof}

%\subsection{Pseudovarieties and Primitive Positive Interpretations}
Pseudo-varieties are linked to primitive positive interpretability from Section~\ref{sect:pp-interpret}. 

\begin{theorem}\label{thm:pp-interpret}
Let $\bC$ be a finite structure with  polymorphism algebra $\fC$. 
Then $\bB \in \PP(\bC)$ if 
and only if there exists ${\bf B} \in \HSPfin({\bf C})$ such that $\Clo(\bf B) \subseteq \Pol(\bB)$. 
\end{theorem}
\begin{proof}
We only prove the `if' part of the statement here; the proof of the `only if' part is similarly easy. 
There exists a finite number $d \geq 1$, a subalgebra $\bf D$ of $\fC^d$,
 and a surjective homomorphism $h$ 
from $\bf D$ to $\bf B$.
We claim that $\bB$ has a primitive positive interpretation $I$
of dimension $d$ in $\bC$. 
All operations of $\fC$ preserve $D$ (viewed as a 
$d$-ary relation over $\bC$),
since ${\bf D}$ is a subalgebra of $\fC^d$.
By Theorem~\ref{thm:inv-pol}, 
this implies that $D$ has a primitive positive definition 
$\delta(x_1,\dots,x_d)$ in $\bC$, which becomes the 
domain formula $\delta_I$ of $I$.
As coordinate map we choose the mapping $h$.
Since $h$ is an algebra homomorphism, the kernel
$K$ of $h$ is a congruence of ${\bf D}$. 
It follows that $K$, viewed as a $2d$-ary
relation over $C$, is preserved by all operations from $\fC$. 
Theorem~\ref{thm:inv-pol} implies that $K$ has a 
primitive positive definition in $\bC$. This definition becomes 
the formula $=_I$.
Finally, let $R$ be a relation of $\bB$ and 
let $f$ be a function symbol from the signature of $\fB$. By assumption, $f^\fB$ preserves
$R$. It is easy to verify that then $f^\fC$ 
preserves $h^{-1}(R)$. 
Hence, all polymorphisms of $\bC$
preserve $h^{-1}(R)$, and the
relation $h^{-1}(R)$ has a primitive positive definition 
in $\bC$ (Theorem~\ref{thm:inv-pol}), which becomes
the defining formula for the atomic formula $R(x_1,\dots,x_k)$ in $I$.
This concludes our construction 
of the primitive positive interpretation $I$ of $\bB$ in $\bC$.
%To see that it is an interpretation, let $R \subseteq
%B^k$ be any relation. 
%If $R$ has a primitive positive
%definition in $\mB$, then it is straightforward to translate this definition into a primitive positive definition of
%$h^{-1}(R)$ in $\mC$. Conversely, suppose that
%$h^{-1}(R)$ is primitively positively definable in $\mC$.
%Then $h^{-1}(R)$ is preserved by $\Clo(\fC)$
%and $R$ is preserved by $\Clo(\fB)$. By assumption
%$\Clo(\fB) = \Pol(\mB)$, and hence $R$ is preserved
%by all polymorphisms of $\mB$ and primitive positive
%definable in $\mB$ by Theorem~\ref{thm:inv-pol}.
\end{proof}

Primitive positive bi-interpretability can also be characterised with the varieties and pseudo-varieties generated by polymorphism algebras.
The following is a special case of Proposition 25 in~\cite{Topo-Birk} (where it is proved for a much larger class of countable structures).

\begin{proposition} 
\label{prop:bi-interpret-pseudo-var}
Let $\bA$ and $\bB$ be structures with finite domains. Then the following are equivalent. 
\begin{itemize}
\item there are polymorphism algebras $\fA$ of $\bA$ and $\fB$ of $\bB$ such that
$\HSPfin(\fA) = \HSPfin(\fB)$;
\item $\bA$ and $\bB$ are primitively positively bi-interpretable.
\end{itemize}
\end{proposition} 
\begin{proof}
For the forward implication, 
we assume that there is a $d_1 \geq 1$,
a subalgebra $\fS_1$ of $\fA^{d_1}$, and a surjective
homomorphism $h_1$ from $\fS_1$ to $\fB$.
Moreover, we assume that there is a $d_2 \geq 1$, 
a subalgebra $\fS_2$ of $\fB^{d_2}$, and a surjective homomorphism $h_2$ from $\fS_2$ to $\fA$.
The proof of Theorem~\ref{thm:pp-interpret} shows that $I_1 := (d_1,S_1,h_1)$
is an interpretation of $\bB$ in $\bA$,
and $I_2 := (d_2,S_2,h_2)$ is an interpretation of $\bA$ in $\bB$. Because the statement is symmetric it
suffices to show that the (graph of the) function $h_1 \circ h_2 \colon (S_2)^{d_1} \to B$ defined by $$(y_{1,1},\dots,y_{1,d_2},\dots,y_{d_1,1},\dots,y_{d_1,d_2}) \mapsto h_1(h_2(y_{1,1},\dots,y_{1,d_2}),\dots,h_2(y_{d_1,1},\dots,y_{d_1,d_2}))$$
(which is undefined whenever the expression above is undefined) 
is primitively positively definable
in $\bB$. 
Theorem~\ref{thm:inv-pol} asserts that this is equivalent to showing that %this relation 
$h_1 \circ h_2$ 
is preserved by all operations $f^\fB$ of $\fB$. 
So let $k$ be the arity of $f^\fB$ and let $b^i = (b^i_1,\dots,b^i_{d_1})$ be elements of
 $(S_2)^{d_1}$, for $1\leq i \leq k$. Then indeed
 \begin{align*}
&  f^{\fB}((h_1 \circ h_2)(b^1),\dots,(h_1 \circ h_2)(b^k)) \\
= \; & 
 h_1\big (f^\fA(h_2(b^1_1),\dots,h_2(b^k_{1})),\dots,f^\fA(h_2(b^1_{d_1}),\dots,h_2(b^k_{d_1}))\big ) \\
%= \; & (h_1 \circ h_2)(f^\fB(b^1_{1,1},\dots,b^k_{1,1}),\dots,f^{\fB} \\
= \; & (h_1 \circ h_2)(f^\fB(b^1,\dots,b^k)) \; .
\end{align*}

For the backwards implication, suppose that $\bA$ and $\bB$ are primitive positive bi-interpretable via an interpretation
$I_1=(d_1,S_1,h_1)$ of $\bB$ in $\bA$
and an interpretation $I_2=(d_2,S_2,h_2)$ of $\bA$ in $\bB$.
Let $\fA$ be a polymorphism algebra of $\bA$. 
The proof of Theorem~\ref{thm:pp-interpret} shows that
$S_1$ 
induces an 
algebra $\fS_1$ in $\fA^{d_1}$ and
$h_1$ is a surjective homomorphism from $\fS_1$ to an algebra $\fB$ satisfying
$\Clo(\fB) \subseteq \Pol(\bB)$ 
(the reverse inclusion is established later in the proof). 
%We may thus find a polymorphism algebra $\fB$ of $\bB$  is a 
%polymorphism algebra of $\bB$ that has the same signature $\tau$ as $\fA$. 
Similarly, $S_2$ is the domain of a subalgebra $\fS_2$ of $\fB^{d_2}$
and $h_2$ is a homomorphism from $\fS_2$ onto an algebra $\fA'$
such that $\Clo(\fA') \subseteq \Pol(\bA)$. 

We claim that 
$\HSPfin(\fA) = \HSPfin(\fB)$. The inclusion `$\supseteq$'
is clear since $\fB \in \HSPfin(\fA)$. 
For the reverse inclusion it suffices to show that $\fA = \fA'$ since $\fA' \in \HSPfin(\fB)$. 
Let $f \in \tau$ be $k$-ary; we show that $f^{\fA} = f^{\fA'}$. Let $a_1,\dots,a_k \in A$. Since $h_2 \circ h_1$ is surjective onto $A$, there are 
$c^i = (c^i_{1,1},\dots,c^i_{d_1,d_2}) \in A^{d_1d_2}$ 
such that $a_i = h_2 \circ h_1(c^i)$.
Then
\begin{align*}
f^{\fA'}(a_1,\dots,a_k) \; = \; & f^{\fA'}(h_2 \circ h_1(c^1),\dots,h_2 \circ h_1(c^k)) \\
= \; &h_2\big (f^{\fB}(h_1(c^1_{1,1},\dots,c^1_{d_1,1}),\dots,h_1(c^k_{1,1},\dots,c^k_{d_1,1})),\dots,\\
 & \quad \;  f^{\fB}(h_1(c^1_{1,d_2},\dots,c^1_{d_1,d_2}),\dots,h_1(c^k_{1,d_2},\dots,c^k_{d_1,d_2}))\big )\\
= \; & h_2 \circ h_1(f^\fA(c^1,\dots,c^k)) \\
= \; & f^\fA(h_2 \circ h_1(c^1),\dots,h_2 \circ h_1(c^k)) \\
 = \; & f^\fA(a_1,\dots,a_k) 
\end{align*}
where the second and third equations hold since $h_2$ and $h_1$ are 
algebra homomorphisms, and the fourth equation holds
because $f^\fA$ preserves $h_2 \circ h_1$,
because $I_2 \circ I_1$ is pp-homotopic to the identity.
\end{proof} 

%\newpage
\begin{exercises}
\exercise[Blau]\label{exe:source-151} Show that an algebra has the empty subuniverse
if and only if the signature does not contain constants
(i.e., function symbols of arity 0).
% Source: selbst, 1.4.23
% Solution sketch: A nullary basic operation is an element of every subuniverse, so the empty set
% cannot be one. If the signature has no constants, closure conditions only concern operations with at
% least one input; they are vacuous on the empty set, which is therefore a subuniverse.
\exercise[Blau]\label{exe:source-153}
Show that the operators $\HS$ and $\SH$ are distinct.
% Solution sketch: Let $\fA=(\{0,1,2,3\};f)$ with rows of the table of $f$ equal to $(1,0,2,0)$,
% $(3,3,3,3)$, $(1,0,3,0)$, and $(3,3,0,3)$. The algebra is simple, and its only proper nonempty
% subalgebras are $\{3\}$, $\{1,3\}$, and $\{0,1,3\}$. On the last one, the partition
% $\{\{0\},\{1,3\}\}$ is a congruence; its two-element quotient is not constant, whereas the only
% two-element subalgebra of $\fA$ is constant. Hence this quotient lies in $\HS(\fA)$ but not in
% $\SH(\fA)$.
\exercise[Blau]\label{exe:source-154} Show that the operators $\SP$ and $\PS$ are distinct.
% Solution sketch: Let $\fA=(\{0,1\};\wedge)$. The set $\{(0,0),(0,1),(1,0)\}$ is a three-element
% subalgebra of $\fA^2$, so it belongs to $\SP(\fA)$. Every product of subalgebras of $\fA$ has
% cardinality a power of two (the factors have size one or two), and therefore no member of $\PS(\fA)$
% is isomorphic to this algebra. Thus $\SP\ne\PS$.
\exercise[Rot]\label{exe:source-152} \label{exe:generation} Let $B$ be a subuniverse of an algebra $\fA$ generated by $S \subseteq A$.
Show that an element $a \in A$ belongs to $B$ if and only if
there exists a term $t(x_1,\dots,x_k)$ and elements $s_1,\dots,s_k \in S$
such that
$a = t^{\fA}(s_1,\dots,s_k)$.
% Solution sketch: The set of all values $t^{\fA}(s_1,\ldots,s_k)$ contains $S$ and is closed under
% every basic operation by term composition, hence contains the generated subuniverse. Conversely,
% induction over terms shows that every such value lies in every subuniverse containing $S$.
\end{exercises}

\subsection{Birkhoff's Theorem}
\label{sect:birkhoff}
Birkhoff's theorem provides a characterisation of varieties in terms of sets of \emph{identities}. 
A sentence in a functional signature $\tau$ is called a ($\tau$-) \emph{identity} 
if it is of the form
$$\forall x_1,\dots,x_n \colon s=t$$
where $s$ and $t$ are $\tau$-terms over the variables $x_1,\dots,x_n$ (such sentences are also called \emph{universally conjunctive}). We follow the usual notation in universal algebra and sometimes write 
such sentences as 
$$ s \approx t.$$
If ${\mathcal K}$ is a class of $\tau$-algebras, then we say that ${\mathcal K}$ \emph{satisfies} $s \approx t$ (or: $s \approx t$ \emph{holds in} ${\mathcal K}$), in symbols 
${\mathcal K} \models s \approx t$, if every algebra in ${\mathcal K}$ satisfies $s \approx t$. 

\begin{theorem}
[Birkhoff~\cite{Bir-On-the-structure}; see e.g.~\cite{HodgesLong} or~\cite{BS}]
\label{thm:birkhoff}
Let $\tau$ be a functional signature, 
 let ${\mathcal K}$ be a class of $\tau$-algebras, and let $\fA$ be a $\tau$-algebra. Then the following are equivalent.
\begin{enumerate}
\item All identities that hold in ${\mathcal K}$ also
hold in $\fA$;
%\item $\fA$ is in the variety generated 
%by $\fB$;
\item $\fA \in \HSP({\mathcal K})$.
\end{enumerate}
If $\fA$ has a finite domain, and ${\mathcal K} = \{\fB\}$ for some algebra 
$\fB$ with a finite domain, then this is also equivalent to 
\begin{enumerate}
\item[3.] $\fA \in \HSPfin(\fB)$.
\end{enumerate}
\end{theorem}

\begin{proof}
To show that
2.\ implies 1., let $s(x_1,\dots,x_n) \approx t(x_1,\dots,x_n)$
 be an identity that holds in ${\mathcal K}$. 
%and suppose that $\fA \in \HSP(\fB)$. 
Then $s \approx t$ is preserved in 
products 
$\fA = \prod_{i \in I} \fB_i$ of algebras $\fB_i \in {\mathcal K}$. To see this, let 
$a_1,\dots,a_n \in A$ be arbitrary.
Since $\fB_i \models s \approx t$ we have
$s^{\fB_i}(a_1[i],\dots,a_n[i]) = t^{\fB_i}(a_1[i],\dots,a_n[i])$ for every 
$i \in I$, and thus $s^{\fA}(a_1,\dots,a_n) = t^{\fA}(a_1,\dots,a_n)$ by the definition of products. 
Since $a_1,\dots,a_n$ were chosen arbitrarily, 
we have $\fA \models s \approx t$. 
Moreover, universal sentences are preserved by taking subalgebras. 
Finally, suppose that $\fB$ is an algebra
that satisfies $s \approx t$, and $\mu$ is a surjective homomorphism from $\fB$ to
some algebra $\fA$. Let $a_1,\dots,a_n \in A$. By the surjectivity
of $\mu$ we can choose $b_1,\dots,b_n$ such that $\mu(b_i) = a_i$ for all $i \leq n$. Then
\begin{align*}
s^{\fB}(b_1,\dots,b_n) = t^{\fB}(b_1,\dots,b_n) 
\Rightarrow & \quad \mu(s^{\fB}(b_1,\dots,b_n)) = \mu(t^{\fB}(b_1,\dots,b_n)) \\
\Rightarrow & \quad t^{\fA}(\mu(b_1),\dots,\mu(b_n)) =  s^{\fA}(\mu(b_1),\dots,\mu(b_n)) \\
\Rightarrow & \quad t^{\fA}(a_1,\dots,a_n) =  s^{\fA}(a_1,\dots,a_n)
\; . 
\end{align*}
%Hence, $\fA \models \phi$. 

%1.\ implies 2.: %
%Since all universal conjunctive $\tau$-sentences that hold in $\fB$ also hold in $\fA$, we have that $s^{\fA}=t^{\fA}$ whenever $s^{\fB} = t^{\fB}$;
%hence, the natural homomorphism 
%$\xi$ from
%$\Clo(\fB)$ onto $\Clo(\fA)$ exists. 
We only show the implication 
from 1.\ to 3.\ (and hence to 2.) if $\fA$ and $\fB$ have finite domains and ${\mathcal K} = \{\fB\}$; the proof of the general case is similar (see Exercise~\ref{exe:full-Birkhoff}). 
Let $a_1,\dots,a_k$ be the elements 
of $\fA$, 
define $m := |B|^k$ and $C := B^k$. Let $c^1,\dots,c^m$ be the elements of $C$; write
$c_i$ for $(c^1_i,\dots,c^m_i)$. 
Let $\fS$ be the smallest subalgebra of $\fB^m$ that contains 
$c_1,\dots,c_k$; so the
elements of $\fS$ are precisely those of the form
$t^{\fB^m}(c_1,\dots,c_k)$, for 
a $k$-ary $\tau$-term $t$. See Figure~\ref{fig:Birkhoff}. 

\begin{figure}
\begin{center}
\includegraphics[scale=0.6]{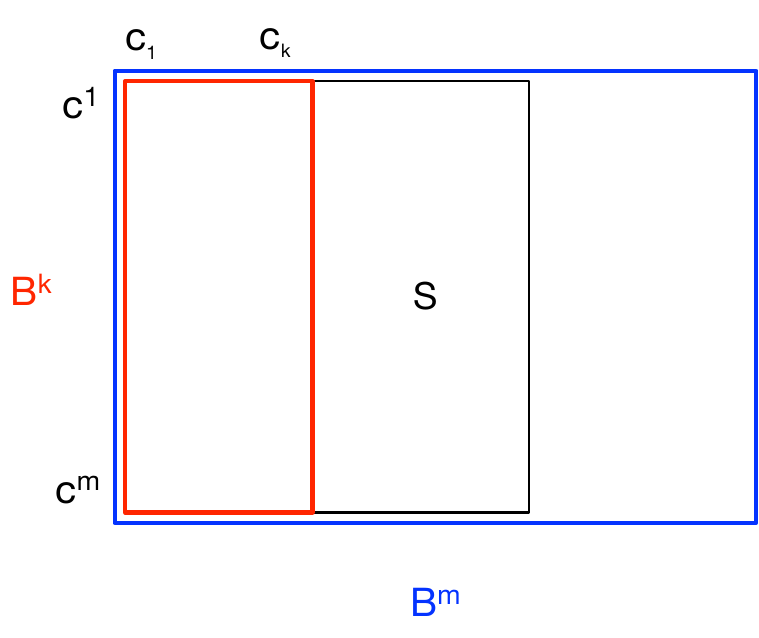} 
\end{center}
\caption{Illustration for the proof of Birkhoff's theorem.}
\label{fig:Birkhoff}
\end{figure}

Define $\mu \colon S \to A$ by 
$$\mu(t^{\fB^m}(c_1,\dots,c_k)) := t^{\fA}(a_1,\dots,a_k).$$ 

 {\bf Claim 1:} $\mu$ is well-defined.
\par\noindent
Since $c^1,\dots,c^m$ enumerate all of $B^k$, the equality
$t^{\fB^m}(c_1,\dots,c_k) = s^{\fB^m}(c_1,\dots,c_k)$ implies $t^\fB = s^{\fB}$.
The assumption therefore gives $t^{\fA}(a_1,\dots,a_k) = s^{\fA}(a_1,\dots,a_k)$. 
 
 {\bf Claim 2:} $\mu$ is surjective. 
For all $i \leq k$, the element $c_i$ is mapped to $a_i$.  
 
 {\bf Claim 3:} $\mu$ is a homomorphism from $\fS$ to $\fA$. 
Let $f \in \tau$ be of arity $n$ and let $s_1,\dots,s_n \in S$. For $i \leq n$, write $s_i = t_i^{\fS}(c_1,\dots,c_k)$ for 
some $\tau$-term $t_i$ (see Exercise~\ref{exe:generation}). Then
	 \begin{align*}
	\mu\big(f^\fS(s_1,\ldots,s_n)\big)= & \, \mu\big (f^{\fS}(t_1^\fS(c_1,\ldots,c_k),\ldots,t_n^\fS(c_1,\ldots,c_k))\big) \\
		= & \, 
		\mu \big (f^\fS(t_1^\fS,\ldots,t_n^\fS)(c_1,\ldots,c_k) \big)\\
		= & \, \mu\big ((f(t_1,\ldots,t_n))^\fS(c_1,\ldots,c_k)\big ) \\
		= & \, 
		\big(f(t_1,\ldots,t_n)\big)^\fA(a_1,\ldots,a_k)\\
		= & \, 
		f^\fA\big (t_1^\fA(a_1,\ldots,a_k),\ldots,t_n^\fA(a_1,\ldots,a_k)\big)\\
		= & \, 
		f^\fA(\mu(s_1),\ldots,\mu(s_n)).
	 \end{align*}
Therefore, $\fA$ is the homomorphic image of the subalgebra $\fS$ of ${\fB}^m$, and so $\fA \in \HSPfin(\fB)$. 
\end{proof}

Theorem~\ref{thm:birkhoff} is 
important for analysing the constraint satisfaction problem for a structure $\bB$, since it can be used to transform the `negative'
statement of not interpreting certain finite structures, which is equivalent to not having a certain finite algebra in the pseudo-variety generated by a polymorphism algebra of $\bB$, into a `positive'
statement of having polymorphisms satisfying non-trivial identities. We will learn several concrete identities that must be satisfied in later sections, e.g., in Section~\ref{sect:taylor}, Section~\ref{sect:6ary}, Section~\ref{sect:cyclic-thm}, and Section~\ref{sect:4ary}. 
%: this will be the content of Section~\ref{sect:abstract-clone}
%and Section~. 

%\ignore{
\subsubsection{The Free Algebra}
%In this section 

In the following we extract an 
important idea from the proof of Birkhoff's theorem and present it in different words which will be useful later. Fix a functional signature $\tau$ and a class of 
$\tau$-algebras ${\mathcal K}$. 

\begin{definition}
Let $\fF$ be a $\tau$-algebra generated by $X \subseteq F$. We say that $\fF$ has the \emph{universal mapping property}
for ${\mathcal K}$ over $X$ if 
for every $\fA \in {\mathcal K}$ and $f \colon X \to A$ there exists a (unique) extension of $f$ to a homomorphism from $\fF$ to $\fA$. 
\end{definition}

\begin{proposition}[Uniqueness]
\label{prop:free-unique}
Suppose that 
$\fF_1,\fF_2 \in {\mathcal K}$ 
have the universal mapping property for ${\mathcal K}$ over $X_i$, for $i \in \{1,2\}$. If $|X_1| = |X_2|$ then 
$\fF_1$ and $\fF_2$ are isomorphic. 
\end{proposition}
\begin{proof}
Fix any bijection between $X_1$ and $X_2$. It extends uniquely to homomorphisms in both directions. Each composite fixes the generators pointwise and is therefore the identity by uniqueness, so the two extensions are inverse isomorphisms. 
\end{proof} 

\begin{proposition}[Existence]
\label{prop:free-subpower}
For every 
%$\tau$-algebra $\fB$ 
class ${\mathcal K}$ 
of $\tau$-algebras 
and for every set $X$ there 
exists a $\tau$-algebra $\fF \in \SP({\mathcal K})$ which has the universal mapping property for  
$\HSP({\mathcal K})$ over $X$. 
\end{proposition}

\begin{definition}[Free algebra]
\label{def:free}
The up to isomorphism unique 
algebra $\fF \in \SP({\mathcal K})$ with the universal mapping property for $\HSP({\mathcal K})$ over $X$ is called
 the  \emph{free algebra for ${\mathcal K}$ over $X$}, and will be denoted by $\fF_{\mathcal K}(X)$. 
\end{definition} 

%\begin{proof}
%\end{proof} 

\begin{lemma}
\label{lem:easy-free}
%Let 
%${\mathcal V}$ be a class of $\tau$-algebras, 
%$s and $t$ be $\tau$-terms, and 
Let $\fF$ be free for ${\mathcal K}$ 
over $\{x_1,\dots,x_n\}$ and let $s(y_1,\dots,y_n)$, $t(y_1,\dots,y_n)$ be $\tau$-terms. Then the following are equivalent. 
\begin{enumerate} 
\item ${\mathcal K} \models s(y_1,\dots,y_n) \approx t(y_1,\dots,y_n)$
%with variables $x_1,\dots,x_n$ 
%holds in every algebra of ${\mathcal K}$;
\item $s(y_1,\dots,y_n) \approx t(y_1,\dots,y_n)$ holds in $\fF$;
\item $s^{\fF} = t^{\fF}$; 
\item $s^{\fF}(x_1,\dots,x_n) = t^{\fF}(x_1,\dots,x_n)$. 
%Then $\fF \models s(x_1,\dots,x_n) \approx t(x_1,\dots,x_n)$ if and only if
%every $\tau$-algebra in ${\mathcal K}$ satisfies $s(x_1,\dots,x_n) \approx t(x_1,\dots,x_n)$. 
\end{enumerate}
\end{lemma}
\begin{proof}
$1. \Rightarrow 2.:$
If $s \approx t$ holds in every algebra of ${\mathcal K}$, then it also holds in products and subalgebras of algebras in ${\mathcal K}$, and hence also in $\fF$. 

$2. \Rightarrow 3.$ and $3. \Rightarrow 4.$ hold trivially. 

$4. \Rightarrow 1$. 
%$\fF \models s(y_1,\dots,y_n) \approx t(y_1,\dots,y_n)$, 
Let $\fA \in {\mathcal K}$. 
If $a_1,\dots,a_n \in A$, then the map that sends $x_i$ to $a_i$ for all $i \in \{1,\dots,n\}$ can be extended to a homomorphism from $\fF$ to $\fA$, and
since $s^{\fF}(x_1,\dots,x_n) = t^{\fF}(x_1,\dots,x_n)$ we have $s^{\fA}(a_1,\dots,a_n) = t^{\fA}(a_1,\dots,a_n)$. Since $a_1,\dots,a_n \in A$ were chosen arbitrarily, this shows that $\fA \models s(y_1,\dots,y_n) \approx t(y_1,\dots,y_n)$. 
\end{proof} 

\begin{remark}\label{rem:hspfin}
Note that if ${\mathcal K} := \HSP(\fB)$, and $B$ and $X$ are finite, then $\fF_{\mathcal K}(X) \leq \fB^{B^X}$ is finite as well. 
\end{remark}

\subsubsection{Equational Theories}
Birkhoff's theorem provides for every class $\mathcal K$ of $\tau$-algebras a characterisation of the class of all $\tau$-algebras that satisfy all identities satisfied by ${\mathcal K}$. Conversely, there is for every set of $\tau$-identities $\Sigma$ a syntactic characterisation of the set of all $\tau$-identities that are satisfied in all algebras that satisfy $\Sigma$ (i.e., we have a proof-theoretic characterisation of the set of all universal conjunctive consequences of $\Sigma$; this can be seen as a special case of the completeness theorem of first-order logic for universally conjunctive sentences in an algebraic signature $\tau$). 

\begin{theorem}\label{thm:Ihringer}
A $\tau$-identity $f \approx g$ is implied by
$\Sigma$ (i.e., holds in all algebras that satisfy $\Sigma$) if and only if $(f,g)$ is contained in the smallest equivalence relation $E$ which 
\begin{itemize}
\item contains $(r,s)$ for every $(r \approx s) \in \Sigma$, 
\item is compatible: if $f \in \tau$ is a function symbol of arity $n$ and $(r_1,s_1),\dots,(r_n,s_n) \in E$,
then $(f(r_1,\dots,r_n),f(s_1,\dots,s_n)) \in E$, 
\item is fully invariant: if $(r(x_1,\dots,x_n),s(x_1,\dots,x_n)) \in E$ and
$t_1,\dots,t_n$ are $\tau$-terms, then 
$(r(t_1,\dots,t_n),s(t_1,\dots,t_n)) \in E$. 
\end{itemize}
\end{theorem}

\begin{exercises}
\exercise[Blau]\label{exe:source-155} Prove Proposition~\ref{prop:free-subpower}.
% Solution sketch: Index a power of $\fA$ by all pairs $(\fB,h)$ under consideration and map each
% generator to its tuple of evaluations. The subalgebra generated by these evaluation tuples is free:
% a map on the generators extends coordinatewise to a homomorphism, and equality of term values in
% this subpower is exactly equality forced by $\fA$. This proves the universal property.
\exercise[Orange]\label{exe:source-156} Show the implication from 1.\ to 2.\ in Birkhoff's theorem in full generality. \label{exe:full-Birkhoff}
% Solution sketch: Let $F$ be the free algebra in $\mathsf V(\fA)$ on the underlying generating set of
% $\fB$. Evaluation gives a surjection $F\to\fB$. The free-subpower proposition embeds $F$ into a
% power of $\fA$; quotienting the corresponding subalgebra by the kernel of evaluation shows
% $\fB\in\HSP(\fA)$, without any finiteness assumption.
\end{exercises}

\subsection{Abstract Clones}
\label{sect:abstract-clone}
Clones (in the literature often \emph{abstract clones}) relate to operation clones in the same way as (abstract) groups relate to permutation groups: the elements of a clone correspond to the operations of an operation clone, and the signature contains composition symbols to code how operations compose. Since an operation clone 
contains operations of various arities, a clone will be formalized 
as a multi-sorted structure, with a sort for each arity. 
%Interessiert keinen: Abstract clones have also been formalized 
%in category theory; we refer to~\cite{Trnkova}. 

\begin{definition}
An (abstract) \emph{clone} $\fC$ is a multi-sorted structure with sorts $\{C^{(i)} \; | \; i \in \mathbb N_{\geq 1}\}$ and the
signature $\{\pi_i^k \; | \; 1\leq i \leq k\}
\cup \{\comp^k_l \; | \; k,l \geq 1\}$. 
For $k \geq 1$, 
the elements of the sort $C^{(k)}$ will be called the \emph{$k$-ary operations} of $\fC$. 
%$$\C = (C;(p_i^k)_{1\leq i \leq k},(\comp^k_l)_{m,l \geq 1})$$ 
%with one sort $C^{(k)} \subseteq C$ for each arity $k \in \mathbb N$.  
%The relationship between operation clones and clones is in perfect analogy to the relationship between permutation groups and groups. Formally, 
We denote a clone by 
$$\fC = (C^{(1)},C^{(2)},\dots;(\pi_i^k)_{1\leq i \leq k},(\comp^k_l)_{k,l \geq 1})$$ and require that
%a clone is a multi-sorted structure
%$\bf C = (C_0,C_1,\dots;(p_i^k)_{1\leq i \leq k},(\comp^k_l)_{m,l \geq 1})$ with sorts indexed by $\omega$ such that 
%\begin{itemize}
%\item 
$\pi_i^k$ is a constant in $C^{(k)}$,
and that  
$\comp_l^k \colon C^{(k)} \times (C^{(l)})^k \to C^{(l)}$ is an
 operation of arity $k+1$.
Moreover, 
for $f \in C^{(k)}$, 
$g_1,\dots,g_k \in C^{(m)}$, and $f_1,\dots,f_k,h_1,\dots,h_m \in C^{(l)}$ it 
holds that
\begin{align}
\comp_k^k(f,\pi_1^k,\dots,\pi_k^k) & = f \label{eq:pin} \\ 
\comp^k_l(\pi_i^k,f_1,\dots,f_k) & = f_i \label{eq:pout} \\
\comp_l^k\big(f,\comp_l^m(g_1,h_1,\dots,h_m),\dots,\comp_l^m(g_k,h_1,\dots,h_m)\big) 
& =  \nonumber  \\
\comp^m_l\big(\comp^k_m(f,g_1,\dots,g_k),h_1,\dots,h_m\big) & \, . \label{eq:comp}
\end{align}
\end{definition}
The final equation generalises associativity in groups and monoids, and we therefore refer to it by \emph{associativity}. 
We also write $f(g_1,\dots,g_k)$ instead
of $\comp^k_l(f,g_1,\dots,g_k)$ when 
$l$ is clear from the context. 
So associativity might be more readable as 
$$f(g_1(h_1,\dots,h_m),\dots,g_k(h_1,\dots,h_m)) = f(g_1,\dots,g_k)(h_1,\dots,h_m) \, .$$

Every operation clone $\cC$ gives rise to an abstract
clone $\fC$ in the obvious way: $\pi^k_i \in C^{(k)}$ denotes the 
$k$-ary $i$-th projection in $\cC$,
and $\comp^k_l(f,g_1,\dots,g_k) \in C^{(l)}$ denotes
the composed operation $(x_1,\dots,x_l) \mapsto f(g_1(x_1,\dots,x_l),\dots,g_k(x_1,\dots,x_l)) \in \cC$.
Conversely, every abstract clone arises from an operation clone - this will follow from Proposition~\ref{prop:clone-variety-clone}. 

\begin{example}
An algebra $\fA$ satisfies 
$f(x_1,x_2) \approx f(x_2,x_1)$ if and only if
\begin{align*}
& \Clo(\fA) \models \comp^2_2(f^\fA,\pi_1^2,\pi_2^2)=\comp^2_2(f^\fA,\pi_2^2,\pi_1^2). \qedhere 
\end{align*}
\end{example}

In the following, we will also use the term `abstract clone' in situations where we want to stress that we are 
working with a clone and \emph{not} with an operation clone.
The notion of a \emph{homomorphism} between clones is just the usual notion of homomorphisms for algebras, adapted to the multi-sorted case. Since we did not  formally introduce homomorphisms for multi-sorted structures, we spell out the definition in the special case of clones. 

\begin{definition}
Let $\fC$ and $\fD$ be clones. A function
$\xi\colon C \to D$ is called a \emph{(clone) homomorphism} if
\begin{enumerate}
\item $\xi$ preserves arities of operations, i.e., $\xi(C^{(i)}) \subseteq D^{(i)}$ for all $i \geq 1$; 
\item $\xi((\pi^k_i)^{\fC}) = (\pi^k_i)^{\fD}$ 
for all $k \geq 1$ and  
$1 \leq i \leq k$;
\item $\xi(f(g_1,\dots,g_n)) = \xi(f)(\xi(g_1),\dots,\xi(g_n))$ for all $n,m \geq 1$, $f \in C^{(n)}$, 
$g_1,\dots,g_n \in C^{(m)}$.
\end{enumerate}
Moreover, $\xi$ is a \emph{(clone) 
isomorphism} if $\xi$ is bijective and
both $\xi$ and $\xi^{-1}$ are a homomorphism. 
\end{definition}

\begin{example}
We write $\Proj$ for the abstract clone
of an algebra with at least two elements all of whose operations are projections; note that any such algebra has the same abstract clone (up to isomorphism), and that $\Proj$ has a homomorphism into any other clone.
\end{example}

\begin{example}
All abstract clones of an algebra on a one-element set are isomorphic, too, but of course not isomorphic to $\Proj$. 
Any clone homomorphically maps to this trivial clone. 
\end{example}

\begin{example}\label{expl:K3Proj}
Using Proposition~\ref{prop:kn-is-projective}, it is easy to see that 
there exists a clone homomorphism from $\Pol(K_3)$ to $\Proj$. 
\end{example}

The following definition plays an important role  throughout the later sections in this text.

\begin{definition}[Star composition]
\label{def:star}
Let $l,m \in \mN$ and $n = lm$.
We write $f * g$ as a shortcut for 
\begin{align*}
\comp^{l}_{n}(f,\comp^{m}_{n}(g,\pi^{n}_1,\dots,\pi^{n}_{m}),\dots,\comp^{m}_{n}(g,\pi^{n}_{(l-1)m+1},\dots,\pi^{n}_{n})) \, .
\end{align*}
\end{definition}
Note that  if $f \colon A^l \to A$ 
and $g \colon A^m \to A$, then 
$f * g$ denotes the operation from
$A^{lm} \to A$ given by 
$$(x_{1,1},\dots,x_{l,m}) \mapsto 
f \big (g(x_{1,1},\dots,x_{1,m}),\dots,g(x_{l,1},\dots,x_{l,m}) \big).$$

\subsection{Clone Formulation of Birkhoff's Theorem}
One can translate back and forth between varieties and abstract clones. 

\begin{definition}[$\Var(\fC)$]
\label{def:var-clone}
For any abstract clone $\fC$, the variety $\Var(\fC)$ is defined as follows. 
We use the elements of $C$ as a functional signature $\tau$, where the elements of $C^{(n)}$ are $n$-ary function symbols. 
We consider the set $\Sigma$ of $\tau$-identities defined as follows. If $f \in C^{(k)}$ and $g_0,g_1,\dots,g_k \in C^{(m)}$ are such that 
$\fC \models (g_0 = \comp^{k}_m(f,g_1,\dots,g_k))$, then 
we add the identity 
$g_0(y_1,\dots,y_m) \approx f(g_1(y_1,\dots,y_m),\dots,g_k(y_1,\dots,y_m))$ to $\Sigma$. 
Moreover, we add the identities
$\pi^n_i(y_1,\dots,y_n) \approx y_i$ to $\Sigma$. 
Then $\Var(\fC)$ denotes the class of $\tau$-algebras that satisfy $\Sigma$. 
\end{definition}

Conversely, to every variety $\mathcal V$
we may associate the clone $\Clo({\mathcal V}) := \Clo(\fF_{\mathcal V}(\{x_1,x_2,\dots\}))$ of the algebra 
$\fF_{\mathcal V}(\{x_1,x_2,\dots\})$ which is free for ${\mathcal V}$ over countably many generators. 

\begin{proposition}\label{prop:clone-variety-clone}
Let $\fC$ be an abstract clone.
Then $\Clo(\Var(\fC))$ is isomorphic to $\fC$. 
\end{proposition}
\begin{proof}
Let ${\mathcal V} := \Var(\fC)$, and let $\Sigma$ be the set of $\tau$-identities that defines ${\mathcal V}$. 
Let $\fF := \fF_{\mathcal V}(\{x_1,x_2,\dots\})$ and let $\fD := \Clo(\fF)$. 

\medskip 
%\noindent
{\bf Claim 1.} The map $\xi$ that sends $f \in C^{(n)}$ to $f^{\fF} \in D^{(n)}$ is a clone homomorphism $\fC \to \fD$: 
\begin{itemize}
\item It clearly preserves arities.
\item If $i \in [n]$ then $\Sigma$ contains
$\pi^n_i(y_1,\dots,y_n) \approx y_i$.
Since $\fF \models \Sigma$ 
we have $(\pi^n_i)^{\fF}(a_1,\dots,a_n) = a_i$ for all $a_1,\dots,a_n \in F$. 
Hence, $\xi((\pi^n_i)^{\fC}) = (\pi^n_i)^{\fF} = (\pi^n_i)^{\fD}$. 
\item If $n,m \geq 1$, $f \in C^{(n)}$, $g_1,\dots,g_n \in C^{(m)}$, and 
$g_0 = f(g_1,\dots,g_n)$, then 
$\Sigma$ contains
$g_0(y_1,\dots,y_m) \approx f(g_1(y_1,\dots,y_m),\dots,g_n(y_1,\dots,y_m))$, and
since $\fF \models \Sigma$ 
it follows that
$\xi(g_0) = \xi(f) (\xi(g_1),\dots,\xi(g_n))$. 
\end{itemize}

\medskip 
{\bf Claim 2.} $\xi$ is surjective. Every element of $D$ is of the form $t^{\fF}$, for some $\tau$-term $t$. It can be shown by induction over the term structure that there exists $s \in \tau$ such that $s^{\fF} = t^{\fF}$, and hence $\xi(s) = t^{\fF}$. 

\medskip 
{\bf Claim 3.} $\xi$ is injective. 
Suppose that $\xi(f) = \xi(g)$ for some $f,g \in C^{(n)}$. 
We construct a
$\tau$-algebra $\fA_n$ with domain $C^{(n)}$. For every function symbol
$h\in C^{(k)}$, define
\[
h^{\fA_n}(q_1,\dots,q_k)
:=
\comp^k_n(h,q_1,\dots,q_k)
\qquad
\text{for }q_1,\dots,q_k\in C^{(n)}.
\]
We claim that $\fA_n\in\mathcal V$, that is, that $\fA_n$ satisfies
$\Sigma$. First, if $q_1,\dots,q_m\in C^{(n)}$, then
\[
(\pi_i^m)^{\fA_n}(q_1,\dots,q_m)
=
\comp^m_n(\pi_i^m,q_1,\dots,q_m)
=
q_i,
\]
so $\fA_n$ satisfies the projection identities from $\Sigma$.
Now let $h\in C^{(k)}$ and $g_0,g_1,\dots,g_k\in C^{(m)}$ be such that
\[
g_0=\comp^k_m(h,g_1,\dots,g_k).
\]
For all $q_1,\dots,q_m\in C^{(n)}$, the associativity axioms of the
abstract clone give
\[
\begin{aligned}
g_0^{\fA_n}(q_1,\dots,q_m)
&=
\comp^m_n(g_0,q_1,\dots,q_m)\\
&=
\comp^k_n\bigl(
h,
\comp^m_n(g_1,q_1,\dots,q_m),
\dots,
\comp^m_n(g_k,q_1,\dots,q_m)
\bigr)\\
&=
h^{\fA_n}\bigl(
g_1^{\fA_n}(q_1,\dots,q_m),
\dots,
g_k^{\fA_n}(q_1,\dots,q_m)
\bigr).
\end{aligned}
\]
Thus $\fA_n$ also satisfies every composition identity from $\Sigma$,
and hence $\fA_n\in\mathcal V$.

The assumption $\xi(f)=\xi(g)$ means that $f^{\fF}=g^{\fF}$ and thus 
\[
\fF\models
f(y_1,\dots,y_n)\approx g(y_1,\dots,y_n).
\]
By Lemma~\ref{lem:easy-free}, this identity holds in every algebra of
$\mathcal V$, and in particular in $\fA_n$. Evaluating it in $\fA_n$
at the elements $\pi_1^n,\dots,\pi_n^n\in C^{(n)}$, we obtain
\[
\comp^n_n(f,\pi_1^n,\dots,\pi_n^n)
=
\comp^n_n(g,\pi_1^n,\dots,\pi_n^n).
\]
The projection axioms of the abstract clone imply that the left-hand
side is $f$ and the right-hand side is $g$. Hence $f=g$, proving that
$\xi$ is injective.
\end{proof}

Proposition~\ref{prop:clone-variety-clone} in particular shows the following, which can be seen as an analog of Cayley's theorem for clones.

\begin{corollary} \label{cor:clone-cayley} 
Every abstract clone is isomorphic to an operation clone.
\end{corollary} 

Proposition~\ref{prop:clone-variety-clone} has a converse; to state it, we need the following definition.

\begin{definition}
Let ${\mathcal V}$, ${\mathcal W}$ be varieties with signatures $\sigma$ and $\rho$, respectively. 
An \emph{interpretation} of ${\mathcal V}$ in ${\mathcal W}$ is a map $I$ from $\sigma$ to $\rho$-terms such that 
symbols of arity $k$ are sent to terms with $k$ free variables, and 
${\mathcal V}$ contains 
%$I(f)$ 
%for $f \in \rho$ of arity $n$ is a $\sigma$-term $t(x_1,\dots,x_n)$ such that 
$\{ I({\fA}) \mid \fA \in {\mathcal W} \}$
where $I({\fA})$ is the $\sigma$-algebra with domain $A$ and the operation $I(f)^{\fA}$ for $f \in \sigma$. 
%$for every ${\fA} \in {\mathcal}$ 
\end{definition}

The following lemma is straightforward from the definitions. 

\begin{lemma}
Let ${\mathcal V}$ and ${\mathcal W}$ be varieties. Then there is an interpretation of ${\mathcal V}$ in ${\mathcal W}$ if and only if there exists a
clone homomorphism from 
$\Clo({\mathcal V})$ to $\Clo({\mathcal W})$. 
\end{lemma}

\begin{proposition}\label{prop:variety-clone-variety}
Let ${\mathcal V}$ be a variety. 
Then $\Var(\Clo({\mathcal V}))$ and ${\mathcal V}$ mutually interpret each other. 
\end{proposition}
\begin{proof}
Let $\sigma$ be the signature of ${\mathcal V}$ and let $\rho$ be the signature of ${\mathcal W} := \Var(\Clo({\mathcal V}))$. 
Let $\fF := \fF_{\mathcal V}(\{x_1,x_2,\dots\})$. 
The identities that hold in every algebra of ${\mathcal V}$ are precisely those that hold in $\fF$ by Lemma~\ref{lem:easy-free}. 
Then the map that sends $f \in \sigma$ of arity $k$ to
$f^{\fF}(x_1,\dots,x_k)$, viewed as a $\rho$-term, is an interpretation of ${\mathcal V}$ in ${\mathcal W}$. 

Conversely, every
$f \in \rho$ of arity $n$ has been introduced for an element of $\fF$ which equals $t^{\fF}(x_{i_1},\dots,x_{i_n})$ for some $i_1,\dots,i_n \in {\mathbb N}$ and some $\sigma$-term $t(y_1,\dots,y_n)$.  
The map $J$ that sends $f$ to $t(y_1,\dots,y_n)$
is an interpretation of ${\mathcal W}$ in ${\mathcal V}$. 
%sends $f \in \sigma$ of arity $n$ to the symbol in $\rho$ that 
%because the algebras in ${\mathcal W}$ are precisely
%the algebras that satisfy the 
%it suffices
%to show that every identity that holds in every algebra of ${\mathcal V}$ also holds in every algebra of ${\mathcal W}$ and vice versa. 
%Let $\Sigma$ be the set of identities involved in the definition of ${\mathcal W}$. 
%And these are precisely the identities in the set $\Sigma$ that defines ${\mathcal W}$. 
\end{proof}

%a clone $\Clo({\mathcal V})$ as follows. 
%\begin{definition}
%Let ${\mathcal V}$ be a variety of $\tau$-algebras. Then $\Clo({\mathcal V})$ denotes the clone 
%\end{definition}
%To every clone $\fC$, we may associate in a canonical way an variety $\Var(\fC)$ as follows: 
The following proposition links the existence of clone homomorphisms with the language of algebras, and in particular identities and (pseudo-) varieties. 

\begin{proposition}\label{prop:clone-homo}
Let $\mathscr C$ and $\mathscr D$ be operation clones on finite sets.  
Then the following are equivalent. 
\begin{enumerate}
\item There is a surjective clone
homomorphism from $\cC$ to $\cD$;
\item there are algebras $\fA$ and $\fB$ 
with the same signature $\tau$ such that 
$\Clo(\fA) = \mathscr D$,
$\Clo(\fB) = \mathscr C$,
and all universal conjunctive $\tau$-sentences
that hold in $\fB$ also hold in $\fA$;
\item there are algebras $\fA$ and $\fB$ 
with the same signature such that 
$\Clo(\fA) = \mathscr D$,
$\Clo(\fB) = \mathscr C$,
and $\fA \in \HSPfin(\fB)$ (equivalently, 
$\fA \in \HSP(\fB)$). 
\end{enumerate}
Moreover, the following are equivalent. 
\begin{itemize}
\item There is a clone isomorphism between ${\mathscr C}$ and ${\mathscr D}$. 
\item there are algebras $\fA$ and $\fB$ 
with the same signature such that 
$\Clo(\fA) = \mathscr D$,
$\Clo(\fB) = \mathscr C$,
and $\HSPfin(\fA) = \HSPfin(\fB)$ (equivalently:
$\HSP(\fA) = \HSP(\fB)$).  
\end{itemize}
\end{proposition}

%\begin{corollary}\label{cor:birk}
%\end{corollary}

%\begin{proposition}\label{prop:clone-homo}
%Let $\fA$ and $\fB$ be $\tau$-algebras. 
%The following are equivalent. 
%\begin{enumerate}
%\item The natural homomorphism from
%$\Clo(\fB)$ to $\Clo(\fA)$ exists;
%\item all universal conjunctive 
%equations that hold in $\fB$ also hold in $\fA$;
%\item $\fA \in \HSP(\fB)$. 
%\end{enumerate}
%\end{proposition}
%\begin{proof}
%The equivalence between (1) and (2) follows from the definitions. The equivalence of (2), (3), and (4) follows
%from Theorem~\ref{thm:birkhoff} applied to ${\cal C} := \{\fB\}$. 
%\end{proof}

In the study of the complexity of CSPs,
the equivalence between (1) and (3) in the above
is the most relevant, since (3) is related to 
our most important tool to prove NP-hardness
of CSPs (because of the link between pseudovarieties and primitive positive interpretations from Theorem~\ref{thm:pp-interpret}), and since (1) is the universal-algebraic property that will be used in the following (see e.g.~Theorem~\ref{thm:taylor} below). 
The following lemma is central for our applications
of abstract clones when studying the complexity of CSPs; it applies to all operation clones $\fF$ on a finite set. 
 
\begin{lemma}\label{lem:clone-compactness}
Let $\fC$ be a clone and let $\fF$ be a clone that has finitely many elements of each sort 
such that there is no clone homomorphism from $\fC$ to $\fF$.
Then there is a primitive positive sentence in the signature $\tau$ of (abstract) clones that
holds in $\fC$ but not in $\fF$. 
\end{lemma}
\begin{proof}
Let $\fE$ be the expansion of $\fC$ by constant symbols such that every element $e$ of $\fE$ is named by a constant $c_e$. 
Let $V$ be the set of atomic sentences that hold in $\fE$. 
Let $U$ be the first-order theory of $\fF$. 
 Suppose for contradiction that $U \cup V$ has 
a model $\fM$. 
%For all $i,n \geq 1$ there is a first-order sentence $\phi^i_{n}$ which holds in a clone if and only if the $i$-th sort has exactly $n$ elements. 
%Since $\fF$ satisfies $\phi^i_{|F^{(i)}|}$, and $\fM$ satisfies $U$,  
%we must have $|M^{(i)}|=|F^{(i)}|$ for all $i \geq 1$. 

For \(n\geq1\), let \(\fF_{\leq n}\) and \(\fM_{\leq n}\) be the $\tau$-reducts of $\fF$ and $\fM$ restricted to the first \(n\) sorts. These are finite (multisorted) structures. Since  \(\fF\) and \(\fM\) satisfy the same first-order theory, \(\fF_{\leq n}\) 
and \(\fM_{\leq n}\) are isomorphic. The isomorphisms between these finite reducts form a finitely branching tree under restriction. By K\H{o}nig's lemma, this tree has an infinite branch. The union of the isomorphisms on this branch is an isomorphism between \(\fF\) and the restriction of 
the $\tau$-reduct of 
\(\fM\) to $\bigcup_{i} M^{(i)}$ (note that there might be elements in $\fM$ outside of $\bigcup_i M^{(i)}$); 
we identify it with $\fF$. Note that for all constants $c_e$ we have that $c_e^{\fM} \in \fF$. 
	Since 
	$\fM$ satisfies all atomic formulas that hold in $\fE$,
	 we have 
	that the mapping $e \mapsto c_e^\fM$, for $e$ an element of $\fE$, 
	is a homomorphism from $\fC$ to $\fF$, in contradiction to our assumptions. 
	
	So $U \cup V$ is unsatisfiable, and by compactness of first-order logic\footnote{See, e.g.,~\cite{BodMathLogic}.} there 
	exists a finite subset $V'$ of $V$ such that $V' \cup U$ 
	is unsatisfiable. 
	Replace each of the new constant symbols in $V'$
	by an existentially quantified variable;
	then the conjunction of the resulting sentences from $V'$ is a primitive positive $\tau$-sentence,
	and it must be false in $\fF$. 
\end{proof}

Similarly as in Definition~\ref{def:var-clone}, 
a primitive positive sentence in the signature of abstract clones that holds in $\Clo(\fA)$ translates into the existence of a set of identities in the signature of $\fA$
that holds in $\fA$. 
A set of identities $\Sigma$ is called \emph{trivial} if there exists an algebra $\fA$ that satisfies $\Sigma$ and $\Clo(\fA)$ is isomorphic to $\Proj$. 

\begin{corollary}\label{cor:identity-compact}
Let $\fA$ be an algebra. If there is no clone homomorphism from $\Clo(\fA)$ to $\Proj$, then there exists a non-trivial finite set of identities that holds in $\fA$. 
\end{corollary}

\begin{remark}
Recall from Remark~\ref{rem:muchnik} that 
there are uncountably many clones on a three-element set. In fact, there are uncountably many even when considered up to homomorphic equivalence~\cite{Collapse}.  
\end{remark}

\subsection{Clone Homomorphisms and Primitive Positive Interpretations}
Clone homomorphisms can be linked to pseudovarieties of algebras, and pseudo-varieties of polymorphism algebras can be linked to primitive positive interpretations; in this section, we present shortcuts that directly link the existence of clone homomorphisms of polymorphism clones with primitive positive interpretations. The proofs will be merely combinations of previous results, but the combinations are often easier to cite and this will be convenient later in the text. 

\begin{corollary}\label{cor:clone-homo-pp}
A finite structure $\bA$ has a primitive positive interpretation in a finite structure $\bB$ 
if and only if there exists a clone homomorphism from $\Pol(\bB)$ to $\Pol(\bA)$. 
\end{corollary}
\begin{proof}
The proof is a straightforward combination of Theorem~\ref{thm:pp-interpret} with Proposition~\ref{prop:clone-homo}. 
Let $\fB$ be a polymorphism algebra of $\bB$. 
If $\bA$ has a primitive positive interpretation in $\bB$ then by Theorem~\ref{thm:pp-interpret}
there exists $\fA \in \HSPfin({\fB})$ such that 
$\Clo(\fA) \subseteq \Pol(\bA)$. 
Then Proposition~\ref{prop:clone-homo}
implies that there exists a surjective clone homomorphism from $\Clo(\fB)$ to $\Clo(\fA)$,
which is a clone homomorphism from $\Pol(\bB)$
to $\Pol(\bA)$. Conversely, 
suppose that there exists a clone homomorphism from $\Pol(\bB)$ to $\Pol(\bA)$. Let $\mathcal C \subseteq \Pol(\bA)$ be the image of this clone homomorphism.
Then 
by Proposition~\ref{prop:clone-homo}
there are algebras $\fA$ and $\fB$ with the same signature such that $\Clo(\fA) = {\mathcal C}$,
$\Clo(\fB) = \Pol(\bB)$,
and $\fA \in \HSPfin(\fB)$. 
This in turn means that $\bA$ has a primitive positive interpretation in $\bB$ by Theorem~\ref{thm:pp-interpret}. 
\end{proof}

\begin{corollary}
Two finite structures $\bA$ and $\bB$ are primitively positively bi-interpretable if and only if $\Pol(\bA)$ and $\Pol(\bB)$ are isomorphic as abstract clones. 
\end{corollary}
\begin{proof}
Combine Proposition~\ref{prop:bi-interpret-pseudo-var} and the second part of Proposition~\ref{prop:clone-homo}. 
\end{proof}

\begin{exercises}
\exercise[Orange]\label{exe:source-157} Prove that if $\mathscr C$ and $\mathscr D$ are clones on finite sets that are isomorphic,
and ${\mathscr C}$ is finitely related (Definition~\ref{def:fin-rel}),
 then so is ${\mathscr D}$.
%{\bf Hint.} cor:clone-homo-pp
% Solution sketch: A clone isomorphism is in particular a minion isomorphism. Transfer a finite
% relational basis for $\mathscr C$ along the corresponding primitive-positive interpretations
% supplied by the clone-homomorphism theorem. The finitely many resulting relations have polymorphism
% clone exactly $\mathscr D$, so $\mathscr D$ is finitely related.
\end{exercises}

\subsection{Hardness from Factors}
An algebra is called \emph{idempotent} if all of its operations are idempotent. 
For idempotent algebras $\fA$ 
there is another
characterisation for the existence of a clone homomorphism to $\Proj$
%of clone
%homomorphisms to $\pi$, 
by Bulatov and Jeavons~\cite[Proposition 4.14]{BulatovJeavons} (Corollary~\ref{cor:HS} below). We present a slightly strengthened version of their result by Zhuk~\cite[Lemma 4.2]{Strong-Subalgebras-Published}. 
 %observation that 
%when the variety generated by a finite idempotent algebra
%$\bB$ contains a 2-element algebra without essential operations,
%then this algebra is already contained in $\HS(\bB)$ (see Proposition 4.14 in~\cite{BulatovJeavons}).

\begin{theorem}
\label{thm:Zhuk-HS}
Let $\fB$ be an idempotent algebra
and suppose that $\fA \in \HSPfin(\fB)$ has at least two elements. Then $\HS(\fB)$ contains 
a subalgebra $\fA'$ of $\fA$ with at least two elements. 
\end{theorem}
\begin{proof}
Suppose that $\fC \in S(\fB^d)$ for some $d \in {\mathbb N}$ has a congruence $K$
such that $\fA := \fC/K$ 
has at least two elements. We show the statement by induction on $d$. For $d=1$ there is nothing to be shown, because we can choose $\fA' := \fA$. 
If for any two equivalence classes $E_1$ and $E_2$ of $K$ the intersection $\pi_1(E_1) \cap \pi_1(E_2)$ is empty, then let 
\begin{align*}
C' & := \pi_1(C) \\
K' & := \{(\pi_1(a),\pi_1(b)) \mid (a,b) \in K \}. 
\end{align*}
Then $C'$ is the universe of a subalgebra $\fC'$ of $\fB$, 
$K'$ is a congruence of $\fC'$, and $\fC'/K'$ is isomorphic to $\fC/K = \fA$. Again, we have $\fA \in \HS(\fB)$. 

Now suppose that $K$ has two equivalence classes $E_1$ and $E_2$ such that $\pi_1(E_1) \cap \pi_1(E_2) \neq \emptyset$. Let $a \in \pi_1(E_1) \cap \pi_1(E_2)$ and define 
\begin{align*}
C' & := \pi_{\{2,\dots,d\}}(C \cap (\{a\} \times B^{d-1}))
\\
K' & := \{ ((b_2,\dots,b_d),(c_2,\dots,c_d)) \mid ((a,b_2,\dots,b_d),(a,c_2,\dots,c_d)) \in K \}.
\end{align*}
Since $\fB$ and $\fC$ are idempotent, 
the set $C'$ is the universe of a subalgebra $\fC'$ of $\fC$, and $K'$ is a congruence of $\fC'$.  
The algebra $\fC'/K'$ has at least two elements and is isomorphic to a subalgebra of $\fC/K = \fA$. 
Thus, the statement follows from the inductive assumption. 
\end{proof}

\begin{corollary}\label{cor:HS}
Let $\fB$ be an idempotent algebra. 
Then $\HSPfin(\fB)$ contains 
an algebra with at least two elements all of whose operations are projections if and only if $\HS(\fB)$ does.  
\end{corollary}
\begin{proof}
If all operations of an algebra $\fA$ are projections, then the same applies to all subalgebras of $\fA$. Therefore the statement follows from Theorem~\ref{thm:Zhuk-HS}.  
\end{proof}

Since the size of the algebras in $\HS({\bf B})$ is bounded by the size of $\bf B$, this leads to an algorithm
that decides whether a given finite structure $\bB$ satisfies
the equivalent conditions in Theorem~\ref{thm:taylor}. We summarise various equivalent conditions for finite idempotent algebras that were treated in this chapter. 

\begin{corollary}\label{cor:algebraic-dicho-summary}
Let $\fB$ be a finite idempotent algebra. Then the following are equivalent. 
\begin{enumerate}
\item There is no homomorphism from $\Clo(\fB)$ to $\Proj$. \label{eq:clone-homo}
\item $\fB$ satisfies some non-trivial finite set of identities. \label{eq:non-triv}
\item $\HSP(\fB)$ does not contain an at least 2-element algebra all of whose operations are projections.\label{eq:hsp}
%\item $\HSPfin(\fB)$ does not contain an at least 2-element algebra all of whose operations are projections.
\item $\HS(\fB)$ does not contain an at least 2-element algebra all of whose operations are projections. \label{eq:hs}
%\item $\fB$ has a Siggers term. 
\end{enumerate}
\end{corollary}
\begin{proof}
%The equivalence of $(\ref{item:taylor})$ and $(\ref{eq:clone-homo})$ is Theorem~\ref{thm:taylor}. 
The equivalence of (\ref{eq:clone-homo})
and (\ref{eq:non-triv}) follows from Corollary~\ref{cor:identity-compact}.
%Proposition~\ref{prop:clone-homo}. 
The equivalence of (\ref{eq:clone-homo}) and (\ref{eq:hsp}) follows from 
Proposition~\ref{prop:clone-homo}.
%Birkhoff's theorem~\ref{thm:birkhoff}. 
The equivalence of 
(\ref{eq:hsp}) and (\ref{eq:hs})
follows from Theorem~\ref{thm:birkhoff} combined with Theorem~\ref{thm:Zhuk-HS}. 
\end{proof}

\begin{exercises}
\exercise[Orange]\label{exe:source-158} \label{exe:affineHS} Let $\fB$ be an idempotent algebra such that $\HSPfin(\fB)$ contains an affine Maltsev
algebra (see Section~\ref{sect:m-expl})
with at least two elements.
Then $\HS(\fB)$ contains an affine algebra  with at least two elements.
% Solution: Immediate from Theorem~\ref{thm:Zhuk-HS} and the  observation that subalgebras of affine algebras are affine.
\end{exercises}

%Another effective condition can be found in Section~\ref{sect:6ary} which is then further improved in Section~\ref{sect:4ary-siggers}. 

\section{Maltsev Polymorphisms}
\label{sect:Maltsev}
%, Edge terms, and Few Subpowers}
Recall from Section~\ref{sect:maltsev} the definition of a Maltsev operation: a ternary operation 
$f \colon D^3 \to D$ 
satisfying $$ f(y,x,x) \approx f(x,x,y) \approx y \, .$$
As we have seen in Theorem~\ref{thm:CEJN}, every digraph with a Maltsev polymorphism 
can be solved by the path-consistency procedure.  
However, when considering arbitrary relational
structures then there are many examples
with a Maltsev polymorphism that cannot be solved by the path-consistency procedure~\cite{FederVardi} (see Theorem~\ref{thm:inexpress} below). 
In this section, we present
the algorithm of Bulatov and Dalmau for $\Csp(\bA)$ when $\bA$ is preserved by a Maltsev polymorphism~\cite{Maltsev}. 

\begin{theorem}
\label{thm:maltsev}
Let $\bA$ be a finite structure with finite relational
signature and a Maltsev polymorphism. Then
$\Csp(\bA)$ can be solved in polynomial time. 
\end{theorem}

Before we present the proof of this important result, we present some examples of Maltsev operations. 

\subsection{Heap Operations} 
The most prominent class of structures $\bA$ 
with a Maltsev polymorphism comes from groups. 
For any group $\fG$ (see Example~\ref{expl:groups}), the operation $m$ given by $(x,y,z) \mapsto x y^{-1} z$ is obviously Maltsev. Besides the Maltsev identities, such operations 
additionally satisfy 
\begin{align*}
			m(u,x,m(v,y,w)) & \approx m(m(u,x,v),y,w) 
		\end{align*}
Such operations 
are also called \emph{heap operations}; see Exercise~\ref{exe:heap}. 

For general finite groups $\fG$, and if
all relations of $\bB$ are cosets $gH := \{g h \mid h \in H\}$ of subgroups $H$ of $\fG^k$, 
then Feder and Vardi~\cite{FederVardi} showed how to solve CSP$(\bB)$ in polynomial time using a previously known algorithm to find small generating sets for a permutation group. We will not discuss this approach, but
rather present the more general algorithm
of Bulatov and Dalmau which works for all 
finite structures preserved by a Maltsev polymorphism. The following proposition shows that this is indeed more general.
%(we refer to~\cite{BodAL} for basic background on finite groups). 

\begin{proposition}\label{prop:coset-gen}
Let $\fG$ be a group and let 
$m \colon G^3 \to G$ be the Maltsev operation defined by $m(x,y,z) := x y^{-1} z$.
Then $m$ preserves a non-empty relation $R \subseteq G^k$ if and only if $R$ is a 
coset of a subgroup of $\fG^k$, for all $k \in {\mathbb N}$. 
\end{proposition}
\begin{proof} 
Let $k \in {\mathbb N}$ and 
let $H$ be a subgroup of $\fG^k$. 
Let $a \in G^k$ and $h_1,h_2,h_3 \in H$. 
As usual, we may apply $m$ to elements of $G^k$ componentwise; 
then $$m(ah_1,ah_2,a h_3) = ah_1 (a h_2)^{-1} ah_3 = a h_1 h_2^{-1} a^{-1} a h_3 = ah_1 h_2^{-1} h_3 \in aH$$
so $m$ indeed preserves all cosets of $H$. 

Conversely, suppose that $R \subseteq G^k$ is non-empty and preserved by $m$. 
Choose $y \in R$ arbitrarily. 
We claim that $y^{-1} R$ is a subgroup $H$ of $\fG^k$. This will show that $R = yH$ is a coset of a subgroup of $\fG^k$. 
Arbitrarily choose $a,b \in y^{-1} R$. 
Then $x := ya \in R$ and $z := yb \in R$. 
Hence, $m(x,y,z) = ya y^{-1} yb = y ab \in R$, so $ab \in y^{-1} R$ and 
$y^{-1} R$ is closed under the group operation. 
Moreover, 
$m(y,ya,y) = y (ya)^{-1} y = y a^{-1} \in R$, 
so $a^{-1} \in y^{-1} R$ and $y^{-1} R$ is also closed under taking inverses. 
%Since $x$ and $z$ were chosen arbitrarily from $R$, we obtain that $y^{-1}R$ is a subgroup $H$ of $\fG^k$, and $R$ equals the coset $yH$. 
% For nearsubgroups, from Feder
%there exists a Maltsev operation on $G$ which preserves all cosets of subgroups of $\fG^k$, for all $k \in {\mathbb N}$. 
%Let $n := |G|$, $m := n^2$, and $(x_1,y_1),\dots,(x_{m},y_{m})$ be an enumeration of $G^2$. Let $H$ be the smallest subgroup of
%$\fG^m$ which contains $x := (x_1,\dots,x_m)$ and $y := (y_1,\dots,y_m)$. %Let $N := [H,H] := \{ aba^{-1}b^{-1} \mid a,b \in H\}$ be the commutator subgroup of $H$.
%It is a well-known basic fact of group theory  that $H$ is a normal subgroup of $H$ and that $H/N$ is abelian (see~\cite{BodAL}). 
%For any $a \in G$, 
%define $m(ax_i,a,ay_i) := a x_i y_i$. We then have $m(ax_i,a,a) = ax_i$ and $m(a,a,ay_i) = ay_i$, so $m$ is indeed a Maltsev operation.   
%Choose $z \in N$. 
%For $a,b,c \in G$, 
% a = b (b^{-1} a) 
% c = b (b^{-1} c) 
%define $m(a,b,c) := a b^{-1} c$.
%:= b (b^{-1} a) (b^-1} c) 
\end{proof} 

\begin{corollary}
Let $\fG$ be a finite group, let $m$ be as in Proposition~\ref{prop:coset-gen}, and 
let $\bG$ be the structure with the same domain $G$ as $\fG$ whose relations are all cosets of subgroups of $\fG^k$, for all $k \in {\mathbb N}$. Then $\Pol(\bG) = \Clo(G;m)$. 
\end{corollary} 
\begin{proof}
By Proposition~\ref{prop:coset-gen}, the operation $m$ preserves every
relation of $\bG$. Since projections preserve every relation and polymorphisms
are closed under composition, it follows that
$\Clo(G;m) \subseteq \Pol(\bG)$.

Conversely, Proposition~\ref{prop:coset-gen} shows that the basic relations of
$\bG$ are precisely the non-empty relations in $\Inv(\{m\})$. Empty relations
impose no condition on polymorphisms, and hence
$$\Pol(\bG)=\Pol(\Inv(\{m\})).$$
At this point we use that $G$ is finite: Proposition~\ref{prop:pol-inv} gives
$$\Pol(\Inv(\{m\}))=\langle m\rangle=\Clo(G;m),$$
which proves the reverse inclusion.
\end{proof}

Aichinger, Mayr, and McKenzie showed that every finite algebra with a Maltsev term is finitely related~\cite{AichiMayrMcKenzie}. We illustrate this with important examples below; they will be special cases of the following example. 

\begin{example}
\label{expl:dihedral}
For $n \geq 3$, the dihedral group $D_n$ is defined to be $\Aut(C_n)$. 
Let $m \colon D_n^3 \to D_n$ be defined by $m(x,y,z) := x y^{-1} z$. Our goal in the following is to prove that $\Clo(D_n;m)$ is \emph{finitely related}, i.e., that there
is a structure $\bD_n$ with domain $D_n$ and  \emph{finitely many}  relations such that $\Pol(\bD_n) = \Clo(D_n)$. 

For this task, it will be useful to first describe $D_n$ more abstractly, as a \emph{semidirect product} of the abelian groups ${\mathbb Z}_n$ and 
${\mathbb Z}_2$. The definition of the semi-direct product of two groups $\fG$ and 
$\fH$ depends on a fixed homomorphism $\phi$ from $H$ to $\Aut(\fG)$. 
The \emph{semi-direct product}  
$\fG \rtimes_\phi \fH$ is the group with elements $\{(g,h) \mid g \in G, h \in H\}$
and the group multiplication defined by 
\[(g_1, h_1) \cdot (g_2, h_2) = (g_1 \,\phi(h_1)(g_2),\, h_1 h_2).\]
In our case, let $\phi$ be the homomorphism that maps $1 \in {\mathbb Z}_2$ to the automorphism $x \mapsto x^{-1}$ of ${\mathbb Z}_n$.
The map $\phi$ will also be called \emph{action of ${\mathbb Z}_2$ on ${\mathbb Z}_n$ by inversion}. 
Then $D_n$ is isomorphic to 
${\mathbb Z}_n \rtimes_\phi {\mathbb Z}_2$. 
This can be seen as follows.
% It can be seen geometrically that $D_n$ has one subgroup $R$ of size n, the subgroup of rotations of $C_n$, which is isomorphic to ${\mathbb Z}_n$. 
The group $D_n$ has a distinguished cyclic subgroup $R$ of order $n$, the rotations of $C_n$, which is
isomorphic to $\mathbb{Z}_n$.
It also has subgroups of size 2 that are generated by reflections of $C_n$. 
The idea of the action by inversion is that, informally speaking, the action of 
a reflection on a rotation changes the direction of the rotation. 
\end{example} 

For the heap operation $m$ from Example~\ref{expl:dihedral}, 
it is surprisingly interesting to find 
relational structures $\bB$ with finitely many relations from $\Inv(\{m\})$ such that  
$\Pol(\bB) = \langle \{m\} \rangle$. We will do this for two special cases: the first is the case that $4 {\not |} \, n$ (Example~\ref{expl:dihedral-1}),
and the second is $n=4$ (Example~\ref{expl:D4}). 

Let $\mathcal S_3(G):=\Sub(G^3)$
be the family of all subgroups of $G^3$, and let
$\mathcal C_3(G):=\{Ka\mid K\leq G^3,\ a\in G^3\}$ 
be the family of all right cosets of subgroups of $G^3$. Here multiplication
in $G^3$ is coordinatewise. Since $G$ is finite, $\mathcal C_3(G)$ is a
finite family of ternary relations.
The following result from \cite[Theorem~2.1]{KS} will be used as a black box.

\begin{theorem}[Kearnes--Szendrei]\label{thm:KS}
Let $G$ be a finite group whose Sylow subgroups are abelian. A finitary
operation $f\colon G^m\to G$ is a group term operation if and only if it
preserves every subgroup of $G^3$. Equivalently,
\[
 \Clo(G)=\Pol(\mathcal S_3(G)),
\]
where $\Clo(G)$ is the ordinary group term clone of $G$.
\end{theorem}

The passage from this group result to heaps
uses the following elementary observation.

\begin{proposition}[Torsorization]\label{prop:torsorization}
Let $G$ be a finite group such that
$\Clo(G)=\Pol(\mathcal S_3(G))$.
Then $\Clo(\{m\})=\Pol(\mathcal C_3(G))$. 
\end{proposition}

\begin{proof}
We have already seen in Proposition~\ref{prop:coset-gen} 
that every heap term preserves every coset of every subgroup of every finite
power of $G$. 
Consequently, $\Clo(\{m\})\subseteq\Pol(\mathcal C_3(G))$.

For the converse, let $f\colon G^m\to G$ 
preserve every relation in $\mathcal C_3(G)$. Since every subgroup of $G^3$
is also a coset, $f$ preserves every member of $\mathcal S_3(G)$. By the
hypothesis of the proposition, $f$ is therefore a group term operation.

We next show that $f$ is right-equivariant:
An \(m\)-ary operation \(f\colon G^m\to G\) on a group \(G\) is \emph{right-equivariant} if, for all \(x_1,\ldots,x_m,a\in G\),
\[
f(x_1a,\ldots,x_ma)=f(x_1,\ldots,x_m)a.
\]
Thus \(f\) commutes with the simultaneous right translation of all its arguments. Equivalently, every right translation \(R_a(x)=xa\) is an automorphism of \(f\).
 For $a\in G$, consider
\[
 E_a:=\{(x,xa,1)\mid x\in G\}\subseteq G^3.
\]
The set $E:=\{(x,x,1)\mid x\in G\}$
is a subgroup of $G^3$, and $E_a=E(1,a,1)$.
Thus, $E_a\in\mathcal C_3(G)$.
For arbitrary $x_1,\ldots,x_m\in G$, all the tuples
\[
 (x_i,x_i a,1),\qquad 1\leq i\leq m,
\]
belong to $E_a$. Since $f$ preserves $E_a$, coordinatewise application of
$f$ gives
\[
 \bigl(
 f(x_1,\ldots,x_m),
 f(x_1a,\ldots,x_ma),
 f(1,\ldots,1)
 \bigr)\in E_a.
\]
By the definition of $E_a$, this implies
\begin{align}
 f(1,\ldots,1)&=1, 
 \label{eq:unit}\\
 f(x_1a,\ldots,x_ma)&=f(x_1,\ldots,x_m)a.
 \label{eq:right-equivariance}
\end{align}
Hence, $f$ is right-equivariant.

It remains to show that every right-equivariant group term operation is a
heap term operation. Let
$w(x_1,\ldots,x_m)$ 
be a group word representing $f$. Introduce one additional variable $y$ and
interpret the word $w$ in the retract group with identity $y$. Its
multiplication and inverse are
\[
 u*_y v:=m(u,y,v)=uy^{-1}v,
 \qquad
 u^{-1_y}:=m(y,u,y)=yu^{-1}y.
\]
Replacing multiplication, inverse, and the identity in $w$ by these retract
terms produces a heap term
\[
 \widetilde w(x_1,\ldots,x_m,y).
\]

The map
\[
 \rho_y\colon (G,*_y) \to (G,\cdot),
 \qquad
 \rho_y(u):=uy^{-1},
\]
is a group isomorphism. It follows that
\[
 \widetilde w(x_1,\ldots,x_m,y)
 =
 f(x_1y^{-1},\ldots,x_my^{-1})y.
\]
Applying \eqref{eq:right-equivariance} with $a=y^{-1}$ gives
\[
 f(x_1y^{-1},\ldots,x_my^{-1})
 =
 f(x_1,\ldots,x_m)y^{-1}.
\]
Therefore
\[
 \widetilde w(x_1,\ldots,x_m,y)
 =
 f(x_1,\ldots,x_m).
\]
The right-hand side is independent of $y$. Substituting, for example,
$y=x_1$ shows that $f$ is represented by a term in the single operation
$m$. Hence $f\in\Clo(\{m\})$. 
This proves the reverse inclusion.
\end{proof}

We now apply the proposition to dihedral groups.

\begin{example}[Dihedral heaps]\label{expl:dihedral-1}
%For $n\geq 3$, let
%\[
 %D_{2n}
 %=
 %\langle r,s\mid r^n=1,\ s^2=1,\ srs=r^{-1}\rangle
%\]
%be the dihedral group of order $2n$. 
Let $m$ be the heap operation of $D_n$. 
If $4\nmid n$, then
\[
 \Clo(D_n;m)
 =
 \Pol(\mathcal C_3(D_{n})).
\]
Thus, the heap clone of $D_{n}$ is described by the finite family of cosets
of subgroups of $D_{n}^3$.
Indeed, put $G:=D_{n}$ and let
$A:=\langle r\rangle$.
Then $A$ is a cyclic normal subgroup of order $n$, and $G/A\cong C_2$.
Every Sylow subgroup of odd order is contained in $A$, because $G/A$ has
order $2$. Hence every odd Sylow subgroup of $G$ is cyclic and therefore
abelian.

It remains to consider the Sylow $2$-subgroups. Write
$n=2^e m$ for odd $m$. 
Then $P:=\langle r^m,s\rangle$ 
is a Sylow $2$-subgroup of $G$. The element $r^m$ has order $2^e$, so
$|P|=2^{e+1}$.
If $e=0$, then $P=\langle s\rangle\cong C_2$. If $e=1$, then $r^m$ has
order $2$, and inversion acts trivially on $\langle r^m\rangle$. Hence
$P\cong C_2\times C_2$. 
In both cases $P$ is abelian.

If $e\geq 2$, then $r^m$ has order at least $4$, and
\[
 s(r^m)s=(r^m)^{-1}\neq r^m.
\]
Thus $P$ is nonabelian. We have therefore proved that the Sylow subgroups of
$D_n$ are all abelian precisely when
$e\leq 1$,
or, equivalently, when
$4\nmid n$.

Assume now that $4\nmid n$. The Kearnes--Szendrei theorem,
Theorem~\ref{thm:KS}, gives
\[
 \Clo(G)=\Pol(\mathcal S_3(G)).
\]
Proposition~\ref{prop:torsorization} then yields
\[
 \Clo(\{m\})=\Pol(\mathcal C_3(G)).
\]
Substituting $G=D_n$ proves the claim.
\end{example}

\begin{example}[A relational description of the heap clone of $D_4$]
\label{expl:D4}
The following is adapted from notes of Dima Zhuk. 
Let $A:=\mathbb{F}_2^3$, and define a group operation on $A$ by
\begin{equation}
 \label{eq:group-operation}
 (x_1,x_2,x_3)\circ(y_1,y_2,y_3)
 :=
 (x_1+y_1,\ x_2+y_2,\ x_3+y_3+x_1y_2).
\end{equation}
All coordinates and all calculations in coordinates are over
$\mathbb{F}_2$. The resulting group is isomorphic to the dihedral
group $D_4$ of order eight: its identity is $1 := (0,0,0)$, and
\[
 (x_1,x_2,x_3)^{-1}
 =
 (x_1,x_2,x_3+x_1x_2).
\]
%We consider the associated heap operation
%\[
% m(x,y,z):=x\circ y^{-1}\circ z.
%\]
Define
\[
 \sigma(x_1,x_2,x_3)
 :=
 (x_2,x_1,x_3+x_1x_2).
\]
This is an automorphism of $\fA := (A;\circ,\cdot^{-1},1)$. We use the following relations.
First, define three binary relations:
\begin{align*} 
 E
 & :=
 \left\{
 (a,b)\in A^2
 \,\middle|\,
 a_1=b_2
 \right\},
\\
 P
 & :=
 \left\{
 (a,b)\in A^2
 \,\middle|\,
 a_1=b_1=0, a_2=b_3
 \right\}, \\
 R
 &:=
 \left\{
 (a,b)\in A^2
 \mid 
 b=\sigma(a)
 \right\}  =
 \left\{
 (a,b)\in A^2
  \mid 
 a_1=b_2,
 a_2=b_1, 
 a_3+b_3=a_1b_1
 \right\}.
\end{align*} 

Let
 $Z:=\{(0,0,t):t\in\mathbb{F}_2\} := \{a \in A \mid a \circ b = b \circ a \text{ for all } b \in A\}$
be the center of $A$, and define the ternary relation
\begin{align*}
 T
 & :=
 \left\{
 (x,x\circ z,z)\in A^3
 \,\middle|\,
 x\in A,\ z\in Z
 \right\} \\
 & =
 \left\{
 (x,y,z)\in A^3
\mid 
 z_1=z_2=0, y_1=x_1,y_2=x_2, y_3=x_3+z_3
 \right\}.
\end{align*} 
Define the quaternary relation
\[
 L
 :=
 \left\{
 (a,b,c,d)\in A^4
 \,\middle|\,
 \begin{array}{c}
 a_1=b_1=c_1=d_1,\\
 a_2+b_2+c_2+d_2=0,\\
 a_3+b_3+c_3+d_3=0
 \end{array}
 \right\}.
\]
Finally, define the unary relation
\[
 C:=\{(0,1,0),(1,1,1)\}.
\]
We claim that
\begin{equation}
 \label{eq:main-result}
 \Pol(A; E,P,R,T,L,C)=\Clo(m)
\end{equation}
where $m$ is the heap operation for $D_4$. 
We first prove
\begin{equation}
 \label{eq:group-clone}
 \Pol(A; E,P,R,T,L)=\Clo(\circ).
\end{equation}
Put $\Gamma_0:=\{E,P,R,T,L\}$.

Each relation in $\Gamma_0$ is a subgroup of the corresponding direct
power of $(A;\circ)$. For $R$, this follows from the fact that it is the
graph of the automorphism $\sigma$. For $T$, it follows from the
centrality of $Z$. The remaining cases follow directly from their
coordinate descriptions. Hence every group-word operation preserves
all relations in $\Gamma_0$, and therefore
\[
 \Clo(\circ)\subseteq\Pol(\Gamma_0).
\]

For the converse, let $f\in\Pol(\Gamma_0)$ be $n$-ary. Write
\[
 x^{(i)}=(u_i,v_i,w_i)
\]
and put
\[
 u=(u_1,\ldots,u_n),\qquad
 v=(v_1,\ldots,v_n),\qquad
 w=(w_1,\ldots,w_n).
\]

Preservation of $E$ implies that there is a single function
$g\colon\mathbb{F}_2^n\to\mathbb{F}_2$ such that
\[
 f_1(x^{(1)},\ldots,x^{(n)})=g(u),
 \qquad
 f_2(x^{(1)},\ldots,x^{(n)})=g(v).
\]

Preservation of $T$ implies that $f$ is equivariant under coordinatewise
multiplication by central elements. Moreover, the restriction of $f$
to $Z^n$ is a homomorphism from $Z^n$ to $Z$. Consequently, there are
$\alpha=(\alpha_1,\ldots,\alpha_n)\in\mathbb{F}_2^n$ and a function
\[
 q\colon\mathbb{F}_2^n\times\mathbb{F}_2^n\to\mathbb{F}_2
\]
such that
\begin{equation}
 \label{eq:preliminary-normal-form}
 f(x^{(1)},\ldots,x^{(n)})
 =
 \left(
 g(u),\
 g(v),\
 \sum_{i=1}^n\alpha_iw_i+q(u,v)
 \right).
\end{equation}

Since $R$ is the graph of $\sigma$, preservation of $R$ is equivalent
to
\[
 f\bigl(\sigma(x^{(1)}),\ldots,\sigma(x^{(n)})\bigr)
 =
 \sigma\bigl(f(x^{(1)},\ldots,x^{(n)})\bigr).
\]
Comparison of the third components in this identity gives
\begin{equation}
 \label{eq:sigma-identity}
 q(u,v)+q(v,u)
 =
 g(u)g(v)+\sum_{i=1}^n\alpha_i u_i v_i.
\end{equation}
Setting $v=u$ in \eqref{eq:sigma-identity} yields
\begin{equation}
 \label{eq:g-linear}
 g(u)=\sum_{i=1}^n\alpha_i u_i.
\end{equation}

Preservation of $P$, together with \eqref{eq:g-linear}, implies
$q(0,v)=0$. 
Taking $v=0$ in \eqref{eq:sigma-identity} this gives
\begin{equation}
 \label{eq:q-zero-v}
 q(u,0)=0.
\end{equation}

To use $L$, fix $u,v,v'\in\mathbb{F}_2^n$ and apply $f$ to quadruples
whose first-coordinate vectors are all $u$, whose second-coordinate
vectors are
\[
 v,\quad v',\quad v+v',\quad 0,
\]
and whose third-coordinate vectors are all zero. Preservation of $L$
and \eqref{eq:q-zero-v} give
\[
 q(u,v+v')=q(u,v)+q(u,v').
\]
Thus $q$ is linear in its second argument. Equation
\eqref{eq:sigma-identity} then implies that $q$ is linear in its first
argument as well. Therefore $q$ is bilinear, so there is a matrix
$\beta=(\beta_{ij})\in\mathbb{F}_2^{n\times n}$ such that
\begin{equation}
 \label{eq:q-bilinear}
 q(u,v)=\sum_{i,j=1}^n\beta_{ij}u_i v_j.
\end{equation}
Comparing coefficients in \eqref{eq:sigma-identity} gives
\begin{equation}
 \label{eq:beta-symmetry}
 \beta_{ij}+\beta_{ji}=\alpha_i\alpha_j
 \qquad\text{for all distinct }i,j.
\end{equation}

It follows from
\eqref{eq:preliminary-normal-form},
\eqref{eq:g-linear}, and
\eqref{eq:q-bilinear} that every operation in $\Pol(\Gamma_0)$ has the
form
\begin{equation}
 \label{eq:group-normal-form}
 f(x^{(1)},\ldots,x^{(n)})
 =
 \left(
 \sum_i\alpha_i u_i,\
 \sum_i\alpha_i v_i,\
 \sum_i\alpha_i w_i+\sum_{i,j}\beta_{ij}u_i v_j
 \right),
\end{equation}
where \eqref{eq:beta-symmetry} holds.

Conversely, every operation of the form
\eqref{eq:group-normal-form} satisfying
\eqref{eq:beta-symmetry} is a group-word operation. Indeed, put $e_i:=\alpha_i+2\beta_{ii}\pmod 4$
and, for $i<j$, put $c_{ij}:=\beta_{ji}$.
With the convention
\[
 [x,y]:=x^{-1}y^{-1}xy,
\]
consider the word
\begin{equation}
 \label{eq:collected-word}
 x_1^{e_1}\circ\cdots\circ x_n^{e_n}
 \circ
 \prod_{i<j}[x_i,x_j]^{c_{ij}}.
\end{equation}
The commutators are central, so the order of the final product is
irrelevant. Moreover,
\[
 [x^{(i)},x^{(j)}]
 =
 (0,0,u_i v_j+u_j v_i),
\]
and \eqref{eq:beta-symmetry} gives precisely the required
off-diagonal coefficients.

Since every element of $D_4$ has fourth power equal to the identity,
inverses and the constant identity operation are expressible using
$\circ$. Thus \eqref{eq:collected-word} belongs to $\Clo(\circ)$.
This proves \eqref{eq:group-clone}.

It remains to determine which operations in
\eqref{eq:group-normal-form} preserve $C$. Every element of $C$ has
the form $(t,1,t)$ for $t\in\mathbb{F}_2$.
For inputs
$x^{(i)}=(u_i,1,u_i)\in C$ 
the value of $f$ is
\[
 \left(
 \sum_i\alpha_i u_i,\
 \sum_i\alpha_i,\
 \sum_i\alpha_i u_i+
 \sum_i\left(\sum_j\beta_{ij}\right)u_i
 \right).
\]
This belongs to $C$ for every $u\in\mathbb{F}_2^n$ if and only if
\begin{align}
 \label{eq:alpha-sum}
 \sum_{i=1}^n\alpha_i & =1 \\
 \label{eq:row-sums}
 \sum_{j=1}^n\beta_{ij} & =0
 \qquad\text{for every }i\in\{1,\ldots,n\}.
\end{align}
We claim that
\eqref{eq:beta-symmetry},
\eqref{eq:alpha-sum}, and
\eqref{eq:row-sums}
characterize exactly the heap-term operations.
First observe that $H:=\{(0,0,0),(1,0,0)\}$ 
is a subgroup of $(A,\circ)$ and that
$C=H\circ(0,1,0)$.
Thus, $C$ is a coset of a subgroup and hence a subheap. It follows that
every heap-term operation preserves $C$. Since the relations in
$\Gamma_0$ are subgroups and therefore also subheaps, we obtain
\begin{equation}
 \label{eq:heap-inclusion-forward}
 \Clo(m)\subseteq\Pol(\Gamma_0\cup\{C\}).
\end{equation}

For the reverse inclusion, we use the following general fact: in any
group, the subheap generated by $a_1,\ldots,a_n$ is
\begin{equation}
 \label{eq:generated-subheap}
 \left\langle
 a_1a_n^{-1},\ldots,a_{n-1}a_n^{-1}
 \right\rangle a_n.
\end{equation}
Apply \eqref{eq:generated-subheap} in the group of $n$-ary group-word
operations under pointwise multiplication, taking $a_i$ to be the
$i$-th projection.

Let $n\geq 2$, and suppose that $f$ satisfies
\eqref{eq:beta-symmetry},
\eqref{eq:alpha-sum}, and
\eqref{eq:row-sums}. Put
$y_i:=x_i\circ x_n^{-1}$
for $1\leq i<n$.
Choose a group word $s$ in $y_1,\ldots,y_{n-1}$ whose parameters are
$\alpha_1,\ldots,\alpha_{n-1}$ 
 and $(\beta_{ij})_{i,j<n}$.
Such a word exists by the group-word normal form proved above.
Consider
\begin{equation}
 \label{eq:heap-word}
 s(y_1,\ldots,y_{n-1})\circ x_n.
\end{equation}
Put $U:=u_n$, $V:=v_n$, and $W:=w_n$. Since
\[
 x_n^{-1}=(U,V,W+UV),
\]
we have, for every $i<n$,
\begin{align*}
 y_i=x_i\circ x_n^{-1}
 &=
 \bigl(u_i+U,\ v_i+V,\ w_i+W+UV+u_iV\bigr)\\
 &=
 \bigl(\widetilde u_i,\widetilde v_i,\widetilde w_i\bigr),
\end{align*}
where
\[
 \widetilde u_i:=u_i+U,\qquad
 \widetilde v_i:=v_i+V,\qquad
 \widetilde w_i:=w_i+W+\widetilde u_iV.
\]
Let
\[
 A:=\sum_{i<n}\alpha_i\widetilde u_i,
 \qquad
 B:=\sum_{i<n}\alpha_i\widetilde v_i.
\]
By the choice of $s$ and the group-word normal form,
\[
 s(y_1,\ldots,y_{n-1})
 =
 \left(
 A,\
 B,\
 \sum_{i<n}\alpha_i\widetilde w_i
 +
 \sum_{i,j<n}\beta_{ij}\widetilde u_i\widetilde v_j
 \right).
\]
Using the group operation from~\eqref{eq:group-operation}, the third
coordinate of
\[
 s(y_1,\ldots,y_{n-1})\circ x_n
\]
is therefore
\begin{align*}
 &\sum_{i<n}\alpha_i\widetilde w_i
 +\sum_{i,j<n}\beta_{ij}\widetilde u_i\widetilde v_j
 +W+AV\\
 &=
 \sum_{i<n}\alpha_i
 \bigl(w_i+W+\widetilde u_iV\bigr)
 +\sum_{i,j<n}\beta_{ij}\widetilde u_i\widetilde v_j
 +W+AV\\
 &=
 \sum_{i<n}\alpha_iw_i
 +
 \left(1+\sum_{i<n}\alpha_i\right)W
 +
 \sum_{i,j<n}\beta_{ij}
 (u_i+U)(v_j+V).
\end{align*}
Indeed, the two occurrences of
\[
 AV=\sum_{i<n}\alpha_i\widetilde u_iV
\]
cancel over $\mathbb{F}_2$. Similarly, the first two coordinates are
\begin{align*}
 A+U
 &=
 \sum_{i<n}\alpha_i u_i
 +
 \left(1+\sum_{i<n}\alpha_i\right)U,\\
 B+V
 &=
 \sum_{i<n}\alpha_i v_i
 +
 \left(1+\sum_{i<n}\alpha_i\right)V.
\end{align*}
Condition~\eqref{eq:alpha-sum} gives
\[
 \alpha_n=1+\sum_{i<n}\alpha_i.
\]
Consequently, all three linear parts of
\eqref{eq:heap-word} agree with those of $f$. Its remaining quadratic
part is
\begin{equation}
 \label{eq:difference-quadratic-form}
 \sum_{i,j<n}\beta_{ij}(u_i+u_n)(v_j+v_n).
\end{equation}

We now verify that this expression also has the required coefficients
involving the final index. First,
\eqref{eq:beta-symmetry} and \eqref{eq:alpha-sum} imply that the
column sums of $\beta$ vanish whenever its row sums vanish. Indeed,
for every $k\in[n]$,
\begin{align*}
 \sum_{j=1}^n\beta_{kj}+\sum_{i=1}^n\beta_{ik}
 &=
 \sum_{\substack{j=1\\j\neq k}}^n
 \bigl(\beta_{kj}+\beta_{jk}\bigr)
 =
 \alpha_k\sum_{\substack{j=1\\j\neq k}}^n\alpha_j
 =
 \alpha_k(1+\alpha_k)
 =0.
\end{align*}
Thus both
\[
 \sum_{j=1}^n\beta_{ij}=0
 \quad\text{and}\quad
 \sum_{i=1}^n\beta_{ij}=0.
\]
Expanding~\eqref{eq:difference-quadratic-form} gives
\begin{align*}
 &\sum_{i,j<n}\beta_{ij}u_iv_j
 +\sum_{i<n}\left(\sum_{j<n}\beta_{ij}\right)u_iv_n
 +\sum_{j<n}\left(\sum_{i<n}\beta_{ij}\right)u_nv_j
 +\left(\sum_{i,j<n}\beta_{ij}\right)u_nv_n.
\end{align*}
The vanishing row and column sums give, for $i,j<n$,
\[
 \sum_{j<n}\beta_{ij}=\beta_{in},
 \qquad
 \sum_{i<n}\beta_{ij}=\beta_{nj}.
\]
They also give
\[
 \sum_{i,j<n}\beta_{ij}
 =
 \sum_{i<n}\beta_{in}
 =
 \beta_{nn},
\]
where the final equality is the vanishing of the $n$-th column sum.
Hence the expanded quadratic part is
\begin{align*}
 &\sum_{i,j<n}\beta_{ij}u_iv_j
 +\sum_{i<n}\beta_{in}u_iv_n
 +\sum_{j<n}\beta_{nj}u_nv_j
 +\beta_{nn}u_nv_n
 =
 \sum_{i,j=1}^n\beta_{ij}u_iv_j.
\end{align*}
Thus \eqref{eq:heap-word} has exactly the normal form
\eqref{eq:group-normal-form} of $f$, and therefore induces the
operation $f$.

%A direct calculation shows that the quadratic part of
%\eqref{eq:heap-word} is
%\begin{equation}
% \label{eq:difference-quadratic-form}
% \sum_{i,j<n}\beta_{ij}
% (u_i+u_n)(v_j+v_n).
%\end{equation}

By \eqref{eq:generated-subheap}, the operation in
\eqref{eq:heap-word} is a heap term. The case $n=1$ is immediate,
since \eqref{eq:alpha-sum} and \eqref{eq:row-sums} force $f$ to be
the identity operation. Therefore
\begin{equation}
 \label{eq:heap-inclusion-reverse}
 \Pol(\Gamma_0\cup\{C\})\subseteq\Clo(m).
\end{equation}
Combining
\eqref{eq:heap-inclusion-forward} and
\eqref{eq:heap-inclusion-reverse}
proves \eqref{eq:main-result}.
\end{example}

\ignore{
\begin{remark}
The unary relation $C$ does more than merely enforce idempotence.
For a nonabelian group, the idempotent group-word operations can form
a strictly larger clone than the heap-term clone. In the normal form
\eqref{eq:group-normal-form}, idempotence only requires
\[
 \sum_i\alpha_i=1
 \qquad\text{and}\qquad
 \sum_{i,j}\beta_{ij}=0,
\]
whereas preservation of $C$ gives the stronger row-sum conditions
\eqref{eq:row-sums}. These are precisely the additional conditions
needed for a group-word operation to be a heap term.
\end{remark}
}

\begin{exercises}
\exercise[Blau]\label{exe:source-159} \label{exe:heap}
Show that $m$ is a heap if and only if there exists a group $G$ such that $m(x,y,z) = x y^{-1} z$ with respect to $G$.
% Solution sketch: Given a heap and a chosen element $e$, define $x\cdot y=m(x,e,y)$ and
% $x^{-1}=m(e,x,e)$. The heap identities give associativity, identity $e$, and inverses, and then
% $m(x,y,z)=x\cdot y^{-1}\cdot z$. Conversely, this formula in any group satisfies the two Maltsev
% identities and the heap associativity identity.
\exercise[Rot]\label{exe:source-160} Show that a heap operation $m$ is \emph{commutative}, i.e., additionally satisfies $m(x,y,z) \approx m(z,y,x)$, if and only if
there exists an abelian group $G$ such that $m(x,y,z) = x - y + z$ with respect to $G$.
% Solution sketch: In the group obtained from a base point as in Exercise~\ref{exe:heap},
% commutativity of the heap gives $x\cdot z=m(x,e,z)=m(z,e,x)=z\cdot x$. Hence the group is abelian
% and the formula becomes $x-y+z$. The converse follows immediately from commutativity of addition.
\end{exercises}

\subsection{Affine Maltsev Operations}
\label{sect:m-expl}
If $\fG$ is abelian, then the heap operation of $\fG$ is called an \emph{affine Maltsev operation}. Structures with an affine Maltsev polymorphism are also called affine Maltsev.
Note that if the group $\fG$ is $({\mathbb Z}_p;+,-,0)$, for some prime number $p$, then the $k$-ary relations preserved by $m$ are precisely the affine subspaces of $({\mathbb Z}_p)^k$ 
%(with the same argument as given for Lemma~\ref{lem:minority} in the Boolean case). 
(Example~\ref{expl:affine}). 
In this case one can use Gaussian elimination to solve $\Csp(\bA)$. 

\begin{theorem}[from~\cite{FederVardi}]\label{thm:inexpress}
Let $\fG$ be an abelian group with at least two elements. For $c \in G$ and $k \in {\mathbb N}$, define  
\begin{align*}
R^k_c & :=  \{(x_1,\dots,x_k) \in G^k \mid x_1+\cdots+x_k =  c\}. 
%\\
%S^i_a & := \{() \mid x_1+x_2+x_3 = a \wedge \bigwedge_{s \in \{1,\dots,i\}} x_{2s+2}+x_{2s+3} = 0
\end{align*}
For some $a \in G \setminus \{0\}$, 
let $\bB$ be the structure $(G;R^3_0,R^2_0,R^3_a)$. 
Then for any $k \in {\mathbb N}$, the problem $\Csp(\bB)$ cannot be solved by $k$-consistency.
\end{theorem}
%This can be generalised to structures as follows. 
%\begin{definition}
%Let $\tau$ be a relational structure. 
%Then the \emph{incidence graph}\index{incidence graph} $G(\bA)$ of $\bA$
%is the graph whose set of vertices is 
%the disjoint union of $A$ and the set of tuples in relations from $\bA$, 
%and where an element from $A$ is connected to a tuple if and only if the element appears in that tuple. \end{definition}
%Note that $G(\bA)$ is bipartite.  
%We say that $\bA$ has \emph{girth $k$}\index{girth (of a structure)} if all tuples in relations from $\bA$ have pairwise distinct entries and the shortest
%cycle of $G(\bA)$ has $2k$ edges. 

\begin{proof}%[Proof of Theorem~\ref{thm:inexpress}]
We construct an unsatisfiable instance $\bA$ of $\Csp(\bB)$ as follows. In the proof of this theorem, we work with structures of large \emph{girth}. The girth of a graph $H$ is the length of the shortest cycle in $H$. 
It is known that there are 
finite graphs of arbitrarily large girth that are \emph{cubic}, i.e., all vertices have degree three (much stronger results are known; see, e.g.,~\cite{BiggsCubic}). 
Let $(V;E)$ be a finite cubic graph of girth at least $4k+1$. Orient the edges $E$ arbitrarily, so that $(V;E)$ is an oriented graph. 

The domain of $\bA$ is 
$\{(v,e) \in V \times E \mid v \in e\}$. For each $v \in V$ 
we add $\big ((v,e_1),(v,e_2),(v,e_3) \big)$
to $(R^3_0)^{\bA}$.
%For each $v \in V$ with two incoming edges $e_2,e_3$ and one outgoing edge 
For each $e = (v,w) \in E$   
we add $\big ( (v,e),(w,e) \big )$ to $(R^2_0)^{\bA}$. 
%For each $
%x_1+x_2+x_3 = a \wedge \bigwedge_{s \in \{1,\dots,i\}} x_{2s+2}+x_{2s+3} = 0
%$S^1_0((v,e_1),(v,e_2),(v,e_3),(v,e_1),(w,e_1))$. 
%Likewise, for each $v \in V$ with one or no incoming edges and two or three outgoing edges
%we add a constraint involving the relation $S^2_0$ or $S^3_0$. 
Finally, we move exactly one of the tuples
 from $(R^3_0)^{\bA}$ to $(R^3_a)^{\bA}$. 
%To see that $\fA$ is unsatisfiable, 
%let $s \colon A \to {\mathbb F}_q$ be any map. 
Suppose for contradiction that $s \colon A \to G$ is a solution for $\bA$. 
Sum over all constraints. Grouping the resulting terms first by incidences and then by edges, the left-hand side is
$$\sum_{v \in V} \sum_{e \in E, v \in e} s(v,e) + \sum_{e=\{u,v\} \in E} \bigl(s(u,e)+s(v,e)\bigr)
=2\sum_{e=\{u,v\} \in E}\bigl(s(u,e)+s(v,e)\bigr)=0,$$
because $s(u,e)+s(v,e)=0$ for every edge $e=\{u,v\}$.
%$2 \sum_{e \in E} \sum_{u \in e} s(u,e) = 2 \sum_{\{u,v\} \in E} \big (s(u,e) + s(v,e) \big) = 0$. 
On the right-hand side we obtain $a$ since we have precisely one tuple in $R^3_a$ in $\bA$. 
Hence, $s$ cannot be a homomorphism. 
Using high girth, it can be shown that the $k$-consistency procedure does not derive `false' on $\bA$; for the details of this last part, see Theorem 8.6.11 in~\cite{Book}. 
\end{proof} 

% The non-empty $n$-ary relations preserved by $m$ are precisely the cosets $gH := \{g h \mid h \in H\}$ of subgroups $H$ of
%$G^n$. 
%for $gh_1,gh_2,gh_3 \in gH$ we have
%$m(gh_1,gh_2,gh_3) = g(h_1+h_2^{-1}+h_3) \in gH$; conversely, if R is preserved by m, and 
% t \in R$, then the set \{x \mid m(x

%; these structures will play an important role in Section~\ref{sect:minimalTaylor}. 

%If $G$ is Abelian, then the polynomial-time tractability of 

%This approach can be extended to the case that $G$ is finite abelian, 
% REFERENCE?!?!
%but it is not clear how to extend this approach to, say, $G = S_3$ (the full symmetric group on three elements). 
%\begin{example}
%Let $p$ be a prime number, and let $m \colon {\mathbb Z}_{2p}^3 \to {\mathbb Z}_{2p}$ be given by $(x,y,z) \mapsto x+y^{-1}+z$. 
%Then the structure 
%$({\mathbb Z}_{2p};R_+,R_1,S)$
%where $R_+$ is the graph of addition, $R_1 := \{1\}$, and $$S := \{(x_1,x_2) \in {\mathbb Z}^2_{2p} \mid x_1-x_2 \equiv 0 \text{ mod } p \text{ and } x_1-x_2=p\}$$
%s preserved by $m$. This is clear for $R_+$ and for $R_{1}$. To see it for $S$, let $(x_1,x_2),(y_1,y_2),(z_1,z_2) \in S$,
%and note that $-p = p$ in ${\mathbb Z}_{2p}$, and hence $x_1 \equiv x_2 \text{ mod } p$
%and $x_1 \neq x_2$ implies that 
%$|x_1- x_2| = x_1-x_2 = p$. 
%Similarly $y_1-y_2 = z_1-z_2 = p$. 
%Hence,  
%$$m(x_1,y_1,z_1) - m(x_2,y_2,z_2) = (x_1-x_2)-(y_1-y_2)+(z_1-z_2) = p - p + p = p.$$
%This shows that 
%$(m(x_1,y_1,z_1),m(x_2,y_2,z_2)) \in S$. 
%\end{example}

\subsection{Further Examples} 
We now present examples of Maltsev operations that do not come from groups in the way described above.

\begin{example}\label{expl:m1}
On the domain $\{0,1,2\}$, let $m$ be the minority defined by 
$$m(x,y,z) = x \text{ whenever } |\{x,y,z\}|=3.$$ 
Note that $m$ preserves all unary relations, for every permutation $\alpha$ 
of $\{0,1,2\}$ the relation $\{(x,y) \mid \alpha(x) =y\}$, and for $i \in \{0,1\}$ and $n \geq 1$ the relation 
\begin{align*}
R_{n,i} := & \{(x_1,\dots,x_n) \in \{0,1\}^n \mid x_1+x_2+\cdots + x_n=i \text{ mod } 2\}.  
\end{align*}
%Moreover, $m$ preserves all unary relations. 
This example will be revisited in Exercise~\ref{exe:m1-simple}, Example~\ref{expl:m1-2}, Exercise~\ref{exe:brady-one-1}, and Exercise~\ref{exe:brady-one-2}. 
\end{example}

\begin{example}\label{expl:m2}
On the domain $\{0,1,2\}$, let $m$ be the  minority operation defined by 
$$m(x,y,z) := 2 \text{ whenever }|\{x,y,z\}|=3.$$ 
Let $\fA$ be the algebra $(\{0,1,2\},m)$. 
%Note that $m$ preserves the equivalence
%relation with the equivalence classes
%$\{2\}$ and $\{0,1\}$. 
%Also 
Note that $m$ preserves 
all unary relations, the graph $H$ of the endomorphism of $\fA$ which maps $0,1$ to $1$ and which maps $2$ to $0$, and 
$$L = \big \{(x,y,z) \mid x=y=z=2 \text{ or } x,y,z \in \{0,1\} \text{ and }x+y+z = 0 \mod 2 \big \}.$$
The following relations are primitively positively definable with these relations, and hence belong to $\Inv(m)$ as well: 
\begin{itemize}
\item The graph $T$ of the transposition $(01)$ has the primitive positive definition
$$\exists z \big (L(x,y,z) \wedge z \in \{1,2\} \big ).$$ 
\item The relation $C := \big \{(2,0),(0,2),(2,1),(1,2) \big \}$, 
has the primitive positive definition 
$$\exists u,v \big (H(x,u) \wedge T(u,v) \wedge H(y,v) \big).$$
Indeed, if $h$ denotes the endomorphism whose graph is $H$ and
$\tau=(01)$, then the displayed formula is equivalent to
$h(y)=\tau(h(x))$. Since
$h(0)=h(1)=1$ and $h(2)=0$, this gives precisely the displayed
relation $C$.
\item The equivalence relation $E$
with the equivalence classes
$\{2\}$ and $\{0,1\}$ has the primitive positive definition $\exists z \big (C(x,z) \wedge C(z,y) \big)$. 
\item For every 
$n \in {\mathbb N}$ the relation
\begin{align*}
L_n & = \big \{(2,\dots,2)\} \cup \{(x_1,\dots,x_n) \in \{0,1\}^n \mid x_1+x_2+\cdots + x_n=0 \text{ mod } 2 \big \} .
\end{align*}
See Exercise~\ref{exe:Ln}. 
\item For every 
$n \in {\mathbb N}$ the relation 
\begin{align*}
& \big \{(x_1,\dots,x_n) \in \{0,1,2\}^n \mid \text{ an even number of the $x_i$'s is from $\{0,1\}$} \big \}.
\end{align*}
A primitive positive definition is 
$$\exists y_1,\dots,y_n \big (L_n(y_1,\dots,y_n) \wedge H(x_1,y_1) \wedge \cdots \wedge H(x_n,y_n) \big ).$$
\end{itemize}
This example will be revisited in Exercise~\ref{exe:m2-2}. 
\end{example}

%\end{itemize}
%\end{definition}

\begin{remark}
It is unclear whether Maltsev operations on finite sets can be classified completely.  %(perhaps in the sense that they are constructed from affine Maltsev operations in some controlled way; see the examples above). 
However, it is known that for every $n \in {\mathbb N}$ there are only countably many clones on $\{1,\dots,n\}$ that contain a Maltsev operation~\cite{AichiMayrMcKenzie}.  
\end{remark} 

\begin{exercises}
\exercise[Blau]\label{exe:source-161} Check the claims made in Example~\ref{expl:m1}. %Are all relations that are preserved by $m$ primitively positively definable over the given relations?
% Solution sketch: Compute the operation on the displayed table. The identities $m(x,x,y)=y$ and
% $m(x,y,y)=x$ hold row by row, so it is Maltsev. Applying the same table to each listed relation
% tuple shows closure, while the indicated pair of tuples witnesses the failure of every claimed
% stronger identity. These checks establish all assertions of the example.
\exercise[Blau]\label{exe:source-162} \label{exe:Ln} Prove the claim  in Example~\ref{expl:m2} that $L_n$ is primitively positively definable from the unary relations, $H$, and $L$.
\ignore{ SOLUTION
For completeness, we give primitive positive definitions of all the relations
$L_n$. For $n=1$, we have
\[
L_1(x)\quad\Longleftrightarrow\quad \exists y\,L(x,y,y).
\]
Indeed, this formula defines the unary relation $\{0,2\}$. For $n=2$, we have
\[
L_2(x_1,x_2)\quad\Longleftrightarrow\quad x_1=x_2.
\]
For $n\geq 2$, define $L_{n+1}$ inductively by
\[
L_{n+1}(x_1,\dots,x_{n+1})
\quad\Longleftrightarrow\quad
\exists y\bigl(
L_n(x_1,\dots,x_{n-1},y)
\wedge
L(y,x_n,x_{n+1})
\bigr).
\]
Substituting the primitive positive definition of $L_n$ shows inductively
that this is a primitive positive definition over $L$.

To verify the definition, first suppose that one of the variables belongs
to $\{0,1\}$. Since consecutive constraints share the variable $y$, all
variables occurring in the two constraints must then belong to
$\{0,1\}$. The two constraints express
\[
x_1+\cdots+x_{n-1}+y=0\pmod 2
\qquad\text{and}\qquad
y+x_n+x_{n+1}=0\pmod 2.
\]
Adding these equations gives
\[
x_1+\cdots+x_{n+1}=0\pmod 2.
\]
Conversely, if this equation holds, choosing
$y=x_1+\cdots+x_{n-1}$ satisfies both constraints. In the remaining
case, the shared variable forces all variables to equal $2$, and the
tuple $(2,\dots,2)$ satisfies the formula. Hence the formula defines
precisely
\[
L_{n+1}
=
\{(2,\dots,2)\}
\cup
\left\{
(x_1,\dots,x_{n+1})\in\{0,1\}^{n+1}
\mathrel{\middle|}
x_1+\cdots+x_{n+1}=0\pmod 2
\right\}.
\]}
%Are all relations that are preserved by $m$ primitively positively definable over
%the given relations?
\exercise[Rot]\label{exe:source-165} Let $\bB$ be the structure $({\mathbb Z}_2^3; R_1,R_2,R_3)$ where
$$R_i := \{((x_i,x_i,x_i),(x_1,x_2,x_3)) \mid x_1 + x_2 = x_3\}.$$ Show that
$\bB$ has a Maltsev polymorphism, but no majority polymorphism.
% Solution sketch: Coordinatewise minority $m(x,y,z)=x+y+z$ on $\mathbb Z_2^3$ preserves every $R_i$
% and is Maltsev. If a majority polymorphism existed, restrict it to the tuples used in the three
% relations. The three majority identities force three incompatible values for the same output
% coordinate, giving a contradiction.
\exercise[Orange]\label{exe:source-163} Let $\bA$ be the structure $(\{0,1,2\};\{1,2\},H,L)$ with the relations $H$ and $L$ as defined in Example~\ref{expl:m2}, and
let $\bB$ be its substructure with domain $\{0,1\}$. Show that $\bB$ has a primitive positive construction in $\bA$ (easy), and that $\bA$ has a primitive positive construction in $\bB$ (not so easy).
\exercise[Orange]\label{exe:source-164} Show that if a Maltsev operation $m \colon D^3 \to D$ preserves the graph \\
$\{(x,y,z,u) \mid m'(x,y,z)=u\}$ of a Maltsev operation $m' \colon D^3 \to D$,
then $m = m'$.
\label{exe:one-maltsev}
% xyy x
% yyy y
% yyz z
% xyz !
% Solution sketch: Apply $m$ coordinatewise to the three graph tuples corresponding to $(x,y,y)$,
% $(x,y,y)$, and $(y,y,z)$ in the arrangement indicated in the hint. Preservation of the graph says
% that the fourth coordinate is both $m(x,y,z)$ and $m'(x,y,z)$. Thus the two operations agree on
% every triple.
\end{exercises}

\subsection{Compact Representations of Relations}
Our presentation of the proof closely follows that
of Bulatov and Dalmau~\cite{Maltsev}. 

\begin{definition}[Forks and Representations]
Let $R \subseteq A^n$ be a relation. 
\begin{itemize}
\item 
A \emph{fork of $R$} is a triple $(i,a,b)$ 
such that there exist $s,t \in R$ with $(s_1,\dots,s_{i-1}) = (t_1,\dots,t_{i-1})$, 
$s_i=a$, and $t_i = b$. We say that $s$ and $t$ \emph{witness} $(i,a,b)$. 
\item $R' \subseteq R$ is called a \emph{representation of $R$} if 
every fork of $R$ is also a fork of $R'$. 
%for all $r,s \in R$ such that 
%$(r_1,\dots,r_{i-1}) = (s_1,\dots,s_{i-1})$ there
%are $r',s' \in R'$ such that 
%\begin{itemize}
%\item $(r'_1,\dots,r_{i-1}') = (s_1,\dots,s_{i-1}')$
%\item $r'_i = r_i$ and $s'_i = s_i$. 
%\end{itemize}
\item A representation $R'$ of $R$ is called \emph{compact} if
its cardinality is at most twice the number of forks of $R$. 
\end{itemize}
\end{definition}

Clearly, every relation has a compact representation. Recall  Exercise~\ref{exe:pp-closure} for the relevance of the following lemma. 

\begin{lemma}
Let $A$ be a finite set and let $m \colon A^3 \to A$ be a Maltsev operation. Let $R \subseteq A^k$ be
a relation preserved by $m$, and let $R'$ be a representation of $R$. Then $R=\langle R' \rangle_m$. 
\end{lemma}
\begin{proof}
We show by induction on $i \in \{1,\dots,k\}$ that
$\pr_{1,\dots,i} (\langle R' \rangle_m) = \pr_{1,\dots,i}(R)$.
Clearly, $\pr_{1,\dots,i} (\langle R' \rangle_m) \subseteq \pr_{1,\dots,i}(R)$, so we only need to show the converse inclusion. 
The case
$i=1$ follows from the fact that $R$ has for every
$t \in R$ the fork $(1,t_1,t_1)$, and since
$R'$ must also have this fork it must contain a
tuple $t'$ such that $t'_1 = t_1$. 

So let us assume that the statement holds for $i < k$. We have to show that for every $t \in R$
we have $(t_1,\dots,t_{i+1}) \in \pr_{1,\dots,i+1} (\langle R' \rangle_m)$. 
By induction hypothesis there exists a tuple $s \in \langle R' \rangle_m$
such that $(s_1,\dots,s_i)=(t_1,\dots,t_i)$.
Then $(i+1,s_{i+1},t_{i+1})$ is a fork of $R$, so there
exist tuples $s',s'' \in R'$ witnessing it. 
Then the tuple $t' := m(s,s',s'') \in \langle R' \rangle_m$ is such that
\begin{align*}
(t'_1,\dots,t'_i,t'_{i+1}) & = (m(t_1,s'_1,s''_1),\dots,m(t_i,s'_i,s''_i),m(s_{i+1},s_{i+1},t_{i+1})) \\
& = (t_1,\dots,t_i,t_{i+1}) 
%& = (t_1,\dots,t_i,t_{i+1})  && \text{(since $s'_{i+1} = r_{i+1}$ and $s'_{i+1} = t_{i+1}$)}
\end{align*}
since $s'_1 = s''_1,\dots,s'_i=s''_i$. 
Hence, $(t_1,\dots,t_i,t_{i+1})$ is a tuple from $\pr_{1,\dots,i+1} (\langle R' \rangle_m)$, as required. 
\end{proof}

\begin{exercises}
\exercise[Rot]\label{exe:source-166} Let $A$ be a finite set.
How many forks does the $n$-ary relation $R := A^n$ have?
Explicitly construct a compact representation
for $R$.
% Note that | A^n | = |A|^n,
% but | R' | \leq 2 |A|^2 n.
% Solution: Fix a \in A,
% and take all tuples of
% the form (a,...,a,d,a,...,a) where d is an
% arbitrary element of A at an arbitrary position, and also take the tuple (a,...,a).
\exercise[Rot]\label{exe:source-167}
Let $R$ be the relation $\{(x,y,z,u) \in \{0,1\}^4 \mid x+y+z = 1 \mod 2\}$.
Find a smallest possible representation $R'$ for $R$.
Explicitly compute $\langle R' \rangle_m$ where
$m$ is the Boolean minority.
\label{exe:maltsev}
% Solution (Example):
% {(1,0,0,0),(0,1,0,0),(0,0,1,0),(0,0,1,1)}.
% (Have |R| = 8)
\end{exercises}

\ignore{
\subsubsection*{The procedure \emph{Closure}}
The procedure \emph{Closure} receives as input 
\begin{itemize}
\item a compact representation $R'$ of a relation $R$ which is preserved by $m$, and 
\item a sequence $i_1,\dots,i_k$ of elements in $[n]$
where $n$ is the arity of $R$. 
\end{itemize}
The output of the procedure is a 
relation $S \subseteq R$ such that
\begin{itemize}
\item $\pr_{i_1,\dots,i_k} S = \pr_{i_1,\dots,i_k} R$. 
\item $|S| \leq |\pr_{i_1,\dots,i_k} R|$. 
\end{itemize}
The running time of the procedure is, for fixed
$\bA$, linear in the input size. 
}

%The output of the procedure is either a tuple 
%$t \in R$ such that $(t_{i_1},\dots,t_{i_k}) \in S$,
%or `No' if no such tuple exists. 

\subsection{The Bulatov-Dalmau Algorithm}
\label{sect:buldalalg}
Let $\exists x_1,\dots,x_n (\phi_1 \wedge \dots \wedge \phi_m)$ be an instance of
$\Csp(\bA)$. % with variables $x_1,\dots,x_n$. 
For $\ell \leq m$, we write $R_\ell$ for the relation
$$\{(s_1,\dots,s_n) \in A^n \mid \bA \models (\phi_1 \wedge \dots \wedge \phi_\ell)(s_1,\dots,s_n) \} .$$
The idea of the algorithm is to inductively construct
a compact representation $R_\ell'$ of $R_\ell$,
adding constraints one by one. Initially, for $\ell = 0$, 
we have $R_\ell = A^n$, and it is easy to come up with a compact representation for this relation. 
Note that when we manage to compute
the compact representation $R'_m$ for $R_m$, we can decide satisfiability
of the instance: it is unsatisfiable if and only if $R'_m$ is empty. 
For the inductive step, we need a procedure called \emph{Next} which is more involved; we first introduce two auxiliary procedures.

\subsubsection*{The procedure \emph{Nonempty}}
\label{sect:nonempty}
The procedure \emph{Nonempty} receives as input 
\begin{itemize}
\item a compact representation $R'$ of a relation $R$ which is preserved by $m$, 
\item a sequence $i_1,\dots,i_k$ of elements in $[n]$
where $n$ is the arity of $R$, and 
\item a $k$-ary relation $S$ which is also preserved by $m$. 
\end{itemize}
The output of the procedure is either a tuple 
$t \in R$ such that $(t_{i_1},\dots,t_{i_k}) \in S$,
or `No' if no such tuple exists. 
The procedure can be found in Figure~\ref{fig:nonempty}. 

\begin{figure}
\begin{center}
\fbox{
\begin{tabular}{l}
Procedure \emph{Nonempty}$(R',i_1,\dots,i_k,S)$. \\
\\
Set $U := R'$. \\
While $\exists r,s,t \in U$ such that $\pr_{i_1,\dots,i_k} (m(r,s,t)) \notin \pr_{i_1,\dots,i_k} (U)$: \\
\hspace{0.5cm} Set $U:= U \cup \{m(r,s,t)\}$ \\
If $\exists t \in U$ such that $(t_{i_1},\dots,t_{i_k}) \in S$ then return $t$ \\
else return `No'. 
\end{tabular}}
\end{center}
\caption{The procedure \emph{Nonempty}.}
\label{fig:nonempty}
\end{figure}

\medskip \noindent 
{\bf Correctness.} 
For the correctness of \emph{Nonempty} we note the following:
\begin{itemize}
\item $R' \subseteq U \subseteq R$: initially we start from $U := R' \subseteq R$, and only add tuples to $U$ obtained by applying $m$ to tuples in $U$, so the added tuples are again in $R$. 
\item It follows that if \emph{Nonempty} returns 
a tuple $t$, then $(t_{i_1},\dots,t_{i_k})$ is indeed from $\pr_{{i_1},\dots,i_k}(R)$ and the output of  the algorithm is correct. 
\item When the algorithm exits the while loop then 
$\pr_{i_1,\dots,i_k} (\langle U \rangle_m) = \pr_{i_1,\dots,i_k}(U)$. 
Since $R' \subseteq U$ we have that $\langle U \rangle_m = R$. Hence, every tuple $t \in \pr_{i_1,\dots,i_k} (R) = \pr_{i_1,\dots,i_k} (\langle U \rangle_m)$ is contained in $\pr_{i_1,\dots,i_k}(U)$, and so the answer of the algorithm is also correct when it returns `No'. 
\end{itemize}

We mention that this procedure does not use the particular properties of a Maltsev polymorphism, but works for any explicitly given polymorphism. 

\medskip 
{\bf Running time.} 
The number of iterations of the while loop can
be bounded by the size $|U|$ of the set $U$ at the 
end of the execution of the procedure. 
Hence, when we want to use this procedure to obtain a polynomial-time running
time, we have to make sure that the size of $U$
remains polynomial in the input size. 
The way this is done in the Bulatov-Dalmau algorithm is to guarantee that at each call of \emph{Nonempty} 
the size $L$ of $\pr_{i_1,\dots,i_k}(R)$ is polynomial in the input size. 
Then $|U|$ is bounded by $L + |R'|$
which is also polynomial. 

We have to test all tuples $r,s,t \in U$; this can be implemented so that $|U|^3$ steps suffice. % (for implementation details, see~\cite{DyerRicherby}). 
In each step 
we have to compute $m(r,s,t)$ and test whether $\pr_{i_1,\dots,i_k} (m(r,s,t)) \in \pr_{i_1,\dots,i_k} (U)$, which can be done in $O(kL)$. 
In the important case that $L$ is bounded by a constant 
in the size of the input $N$, the running
time of \emph{Nonempty} is in $O(N^4)$. 

%\paragraph{Exercises.}
%\begin{enumerate}
%\setcounter{enumi}{\value{mycounter}}
%\item Let $R$ be the
%4-ary relation $\{(x,y,z,u) \in \{0,1\}^4 \mid x+y+z = 1 \mod 2\}$, and let $R'$ be a representation
%of $R$ (see Exercise~\ref{exe:maltsev}).
%Let $S$ be the ternary relation 
%$\{(x,y,z) \in \{0,1\}^3 \mid x+y+z = 1 \mod 2\}$. 
%Compute \emph{Nonempty}$(R',2,3,4,S)$. 
% Solution: (starting from the compact representation
% R' := {(1,0,0,0),(0,1,0,0),(0,0,1,0),(0,0,1,1)}. 
% Then (1,0,0,0)+(0,1,0,0)+(0,0,1,1)=(1,1,1,1),
% and (1,1,1) \in S, 
% so t = (1,1,1) can be returned. 
%\setcounter{mycounter}{\value{enumi}}
%\end{enumerate}

\ignore{
{\bf No real improvement:}
\begin{figure}
\begin{center}
\fbox{
\begin{tabular}{l}
Procedure \emph{New-Nonempty}$(R',i_1,\dots,i_k,S)$. \\
\\
Set $U:= \emptyset$. \\
For each $u \in S$: \\
\hspace{0.5cm} If $\exists r,s,t \in R'$ such that $u = \pr_{i_1,\dots,i_k} m(r,s,t)$: \\
\hspace{1cm} Set $U:= U \cup \{u\}$ \\
If $U=\emptyset$ then return `No' \\
else return an element $u$ from $U$. 
\end{tabular}}
\end{center}
\caption{The procedure \emph{Nonempty}.}
\label{fig:nonempty}
\end{figure}}

\subsubsection*{The procedure \emph{Fix-values}}
\label{sect:fix-values}
The procedure \emph{Fix-values} receives as input
\begin{itemize}
\item a compact representation $R'$ of an $n$-ary relation $R$ preserved by $m$, and
\item a sequence $c_1,\dots,c_k \in A$ for $k \leq n$. 
\end{itemize}
The output of \emph{Fix-values} is a compact representation of the relation 
$$ R \cap (\{c_1\} \times \cdots \times \{c_k\} \times A \times \cdots \times A) . $$
The procedure can be found in Figure~\ref{fig:fix-values}. 
The algorithm computes inductively 
a compact representation $U_j$ of the relation 
$$R_j = R \cap (\{c_1\} \times \cdots \times \{c_j\} \times A \times \cdots \times A) \;$$
This is immediate for $U_0 = R'$, and the set $U_k$
is the relation that we have to compute. 

\ignore{
\begin{figure}
\begin{center}
\fbox{
\begin{tabular}{l}
Procedure \emph{Fix-values}$(R',c_1,\dots,c_k)$. \\
\\
Set $j := 0$; $U_j := R'$. \\
While $j < k$ do: \\
\hspace{0.5cm} Set $U_{j+1} := \emptyset$. \\
%\hspace{0.5cm} For each $i \in [n], a \in A$ \\
%\hspace{1cm} compute $t_{i,a}$ := Nonempty$(U_j,j+1,i,\{(c_{j+1},a)\})$. 
\hspace{0.5cm} For each $(i,a,b) \in [n] \times A^2$: \\
\hspace{1cm} If $\exists s,t \in U_j$ witnessing $(i,a,b)$ (assuming $s=t$ if $a=b$): \\
%\hspace{1.5cm} If $a \neq b$ or $s=t$: \\
\hspace{1.5cm} If 
$r := $ Nonempty$(U_j,j+1,i,\{(c_{j+1},a)\}) \neq$ `No' \\
\hspace{2cm} If $(i>j+1)$ or $(a=b=c_i)$: \\
\hspace{2.5cm} Set $U_{j+1} := U_{j+1} 
\cup \{r,m(r,s,t)\}$ \\
\hspace{0.5cm} Set $j := j+1$. \\
Return $U_k$. 
\end{tabular}}
\end{center}
\caption{The procedure \emph{Fix-values}.}
\label{fig:fix-values}
\end{figure}
}
\begin{figure}
\begin{center}
\fbox{
\begin{tabular}{l}
Procedure \emph{Fix-values}$(R',c_1,\dots,c_k)$. \\
\\
Set $j := 0$; $U_j := R'$. \\
While $j < k$ do: \\
\hspace{0.5cm} Set $U_{j+1} := \emptyset$. \\
\hspace{0.5cm} For each $(i,a,b) \in [n] \times A^2$: \\
\hspace{1cm} If $\exists s,t \in U_j$ witnessing $(i,a,b)$ (assuming $s=t$ if $a=b$): \\
%\hspace{1.5cm} If $a \neq b$ or $s=t$: \\
\hspace{1.5cm} If 
$r := $ Nonempty$(U_j,j+1,i,\{(c_{j+1},a)\}) \neq$ `No' \\
\hspace{2cm} If $(i>j+1)$ or $(a=b=c_i)$: \\
\hspace{2.5cm} Set $U_{j+1} := U_{j+1} 
\cup \{r,m(r,s,t)\}$ \\
\hspace{0.5cm} Set $j := j+1$. \\
Return $U_k$. 
\end{tabular}}
\end{center}
\caption{The procedure \emph{Fix-values}.}
\label{fig:fix-values}
\end{figure}
For its correctness, suppose inductively that $U_j$ is a compact representation of $R_j$. We have to show that 
the set $U_{j+1}$ computed by the procedure is a compact representation of $R_{j+1}$: 
\begin{enumerate}
%To show that $U_{j+1}$ is a representation of $R_{j+1}$, we 
\item $U_{j+1} \subseteq R_{j+1}$. 
Suppose that the procedure adds $\{r,m(r,s,t)\}$ to $U_{j+1}$,
where $s$ and $t$ witness the fork $(i,a,b)$ of $U_j$ processed in the for-loop of the procedure. 
Note that $r \in R_{j+1}$ since $r \in \langle U_j \rangle_{m} \subseteq R_j$ and $r_{j+1} = c_{j+1}$. Since $m$ preserves $R$ and is idempotent, it also preserves $R_j$, and since $r,s,t \in R_j$
it follows that $m(r,s,t) \in R_j$. To show that $m(r,s,t) \in R_{j+1}$ it suffices to show that $s_{j+1} = t_{j+1}$ because then $m(r,s,t)_{j+1} = r_{j+1} = c_{j+1}$ since $m$ is Maltsev. If $i > j+1$ then we have that $s_{j+1} = t_{j+1}$
since $s,t$ witness $(i,a,b)$. Otherwise, we must have
$a=b=c_i$ because of the innermost if-clause of the procedure.
But then $s=t$ by the stipulation of the algorithm on the choice of $s$ and $t$. 
\item All forks $(i,a,b)$ of $R_{j+1}$ are forks of $U_{j+1}$. If $R_{j+1}$ has  
the fork $(i,a,b)$, 
then by inductive assumption $U_j$ must contain witnesses $s,t$ for $(i,a,b)$. 
Therefore, the first if-clause of the procedure is positive. 
Moreover, $s_{j+1} = c_{j+1}$ and $s_i = a$,
so $r :=$ \emph{Nonempty}$(U_j,j+1,i,\{(c_{j+1},a)\}) \neq$ `No'.  
Also note that if $i \leq j+1$, then $a = s_i = c_i = t_i = b$.
So all the if-clauses of the procedure are positive, and 
the procedure adds $r$ and $m(r,s,t)$ to 
$U_{j+1}$. The tuples $r$ and $m(r,s,t)$ witness $(i,a,b)$.
Since $s,t$ witness $(i,a,b)$ we have that $(s_1,\dots,s_{i-1}) = (t_1,\dots,t_{i-1})$. Hence, $\pr_{1,\dots,i-1}(m(r,s,t)) = (r_1,\dots,r_{i-1})$. Furthermore, 
we have that $\pr_i(m(r,s,t)) = m(a,a,b) = b$. 
\item The representation $U_{j+1}$ of $R_{j+1}$ is compact
 since at most two tuples are added to $U_{j+1}$ for each fork of $R_{j+1}$. 
\end{enumerate}

{\bf Running time.} 
The while loop is performed $k \leq n$ times; the inner for-loop is executed for each $(i,a,b) \in [n] \times A^2$, which is linear for fixed $\bA$. 
The cost of each iteration is dominated by the cost of calling the procedure \emph{Nonempty}. 
Note that when calling \emph{Nonempty}, the size of
$\pr_{j+1,i} (U_j)$ is polynomial in the input size (even constant size when $\bA$ is fixed), so the cost of \emph{Nonempty} is in $O(N^4)$ where $N$ is the size of the input. 
% as explained at the end of Section~\ref{sect:nonempty}. 
Therefore, the total time complexity of the procedure \emph{Fix-values} is polynomial in the input size (for fixed $\bA$ it is in $O(N^5)$). 

\subsubsection*{The procedure \emph{Next}}
Now comes the heart of the algorithm, which is the procedure \emph{Next} that updates a compact representation of the solution space when constraints are added one by one.  
The input of the procedure is 
\begin{itemize}
\item a compact representation $R'$ of a relation $R \subseteq A^n$ that is preserved by $m$, 
\item a sequence $i_1,\dots,i_k$ of elements from $[n]$, 
\item a $k$-ary relation $S$ which is also preserved by $m$. 
\end{itemize}
The output of the procedure is a compact representation of the relation $$R^* := \{t \in R \mid (t_{i_1},\dots,t_{i_k}) \in S\}.$$

%We first present a procedure, called \emph{Next$^\beta$}, that achieves this, but
%which does not have a polynomial running time. 
%However, there is a way to use \emph{Next$^\beta$}
%in order to obtain a polynomial procedure \emph{Next}. 
%The procedure  \emph{Next$^\beta$}
% can be found in Figure~\ref{fig:next-beta}. 
The procedure  \emph{Next}
 can be found in Figure~\ref{fig:next}. 
\begin{figure}
\begin{center}
\fbox{
\begin{tabular}{l}
Procedure \emph{Next}$(R',i_1,\dots,i_k,S)$. \\
\\
Set $U := \emptyset$. \\
For each $(i,a,b) \in [n] \times A^2$: \\
\hspace{0.5cm} If \emph{Nonempty}$(R',i_1,\dots,i_k,i,S \times \{a\}) =: t \neq$ `No': \\
\hspace{1cm} If \emph{Nonempty}$($\emph{Fix-values}$(R',t_1,\dots,t_{i-1}),i_1,\dots,i_k,i,S \times \{b\}) =: t' \neq$ `No': \\
\hspace{1.5cm} Set $U := U 
\cup \{t,t'\}$.  \\
Return $U$. 
\end{tabular}}
\end{center}
\caption{The procedure \emph{Next}.}
\label{fig:next}
\end{figure}
Observe that 
\begin{itemize}
\item the condition \emph{Nonempty}$(R',i_1,\dots,i_k,i,S \times \{a\}) \neq$ `No' from the first if-clause is satisfied if and only if there exists a tuple $t \in R$ such that $(t_{i_1},\dots,t_{i_k}) \in S$ and $t_i = a$. Hence,
if such a tuple does not exist, then $(i,a,b)$ cannot be a fork of $R^*$, and nothing needs to be done. 
\item the condition \emph{Nonempty}$($\emph{Fix-values}$(R',t_1,\dots,t_{i-1}),i_1,\dots,i_k,i,S \times \{b\}) \neq$ `No'
from the second if-clause is satisfied if and only if there exists
a tuple $t' \in R$ such that 
\begin{itemize}
\item $(t'_1,\dots,t'_{i-1}) = (t_1,\dots,t_{i-1})$, 
\item $(t'_{i_1},\dots,t'_{i_k}) \in S$,  and 
\item $t'_i = b$.
\end{itemize}
If this condition holds, and since $t_i=a$, we have that 
$t$ and $t'$ witness $(i,a,b)$. It only remains to show that if $(i,a,b)$ is a fork of $R^*$, then such a tuple $t'$ must exist. 
So let $r$ and $s$ be witnesses for $(i,a,b)$ in $R^*$. 
Then the tuple $t' := m(t,r,s)$ has the desired properties:
\begin{itemize}
\item for $j < i$ we have that $t'_j = m(t_j,r_j,s_j) = t_j$;
\item $t' \in R^*$, because $(r_{i_1},\dots,r_{i_k}),(s_{i_1},\dots,s_{i_k}),(t_{i_1},\dots,t_{i_k}) \in S$ and $m$ preserves $S$. 
\item $t'_i = m(t_i,r_i,s_i) = m(a,a,b) = b$. 
\end{itemize}
\item The cardinality of $U$ is bounded by twice the number of forks of $R^*$, so the representation computed by the algorithm is compact. 
\end{itemize}
{\bf Running time.}
The for-loop of the procedure \emph{Next} is performed $n |A|^2$ times and the cost of each iteration is polynomial in the cost of \emph{Nonempty} and \emph{Fix-values}. Also note that $k$ is bounded by the maximal arity of the relations in $\bA$, so
constant for fixed $\bA$. It follows that 
$\pr_{i_1,\dots,i_k,i}(R)$ is polynomial, 
so the running time of the calls to \emph{Nonempty} are polynomial. For fixed $\bA$, the global running time of the procedure \emph{Next}
is in $O(N^6)$ where $N$ is the size of the input.  

\begin{proof}[Proof of Theorem~\ref{thm:maltsev}]
Starting from an empty list of constraints, we add constraints on the variables $x_1,\dots,x_n$ one by one, and maintain a compact representation 
of the $n$-ary relation defined by the constraints considered so far. Initially, we start with a compact representation of the full relation $A^n$. 
In later steps, we use the procedure \emph{Next} to compute a compact representation when a constraint is added, in $O(N^6)$ for fixed $\bA$ and $N$ the size of the input. The instance is unsatisfiable
if and only if at the final stage we end up with an empty representation. The entire running time
of the algorithm is in $O(N^7)$. 
\end{proof}

% HERE, GLOBAL MALTSEV ALGORITHM 
% DESCRIPTION. 

% SIMPLIFICATION IN PRESENTATION POSSIBLE
% FOR NON-UNIFORM CSP!

\begin{exercises}
\exercise[Orange]\label{exe:source-168} Let $\bA$ be the structure $(\{0,1\};L_0,L_1)$
where $L_i := \{(x,y,z) \mid x+y+z=i \mod 2\}$,
which has the Boolean minority $m$ as polymorphism. Consider the instance
$$\exists x_1,\dots,x_5 \, \big (L_1(x_1,x_2,x_3) \wedge L_1(x_2,x_3,x_4) \wedge L_1(x_3,x_4,x_5) \wedge
L_0(x_1,x_3,x_5) \big )$$
Compute compact representations $R_\ell'$ of $R_\ell$, for $\ell \in \{1,2,3,4\}$.
%(a `non-degenerate' fork is in such an example uniquely given
%by a value in $\{1,\dots,5\}$).
% Solution:
% * We already know that R'_4 = R_4 = \emptyset
% since obviously there is no solution (sum up
% all the equations!)
% * $R'_1$ can for example be:
% (1,0,0,0,0),
% (0,1,0,0,0),
% (0,0,1,0,0).
% (1,0,0,1,0),
% (1,0,0,0,1),
% Note that there is no fork 3
% * For $R'_2$, we find for the fork 1
% the witnesses t=(0,1,0,0,0) and t' = (1,0,0,1,0)
% and for the fork 2 we find the witnesses
% t=(0,0,1,0,0) and t'=(0,1,0,0,0)
% Note that 3 and 4 are no forks.
% For the fork 5 we get the witnesses
% t=(1,0,0,1,0) and t'=(1,0,0,1,1)
% = m((1,0,0,1,0),(1,0,0,0,0),(1,0,0,0,1))
% So this gives for R'_2:
% (1,0,0,1,0)
% (0,1,0,0,0),
% (0,0,1,0,0),
% (1,0,0,1,0)
% (1,0,0,1,1)
% * For the computation of $R'_3$, we even loose
% the fork 5.
% For the fork 1 we have the witnesses
% t=(0,0,1,0,0) and t'=(1,0,0,1,0)
% For the fork 2 we have the witnesses
% t=(0,0,1,0,0) and t'=(0,1,0,0,1)=m((0,1,0,0,0),(1,0,0,1,0),(1,0,0,1,1)). This gives:
% (0,0,1,0,0),
% (1,0,0,1,0),
% (0,1,0,0,1).
% Now the call Nonempty(R'_3,1,3,5,L_0) yields
% `No' for all numbers i \in [n] (there are no forks).
\exercise[Schwarz]\label{exe:source-169} Let $\bB$ be a structure with
a Maltsev polymorphism $f$ and an \emph{infinite} relational signature.
Note that we have defined
$\Csp(\bB)$ only if $\bB$ has a finite signature.
%It is standard to define $\Csp(\bB)$ as the
%problem of deciding whether a finite
%structure with the same signature as $\bB$ has a homomorphism to $\bB$.
If we want to define $\Csp(\bB)$ also for structures $\bB$ with an infinite signature, it is important \\ to
discuss how the relation symbols in the
signature of $\bB$
are represented
in the input. We choose to represent a relation symbol $R$ from $\bB$ by listing
the tuples in $R^{\bB}$.
Adapt the Dalmau algorithm
such that it can solve
$\Csp(\bB)$
in polynomial time for this choice of
representing the relations in $\bB$.
% Solution sketch: Store only the relations that actually occur in the finite input, each by its
% explicit tuple list. Dalmau's compact representations and all closure tests are then performed by
% scanning those lists; no step enumerates the infinite signature. Since the template domain is fixed,
% every operation is polynomial in the total encoded input length.
\exercise[Schwarz]\label{exe:source-170} The \emph{graph isomorphism problem (GI)} is a famous computational problem that is neither known to be solvable in polynomial time, nor expected to be NP-hard. An instance of GI consists of two graphs $G$
and $H$, and the question is to decide
whether $G$ and $H$ are isomorphic.
Consider the variant of the
graph-isomorphism problem where
the vertices are coloured, each color
appears at most $k$ times for
some constant $k$, and the isomorphism between $H$ and $G$ that we are
looking for is required to additionally  preserve the colours. Show that
this problem can
be solved in polynomial time using
Dalmau's algorithm
(use Exercise~\ref{exe:source-169}).
% Here is the solution from Feder.
% "For simplicity, assume that each
% color occurs in $k$ of the vertices of
% each of the two graphs.
% Let G be the group of permutations
% on 2k elements, corresponding to
% the 2k vertices of the same color in the
% two graphs. The constraint that vertices
% in one graph map to vertices in the
% other is a coset in G while the
% constraint that adjacent vertices map
% to adjacent vertices is a subgroup
% of G for adjacent vertices of the same
% color, and of G^2 for adjacent vertices
% of different colors. Thus labeled graph
% isomorphism can be viewed as a
% subgroup CSP.
% Frage 11.11.22: geht das nicht auch mit endlicher Sprache?
\end{exercises}

%\subsection{The Dyer-Richerby Algorithm}

% END INLINED SOURCE: basics.tex
% BEGIN INLINED SOURCE: Minions.tex
% !TEX root = GH-UA.tex
\section{Minions}
\label{sect:minions}
(Abstract) minions generalise (abstract) clones,  
and function minions generalise operation clones. 
The name has been introduced in 2018
 when it became clear that function minions over finite domains capture the complexity of so-called \emph{promise CSPs}, which generalise CSPs~\cite{BartoBKO21}. 
This text does not cover promise CSPs; however, we still introduce minions, because  minion homomorphisms play an important role when studying clones as well (see, e.g.,~\cite{wonderland,smooth-digraphs,Collapse}). In particular, minion homomorphisms can be used to characterise 
the operator $\HI$ from the second formulation of the tractability theorem, Theorem~\ref{thm:tractability-2}. 

\subsection{Minors and Minions} 
\label{sect:minors}
Let $A,B$ be sets and $k,m \in {\mathbb N}_{\geq 1}$. Let $f \colon A^k \to B$ be a function and let $\alpha \colon [k] \to [m]$. Then $f_\alpha$ denotes the function $g \colon A^m \to B$ given by $g(x_1,\dots,x_m) := f(x_{\alpha(1)},\dots,x_{\alpha(k)})$. 
A \emph{minor} of $f$ is a function of the form $f_{\alpha}$, for some $\alpha \colon [k] \to [m]$. 
Note that 
\begin{itemize}
\item if $\alpha \colon [k] \to [m]$ and $\beta \colon [m] \to [n]$,
then $$(f_\alpha)_\beta = f_{\beta \circ \alpha},$$
so the minor relation is transitive. 
\item if $\alpha \colon [k] \to [m]$ 
then $(\pi^k_i)_\alpha = \pi^{m}_{\alpha(i)}$. 
\item if $f$ and $g$ are operations on $A$, then 
 for all $\alpha_1,\dots,\alpha_k \colon [m] \to [n]$ there exists $\beta \colon [km] \to [n]$ such that 
\begin{align*}
f(g_{\alpha_1},\dots,g_{\alpha_k}) & = (f * g)_{\beta}
&& \text{(recall Definition~\ref{def:star}).}
\end{align*}
\end{itemize}

\begin{definition}
A \emph{function minion} is a subset ${\mathscr M}$ of  $\bigcup_{k \geq 1} B^{A^k}$, where $A$ and $B$ are sets, which is closed under taking minors.
\end{definition}

Note that every operation clone is a function minion where $A = B$ and where we additionally require the presence of the projections and closure under composition. 
A \emph{minion} is the abstract version of a function minion, analogously as clones can be viewed as the abstract version of operation clones. %Again, they can be viewed as a multi-sorted algebra. 

\begin{definition}
An \emph{(abstract) minion} is a multi-sorted algebra ${\fM}$ with sorts $M^{(1)},M^{(2)},\dots$ and for each $\alpha \colon [n] \to [m]$ the operation ${ }_{\alpha} \colon  
M^{(n)} \to M^{(m)}$ such that
%\begin{itemize}
%\item 
for every $f \in M^{(n)}$ we have $f_{\id_{[n]}}=f$, and for every $\sigma \colon [n] \to [m]$ and $\rho \colon [m] \to [k]$ we have
$$ (f_\sigma)_\rho = f_{\rho \circ \sigma}.$$
%\item for every injective $\alpha \colon [n] \to [m]$ the operation $f \mapsto f_\alpha$ from $M^{(n)}$ to $M^{(m)}$ is injective, 
%\item for every surjective $\alpha \colon [n] \to [m]$ the operation $f \mapsto f_\alpha$ from $M^{(n)}$ to $M^{(m)}$ is surjective. 
%\end{itemize}
\end{definition} 

Clearly, every function minion gives rise to a minion in the obvious way. 
This statement has a converse, which is analogous to Cayley's theorem for groups (see Proposition~\ref{prop:clone-variety-clone} for the corresponding statement for clones); 
see Exercise~\ref{exe:cayley-minion}. 
%we will prove this in Section~\ref{sect:h1-birkhoff} (Proposition~\ref{prop:cayley-minion}).  

\begin{definition}
Let $\fM$ and $\fN$ be minions.  
\begin{itemize}
\item A \emph{minion homomorphism} from $\fM$ to $\fN$ is a map $\xi \colon M \to N$ such that for every $n \in {\mathbb N}$ and $f \in M^{(n)}$ we have $\xi(f) \in N^{(n)}$ and for every $m,n \in {\mathbb N}_{\geq 1}$ and $\alpha \colon [n] \to [m]$ and $f \in M^{(n)}$ we have $\xi(f_\alpha) =\xi(f)_\alpha$. 
\item A \emph{minion isomorphism} is a bijective minion homomorphism $\xi \colon \fM \to \fN$ such that  $\xi^{-1}$ is a minion homomorphism as well. In this case, $\fM$ and $\fN$ are called \emph{isomorphic}. 
\end{itemize}
\end{definition}

Similarly as for clones in Lemma~\ref{lem:clone-compactness}, we may apply the compactness theorem of first-order logic\footnote{See, e.g.,~\cite{BodMathLogic}.} to characterise the existence of minion homomorphisms to function minions on finite sets. 
%with finitely many elements in each sort. 

\begin{lemma}\label{lem:minion-compactness}
Let $\fM$ be a minion and let $\fF$ be a minion with finitely many elements of each sort. 
%be isomorphic to a function minion on finite sets. 
If there is no minion homomorphism from $\fM$ to $\fF$, then there exists a primitive positive sentence over the signature of (abstract) minions that holds in $\fM$ but not in $\fF$. 
%is a primitive positive sentence in the language $\tau$ of (abstract) clones that
%holds in $\fC$ but not in $\fF$. 
\end{lemma}

\begin{exercises}
\exercise[Blau]\label{exe:source-171} Let $\fC$ and $\fD$ be clones.
Show that $\xi \colon C \to D$ is a minion homomorphism if and only if
\begin{itemize}
\item $\xi$ preserves arities, i.e.,
$\xi(C^{(i)}) \subseteq D^{(i)}$ for all $i \in {\mathbb N}$, and
\item $\xi$ preserves composition with projections, that is, for all $n,k \geq 1$
and $f \in C^{(k)}$
$$ \xi(f \big ((\pi^n_{i_1})^{\fC},\dots,(\pi^n_{i_k})^{\fC} \big)) = \xi(f) \big ((\pi^n_{i_1})^{\fD},\dots,(\pi^n_{i_k})^{\fD} \big).$$
\end{itemize}
%\newpage
%\vspace{-.2cm}
%\newpage
% Solution sketch: A minor $f_\alpha$ is exactly a composition of $f$ with projections. Hence
% preservation of arity and of such compositions implies $\xi(f_\alpha)=\xi(f)_\alpha$, the definition
% of a minion homomorphism. The converse follows by taking the map $\alpha$ that lists the chosen
% projections.
\exercise[Blau]\label{exe:source-172} Let $\alpha \colon [k] \to [m]$. Write down a primitive positive formula
$\phi(x,y)$
in the language of clones such that for every operation clone ${\mathscr C}$ and
all $f,g \in {\mathscr C}$ we have ${\mathscr C} \models \phi(f,g)$ if and only if $g = f_{\alpha}$.
%\item Show that $f \colon A^n \to A$ is a Taylor operation if and only if \\ for every $i \in [n]$
%there are $\alpha,\beta \colon [n] \to [k]$ such that $t_\alpha = t_{\beta}$ and $\alpha(i) \neq \beta(i)$.
\exercise[Rot]\label{exe:source-173} Let $\fM$ be a minion. Show that for every injective $\alpha \colon [n] \to [m]$
the operation $f \mapsto f_\alpha$ from $M^{(n)}$ to $M^{(m)}$ is injective, and for every
surjective $\alpha \colon [n] \to [m]$ the operation $f \mapsto f_\alpha$ from $M^{(n)}$ to $M^{(m)}$
is surjective.
% Solution sketch: If $\alpha$ is injective, choose a left inverse $\beta$; then $(f_\alpha)_\beta=f$,
% proving injectivity. If $\alpha$ is surjective, choose a right inverse $\beta$; for any $g\in
% M^{(m)}$, the operation $f=g_\beta$ satisfies $f_\alpha=g$, proving surjectivity.
\exercise[Schwarz]\label{exe:source-174} \label{exe:functor} (For readers familiar with basic category theory)
Let {\bf Set} denote the category
of all sets and mappings between them, and let
{\bf FinOrd} denote the category of finite ordinals and mappings between them.
\begin{itemize}
\item
Explain how minions can be viewed as
functors from ${\bf FinOrd}$ to ${\bf Set}$. Show that natural transformations between such functors are minion homomorphisms.
\item Explain how minions can be viewed
as {\bf Set}-endofunctors $F$
such that $F(\emptyset) = \emptyset$ and
$F$ is \emph{finitary}:
that is, if $X$ is a set and $x \in F(X)$, there exists a finite subset $Y \subseteq X$ and $y \in F(Y)$ such that $F(f)(y) = x$,
where $f \colon Y \to X$ is the inclusion map.
%, then $F(X)$ is the union of the sets $\{F(i)(u) \mid u \in U\}$, where $U$ is a finite subset of $X$ and $i \colon U \to X$ is the inclusion map.
Show that natural transformations between
such {\bf Set}-endofunctors correspond to  minion homomorphisms.
\end{itemize}
% Sei M Minion.
% Betrachten Kategorie N:
% Objekten := natuerliche Zahlen
% als Mengen gedacht
% Morphismen: Beliebige Abbildungen zwischen solchen Mengen.

% F ist Funktor von N nach Set.
% definiert auf Objekten wie folgt:
% n \mapsto M^{(n)}
% (f: n->m) \mapsto
% die Abbildung M^{(n)} -> M^{(m)}
% die gegeben ist durch g \mapsto g_f

% Fine with natural transformations.

% But not yet an endofunctor!
% Composition of endofunctors is
% endofunctors: which one?

\end{exercises}

%All the minions that appear in this text 

% TODO: Cayley for minions? 

%\begin{definition}
%Let $\fC$ and $\fD$ be minions. 
%A function $\xi \colon C \to D$ is
%called a \emph{(minion) homomorphism} (also: \emph{minor preserving}) 
%(also: \emph{minion homomorphism}) 
%if 
%\end{definition}
%Let ${\mathscr C}$ and ${\mathscr D}$ be operation clones. Note that 
%\begin{itemize}
%\item an arity-preserving function $\xi \colon {\mathscr C} \to {\mathscr D}$ is minor-preserving
%if and only if for all 
%$f \in {\mathscr C}^{(k)}$,
%$g \in {\mathscr C}^{(m)}$, and $\alpha \colon [k] \to [m]$  
%such that $g = f^\alpha$ we have  
%$\xi(g) = \xi(f)^{\alpha}$. 
%\item 
%\end{itemize}

\subsection{Reflections}
\label{sect:reflections}
In Section~\ref{sect:pseudo-var} we have
seen that the $\HSPfin$ operator is the algebraic counterpart to full 
primitive positive interpretations. This section treats a relatively new universal-algebraic operator, for forming \emph{reflections} (introduced in~\cite{wonderland}), which can be used to characterise the structure-building operator $\HI$.
Recall from Section~\ref{sect:wonderland} 
that $\HI$ is the operator that is most relevant for  constraint satisfaction. 

\begin{definition}\label{def:refl}
Let $\fB$ be a $\tau$-algebra, let $A$ be a set,
and let $h \colon B \to A$ and $g \colon A \to B$ be two maps. Then the \emph{reflection}\index{reflection} of $\fB$ with respect to $(h,g)$ is the $\tau$-algebra $\fA$ with domain $A$ where for all $a_1,\dots,a_n \in A$
and $f \in \tau$ of arity $n$ we define 
$$f^{\fA}(a_1,\dots,a_n) :=  h(f^{\fB}(g(a_1),\dots,g(a_n))) \, .$$
The class of reflections of a class of $\tau$-algebras ${\mathcal C}$ is denoted by 
$\R({\mathcal C})$.
\end{definition}

As for the other operators on algebras, we write
$\R(\fB)$ instead of $\R(\{\fB\})$. 
The following is an analog to the $\HSP$-lemma (Lemma~
\ref{lem:hsp}). %and the $\HSPfin$ 

\begin{lemma}[from~\cite{wonderland}]\label{lem:wonderland}
Let $\mathcal C$ be a class of $\tau$-algebras. If the empty $\tau$-algebra exists, assume that it belongs to $\mathcal C$.
\begin{itemize}
\item The smallest class of
$\tau$-algebras that contains $\mathcal C$ and is closed under $\R$, $\Hom$, $\S$, and $\P$ 
equals $\RP({\mathcal C})$. 
\item The smallest class of
$\tau$-algebras that contains $\mathcal C$ and is closed under $\R$, $\Hom$, $\S$, and $\Pfin$ 
equals $\RPfin({\mathcal C})$. 
\end{itemize}
\end{lemma}
\begin{proof}
For the first statement, it suffices to prove that 
$\RP({\mathcal C})$
is closed under $\R$, $\Hom$, $\S$, $\P$,
and for the second that $\RPfin({\mathcal C})$
is closed under $\R$, $\Hom$, $\S$, 
$\P^{\fin}$. 
For the operator $\R$ this follows 
from the simple fact that $\R \R({\mathcal K}) = \R({\mathcal K})$ for any class
$\mathcal K$. 
% To see this, observe that if
%$\bC$ is a reflection of $\bB$ witnessed by $h_1 \colon C \to B$ and $g_1 \colon B \to C$,
%and that 

To prove that $\RP({\mathcal C})$ and 
$\RPfin({\mathcal C})$ are closed under $\Hom$,
we show that $\Hom({\mathcal K}) \subseteq \R({\mathcal K})$ for any class ${\mathcal K}$. 
Let  $\fB \in {\mathcal K}$ and $h \colon B \to A$
be a surjective homomorphism to an algebra $\fA$.
Pick any function $g$ such that $h \circ g$ is the identity on $A$. Then $h$ and $g$ witness that $\fA$ is a reflection of $\fB$ since
\begin{align*}
h(f^{\fB}(g(x_1),\dots,g(x_n))) & = f^{\fA}(h \circ g(x_1),\dots,h \circ g(x_n)) && \text{(since $h$ is a homomorphism)} \\
& = f^{\fA}(x_1,\dots,x_n) && \text{(by the choice of $g$).}
\end{align*}
To prove that $\RP({\mathcal C})$ and $\RPfin({\mathcal C})$ are closed under $\S$,
we show that $\S({\mathcal K}) \subseteq \R({\mathcal K})$ for any class ${\mathcal K}$. 
Let $\fB \in {\mathcal K}$ and suppose that $\fA$ is a subalgebra of $\fB$. Let $g \colon A \to B$ be the identity on $A$, and $h \colon B \to A$ be any 
extension of $g$ to $B$. Then $h$ and $g$ show that $\fA$ is a reflection of $\fB$ since 
$$h(f^{\fB}(g(x_1),\dots,g(x_n))) = f^{\fB}(x_1,\dots,x_n) = f^{\fA}(x_1,\dots,x_n) \, .$$

Let $I$ be an arbitrary set, $(\fB_i)_{i \in I}$ be algebras from $\P({\mathcal C})$, and suppose that $\fA_i$ is a reflection of $\fB_i$ for every $i \in I$, witnessed by functions $h_i \colon B_i \to A_i$ and $g_i \colon A_i \to B_i$. Then the map $h \colon \prod_{i \in I} B_i \to \prod_{i \in I} A_i$ that sends
$(b_i)_{i \in I}$ to $(h_i(b_i))_{i \in I}$ and the map 
$g \colon \prod_{i \in I} A_i \to \prod_{i \in I} B_i$ that sends
$(a_i)_{i \in I}$ to $(g_i(a_i))_{i \in I}$ witness
that $\prod_{i\in I} \fA_i$ is a reflection of $\prod_{i\in I} \fB_i$. This shows that $\P(\RP({\mathcal C})) \subseteq \RP({\mathcal C})$
and likewise that $\P^{\fin}(\RPfin({\mathcal C})) \subseteq \RPfin({\mathcal C})$. 
\end{proof}

\begin{theorem}\label{thm:hi-erpfin}
Let $\bB,\bC$ be finite relational structures and let $\fC$ be a polymorphism algebra of $\bC$. Then 
\begin{enumerate}
\item $\bB \in \Hom(\bC')$ for some 
structure $\bC'$ which is primitively positively  definable in $\bC$
if and only if there is an algebra $\fB \in \R(\fC)$ such that $\Clo(\fB) \subseteq \Pol(\bB)$. 
%\item $\bB \in \H(\bC)$ if and only if there is an
%algebra $\fB \in \R(\fC)$ such that $\Clo(\fB) = \Pol(\bB)$. CAN'T BE TRUE, SINCE THE SECOND IS SIGNATURE-INDEPENDENT, WHILE THE FIRST IS NOT.  TO FIX IT, ONE COULD
% INTRODUCE A PP-interdefinability operator,
% which needs to be applied first. 
% But even then, the statement needs proof
% and below I didn't manage to finish it. 
%\item $\bB \in \H \FI(\bC)$  if and only if 
%there is an algebra $\fB \in \RPfin (\fC)$ such that $\Clo(\fB) = \Pol(\bB)$.
\item $\bB \in \HI(\bC)$ if and only if there is an algebra $\fB \in \RPfin (\fC)$ such that $\Clo(\fB) \subseteq \Pol(\bB)$.
\end{enumerate}
\end{theorem}
\begin{proof}
To show $(1)$, first suppose that $\bB \in \Hom (\bC')$ for some $\bC'$ which is pp definable in $\bC$; let $h \colon \bC' \to \bB$ and $g \colon \bB \to \bC'$ be homomorphisms witnessing homomorphic equivalence of $\bB$ and $\bC'$. 
Let $\fC'$ be an expansion of $\fC$ which is a polymorphism algebra 
of $\bC'$. 
Let $\fB'$ be the reflection of $\fC'$ with respect to $(h,g)$. Every operation of $\Clo(\fB')$ is obtained as a composition of homomorphisms, so preserves all the relations of $\bB$, so $\Clo(\fB') \subseteq \Pol(\bB)$. Let $\fB$ be the reduct of $\fB'$ where we only keep the operations for the signature of $\fC$, and note that $\fB \in \R (\fC)$ is such that $\Clo(\fB) \subseteq \Pol(\bB)$.

Conversely, suppose that 
the reflection $\fB$ of $\fC$ at $h \colon C \to B$ and $g \colon B \to C$ is such that $\Clo(\fB) \subseteq \Pol(\bB)$. Let $\bC'$ be the structure
with domain $C$ and the same signature as $\bB$  which contains for every $k$-ary relation symbol $R$ of $\bB$ the relation
\begin{align*}
R^{\bC'} :=  \{ & (f(g(b^1_1),\dots,g(b^{\ell}_1)),\dots,f(g(b^1_k),\dots,g(b^{\ell}_k))) \\
& \mid \ell \geq 1, f \in \Pol(\bC)^{(\ell)}, (b^1_1,\dots,b^1_k), \dots (b^{\ell}_1,\dots,b^{\ell}_k) \in R^{\bB}\}.
\end{align*}
These relations are preserved by $\Pol(\bC)$, so they are pp definable in $\bC$ by Theorem~\ref{thm:inv-pol},
and hence $\bC'$ is primitively positively definable in $\bC$. Clearly, $g$ is a homomorphism from $\bB$ to $\bC'$. We claim that $h$ is a homomorphism from $\bC'$ to $\bB$. 
Let $R$ be a $k$-ary relation symbol and let
$(c_1,\dots,c_k)\in R^{\bC'}$. By the definition of $R^{\bC'}$,
there exist $\ell\geq 1$, an operation
$f\in\Pol(\bC)^{(\ell)}$, and tuples
\[
\bar b^j=(b^j_1,\dots,b^j_k)\in R^{\bB}
\qquad (j\in[\ell])
\]
such that
\[
c_i=f\bigl(g(b^1_i),\dots,g(b^\ell_i)\bigr)
\qquad (i\in[k]).
\]
%The operation of the reflection $\fB$ corresponding to $f$ is
%\[
%f^{\fB}(x_1,\dots,x_\ell)
%:=
%h\bigl(f(g(x_1),\dots,g(x_\ell))\bigr).
%\]
Since $\Clo(\fB)\subseteq\Pol(\bB)$, the operation $f^{\fB}$
preserves $R^{\bB}$. Applying it coordinatewise to
$\bar b^1,\dots,\bar b^\ell\in R^{\bB}$ therefore gives
\begin{align*}
\bigl(h(c_1),\dots,h(c_k)\bigr)
&= \big (h(f(g(b^1_1),\dots,g(b^{\ell}_1))),\dots,h(f(g(b_k^1),\dots,g(b^{\ell}_k))) \big)
\\
& = \bigl(
 f^{\fB}(b^1_1,\dots,b^\ell_1),\dots,
 f^{\fB}(b^1_k,\dots,b^\ell_k)
\bigr)\\
&=
f^{\fB}(\bar b^1,\dots,\bar b^\ell)
\in R^{\bB}.
\end{align*}
Thus, $h$ preserves every relation of $\bC'$ and  $h$ is a
homomorphism from $\bC'$ to $\bB$.
%We claim that $h$ is a homomorphism from $\bC'$ to $\bB$.
%Indeed, if $b_1,\dots,b_k \in B^$ are such that $(f(g(b_1),\dots,g(b_k))) \in R^{\bC'}$,
%then $h(f(g(b_1),\dots,g(b_k))) \in R^{\bB}$ because
%the operation $(x_1,\dots,x_k) \mapsto h(f(g(x_1),\dots,g(x_k)))$ is an operation of $\fB' \in \R(\fC)$
%and hence a polymorphism of $\bB$
%since $\Clo(\fB') \subseteq \Pol(\bB)$. 
Therefore, $\bB \in \Hom(\bC')$.

%\medskip 
%To show $(2)$, suppose first that 
%$\bB \in \H \FI(\bC)$. Then there exists a structure
%$\bD \in \FI(\bC)$ such that $\bB \in \H (\bD)$. 
%By Theorem~\ref{thm:pp-interpret} there is an algebra $\fD \in \HSPfin(\fC)$ such that $\Clo(\fD) = \Pol(\bD)$. Let $g \colon \bD \to \bB$ and $h \colon \bB \to \bD$ be homomorphisms witnessing homomorphic equivalence of $\bB$ and $\bD$. 
%Let $\fB$ be the reflection of $\fD$ with respect to $g$ and $h$. 
%Every operation of $\Clo(\fB)$ is obtained as a composition of homomorphisms, so preserves all the relations of $\bB$, so $\Clo(\fB) \subseteq \Pol(\bB)$. To prove that $\Pol(\bB) \subseteq \Clo(\fB)$, let $f \in \Pol(\bB)$ be $k$-ary. Then
%$(x_1,\dots,x_k) \mapsto h(f(g(x_1),\dots,g(x_k)))$is a polymorphism of $\bD$, and hence in $\Clo(\fD)$. Therefore, 

\medskip 
Item $(2)$ is a combination of 
item $(1)$ with Theorem~\ref{thm:pp-interpret}: 
First suppose that 
$\bB \in \HI(\bC)$.
Then there exists a structure $\bD \in I(\bC)$ 
such that $\bB \in \Hom(\bD)$. 
%$\bD \in \FI(\bC)$ such that $\bB \in \Hom \Red(\bD)$. 
By Theorem~\ref{thm:pp-interpret} there is an algebra $\fD \in \HSPfin(\fC)$ such that $\Clo(\fD) \subseteq \Pol(\bD)$, and by item $(1)$ there is an 
algebra $\fB \in \R(\fD)$ such that $\Clo(\fB) \subseteq \Pol(\bB)$. This proves the statement since 
\begin{align*}
\fB \in \R(\fD) & \subseteq \R \HSPfin(\fC) \\
& = \RPfin(\fC) && \text{(by Lemma~\ref{lem:wonderland}).} 
\end{align*}
Conversely, suppose that there exists
$\fB \in \RPfin(\fC)$ such that $\Clo(\fB) \subseteq \Pol(\bB)$. Then there 
exists $\fD \in \Pfin(\fC)$ such that $\fB \in \R(\fD)$. Let $\bD$ be the structure with the same domain as $\fD$ which contains all the relations preserved by all operations
of $\Clo(\fD)$. Then $\bD \in \I(\bC)$.
Moreover, by item (1) there exists $\bD'$ which is pp definable in $\bD$ such that
$\bB \in \Hom(\bD')$. Clearly, $\bD' \in \I(\bC)$ and hence $\bB \in \HI(\bC)$. 
%Since $\fB \in \R(\bD)$, we get  
%$\bB \in \H(\bD)$. 
\end{proof}

%The following hardness condition for CSPs is stronger than the one presented 
%in~\ref{thm:pseudo-var-projective}: there are situations where the conditions of the following corollary applies, but where the conditions of 
%Theorem~\ref{thm:pseudo-var-projective} do not. 

We now characterise in many different ways 
the hardness condition from the second formulation of the tractability theorem (Theorem~\ref{thm:tractability-2}); quite remarkably, we do not need to assume that the involved polymorphism algebra is idempotent. 

\begin{corollary}\label{cor:hi-hard}
Let $\bB$ be a structure with a finite domain and let $\fB$ be a polymorphism algebra of $\bB$. Then the following are equivalent.
\begin{enumerate}
\item $\HI(\bB)$ contains $K_3$; 
\label{eq:hi-k3}
\item $\HI(\bB)$ contains all finite structures;
\label{eq:hi-all}
\item $\HI(\bB)$ contains $(\{0,1\};\NAE)$; 
\label{eq:hi-oit}
\item 
\label{eq:rp1}
$\R \Pfin(\fB)$ contains an algebra of size at least 2 all of whose operations are projections.

\item $\R \Pfin(\fB)$ contains for every finite set $A$ an algebra on $A$ all of whose operations are projections.
\label{eq:rp2}
\end{enumerate}
If these conditions apply then $\bB$ has a finite-signature reduct with an NP-hard $\Csp$. 
\end{corollary}
\begin{proof}
The implication from $\ref{eq:hi-k3}.$ to 
$\ref{eq:hi-all}.$ follows from the
fact that $\I(K_3)$ contains all finite structures (Theorem~\ref{thm:k3-interprets}),
and that $\I \HomE \I(\bB) \subseteq \HI(\bB)$
by Theorem~\ref{thm:wonderland}. 
The implication from $\ref{eq:hi-all}.$ to $\ref{eq:hi-oit}.$ is trivial.
The equivalence of $\ref{eq:hi-oit}.$ 
and $\ref{eq:rp1}.$ follows
from the fact that all polymorphisms
of $(\{0,1\};\NAE)$ are essentially unary (Exercise~\ref{exe:NAE}) 
and Theorem~\ref{thm:hi-erpfin}. 
We leave the proof of the equivalence of 
$\ref{eq:rp2}.$ to the reader. 
%For the implication from 
%$(\ref{eq:rp1})$ 
%to $(\ref{eq:rp2})$, 
%suppose that $\R \Pfin(\fB)$ contains an algebra 
%$\fA$ of size at least 2 all of whose operations are projections. By Theorem~\ref{thm:pseudo-var-projective},
%$\HSPfin(\fA)$ contains for every finite set $S$ an algebra on $S$ all of whose operations are projections. 
%The statement follows since $\HSPfin(\fA) \subseteq \HSPfin(\HI(\bB)) \subseteq \HI(\bB)$
%by Theorem~\ref{thm:wonderland}. 
%To show that  
%$(\ref{eq:rp2})$ implies $(\ref{eq:hi-k3})$,
%et $\fA \in \HomE(\bB)$ be such that $A = \{0,1,2\}$ and all operations of $\fA$ are projections. Then $\Pol(\fA) \subseteq \Pol(K_3)$
%and hence $K_3 \in \HI (\bB)$ by Theorem~\ref{thm:hi-erpfin}. 
The final statement follows from Corollary~\ref{cor:pp-interpret-hard}.
\end{proof}

\begin{exercises}
\exercise[Blau]\label{exe:source-175} Prove the equivalence of \ref{eq:rp2}.~ with the other items of Corollary~\ref{cor:hi-hard}.
% Solution sketch: Translate item~\ref{eq:rp2} into the existence of the corresponding minion
% homomorphism by the pp-construction theorem. Compose it with the minion homomorphism from the hard
% three-colour template and use transitivity of pp-constructions. Reversing these steps yields each of
% the other items of Corollary~\ref{cor:hi-hard}.
\end{exercises}

\subsection{Birkhoff's Theorem for Height-One Identities}
\label{sect:h1-birkhoff}
A \emph{height-one identity}
is an identity $s \approx t$ where the involved terms $s$ and $t$ have \emph{height one}, i.e., each term involves exactly one function symbol.  
Three examples of properties that
can be expressed by finite sets of height-one identities are listed below. 
\begin{align*}
f(x,y) & \approx f(y,x) && \text{($f$ is symmetric)} \\
f(x,x,y) & \approx f(x,y,x) \approx f(y,x,x) \approx f(x,x,x) && \text{($f$ is quasi majority)} \\
f(x,x,y) & \approx f(y,x,x) \approx f(y,y,y) && \text{($f$ is quasi Maltsev)} 
\end{align*}
A non-example is furnished by the Maltsev identities 
$f(x,x,y) \approx f(y,x,x) \approx y$ because the term $y$ involves no function symbol. Identities where each term involves \emph{at most} one function symbol are called \emph{linear}; so the Maltsev identities are an example of a set of linear identities. 
An example of a non-linear identity is the associativity law
$$ f(x,f(y,z)) \approx f(f(x,y),z).$$

If $\fA$ is a $\tau$-algebra, then 
we write $\Minion(\fA)$ for the smallest function minion 
that contains $\{f^{\fA} \mid f \in \tau\}$. 
If $\fA$ and $\fB$ are $\tau$-algebras 
then 
there exists a minion homomorphism $\xi \colon \Minion(\fB) \to \Minion(\fA)$ 
that maps $f^{\fB}$ to $f^{\fA}$ 
if and only if for all 
$f,g \in \tau$ of arity $k$ and $l$ and all $m$-ary projections $p_1,\dots,p_k,q_1,\dots,q_l$ we have that 
$f^\fA(p_1,\dots,p_k) = g^\fA(q_1,\dots,q_l)$ whenever $f^\fB(p_1,\dots,p_k)= g^\fB(q_1,\dots,q_l)$. 
If this map exists it 
must be surjective and we call it the \emph{natural minor-preserving map from
$\Minion(\fB)$ to $\Minion(\fA)$}.   
The following theorem is a variant of Birkhoff's theorem (Theorem~\ref{thm:birkhoff}) 
 for height-one identities. 

 \begin{theorem}[cf.~Proposition 5.3~of~\cite{wonderland}]
 \label{thm:h1-birkhoff}
Let $\fA$ and $\fB$ be $\tau$-algebras
such that $\Minion(\fA)$ and $\Minion(\fB)$ are
operation clones. Then the following are equivalent. 
\begin{enumerate}
\item The natural minor-preserving map from 
$\Minion(\fB)$ to $\Minion(\fA)$ exists. 
\item All height-one identities that hold in $\fB$ also hold in $\fA$. 
\item $\fA \in \RP(\fB)$. 
\end{enumerate}
Moreover, if $A$ and $B$ are finite then we can add the following to the list: 
\begin{enumerate}
\item[4.] $\fA \in \RPfin(\fB)$. 
\end{enumerate}
\end{theorem}

\begin{proof}
The equivalence of $1.$ and $2.$ is straightforward from the definitions, as in the proof of Theorem~\ref{thm:birkhoff}. 

The proof that $2.$ implies $3.$ is similar to the proof of Theorem~\ref{thm:birkhoff}. 
%Let $\fC$ be the algebra $\fB^{B^A}$. 
For every $a \in A$, let $\pi^A_a \in 
%C =  
B^{B^A}$
be the function that maps every tuple in $B^A$ to its $a$-th entry. Let $\fS$ be the 
subalgebra of $\fB^{B^A}$ generated by
$\{\pi^A_a \mid a \in A\}$. 
Define $h \colon S \to A$ as 
$$h(f^{\fB}(\pi^A_{a_1},\dots,\pi^A_{a_n})) :=  f^{\fA}(a_1,\dots,a_n).$$
Similarly as in the proof of Theorem~\ref{thm:birkhoff} one can show that
$h$ is well defined using that all height-one identities that hold in $\fB$ also hold in $\fA$. 
Note that $h$ is defined on all of $S$ 
because $\Minion(\fB)$ is an operation clone. 

Let $g \colon A \to S$ be the mapping which sends every $a \in A$ to $\pi^A_a$. Then $h$ and $g$
show that $\fA \in \R(\fS) \subseteq \R \S \P(\fB) = \RP(\fB)$: for all $a_1,\dots,a_n \in A$ 
\begin{align*}
f^\fA(a_1,\dots,a_n) = h(f^{\fB}(g(a_1),\dots,g(a_n))). 
\end{align*}
If $A$ and $B$ are finite, then $B^A$ is finite
and hence $\fS \in \S \Pfin(\fB)$, so the proof implies that $\fA \in \RPfin(\fB)$. 

$3.$ implies $2$. 
If $\fA \in \P(\fB)$ then the statement follows from Theorem~\ref{thm:birkhoff}. Now 
suppose that $\fA$ is a reflection of $\fB$ via the maps $h \colon B \to A$ and
$g \colon A \to B$. 
%Let $f \in \tau$; then 
%\begin{align*}
%(x_{i_1},\dots,x_{i_n})) = \xi(
Let $\phi$ be the identity $\forall x_1,\dots,x_n \colon f_1(x_{i_1},\dots,x_{i_k})=f_2(x_{j_1},\dots,x_{j_l})$ for $f_1,f_2 \in \tau$  
% $i_1,\dots,i_k,j_1,\dots,j_l \in \{1,\dots,n\}$. 
and suppose that $\fB \models \phi$. 
For all $a_1,\dots,a_n \in A$ we have
\begin{align*}
f^{\fA}_1(a_{i_1},\dots,a_{i_k}) & = h(f^{\fB}_1(g(a_{i_1}),\dots,g(a_{i_k})))  \\
& = h(f^{\fB}_2(g(a_{j_1}),\dots,g(a_{j_l}))) = f^{\fA}_2(a_{j_1},\dots,a_{j_l}) .
\end{align*}
Since $a_1,\dots,a_n$ were chosen arbitrarily, we have that $\fA \models \phi$. 
\end{proof}

\ignore{
\begin{exercises}
\exercise[Rot]\label{exe:source-176} Let $\fA$ be an idempotent algebra with finite signature. Then there exists
a single $f \in \Clo(\fA)^{(n)}$ such that for every operation $g$ of $\fA$ there exists $\alpha \colon [n] \to $.
% Solution sketch: List the finitely many basic operations. Raise their arities to one common arity by
% adding dummy variables, and combine them into a single term operation using the idempotent
% composition available in the proof preceding the exercise. Each original basic operation is then
% obtained as a minor $f_\alpha$. The missing codomain in the statement should read $[m]$, where $m$
% is the arity of the relevant operation.
\end{exercises}
}

%\begin{proposition}\label{prop:cayley-minion}
%Let $\fM$ be a minion. Then there exists a function minion which is isomorphic to $\fM$. 
%\end{proposition}
%\begin{proof}
%\end{proof} 

\begin{exercises}
\exercise[Schwarz]\label{exe:source-177} \label{exe:cayley-minion}
Prove a minion version of Cayley's theorem: show that every minion is isomorphic to a function minion.
% This exercise becomes an easy exercise when
% all the setting is there. But here we don't have the tools for it, this is why the exe is black:
% we are missing on the algebraic side PAIRS of algebras (which then correspond to promise CSP templates)
\exercise[Gelb]\label{exe:source-178} \label{exe:self-red}
Let $\Sigma$ be a finite set of height-one identities.
\\
Assume that there exists an algorithm with the following properties:
\begin{itemize}
\item it takes as input two finite $\tau$-structures $\bA$ and $\bB$;
\item if the algorithm returns `no' then $\bA \not\to \bB$;
\item it runs in polynomial time in the size of $\bA$ and $\bB$;
\item if the polymorphisms of $\bB$ satisfy $\Sigma$, and the algorithm returns `yes', then $\bA \to \bB$.
\end{itemize}
Show that:
\begin{enumerate}
\item if $\Sigma$ expresses the existence of a majority operation, then the
path-consistency procedure $\PC_H$ provides an example for such an algorithm $A$ (viewing the graph $H$ as part of the input of $\PC_H$);
\item if there is such an algorithm $A$ for $\Sigma$, then there exists a polynomial-time algorithm that decides for
a given finite $\tau$-structure $\bB$ whether $\bB$ has polymorphisms that satisfy $\Sigma$.
\end{enumerate}
% Solution sketch: For majority identities, the standard path-consistency correctness proof gives the
% first item uniformly in the input template. For the second, construct the finite canonical test
% structure whose variables are the operation-table entries occurring in $\Sigma$. Run the assumed
% algorithm on this structure and the candidate template; it returns yes exactly when those entries
% can be completed to polymorphisms satisfying $\Sigma$. The construction and the call are polynomial
% in the template size.
\end{exercises}

\subsection{Minion Homomorphisms and Primitive Positive Constructions} 
We have characterised primitive positive constructions in terms of polymorphism algebras and the reflection operator, and then we have characterised varieties that are additionally closed under reflection in terms of minion homomorphisms. In this section we present straightforward combinations of these links that are elegant and convenient for later use. 
The following is analogous to Corollary~\ref{cor:clone-homo-pp} for primitive positive constructions and clone homomorphisms. 

\begin{corollary}\label{cor:pp-minion}
Let $\bB$ and $\bA$ be finite structures. Then 
$\bA \in \HI(\bB)$ 
if and only if there exists a minion homomorphism from $\Pol(\bB)$ to $\Pol(\bA)$. 
\end{corollary}
\begin{proof}
Combine Theorem~\ref{thm:hi-erpfin} with 
Theorem~\ref{thm:h1-birkhoff}. 
%Theorem~\ref{thm:pp-interpret} with 
%Proposition~\ref{prop:clone-homo}. 
\end{proof}

\begin{remark}\label{rem:forward-lin-birk}
By inspection of the proof, one can see that the forward implication remains true even if $\bA$ and $\bB$ have infinite domains. 
\end{remark}

\begin{corollary}\label{cor:minion-hard}
Let $\bB$ be a finite structure. Then there exists a minion homomorphism from $\Pol(\bB)$ to $\Proj$ if and only if 
$K_3 \in \HI(\bB)$. 
%In these cases, $\bB$ has a finite signature reduct whose CSP is NP-hard. 
\end{corollary} 
\begin{proof}
Let $\fB$ be an algebra such that $\Minion(\fB) = \Pol(\bB)$. Corollary~\ref{cor:hi-hard}
states that $K_3 \in \HI(\bB)$ if and only if 
$\R \Pfin(\fB)$ contains an algebra of size at least 2 all of whose operations are projections, and whose clone is therefore $\Proj$. 
Theorem~\ref{thm:h1-birkhoff} implies that this is equivalent to the existence of a minor-preserving map to $\Proj$. 
\end{proof}

%\begin{example}
%Let $\fA$ be a module over a finite ring $\fR$ (Example~\ref{expl:module}). Then $\Clo(\fA)$ 
%has a minion homomorphism to 
%Every module has a minion homomorphism
%\end{example}

%\subsection{The Tractability Conjecture Revisited} 

\begin{figure}
\begin{center}
\begin{tabular}{l|l}
(Abstract) Clone & (Abstract) Minion \\
\hline
Operation Clone & Function Minion \\
\hline
Identity & Height-one Identity \\
\hline
Clone Homomorphism & Minion Homomorphism \\
\hline
$\HSP$ & $\RP$ \\
\hline
$\HSPfin$ & $\RPfin$ \\
\hline
Primitive Positive Interpretation & Primitive Positive Construction \\
\hline
Corollary~\ref{cor:clone-cayley} & 
Exercise~\ref{exe:cayley-minion}
\end{tabular} 
\end{center}
\caption{A dictionary between corresponding notions.}
\end{figure}

\begin{proposition}\label{lem:idempotent-reduct}
For every operation clone $\mathscr C$ on a finite set $C$ there exists an idempotent operation clone $\mathscr D$
on a finite set $D$, with $|D| \leq |C|$, 
such that there exists a minion homomorphism from $\mathscr C$ to $\mathscr D$ and from $\mathscr D$ to $\mathscr C$. 
\end{proposition}
\begin{proof}
Let $\bB$ be a structure such that $\Pol(\bB) = {\mathscr C}$ (such a $\bB$ exists by  Proposition~\ref{prop:pol-inv}). Let $\bC$ be the core of $\bB$ (which exists by the generalisation of Proposition~\ref{prop:core} to relational structures). Let $\bD$ be the expansion of $\bC$ by all unary singleton relations.
%note that $\mathscr D := \Pol(\bD)$ is idempotent. 
We have that
$\bD \in \HI(\bB)$ by Proposition~\ref{prop:wonderland-constants}. 
It follows from Corollary~\ref{cor:pp-minion} that there exists a minion homomorphism from ${\mathscr C} = \Pol(\bB)$ to the idempotent clone ${\mathscr D} := \Pol(\bD)$. Conversely, we have that
${\mathscr D} \subseteq \Pol(\bC)$,
and $\Pol(\bC)$ has a minion homomorphism to
$\Pol(\bB) = {\mathscr C}$ since  $\bB \in \HomE(\bC)$ and again by Corollary~\ref{cor:pp-minion}. 
\end{proof}

\begin{exercises}
\exercise[Blau]\label{exe:source-180} Prove the claim in Remark~\ref{rem:forward-lin-birk}.
% Solution sketch: Use the free algebra on the variables occurring in the finite linear identity
% system. Map each generator to its coordinate column in a suitable finite power. Satisfaction of the
% forward identities makes this map well-defined, and its image is a subalgebra; quotienting by the
% equality of prescribed columns gives exactly the required algebra. This is the forward linear form
% of Birkhoff's argument.
\exercise[Orange]\label{exe:source-179} Show that every finite structure $\bB$ with totally symmetric polymorphisms of all arities can be pp-constructed in $(\{0,1\};\{0,1\}^3 \setminus \{(1,1,0)\},\{0\},\{1\})$.

\medskip
\par\noindent {\bf Hints:} Exercise~\ref{exe:horn}, Lemma~\ref{lem:horn}, Corollary~\ref{cor:pp-minion}, Theorem~\ref{thm:semilattice}, Theorem~\ref{thm:totally-symmetric}.
%\vspace{-0.2cm}
% Solution sketch: By Theorem~\ref{thm:totally-symmetric}, AC solves $\Csp(\bB)$. The
% minion/pp-construction correspondence reduces $\bB$ to the Boolean semilattice template.
% Exercise~\ref{exe:horn} and Lemma~\ref{lem:horn} give a finite primitive-positive presentation of
% all invariant relations needed for the interpretation, proving the claim.
\end{exercises}

%\begin{lemma}\label{lem:idempotent-reduct}
%Let $\fA$ be a finite algebra. 
%Then there exists a finite idempotent algebra
%$\fB$ with $|B| \leq |A|$ that satisfies the same height one identities as $\fA$. 
%\end{lemma}
% Stimmt das wirklich? 

\subsection{Taylor Terms}
\label{sect:taylor}
The following
goes back  to Walter Taylor~\cite{Taylor}.
We slightly deviate from the historical  definition in that we do not require idempotence -- this allows us to give stronger formulations of several results in the following. 

\begin{definition}[Taylor operations]
A function $f \colon B^n \to B$, for $n \geq 2$, 
is called a \emph{Taylor operation}
if for every $i \in [n]$ there are $\alpha,\beta \colon [n] \to [2]$ such that $f_\alpha = f_\beta$ and $\alpha(i) \neq \beta(i)$. 
%for each $1 \leq i \leq n$ there are variables $z_1,\dots,z_n,z'_1,\dots,z'_n \in \{x,y\}$ with $z_i \neq z'_i$ such that  for all $x,y \in B$
%\begin{align*}
%f(z_1,\dots,z_n) = f(z'_1,\dots,z'_n) \, .
%\end{align*}
\end{definition}

Examples for Taylor operations are binary commutative operations, 
majority operations, and Maltsev operations. Since we do not insist on idempotence, also quasi majority operations (Exercise~\ref{exe:qmaj}) are examples of Taylor operations. 

%\end{definition}
A \emph{Taylor term} of a $\tau$-algebra $\fB$ is a $\tau$-term $t(x_1,\dots,x_n)$, for $n \geq 2$, such that $t^{\fB}$ is a Taylor operation.
Note that $t$ is a Taylor term  
if and only if
it satisfies a set of $n$ height-one identities that can be written as 
\begin{align}
t \begin{pmatrix} x & ? & ? & \cdots & ? \\
? & x & ? &  & \vdots \\
\vdots & ? & \ddots & \ddots & \vdots \\
\vdots & & \ddots & x & ? \\
? & \cdots & \cdots & ? & x 
\end{pmatrix}
\approx t \begin{pmatrix}
 y & ? & ? & \cdots & ? \\
? & y & ? &  & \vdots \\
\vdots & ? & \ddots & \ddots & \vdots \\
\vdots & & \ddots & y & ? \\
? & \cdots & \cdots & ? & y \\
\end{pmatrix}
\label{eq:taylor}
\end{align}
where $t$ is applied row-wise and ? stands for either $x$ or $y$. 

Walter Taylor did not just introduce Taylor operations, but he also found a
beautiful statement about their existence (Theorem~\ref{thm:taylor}). In the proof of this statement, we need the following important observation about the star composition in idempotent algebras.

\begin{lemma}\label{lem:combine}
For $n \in {\mathbb N}$, let $(A;f_1,\dots,f_n)$ be an idempotent algebra. 
Then there exists $g \in \Clo(\fA)$ such that 
for every $f \in \{f_1,\dots,f_n\}$ there exists
$\alpha$ such that $g_\alpha = f$. 
\end{lemma}
\begin{proof}
The statement is clear if $n \leq 1$. 
First consider the case that $n=2$. 
Let $m$ be the arity of $f_1$ and let $l$ be the arity of $f_2$. 
Note that
\begin{align}
\Clo(\fA) \models & \; 
\big (f_1 = \comp^{ml}_m(f_1 * f_2,\underbrace{\pi^m_1,\dots,\pi^m_1}_{l \text{ times}},\underbrace{\pi^m_2,\dots,\pi^m_2}_{l \text{ times}},\dots,\underbrace{\pi^m_m,\dots,\pi^m_m}_{l \text{ times}}) \big) \label{eq:expr-y} \\
\text{ and } \Clo(\fA) \models & \; \big(f_2 = \comp^{ml}_l(f_1 * f_2,\underbrace{\pi^l_1,\dots,\pi^l_l}_{m \text{ times}}, \underbrace{\pi^l_1,\dots,\pi^l_l}_{m \text{ times}},\dots,\underbrace{\pi^l_1,\dots,\pi^l_l}_{m \text{ times}})\big) \label{eq:expr-x} 
\end{align}
since $\fA$ is idempotent. The general case can be shown easily by induction on $n$. 
\end{proof}

%The following is a slightly expanded presentation of the proof of Lemma 9.4 in~\cite{HobbyMcKenzie}.

%The following proposition is a special case of Proposition 5.6 in~\cite{Topo} for idempotent clones. 
%\begin{theorem}
%\label{thm:clone-homo-proj}
%Let $\fC$ be a clone and let 
%$\xi \colon \fC \to \Proj$ be a minor-preserving map. 
%Then $\xi$ is a clone
%homomorphism. 
%\end{theorem}
%\begin{proof}

\begin{theorem}
\label{thm:taylor}
Let $\fB$ be an idempotent algebra. Then  
the following are equivalent. 
\begin{enumerate}
\item[(1)] $\bf B$ has a Taylor term $t$.
\item[(2)] there is no minion homomorphism
from $\Clo(\fB)$ to ${\bf Proj}$. 
%\item[(3)] there is no clone homomorphism 
%from $\Clo(\fB)$ to ${\bf Proj}$. 
\end{enumerate}
\end{theorem}

%Before we show the theorem, we first mention
%yet another equivalent characterisation that will be used in the proof. 

\begin{proof}
To show that (1) implies (2), 
suppose for contradiction that there is a minion homomorphism
$\xi$ from $\Clo(\fB)$ to ${\bf Proj}$. 
%Let $f$ be the element
%of $\Clo(\fB)$ that is denoted by $t^{\fB}$. 
By definition of ${\bf Proj}$ we have $\xi(t^{\fB}) = \pi^n_l$ for some $l \leq n$. 
By assumption, there are $\alpha,\beta \colon [n] \to [2]$ such that $(t^{\fB})_\alpha = (t^{\fB})_\beta$ and $\alpha(l) \neq \beta(l)$. 
 %By assumption, $\fB$ satisfies 
%\begin{align}
%t(z_1,\dots,z_n) \approx  t(z'_1,\dots,z'_n)
%\label{eq:algebra-eq}
%\end{align}
%for $z_1,\dots,z_n,z'_1,\dots,z'_n \in \{x,y\}$ such that $z_l \neq z'_l$. Then $\Clo(\fB)$ satisfies
%\begin{align}
%\comp^n_2(f,\pi^2_{i_1},\dots,\pi^2_{i_n}) = \comp^n_2(f,\pi^2_{j_1},\dots,\pi^2_{j_n}) \label{eq:to-clone}
%\end{align}
%for $i_1,\dots,i_n,j_1,\dots,j_n \in \{1,2\}$ such that
%$i_l = 1$ if and only if $z_l = x$, 
%$j_l = 1$ if and only if $z'_l = x$, 
%and $i_l \neq j_l$. 
Since $\xi(t^{\fB}) = \pi^n_l$ and $\xi$ is a minion homomorphism, we therefore obtain 
that $\pi^2_1 = \pi^2_2$, which does not hold in ${\bf Proj}$, a contradiction. 

To show the converse implication, 
suppose that $\fB$ does not have a Taylor term. 
We have to show that $\Clo(\fB)$ has a minion homomorphism to $\Proj$. 
By Lemma~\ref{lem:minion-compactness}, it suffices to show that every primitive positive sentence $\phi$ in the language of minions that holds in $\Clo(\fB)$ also holds in $\Proj$. If $g_1,\dots,g_m$ are the existentially quantified variables in $\phi$, then by Lemma~\ref{lem:combine} there exists $g \in \Clo(\fB)^{(n)}$, for some $n$, such that every $g_i$, $i \in [m]$, is a minor $g_{\alpha_i}$ of $g$. 
Hence, it suffices to define a minion homomorphism from $\Minion((B;g))$ to $\Proj$. 
By assumption, $g$ is not a Taylor term, so there exists an argument $i$ 
such that for all $\alpha,\beta \colon [n] \to [2]$ 
we have $\alpha(i) = \beta(i)$ or $g_\alpha \neq g_\beta$. For $\alpha \colon [n] \to [k]$, define $\xi(g_\alpha) := \pi^k_{\alpha(i)}$. This map is well-defined because if $g_\alpha = g_\beta$ for $\alpha,\beta \colon [n] \to [k]$,
then $(g_{\alpha})_\gamma = (g_{\beta})_\gamma$ for all $\gamma \colon [k] \to [2]$, and hence $\gamma \circ \alpha(i) = \gamma \circ \beta(i)$ for all $\gamma \colon [k] \to [2]$, 
which implies that $\alpha(i) = \beta(i)$. Moreover, $\xi$ is a minion homomorphism because $\xi(g_\alpha) = \pi^k_{\alpha(i)} = (\pi^n_i)_{\alpha} = \xi(g)_{\alpha}$. 
%Note that for every $f \in \Clo(\fB)$ there 
\end{proof}

\begin{remark}
The original statement of Taylor is 
Theorem~\ref{thm:taylor} with \emph{clone homomorphism} instead of minion homomorphism; 
the version with minion homomorphisms in Theorem~\ref{thm:taylor} is the statement we really care about in this course and leads to an easier proof. 
\end{remark} 

The following lemma should be clear from the results that we have already seen. 

\begin{lemma}\label{lem:homo-equiv-taylor}
Let $\bB$ and $\bC$ be homomorphically
equivalent structures.  
Then $\bB$ has a Taylor polymorphism
if and only if 
$\bC$ has a Taylor polymorphism. 
\end{lemma}
\begin{proof}
%This follows from results that we have already established, but since the proof is so short we prefer to state it. 
%But since the proof is so short 
Let $h$ be a homomorphism from $\bB$
to $\bC$, and $g$ be a homomorphism from
$\bC$ to $\bB$. 
Suppose that $f$ is a Taylor polymorphism for $\bB$ of arity $n$.
%, that is,
%$\Pol(\bB) \models \Phi_n(f)$
%for the formula $\Phi_n$ given in $(\ref{eq:taylor})$ above. 
%That is, there are $v,v' \in \{p^2_1,p^2_2\}^{n \times n}$ such that BLA.
%It suffices to show that the operation 
%$f'$ given by 
Then $(x_1,\dots,x_n) \mapsto h(f(g(x_1),\dots,g(x_n)))$ is a Taylor polymorphism of $\bC$. %Indeed, for all $i \leq n$
%we have that
%\begin{align*}
%f'(v_{1,i},\dots,v_{n,i}) = & \; 
%h\big (f(g(v_{1,i}),\dots,g(v_{n,i}))\big ) \\
%= & \; h\big (f(g(v'_{1,i}),\dots,g(v'_{n,i})) \big ) \\
%= & \; f'(v'_{1,i},\dots,v'_{n,i}) \qedhere
%\end{align*}
%This can either be shown directly or derived from the general facts that we have already seen. 
\end{proof}

\begin{corollary}\label{cor:Taylor-minion}
%Let $\fA$ be an algebra with a Taylor term. Then $\Clo(\fA)$ does not have a minion homomorphism to $\Proj$. If $\fA$ has a finite domain, then the converse is true as well. 
Let $\bB$ be a finite structure. Then the following are equivalent. 
\begin{enumerate}
\item $K_3 \notin \HI(\bB)$. 
\item $\bB$ has a Taylor polymorphism. 
\item $\Pol(\bB)$ has no minion homomorphism to $\Proj$. 
\end{enumerate}
If these conditions do not hold then $\bB$ has a finite-signature reduct with an NP-hard $\Csp$. 
\end{corollary}

\begin{proof}
The equivalence of 1.\ and 3.\ is Corollary~\ref{cor:minion-hard}. 
Now suppose that $K_3 \notin \HI(\bB)$. If $\bB'$ is the core of $\bB$, and $\bC$ is the expansion of $\bB'$ by all unary singleton relations,
then $\CC(\HH(\bB)) \subseteq \HI(\bB)$ implies that 
$K_3$ is not pp constructible in $\bC$.
Hence, the idempotent clone $\Pol(\bC)$ does not have a minion homomorphism to $\Proj$. 
Theorem~\ref{thm:taylor} shows that $\bC$ and thus also $\bB'$ must
have a Taylor polymorphism. 
Lemma~\ref{lem:homo-equiv-taylor} implies that $\bB$ has a Taylor polymorphism. 

Note that the existence of Taylor polymorphisms is preserved by minion homomorphisms, and since $\Proj$ does not have a Taylor operation we have that $2.$ implies $3$.

%If $\bB$ has a Taylor polymorphism,
%then so has any structure in 
%$\PP(\bB)$, and so has any structure
%in $\HI(\bB)$ by Lemma~\ref{lem:homo-equiv-taylor}. But all polymorphisms of $K_3$ are essentially unary (Proposition~\ref{prop:kn-is-projective}), so $K_3 \notin \HI(\bB)$. 

The final statement follows from Corollary~\ref{cor:pp-interpret-hard}.
\end{proof}

%\begin{corollary}\label{cor:taylor-or-npc}
%Let $\bB$ be a finite structure. Then $\bB$ has a Taylor polymorphism, or $\Csp(\bB)$ is NP-hard. 
%\end{corollary}

\begin{theorem}[Tractability Theorem, Version 3]
\label{thm:tractability-3}
Let $\bB$ be a relational structure with finite domain and finite signature. 
If $\bB$ has a Taylor polymorphism, then $\Csp(\bB)$ is in P. Otherwise, 
$\Csp(\bB)$ is NP-complete.
\end{theorem}

\begin{proof}
An immediate consequence of Corollary~\ref{cor:Taylor-minion} and Theorem~\ref{thm:tractability-2}. 
%If $\bB$ does not have a Taylor polymorphism, then $K_3 \in \HI(\bB)$ by Corollary~\ref{cor:hpp},
%and $\Csp(\bB)$ is NP-hard by 
%Corollary~\ref{cor:pp-interpret-hard}. 
%Otherwise,  
%$\bB$ has a Taylor polymorphism by Corollary~\ref{cor:hpp}. 
\end{proof}

A clone is said to be \emph{Taylor} if it has a Taylor operation, and an algebra is called \emph{Taylor} 
if it has a Taylor term operation. 

\begin{remark}\label{rem:Taylor}
We will from now on often use the formulation `let $\fA$ be a finite Taylor algebra' instead of `let $\fA$ be a finite algebra such that $\Clo(\fA)$ does not have a minion homomorphism to $\Proj$', even if we can avoid in the proofs the use of Taylor operations altogether. The reason is that the assumption is shorter to state, and equivalent by the results of this section (see Exercise~\ref{exe:taylor-idempotent}). 
\end{remark}

\begin{exercises}
\exercise[Rot]\label{exe:source-181} \label{exe:taylor-idempotent}
Show that the assumption in Theorem~\ref{thm:taylor} that $\fB$ is idempotent can be replaced
by the assumption that its domain is finite.
% Solution sketch: For a finite template, take an idempotent power of the diagonal unary map of every
% polymorphism and pass to a core image. On that image the diagonal maps are automorphisms; composing
% with their inverses idempotentises the polymorphisms without changing the relevant height-one
% identities. Apply the idempotent Taylor theorem there and transport the resulting polymorphism back
% along the homomorphic equivalence.
\exercise[Rot]\label{exe:source-182} Show that a finite structure without a Taylor polymorphism
pp-constructs all finite structures.
% From Corollary {cor:Taylor-minion}
% we get that K_3 is in HI(B).
% and hence all finite structures by
%{cor:hi-hard}
\label{exe:pp-construct-all-finite}
% Solution sketch: By the Taylor-minion corollary, absence of a Taylor polymorphism gives a minion
% homomorphism to the projection clone, equivalently a pp-construction of $K_3$. The hard-template
% corollary says that $K_3$ pp-constructs every finite structure. Transitivity of pp-constructions
% proves the assertion.
\exercise[Schwarz]\label{exe:source-183} \label{exe:HAaffineTaylor}
Show that if $\fA$ is a finite idempotent algebra such that $\HS(\fA)$
does not
contain an abelian algebra with at least two elements, then $\fA$ has a Taylor term.

\par\noindent {\bf Hint.} 
%Use Exercise~\ref{exe:affineHS}.
Theorem~\ref{thm:Zhuk-HS} and Corollary~\ref{cor:Taylor-minion}. 
% Hints:
%This follows by combining theorems that we have already shown: if $\fA$ has no Taylor term,
%then the structure with relations $\Inv(\Clo(\fA))$
%pp-constructs all finite structures, and in particular $({\mathbb Z}_2;+,1)$ by TODO Since $\fA$ is idempotent, we even have a pp interpretation of
%$\Pol({\mathbb Z}_2;+,1)$ by TODO.
% If B is the polymorphism algebra of ({\mathbb Z}_2;+,1), we find B in HSP(A) by TODO.
% Then use Exercise~\ref{exe:HSaffine}
% to find is in HS.

% Solution sketch: Assume that $\fA$ has no Taylor term. The invariant relational structure of $\fA$
% then pp-interprets the affine two-element algebra. Birkhoff's theorem places its polymorphism
% algebra in $\HSP(\fA)$; the affine-factor lemma moves a nontrivial affine member into $\HS(\fA)$,
% contradicting the hypothesis. Hence $\fA$ has a Taylor term.
\exercise[Schwarz]\label{exe:source-184} \label{exe:cube} A \emph{cube operation} is an operation $c$ of arity $n \geq 3$ which satisfies for every $i \in \{1,\dots,n\}$ an identity of the form $c(z_1,\dots,z_n) \approx y$ where $z_1,\dots,z_n \in \{x,y\}$
such that $z_i = x$.
%Motivation for this terminology will be provided in  Exercise~\ref{exe:cube-blocker}.
\begin{itemize}
\item Present the definition of cube operations in a similar way as Taylor operations are presented in~\eqref{eq:taylor}.
\item Give two different ternary cube  operations on a two-element set. 
\item Show that every operation $c$ that satisfies
\begin{align*}
c(x,x,x,x,y,y,y) & \approx y \\
c(x,x,y,y,x,x,y) & \approx y \\
c(x,y,x,y,x,y,x) & \approx y
\end{align*}
is a cube operation.
\item Show that the clone generated by a cube operation contains an operation $f$
of arity $m := 2^k-1$, for some $k \geq 2$,
satisfying $
f(z_{1,j},\dots,z_{m,j}) \approx y
 $
 for every $j \in \{1,\dots,k\}$, where $z_{1,1},\dots,z_{m,k} \in \{x,y\}$ are such that
$(z_{1,1},\dots,z_{1,k}),\dots,(z_{m,1},\dots,z_{m,k})$ enumerates $\{x,y\}^k \setminus \{(y,\dots,y)\}$.
% Proof: take minimal subset of the cube term equations satisfied by the cube term,
% identify arguments of operation if columns repeat; let k be the resulting arity. Add dummy arguments to fill up to right arity 2^k-1: this will be f.
\end{itemize}
\end{exercises}

%\begin{proof}
%The identities that define Taylor terms have height one, and hence are preserved by minion homomorphisms. $\Proj$ does not satisfy these identities. TODO
%$\end{proof}

\subsection{Arc-consistency Revisited} 
\label{sect:ACrevisited}
In this section we revisit the arc-consistency procedure from Section~\ref{sect:AC}, generalised to arbitrary relational structures with finite domain and finite signature, in the light of minions and primitive positive constructions. 

\begin{definition}\label{def:mac}
The minion ${\bf M}_{\text{AC}}$ is defined as follows. 
For $n \geq 1$, the set $M_{\text{AC}}^{(n)}$ consists of the set of all non-empty subsets of $\{1,\dots,n\}$.
 For $\alpha \colon [n] \to [m]$ and $f \in M_{\text{AC}}^{(n)}$, we define
$f_\alpha$ to be $\big \{ \alpha(a) \mid a \in f \}$. 
% define $_\alpha \colon H^{(n)}
%to H^{(m)}$ by setting for 
\end{definition}

\begin{remark}\label{rem:mac-horn}
${\bf M}_{\text{AC}}$ is isomorphic to 
$\Pol(\{0,1\}; \{0\},\{1\},\{0,1\}^3 \setminus \{(1,1,0)\})$ (see Exercise~\ref{exe:Horn}). 
\end{remark}

\begin{definition}[$\AC_{\bB}$]\label{def:ACB}
For a structure $\bB$ with finite domain and finite relational signature $\tau$, 
we write $\AC_{\bB}$ for the generalisation 
of the arc-consistency procedure $\AC_H$ from digraphs $H$ to general relational structures $\bB$ (Exercise~\ref{exe:gac}).
\end{definition}

We also generalise the concept of \emph{arc-consistency} for digraphs from Section~\ref{sect:AC}.

\begin{definition}[Arc-consistency]
\label{def:ac}
Let $\bB$ be a structure with the relational signature $\tau$.  
A finite $\tau$-structure $\bA$ together with a set $L(a) \subseteq B$ for every $a \in A$ is called \emph{arc-consistent (with respect to $\bB$)} if for every 
$R \in \tau$ of arity $k$, $(a_1,\dots,a_k) \in R^{\bA}$, $i \in [k]$, $c \in L(a_i)$
there exists $(b_1,\dots,b_k) \in R^{\bB}$
such that $b_i = c$ and $b_j \in L(a_j)$ for all $j \in [k]$. 
\end{definition} 

Note that for any two finite $\tau$-structures $\bA$ and $\bB$, if $L(a) \subseteq B$ is the list computed for $a \in A$ by $\AC_{\bB}$ at the final stage of the algorithm, 
then $\bA$ together with these sets $L(a)$ is arc-consistent. 

\begin{theorem}\label{thm:ac-rev}
Let $\bB$ be a finite structure. Then the following are equivalent. 
\begin{enumerate}
\item $\Csp(\bB)$ is solved by $\AC_{\bB}$.
\item $P(\bB) \to \bB$. 
\item $\Pol(\bB)$ has totally symmetric polymorphisms of all arities.   
\item ${\bf M}_{\AC}$ has a minion homomorphism to $\Pol(\bB)$. 
\item $\bB$ has a primitive positive construction in $(\{0,1\};\{0\},\{1\},\{0,1\}^3 \setminus \{(1,1,0)\})$. 
\end{enumerate} 
\end{theorem}
\begin{proof}
$1 \Leftrightarrow 2$ was already shown for digraphs in Theorem~\ref{thm:totally-symmetric} and generalised to general structures in Exercise~\ref{exe:gac}. 

$2 \Leftrightarrow 3$ was already shown for digraphs in Theorem~\ref{thm:totally-symmetric} and generalised to general structures in Exercise~\ref{exe:tsn}. 

\ignore{2 -> 3
% Incorrect proof: 
 let $s_k \in \Pol(\bB)$ 
 be a totally symmetric operation of arity $k$. 
Then the map which sends for every $n \in {\mathbb N}$ and $k \leq n$ the element 
$\{i_1,\dots,i_k\} \in M^{(n)}_{\text{AC}}$ to the operation
$(x_1,\dots,x_n) \mapsto s_{k}(x_{i_1},\dots,x_{i_k})$
is a minion homomorphism $\xi \colon {\bf M}_{\text{AC}} \to \Pol(\bB)$: 
if $\alpha \colon [n] \to [m]$ for some $n,m \in {\mathbb N}$, then 
\begin{align*}
\xi(\{i_1,\dots,i_k\}_\alpha) & = \xi(\{\alpha(i_1),\dots,\alpha(i_k)\}) \\
& =  ((x_1,\dots,x_m) \mapsto s_{k}(x_{\alpha(i_1)},\dots,x_{\alpha(i_k)}) \\
% This is the incorrect step for variable identification. 
& = ((x_1,\dots,x_n) \mapsto s_{k}(x_{i_1},\dots,x_{i_k})_{\alpha} \\
& = \xi(\{i_1,\dots,i_k\})_{\alpha}. 
\end{align*}}

$2 \Rightarrow 4$. Suppose that
there exists a homomorphism
$h\colon P(\bB)\to\bB$ 
from the powerset structure of $\bB$ to $\bB$. For every $n\geq 1$ and every nonempty set $S\subseteq[n]$, define
the $n$-ary operation
\[
\xi(S)(x_1,\dots,x_n)
:=
h\bigl(\{x_i\mid i\in S\}\bigr).
\]
We first verify that $\xi(S)$ is a polymorphism of $\bB$. Let $R$ be
a $k$-ary relation symbol and let
\[
\bar b^r=(b^r_1,\dots,b^r_k)\in R^{\bB}
\qquad(r\in[n]).
\]
For every $j\in[k]$, put
\[
U_j:=\{b^r_j\mid r\in S\}.
\]
Then $(U_1,\dots,U_k)\in R^{P(\bB)}$. Indeed, for every $j\in[k]$
and every $c\in U_j$, there exists an $r\in S$ such that
$c=b^r_j$, and the tuple $\bar b^r\in R^{\bB}$ witnesses that $c$
has a support whose $\ell$-th coordinate belongs to $U_\ell$ for
every $\ell\in[k]$. Since $h$ is a homomorphism, it follows that
\[
\bigl(h(U_1),\dots,h(U_k)\bigr)\in R^{\bB}.
\]
This tuple is obtained by applying $\xi(S)$ coordinatewise to
$\bar b^1,\dots,\bar b^n$. Hence $\xi(S)\in\Pol(\bB)$.

It remains to verify preservation of minors. Let
$\alpha\colon[n]\to[m]$. Since
\[
S_\alpha=\{\alpha(i)\mid i\in S\},
\]
we have, for all $x_1,\dots,x_m\in B$,
\begin{align*}
\xi(S_\alpha)(x_1,\dots,x_m)
&=
h\bigl(\{x_j\mid j\in S_\alpha\}\bigr)\\
&=
h\bigl(\{x_{\alpha(i)}\mid i\in S\}\bigr)\\
&=
\xi(S)(x_{\alpha(1)},\dots,x_{\alpha(n)})\\
&=
\xi(S)_\alpha(x_1,\dots,x_m).
\end{align*}
Therefore $\xi\colon{\bf M}_{\mathrm{AC}}\to\Pol(\bB)$ is a minion
homomorphism. This proves $2\Rightarrow 4$.

$4 \Rightarrow 3$: let $\xi$ be the minion homomorphism from ${\bf M}_{\text{AC}}$ to $\Pol(\bB)$. Then the operation $s_n := \xi(\{1,\dots,n\}) \in \Pol(\bB)$ is totally symmetric and of arity $n$. Indeed, suppose that $a_1,\dots,a_n,b_1,\dots,b_n \in B$ 
are such that $\{a_1,\dots,a_n\} = \{b_1,\dots,b_n\} = \{c_1,\dots,c_m\}$
for some $m \leq n$ 
such that $c_1,\dots,c_m$ are pairwise distinct. 
%We have to show that $s_n(a_1,\dots,a_n) = s_n(b_1,\dots,b_n)$. 
Then there are $\alpha_1,\alpha_2 \colon [n] \to [m]$ such that $a_i=c_{\alpha_1(i)}$ and $b_i=c_{\alpha_2(i)}$ for every $i \in [n]$. Since every $c_j$ occurs in both tuples, $[n]_{\alpha_1}=[m]=[n]_{\alpha_2}$ in ${\bf M}_{\mathrm{AC}}$. Hence, using $s_n=\xi([n])$ and preservation of minors,
$$s_n(a_1,\dots,a_n)=(s_n)_{\alpha_1}(c_1,\dots,c_m)=\xi([m])(c_1,\dots,c_m)=(s_n)_{\alpha_2}(c_1,\dots,c_m)=s_n(b_1,\dots,b_n).$$ 
%Define $\alpha \colon [2n] \to [m]$ by
%setting $\alpha(i) := \alpha_1(i)$ if $i \leq n$
%and  setting $\alpha(i) := \alpha_2(i)$ if $i \geq n+1$. Note that in ${\bf M}_{\text{AC}}$ we have $[n]_{\alpha_1} = [m] = [2n]_{\alpha} = [n]_{\alpha_2}$. 
% then there exists $\alpha \colon [2n] \to [m]$ for some $m \leq n$ such that $a_i = a_{\alpha(i)} = b_{\alpha(n+i)} = b_i$ for every $i \in \{1,\dots,n\}$. 
%Hence, 
%\begin{align*}
%s_n(a_1,\dots,a_n)  = \xi([n])(a_{1},\dots,a_{n}) 
%&  = \xi([n])_{\alpha_1}(c_1,\dots,c_m) \\
%& = \xi([2n])_{\alpha}(c_1,\dots,c_m) \\
%& = \xi([n])_{\alpha_2}(c_1,\dots,c_m)  = \xi([n])(b_1,\dots,b_n) = s_n(b_1,\dots,b_n). 
%\end{align*} 
%$\xi(\{a_1,\dots,a_n\}) = \xi(\{b_1,\dots,b_n\}) = $.
 
$4 \Leftrightarrow 5$: Corollary~\ref{cor:pp-minion} and Remark~\ref{rem:mac-horn}.  
\end{proof} 

Later, Theorem~\ref{thm:ac-rev} will be the blueprint for characterisations of various other algorithms. To prove these characterisations, we need to generalise 
$P(\bB)$ 
%one important concept that was relevant in the proof of Theorem~\ref{thm:ac-rev} (since it uses Theorem~\ref{thm:totally-symmetric}): the concept of $P(H)$.
%Obviously, this concept can be generalised to relational structure $\bB$ instead of $H$.
%Thus we get that $P(\bB)$ has a homomorphism to $\bB$ if and only if ${\bf M}_{\AC}$ has a homomorphism to $\Pol(\bB)$ (combining Theorem~\ref{thm:ac-rev} with Theorem~\ref{thm:totally-symmetric}). 
%Less obviously, this can also be generalised
from ${\bf M}_{\AC}$ to arbitrary minions ${\bf M}$, and this is the content of the next section. 
%For the minion ${\bf M}_{\AC}$, the definition below specialises to $P(\bB)$. 

\begin{exercises}
\exercise[Rot]\label{exe:source-185} \label{exe:ac-pp}
Let $\bA$ and $\bB$ be finite structures with finite relational signature $\tau$ and let $\fB$ be an algebra such that
$\Clo(\fB) = \Pol(\bB)$.
Show that if
$L(a)$ is the list computed by $\AC_{\bB}$ on input $\bA$ for $a \in A$, then $L(a) \leq \fB$.
% Solution sketch: Initially each list is a subuniverse because it is the full universe. If a value
% has supporting tuples for several list elements, applying any term operation of $\fB$ coordinatewise
% to those supports gives a support for the term value. Therefore every deletion step leaves a
% subuniverse, and so does the fixed-point list $L(a)$.
\end{exercises}

\subsection{The Free Structure} 
The definition of free structures first appeared in `the wonderland paper'~\cite{wonderland}, and also plays a fundamental role for the theory of promise CSPs~\cite{BartoBKO21};
 we essentially follow the presentation there (Definition 4.1). However, unlike there, we use the following notational convention: we allow arbitrary finite sets $A$ as index sets for tuples, rather than only sets of the form $[n] = \{1,\dots,n\}$. 
Since $n$-tuples from a set $A$ can be formalised as functions from $[n]$ to $A$,
this is just a small stretch of the usual notation. We already used the notation $B^A$ for the set of all functions from $A$ to $B$, i.e., for a power of $B$ where the factors are indexed by $A$. 
Likewise, $B^A \to C$ are functions from $B^A$ to $C$ where the arguments are indexed by $A$.
%, and $B^A \to B$ are  operations on $B$ (functions from $B^A \to B$) where the arguments are indexed by $A$. 
This notation extends 
\begin{itemize}
\item to minors, which are now specified by maps $\alpha \colon A \to B$ between arbitrary finite sets $A$ and $B$ (rather than just maps of the form $\alpha \colon [k] \to [m]$),
which means in particular that we may use tuples $d \in C^r$ to specify a minor $f_d \colon B^C \to B$ of an operation $f \colon B^r \to B$, since $d$ is a function from $[r]$ to $C$, and 
\item to 
minions, where we are now allowed to write $M^{(A)}$ for the elements of $M^{(|A|)}$
(in other words, we now treat minions as functors from {\bf FinSet} to {\bf Set}, rather than functors from {\bf FinOrd} to {\bf Set}; see Exercise~\ref{exe:functor}). 
\end{itemize}
Note that if $f \colon B^r \to B$, $\alpha \colon [r] \to [s]$, $d \in B^s$, then 
\begin{align}
f_\alpha(d) & = f(d \circ \alpha). \label{eq:minor}
\end{align}

\begin{definition}[the free structure]
\label{def:free-struct} 
Let ${\bf M}$ be a minion and $\bB$ be a structure with relational signature $\tau$. Then $\bF_{\bf M}(\bB)$, the \emph{free structure of ${\bf M}$ generated by $\bB$}, is the $\tau$-structure with the domain $M^{(B)}$. For $R \in \tau$ of arity $k$, the relation
 $R^{\bF_{\bf M}(\bB)}$ is defined as follows. 
 %Let $m := |R^{\bB}|$.
 % and $R^{\bB} = \{r_1,\dots,r_m\}$. 
Below we will use $\pi_i$, for $i \in [k]$, as a shortcut for the restriction of $\pi^k_i$ to 
$R^{\bB}$, which is a function from $R^{\bB}$ to $B$, and hence can be used to form minors of functions in $M^{(R^{\bB})}$. 
If $R^{\bB}=\emptyset$, define $R^{\bF_{\bf M}(\bB)}:=\emptyset$. Otherwise, the relation $R^{\bF_{\bf M}(\bB)}$
 consists of the set of all $k$-tuples $(f_1,\dots,f_k) \in (M^{(B)})^k$ 
 such that there exists $g \in M^{(R^{\bB})}$ satisfying $g_{\pi_i} = f_i$ for every $i \in [k]$.
 %, where 
 %$\pi_i \colon R^{\bB} \to B$ is 
 \end{definition} 
% IMPOSSIBLE TO TEACH!!! 24.9.25

For readers who need some intuition for Definition~\ref{def:free-struct}, but 
are already comfortable with the material in Section~\ref{sect:ts}, we strongly recommend doing Exercise~\ref{exe:power-set-struct-free}. 

\begin{proposition}\label{prop:free-struct} 
Let $\bB$ be a structure with a finite domain and a relational signature $\tau$. Let $\fM$ be a minion. 
Then the following are equivalent. 
\begin{itemize}
\item ${\bF}_{\fM}(\bB) \to \bB$.
\item $\fM$ has a homomorphism to $\Pol(\bB)$. 
\end{itemize} 
\end{proposition}
\begin{proof}
Let 
$B = \{b_1,\dots,b_n\}$. First suppose that 
$h \colon {\bF}_{\bf M}(\bB) \to \bB$ is a homomorphism. 
To define a minion homomorphism $\mu \colon \fM \to \Pol(\bB)$, let $f \in M^{(r)}$ for some $r \in {\mathbb N}$. 
For $c \in B^r$, 
%let $\alpha \colon [r] \to B$ be defined
%by $\alpha(j) := c_j$, and 
define 
$$\mu(f)(c) := h(f_c).$$
To see that $\mu$ is a minion homomorphism, suppose that 
$\beta \colon [r] \to [s]$ is a function; we need to verify that $\mu(f_\beta) = \mu(f)_\beta$. 
Let $d \in B^s$. 
%, and 
%let $\gamma \colon [s] \to B$ be defined by $\gamma(j) := d_j$. 
%Let $\delta \colon [r] \to B$ be defined by
%$\delta(j) := d_{\beta(j)}$. 
Then 
\begin{align*}
\mu(f_\beta)(d) 
& = h((f_\beta)_d) && \text{(Definition of $\mu$)} \\
& = h(f_{d \circ \beta}) && \text{($\fM$ is a minion)}\\
%& = h(f_{\alpha})(b_1,\dots,b_n) \\
& = \mu(f)(d \circ \beta) && \text{(Definition of $\mu$)} \\ 
& = \mu(f)_\beta(d) && \eqref{eq:minor}. 
\end{align*}
We also have to verify that $\mu(f) \in \Pol(\bB)$, i.e., that $\mu(f)$ preserves $R^{\bB}$ for every $R \in \tau$.
Let $t^1,\dots,t^r \in R^{\bB}$; we need to show that $\mu(f)(t^1,\dots,t^r) \in R^{\bB}$. 
Let $k$ be the arity of $R$. 
For every $i \in [k]$ let $t_i := (t^1_i,\dots,t^r_i)$. 
%Then 
%$\mu(f)(t_i) = h(f_{t_i})$.
Let $t$ be the function from $[r]$ to $R^{\bB}$ 
%to $[k]$ 
that maps $j \in [r]$ to $t^j$, 
and let 
$g := f_t$. 
By the definition of $\bF_{\fM}(\bB)$,  
we have $(f_{t_1},\dots,f_{t_k}) \in R^{\bF_{\fM}(\bB)}$, since $g_{\pi_i} = (f_t)_{\pi_i} = f_{\pi_i \circ t} = f_{t_i}$ for every $i \in [k]$. 
Then 
\begin{align*}
\mu(f)(t^1,\dots,t^r) & = (\mu(f)(t_1),\dots,\mu(f)(t_k)) && \text{(by the definition of $t_1,\dots,t_k$)} \\
& = (h(f_{t_1}),\dots,h(f_{t_k})) && \text{(by the definition of $\mu(f)$)} \\
& \in R^{\bB}
&& \text{(since $h$ is a homomorphism).}
\end{align*} 

For the converse direction, 
suppose that $\mu \colon {\bf M} \to \Pol(\bB)$ is a (minion) homomorphism. 
To define the homomorphism $h \colon {\bF}_{\bf M}(\bB) \to \bB$, 
let $f \in M^{(B)}$ be an element of 
${\bF}_{\bf M}(\bB)$, and define $h(f) := \mu(f)(b_1,\dots,b_n)$. 
To see that this is a homomorphism, let 
$R \in \tau$ have arity $k$.
If $R^{\bB}=\emptyset$, there is nothing to prove. Otherwise, write
$R^{\bB}=\{r_1,\dots,r_m\}$, 
for $r_j=(r_{j,1},\dots,r_{j,k})$. 
Suppose that $(f_1,\dots,f_k) \in R^{{\bF}_{\bf M}(\bB)}$. By the definition of
the free structure, there exists $g \in M^{(R^{\bB})}$ such that $g_{\pi_i} = f_i$ for every $i \in [k]$. 
Then
\begin{align*} (h(f_1),\dots,h(f_k)) & = (\mu(f_1)(b_1,\dots,b_n),\dots,\mu(f_k)(b_1,\dots,b_n)) && \text{(by definition of $h$)} \\
& = (\mu(g_{\pi_1})(b_1,\dots,b_n),\dots,\mu(g_{\pi_k})(b_1,\dots,b_n)) &&  \text{(by assumption)} \\
&=
\bigl(
 \mu(g)(r_{1,1},\dots,r_{m,1}),\dots,
 \mu(g)(r_{1,k},\dots,r_{m,k})
\bigr)
&& \text{($\mu$ preserves minors)} 
\\
& = \mu(g)(r_1,\dots,r_m)  
%&& \text{(hint: read bottom-up)}
%(see footnote\footnote{Read this equality from bottom to top.})} 
 \\
& \in R^{\bB} && \text{since $\mu(g) \in \Pol(\bB)$,}
\end{align*}  
which shows that $h$ is a homomorphism. 
\end{proof} 

\begin{exercises}
\exercise[Blau]\label{exe:source-186} Verify that ${\bf M}_{\text{AC}}$ (Definition~\ref{def:mac}) is indeed a minion.
% Solution sketch: The prescribed sets are nonempty in every arity, and taking a minor merely
% identifies, permutes, or introduces coordinates in the same family. Identity minors and composition
% of coordinate maps hold by direct substitution. Hence all minion axioms are satisfied.
\exercise[Blau]\label{exe:source-187} Let $H$ be a digraph. Prove that $\bF_{{\bf M}_{\AC}}(H)$  equals $P(H)$.
\label{exe:power-set-struct-free}
% Solution sketch: An element of the free structure is a nonempty subset of $V(H)$. Unfolding the
% definition of a relation in $\bF_{{\bf M}_{\AC}}(H)$ says exactly that every element on either side
% has a compatible neighbour on the other side. Those are precisely the two support conditions
% defining the edge relation of $P(H)$.
\end{exercises}

\section{Undirected Graphs}
\label{sect:HN}
This section contains a proof of the dichotomy for finite undirected graphs of Hell and Ne\v{s}et\v{r}il, Theorem~\ref{thm:HN}. We prove something stronger, 
namely that
the tractability theorem (Theorem~\ref{thm:tractability-2}) is true
for finite undirected graphs $\bB$~\cite{BulatovHColoring}. 
More specifically, the following is true.

\begin{theorem}\label{thm:full-HN}
Let $\bB$ be a finite undirected graph. 
Then either 
\begin{itemize}
\item $\bB$ is \emph{bipartite} (i.e., homomorphic to $K_2$) or has a loop, 
%and hence has a quasi majority polymorphism (cf.~Exercise~\ref{exe:qmaj}) 
or 
\item $\HI(\bB)$ contains all finite structures. 
%every finite structure is homomorphically
%equivalent to a structure with a primitive positive interpretation in $\bB$. 
\end{itemize}
\end{theorem}
Note that in combination with 
Corollary~\ref{cor:pp-interpret-hard}, 
this theorem implies the tractability theorem (Theorem~\ref{thm:tractability-2}) for the special case of finite undirected graphs. 
This theorem also has a remarkable 
consequence in universal algebra, 
whose significance goes beyond the study 
of the complexity of CSPs, and 
%independent from the study of CSP, 
which provides a strengthening of
Taylor's theorem (Theorem~\ref{thm:taylor}), discovered by Siggers in 2010 (see Section~\ref{sect:6ary}). 

\subsection{The Hell-Ne\v{s}et\v{r}il Theorem}
The graph $K_4-\{0,1\}$ (a clique with four vertices where one  edge is missing) is called a \emph{diamond}. 
A graph is called \emph{diamond-free}
if it does not contain a copy of a diamond as a (not necessarily induced) subgraph. For every $\ell \in {\mathbb N}$, the graph $(K_3)^{\ell}$ is an example of a diamond-free graph. 

\begin{lemma}\label{lem:diamond}
Let $\bB$ be a finite undirected loopless graph
which is not bipartite. 
Then $\bB$ pp-constructs a diamond-free core containing a triangle.
\end{lemma} 
\begin{proof}
We may assume that 
\begin{enumerate}
\item $\HI(\bB)$ 
does not contain 
a non-bipartite loopless graph with fewer vertices than $\bB$, since otherwise
we could replace $\bB$ by this graph. In particular, $\bB$ must then be a core. 
\item $\bB = (V;E)$
contains a triangle: if the length of the shortest odd cycle in $\bB$ is $k$, then $(B;E^{k-2})$
%, 
%where $E^{k-2}$ is defined via the usual composition of binary relations and so primitive positive definable in $\bB$, 
is a graph and 
contains
a triangle, so it can replace $\bB$. 

%\item Every edge of $\bB$ is contained in a triangle: otherwise, the primitive positive formula 
%$$\exists z \big(E(x,y) \wedge E(x,z) \wedge E(y,z) \big)$$ 
%defines a new undirected binary relation in $\bB$ that still contains a triangle and does not contain loops, so we may replace $E$ by this relation. 
\end{enumerate}

{\bf Claim 1.} Every vertex
of $\bB$ is contained in a triangle: 
Otherwise, we can replace $\bB$
by the subgraph of $\bB$ induced 
by a set defined by
the primitive positive formula $$\exists u,v \, \big (E(x,u) \wedge E(x,v) \wedge E(u,v) \big)$$
which still contains a triangle, contradicting our first assumption. 

{\bf Claim 2.} $\bB$ does not contain
a copy of $K_4$. Otherwise, if $a$ is an element from a copy of $K_4$, then 
the subgraph of $\bB$ induced by
the set defined by the primitive positive formula $E(a,x)$ 
is a non-bipartite 
graph $\bA$, which has strictly fewer vertices than $\bB$ because $a \notin A$. Moreover, Theorem~\ref{thm:wonderland} implies that  expansions of cores by constants can be pp-constructed, and hence that $\bB$ pp-constructs
$\bA$, contrary to our initial assumption. 

{\bf Claim 3.} The graph $\bB$ must also be diamond-free. 
To see this, let $R$ be the binary relation with the
primitive positive definition 
$$R(x,y) : \Leftrightarrow \exists u,v \, \big(
E(x,u) \wedge E(x,v) \wedge E(u,v) \wedge E(u,y) \wedge E(v,y) \big)$$
and let $T$ be the transitive closure of $R$.
The relation $T$ is clearly symmetric, and 
since every vertex of $\bB$ is contained
in a triangle, it is also reflexive, and hence
an equivalence relation of $\bB$. 
Since $B$ is finite, for some $n$
the formula 
$\exists u_1,\dots,u_n \, \big(R(x,u_1) \wedge 
R(u_1,u_2) \wedge \cdots \wedge 
R(u_n,y) \big)$ defines $T$, showing that 
$T$ is 
primitively positively definable in $\bB$. 

\begin{figure}
\begin{center}
\includegraphics[scale=0.6]{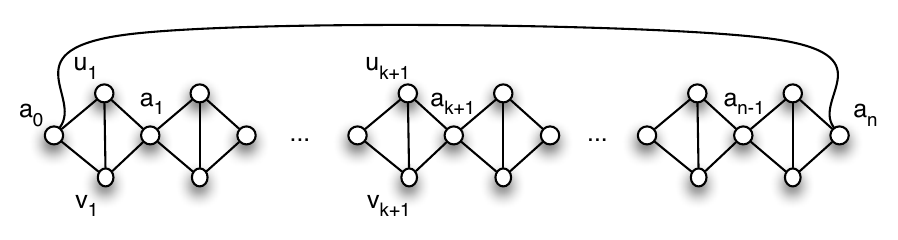} 
\end{center}
\caption{Diagram for the proof of Lemma~\ref{lem:diamond}.}
\label{fig:bul-chain}
\end{figure}

We claim that the graph $\bB/T$ (see Example~\ref{expl:fact}) does not contain loops. It suffices to show that $T \cap E = \emptyset$. Otherwise, let $(a,b) \in T \cap E$. Choose 
$(a,b)$ in such a way that
the shortest sequence $a=a_0,a_1,\dots,a_n=b$ with $R(a_0,a_1)$, $R(a_1,a_2)$, \dots, $R(a_{n-1},a_n)$ in $\bB$ is shortest possible;
see Figure~\ref{fig:bul-chain}. 
This chain cannot have 
%the form $R(a_0,a_1)$ 
length one, because $\bB$ does not contain $K_4$ subgraphs. 
For every $i\in\{1,\dots,n\}$, choose vertices $u_i,v_i$
witnessing $R(a_{i-1},a_i)$; thus
\[
E(a_{i-1},u_i),\ E(a_{i-1},v_i),\ E(u_i,v_i),\
E(u_i,a_i),\ E(v_i,a_i)
\]
hold. For $m\geq 0$, let $R^m$ denote the $m$-fold relational
composition of $R$, where $R^0$ is the equality relation.

Suppose first that $n=2k$ is even. Define $S$ by the primitive
positive formula
\[
S(x)\ \Leftrightarrow\
\exists y,z\,
\bigl(
E(u_{k+1},y)\wedge E(v_{k+1},y)
\wedge R^{k-1}(y,z)\wedge E(z,x)
\bigr).
\]
This formula also covers $k=1$, since then $R^{k-1}=R^0$.

The vertices $a_0,u_1,v_1$ belong to $S$ and form a triangle.
Indeed, for $a_0$ we may choose $y=a_{k+1}$ and $z=a_n$,
whereas for $u_1$ and $v_1$ we may choose $y=a_k$ and $z=a_1$.
We claim that $a_n\notin S$. Otherwise, let $y,z$ witness
$a_n\in S$. Since $y$ and $a_{k+1}$ are both adjacent to
$u_{k+1}$ and $v_{k+1}$, we have $R(y,a_{k+1})$. Using the
symmetry of $R$, we obtain
\[
R^{k-1}(z,y),\qquad
R(y,a_{k+1}),\qquad
R^{k-1}(a_{k+1},a_n).
\]
Consequently, $R^{2k-1}(z,a_n)$, while $E(z,a_n)$ holds by the
definition of $S$. Thus, the endpoints of an edge are joined by an
$R$-chain of length at most $2k-1=n-1$, contradicting the minimal
choice of $n$. Hence $a_n\notin S$. The graph induced by $S$ is
therefore non-bipartite and has strictly fewer vertices than $\bB$,
contrary to our initial assumption.

Now suppose that $n=2k+1$ is odd. Since $n\neq 1$, we have $k\geq 1$.
Define $S$ by
\[
S(x)\ \Leftrightarrow\
\exists z\,\bigl(R^k(a_{k+1},z)\wedge E(z,x)\bigr).
\]
Again $a_0,u_1,v_1\in S$: for $a_0$ choose $z=a_n$, and for
$u_1$ and $v_1$ choose $z=a_1$. Thus $S$ contains a triangle.
If $a_n\in S$, witnessed by $z$, then symmetry gives
$R^k(z,a_{k+1})$. Together with
$R^k(a_{k+1},a_n)$ this yields
\[
R^{2k}(z,a_n)=R^{n-1}(z,a_n).
\]
Since also $E(z,a_n)$, this again contradicts the minimal choice of
$n$. Hence $a_n\notin S$, and the same smaller-graph contradiction
follows.

So we conclude that $\bB/T$ does not contain loops. It also follows that
$\bB/T$ contains a triangle, because $\bB$ contains a triangle. 
Thus, the initial assumption on $\bB$ then
implies that $T$ must be the equality relation on $B$, which in turn implies
that $\bB$ does not contain any diamonds. 
\end{proof}

\begin{figure}
\begin{center}
\includegraphics[scale=0.6]{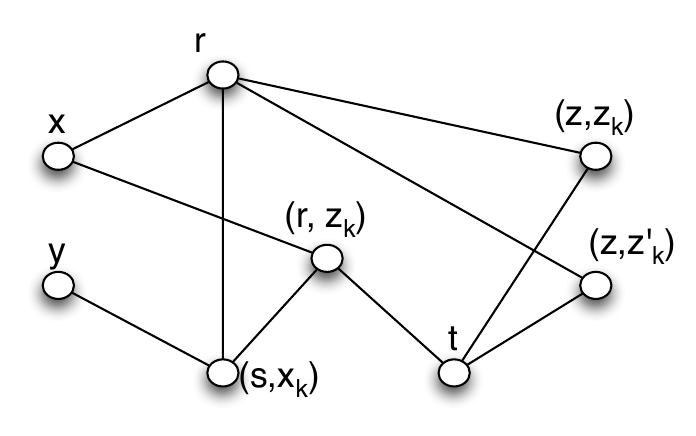} 
\end{center}
\caption{Diagram for the proof of Lemma~\ref{lem:k3k-image}.}
\label{fig:Bul}
\end{figure}

\begin{lemma}[from~\cite{BulatovHColoring}]\label{lem:k3k-image}
Let $\bB$ be a diamond-free undirected graph and let $h \colon (K_3)^k \to \bB$ 
be a homomorphism. Then the image
of $h$ is isomorphic to $(K_3)^m$ for some
$m \leq k$. 
\end{lemma}
\begin{proof}
Let $I \subseteq \{1,\dots,k\}$ be maximal
such that $\ker(h) \subseteq \ker(\pr_I)$.
Note that $\pr_I$ is defined even if $I = \emptyset$ (Definition~\ref{def:proj}).  
Such a set exists, because $\ker(\pr_{\emptyset})$ is the total relation. 
We claim that $\ker(h) = \ker(\pr_I)$;
this clearly implies the statement. 

By the maximality of $I$, for every 
$j \in \{1,\dots,k\} \setminus I$ there
are $x,y \in (K_3)^k$ 
such that $h(x) = h(y)$ and $x_j \neq y_j$. 
%We have to show that for all $z_1,\dots,z_k,z_j' \in \{0,1,2\}$ 
%$$h(z_1,\dots,z_j, \dots, z_k) = h(z_1,\dots,z_{j-1},z_j',z_{j+1},\dots,z_k).$$
For fixed $z_i\in\{0,1,2\}$, $i\neq j$, and $a\in\{0,1,2\}$, write
\[
z^{(a)}
:=
(z_1,\dots,z_{j-1},a,z_{j+1},\dots,z_k).
\]
We have to show that $h(z^{(a)})=h(z^{(b)})$ 
for all $a,b\in\{0,1,2\}$. It suffices to prove that
$h(z^{(a)})=h(z^{(x_j)})$
for every $a\neq x_j$. Indeed, once this has been proved, both
$h(z^{(a)})$ and $h(z^{(b)})$ are equal to $h(z^{(x_j)})$; if
$a=x_j$ or $b=x_j$, the corresponding equality is trivial.

Fix, therefore, $a\neq x_j$. Renaming the coordinate positions
throughout, we may assume that $j=k$. Put $z_k:=a$ and
$z_k':=x_k$. Thus $z_k\neq x_k$ and $z_k'=x_k$, and it remains
to prove
\[
h(z_1,\dots,z_{k-1},z_k)
=
h(z_1,\dots,z_{k-1},z_k').
\]
%We may suppose that $z_j \neq x_j$ and $z'_j = x_j$. 
%To simplify notation, we assume that $j=k$. 
As we have seen in Exercises~\ref{exe:one-common-neighbour} and \ref{exe:two-common-neighbours}, any two vertices
in $(K_3)^k$ have a common neighbour. 
\begin{itemize}
\item Let $r$ be a common neighbour of
$x$ and $(z,z_k) := (z_1,\dots,z_k)$. 
Note that $r$ and $(z,z_k')$ are adjacent, too. 
\item 
For all $i \neq k$ we choose an element $s_i$
of $K_3$ that is distinct from both $r_i$
and $y_i$. Since $x_k$ is distinct 
from $r_k$ and $y_k$
we have that 
$(s,x_k) := (s_1,\dots,s_{k-1},x_k)$ 
is a common neighbour of $r$ and $y$. 
\item The tuple
$(r,z_k) := (r_1,\dots,r_{k-1},z_k)$
is a common neighbour of both $x$ and $(s,x_k)$. 
\item Finally, for $i \neq k$ choose $t_i$
to be distinct from $z_i$ and $r_i$,
and choose $t_k$ to be distinct
from $z_k$ and from $z_k'$. 
Then $t := (t_1,\dots,t_{k-1},t_k)$
is a common neighbour of $(z,z_k)$,
of $(z,z_k')$, and
of $(r,z_k)$. 
\end{itemize}
The situation is illustrated in Figure~\ref{fig:Bul}. 
Since $\bB$ is diamond-free, 
$h(x) = h(y)$ implies that $h(r) = h(r,z_k)$
and for the same reason $h(z,z_k)=h(z,z_k')$
which completes the proof. 
\end{proof}

\begin{lemma}[from~\cite{BulatovHColoring}]
\label{lem:k3k-interpret}
If a finite diamond-free graph $\bB$
contains a triangle, then for some 
$k \in {\mathbb N}$ 
there is a primitive positive
interpretation of 
$(K_3)^k$ with constants in $\bB$. 
\end{lemma}
\begin{proof}
We construct a strictly increasing sequence
of subgraphs $G_1 \subset G_2 \subset \cdots$ of $\bB$ such that $G_i$ is isomorphic to
$(K_3)^{k_i}$ for some $k_i \in {\mathbb N}$. 
Let $G_1$ be any triangle in $\bB$. 
Suppose now that $G_i$ has already been 
constructed. If the domain of $G_i$ is
primitively positively definable in $\bB$
with constants, then we are done. 
Otherwise, there exists $\ell \in {\mathbb N}_{\geq 1}$, an idempotent $f \in \Pol(\bB)^{(\ell)}$, and $v_1,\dots,v_{\ell} \in G_i$
such that $f(v_1,\dots,v_{\ell}) \notin G_i$.
The restriction of $f$ to $(G_i)^{\ell}$ is a homomorphism from
$(G_i)^{\ell} \simeq (K_3)^{k_i\ell}$ to
the diamond-free graph $\bB$.
Lemma~\ref{lem:k3k-image} shows
that $G_{i+1} := f((G_i)^{\ell})$ induces a copy
of $(K_3)^{k_{i+1}}$ for some $k_{i+1} \leq k_i \cdot \ell$. 
Since $f$ is idempotent, we have
that $G_i \subseteq G_{i+1}$, and by the choice of $f$ the containment is strict. 
Since $\bB$ is finite, for some $m$ the set $G_m$ must have a primitive positive definition in $\bB$ with constants. 
\end{proof}

\begin{proof}[Proof of Theorem~\ref{thm:full-HN}]
Let $\bB$ be a finite undirected graph that is not bipartite and loopless. Then 
$\bB$ pp-constructs 
a diamond-free core $\bC$ containing a triangle, by Lemma~\ref{lem:diamond}. We may now apply Lemma~\ref{lem:k3k-interpret} to $\bC$ and obtain 
that for some $k \in {\mathbb N}$ there
is a primitive positive interpretation of 
$(K_3)^k$ with constants in $\bC$.
%Since $\bC$ is a core, and since $
The structure $(K_3)^k$ is
homomorphically equivalent to $K_3$, 
%it follows
%that there is a primitive positive %interpretation
%of a structure that is homomorphically equivalent to $K_3$ in $\bC$.
which interprets all finite structures primitive positively (Theorem~\ref{thm:k3-interprets}). 
So Theorem~\ref{thm:wonderland} implies
that $\HH(\PP(\bB))$ contains all finite structures. 
\end{proof}

\subsection{Siggers Terms of Arity 6}
\label{sect:6ary}
% TODO: mention "Siggers term" somewhere
An operation $s \colon B^6 \to B$
is called a \emph{Siggers operation} (of arity six\footnote{We stress the arity here since there is also a notion of Siggers operations for arity four, which is a similar but stronger result, see Section~\ref{sect:4ary}.}) 
if for all $x,y,z \in B$ 
$$s(x,y,x,z,y,z) = s(y,x,z,x,z,y).$$
As usual, if $\fA$ is an algebra and $t$ is a term such that $t^{\fA}$ is a Siggers operation, we call $t$ a \emph{Siggers term}. 

\begin{theorem}[from~\cite{Siggers}]
\label{thm:siggers}
Let $\bB$ be a finite structure. Then
either $\bB$ primitively positively interprets all finite structures
up to homomorphic equivalence, 
%the core of $\bB$ interprets all finite
%structures with constants, 
or $\bB$ 
has a Siggers polymorphism. 
\end{theorem}
\begin{proof}
Pick $k \geq 1$ and $a,b,c \in B^k$ such
that $\{(a_i,b_i,c_i) \mid i \leq k\} = B^3$. 
Let $R$ be the binary relation on $B^k$
such that $(u,v) \in R$ if and only if there exists a 6-ary $s \in \Pol(\bB)$ such that 
$u = s(a,b,a,c,b,c)$ and $v = s(b,a,c,a,c,b)$. 
We make the following series of observations.
\begin{itemize}
\item The relation $R$ (as a $2k$-ary relation over $B$) is preserved by $\Pol(\bB)$, and hence
$(B^k;R)$ has a primitive positive interpretation in $\bB$. 
To see this, let
$p\in\Pol(\bB)^{(\ell)}$ and let
$(u^j,v^j)\in R$, for $j\in[\ell]$, be witnessed by
$s_j\in\Pol(\bB)^{(6)}$. Define
\[
s(x_1,\dots,x_6)
:=
p\bigl(s_1(x_1,\dots,x_6),\dots,s_\ell(x_1,\dots,x_6)\bigr).
\]
Then $s\in\Pol(\bB)$, and
\begin{align*}
s(a,b,a,c,b,c)
&=p(u^1,\dots,u^\ell),\\
s(b,a,c,a,c,b)
&=p(v^1,\dots,v^\ell).
\end{align*}
Thus $s$ witnesses that
$\bigl(p(u^1,\dots,u^\ell),p(v^1,\dots,v^\ell)\bigr)\in R$. 
\item The vertices $a,b,c \in B^k$ induce in $(B^k;R)$ a copy of $K_3$: each of the six edges of $K_3$ is witnessed by one of the six 6-ary projections from $\Pol(\bB)$. 
\item The relation $R$ is symmetric: 
Suppose that $(u,v) \in R$ and let $s \in \Pol(\bB)$ be such that 
$u = s(a,b,a,c,b,c)$ and $v = s(b,a,c,a,c,b)$.
Define $s' \in \Pol(\bB)$ by $s'(x_1,\dots,x_6) := s(x_2,x_1,x_4,x_3,x_6,x_5)$; then 
\begin{align*}
v & = s(b,a,c,a,c,b) = s'(a,b,a,c,b,c) \\
u & = s(a,b,a,c,b,c) = s'(b,a,c,a,c,b) 
\end{align*}
and hence $s'$ witnesses that $(v,u) \in R$. 
\item If the graph $(B^k;R)$ contains a loop
$(w,w) \in R$, then there exists a 6-ary
$s \in \Pol(\bB)$ such that 
$$s(a,b,a,c,b,c) = w = s(b,a,c,a,c,b) \, .$$
The operation $s$ is Siggers:
for all $x,y,z \in B$ there exists an $i \leq k$
such that $(x,y,z) = (a_i,b_i,c_i)$,
and the above implies that
$$s(a_i,b_i,a_i,c_i,b_i,c_i) = s(b_i,a_i,c_i,a_i,c_i,b_i)$$
and we are done in this case.
\end{itemize}
So we may assume in the following
that $(B^k;R)$ is a simple (i.e., undirected and loopless) graph that contains a copy of $K_3$. 
By Theorem~\ref{thm:full-HN} applied to the undirected graph $(B^k;R)$, 
there is a primitive positive interpretation in $(B^k;R)$ of all finite structures up to homomorphic equivalence, and hence also
in $\bB$, and this concludes the proof. 
%In other words, $R$ is the smallest
%relation that contains 
\end{proof}

\begin{theorem}[Tractability Theorem, Version 4]
\label{thm:tractability-4}
Let $\bB$ be a relational structure with finite domain and finite signature. 
If $\bB$ has a Siggers polymorphism, then $\Csp(\bB)$ is in P. Otherwise, 
$\Csp(\bB)$ is NP-complete.
\end{theorem}

\begin{proof}
If $K_3\in\HI(\bB)$, then the NP-hardness of $\Csp(\bB)$ follows from 
Corollary~\ref{cor:pp-interpret-hard}.
Otherwise, Theorem~\ref{thm:siggers}
implies that $\bB$ has a Siggers polymorphism. 
Moreover, 
no projection satisfies the Siggers identity, so $\Pol(\bB)$ has no minion homomorphism to $\Proj$. 
Hence, 
$K_3 \notin \mathrm{HI}(\mathfrak{B})$ by Corollary~\ref{cor:minion-hard}. 
Theorem~\ref{thm:tractability-2} then implies that $\Csp(\bB)$ is in P. 
\end{proof}

\begin{exercises}
\exercise[Rot]\label{exe:source-188} A Taylor clone $\bf C$ is called \emph{minimal Taylor} if every proper subclone of $\bf C$
is not Taylor.
Show that every Taylor clone on a finite set contains a minimal Taylor clone.
% Solution sketch: Consider the set of Taylor subclones ordered by inclusion. A descending chain has a
% Taylor lower bound: on a finite domain only finitely many operations of each bounded arity occur,
% and the finite Taylor identity system stabilises along the chain. Zorn's lemma therefore gives an
% inclusion-minimal Taylor subclone.
\end{exercises}

% END INLINED SOURCE: UndirectedGraphs.tex
% BEGIN INLINED SOURCE: congruences.tex
% !TEX root = GH-UA.tex

\section{Congruences}
\label{sect:congruences}
The study of congruences of algebras and varieties is one of the central topics in universal algebra. In Section~\ref{sect:congr-lattice} we present some basic facts about congruences that will be used later in the text. Section~\ref{sect:congr-perm} about congruence permutability 
and Section~\ref{sect:congr-distr} about congruence distributivity 
cover material that is central in universal algebra, but will not be used later in the text and can be skipped by the hasty reader. 
%We revisit classical concepts and results that will then become relevant in later chapters, either because they will be directly used, or because they provide the intuition for the more abstract general results later. 

\subsection{The Congruence Lattice}
\label{sect:congr-lattice}
Let $\fA$ be a $\tau$-algebra. We write \emph{$\Con(\fA)$} for the set of all congruences of $\fA$ (Definition~\ref{def:congruence}). 
Clearly, $\Con(\fA)$ is closed under arbitrary intersections. 
On the other hand, the union of two congruences is in general not a congruence. 

%\begin{definition}
%Let 
%$R$ and $S$ be binary relations.
%Then the \emph{composition} of $R$ and $S$ is the binary relation $R \circ S := \{(x,z) \mid \exists y \, (R(x,y) \wedge R(y,z))\}$. 
%\end{definition}

\begin{lemma}
Let $\fA$ be an algebra. Then 
$(\Con(\fA),\subseteq)$ is a 
complete % (don't need, but would be strange to not do the full proof)
lattice (Example~\ref{expl:lattice}). 
\end{lemma}
\begin{proof}
Let $(E_i)_{i \in I}$ be a sequence of congruences of $\fA$. 
Intersections of congruences are congruences, and this
intersection is clearly the largest congruence contained in every
$E_i$.
Define (recall the definition of the relational product, Definition~\ref{def:rel-prod}) 
$$J := \bigvee_{i \in I} E_i := \bigcup \big \{E_{i_1} \circ \cdots \circ E_{i_k} \mid i_1,\dots,i_k \in I, k \in {\mathbb N} \big \};$$
if $k=0$, then $E_{i_1} \circ \cdots \circ E_{i_k}$ stands for $\id_{A}$. 
The relation $J$ is an
equivalence relation: reflexivity follows from the case $k=0$,
symmetry follows by reversing a word, and transitivity follows by
concatenating words.

We claim that this is the smallest (with respect to inclusion) equivalence relation that contains all the $E_i$. 
Let $f\in\tau$ be $n$-ary and suppose that
$(a_j,b_j)\in\bigvee_{i\in I}E_i$ for every $j\in\{1,\dots,n\}$.
For each $j$, choose a word in the congruences $E_i$ and a corresponding
chain from $a_j$ to $b_j$. Concatenate these finitely many words. Since
every congruence is reflexive, each of the chains can be padded by
constant steps and hence regarded as a chain for this common word.
Thus, writing the common word as $E_{i_1}\circ\cdots\circ E_{i_k}$,
there are elements
\[
a_j=c_{j,0},c_{j,1},\dots,c_{j,k}=b_j
\]
such that $(c_{j,\ell-1},c_{j,\ell})\in E_{i_\ell}$ for all $j$ and
$\ell$. Since $E_{i_\ell}$ is a congruence,
\[
\bigl(f(c_{1,\ell-1},\dots,c_{n,\ell-1}),
      f(c_{1,\ell},\dots,c_{n,\ell})\bigr)\in E_{i_\ell}.
\]
Consequently,
\[
\bigl(f(a_1,\dots,a_n),f(b_1,\dots,b_n)\bigr)
 \in E_{i_1}\circ\cdots\circ E_{i_k}
 \subseteq \bigvee_{i\in I}E_i.
\]
Hence, $\bigvee_{i\in I}E_i$ is a congruence.
Moreover, $E_i\subseteq J$ for every $i\in I$, by taking words of
length one. If $C$ is any congruence containing every $E_i$, then
transitivity of $C$ implies
\[
E_{i_1}\circ\cdots\circ E_{i_k}\subseteq C
\]
for every finite word $i_1 \dots i_k$. Therefore, $J\subseteq C$ and 
$J$ is the least congruence containing all the $E_i$. Thus, $\bigvee_{i\in I}E_i=J$.
%Let $f \in \tau$ be $n$-ary and 
%$(a_1,b_1),\dots,(a_n,b_n) \in E$. 
%Then there are $i_1,\dots,i_k \in I$ such that for all $j \in \{1,\dots,n\}$ 
%$$(a_j,b_j) \in E_{i_1} \circ \cdots \circ E_{i_k}.$$
%Hence, 
%$(f(a_1,\dots,a_n),f(b_1,\dots,b_n)) \in E_{i_1} \circ \cdots \circ E_{i_k}$ and $E \in \Con(\fA)$. 

We have constructed arbitrary meets and joins in
$(\Con(\fA),\subseteq)$, so it is a complete lattice. In particular,
the associativity, commutativity, idempotence, and absorption
identities for the binary meet and join follow from their
characterisation as greatest lower bounds and least upper bounds.
\end{proof}

Every algebra has the following two congruences. 
\begin{itemize}
\item \emph{$\Delta_A$}: the \emph{diagonal relation} 
$\{(a,a) \mid a \in A\}$. 
\item \emph{$\nabla_A$}: the \emph{universal relation} $A^2$. 
\end{itemize}
%Congruences that are different from $\Delta_A$
%are called non-trivial, 
%and 
Congruences that are different from 
$\nabla_A$ and $\Delta_A$ are called \emph{proper}. 
%and congruences that are different from $\Delta_A$ are called \emph{non-trivial}. 

\begin{definition}\label{def:simple} 
Algebras $\fA$ without proper congruences are called  
%where $\Delta_A$ and $\nabla_A$
%are the only congruences are called 
\emph{simple}. 
\end{definition}

\begin{example} 
Groups that are simple in the sense of group theory are simple in the more general sense of
Definition~\ref{def:simple}. 
\end{example}

\begin{example}
Let $G$ be a permutation group on the set $A$. 
Let $\fA$ be the algebra with domain $A$ and signature $G$, and define 
$g^{\fA} := g$ for all $g \in G$. 
Then $\fA$ is simple if and only if $G$ is primitive 
as a permutation group. 
\end{example}

\begin{definition}\label{def:polynomial}
Let $\fA$ be an algebra and let $\fB$ be the expansion of $\fA$ by all constant operations. 
A \emph{polynomial over $\fA$} is a term in the signature of $\fB$. A \emph{polynomial operation of $\fA$} is a term operation of $\fB$. 
\end{definition}

\begin{remark}
In these notes, all term and polynomial operations have positive arity; in particular, nullary polynomial operations are not included. Constant operations of every positive arity are polynomial operations. 
\end{remark}

\begin{definition}\label{def:poly-equiv}
Two algebras $\fA_1,\fA_2$ with the same domain $A$ are called \emph{polynomially equivalent}
if they have the same polynomial operations. 
\end{definition}

%\begin{lemma}
Note that polynomially equivalent algebras 
% is polynomially equivalent to $\fA$,
%then $\fA$ and $\fB$ 
have the same congruences. 
%\end{lemma}
%\begin{proof}
%Exercise~\ref{}. 
%\end{proof}

\begin{lemma}\label{lem:cg}
Let $\fB$ be an algebra and $X \subseteq B^2$. 
Then the smallest congruence of $\fB$ that contains $X$, denoted by $\Cg_{\fB}(X)$,  equals the reflexive symmetric transitive  closure of
\begin{align}
T := \{(p(a),p(b)) \in B^2 \mid (a,b) \in X, p \text{ a unary polynomial operation of } \fB \}. \label{eq:cg}
\end{align}
\end{lemma} 
\begin{proof}
Let $C$ be the reflexive symmetric transitive closure of $T$. Clearly, if $(a,b) \in X$ and $p$ is a unary polynomial operation of $\fB$, then $(p(a),p(b)) \in \Cg_{\fB}(X)$ since congruences are preserved by term operations and are reflexive. 
Since $\Cg_{\fB}(X)$ is reflexive, symmetric, and transitive, we obtain that $C \subseteq \Cg_{\fB}(X)$. To prove that  $\Cg_{\fB}(X) \subseteq C$ it suffices to prove that $C$ is a congruence of $\fB$ that contains $X$. By definition, $C$ is reflexive, symmetric, and transitive,
and we clearly have that $X \subseteq C$. 
We are left with the verification that $C$ is a congruence. If $X=\emptyset$, then $C$ is equality and there is nothing to prove, so assume that $X\neq\emptyset$. Then $T$ is reflexive, since constant unary polynomial operations show that $(a,a)\in T$ for every $a\in B$. If $(u_1,v_1),\dots,(u_k,v_k) \in C$
 and $f$ is an operation of $\fB$ of arity $k$, 
 then for each $i \in \{1,\dots,k\}$ there exists a path $P_i = (a_{i,1},\dots,a_{i,r})$ whose edges are forward or backward edges in $T$ and that starts in $u_i$ and ends in $v_i$. By the reflexivity of $T$ we may duplicate elements on these paths such that 
 \begin{itemize}
\item all these paths $P_i$ have the same length $r$; 
\item for every $\ell \in \{1,\dots,r-1\}$ 
there is at most one index $i(\ell)$ such that the $\ell$-th edge of $P_{i(\ell)}$ is not of the
form $(a,a)$.
\end{itemize} 
Hence, for every $\ell \in \{1,\dots,r-1\}$ 
the unary polynomial 
$$p_l(x) := f(a_{1,\ell},\dots,a_{i(\ell)-1,\ell},x,a_{i(\ell)+1,\ell},\dots,a_{k,\ell})$$ shows that 
there is a forward or backward edge between the elements 
$f(a_{1,\ell},\dots,a_{k,\ell})$ and $f(a_{1,\ell+1},\dots,a_{k,\ell+1})$.
By symmetry and transitivity, we therefore obtain a path from $u_0 := f(u_1,\dots,u_k)$ to $v_0 := f(v_1,\dots,v_k)$ whose edges are forward or backward edges in $T$, showing that $(u_0,v_0) \in C$, and we are done. 
% Solution: combines three ideas:
% constants correspond to reflexivity,
% and: Reflexivitaet von Kongruenzen ausnutzen um
% f(a1,...an) = f(b1,...,bn) auseinanderziehen
\end{proof}

\begin{exercises}
\exercise[Blau]\label{exe:source-189} Show that the algebra from Example~\ref{expl:m1} is simple.
\label{exe:m1-simple}
%\item Is the algebra from Example~\ref{expl:m2} simple?
%\label{exe:m2-not-simple}
% Solution sketch: Start with any nontrivial congruence and one related pair. Compatibility with the
% operation table in Example~\ref{expl:m1} successively relates that pair to every other element, so
% the congruence is universal; hence the algebra is simple. For Example~\ref{expl:m2}, the partition
% displayed by the two fibres of $H$ is a proper compatible equivalence, so that algebra is not
% simple.
\exercise[Blau]\label{exe:source-190} Show that if $\fA$ is an idempotent algebra
and $C$ a congruence of $\fA$,
then every congruence class of $C$ is a subalgebra of $\fA$.
% Solution sketch: If $a_1,\ldots,a_k$ lie in the congruence class of $c$, compatibility gives
% $f(a_1,\ldots,a_k)\equiv f(c,\ldots,c)=c$. Idempotence is used in the last equality. Thus every
% basic operation maps the class into itself.
\exercise[Blau]\label{exe:source-191} Prove the remark after Definition~\ref{def:poly-equiv}.
%\item \label{exe:cg} Let $\fB$ be an algebra and $X \subseteq B^2$. Prove that the smallest congruence of $\fB$ that contains $X$, denoted by $\Cg_{\fB}(X)$,  equals the symmetric transitive  closure of
%$$\{(p(a),p(b)) \mid (a,b) \in X, p \text{ a unary polynomial operation of } \fB \}.$$
\exercise[Blau]\label{exe:source-195} Show that if a finite abelian group  is polynomially complete, then it has just one element.
% Quelle: Ihringer, 7.1.3.
% Hinty: must be simple, thus Z_p.
% Solution sketch: Polynomial completeness implies simplicity, so a finite abelian group must be
% cyclic of prime order. But every polynomial operation of such a group is affine and therefore
% preserves the nontrivial relation $x-y+z-w=0$. A polynomially complete algebra preserves only
% relations forced by equality, a contradiction unless the group has one element.
\exercise[Blau]\label{exe:source-196} Show that if a lattice is polynomially  complete, then it has just one element.
% Quelle: Ihringer, 7.1.3.
% Hint: use monotonicity property.
% Solution sketch: Every lattice polynomial is monotone in each variable. On a lattice with two
% distinct comparable elements, a unary map interchanging them is not monotone and hence is not a
% polynomial operation. Polynomial completeness would supply every unary map, so only the one-element
% lattice can be polynomially complete.
\exercise[Rot]\label{exe:source-192} Let $\fA$ be an algebra. Show that an equivalence relation $R \subseteq A^2$ is
a congruence of $\fA$
%preserved by $\Clo(\fA)$
if and only if it is preserved by all \emph{unary} polynomials of $\fA$.
%Show that the same is true more
Which properties of an equivalence relation do you need in the proof?
% Solution sketch: Necessity is immediate. Conversely, suppose all unary polynomials preserve $R$. For
% a basic $k$-ary operation, replace the coordinates of one related input tuple by those of the other
% one at a time. Each replacement is a unary polynomial with all other arguments fixed; transitivity
% of $R$ chains the resulting pairs. Only reflexivity, symmetry, and transitivity of the equivalence
% relation are used.
\exercise[Rot]\label{exe:source-193} Let $\fA$ be an algebra. Then a relation $R \subseteq A^n$ is called
\begin{itemize}
\item \emph{reflexive} if $(a,\dots,a) \in R$ for every $a \in A$.
\item \emph{transitive} if for all $(a_{ij})_{i,j \in \{1,\dots,n\}} \in A^{n \times n}$ whose rows and columns are from $R$, we have that
 $(a_{11},\dots,a_{nn}) \in R$.
\end{itemize}
Show that a reflexive and transitive $R$ is preserved by $\Clo(\fA)$ if and only if it is preserved by all unary polynomial operations of $\fA$.
% Solution sketch: Use the same one-coordinate-at-a-time replacement as in
% Exercise~\ref{exe:congruence-polynomial}, but arrange the intermediate tuples as the rows and
% columns of an $n\times n$ matrix. Reflexivity supplies constant rows and columns and the stated
% transitivity property yields the diagonal tuple. This proves preservation by every term operation;
% the converse follows because unary polynomials are obtained by fixing arguments.
\exercise[Rot]\label{exe:source-197} \label{exe:fin-congruence-generation}
Let $\fA$ be an algebra on a finite set and $R \leq \fA^2$ be subdirect.
Then $\bigcup_{i \in {\mathbb N}} (R^{-1} \circ R)^i$
is a congruence of $\fA$.
% Source: selbsterfunden, wird gebraucht in Zhuk's proof
\exercise[Orange]\label{exe:source-194} \label{exe:PC}
Let $\fA$ be a finite idempotent algebra. Show that the following are equivalent.
\begin{enumerate}
\item $\fA$ is \emph{polynomially complete}, i.e., the clone of polynomial operations of $\fA$ equals the set of all operations on $A$.
\item The \emph{discriminator operation} $d \colon A^3 \to A$ is a polynomial operation of $\fA$. The discriminator operation is defined as follows:
\begin{align*}
d(x,y,z) := \begin{cases} z & x=y \\
x & x \neq y.
\end{cases}
\end{align*}
\item if $R \leq \fA^2$ strictly contains $=_A$, then $R = A^2$, and if $R \leq \fA^3$ is subdirect such that $\pi_{i,j}(R) = A^2$ for all $\{i,j\} \in \binom{[3]}{2}$, then $R=A^3$.
\end{enumerate}
\end{exercises}

%\subsection{Congruence Lattice Identities}
%\begin{center}
%\begin{tabular}{l|l|l}
%Term Condition & Congruence Lattice Property & Finitely Related Collapse \\
%\hline
%Maltsev & Congruence Permutable & No Collapse \\
%Jonsson chain & Congruence Distributive & Near Unanimity Terms \\
%Gumm chain & Congruence Modularity & Edge Terms \\
%\end{tabular} 
%\end{center} 

\subsection{Congruence Permutability}
\label{sect:congr-perm}
Two congruences $C_1,C_2 \in \Con(\fA)$ \emph{permute} if 
$$C_1 \circ C_2 = C_2 \circ C_1.$$ 
An algebra $\fA$ is called \emph{congruence permutable} 
if all pairs of congruences of $\fA$ permute. 
In the following, it will be convenient to write $a_0 C_1 a_1 C_2 a_2 \cdots a_{n-1} C_n a_n$
as a shortcut for 
$a_0 C_1 a_1$ and $a_1 C_2 a_2$ and 
$\cdots a_{n-1} C_{n} a_n$.

\begin{lemma}\label{lem:Maltsev}
Let $\fA$ be an algebra such that $\Clo(\fA)$ contains a Maltsev operation $p$. 
Then $\fA$ is congruence permutable. 
\end{lemma}
%eine exakte Charakterisierung der Existenz
%von Maltsevtermen mit Hilfe von Kongruenzen. 
\begin{proof}
Let $C,E \in \Con(\fA)$ 
%Es gen\"ugt zu zeigen: $C \circ E \subseteq E \circ C$. \\
and let $(a,b) \in C \circ E$. Then there exists $c \in A$ with 
$(a,c) \in C$ and $(c,b) \in E$. 
Note that $$ b=p^{\fA}(c,c,b) \; C \; p^{\fA}(a,c,b) \; E \; p^{\fA}(a,b,b) = a $$
and thus $(b,a) \in C \circ E$ and $(a,b) \in E \circ C$.
\end{proof}

Note that most classical algebras, such as groups, rings, fields, etc., do have a Maltsev term operation, and hence are congruence permutable. 
Lemma~\ref{lem:Maltsev} has a converse; however, for the converse we have to work with varieties of algebras. 
A variety is called \emph{congruence permutable} if all algebras in the variety are congruence permutable. 
We will show that an algebra $\fA$ has a Maltsev term if and only if the variety generated by $\fA$ is congruence permutable (Corollary~\ref{cor:congr-perm}). 

\ignore{
\begin{theorem}[Maltsev]
Let ${\mathcal K}$ be a class of  
$\tau$-algebras. 
Then $\HSP({\mathcal K})$ is congruence permutable if and only if there exists a $\tau$-term $t(x,y,z)$ such that every algebra in ${\mathcal K}$ satisfies 
$t(y,x,x) \approx t(x,x,y) \approx y$.  
\end{theorem}
\begin{proof}
``$\Leftarrow$''. If every algebra in ${\mathcal K}$ has a Maltsev term operation, then so does $\HSP({\mathcal K})$, and hence the statement follows from  Lemma~\ref{lem:Maltsev}. 

``$\Rightarrow$''. 
Let $\fF := F_{\mathcal K}(\{x,y,z\})$. 
For $\fF := \fF_{\mathcal K}(X)$ und $u,v \in \{x,y,z\}$ we write $C(u,v)$ for the smallest 
congruence of $\fF$ that contains $(u,v)$.
Let $C_1 := C(x,y) \in \Con(\fF)$  and $C_2 := C(y,z) \in \Con(\fF)$. 
Since $(x,z) \in C_1 \circ C_2 = C_2 \circ C_1$ 
there exists $b \in F$ with $(x,b) \in C_2$
and $(b,z) \in C_1$. Since $\fF$ is generated by  
$\{x,y,z\}$, there is a $\tau$-term $p(x,y,z)$ with $b = p^{\fF}(x,y,z)$. We will show that ${\mathcal K} \models \forall x,y. \, p(x,x,y)=y$. 
Let $\fA \in {\mathcal K}$ and $u,v \in A$. 
Let $f \colon \fF \to \fA$ be a homomorphism with $f(x) = u$, $f(y) = u$, and $f(z) = v$. Then $f(p^{\fF}(x,y,z)) = p^{\fA}(u,u,v)$. 
Since $(x,y) \in \Ker(f)$ we have $C_1 \subseteq \Ker(f)$. Thus, $(b,z) \in \Ker(f)$ and $$v = f(z) = f(b) = f\big (p^{\fF}(x,y,z) \big) = p^{\fA}(u,u,v) .$$ 
${\mathcal K} \models p(y,x,x) \approx y$ can be shown similarly. 
\end{proof}
}

In our proof of this result, we will use 
the concept of rectangularity, which we have already encountered 
 in Section~\ref{sect:maltsev},
 where we proved that digraphs with a Maltsev polymorphism are rectangular.
 %, and in Theorem~\ref{thm:CEJN} we characterised the existence of Maltsev polymorphisms of digraphs using total rectangularity. The following corollary clarifies the connection between rectangularity and Maltese  terms. 

\ignore{
\begin{proposition}\label{prop:rectangularity-variety}
Let $\fA$ be an algebra. 
Then $\fA$ has a Maltsev term if and only if 
every $R \leq \fB^2$, 
for every algebra $\fB \in \HSP(\fA)$, is rectangular. 
\end{proposition}
\begin{proof}
Clearly, if $\fA$ has a Maltsev term, then every algebra $\fB$ in $\HSP(\fA)$ has a Maltsev term, and hence every $R \leq \fB^2$ is rectangular. Conversely, let
$\fB \in \HSP(\fA)$ be the free algebra for 
$\HSP(\fA)$ over $\{x,y\}$, and 
let $R$ be the subuniverse of $\fB^2$ generated by 
$\{(x,x),(x,y),(y,y)\}$. Since $R$ is rectangular, we have that $(x,y) \in R$. 
Hence, there exists a term $t$ such that $f((x,x),(x,y),(y,y)) = (y,x)$. Then $t$ satisfies 
$t(x,x,y) \approx y$ and $t(x,y,y) \approx x$
in every algebra of $\HSP(\fA)$, 
(Lemma~\ref{lem:easy-free}), and hence $\fA$ has a Maltsev term. 
\end{proof}
}

\begin{theorem}[Maltsev]
\label{prop:rectangularity-variety}
Let ${\mathcal K}$ be a class of  
$\tau$-algebras. Then the following are equivalent. 
\begin{enumerate}
\item $\HSP({\mathcal K})$ is congruence permutable.
\item For every algebra $\fB \in \HSP({\mathcal K})$, 
every subalgebra $R \leq \fB^2$ is rectangular. 
\item There exists a $\tau$-term
which denotes a Maltsev operation 
in every algebra in ${\mathcal K}$.  
\end{enumerate} 
\end{theorem}
\begin{proof}
3.\ implies 1.: 
If every algebra in ${\mathcal K}$ has a Maltsev term operation, then so does $\HSP({\mathcal K})$, and hence the statement follows from  Lemma~\ref{lem:Maltsev}. 

1.\ implies 2.: Apply congruence permutability to the two projection congruences of $R \leq \fB^2$: these are the congruences 
\begin{align*} 
C_1 & := \{((a,b),(c,d)) \mid (a,b),(c,d) \in R, a=c\} \\ 
C_2 & := \{((a,b),(c,d)) \mid (a,b),(c,d) \in R, b=d\}.
\end{align*} 

2.\ implies 3.: Let
%$\fB \in \HSP({\mathcal K})$ 
$\fF := \fF_{\mathcal K}(\{x,y\})$ 
be the free algebra for 
${\mathcal K}$ over $\{x,y\}$, and 
let $R$ be the subuniverse of $\fF^2$ generated by 
$\{(x,x),(x,y),(y,y)\}$. Since $R$ is rectangular, we have that $(y,x) \in R$. 
Hence, there exists a term $t$ such that $t^{\fF^2}((x,x),(x,y),(y,y)) = (y,x)$. Then $t$ satisfies 
$t(x,x,y) \approx y$ and $t(x,y,y) \approx x$
in every algebra of $\HSP({\mathcal K})$  
(Lemma~\ref{lem:easy-free}). 
\end{proof}

\begin{corollary}\label{cor:congr-perm}
Let $\fA$ be an algebra. Then the following are equivalent. 
\begin{enumerate}
\item Every algebra in $\HSP(\fA)$ is congruence permutable. 
\item For every algebra $\fB \in \HSP(\fA)$,  every $R \leq \fB^2$ is rectangular.
\item $\fA$ has a Maltsev term. 
\end{enumerate}
\end{corollary}

\subsection{Congruence Distributivity}
\label{sect:congr-distr}
A lattice $\fL = (L;\wedge,\vee,0,1)$ 
is called \emph{distributive} if it satisfies
\begin{align*}
(x \wedge y) \vee z & \approx (x \vee z) \wedge (y \vee z) \\
\text{ and } \quad  (x \vee y) \wedge z & \approx (x \wedge z) \vee (y \wedge z). 
\end{align*}
%{\bf Examples:} 
%\begin{itemize}
%\item 
An example of a distributive lattice is 
the set of subsets of a set $S$, equipped with the operations of intersection and union. 
%: $({\mathcal P}(S);\subseteq)$. 
%\item Der Teilerverband
%\item Idealverb\"ande (Darstellungssatz von Birkhoff);
%\item Every lattice that does not contain the lattice $M_3$ or the lattice $N_5$ as a sublattice (see TODO). 
%\end{itemize}
%TODO: Draw $M_3$. 
If the congruence lattice of an algebra $\fA$ is distributive, we call $\fA$ \emph{congruence distributive}. 

\begin{lemma}
Every algebra with a majority term operation is congruence distributive.  
\end{lemma}

\begin{proof}
Let $C,D,E \in \Con(\bA)$ and $(a,b) \in C \wedge (D \vee E)$. 
Then there are  
$c_1,\dots,c_n$ such that 
$$a D c_1 E c_2 D c_3 \dots c_n E b.$$
Since $(b,a) \in C$, we have for all $c \in A$ that 
\begin{align*}
m^{\bA}(a,c,b) \, C \, m^{\bA}(a,c,a) = a. 
%\label{eq:maj-sd}
\end{align*}
Thus, for all $c_1,c_2 \in A$ we obtain 
\begin{align}
m^{\bA}(a,c_1,b) \, C \, a \, C\, m^{\bA}(a,c_2,b). \label{eq:maj-sd}
\end{align}
Therefore, 
\begin{align*} a = m^{\bA}(a,a,b)
& (C \wedge D) m^{\bA}(a,c_1,b)  
&& \text{(by~\eqref{eq:maj-sd})} \\
& (C \wedge E) m^{\bA}(a,c_2,b) \\
& \cdots \\
& (C \wedge D) m^{\bA}(a,c_n,b) \\
& (C \wedge E) m^{\bA}(a,b,b) = b. 
\end{align*}
We conclude that $(a,b) \in (C \wedge D) \vee (C \wedge E)$. 
\end{proof}

%\begin{example}
%Sei $A$ beliebige Menge. 
%Betrachte $\bA := (A;f)$ mit
%\begin{align*} f^{\bA}(x,y,z) := \begin{cases}
%z & \text{ falls } y=z \\
%x & \text{ sonst } 
%\end{cases} & \qedhere
%\end{align*}
%\end{example}

As in the case of congruence permutability, 
there is even an equational characterisation of congruence distributivity of varieties. 

\begin{theorem}[J\'onsson]
\label{thm:jonsson}
$\HSP({\mathcal K})$ 
is congruence distributive if and only if there 
exists $n \in {\mathbb N}$ and 
$\tau$-terms $p_0,\dots,p_n$ such that 
${\mathcal K}$
satisfies the following identities. 
%f\"ur alle $i \in \{0,1,\dots,n\}$
\begin{align*}
p_i(x,y,x) & \approx x && \text{ for } i \in \{1,\dots,n\} \\
p_0(x,y,z) &  \approx x \\
p_i(x,x,y) &  \approx p_{i+1}(x,x,y)  && \text{ for $i$ even}  \\
p_i(x,y,y)  &  \approx p_{i+1}(x,y,y)  && \text{ for $i$ odd} \\
p_n(x,y,z)  &  \approx z
\end{align*}
\end{theorem}
\begin{proof}
``$\Rightarrow$''. Let $\fF := \fF_{\mathcal K}(\{x,y,z\})$ be the free algebra for ${\mathcal K}$ over $\{x,y,z\}$. 
By freeness, the following three substitutions
\begin{align}
(x,y,z) \mapsto(x,x,z) ,\qquad
(x,y,z) \mapsto(x,y,y) ,\qquad
(x,y,z) \mapsto(x,y,x) 
\end{align}
 extend uniquely to endomorphisms 
 $\sigma_{xy}$, $\sigma_{yz}$, and $\sigma_{xz}$ of $\fF$. 
Define
$$
C_{xy}:=\ker(\sigma_{xy}),\qquad
C_{yz}:=\ker(\sigma_{yz}),\qquad
C_{xz}:=\ker(\sigma_{xz}).
$$
%Let $C_{xy}$ be the congruence of 
%$\fF$ obtained by identifying the variables %$x$ and $y$, i.e., 
%For $u,v \in \{x,y,z\}$, define
%$$C_{xy} := \{(r,s) \in \fF^2 \mid \fF \models r(x,x,z) \approx s(x,x,z) \}.$$
%The congruences $C_{yz}$ and $C_{xz}$ are %defined analogously.  
We have 
$$C_{xz} \wedge \big(C_{xy} \vee C_{yz} \big) = \big(C_{xz} \wedge C_{xy}\big) \vee \big (C_{xz} \wedge C_{yz} \big)$$
and hence $(x,z) \in \big(C_{xz} \wedge C_{xy}\big) \vee \big (C_{xz} \wedge C_{yz} \big)$ in $\fF$. Thus, there are $p_1,\dots,p_{n-1} \in \fF$ such that each of $(x,p_1),(p_1,p_2),\dots,(p_{n-1},z)$ is in $C_{xz} \wedge C_{xy}$
or in $C_{xz} \wedge C_{yz}$. By padding the sequence $p_1,\dots,p_{n-1}$ with repeated entries, we may even suppose that 
%\begin{align*}
%x & \big(C(x,z) \wedge C(x,y)\big) p_1 & x = \; & p_1(x,y,x) = p_1(x,x,y), \\
%p_1 & \big(C(x,z) \wedge C(y,z)\big) p_2 & &  p_1(x,y,x) = p_2(x,y,x), \\
%& & & p_1(x,y,y) = p_2(x,y,y), \\
%& \vdots \\
%p_{n-1} & \big(C(x,z) \wedge C(y,z)\big) z  & & p_{n-1}(x,y,x) = p_{n-1}(x,y,y) = z
%\end{align_{xy}*}
\begin{align}
x & \big(C_{xz} \wedge C_{xy}\big) p_1 \label{eq:j1} \\
p_1 & \big(C_{xz} \wedge C_{yz}\big) p_2 \label{eq:j2}  \\
& \quad \vdots \nonumber \\
p_{n-1} & \big(C_{xz} \wedge C_{yz}\big) z  
\label{eq:jn}
\end{align}
and that in particular $n$ is odd. 
From~\eqref{eq:j1} we obtain that 
$$x \approx p_1(x,y,x)  \approx p_1(x,x,y),$$
from~\eqref{eq:j2} we obtain 
\begin{align*}
p_1(x,y,x) &  \approx p_2(x,y,x) \\
\text{and} \quad p_1(x,y,y) &  \approx p_2(x,y,y), 
\end{align*}
and from~\eqref{eq:jn} that 
$p_{n-1}(z,x,z)  \approx p_{n-1}(x,z,z)  \approx z$. 
Similarly, all other identities from the statement can be derived from the sequence
\eqref{eq:j2}-\eqref{eq:jn}. 

``$\Leftarrow$''. Let $\fA \in \HSP({\mathcal K})$ and let
$C_1,C_2,C_3 \in \Con(\fA)$. Put
\[
Q := (C_1\wedge C_2)\vee(C_1\wedge C_3).
\]
It suffices to prove that
\[
C_1\wedge(C_2\vee C_3)\subseteq Q,
\]
since the converse inclusion holds in every lattice. Let $(a,b)\in C_1\wedge(C_2\vee C_3)$. Since
$(a,b)\in C_2\vee C_3$, there exist
\[
a=d_0,d_1,\dots,d_m=b
\]
such that, for every $j\in\{1,\dots,m\}$, either
$d_{j-1}\mathrel{C_2}d_j$ or $d_{j-1}\mathrel{C_3}d_j$.
For $i\in\{0,\dots,n\}$, put
\[
L_i:=p_i(a,a,b)
\qquad\text{and}\qquad
R_i:=p_i(a,b,b).
\]

We claim that
\[
L_i\mathrel{Q}R_i
\qquad\text{for every }i\in\{1,\dots,n\}.
\]
Indeed, fix such an $i$. Since $(a,b)\in C_1$ and $C_1$ is
compatible with the term operation $p_i$, for every
$j\in\{0,\dots,m\}$ we have
\[
p_i(a,d_j,b)\mathrel{C_1}p_i(a,d_j,a)=a.
\]
Moreover, if $d_{j-1}\mathrel{C_\ell}d_j$, where
$\ell\in\{2,3\}$, then compatibility with $C_\ell$ gives
\[
p_i(a,d_{j-1},b)\mathrel{C_\ell}p_i(a,d_j,b).
\]
Consequently,
\[
p_i(a,d_{j-1},b)
\mathrel{C_1\wedge C_\ell}
p_i(a,d_j,b),
\]
and hence
\[
p_i(a,d_{j-1},b)\mathrel{Q}p_i(a,d_j,b).
\]
Following the sequence $d_0,\dots,d_m$, and using the transitivity
of $Q$, yields
\[
L_i=p_i(a,d_0,b)\mathrel{Q}p_i(a,d_m,b)=R_i,
\]
as claimed.

It remains to concatenate these relations for the successive
Jonsson terms. The linking identities imply
\[
L_i=L_{i+1}\quad\text{if $i$ is even},
\qquad
R_i=R_{i+1}\quad\text{if $i$ is odd}.
\]
Furthermore, the endpoint identities give
\[
L_0=p_0(a,a,b)=a
\qquad\text{and}\qquad
L_n=R_n=b.
\]
Thus, using also the symmetry of $Q$, the relations
$L_i\mathrel{Q}R_i$ concatenate as
\[
a=L_0=L_1\mathrel{Q}R_1=R_2
\mathrel{Q}L_2=L_3
\mathrel{Q}R_3=R_4
\mathrel{Q}\cdots\mathrel{Q}b.
\]
Therefore $(a,b)\in Q$, and hence
\[
C_1\wedge(C_2\vee C_3)
\subseteq
(C_1\wedge C_2)\vee(C_1\wedge C_3).
\]
This proves that $\Con(\fA)$ is distributive.
\end{proof}

A sequence of operations $(p_0,\dots,p_n)$ which satisfy the identities from Theorem~\ref{thm:jonsson} is called a J\'onsson chain; if $\fA$ is a $\tau$-algebra and $t_0,\dots,t_n$ are $\tau$-terms such that $(t_0^{\fA},\dots,t_n^{\fA})$ is a J\'onsson chain, then $t_0,\dots,t_n$ are called \emph{J\'onsson terms}. 

\begin{example}
Let $\fA = (\{0,1\};p)$ be the algebra where $p$ is the ternary operation given by $p(x,y,z) := x \wedge (y \vee z)$. 
Then 
$$\underbrace{p(x,x,x)}_{=:p_0},\underbrace{p(x,y,z)}_{=:p_1},\underbrace{p(x,z,z)}_{=:p_2},\underbrace{p(z,y,x)}_{=:p_3},
% Different from Albert, dort steht p(z,x,y),
% ist aber egal. 
\underbrace{p(z,z,z)}_{=:p_4}$$
are J\'onsson terms: for any $x,y,z \in \{0,1\}$ we have 
\begin{itemize}
\item $p^{\fA}_i(x,y,x) = x \wedge (y \vee x) = x$ for every $i \in \{1,\dots,4\}$, 
\item $p^{\fA}_0(x,y,z) = x$, 
\item $p^{\fA}_1(x,y,y) = p(x,y,y) = p^{\fA}_2(x,y,y)$, 
\item $p^{\fA}_2(x,x,y) = p(x,y,y) = (x \wedge y) = (y \wedge x) = p(y,x,x) = p^{\fA}_3(x,x,y)$, 
\item $p^{\fA}_3(x,y,y) = p(y,x,y) = y = p^{\fA}_4(x,y,y)$,  
\item $p^{\fA}_4(x,y,z) = p(z,z,z) = z$. 
\qedhere 
\end{itemize} 
\end{example} 

%\subsection{J\'onsson Chains}
%\label{sect:jonsson}
%J\'onsson chains have been discovered
%by Bjarni J\'onsson; they provide an equivalent characterisation of congruence distributive  varieties. 

If the variety is generated by the polymorphism clone of a finite structure $\bB$ with finite relational signature, the term condition in Theorem~\ref{thm:jonsson} has drastic consequences
for $\Csp(\bB)$, similarly as in the previous section for congruence permutable varieties. 
Barto~\cite{barto-cd} proved that in this case $\bB$ must also have a near unanimity polymorphism and hence can be solved in polynomial time (see Exercise~\ref{exe:nu-kcons}, or Corollary~\ref{cor:nu-sac}).

%\medskip
%\newpage
% \noindent
%{\bf Exercises.}
\begin{exercises}
\exercise[Blau]\label{exe:source-199}
Let $f \colon \fA \to \prod_{i \in I} \fA_i$
be a homomorphism. Show that
$$\Ker(f) = \bigcap_{i \in I} \Ker( \pi_i \circ f)$$
% Solution sketch: Two elements have the same image under $f$ exactly when all coordinates of their
% images agree. Coordinate $i$ agrees exactly when the pair belongs to $\ker(\pi_i\circ f)$. Taking
% the conjunction over all $i$ gives the displayed intersection.
\exercise[Blau]\label{exe:source-200} Show that the variety of all lattices is congruence
distributive, but not congruence permutable.
% Solution sketch: The Funayama--Nakayama theorem gives distributivity of every lattice congruence
% lattice; equivalently, one may use the standard Jonsson terms built from $\wedge$ and $\vee$.
% Permutability fails already for the three-element chain: its two nontrivial congruences, collapsing
% the bottom or the top adjacent pair, do not commute under relational composition.
\exercise[Blau]\label{exe:source-201} Show that the variety of Boolean algebras
is both
congruence permutable and congruence distributive.
% Solution sketch: Boolean algebras have the Maltsev term
% $p(x,y,z)=x\mathbin{\triangle}y\mathbin{\triangle}z$, so their congruences permute. The lattice
% reduct is distributive and its standard Jonsson terms are Boolean terms, so their congruence
% lattices are distributive as well.
\exercise[Blau]\label{exe:source-205}
% Source: Brady notes
% Pixley 2: A simple and Pixley: -> PC.
Show that if $\fA$ is a finite simple algebra and it has a Pixley term, then it is polynomially complete (see Exercise~\ref{exe:PC}).
% Solution sketch: Pixley's characterisation says that in a variety with a Pixley term every
% compatible relation is determined by congruences. In a simple algebra the only congruences are
% equality and the universal relation, so every operation preserving the basic operations is a
% polynomial operation. Equivalently, every finitary operation is polynomial, and the algebra is
% polynomially complete.
\exercise[Blau]\label{exe:source-206}
Show that the algebra $\fA = (\{0,1\};f_k)$ from Example~\ref{expl:bool-nu} generates a congruence distributive variety.
% Source: me
% Solution: the following operation
\exercise[Blau]\label{exe:source-207} Show that every algebra with a near unanimity term also has J\'onsson terms.
% Source: folklore.
% Generalises the Boolean example.
% Solution sketch: Let $u$ be an $n$-ary near-unanimity term. Define successive ternary terms by
% filling an increasing initial block of arguments of $u$ with $x$, the next block with $y$, and the
% remaining arguments with $z$. Adjacent terms agree on the alternating Jonsson substitutions by the
% near-unanimity identities, giving a Jonsson chain.
\exercise[Rot]\label{exe:source-198} Show that $\SH({\mathcal K}) \subseteq \HS({\mathcal K})$,
$\PS({\mathcal K}) \subseteq \SP({\mathcal K})$, and $\PH({\mathcal K}) \subseteq \HP({\mathcal K})$. \label{exe:operators}
% Solution sketch: For $\SH\subseteq\HS$, restrict a quotient map to a subalgebra and take its image.
% For $\PS\subseteq\SP$, a product of subalgebras is the corresponding subalgebra of the product. For
% $\PH\subseteq\HP$, the product of quotient maps is a quotient map from the product. These
% constructions give the three inclusions.
\exercise[Rot]\label{exe:source-202} Show that a lattice satisfies the two identities
\begin{align*}
x \wedge (y \vee z)  \approx (x \wedge y) \vee (x \wedge z) \\
x \vee (y \wedge z)  \approx (x \vee y) \wedge (x \vee z)
\end{align*}
if and only if it satisfies one of those identities.
\label{exe:lattices}
%\item Show that every algebra with a majority term operation has J\'onsson terms as in Theorem~\ref{thm:jonsson}.
% Solution sketch: Order duality interchanges $\wedge$ and $\vee$. More explicitly, substitute the
% first distributive law into itself after interchanging suitable variables and use absorption to
% derive the second; the reverse derivation is its dual. Hence either identity implies the other.
\exercise[Rot]\label{exe:source-203} \label{exe:Pixley}
Let $\fA$ be an algebra.
A term $t$ is called a \emph{Pixley term} if it  satisfies
\begin{align*}
t(x,y,y) & \approx x \\
t(x,y,x) & \approx x \\
t(y,y,x) & \approx x.
\end{align*}
Show that if $t$ is a Pixley term, then
$t(x,t(x,y,z),z)$ is a majority term.
% Solution sketch: Put $m(x,y,z)=t(x,t(x,y,z),z)$. Substituting $(x,x,y)$, $(x,y,x)$, and $(y,x,x)$
% and applying the three Pixley identities from the inside out gives respectively $x,x,x$. Thus $m$
% satisfies all majority identities.
\exercise[Schwarz]\label{exe:source-204} Show that an algebra $\fA$ generates a variety which is both congruence permutable and congruence distributive if and only if it has a Pixley term.
% Source: Brady notes
% Solution sketch: A Pixley term yields a Maltsev term directly and, by Exercise~\ref{exe:Pixley}, a
% majority term; hence the generated variety is congruence permutable and distributive. Conversely,
% combine a Maltsev term with a chain of Jonsson terms, successively correcting the three exceptional
% input patterns. The resulting composite satisfies the three Pixley identities.
\end{exercises}

% END INLINED SOURCE: congruences.tex
% BEGIN INLINED SOURCE: Abelian.tex
% !TEX root = GH-UA.tex
\section{Abelian Algebras}
\label{sect:abelian}
%We have seen a strong analysis of finite conservative
%algebras $\fA$ based on their 2-element subalgebras in Section~\ref{sect:list}. 
%In order to generalise this approach to 
%finite idempotent algebras $\fA$, 
%the assumption that $\fA$ is minimal Taylor is very fruitful. One of the ingredients for this approach is a good understanding of finite idempotent algebras with a Taylor term that are additionally \emph{abelian}. 
%The concept of \emph{abelian} algebras allows to generalise several ideas that are well-known from group theory to a much wider class of algebras. 
Modules (Example~\ref{expl:module}) have many strong properties and are very well understood. 
Affine algebras are `essentially' modules and introduced in Section~\ref{sect:affine}. 
The relevance of affine algebras in the context of constraint satisfaction is that core structures with more than one element and whose polymorphism algebra is idempotent and affine can 
pp-construct $({\mathbb Z}_p;+,1)$, for some prime $p$ (Section~\ref{sect:aff-pp}). 
We have seen in Theorem~\ref{thm:inexpress}
%We will see in Section~\ref{sect:bounded-width}
that the CSP of such structures cannot be solved by $k$-consistency, for any $k$. 
% (i.e., not by local consistency techniques). 

\emph{Abelian} algebras are defined more abstractly (Section~\ref{sect:term-cond}). 
It turns out that under fairly general conditions, abelian algebras must be affine; this is the content of the fundamental theorem of abelian algebras which will be presented in Section~\ref{sect:term-cond} (Theorem~\ref{thm:abelian}) and generalised later in Section~\ref{sect:abelian-revisited}. Section~\ref{sect:congr-cond} presents other useful characterisations of abelian algebras in terms of congruences.  

%A guiding question of this section is 
% DON'T NEED
%\begin{definition}\label{def:abelian}
% Explicit link with other defs: 
% Taylor, McNulty, McKenzie
%An algebra $\fA$ is called \emph{abelian} if $\Delta_{A} := \{(a,a) \mid a \in A\}$ is a congruence class of a congruence of $\fA^2$. 
%\end{definition}

%\begin{example}\label{expl:affine-m}
%For $n \in {\mathbb N}$, 
%the algebra $(\{0,\dots,n-1\};m)$, where $m \colon \{0,\dots,p-1\}^3 \to \{0,\dots,p-1\}$ is defined by $m(x,y,z) := x-y+z$, is abelian, witnessed by the congruence $\theta$ defined as follows:
%\begin{align*}
%& ( (x_1,x_2), (y_1,y_2) ) \in \theta \; \Leftrightarrow \; (x_1-x_2 = y_1 - y_2).  \qedhere
%\end{align*}
%\end{example}
%With the following definition we generalise Example~\ref{expl:affine-m}.

\subsection{Affine Algebras}
\label{sect:affine}
In the study of the CSP, we may assume without loss of generality that our template contains all unary singleton relations (Section~\ref{sect:cores-constants}). The  polymorphism clone of the resulting template is then idempotent. 
In this section, in order to define affine algebras, we do the opposite: we first add all constant operations to the polymorphism clone of our structure! This might appear strange from the CSP perspective, because templates with constant polymorphisms have a trivial CSP. However, miraculously we can still use the resulting definitions and theorems about algebras and interpret them in the original idempotent setting. 

%\begin{definition}\label{def:affine}
An algebra $\fA$ is called \emph{affine} if $\fA$ is polynomially equivalent (Definition~\ref{def:poly-equiv}) to a module (Example~\ref{expl:module}). 
%\end{definition}
Clearly, every module $\fM$ has a Maltsev term operation, namely $(x,y,z) \mapsto x-y+z$, so every affine algebra has a Maltsev polynomial operation. Something stronger holds.

\begin{lemma}\label{lem:m-unique}
If $\fA$ is affine, then 
$(x,y,z) \mapsto x-y+z$ is the \emph{unique}
Maltsev polynomial operation of $\fA$.  
\end{lemma}
\begin{proof}
Suppose that $\fA$ is polynomially equivalent to a module $\fM$ over the ring $\fR$. 
Let $m(x,y,z) = \alpha x + \beta y + \gamma z + d$, for $\alpha,\beta,\gamma \in R$ and $d \in M$, be a Maltsev polynomial operation of $\fA$. Since $m(0,0,0) = 0$ we must have $d = 0$. Moreover, 
for all $x \in M$ we have $x = m(x,0,0)=\alpha x$ 
and analogously we obtain $x = \gamma x$. Finally, $m(x,x,0) = x + \beta x = 0$, and therefore 
$\beta x = -x$. 
We conclude that $m(x,y,z) = x-y+z$. 
\end{proof}

\begin{remark}
The operation $(x,y,z) \mapsto x-y+z$ is an affine Maltsev operation as defined in Section~\ref{sect:m-expl}. 
An algebra whose 
term operations are generated by 
an affine Maltsev operation is called an \emph{affine Maltsev algebra} (also see Exercise~\ref{exe:affine-maltsev-terms}). 
Note that affine Maltsev algebras 
are affine in the sense defined above. 
%of Definition~\ref{def:affine}. 
\end{remark}

\begin{example} 
Every abelian group is affine: let $\fR$ be the ring of integers ${\mathbb Z}$, and define scalar multiplication as follows
\begin{align*} 
n \cdot x & := \begin{cases} \underbrace{x+\cdots+x}_{n \text{ times}} &  \text{ if } n>0 \\
0 & \text{ if } n=0 \\
- \underbrace{x+\cdots+x}_{-n \text{ times}} &  \text{ if } n < 0 . 
\end{cases} \qedhere
\end{align*} 
% Source: comment in kompatscher script;
\end{example} 

\begin{definition}\label{def:central}
Let $\fA$ be an algebra. An operation $m \colon A^k \to A$ is 
called \emph{central in $\fA$} if 
$m$ is a homomorphism from $\fA^k$ to $\fA$. 
\end{definition}

\begin{remark}
Note that $m$ is central in $\fA$ if and only if
every operation of $\fA$ preserves the graph of $m$ (see Exercise~\ref{exe:graphof}). Hence, if $\Clo(\fA) = \Pol(\bA)$ for some finite structure $\bA$, then $m$ is central in $\fA$ if and only if its graph has a primitive positive definition in $\bA$. 
\end{remark}

\begin{lemma}\label{lem:m-central}
Let $\fA$ be affine. Then the operation $(x,y,z) \mapsto x-y+z$ is central. 
\end{lemma}
\begin{proof}
Let $f$ be a basic operation of $\fA$ of arity $n$.  
Since $\fA$ is affine we can write $f$ as $\sum_{i = 1}^n \alpha_i x_i + c$. Then
\begin{align*} 
f(\bar x)-f(\bar y)+f(\bar z) & = \sum_{i \in \{1,\dots,n\}} \alpha_i x_i + c - \left (\sum_{i  \in \{1,\dots,n\}} \alpha_i y_i + c \right ) + \sum_{i  \in \{1,\dots,n\}} \alpha_i z_i + c \\
& = \sum_{i \in \{1,\dots,n\}} \alpha_i(x_i-y_i+z_i) + c \\
& = f(x_1-y_1+z_1,\dots,x_n-y_n+z_n).  
\qedhere
\end{align*}
\end{proof}

\begin{exercises}
\exercise[Orange]\label{exe:source-208} \label{exe:cons} Let $\fA$ be an affine conservative algebra with $|A| \geq 2$. Show that $|A| = 2$.
%\vspace{-2cm}
%\begin{flushright}
%\includegraphics[scale=.3]{Orange.jpg}
%\hspace{1cm}{ }
%\end{flushright}
%\vspace{-.5cm}
% Solution sketch: An affine idempotent term has the form $\sum_i r_i x_i$ with $\sum_i r_i=1$.
% Conservativity of the Maltsev term $x-y+z$ forces $x-y+z\in\{x,y,z\}$ for all triples. Taking
% $x=z=0$ shows that every element equals its inverse, and taking distinct nonzero elements then gives
% a contradiction. Thus there is at most one nonzero element and $|A|=2$ for a nontrivial algebra.
\exercise[Schwarz] 
Find structures $\bB$ and $\bC$ with finite relational signatures such that
%\begin{itemize}
%\item 
$\Pol(\bC)$ equals the set of all idempotent operations generated by $\Pol(\bB) \cup \{x \mapsto b \mid b \in B\}$, and $\Csp(\bB)$ is NP-hard and $\Csp(\bC)$ is in P. 
\end{exercises}

\subsection{Structures with an Idempotent Affine Polymorphism Clone}
\label{sect:aff-pp}
In this section we show that if a finite structure of size at least two has an \emph{idempotent} affine polymorphism algebra, it can simulate systems of linear equations over a finite field. 

\begin{proposition}\label{prop:pp-construct-finite-field}
Let $\bB$ be a finite structure with at least two elements and let 
$\fA$ be an affine idempotent algebra such that $\Clo(\fA) = \Pol(\bB)$. Then there exists a prime number $p$ such that $\bB$ pp-constructs 
$({\mathbb Z}_p;+,1)$. 
\end{proposition}
\begin{proof}
Let $R_+$ 
be the graph of addition in the abelian group underlying $\fA$. Since
$m(x,y,z)=x-y+z$ is central in $\fA$
(Lemma~\ref{lem:m-central}) and $\fA$ is idempotent, the operation
$(x,y)\mapsto m(x,0,y)=x+y$ is central as well. Hence, $R_+$ is
preserved by $\Pol(\bB)=\Clo(\fA)$ and has a  primitive positive
definition $\alpha_+$ in $\bB$ by Theorem~\ref{thm:inv-pol}. 
Every operation in $\Pol(\bB)=\Clo(\fA)$ is idempotent and therefore
preserves the singleton relation $\{b\}$. Again by
Theorem~\ref{thm:inv-pol}, there is a primitive positive formula
$\beta_b(x)$ defining $\{b\}$ in $\bB$.

Since $A$ has at least two elements, the
finite abelian group $(A;+,-,0)$ contains an element of prime order.
Fix such a prime $p$. The formula
\begin{align*}
\phi_p(x):=\exists s_2,\dots,s_p\bigl(&\alpha_+(x,x,s_2)
 \wedge \bigwedge_{j=2}^{p-1}\alpha_+(s_j,x,s_{j+1}) \wedge \beta_0(s_p) \bigr)
\end{align*}
defines $A_p:=\{x\in A\mid px=0\}$,
which is a non-trivial subgroup of $(A;+,-,0)$ and an abelian $p$-group. So by elementary group theory (see, e.g., Theorem 5 in Chapter 5 of~\cite{DummitFoote}) 
there exist $k\geq1$ and a group isomorphism
\[
i\colon A_p\to({\mathbb Z}_p)^k.
\]
Put
\[
b:=i^{-1}(1,0,\dots,0).
\]
Let $\bC$ be the relational structure with domain $A_p$, with the
restriction of $R_+$ to $(A_p)^3$, and with the unary relation
$\{b\}$. The structure $\bC$ has a one-dimensional primitive positive
interpretation in $\bB$: use $\phi_p(x)$ as the domain formula,
\[
\phi_p(x) \wedge \phi_p(y)\wedge \phi_p(z)\wedge \alpha_+(x,y,z)
\]
as the formula for its addition relation, and $\beta_b(x)$ as the
formula for its singleton relation.

It remains to observe that $\bC$ is homomorphically equivalent to
$({\mathbb Z}_p;+,1)$, where $+$ denotes the graph of addition and
$1$ denotes the singleton relation $\{1\}$. A homomorphism from
$\bC$ to $({\mathbb Z}_p;+,1)$ is
\[
\pr_1 \circ i,
\]
and a homomorphism in the other direction is
\[
x\mapsto i^{-1}(x,0,\dots,0).
\]
Both maps preserve addition, and they map $b$ to $1$ and $1$ to $b$,
respectively. Hence, $({\mathbb Z}_p;+,1)$ is homomorphically
equivalent to a structure with a primitive positive interpretation
in $\bB$. Therefore, $\bB$ pp-constructs
$({\mathbb Z}_p;+,1)$.
\end{proof}

\subsection{The Term Condition}
\label{sect:term-cond}
We will see that affine algebras 
satisfy a general `universal-algebraic' condition, abelianness. 

%are abelian (Corollary~\ref{cor:aff-ab}); this follows easily from the following characterisation of abelianess. 
%We  a general `universal-algebraic' definition. 

\begin{definition}\label{def:abelian}
An algebra $\fA$ is \emph{abelian} if 
%and only if 
it \emph{satisfies the term condition}, i.e., for
every term $t$ of arity $k+1$, 
all $a,b \in A$ and tuples $c,d \in A^k$
$$t(a,c) = t(a,d) \Rightarrow t(b,c) = t(b,d).$$ 
\end{definition}
We also say that in Definition~\ref{def:abelian} the term condition is \emph{applied to the first argument of $t$}; 
since we can permute arguments of $t$ it is clear what is meant by applying the term condition to other arguments of $t$. 

\begin{example}
Every algebra all of whose operations are essentially unary is abelian. 
\end{example}

%\begin{proof}
%TODO
%\end{proof}

\begin{example}\label{expl:abelian}
A group $\fG = (G;\circ,^{-1},e)$ (Example~\ref{expl:groups}) is abelian if and only if multiplication is commutative, i.e., $\fG$ satisfies $x \circ y \approx y \circ x$. 
Let us consider the term operation 
$$[z_1,z_2] := z_1^{-1} \circ z_2^{-1} \circ z_1 \circ z_2$$ (the \emph{commutator} from group theory)
and let $x,y \in G$. Then $[e,y] = e = [e,e]$. The term condition implies that we can exchange $e$ in the first argument of the term by $x$, and obtain $[x,y]=[x,e] = e$. Thus, $ [x,y] = x^{-1} y^{-1} x y = e$
which implies that $xy = yx$.   
The converse direction follows from Lemma~\ref{lem:aff-ab}. 
\end{example}

\begin{lemma}\label{lem:aff-ab}
Every affine algebra $\fA$ is abelian. 
\end{lemma}
\begin{proof}
To verify the term condition of $\fA$, let $t$ be a term operation. By assumption, $t$ can be written as $t(x,y_1,\dots,y_n) = \alpha_0 x + \sum_{i \in \{1,\dots,n\}} \alpha_i y_i + c$. Now, if $a,b \in A$ and $u,v \in A^n$ then
\begin{align*}
t(a,u) = t(a,v) & \Leftrightarrow \sum_{i \in \{1,\dots,n\}} \alpha_i u_i = \sum_{i \in \{1,\dots,n\}} \alpha_i v_i \\
& \Leftrightarrow t(b,u) = t(b,v) . \qedhere
\end{align*}
\end{proof}

\begin{lemma}\label{lem:term-condition-polynomials}
Let $\fA$ be an algebra. Then $\fA$ satisfies the term condition if and
only if, for every $(k+1)$-ary polynomial operation
$p\colon A^{k+1}\to A$, all $a,b\in A$, and all $c,d\in A^k$, we have
\[
p(a,c)=p(a,d)\quad\Longrightarrow\quad p(b,c)=p(b,d).
\]
\end{lemma}

\begin{proof}
Suppose first that $\fA$ satisfies the term condition, and let
$p\colon A^{k+1}\to A$ be a polynomial operation. Then there exist a
term
\[
t(x,y_1,\dots,y_k,z_1,\dots,z_\ell)
\]
and parameters $e_1,\dots,e_\ell\in A$ such that
\[
p(x,y_1,\dots,y_k)
=
t^{\fA}(x,y_1,\dots,y_k,e_1,\dots,e_\ell)
\]
for all $x,y_1,\dots,y_k\in A$.
Let $a,b\in A$ and $c,d\in A^k$, and suppose that $p(a,c)=p(a,d)$.
Writing $e:=(e_1,\dots,e_\ell)$, this means that
\[
t^{\fA}(a,c,e)=t^{\fA}(a,d,e).
\]
Applying the term condition to the first argument of $t$ and to the
tuples $(c,e)$ and $(d,e)$ gives
\[
t^{\fA}(b,c,e)=t^{\fA}(b,d,e).
\]
Hence $p(b,c)=p(b,d)$.

Conversely, every term operation is a polynomial operation, obtained
by taking no parameters. Therefore, if the displayed condition holds
for all polynomial operations, then it holds in particular for all
term operations, and thus $\fA$ satisfies the term condition.
\end{proof}

The following result was found by H.\ P.\ Gumm and, independently, J.\ D.\ H.~Smith; 
note that it is \emph{not} restricted to algebras with a finite domain. 

\begin{theorem}[Fundamental theorem of abelian algebras; see~\cite{McKenzieMcNultyTaylor}]
\label{thm:abelian}
Let $\fA$ be an algebra.
% with a Maltsev term $m$. 
Then the following are equivalent.
\begin{enumerate}
\item[$(1)$] $\fA$ is abelian and has a Maltsev term $m$. 
%$p^{\fA}$ is a homomorphism from $\fA^3 \to \fA$. 
\item[$(2)$] $\fA$ is affine.  
\item[$(3)$] There exists an 
abelian group $(A;+,-,0)$ such that the operation $(x,y,z) \mapsto x-y+z$ belongs to $\Clo({\fA})$ and is central in $\fA$. 
%\item If $p(x,y,z)$ is a Maltsev term in $\fA$, then 
%$p$ is central in $\fA$. 
\item[$(4)$] $\fA$ has a central Maltsev term $m$. 
\end{enumerate}
\end{theorem}
\begin{proof}
%We have already seen that $(2)$ implies $(1)$ (Lemma~\ref{lem:aff-ab})
%and that $(2)$ implies $(4)$ Lemma~\ref{lem:m-central}). 
% STIMMT GAR NICHT, m-central geht ja um x-y+z, aber das gegebene m ist ja gar nicht unbedingt x-y+z! Dafuer brauchen wir noch Lemma 12.1...
%We first prove that (4) implies (1), then that (1) implies (2), and finally that (3) implies (4). 
We prove these implications in cyclic order. 

For the implication from (1) to (2), we need to construct a module $\fM$ that is polynomially equivalent to $\fA$. Arbitrarily fix $0 \in A$. Define $x +^{\fM} y := m(x,0,y)$ and $-^{\fM} x := m(0,x,0)$. 
The scalar ring $\fR$ has domain
\[
R:=\{r\in A^A\mid r\text{ is a unary polynomial operation of }\fA
\text{ and }r(0)=0\}.
\]
For $r,s\in R$, define
\begin{itemize}
\item $rs$ by $(rs)(x):=r(s(x))$ for all $x \in A$;
\item $r+s$ by $(r+s)(x):=m(r(x),0,s(x))$ for all $x \in A$;
\item $-r$ by $(-r)(x):=m(0,r(x),0)$ for all $x \in A$;
\item $0^{\fR}$ to be the constant-zero operation;
\item $1^{\fR}$ to be the identity operation.
\end{itemize}
These operations are well-defined on $R$. Indeed, polynomial
operations are closed under composition, and the displayed operations
are polynomial operations of $\fA$. Moreover, each of them maps $0$
to $0$, since every $r\in R$ satisfies $r(0)=0$ and
$m(0,0,0)=0$.

For $r\in R$, define scalar multiplication on $M:=A$ by
\[
ra:=r(a).
\]
We first verify that $(M;+,-,0)$ is an abelian group.
\begin{itemize}
\item For every $a\in A$,
\[
a+0=m(a,0,0)=a=m(0,0,a)=0+a,
\]
so $0$ is a neutral element.

\item For associativity, consider the polynomial operation
\[
t(x_1,x_2,x_3,x_4):=(x_1+x_2)+(x_3+x_4).
\]
For all $b,c\in A$, we have
$t(0,0,b,c)=t(0,b,0,c)$. Applying the term condition for polynomial
operations, Lemma~\ref{lem:term-condition-polynomials}, to the first
argument gives
\[
t(a,0,b,c)=t(a,b,0,c)
\]
for every $a\in A$. Hence,
\[
a+(b+c)=(a+b)+c.
\]

\item For all $a,b\in A$, the Maltsev identities give
\[
m(a,a,b)=b=m(b,a,a).
\]
Applying the term condition to the middle argument yields
\[
m(a,0,b)=m(b,0,a),
\]
and therefore $a+b=b+a$.

\item To see that $-a=m(0,a,0)$ is the inverse of $a$, consider the
polynomial operation
\[
t(x,u_1,u_2):=u_1+m(x,u_2,0).
\]
For every $a\in A$, we have $t(a,a,a)=t(a,0,0)$. Applying the term
condition to the first argument gives
\[
t(0,a,a)=t(0,0,0),
\]
and hence $a+(-a)=0$.
\end{itemize}

Next, every $r\in R$ is an endomorphism of the abelian group
$(M;+,-,0)$. To see this, consider the polynomial operation
\[
t(x,y):=r(x+y)-r(x)-r(y).
\]
For every $b\in A$, we have $t(0,b)=0=t(0,0)$. The term condition
applied to the first argument gives $t(a,b)=t(a,0)=0$ for all
$a,b\in A$. Thus,
\[
r(a+b)=r(a)+r(b).
\]
Together with $r(0)=0$, this also implies $r(-a)=-r(a)$.
The closure observations above show that $R$ is a unital subring of $\operatorname{End}(M)$. Hence,  evaluation makes $M$ a unitary $R$-module.

To show that $\fA$ and $\fM$ are polynomially equivalent, first observe that 
every operation of $\fM$ has been defined by a polynomial over $\fA$. Conversely, let $p$ be an $n$-ary polynomial operation of $\fA$.  
We prove by induction on $n$ that $p$ is a polynomial operation of $\fM$. 
If $n = 1$ then consider the unary polynomial operation $r(x) := p(x) - p(0)$.
We have $r \in R$, and thus see that $p(x) = rx + p(0)$
is indeed a polynomial operation of $\fM$. 
If $n > 1$, let $t(x_1,\dots,x_n)$ be the polynomial 
$$p(x_1,x_2,\dots,x_n) - p(0,x_2,\dots,x_n) - p(x_1,0,\dots,0) + p(0,0,\dots,0).$$
We have $t(0,a_2,\dots,a_n) = 0 = t(0,0,\dots,0)$ for all $a_2,\dots,a_n \in A$, 
and by the term condition we get that
$t(a_1,a_2,\dots,a_n) = t(a_1,0,\dots,0) = 0$. So $$p(x_1,x_2,\dots,x_n) = p(0,x_2,\dots,x_n) + p(x_1,0,\dots,0) - p(0,0,\dots,0);$$
the three polynomials on the right have fewer variables and by the induction hypothesis can be written as polynomials over $\fM$, which shows that $p$ can be written as a polynomial over $\fM$ as well.

$(2)$ implies $(3)$. The assumptions imply that there exists an abelian group $(A;+,-,0)$ such that in particular the operation 
$m \colon (x,y,z) \mapsto x-y+z$ is a 
  polynomial operation in $\fA$. 
% By Lemma~\ref{lem:m-unique}, 
%this operation equals $m$. 
   We have to show that $m$ is not only a polynomial operation, but 
	even a term operation of $\fA$. 
Let $t$ be a term such that $t^{\fA}(x,y,z,a_1,\dots,a_n) = x-y+z$ for some constants $a_1,\dots,a_n \in A$. 
Since $\fA$ is affine, there are $\lambda_1,\dots,\lambda_n \in R$ and $c \in A$ such that $t^{\fA}$ can be written in the form
$$(x,y,z,w_1,\dots,w_n) \mapsto x-y+z + \sum_{i=1}^n \lambda_i w_i + c.$$
Now consider the term 
	$$s(x,y,z) := t(x,t(y,y,y,y,\dots,y),z,y,\dots,y).$$ 
	Note that 
	\begin{align*}
	s^{\fA}(x,y,z) & = x - (y + \sum \lambda_i y + c) + z + \sum \lambda_i y + c 
	\\
	& = x - y + z.
	 \end{align*}  
	 so indeed $m \in \Clo(\fA)$. 
Lemma~\ref{lem:m-central} states that $m$ is central.

$(3)$ implies $(4)$. The operation \(m\colon(x,y,z)\mapsto x-y+z\) belongs to \(\Clo(\fA)\), is a Maltsev operation, and is central in \(\fA\). Hence \(m\) is a central Maltsev term operation of \(\fA\).

$(4)$ implies $(1)$. 
Suppose that $m$ is central. We verify that $\fA$ satisfies the term condition. Let $t$ be a term operation of $\fA$ and 
let $x,y \in A$ and $u,v \in A^n$ be such that 
$t(x,u) = t(x,v)$. We have to show that
$t(y,u) = t(y,v)$. And indeed, 
\begin{align*} 
t(y,u) & = m(t(y,u),t(x,u),t(x,v)) && \text{(since $m$ is Maltsev)} \\
& = t(m(y,x,x),m(u_1,u_1,v_1),\dots,m(u_n,u_n,v_n)) && \text{(centrality)} \\
& = t(y,v). && \qedhere 
\end{align*}
%and use Lemma~\ref{lem:abelian}. 
\end{proof}

\begin{exercises}
\exercise[Blau]\label{exe:source-210} \label{exe:abelian-subalgebra} Show that subalgebras of abelian algebras are abelian.
% Suppose A is abelian and B \leq A.
% Then B satisfies the term condition because A
% does (using lemma: if B is subalgebra of A,
% then term operations of B are restrictions of term operations  of A)
\exercise[Blau]\label{exe:source-211} Show that a semilattice $(L;\wedge)$ (Example~\ref{expl:semilattice}) is abelian if and only if $|L|=1$.
% source: Kompatscher script.
% If |L| = 1 then L is clearly abelian (holds for all 1-element algebras). Conversely: let a,b \in L be distinct.
%Then a \neq (a \wedge b) or b \neq (a \wedge b). wlog the former.
% (a \wedge b) \wedge a = a \wedge b = (a \wedge b) \wedge b
% Term condition:
% a \wedge a = a \wedge b, contradiction.
\exercise[Blau]\label{exe:source-213} A ring $\fR$ (Example~\ref{expl:ring}) is abelian in the sense of Definition~\ref{def:abelian}
 if and only if for all $x,y \in R$ we have $x \cdot y = 0$.
% Source: Libor's notes.
% Solution from Michael K's notes:
% Consider term t(z_1,z_2) := z1 \cdot z2
% and let
%$x,y \in R. Then t(0,y) = t(0,0) = 0.
% Applying term condition:
% t(x,y) = t(x,0) = 0.
% Backwards: may remove all multiplications from any term in R. The resulting terms are linear and hence satisfy the term condition.
	\exercise[Blau]\label{exe:source-216} Let $(A,+,-,0)$ be a group. Show that $(x,y,z) \mapsto x-y+z$ is central in $\fA$  if and only if
	$\{(x,y,u,v) \in A^4 \mid x+y = u+v\}$ is a subalgebra of ${\bf A}^4$.
	% Appears in Brady's notes as "quasi affine".
	% x+y=u+v \leq A^4
	% iff
	% for every f \in Clo(A) and (x,y,u,v) \in A^4n with x+y=u+v, we have f(x)+f(y)=f(u)+f(v)
	% iff
	% for every f \in Clo(A) and a,b,c \in A^n
	% we have f(a-b+c) = f(a)-f(b)+f(c)
	% Last iff: forward:
	% have a+c = b + (a+c-b)
	% and hence f(a)+f(c) = f(b) + f(a-b+c)
	% thus f(a-b+c) = f(a) - f(b) + f(c)
	% backward:
	% if x+y=u+v, then x+y-u=v and thus
	% f(v) = f(x+y-u) = f(x) + f(y) - f(u)
	% and the claim follows.
\exercise[Rot]\label{exe:source-209} Show that if $(A;+,-,0)$ is a group, and
$\fA = (A;m)$ where
 $m$ is the
Maltsev operation given by $(x,y,z) \mapsto x-y+z$,
then
$\fA$ is abelian
if and only if the group
$(A;+,-,0)$ is abelian.
%\newpage
% Proof: if the group is Abelian, then it is
% abelian in the universal algebraic sense (Example~\ref{expl:abelian},
% so in particular the Maltsev term operation satisfies the term condition, and hence $A$ is abelian.
% Conversely, if A is Abelian, then by (2) of the fundamental theorem, A is affine, and
% by Lemma 12.1 m is the unique Maltsev
% polynomial operation over the group,
% and so the group must be Abelian as well.
\exercise[Rot]\label{exe:source-212} Show that subalgebras of affine algebras are affine. \label{exe:affine}
% Sei A affine und B \leq A. Dann hat A Maltsev und ist Abelsch. Damit ist auch B Abelsch nach Uebung und hat ebenfalls einen Maltsev term (den selben). Nach dem Satz der VL ist B also affin.
%\newpage
\exercise[Rot]\label{exe:source-215} Show that $({\mathbb Q};(x,y) \mapsto \frac{x+y}{2})$ is idempotent abelian,
 but has no Maltsev polynomial.
\label{exe:affine-Q}
% Abelian: every term t(x1,...,xn) can be rewritten into
% 1/2^{-n} k_i x_i. Now just verify the term condition.
% use \leq to show that no Maltsev.
% Source: Michael K.'s notes.  Libor's notes.
%\newpage
	\exercise[Rot]\label{exe:source-217} Let $(A,+,-,0)$ be a group. Show that
	$(x,y,z) \mapsto x-y+z$ is central in $\fA$
	if and only if  for every $f \in \Clo({\bf A})$ there exist $a \in A$ and endomorphisms $e_1,\dots,e_n$
	of $(A;+,-,0)$
	such that for all $x_1,\dots,x_n \in A$
	$$f(x_1,\dots,x_n) = \sum_{i=1}^n e_i(x_i) + a.$$
	% Solution:
	% 2->1: f(x-y+z) = \sum_{i=1}^n e_i(x_i - y_i + z_i) + a = f(x) - f(y) + f(z)
	% 1->2: define a := f(0,...,0) and
	% e_i (x_i) = f(0,...,0,x_i,0,...,0) - a
	% By assumption:
	% f(x1,...,xn) = f(x1,0,...,0)-f(0,...,0) + f(0,x2,...,xn) = e1(x_1) + f(0,x2,...,xn)
	% and by induction
	% f(x1,...,xn) = \sum e_i(x_i) + a.
	% e_n(0) = 0
	% e_n(x) + e_n(y) = e_n(x+y)
	% x_n(-x) = -e_n(x)
	% so e_n is indeed endo of A.
%\item \label{exe:abelian}
% Vom Besuch von Armin Weiss:
%	Let $(A;+,-,0)$ be a group. Show that $(x,y,z) \mapsto x-y+z$ preserves the relation $\{(a,b,c) \in A^3 \mid a+b=c\}$ if and only if the group is abelian.
	% Solution:
	% if A is Abelian, then simply verify that
	% x1-x2+x3 + y1-y2+y3 = (x1+y1) - (x2+y2) + (x3+y3).
	% Conversely, if we have this identity,
	% then -x2+x3+y1-y2 = y1-x2+y2+x3
	% Ja und?
\exercise[Rot]\label{exe:source-218} \label{exe:armin}
	Let $\fA$ be a group. Show that the structure $(A;\{(a,b,ab) \mid a,b \in A\})$ has a Maltsev polymorphism if and only if $\fA$ is abelian (see Example~\ref{expl:abelian}). 
	% Solution:
	% If \fA=(A;+) is abelian, $+$ is commutative (Exercise), and the operation
	% (x,y,z) \mapsto x y^{-1} z is a Maltsev operation that preserves the given relation.
	% Conversely, suppose that m is a
	% Maltsev operation and preserves
	% R := {(a,b,ab): a,b \in A}.
	% Applying m to (a,1),(1,1),(1,b)
	% we obtain (a,b).
	% Since (a,1,a),(1,1,1),(1,b,b) \in R,
	% we thus have ab = m(a,1,b).
	%
	% Applying m to (1,a),(1,1),(b,1)
	% we also obtain (b,a).
	% Since (1,a,a),(1,1,1),(b,1,b) \in R,
	% we thus have ba = m(a,1,b),
	% hence ab = ba. Now use again the exercise. 
	\exercise[Rot]\label{exe:source-219} Let $\fA$ be an algebra. Show that if $\fA$ has a Maltsev polymorphism (here we study polymorphisms of an algebra, not a relational structure!), then $\fA$ is abelian.
	% Source: Larose Zadori Lemma 3.8
	% in `complexity of polynomial equations'
%	Proof. Let m denote the Mal?tsev operation that commutes with the basic operations of B and let t be any term of B. Let u,v,x2,...,xn,y1,...,yn be such that t(u,x2,...,xn) = t(u,y2,...,yn). Consider the following 3 \ensuremath{\times} 2(n + 1) matrix: (u x2 \ensuremath{\cdot}\ensuremath{\cdot}\ensuremath{\cdot} xn u y2 \ensuremath{\cdot}\ensuremath{\cdot}\ensuremath{\cdot} yn? ?uu\ensuremath{\cdot}\ensuremath{\cdot}\ensuremath{\cdot}uuu\ensuremath{\cdot}\ensuremath{\cdot}\ensuremath{\cdot}u?.vu\ensuremath{\cdot}\ensuremath{\cdot}\ensuremath{\cdot}uvu\ensuremath{\cdot}\ensuremath{\cdot}\ensuremath{\cdot}u
%By hypothesis, if we apply the term t to the first n + 1 entries of any row we get
%the same result as applying it to the last n + 1 entries:
%t(u,x2,...,xn) = t(u,y2,...,yn) t(u,u,...,u) = t(u,u,...,u) t(v,u,...,u) = t(v,u,...,u).
%Since m and t commute, applying the operation m to the columns of the matrix of the entries will yield
%t(v,x2,...,xn) = t(m(u,u,v),m(x2,u,u),...,m(xn,u,u))
%= m(t(u,x2,...,xn),t(u,u,...,u),t(v,u,...,u)) = m(t(u,y2,...,yn),t(u,u,...,u),t(v,u,...,u))
%= t(m(u,u,v),m(y2,u,u),...,m(yn,u,u)) = t(v,y2,...,yn).
\exercise[Rot] % Source: myself
Show that the converse of the previous exercise fails: 
there are Abelian algebras that do not have a Maltsev polymorphism. 
\ignore{ GPT 5.6 solution, 2.9.26
Take the unary algebra
\[
\mathbf A=(\{0,1,2\};f,g),
\]where
\[
\begin{array}{c|ccc}
x&0&1&2\\ \hline
f(x)&0&0&1\\
g(x)&0&1&1
\end{array}
\]Every term operation of \(\mathbf A\) is essentially unary, so \(\mathbf A\) is abelian.
Suppose that \(m\) were a Maltsev polymorphism of \(\mathbf A\), and put \(a:=m(0,1,2)\). Since \(m\) commutes with \(f\),
\[
f(a)=m(f(0),f(1),f(2))=m(0,0,1)=1.
\]The only element mapped to \(1\) by \(f\) is \(2\), so \(a=2\). But commutation with \(g\) gives
\[
g(a)=m(g(0),g(1),g(2))=m(0,1,1)=0.
\]On the other hand, \(a=2\) implies \(g(a)=g(2)=1\), a contradiction.
Thus \(\mathbf A\) is abelian but has no Maltsev polymorphism.}
\exercise[Rot]\label{exe:source-220} Show that if $m$ is a commutative heap operation on $A$, then $(A;m)$ is abelian in the universal-algebraic sense (i.e., it satisfies the term condition).
% Source: also showed up in the Prag 26
% POCOCOP meeting, but not for heaps, but for Maltsev in general. With Armin?
% proof: comes from a group since it is a heap, as in Exercise~\ref{exe:source-219}. Then this group is also commutative, and hence
% abelian in the universal-algebraic sense.
Conversely, show that if $m$ is a heap operation on $A$ such that $(A;m)$ is abelian in the universal-algebraic sense, then $m$
must be commutative.
%(proof: if heap is abelian in the universal-algebraic sense,
% then by Thm 12.12 there exists abelian group s.t. x-y+z is central, and unique Maltsev, so equal to m, and hence m is commutative.
\exercise[Orange]\label{exe:source-214} We have seen in Lemma~\ref{lem:term-condition-polynomials} that in definition of the term condition
(Definition~\ref{def:abelian}),
we could have equivalently phrased the condition for polynomial operations instead of term operations $t$. However, show that it is not sufficient to require the condition only for the operations of the algebra.
% Source: Kompatscher Skript
% Hint for second part: "groups." (signature?)
%\vspace{-1cm}
%\newpage
% Solution sketch: Take the nonabelian group $S_3$ in its usual signature $(\cdot,{}^{-1},1)$.
% Cancellation shows that multiplication satisfies the term-condition implication when tested as a
% basic operation; inverse and the constant do so trivially. Nevertheless the group algebra is not
% Abelian: a commutator term evaluated at two noncommuting permutations violates the term condition.
% Thus testing only the basic operations is insufficient.
\exercise[Schwarz]\label{exe:source-221} % Source: me, inspired by POCOCOP meeting 26.
Find a Maltsev operation $m$ on a finite set which satisfies $m(x,y,z) \approx m(z,y,x)$
% Solution sketch: On $A=\{0,1,2,3\}$ put $m(x,y,z)=z$ if $x=y$, put $m(x,y,z)=x$ if $y=z$, and put
% $m(x,y,z)=y$ otherwise. This is a symmetric Maltsev operation. Heap associativity fails at
% $(x,y,z,u,v)=(0,1,0,0,2)$, whereas every affine Maltsev operation is a heap; hence this operation is
% not affine.
\end{exercises}

\subsection{The Congruence Condition}
\label{sect:congr-cond}
We close with a relational characterisation of abelianness (which for some authors is the official definition of abelianness).
Recall that $\Delta_A$ denotes $\{(a,a) \mid a \in A\}$, and that $\Cg_{\fA}(X)$ denotes the smallest congruence of $\fA$ that contains $X$ (see Lemma~\ref{lem:cg}). 

\begin{theorem}\label{thm:congr-ab}
% Source: Libor Notes! But also in Ihringer and in Jezek. 
Let $\fA$ be an algebra and $\Delta := \Delta_A$. Then the following are equivalent. 
\begin{enumerate}
\item $\fA$ is abelian. 
\item $\Delta$ is a congruence class of $\Cg_{\fA^2}(\Delta^2)$. 
\item $\Delta$ is a congruence class of a congruence of $\fA^2$. 
\end{enumerate}
\end{theorem}
\begin{proof}
The implication $(2) \Rightarrow (3)$ is trivial.
For the implication $(3) \Rightarrow (2)$, 
suppose that $C$ is a congruence of $\fA^2$ where $\Delta$ is a congruence class. Since $C' :=\Cg_{\fA^2}(\Delta^2)$
contains $\Delta^2$ we have that $\Delta$ is contained in a congruence class of $C'$. But since $C' \subseteq C$, 
this congruence class must be $\Delta$. 

For the equivalence between $(1)$ and $(2)$, recall from Lemma~\ref{lem:cg} that $\Cg_{\fA^2}(\Delta^2)$
equals the reflexive symmetric transitive closure of
%\begin{align*}
$$\{(p(u),p(v)) \mid u,v \in \Delta, p \text{ a unary polynomial operation of } \fA^2\}.$$
Note that every unary polynomial operation $p(x)$ of $\fA^2$ can be written as
$$f(x,{c_1 \choose d_1},\dots,{c_n \choose d_n})$$
for some $c_1,d_1,\dots,c_n,d_n \in A, n \in {\mathbb N}$ and $f \in \Clo(\fA)$. 
Hence, $\Delta$ is a congruence class of $\Cg_{\fA^2}(\Delta^2)$ if and only if for all $a,b \in A,c_1,d_1,\dots,c_n,d_n \in A, n \in {\mathbb N}$ and every term $t$
% f \in \Clo(\fA^2)^{(n)}$ 
%we have
$$t^{\fA^2}({a \choose a},{c_1 \choose d_1},\dots,{c_n \choose d_n}) \in \Delta \text{ if and only if }
t^{\fA^2}({b \choose b},{c_1 \choose d_1},\dots,{c_n \choose d_n}) \in \Delta;$$
this is exactly the term condition for $\fA$ applied to the first argument of $t$. 
%note that $\Delta_A$ is a congruence class of
%$\Cg_{\fA^2}(\Delta^2)$
%if and only if it is the transitive symmetric closure for every unary polynomial operation $f$ of $\fA^2$ we have that
%$f((a,a)) \in \Delta_A$ 
\end{proof} 

\begin{example}\label{expl:affine-m}
Let $n \geq 1$. Let $\fA$ be 
the algebra $({\mathbb Z}_n;m)$, where $m \colon {\mathbb Z}_n^3 \to {\mathbb Z}_n$ is given by $(x,y,z) \mapsto x-y+z$. Then $\fA^2$ has the congruence $C := \{ ( (x_1,x_2), (y_1,y_2) ) \mid  (x_1-x_2 = y_1 - y_2) \}$ 
and clearly $\{(a,a) \mid a \in A\}$ is a congruence class of $C$. 
\end{example}
%With the following definition we generalise Example~\ref{expl:affine-m}.

We now present practical conditions for ternary subpowers that imply that $\fA$ is abelian. 

\begin{proposition}\label{prop:prove-abelian}
% Source?!
Let $\fA$ be an algebra 
with %a subdirect 
$R \leq {\fA}^3$ such that 
for every $a \in A$ and $i \in \{1,2,3\}$ the binary relation defined by $\exists x_i (R(x_1,x_2,x_3) \wedge x_i = a)$ is the graph of an automorphism of $\fA$. 
Then $\fA$ is abelian. 
\end{proposition}
\begin{proof}
Note that the assumptions imply that $R$ is the graph of a surjective binary operation $f \colon A^2 \to A$. 
Also note that $f$ is central, i.e., $f \colon \fA^2 \to \fA$ is a homomorphism, because $R \leq \fA^3$. 
Arbitrarily pick $a \in A$. Then $f^{-1}(a)$ is the graph of an automorphism 
$\alpha$ of $\fA$. 
Hence, $(x,y) \mapsto  f(x,\alpha(y))$ is central, and its kernel $C$ is a congruence of $\fA^2$. 
The congruence class $(a,a)/C$ equals $\Delta_A$ and the statement follows from Theorem~\ref{thm:congr-ab}. 
\end{proof} 

A relation $R \subseteq A^3$ is called \emph{strongly functional} if
\begin{itemize}
\item for distinct $i,j \in \{1,2,3\}$ we have $\pi_{i,j}(R) = A^2$,  and
\item for every $a,b \in A$ there exists precisely one $c \in A$ such that $(a,b,c) \in R$. 
\end{itemize} 

The following is from~\cite{theoretics:11361}. 

\begin{proposition}
Let $\fA$ be a finite idempotent algebra
and let $R \leq \fA^3$ be strongly functional.  Then $\fA$ is abelian. 
\end{proposition}
\begin{proof}
By strong functionality, there is a binary operation
$q\colon A^2\to A$ such that
\[
R=\{(x,y,q(x,y))\mid x,y\in A\}.
\]
For every $x\in A$, define $\lambda_x\colon A\to A$ by
$\lambda_x(y):=q(x,y)$. Since $\pi_{1,3}(R)=A^2$, the map
$\lambda_x$ is surjective. Since $A$ is finite, it is bijective.
Similarly, for every $y\in A$, the map $\rho_y\colon A\to A$ given by
$\rho_y(x):=q(x,y)$ is surjective, because
$\pi_{2,3}(R)=A^2$, and hence it is bijective.

We verify the assumptions of Proposition~\ref{prop:prove-abelian}.
Let $a\in A$. The binary relation obtained by fixing the first
coordinate of $R$ to $a$ is
\[
S_a^1:=\{(y,z)\in A^2\mid (a,y,z)\in R\}
      =\{(y,\lambda_a(y))\mid y\in A\},
\]
so it is the graph of the bijection $\lambda_a$. Similarly,
\[
S_a^2:=\{(x,z)\in A^2\mid (x,a,z)\in R\}
      =\{(x,\rho_a(x))\mid x\in A\}
\]
is the graph of the bijection $\rho_a$.
Finally, consider
\[
S_a^3:=\{(x,y)\in A^2\mid (x,y,a)\in R\}.
\]
For every $x\in A$, the bijectivity of $\lambda_x$ gives a unique
$y\in A$ such that $q(x,y)=a$. Hence, $S_a^3$ is the graph of the
function
\[
\sigma_a\colon A\to A,\qquad
\sigma_a(x):=\lambda_x^{-1}(a).
\]
For every $y\in A$, the bijectivity of $\rho_y$ gives a unique
$x\in A$ such that $q(x,y)=a$. It follows that $\sigma_a$ is
bijective, with inverse 
$\sigma_a^{-1}(y)=\rho_y^{-1}(a)$.
Clearly, each of $S_a^1,S_a^2,S_a^3$ is a subalgebra of
$\fA^2$, using idempotence of $\fA$. 
%Let $S_a^i$ be one of these relations, let $f$ be a basic
%operation of $\fA$ of arity $k$, and let
%\[
%(u_1,v_1),\dots,(u_k,v_k)\in S_a^i.
%\]
%The corresponding triples in $R$ all have $a$ in their $i$-th
%coordinate. Since $R\leq\fA^3$, applying $f$ coordinatewise produces
%another tuple in $R$. Its $i$-th coordinate is $f(a,\dots,a)=a$ by idempotence. Therefore
%\[
%\bigl(f(u_1,\dots,u_k),f(v_1,\dots,v_k)\bigr)\in S_a^i,
%\]
%which proves that $S_a^i\leq\fA^2$.
If the graph of a function $\alpha\colon A\to A$ is a subalgebra of
$\fA^2$, then $\alpha$ is a homomorphism from $\fA$ to $\fA$.
Moreover, if $\alpha$ is bijective, then
% the graph of $\alpha^{-1}$
%is obtained from the graph of $\alpha$ by exchanging the two
%coordinates. Hence it 
%is also a subalgebra of $\fA^2$, and
%then $\alpha^{-1}$ is a homomorphism as well. Thus 
$\alpha$ is an
automorphism of $\fA$.
Consequently, for every $a\in A$ and every $i\in\{1,2,3\}$, the
binary relation obtained by fixing the $i$-th coordinate of $R$ to
$a$ is the graph of an automorphism of $\fA$. The statement now
follows from Proposition~\ref{prop:prove-abelian}.
\end{proof}
% END INLINED SOURCE: Abelian.tex
% BEGIN INLINED SOURCE: absorption.tex
% !TEX root = GH-UA.tex
% !TEX root = Graph-Homomorphisms.tex
\section{Absorption}
\label{sect:absorption}
\begin{flushright}
\emph{``The notion of absorption is, in a sense, complementary to abelianness''} \\
(Barto and Kozik~\cite{absorption})
\end{flushright}
Absorption theory is an important topic in universal algebra, developed by Marcin Kozik and Libor Barto, which has powerful applications for the study of homomorphism problems. It can be seen
as a tool to show the existence of certain solutions
in instances of a CSP. This section covers material that stems from~\cite{cyclic,BartoKozikStanovsky,theoretics:11361,Strong-Subalgebras-Published}. 

Throughout, $\fA$ denotes an algebra with a not necessarily finite domain and signature $\tau$.

\begin{definition}[Absorbing subalgebras]
\label{def:absorb}
Let $\fB$ be an algebra and $t(x_1,\dots,x_n)$ a $\tau$-term.  
%\in \Clo(\fA)$ of arity $n$. 
A subalgebra $\fA$ of $\fB$
is called an \emph{absorbing subalgebra of $\fB$ with respect to $t$}, in symbols $\fA \abs_t \fB$, if for all $i \in \{1,\dots,n\}$ 
%$$t^{\fB}(A \times A \times \underbrace{B}_i \times A \times \cdots \times A) \subseteq A,$$
$$t^{\fB}(A^{i-1} \times B \times A^{n-i}) \subseteq A,$$
i.e., if for all $b_1,\dots,b_n \in B$ we have 
$t^{\fB}(b_1,\dots,b_n) \in A$ whenever all
but at most one out of $b_1,\dots,b_n$ 
are from $A$.\footnote{We also use the notation $\abs_{t^{\fB}}$ instead of $\abs_t$.}
If such a $t$ exists
we say that $\fA$ \emph{absorbs} $\fB$,
and write $\fA \abs \fB$. 
A subalgebra $\fA$ of $\fB$ is called \emph{n-absorbing}
if $\fA \abs_t \fB$ for some $t$
with $n$ variables. 
\end{definition}

Since subalgebras are uniquely determined 
by their domain, we also use the notation
$A \abs \fB$ if $A$ is the domain of
an absorbing subalgebra $\fA$ of $\fB$. 
Note that if $\fB$ is idempotent, then $A \subseteq B$ is $1$-absorbing if and only if $A=B$.
 
\begin{remark}\label{rem:abs-arity} If $A$ is $n$-absorbing, then it is also $m$-absorbing for all $m \geq n$. 
\end{remark}
Hence, if we ignore $1$-absorption, which is uninteresting as we have seen above, then $2$-absorption is the strongest form of absorption. 
Subalgebras that are 2-absorbing are also called \emph{binary absorbing}. 
We say that $\fB$ is \emph{absorption-free} if $\fB$ has no proper absorbing subuniverse. 

\begin{remark} 
Impatient readers who would like to see algorithmic applications of the concept of absorption can consult Section~\ref{sect:rrr}. 
\end{remark} 

\begin{example}\label{expl:wedge-absorption}
The subuniverse $\{0\}$ of the algebra $(\{0,1\};\wedge)$ is absorbing with respect to the operation $\wedge$. 
More generally, if $\fB = (B;\wedge)$ where $B$ is finite and non-empty and $\wedge$ is a semilattice operation, then 
$\{\bigwedge B\} \abs \fB$.  
\end{example} 

\begin{example}\label{expl:affine-not-absorbing} 
Let $p$ be a prime, and let $\fB = (\{0,\dots,p-1\};m)$ be the algebra where $m(x,y,z) := x-y+z$ and where $+$ and $-$ are the usual addition and subtraction modulo $p$. Then the only absorbing subuniverses are $\emptyset$ and $B$, so $\fB$ is absorption-free. 
Indeed, the only proper subuniverses of $\fB$ are of the form $\{b\}$, for $b \in B$ (Exercise~\ref{exe:subalgebras-Ap}). 
Every term operation $t$ of $\fB$ has the form $t(x_1,\dots,x_n)=\sum_{i=1}^n\alpha_i x_i$ with $\sum_{i=1}^n\alpha_i=1$ (Example~\ref{expl:affine}). If $\{b\}$ were absorbing with respect to a term $t$, then for every coordinate $i$ and every $a\in B$ we would have $$b = t(b,\dots,b,a,b,\dots,b) = \alpha_i a + \sum_{j \in [n] \setminus \{i\}} \alpha_j b = \alpha_i a+(1-\alpha_i)b = b + (a-b) \alpha_i.$$ Hence, $\alpha_i=0$ for every $i$, contradicting $\sum_i\alpha_i=1$. Thus, no singleton is absorbing. 
\end{example}

\begin{example}\label{expl:nu-absorption} 
If $\fB = (\{0,1\};\majority)$, then both $\{0\} \abs \fB$ 
and $\{1\} \abs \fB$. 
More generally, in any algebra $\fB$ with a near unanimity term $t$ (see Section~\ref{sect:nu}), every one-element subalgebra is absorbing with respect to $t^{\fB}$. 
\end{example}

%Following the presentation in~\cite{absorption}, we will prove in 

Proposition~\ref{prop:nu-absorption} below is a converse to the statement from Example~\ref{expl:nu-absorption}. 
%The following lemma justifies the notation 
%$\abs$ instead of $\abs_f$. 

%in the proof, we need a useful definition. 
%We start with some warm-up statements to strengthen the intuition about absorption. 
%\begin{definition}\label{def:star}
%If $s,t$ are operations (or terms) of arity $p$ and $q$, respectively, then the \emph{star composition} of $s$ and $t$ is the operation $s*t$ (or term) of arity $pq$ 
%defined by 
%$$(x_{1,1},\dots,x_{p,q}) \mapsto 
%%s \big (t(x_{1,1},\dots,x_{1,q}),\dots,t(x_{p,1},\dots,x_{p,q}) \big).$$
%\end{definition}

\begin{lemma}\label{lem:combine-abs}
If $\fB,\fC \abs \fA$ then $\fA$ has a
term operation $f$ such that 
$\fB,\fC \abs_f \fA$. 
\end{lemma}
\begin{proof}
If $\fB \abs_s \fA$
 and 
$\fC \abs_t \fA$ 
for some $s,t \in \Clo(\fA)$, choose $f := s*t$ (Definition~\ref{def:star}). 
%$\f \abs_{s * t} \fA$. 
\end{proof} 

%\begin{remark}
%If $B$ is $n$-absorbing, then it is also $n+1$-absorbing. 
%\end{remark}

\begin{proposition}\label{prop:nu-absorption}
Let $\fB$ be a finite algebra. If every one-element subset is the domain of an absorbing subalgebra
of $\fB$, then $\fB$ has a near unanimity term. 
\end{proposition}
\begin{proof}
%If $\fB \abs_f \fA$ and $\fC \abs_g \fA$, where $f$ is $n$-ary and $g$ is $m$-ary, then $\fB \abs_{f*g} \fA$ and $\fC \abs_{f*g} \fA$. 
If $|B|\leq 1$, then every ternary projection is a near unanimity term, so suppose that $|B|\geq 2$.
Since $\fB$ is finite we can use Lemma~\ref{lem:combine-abs} to construct a single term operation 
$h$ such that $\fA \abs_h \fB$ for every
one-element subalgebra $\fA$ of $\fB$. Note that $h$ must have arity $\ge 3$: a binary $h$ with $\{0\} \lhd_h \mathbf{B}$ and $\{1\} \lhd_h \mathbf{B}$ would
force $h(0,1) \in \{0\}\cap\{1\}$. But then $h$ must be a near unanimity operation. 
\end{proof}

An operation $f \colon B^n \to B$ is called \emph{(fully) symmetric} if $f(x_1,\dots,x_n) \approx f(x_{\rho(1)},\dots,x_{\rho(n)})$ for every permutation $\rho \in \Sym(\{1,\dots,n\})$.  
The following example from~\cite{KunSzegedy} shows a structure 
which has fully symmetric polymorphisms of all arities, but not totally symmetric polymorphisms of arity 3. 

\begin{example}\label{expl:kun} 
Let $\fB$ be the algebra on the domain $B = \{-1,0,1\}$ which has for every $k \geq 1$ the $k$-ary operation $s_k$ defined as follows:
$$
s_k(x_1,\dots,x_k) = \begin{cases} 
0 & \text{ if }
\frac{x_1+\cdots+x_k}{k} \in (-\frac{1}{3},\frac{1}{3}) \\
1 & \text{ if }
\frac{x_1+\cdots+x_k}{k} \geq \frac{1}{3} \\
-1 & \text{ otherwise.}
\end{cases}
$$
Note that $s_k$ is fully symmetric and idempotent. Note that for every $k \geq 1$,
the operation $s_k$ preserves the following two relations
\begin{align*}
R_+ & := \{(a_1,a_2,a_3) \in B^3 \mid a_1+a_2+a_3 \geq 1 \} \\
R_- & := \{(a_1,a_2,a_3) \in B^3 \mid a_1+a_2+a_3 \leq -1 \} 
\end{align*}
Suppose for contradiction that there exists 
a totally symmetric operation $t \colon B^3 \to B$ which preserves $R_+$ and $R_-$. 
Then $$t(1,1,-1) = t(1,-1,1) = t(-1,1,1) = t(1,-1,-1) = t(-1,1,-1) = t(-1,-1,1) =: c.$$
If we apply $t$ to the three tuples $(1,1,-1),(1,-1,1),(-1,1,1) \in R_+$ we obtain 
$(c,c,c) \in R_+$, hence $c = 1$. The same argument applied to $R_-$ instead of $R_+$ shows that $c = -1$, a contradiction. 

Note that $\fB$ has the proper subalgebras $\{-1\},\{0\},\{1\},\{-1,0\},\{0,1\}$. All of them are absorbing, witnessed by $s_k$ for some large enough $k$. 
\end{example} 

\subsection{Absorption Transfer}
Many natural algebraic constructions preserve the absorption property. 

\begin{lemma}\label{lem:abs-trans}
If $\fC \abs \fB$ and $\fB \abs \fA$, then $\fC \abs \fA$. 
\end{lemma}
\begin{proof}
If $\fB \abs_t \fA$ for some $t \in \Clo(\fA)$ and 
$\fC \abs_s \fB$ for some $s \in \Clo(\fB)$, then $\fC  \abs_{s * t} \fA$. 
\end{proof}

\begin{corollary}\label{cor:abs-intersect}
If $\fB \abs \fA$ and $\fC \abs \fA$
then $(B \cap C) \abs \fA$. 
\end{corollary}
\begin{proof}
Note that $(B \cap C) \abs \fC$ with respect 
to the same term as $B \abs \fA$. 
Now the statement follows from 
Lemma~\ref{lem:abs-trans}. 
\end{proof} 

The following transfer principles also preserve the term that witnesses absorption. 

\begin{lemma}\label{lem:abs-hom}
Let $t(x_1,\dots,x_n)$ be a $\tau$-term, let $\sim$ be a congruence of $\fB$, and suppose that $A \abs_t \fB/{\sim}$. Then $\bigcup A \abs_t \fB$.
\end{lemma}
\begin{proof}
%If $B \abs_{t^{\fA/{\sim}}} (\fA/{\sim})$ for some term $t(x_1,\dots,x_n)$ and 
If $b_1,\dots,b_n \in B$ are such that all but one are from $\cup A$, then
$t^{\fB}(b_1,\dots,b_n)/_{\sim} = t^{\fB/\sim}(b_1/_{\sim},\dots,b_n/_{\sim}) \in A$. Hence,  
$\bigcup A \abs_{t} \fB$.
\end{proof}

\begin{lemma}\label{lem:abs-trans-rel}
Let $\fB_1,\dots,\fB_n$ be $\tau$-algebras and let $t$ be a $\tau$-term. Let $R \leq \fB_1 \times \cdots \times \fB_n$ and $A_i \abs_t \fB_i$ for every $i \in [n]$. Then $R \cap (A_1 \times \cdots \times A_n) \abs_t \fR$. 
\end{lemma} 

\begin{example}\label{expl:m1-2}
Consider the Maltsev operation $m$ on $\{0,1,2\}$ from
Example~\ref{expl:m1}. We claim that the algebra
$(\{0,1,2\};m)$ is absorption-free.

First, we show that $\{1\}$ is not absorbing. The set $\{0,1\}$ is a
subuniverse, and the restriction of $m$ to this set is the Boolean
minority operation, that is, addition modulo $2$.
The restriction of every $n$-ary term
operation $t$ to $\{0,1\}$ has the form
\[
t(x_1,\dots,x_n)=\sum_{j=1}^n\alpha_jx_j
\]
for some $\alpha_1,\dots,\alpha_n\in{\mathbb Z}_2$ satisfying
$\sum_{j=1}^n\alpha_j=1$ (Example~\ref{expl:affine}). 
Suppose for contradiction that $t$ witnesses that $\{1\}$ absorbs
$(\{0,1,2\};m)$. For every $i\in[n]$, absorption and the description
of the restriction of $t$ give, with all sums computed in
${\mathbb Z}_2$,
\[
1
=t(1,\dots,1,0,1,\dots,1)
=\sum_{j\neq i}\alpha_j
=1-\alpha_i.
\]
Hence $\alpha_i=0$ for every $i\in[n]$, contradicting
$\sum_{j=1}^n\alpha_j=1$. Thus, $\{1\}$ is not absorbing, and by
symmetry no singleton is absorbing.

Finally, suppose that a two-element subuniverse were absorbing.
Every permutation of $\{0,1,2\}$ is an automorphism of the algebra, so
another two-element subuniverse intersecting it in a singleton would
also be absorbing. Corollary~\ref{cor:abs-intersect} would then imply
that this singleton is absorbing, a contradiction. Therefore,
$(\{0,1,2\};m)$ is absorption-free.
\end{example}

% Lecture notes of Libor stop here,
% the following stuff is only from some earlier papers. The next lemma is from the notes
% of Brady, and needed in the proof of Zhuk's cases. 

\begin{lemma}\label{lem:abs-prod-1}
Let $\fR$ be a subalgebra of $\fA \times \fB$, let $t(x_1,\dots,x_n)$ be a $\tau$-term, and let $S \abs_t \fR$. 
Then $\pi_1(S) \abs_t \pi_1(R)$. 
\end{lemma} 
\begin{proof}
Suppose that $a_1,\dots,a_n \in \pi_1(S)$, $c \in \pi_1(R)$, and $i \in \{1,\dots,n\}$. 
Then there are $b_1,\dots,b_n \in B$ and $d \in B$ such that $(a_1,b_1),\dots,(a_n,b_n) \in S$ and $(c,d) \in R$. Since $S \abs_t \fR$
we have $$t^{\fR}((a_1,b_1),\dots,(a_{i-1},b_{i-1}),(c,d),(a_{i+1},b_{i+1}),\dots,(a_n,b_n)) \in S,$$ 
and hence $t^\fA(a_1,\dots,a_{i-1},c,a_{i+1},\dots,a_n) \in \pi_1(S)$, which proves that 
$\pi_1(S) \abs_t \pi_1(R)$. 
% the statement of the lemma. 
\end{proof} 

\begin{lemma}\label{lem:abs-prod-2}
Let $\fB$ be an idempotent algebra and suppose that $\fB^2$ has a proper $n$-absorbing subalgebra. Then $\fB$ has a proper $n$-absorbing subalgebra as well. 
\end{lemma}
\begin{proof}
Suppose that $A$ is a proper non-empty subset of $B^2$ such that 
 $A \abs_f \fB^2$ for some $f \in \Clo(\fB)^{(n)}$. 
Note that $\pi_1(A) \abs_f \fB$ by Lemma~\ref{lem:abs-prod-1}, and that $\pi_1(A)$ is non-empty. Hence, 
if $\pi_1(A) \neq B$ then we are done, so suppose that $\pi_1(A) = B$. 
Since $A \neq B^2$, there exists a $b \in B$ such that 
$$A' := \pi_2 \big (A \cap (\{b\} \times B) \big) = \{a \in B \mid (b,a) \in A \} \neq B.$$ 
%We have that $B'' := B \times (\{a\} \times A) \leq \fA^2$ since $\fA$ is idempotent,
%and $B' \abs_f B''$ by 
Since $\pi_1(A) = B$ we have that $A' \neq \emptyset$.
Clearly, $A' \leq \fB$ by the idempotence of $\fB$.
So it suffices to show that $A' \abs_f \fB$. Let $a_1,\dots,a_n \in A'$, $i \leq n$, and $c \in B$. We have to show that
$d := f(a_1,\dots,a_{i-1},c,a_{i+1},\dots,a_n) \in A'$. By the definition of $A'$ we have 
$(b,a_1),\dots,(b,a_n) \in A$. Then 
\begin{align*}
& f((b,a_1),\dots,(b,a_{i-1}),(b,c),(b,a_{i+1}),\dots,(b,a_n)) \\
= \; & (f(b,\dots,b),f(a_1,\dots,a_{i-1},c,a_{i+1},\dots,a_n)) = (b,d) \in A
\end{align*} since $A \abs_f \fB^2$, hence $d \in A'$. 
\end{proof}

\begin{exercises}
\exercise[Blau]\label{exe:source-222} \label{exe:abs-fact}
Let $\fB$ be an algebra and $\sim$ a congruence of $\fB$. If $A \abs_f \fB$,
then $A/_{\sim} \abs_f \fB/_{\sim}$ (see Remark~\ref{rem:congnot}).
% Source: Dima Strong Subalgebras, without proof. But easy!
\exercise[Blau]\label{exe:source-223} \label{exe:abs-dummy}
Let $\fA$ and $\fB$ be algebras and $\fC \abs \fA$. Then $\fC \times \fB \abs \fA \times \fB$.
% Source: Dima Strong Subalgebras 6.1.
% There, he uses the pp formula lemma.
% Solution: hypereasy.
% Let t be the term that witnesses
%\fC \abs \fA, say of arity k. Let c1,..,ck-1 \in C, ck \in A, and b_1,...,bk \in B.
% Then t((c1,b1),...,(ck-1,bk-1),(ck,bk)) = (c0,b0) for some c0 \in C and b0 \in B.
%\item \label{exe:abs-product}
%Show that if $\fA$ has no proper $n$-absorbing subuniverse, then so has $\fA^2$.
% IS THIS TRUE?!
\exercise[Blau]\label{exe:source-224} \label{exe:walk}
Let $\fA$ and $\fB$ be two algebras of the same signature,
and let $R$ be the domain of a subalgebra of $\fA \times \fB$. For $X \subseteq A$ %and $Y \subseteq B$
we define
%the \emph{neighborhoods} of $X \subseteq A$ and
%$Y \subseteq B$
%we define
\begin{align*}
X+R & := \{b \in B \mid \exists a \in X \colon R(a,b) \} .
%\\
%Y-R & := \{a \in A \mid \exists b \in Y \colon R(a,b) \}. %= Y + R^{-}.
\end{align*}
%Let $R$ be a subalgebra of $\fA \times \fB$.
Prove that
\begin{itemize}
%\item $Y-R = Y + R^{-1}$;
\item if $X \leq \fA$ then $(X+R) \leq \fB$;
\item if
$R \leq \fA \times \fB$ is subdirect and $X \abs_f \fA$, then  $(X+R) \abs_f \fB$.
\end{itemize}
% Source? Solution?
% Proof of second item, for X+R \abs B:
% Suppose f is n-ary and proves absorption for X.
% Then f proves absorption for X+R:
% if c1,...,cn \in X+R, i in [n], and b \in B,
% then there are x1,...,xn \in X such that
% (xi,ci) \in R, and
% by subdirectness there is a \in A such that
% (a,b) \in R. Then a' := f(x1,...a,...,xn) \in X by absorption and (a',f(c1,...,b,...,cn)) \in R
%
%\item \label{exe:proj-abs-trans}
%If $\fR \leq \fA \times \fB$ and $S \abs \fR$,
%then $\pi_1(S) \abs \pi_1(R)$.
%\item \label{exe:product-abs-trans}
%If $\fA$ has a proper absorbing subalgebra, then so has $\fA^2$.
%If $\fA^2$ has no proper absorbing subalgebras, then neither has $\fA$.
%\vspace{-2cm}
%\begin{flushright}
%\includegraphics[scale=.3]{Blau.jpg}
%\hspace{1cm}{ }
%\end{flushright}
%\vspace{-.7cm}
%\item \label{exe:product-abs-trans-2} If $\fC$ is a proper subalgebra of $\fA^2$ and
%$\fC \abs_f \fA$, then $\fA$ has a proper \\
%subalgebra $\fB$ with $\fB \abs_f \fA$.
%\vspace{-2.2cm}
%\begin{flushright}
%\includegraphics[scale=.3]{Rot.jpg}
%\end{flushright}
%\vspace{-.2cm}
% Zaneta's solution:
%Since C is non-empty and proper, there is (c,d) in C such that at least one of these applies:
%(1) there is c' \in A s. t. (c' d) \not \in C
%(2) there is d' \in A s. t. (c, d') \not \in C
%The two cases are symmetric, so let's go with (1). Define B:= {a \in A | (a,d) \in C}, then B is non-empty and proper. If a_1, .., a_n are all but at most one from B, then
%f( (a_1, d), ... (a_n, d)) lies since C is absorbing, %which also proves that B is absorbing.
\exercise[Rot]\label{exe:source-227} Show that a binary absorbing subalgebra of a binary absorbing subalgebra
might not be binary absorbing.
% Source: Brady, comments after Prop 3.2.17 in ArXiv Version 2.
% Solution: {(0,1)} \abs_2 {(0,1),(1,)} \abs_2 {0,1}^2 wrt \wedge, vee.
% GPT proposed Solution: In the algebra $(\{0,1\}^2;\wedge,\vee)$, with the
% operations acting coordinatewise, we have
% $\{(0,1)\}\abs_\wedge\{(0,1),(1,1)\}
% \abs_\vee\{0,1\}^2$, but $\{(0,1)\}$ is not binary absorbing
% in $\{0,1\}^2$.
\exercise[Rot]\label{exe:source-225} \label{exe:abs-pp-trans}
Let $\bB$ be a relational $\tau$-structure
and $\fA$ an algebra such that
$\Pol(\bB) = \Clo(\fA)$.
%Recall that if $\Clo(\fA) = \Pol(\bB)$, then $R \subseteq A^n$ is primitive positive
%definable in $\bB$ if and only if $R$ is the domain of a subalgebra of $\fA^n$.
%defining the relation $S \subseteq A^n$.
Let $\bB'$ be a $\tau$-structure with the same domain $B$
such that $R^{\bB'} \abs R^{\bB} \leq \fA^{\ar(R)}$
for every $R \in \tau$.
If a primitive positive \\
$\tau$-formula $\phi$ defines $S$ in $\bB$ and defines $S'$ in $\bB'$, then $S' \abs S$.
 If $R^{\bB'} \abs_f R^{\bB}$ for the same operation $f$ for all $R \in \tau$,
 then $S' \abs_f S$.
% Should also work for infinite domains:
% use * product to combine all the
% absorption terms into one.
% then verify that this term also does it for
% the relation defined by \phi.
% Problem: arity changes!
%\item Let $\fA$ be a finite algebra and $B \leq \fA$. Prove that $B$ $n$-absorbs $\fA$ if an only if there is no $R \leq \fA^n$ such that $R \cap B^n = \emptyset$ and for every $i \in \{1,\dots,n\}$
%$$R \cap (B \times \cdots \times B \times \underbrace{A}_{i} \times B \times \cdots \times B) \neq \emptyset.$$
% Source:  Kazda+Barto's deciding absorption, Prop 16 in ArXiv version.
\exercise[Rot]\label{exe:source-226} Suppose that $\fA \abs \fB \leq \fC$. Do we then have $\fA \abs \fC$?
% no: take paper-scissor stone. It is absorption-free, but has 2-element semilattice subalgebras.
% Solution sketch: No. In the rock--paper--scissors algebra every winning pair forms a two-element
% semilattice subalgebra and therefore has an absorbing singleton. The full three-element algebra is
% absorption-free. Thus one may have $\fA\abs\fB\leq\fC$ without $\fA\abs\fC$.
\exercise[Orange]\label{exe:source-228} Let $f \colon \{0,1,2\}^3 \to \{0,1,2\}$ be the majority operation given by
$f(x,y,z) = x$ if $x,y,z$ are pairwise distinct.
Show that $\{0,1\}$ is a 4-absorbing, but not a 3-absorbing subuniverse.
\end{exercises}

\subsection{Essential Relations}
This section presents a relational characterisation of absorption that will be needed in Section~\ref{sect:ternary-abs} and in Section~\ref{sect:cyclic-thm}.
%~\ref{sect:zhuk}. 
The following 
%concepts have been developed in
material is mostly from Barto and Kazda~\cite{BartoKazda}.
Throughout, $\fA$ denotes an algebra with a not necessarily finite domain. 
%; we follow the presentation in~\cite{BradyNotes}. 

\begin{definition}\label{def:B-essential}
Let $A \leq \fB$ and $n \geq 1$. 
% $n \in {\mathbb N}$. 
Then $R \leq \fB^n$ is \emph{$A$-essential} if for every $i \in \{1,\dots,n\}$ 
\begin{align*} 
%R \cap (A \times \dots \times A \times \underbrace{B}_i \times A \times \cdots \times A) & \neq \emptyset \\
R \cap (A^{i-1} \times B \times A^{n-i}) & \neq \emptyset \\
\text{ and } \quad R \cap A^n & = \emptyset.
\end{align*}
\end{definition}

\begin{example}
Let $\fB = (\{0,1\};f_k)$ be the algebra with the near unanimity operation $f_k$ of arity $k \geq 3$ which returns $0$ iff at least two of its arguments are 0 (Example~\ref{expl:bool-nu}). 
Then $A := \{1\} \abs \fB$ is absorbing and 
the relation $A_n := \{0,1\}^n \setminus \{(1,\dots,1)\} \leq \fB^n$ is $A$-essential 
if and only if $n < k$. 
\end{example} 

Note that if $A=B$, then there are no $A$-essential relations. If $A = \emptyset$, then every nonempty subalgebra of $\fB$ is $A$-essential. 
%Note that $R \leq \fA$ is $B$-essential if $R \neq \emptyset$ and $R \cap B = \emptyset$; hence, 
Also note that if 
$A \leq \fB$ is a proper subuniverse and $\fB$ is idempotent, then $\{b\}$ is $A$-essential for every 
$b \in B \setminus A$.

\begin{lemma}\label{lem:essential-down}
Let $A \leq \fB$. If there is no $A$-essential $R \leq \fB^m$, then for every $n \geq m$ there 
is no $A$-essential $S \leq \fB^n$. 
\end{lemma}
\begin{proof}
If $R \leq \fB^n$ is $A$-essential, 
then $\pi_{1,\dots,n-1}(R \cap (B^{n-1} \times A))$ is $A$-essential. 
\end{proof}

\begin{lemma}\label{lem:essential-abs} 
Let $t$ be a term operation of $\fB$ of arity $m$ such that 
$A \abs_t \fB$. Then there are no $A$-essential relations $R \leq \fB^m$. 
\end{lemma}
\begin{proof}
Suppose for contradiction that $R \leq \fB^m$ is $A$-essential. 
Then there are $b^1,\dots,b^m \in R$ such that  
$\{b^i_1,\dots,b^i_m\} \setminus \{b^i_i\} \subseteq A$ for every $i \in \{1,\dots,m\}$. Therefore, $t(b^1,\dots,b^m) \in R \cap A^m$, because $A \abs_t \fB$, contrary to our assumptions. 
%Let $b_{1,1},\dots,b_{m,m} \in B$ and $a_1,\dots,a_m \in A$ be such that 
\end{proof}

\begin{proposition}\label{prop:essential}
Let $A \leq \fB$ and $R \leq \fB^n$ for $n \geq m-1$.
Suppose that $\fB$ has no $A$-essential relation of arity $m$ and for every $I \in {\{1,\dots,n\} \choose m-1}$ we have $\pi_I(R) \cap A^{m-1} \neq \emptyset$.  
Then 
$$ R \cap A^n \neq \emptyset. $$
%$$ \pi_{[n] \setminus \{i\}}
\end{proposition} 
\begin{proof}
The proof is by induction on $n \geq m-1$. The base case $n = m-1$ is immediate by the assumption applied for $I = \{1,\dots,n\}$. For the inductive step, suppose that $n \geq m$. 
For every $i \in \{1,\dots,n\}$ define 
$$R_i := \pi_{[n] \setminus \{i\}}(R) \leq \fB^{n-1}$$ 
and note that $\pi_I(R_i) \cap A^{m-1} \neq \emptyset$
 for every $I \in {\{1,\dots,n\} \setminus \{i\} \choose m-1}$. 
Hence, by the inductive assumption 
we have that 
$$ R_i \cap A^{n-1} \neq \emptyset.$$ 
Since $R$ is not essential by Lemma~\ref{lem:essential-down} 
% since $R$ has arity $n \geq m$. 
we therefore must have
$R \cap A^n \neq \emptyset$. 
\end{proof}

\begin{corollary}\label{cor:essential}
Let $t \in \Clo(\fB)^{(m)}$ be such that 
$A \abs_t \fB$ and let $R \leq \fB^n$ for $n \geq m-1$.
Suppose that for every $I \in {\{1,\dots,n\} \choose m-1}$ we have $\pi_I(R) \cap A^{m-1} \neq \emptyset$. Then 
$$ R \cap A^n \neq \emptyset. $$
\end{corollary} 
\begin{proof}
Combine 
 Proposition~\ref{prop:essential} with
 Lemma~\ref{lem:essential-abs}. 
\end{proof}

Lemma~\ref{lem:essential-abs} has a converse (Proposition 16 in~\cite{BartoKazda}); also see~\cite{BradyNotes,Strong-Subalgebras-Published}.
It will be needed in the proof of the important Theorem~\ref{thm:centrally-3abs} later. 

% Backwards direction used 
\begin{theorem}\label{thm:essential-converse}
% Source? 
Let $m \geq 1$. 
A subalgebra 
$\fA \leq \fB$ $m$-absorbs $\fB$ if and only if there are no $A$-essential relations $R \leq {\fB}^m$. 
%no $R \leq \fA^n$ such that $R \cap B^n = \emptyset$ and for every $i \in \{1,\dots,n\}$ 
%$$R \cap (B \times \cdots \times B \times \underbrace{A}_{i} \times B \times \cdots \times B) \neq \emptyset.$$ 
\end{theorem}
\begin{proof}
The forward implication is Lemma~\ref{lem:essential-abs}. For the converse, the statement is trivial if $\fA$ is not proper, so suppose that $\fA$ is a proper subalgebra of $\fB$. 
Suppose that $\fB$ has no $A$-essential relations of arity $m$. Let $\fF \leq \fB^{B^m}$ be the free algebra generated by $x_1,\dots,x_m$ in $\HSP(\fB)$ 
(see Section~\ref{sect:birkhoff}). 
For $i \in \{1,\dots,m\}$, let $X_i := A^{i-1} \times (B \setminus A) \times A^{m-i}$, and let $X :=  X_1 \cup \cdots \cup X_m$, 
which will be used as index set of the relation $R$ defined as follows: 
$$ R := \pi_{X} (F) \leq \fB^{X_1} \times \cdots \times \fB^{X_m}.$$ 
%For every $i \in \{1,\dots,m\}$, since $\fF$ contains $\pi^m_i$, we have 
%Let $n := |X_1| + \cdots + |X_m|$ be the arity of $R$ and let $I \in { \{1,\dots,n\} \choose m-1 }$. 
Let $I \in {X \choose m-1}$. 
We claim that $\pi_I(R) \cap A^{m-1} \neq \emptyset$:
indeed, 
by the pigeon-hole principle there exists $i \in \{1,\dots,m\}$ such that
$I \cap X_i = \emptyset$. 
Since $\fF$ contains $\pi^m_i$
we have 
$$R \cap (A^{X_1} \times \cdots \times  A^{X_{i-1}} \times B^{X_i} \times A^{X_{i+1}} \times \cdots \times A^{X_m}) \neq \emptyset$$
which shows the claim. 

%If $R \cap B^n = \emptyset$, then there is a $B$-essential relation of arity $m$: TODO. 
%If the relation $R$ is not $B$-essential, 
%then we must have 
%So we may suppose that 
Therefore, Proposition~\ref{prop:essential} implies that 
$R \cap A^X \neq \emptyset$. By definition, any element of $R \cap A^{X}$ can be extended to an element $t \in \fF$, and any such $t$ is a term operation of $\fB$ of arity $m$ which absorbs $A$. 
\end{proof} 

\begin{exercises}
\exercise[Rot]\label{exe:source-229} Find for every $k \geq 3$ a finite idempotent algebra with a $k$-absorbing proper subuniverse which is not $k-1$-absorbing.
% Source: me
% Solution: take the Boolean NU algebras.
\end{exercises}

\subsection{The Absorption Theorem}
\label{sect:absorb}
The presentation of this section is based on  lecture notes of Libor Barto. The goal of this section is to show that finite idempotent 
Taylor algebras (see Remark~\ref{rem:Taylor})
%whose clone of term operations has no minion homomorphism to $\Proj$ 
must have some form of absorption; this idea will be formalised in the absorption theorem, Theorem~\ref{thm:absorb} below, which is from~\cite{cyclic}.

\begin{definition}\label{def:cube-term-blocker}
Let $\fA$ be an algebra. A subset 
$B \subseteq A$ is called \emph{projective in $\fA$} (in some papers called \emph{cube term blocker}~\cite{MarkovicMarotiMcKenzie}) if for every 
$f \in \Clo(\fA)$ of arity $n$ there exists $i \in \{1,\dots,n\}$
such that 
\begin{align*}
f(A,A,\dots,A,\underbrace{B}_{\text{position $i$}},A,\dots,A) \subseteq B;
\end{align*}
as usual, the term on the left stands for $\{f(x_1,\dots,x_n) \mid x_1,\dots,x_n \in A, x_i \in B\}$. 
\end{definition}

Note that subsets of $A$ that are projective in $\fA$ are subuniverses of $\fA$. 
Recall the definition of minion homomorphisms from Section~\ref{sect:minors}. 
Our starting point is the following theorem. 

\begin{theorem}\label{thm:2-abs}
Let $\fA$ be an algebra such that there is no minion homomorphism from $\Clo(\fA)$ to $\Proj$ and let $B \subseteq A$ be projective in $\fA$. Then $B$ 2-absorbs $\fA$. 
\end{theorem}

\begin{proof}
Since $B$ is projective in $\fA$, 
for every $f \in \Clo(\fA)$
of arity $n$ there exists $i_f \in [n]$ such that 
\begin{align}
f(A,\dots,A,\underbrace{B}_{i_f},A,\dots,A) \subseteq B.
\label{eq:uniqueness-minionhom}
\end{align}
If $i_f$ is \emph{unique} for every $f \in \Clo(\fA)$, 
then $\Clo(\fA) \to \Proj$
given by $f \mapsto \pi^n_{i_f}$ is a minion homomorphism: 
indeed, let $\alpha \colon [n] \to [k]$ and $f \in \Clo(\fA)$ 
and suppose that there exists a unique $i \in [n]$ such that~\eqref{eq:uniqueness-minionhom} holds. 
Then $f_\alpha$ is an operation of arity $k$ such that 
$$f_{\alpha}(A,\dots,A,\underbrace{B}_{\alpha(i)},A,\dots,A) \subseteq f(A,\dots,A,\underbrace{B}_{i},A,\dots,A) \subseteq B$$
and by assumption $\alpha(i)$ is the only 
index $j \in [k]$ such that $f_{\alpha}(A,\dots,A,\underbrace{B}_{j},A,\dots,A) \subseteq B$. Hence, 
$$\xi(f_\alpha) = \pi^k_{\alpha(i)} =  (\pi^n_i)_\alpha = \xi(f)_\alpha$$ and $\xi$ is a minion homomorphism. 
So there exists $f \in \Clo(\fA)$ and $i \neq j$ such that 
\begin{align*}
f(A,\dots,A,\underbrace{B}_i,A,\dots,A) & \subseteq B 
& \text{ and } 
\quad  f(A,\dots,A,\underbrace{B}_j,A,\dots,A)  & \subseteq B.
\end{align*}
Define $r(x,y) := f(x,\dots,x,\underbrace{y}_j,x,\dots,x)$
and observe that 
$B \abs_r \fA$. 
\end{proof}

If $\fA$ does not have proper projective subuniverses, 
then in exchange it must have 
a term operation satisfying the following strong condition. 

\begin{definition}
An operation $t \colon A^n \to A$ is called \emph{transitive} if for every $a \in A$ and $i \in \{1,\dots,n\}$ we have
$$t(A,\dots,A,\underbrace{\{a\}}_i,A,\dots,A) 
= A.$$
\end{definition}
Clearly, if $|A| > 1$, then a transitive operation must have arity at least two. 
The following theorem for the first time depends on $A$ being finite. 

\begin{theorem}\label{thm:transitive}
Let $\fA$ be a finite idempotent algebra without proper  projective subuniverses. 
Then $\Clo(\fA)$ contains a transitive operation. 
\end{theorem}
\begin{proof}
By assumption, for every proper subset $B$ of $A$ there exists $t_B \in \Clo(\fA)$ of arity $n$ such that for every $i \in \{1,\dots,n\}$ 
$$t_B (A,\dots,A,\underbrace{B}_i,A,\dots,A) \not \subseteq B.$$
Using the star composition and the idempotence of $\fA$, Lemma~\ref{lem:combine} implies that we may suppose that there exists a single term $t$ that works for all proper  $B \subset A$. 
Then 
$$u := \underbrace{t * \cdots * t}_{|A| \text{ times }}$$ is transitive, because for every $a \in A$ and $j \in \{1,\dots,|A|\}$
$$|\underbrace{t * \cdots * t}_{j \text{ times}}(A,\dots,A,\underbrace{\{a\}}_{i},A,\dots,A)| \geq j$$  and hence
$u(A,\dots,A,\underbrace{\{a\}}_{i},A,\dots,A) = A$. 
\end{proof} 

\begin{corollary}\label{cor:trans}
Let $\fA$ be a finite idempotent 
Taylor algebra.
% such that
%there is no minion. 
Then $\fA$ has a proper 2-absorbing subuniverse, or $\Clo(\fA)$ contains a transitive operation. 
\end{corollary}
\begin{proof}
Since $\fA$ is Taylor, there is no minion  homomorphism from $\Clo(\fA)$ to $\Proj$
(Theorem~\ref{thm:taylor}; Remark~\ref{rem:Taylor}). 
If $\fA$ has a proper projective subuniverse $B$, then $B$ 2-absorbs $\fA$
by Theorem~\ref{thm:2-abs} and we are done. 
Otherwise, all proper  subuniverses of $\fA$ are not projective, and $\Clo(\fA)$ contains a 
transitive operation by Theorem~\ref{thm:transitive}.
\end{proof} 

We will now explore consequences of having a transitive term operation
for the existence of proper absorbing subuniverses. 
Let $\fA$ and $\fB$ be algebras,
and let $R \subseteq A \times B$ be a relation.

\begin{figure}
\begin{center}
\includegraphics[scale=.5]{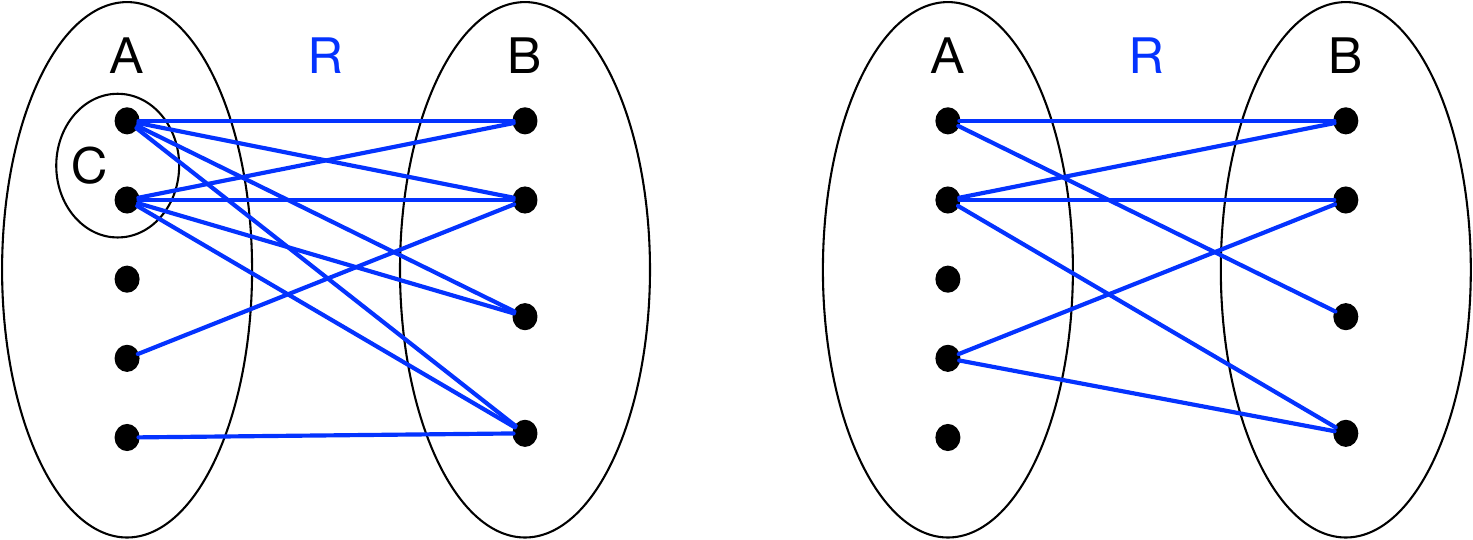}
\end{center}
\caption{Illustrations of $R \leq \fA \times \fB$ with non-empty left center $C$ (left side), and of $R \leq \fA \times \fB$ which is linked (right side); none of the two examples is subdirect.}
\label{fig:linked}
\end{figure}

\begin{definition}[left centre\footnote{There is no connection with the notion of centrality from Definition~\ref{def:central}.}]
\label{def:center}
%Let $R \subseteq A \times B$. 
The \emph{left centre of $R$} 
is the set 
$$\{a \in A \mid (a,b) \in R \text{ for every } b \in B\}.$$
\end{definition}

See Figure~\ref{fig:linked}, left side. 

\begin{proposition}\label{prop:absorb-left-center}
Let $\fA$ and $\fB$ be idempotent algebras with the same signature and let 
$R \leq \fA \times \fB$ 
with left centre $C$ be such that $\pi_1(R) = A$. 
%for every $a \in A$ there exists $b \in B$ such that $(a,b) \in R$. 
% Where do we need that subdirect is relaxed
% to this one-sided condition? In the proof of Zhuk's cases when stepping to higher-ary relations
% like R \leq A^m = A \times A^{n-1}:
% have that R is subdirect, but not as a subalgebra
% of $A \times A^{n-1}. 
If there exists a term $t$ such that $t^{\fB}$ is transitive, then $C \abs_{t^{\fA}} \fA$. 
\end{proposition}
\begin{proof}
If $C$ is empty, then the statement is trivial, so suppose that $C$ is non-empty. 
Since $\fB$ is idempotent, $C \leq \fA$. Let $i \in [n]$, $z_1,\dots,z_{i-1},z_{i+1}, \dots,z_n \in C$, $a \in A$, and $$a' := t^{\fA}(z_1,\dots,z_{i-1},a,z_{i+1}, \dots,z_n).$$
To show that $C \abs_{t^{\fA}} \fA$ we need to show that $a' \in C$, i.e., 
 $(a',b) \in R$ for every 
 $b \in B$. Arbitrarily choose $b \in B$. 
By assumption, there exists $c \in B$ such that $(a,c) \in R$. 
By the transitivity of $t^{\fB}$ there are $d_1,\dots,d_{i-1},d_{i+1},\dots,d_n \in B$ such that 
$t^{\fB}(d_1,\dots,d_{i-1},c,d_{i+1},\dots,d_n) = b$. 
Note that $(z_1,d_1),\dots,(z_{i-1},d_{i-1}),(z_{i+1},d_{i+1}),\dots,(z_n,d_n) \in R$ since $z_1,\dots,z_{i-1},z_{i+1},\dots,z_n$ are from the left center of $R$. 
Since $R \leq \fA \times \fB$, 
we have that $$(a',b) = \big (t^{\fA}(z_1,\dots,z_{i-1},a,z_{i+1}, \dots,z_n), t^{\fB}(d_1,\dots,d_{i-1},c,d_{i+1},\dots,d_n) \big) \in R$$
and the proof is complete. 
\end{proof}

\begin{corollary}\label{cor:2-abs-centre}
Let $\fA$ and $\fB$ be finite idempotent algebras with the same signature 
such that $\fB$ is Taylor. 
%there is no minion homomorphism from $\Clo(\fB)$ to $\Proj$. 
Let $R \leq \fA \times \fB$ with left centre $C$ be such that $\pi_1(R) = A$. 
%for every $a \in A$ there exists $b \in B$ such that $(a,b) \in R$. 
Then $\fB$ has a proper 2-absorbing subuniverse or $C \abs \fA$. 
\end{corollary}
\begin{proof}
Suppose that $\fB$ does not have a proper 2-absorbing subuniverse. 
Then Corollary~\ref{cor:trans} implies that $\fB$ has a transitive term operation $t$. Hence, Proposition~\ref{prop:absorb-left-center} implies that $C \abs_{t} \fA$. 
\end{proof} 

The relation $R$ can be viewed as the edge relation of a bipartite graph $G_R$ with color classes $A$ and $B$ (this perspective was already presented in Section~\ref{sect:alg-csp}). 

\begin{definition}[Linked relations]
A relation $R \subseteq A \times B$ is called \emph{linked}
if $G_R$ is connected after removing
isolated vertices.
\end{definition}

% TODO: find arguments why we define this for
% not necessarily subdirect relations. 

See Figure~\ref{fig:linked} on the right. Note that if $R$ is a subdirect subalgebra of $\fA \times \fB$ (Definition~\ref{def:subdirect})
then $G_R$ has no isolated vertices. Also note that if $R$ has a non-empty left centre, then it is linked (but not necessarily subdirect). 
Recall the definition of $R^{-1}$ from Exercise~\ref{exe:converse}. 

\begin{proposition}\label{prop:linked-gives-left-centre}
Let $\fA$ and $\fB$ be finite idempotent algebras 
with the same signature 
and let $R \leq \fA \times \fB$ have
%proper, subdirect, and linked,
empty left centre and such that $R^{-1}  \circ R = B^2$. 
Then there exists a subdirect 
$R' \leq \fB^2$ whose left centre is a proper  subuniverse of $\fB$. 
\end{proposition}

Before we go into the proof we consider an example. 

\begin{example}
Suppose that $\fB$ has domain $B = \{1,2,3\}$ and 
$\fB^2$ has the subuniverse $R := \{(u,v) \in B^2 \mid u \neq v\}$. 
Clearly, the left centre of $R$ is empty. Then 
$$R' := \{ (x,y) \mid \exists a (R(a,x) \wedge R(a,y) \wedge R(a,1))\} 
%= B^2 \setminus \{(1,1)\}$$
= B^2 \setminus \{(2,3),(3,2)\}$$
is a subuniverse of $\fB^2$ (we use that $\{1\}$ is a subuniverse), is subdirect, 
and has the left centre $\{1\}$. 
\end{example}

\begin{figure}
\begin{center}
\includegraphics[scale=.5]{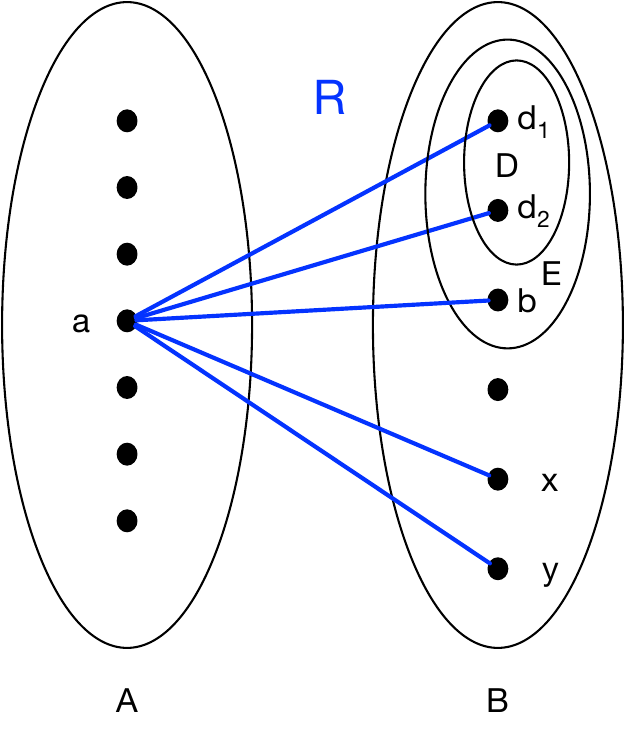}
\end{center}
\caption{An illustration for the definition of $S_E$ in the proof Proposition~\ref{prop:linked-gives-left-centre}.}
\label{fig:absorption}
\end{figure}

\begin{proof}[Proof of Proposition~\ref{prop:linked-gives-left-centre}]
For $X = \{d_1,\dots,d_k\} \subseteq B$ 
define
$$S_X := \big \{(x,y) \in B^2 \mid \exists a \big (R(a,x) \wedge R(a,y) \wedge R(a,d_1) \wedge \cdots \wedge R(a,d_k) \big ) \big \}.$$
Then  
\begin{itemize}
\item $S_X \leq \fB^2$ by the idempotence of $\fB$, 
\item $S_\emptyset = R^{-1} \circ R = B^2$ by assumption, 
and 
\item $S_B = \emptyset$ because the left centre of $R$ is empty. 
%$Z = \emptyset$. 
\end{itemize}
Choose $D\subseteq B$ maximal subject to $S_D=B^2$. Such a $D$
exists because $S_\emptyset=B^2$. Since $S_B=\emptyset$, we have
$D\neq B$. Choose $b\in B\setminus D$ and put
$E:=D\cup\{b\}$. By the maximality of $D$, $S_E\neq B^2$.
%OLD, buggy: Let $D$ be maximal such that $S_D = B^2$, and 
%let $E \subseteq B$ and $b \in B$ be a set such 
%that $E \setminus D = \{b\}$. 
See Figure~\ref{fig:absorption}. 
Let $C$ be the left centre 
of $S_E$. 
\begin{itemize}
\item $C$ contains $b$ and hence is non-empty:
%for $x \in Z'$ 
indeed, for any $y \in B$ there exists
$a \in A$ witnessing that $(b,y) \in S_D = B^2$, i.e., $R(a,b)$,  $R(a,y)$, and $R(a,d)$ for every $d \in D$. 
Hence, $(b,y) \in S_E$ and $b \in C$. 
\item Since \(C\) is non-empty, choose \(c\in C\). By the definition of the left centre, \((c,y)\in S_E\) for every \(y\in B\). Since \(S_E\) is symmetric, also \((y,c)\in S_E\). Hence both projections of \(S_E\) equal \(B\), so \(S_E\) is subdirect.
%Old, buggy: $S_E \leq \fB^2$ is subdirect: since $C$ is non-empty, for every $y \in B$ there is $c \in C$ such that $(y,c) \in S_E$ and $(c,y) \in S_E$. 
\item Finally, \(C\neq B\): otherwise, by the definition of the left centre, \((x,y)\in S_E\) for all \(x,y\in B\), and hence \(S_E=B^2\), a contradiction. 
%Old, buggy: $C$ is a proper subset of $B$. Otherwise,  the centrality of $C$ would imply that $S_E = B^2$, contrary to the choice of $D$ and $E$. 
\end{itemize} 
Therefore, $R' := S_E$ meets the requirements.
\end{proof} 

\begin{proposition}\label{prop:new-absorption}
Let $\fA$ and $\fB$ be finite idempotent algebras with the same signature and let $R \leq \fA \times \fB$ be subdirect and linked such that $R \neq A \times B$. 
Then at least one of the following cases applies.
\begin{itemize}
\item %there exists $R' \leq \fA \times \fB$ with non-empty left centre.
%, and $R' \neq A \times B$. 
$R$ has a non-empty left centre. 
\item there exists a subdirect $R' \leq \fB^2$ whose left centre 
is a proper subuniverse of $\fB$. 
%such that $R' \neq B^2$. 
\end{itemize} 
\end{proposition}
\begin{proof}
Suppose that the left centre $C$ of $R$ is empty. 
%Let $C$ be the left centre of $R$. 
%Then $C$ is a proper subset of $A$ because $R$ is a proper subset of $A \times B$ and subdirect. Hence, 
%If $C$ is non-empty,  then the first item of the proposition applies and we are done. 
%we can choose $R' = R$ and are done. 
%So suppose  in the following that $C=\emptyset$. 
If $R^{-1} \circ R = B^2$ then the statement follows from Proposition~\ref{prop:linked-gives-left-centre}. 
Otherwise,  
$R^{-1} \circ R \leq \fB^2$ is subdirect, proper, and linked (Exercise~\ref{exe:linked}), so we may replace $\fA$ by $\fB$ and $R$ by 
$R^{-1} \circ R$. 
Put $S:=R^{-1} \circ R$. Since $R$ is linked and subdirect, $S^n=B^2$ for some $n \in {\mathbb N}$. The relation $S$ is reflexive and symmetric, so $S^m=B^2$ for every $m \geq n$. Moreover, every repetition replaces the current symmetric relation by its square. Thus the relations tested in successive repetitions are $S^2,S^4,S^8,\dots$. If $j$ is least such that $S^{2^j}=B^2$, then $S^{2^{j-1}}$ is still proper and its inverse composite is $B^2$. Hence we eventually find a proper, subdirect, and linked subuniverse $R$ of $\fB^2$ such that $R^{-1} \circ R=B^2$.
If the left centre of $R$ is non-empty it is a proper 
subalgebra of $\fB$ and we are done. 
Otherwise, the statement again follows from Proposition~\ref{prop:linked-gives-left-centre}. 
%Note that $R \circ R^{-1}$ is a subuniverse of $\fB^2$ which is subdirect and linked. If $R \circ R^{-1}$ is a proper subset of $\fB^2$ then Lemma~\ref{prop:linked-gives-left-centre} implies the second item of the statement of the proposition. 
%Suppose that $R$ has an empty left centre. 
\end{proof} 

% Only need Taylor assumption for $\fB$!
\begin{theorem}[Absorption theorem~\cite{cyclic}]
\label{thm:absorb}
Let $\fA$ and $\fB$ be finite idempotent algebras with the same signature such that $\fB$ is Taylor. 
%$\Clo(\fB)$ has no minion homomorphism to $\Proj$. 
Then 
%$\fA,\fB \in \HSP(\fC)$ and any
for every linked and subdirect $R \leq \fA \times \fB$ one of the following is true: 
\begin{enumerate}
\item $R = A \times B$; 
\item $\fA$ has a proper absorbing subuniverse;
\item $\fB$ has a proper absorbing subuniverse. 
\end{enumerate}
\end{theorem}
\begin{proof}
%Suppose that $\fB$ has a proper projective subuniverse $B'$;
%then Theorem~\ref{thm:2-abs} implies that $B'$ is 2-absorbing. Hence, item 3 of the theorem is true and we are done. Otherwise, 
%Theorem~\ref{thm:transitive} implies
%that $\fB$ has a transitive term operation. 
Suppose that 
$R \neq A \times B$ because otherwise item 1 of the theorem holds and we are done. 
Let $C$ be the left centre of $R$.
Note that $C \neq A$ because $R \neq A \times B$. 

Suppose also that $\fB$ is absorption-free, because otherwise item~3 of the theorem holds. 
Corollary~\ref{cor:2-abs-centre} then implies that  $C \abs \fA$.
If $C$ is non-empty,  then we have found a proper absorbing subuniverse of $\fA$ and item 2 of the theorem holds. 
Otherwise, Proposition~\ref{prop:new-absorption} implies that there 
exists a subdirect $R' \leq \fB^2$ whose left centre is a proper subuniverse of $\fB$. 
In this case, $\fB$ has a proper absorbing subuniverse by Corollary~\ref{cor:2-abs-centre}, in contradiction to the assumption above. 
\end{proof}

We will prove a strengthening of this result in Theorem~\ref{thm:3-absorb}.

\begin{exercises}
\exercise[Rot]\label{exe:source-231} Let $\fA$ and $\fB$ be idempotent algebras and $R \leq \fA \times \fB$. Show that
the left center of $R$ is a subalgebra of $\fA$.
% Solution sketch: Let $a_1,\ldots,a_k$ lie in the left center and choose witnesses $b_i$ with
% $(a_i,b_i)\in R$ and the required full right fibres. For any basic operation $f$, closure of $R$
% gives $(f(\bar a),f(\bar b))\in R$; applying $f$ coordinatewise to arbitrary fibre witnesses shows
% that $f(\bar a)$ again has a full fibre. Hence the center is a subalgebra.
\exercise[Rot]\label{exe:source-232} \label{exe:linked}
Show that if $R \leq \fA \times \fB$ is linked, then $R^{-1} \circ R \leq \fB^2$ is linked, too.
% Proof: first note that R \circ R^{-1} contains diagonal (and is symmetric).
% let a,b \in B. Take a path from a to b in G_R: must have even length. Then this path translates to a path linking a with b in G_{R \circ R^{-1}},
\exercise[Rot]\label{exe:source-233} \label{exe:simple-linked}
Suppose that $\fA$ is a simple algebra. Then every subdirect
$R \leq \fA^2$
is linked or the graph of an automorphism of $\fA$.
\par\noindent {\bf Hint.} Consider $\bigcup_{i \in {\mathbb N}} (R \circ R^{-1})^i$.
% From reading the notes of Brady,
% in particular the proof of the pss expl.
% Solution: if R is not linked,
% then B := lim_n (R^{-1} \circ R)^n is not everything.
% but it is clearly
% reflexive (from subdirect)
% symmetric (by construction)
% transitive (by construction),
% hence an equivalence relation.
% Moreover, B = (R^{-1} \circ R)^{n0}, for some $n0$, which is pp-definable, so we even
% have a congruence of $A$. Since $A$ is simple,
% and B \neq A^2, the congruence has singleton classes. It follows that
% R is the graph of a permutation.
% But it is also preserved by A, so
% the permutation commutes with all operations in A, so it is an automorphism of A.
\exercise[Rot]\label{exe:source-235} Let $\fA$ and $\fB$ be algebras and let $R \leq \fA \times \fB$ be subdirect. Let $\theta_A$ be the kernel of $\pi_1 \colon R \to A$ and let $\theta_B$ be the kernel of $\pi_2 \colon R \to B$.
Show that $R$ is linked if and only if $\theta_A \vee \theta_B = 1_R$.
% Solution sketch: Edges in the bipartite link graph of $R$ alternately change only the first or only
% the second coordinate. These two moves are precisely the pairs in $\theta_B$ and $\theta_A$. Thus
% two elements of $R$ lie in one link component exactly when they are related by the congruence
% generated by $\theta_A\cup\theta_B$, namely $\theta_A\vee\theta_B$. The graph is connected exactly
% when this join is $1_R$.
\exercise[Rot]\label{exe:source-236} \label{exe:connectedness-absorption}
Let $R \leq \fA^2$, $S \lhd R$, and $(a,a),(b,b) \in S$ be such that $(a,b) \in R$.
Then there exists $k \in {\mathbb N}$ such that $(a,b) \in S^k$.
% Quelle: Brady Seite 181.
% Let $f$ be the term that witnesses that
% S absorbs R. Feed (a,a),(a,a),...,(a,b) into
% this term and we obtain a pair in S, whose
% first component is a, the second is c
% and (a,c)
% Do the same with (a,a),...,(a,a),(a,b),(b,b)
% Again, obtain a pair in S,  whose first pair is c, and second is c'. Continue like this.
\exercise[Rot]
\label{exe:cube-blocker}
A \emph{cube term blocker} in an algebra $\fA$ is a pair $(C,B)$
such that $\fB\leq\fA$, $\emptyset\neq C\subsetneq B$, and $C$
is projective in $\fB$ (Definition~\ref{def:cube-term-blocker}).
Show that a finite idempotent algebra with a cube term
(Exercise~\ref{exe:cube}) has no cube term blocker.
The converse holds as well, but its proof requires substantial
work~\cite[Theorem 2.1]{MarkovicMarotiMcKenzie}.
%\label{exe:source-237} Show that a finite idempotent algebra $\fA$ with a cube term (see Exercise~\ref{exe:cube}) has no cube term blocker (Definition~\ref{def:cube-term-blocker}; the converse is true as well, but requires substantial work~\cite[Theorem 2.1]{MarkovicMarotiMcKenzie}).
%\label{exe:cube-blocker}
% Solution: Theorem 2.1 in the above cited paper.
\exercise[Orange]\label{exe:source-234} Let $\fA$ be a finite algebra
% TODO: check finite.
and $B \subseteq A$. Then
$B$ is projective in $\fA$
if and only if $\Clo(\fA)$ preserves
for every $n$
the relation $B[n] := A^n \setminus (A \setminus B)^n.$
% Source: minimal Taylor in Prop 3.8 refers to Lemma 3.2 in CubeTermBlockers.
% Solution: Suppose B is projective,
% and let f \in Clo(A)^k,
% n \in N and a1,...,ak be from B[n]. Then there exists i \in {1,\dots,n}$
% such that for all j, if ai_j \in B
% then f(a1_j,...,an_j) \in B.
%Note that for some j_0 we must have
% ai_j_0 \in B since ai \in B.
% therefore, f(a1,...,ak)_j_0 = f(a1_j_0,...,an_j_0) \in B
% and f(a1,..,ak) \in B[n]. So Clo(A)
% preserves B[n].
% Conversely, we prove the contraposition. Let
% f \in Clo(\fA) be of arity n.
% Suppose that for every i \leq n
% there exists a^i \in A^n with a^i_i \in B such that f(a^i) \notin B.
% Then (a^1_1,\dots,a^n_1),\dots,(a^n_1,\dots,a^n_n) \in B[n]$,
% but $f(a^1,\dots,a^n) \in (A \setminus B)^n, hence $f$ violates B[n]$.
\exercise[Schwarz]\label{exe:source-230} Find a counterexample to the absorption theorem if we drop the assumption that $R$ is linked.
% Solution sketch: Let $\fA$ be any nontrivial simple idempotent absorption-free algebra and take the
% diagonal $R=\{(a,a):a\in A\}\leq\fA\times\fA$. It is subdirect but its link congruences are
% equality, so it is not linked. Neither factor has a proper absorbing subuniverse, yet $R$ is not the
% full product, contradicting the theorem's conclusion without linkedness.
\end{exercises}

\subsection{Abelianness Revisited}
\label{sect:abelian-revisited}
The fundamental theorem of abelian algebras (Theorem~\ref{thm:abelian}) implies that every abelian algebra with a Maltsev term is affine. 
In this section we considerably strengthen this theorem for finite idempotent algebras by replacing the assumption of having a Maltsev term by having a Taylor term (Corollary~\ref{cor:abelian-affine}).
This result follows from tame congruence theory (Hobby and McKenzie~\cite{HobbyMcKenzie}; see the discussion in \cite{BartoKozikStanovsky});
the new proof based on absorption that we present here is from~\cite{BartoKozikStanovsky}; the presentation follows lecture notes of Libor Barto. 

\begin{definition}
An algebra $\fA$ is called \emph{hereditarily absorption-free (HAF)} if no subalgebra of $\fA$ has a proper absorbing subalgebra, i.e., whenever $\fC$ is a non-empty absorbing subalgebra of a subalgebra $\fB$ of $\fA$, then $C=B$. 
\end{definition}

We will first prove that `HAF and Taylor implies Maltsev' (Theorem~\ref{thm:haf-maltsev}),
and then that `abelian implies HAF' (Theorem~\ref{thm:abelian-haf}).
First we prove that HAF is closed under taking direct products. 

\begin{lemma}\label{lem:product-HAF}
Let $\fA,\fB$ be idempotent hereditarily absorption-free algebras with the same signature. Then 
$\fA \times \fB$ is hereditarily absorption-free. 
\end{lemma}
\begin{proof}
Suppose that $S \abs R \leq \fA \times \fB$ is non-empty. We have to show that 
$S=R$. 
Let $(a,b) \in R$ and consider the subalgebras $\fD \leq \fC \leq \fB$ given by 
 \begin{align*}
C & := \{b' \mid (a,b') \in R \} \leq \fB && \text{(since $\fB$ is idempotent)} \\
\text{ and } D & := \{b' \mid (a,b') \in S \} \leq \fC && \text{(since $\fB$ is idempotent)} .\end{align*}

{\bf Claim 1.} $D \neq \emptyset$. 
Note that $\pi_1(S) \leq \pi_1(R) \leq \fA$ (for the notation, see the comments after Definition~\ref{def:proj}). 
We even have $\pi_1(S) \abs \pi_1(R)$ (Lemma~\ref{lem:abs-prod-1}). 
 Since $\fA$ is HAF, we get $\pi_1(S) = \pi_1(R)$.
 Since $a \in \pi_1(R)$, there must be $b' \in B$ such that $(a,b') \in S$. Therefore, $b' \in D \neq \emptyset$. 

{\bf Claim 2.} $D \abs \fC$. 
By assumption, there exists a term operation $f \in \Clo(\fR)^{(n)}$ such that $S \abs_f \fR$. 
Let $b_1,\dots,b_n \in C$ be such that all but one of them are from $D$. Then $f((a,b_1),\dots,(a,b_n)) = (a,f(b_1,\dots,b_n)) \in S$ since $\fA$ is idempotent and $S \abs_f \fR$. It follows that $f(b_1,\dots,b_n) \in D$. 

\medskip 
By the assumption that $\fB$ is HAF, we must have $D=C$ and hence 
 $(a,b) \in S$. Since $(a,b) \in R$ was chosen arbitrarily, this implies that $R=S$. 
\end{proof}

\begin{corollary}\label{cor:abs-pvar}
The class of idempotent HAF algebras of fixed signature $\tau$ forms a pseudo-variety.  
\end{corollary}
\begin{proof}
By definition of HAF, the class is closed under subalgebras. Closure under finite products has been established in Lemma~\ref{lem:product-HAF}. 
%Closure under homomorphic images is by Lemma~\ref{lem:abs-hom}. 
%Exercise~\ref{exe:HAF-Hom}. 

For closure under homomorphic images, let
$h\colon\fA\to\fB$ be a surjective homomorphism, where $\fA$ is HAF,
and suppose that $\emptyset\neq D\abs_t\fC\leq\fB$. Then $h^{-1}(C)$ is a subalgebra of $\fA$, and
$h^{-1}(D)$ is a non-empty subalgebra of $h^{-1}(C)$. Applying
Lemma~\ref{lem:abs-hom} to the restriction
$h|_{h^{-1}(C)}\colon h^{-1}(C)\to C$
gives
$h^{-1}(D)\abs_t h^{-1}(C)$. 
Since $\fA$ is hereditarily absorption-free, the subalgebra
$h^{-1}(C)$ is absorption-free. Hence
$h^{-1}(D)=h^{-1}(C)$, and applying $h$ yields $D=C$. Thus $\fB$ is
HAF. Idempotence is also preserved under homomorphic images, so the
class is closed under homomorphic images.
\end{proof}

\begin{theorem}[Theorem 1.4 in~\cite{BartoKozikStanovsky}]
%; presentation from~\cite{BradyNotes}]
% Presentation is clear!
\label{thm:haf-maltsev}
Let $\fA$ be a finite idempotent 
Taylor algebra.
% such that there is no minion homomorphism from $\Clo(\fA)$ to $\Proj$. 
If $\fA$ is hereditarily absorption-free, then $\fA$ has a Maltsev term. 
\end{theorem}
\begin{proof}
Let $\fF \in \HSPfin(\fA)$ be the free algebra 
over two generators $x,y$ (see Section~\ref{sect:birkhoff}). It follows from Corollary~\ref{cor:abs-pvar} that  $\fF$ is HAF. 
Let $\fR$ be the subalgebra of $\fF^2$ generated by $(x,y)$, $(x,x)$, and $(y,x)$. 

\medskip 
{\bf Claim 1.} $R \leq \fF^2$ is subdirect. 
Let $u\in F$. Since $\fF$ is generated by $x$ and $y$, there is a
binary term $t$ such that $u=t^{\fF}(x,y)$. Since
$(x,x),(y,x)\in R$ and $\fR\leq\fF^2$, we have
\[
t^{\fR}\bigl((x,x),(y,x)\bigr)
=
\bigl(t^{\fF}(x,y),t^{\fF}(x,x)\bigr)
=
(u,x)\in R,
\]
where the last equality uses the idempotence of $\fF$. Hence
$u\in\pi_1(R)$. Similarly, since $(x,x),(x,y)\in R$, we have
\[
t^{\fR}\bigl((x,x),(x,y)\bigr)
=
\bigl(t^{\fF}(x,x),t^{\fF}(x,y)\bigr)
=
(x,u)\in R.
\]
Hence $u\in\pi_2(R)$. Since $u\in F$ was arbitrary, both projections
of $R$ are equal to $F$, and therefore $\fR$ is subdirect in
$\fF^2$.

% Old proof, buggy: 
%Every element of $F$ can be written as $t(x,y)$ for some term $t$, and
%since $(x,y) \in R$ and $(y,x) \in R$ we have that
%$(t^{\fF}(x,y),t^{\fF}(y,x)) \in R$. A similar statement holds for the second argument of $R$. This shows
%that $\fR$ is 
%a subdirect subalgebra of $\fF^2$.

{\bf Claim 2.} $R$ is linked. 
Every element of 
$R$ can be written as 
$$s^{\fR}((x,y),(x,x),(y,x)) = (s^{\fF}(x,x,y),s^{\fF}(y,x,x))$$ for some term $s$. 
Since $(x,x),(y,x) \in R$ we 
have that 
$(s^{\fF}(x,x,y),s^{\fF}(x,x,x)) \in R$  
%$(t^{\fF}(x,x,x),t^{\fF}(x,x,x)) \in R$, 
and since $(x,y),(x,x) \in R$ we have 
$(s^{\fF}(x,x,x),s^{\fF}(y,x,x)) \in R$. 
Note that $s^{\fF}(x,x,x) = x$ by the idempotence of $\fA$ and $\fF$, and thus between any two elements of $F$ there is a path of length at most three in the bipartite graph $G_R$ of $R$, which proves the claim. 

Since $\fF \in \SP(\fA)$ (Proposition~\ref{prop:free-subpower}) and $\fA$ has a Taylor term, so has 
%$\Clo(\fA)$ has no minion homomorphism to $\Proj$, neither has 
$\fF$. % (Proposition~\ref{prop:clone-homo}). 
Since $\fF$ has no proper non-empty absorbing subalgebra, 
Theorem~\ref{thm:absorb}
implies that $R = F \times F$. 
Let $m$ be a term such that 
$m^{\fF}((x,y),(x,x),(y,x)) = (y,y)$. Then $m^{\fA}$ is a Maltsev operation. 
\end{proof}

The following is a special case of Lemma 4.1 in~\cite{BartoKozikStanovsky}.

\begin{theorem}\label{thm:abelian-haf}
Let $\fA$ be a finite idempotent algebra. If $\fA$ is abelian, then $\fA$ is hereditarily absorption-free. 
\end{theorem}
\begin{proof}
Since every subalgebra of an abelian algebra is abelian (Exercise~\ref{exe:abelian-subalgebra}), it suffices to show that if $\fB \abs \fA$, then $B = A$. 
We will show that if $\fB \abs_t \fA$ for some $n$-ary term operation $t \in \Clo(\fA)$, for $n \geq 2$, then $\fB \abs_s \fA$ for some $n-1$-ary $s$. This is enough, because if $\fB \abs_s \fA$ and $s$ is unary, then $s$ must be the identity by the idempotence of $\fA$, hence $B = A$. 
%For notational simplicity, we present the proof 
%for $n = 2$. 
Define the term $t_m(\bar x,y)$, where $\bar x = (x_1,\dots,x_{n-1})$, as follows 
\begin{align*}
t_1(\bar x,y) &  :=t(\bar x,y) \\
t_{m+1}(\bar x,y) & :=t(\bar x,t_m(\bar x,y)) 
\end{align*}
% with $m$ occurrences of $t$, of the form 
%$$t_m(\bar x,y) := \underbrace{t(\bar x,t(\bar x,\dots,t(\bar x}_{m \text{ times}},y))).$$
Note that $\fB \abs_{t_m} \fA$ for every $m \geq 1$. 

\medskip 
{\bf Claim 1.} For $m = |A|!$ we have  \begin{align}
t_m(\bar x,t_m(\bar x,y)) = t_m(\bar x,y). \label{eq:idemp}
\end{align}
To see this, define $r_{\bar x} \colon A \to A$ by $r_{\bar x}(y) := t(\bar x,y)$. Then note that $$t_m(\bar x,y) = r_{\bar x}^m(y) = \underbrace{r_{\bar x} \circ \cdots \circ r_{\bar x}}_{m \text{ times}}(y)$$ and observe that (see Exercise~\ref{exe:unary})
$$\underbrace{r_{\bar x} \circ \cdots \circ r_{\bar x}}_{ 2m \text{ times}} (y) = 
\underbrace{r_{\bar x} \circ \cdots \circ r_{\bar x}}_{ m \text{ times}}(y).$$ 

\medskip 
{\bf Claim 2.} $\fB \abs_s \fA$ for $s \colon A^{n-1} \to A$ defined by 
$$s(x_1,\dots,x_{n-1}) := t_m(x_1,\dots,x_{n-1},x_{n-1}).$$
Let $a \in A$ and $b_1,\dots,b_{n-1} \in B$. 
Clearly, $s(b_1,\dots,b_{i-1},a,b_{i+1},\dots,b_{n-1}) \in B$ for all $i < n-1$,  since $B \abs_{t_m} \fA$. 
We have to verify that 
$$s(b_1,\dots,b_{n-2},a) = t_m(b_1,\dots,b_{n-2},a,a) \in B.$$ 
From \eqref{eq:idemp} we obtain that
$$ t_m(b_1,\dots,b_{n-2},b_{n-1},a)  = t_m(b_1,\dots,b_{n-2},b_{n-1},
t_m(b_1,\dots,b_{n-2},b_{n-1},a))$$ 
and since $\fA$ is abelian, we may apply the term condition to the term $t_m$ at the $(n-1)$-st argument and obtain 
$$t_m(b_1,\dots,b_{n-2},a,a) = t_m(b_1,\dots,b_{n-2},a,
t_m(b_1,\dots,b_{n-2},b_{n-1},a)).$$
The left hand side of this equation equals $s(b_1,\dots,b_{n-2},a)$, and the right hand side is contained in $B$ since $B \abs_{t_m} \fA$. 
\end{proof}

\begin{corollary}\label{cor:abelian-affine}
Let $\fA$ be a finite idempotent abelian Taylor algebra. 
%such that there is no minion homomorphism from $\Clo(\fA)$ to $\Proj$. 
%with a Taylor term. 
Then $\fA$ is affine. 
%has an affine Maltsev term. 
\end{corollary}
\begin{proof}
If $\fA$ is abelian, then by Theorem~\ref{thm:abelian-haf} 
it is hereditarily absorption-free. 
%Since $\fA$ has a Taylor term, 
%$\Clo(\fA)$ has no minion homomorphism to $\Proj$ (Corollary~\ref{cor:Taylor-minion}) and therefore 
Theorem~\ref{thm:haf-maltsev}
implies that $\fA$ has a Maltsev term $m$. 
%It follows that 
%$\fA$ generates a congruence permutable, and in particular congruence modular variety, so by 
Then Theorem~\ref{thm:abelian} implies that 
$\fA$ is affine. 
\end{proof}

\begin{exercises}
\exercise[Rot]\label{exe:source-240} Show that the assumption that
$\fA$ is finite is necessary in Corollary~\ref{cor:abelian-affine}.
\par\noindent {\bf Hint.} Consider $\fA = ([0,1];(x,y) \mapsto \frac{x+y}{2})$.
% Factor with classes [0,1) {1} has semilattice and no Maltsev.
% Solution sketch: All terms of the midpoint algebra are dyadic convex combinations, so no term can be
% a Maltsev term. Nevertheless the algebra is abelian in the term-condition sense. The quotient with
% classes $[0,1)$ and $\{1\}$ is a two-element semilattice, showing directly why the finite proof of
% affine representability fails. Hence finiteness in the corollary is essential.
	\exercise[Schwarz]\label{exe:source-238} \label{exe:GoldmannRussell}
	Let $\fA := (A,+,-,0)$ be a group
	with domain $A = \{a_1,\dots,a_n\}$.
	Show that
	$$\Csp(A;\{(a,b,c) \mid a+b=c\},\{a_1\},\dots,\{a_n\})$$
	is in P if $\fA$ is abelian, and is
	NP-hard otherwise.\footnote{This result is due to Goldmann and Russell~\cite{GoldmannRussell}; it can be derived relatively easily from the results in this text.
	\par\noindent Hint: Combine Corollary~\ref{cor:Taylor-minion},
	Proposition~\ref{prop:prove-abelian}, Corollary~\ref{cor:abelian-affine}, and
	Theorem~\ref{thm:maltsev}.}
%	Combine Exercise~\ref{exe:abelian}, Theorem~\ref{thm:maltsev}, Proposition~\ref{prop:prove-abelian}, Corollary~\ref{cor:Taylor-minion}, and Corollary~\ref{cor:abelian-affine}.}
	% Solution: If the group is Abelian, then
	% by Exercise~\ref{exe:abelian} we have a Maltsev, and are in P by Theorem~\ref{thm:maltsev}.
	% So suppose that the group is not Abelian.
	% Note, however, that the polymorphism algebra is abelian by prop:prove-abelian!
%by cor:abelian-affine, either
	% there is no Taylor, in which case it is
	% hard by Theorem~\ref{cor:Taylor-minion}, or the algebra has a Maltsev.
%Then by Exercise~\ref{exe:abelian} (this time the interesting direction) we have
% a contradiction to the assumption that
% G is not abelian.
\exercise[Schwarz]\label{exe:source-239} Prove that the converse of Theorem~\ref{thm:abelian-haf} is false:
i.e., find an example of a finite idempotent algebra which is hereditarily absorption-free, but not abelian.
% D_4: not abelian, but no absorption?
% Solution sketch: Let $G=S_3$ and take the idempotent heap algebra $(G;m)$ with $m(x,y,z)=xy^{-1}z$.
% Every subalgebra is a coset of a subgroup and again has a Maltsev term, so it has no proper
% absorbing subuniverse; the algebra is hereditarily absorption-free. It is not abelian as a universal
% algebra because the underlying group is nonabelian and violates the term condition.
\end{exercises}

%\subsection{Cube Terms}
% Move to somewhere later. Incorporate algorithm. 

%\paragraph{Exercises.}
%\begin{enumerate}
%\setcounter{enumi}{\value{mycounter}}
% this is the property that Libor, David, and Marcin use when stating the absorption theorem. 
%\item \label{exe:HAF-Hom} Show that if $\fA$ is a HAF algebra with a congruence $\theta$, then $\fA/\theta$ is also HAF.
% Source: libor's notes. 
% Solution: If C absorbs A/theta via t^{A/theta},
% then $the union of all classes in C absorbs A via t. 
%\setcounter{mycounter}{\value{enumi}}
%\end{enumerate}

%\subsection{$n$-Absorption}
%We define a refinement of absorption which
%takes the arity of the function into account. 
%\begin{definition}
%Let $\fA$ be an algebra and $B \subseteq A$.
%Then $B$ is called \emph{$n$-absorbing} if there exists a term $t$ such that $t^{\fA}(a) \in B$ whenever $a \in A^n$ and $|\{i \mid a_i \in B\} | \geq n-1$. If $B$ is the domain of a subalgebra 
%$\fB$ of $\fA$, we write $\fB \abs_n \fA$. 
%\end{definition}
%Clearly, $\fB \abs \fA$ (according to Definition~\ref{def:absorb}) if there exists an $n \in {\mathbb N}$ such that $\fB \abs_n \fA$. 

%Important conditions that implies hereditary absorption-freeness will be treated in Section~\ref{sect:minimalTaylor}. 

% END INLINED SOURCE: absorption.tex
% BEGIN INLINED SOURCE: Zhuk.tex
% !TEX root = GH-UA.tex

\subsection{Central Absorption}
\label{sect:ternary-abs}
It will be convenient to work with ternary absorbing (i.e., 3-absorbing) subalgebras instead of absorbing subalgebras with respect to terms of unbounded arity; this will in particular help in the applications of the absorption theorem in Section~\ref{sect:cyclic}
and Section~\ref{sect:bounded-width}. 
In fact, we can often guarantee the existence of subalgebras with an even stronger form of absorption, namely \emph{central absorption}, which implies ternary absorption. 
The results in this section are from~\cite{Strong-Subalgebras-Published} and parts of the presentation are inspired by~\cite{BradyNotes}. 

\begin{definition}\label{def:central-abs}
Let $\fA$ be an algebra. We say that $C \abs \fA$ is \emph{centrally absorbing}\footnote{This terminology is motivated by the concept of a center from Definition~\ref{def:center}, but not with the notion of centrality from Definition~\ref{def:central}.}, written 
$C \cabs \fA$, if  for every $a \in A \setminus C$ 
$$ (a,a) \notin \langle (\{a\} \times C) \cup (C \times \{a\}) \rangle_{\fA^2}.$$
\end{definition}

\begin{figure}
\begin{center}
\includegraphics[scale=.5]{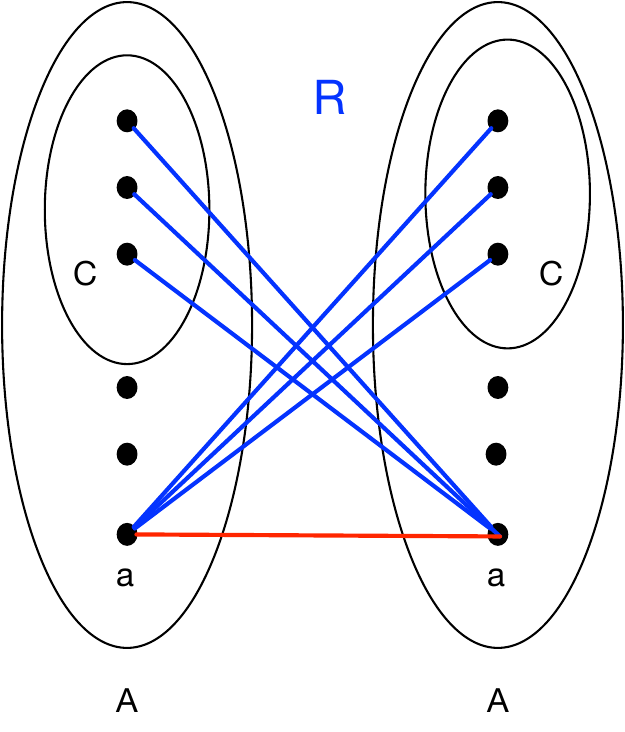}
\end{center}
\caption{An illustration for the definition of central absorption (Definition~\ref{def:central-abs}).}
\label{fig:central-abs}
\end{figure}

See Figure~\ref{fig:central-abs}. 

\begin{example}\label{expl:cabs}
If $\fA = (A;m)$ where $m$ is a majority operation, then $\{c\} \cabs \fA$ for every $c \in A$, because for every $a \in A \setminus C$ 
%the subalgebra $\{a\}$ is centrally 
%The absorbing subuniverse $\{0\}$ of
%$\fA := (\{0,1\};\majority)$ (Example~\ref{expl:wedge-absorption}) is centrally absorbing, because 
$$(a,a) \notin \langle (c,a),(a,c) \rangle_{\fA^2} = \{(c,a),(a,c)\}.$$
The absorbing subuniverse $\{0\}$ of $\fB := (\{0,1\};\wedge)$ (Example~\ref{expl:nu-absorption}) is centrally absorbing, because $(1,1) \notin \langle (0,1),(1,0) \rangle_{\fB^2} = \{(0,0),(0,1),(1,0)\}$. 
\end{example} 

The next example shows an algebra with an absorbing subuniverse which is \emph{not} centrally absorbing. 

\begin{example}
Let $\fA := (\{0,1\};\wedge,\vee)$ be the 2-element lattice (Example~\ref{expl:lattice}).  
Note that $\{0\} \leq \fA$ is absorbing with respect to $\wedge$. However, $\langle \{(0,1),(1,0)\} \rangle = \{0,1\}^2$ contains $(1,1)$,
 so $\{0\}$ is not centrally absorbing. 
\end{example}

The following lemma is the key for proving  several transfer principles for centrally absorbing subalgebras. 

\begin{lemma}[Avoiding a subalgebra square]
\label{lem:central-avoiding-square}
Let $\fA$ be a finite idempotent algebra, let $C \cabs \fA$, and
let $\emptyset\neq E\leq\fA$ satisfy $E\cap C=\emptyset$. Then there
exists $e\in E$ such that
\[
L:=
\left\langle
(\{e\}\times C)\cup(C\times\{e\})
\right\rangle_{\fA^2}
\]
satisfies $L\cap E^2=\emptyset$.
\end{lemma}

\begin{proof}
Since $\fA$ is finite, we may choose $e\in E$ such that $D:=\langle C\cup\{e\}\rangle_{\fA}$ 
is minimal with respect to inclusion among the subalgebras
$\langle C\cup\{e'\}\rangle_{\fA}$, where $e'\in E$. Let
\[
L:=
\left\langle
(\{e\}\times C)\cup(C\times\{e\})
\right\rangle_{\fA^2}.
\]
Suppose, for a contradiction, that $(e_1,e_2)\in L\cap E^2$. Define
\[
P:=\{y\in D\mid (x,y)\in L\text{ for some }x\in E\}.
\]
Since $E\leq\fA$ and $L\leq\fA^2$, the set $P$ is a subalgebra of
$\fA$. Moreover, $P\leq D$, since $L\leq D^2$. For every $c\in C$,
we have $(e,c)\in L$, and hence $c\in P$. Also $e_2\in P$, since
$(e_1,e_2)\in L$. Consequently,
\[
\langle C\cup\{e_2\}\rangle_{\fA}\leq P\leq D.
\]
By the choice of $e$, this implies
$\langle C\cup\{e_2\}\rangle_{\fA}=D$, and therefore $e\in P$.
Thus, there exists $e_3\in E$ such that $(e_3,e)\in L$.
Now define
\[
Q:=\{x\in D\mid (x,e)\in L\}.
\]
Since $\fA$ is idempotent, $Q$ is a subalgebra of $\fA$. It contains
$C$, because $(c,e)\in L$ for every $c\in C$, and it contains $e_3$.
Hence
\[
\langle C\cup\{e_3\}\rangle_{\fA}\leq Q\leq D.
\]
Again, the choice of $e$ implies
$\langle C\cup\{e_3\}\rangle_{\fA}=D$. In particular, $e\in Q$, so
$(e,e)\in L$. Since $e\notin C$, this contradicts $C\cabs\fA$.
Therefore, $L\cap E^2=\emptyset$.
\end{proof}

We have the following transfer principles for centrally absorbing subalgebras.

\begin{lemma}
\label{lem:central-abs-trans}
% Source? 
% both parts should be in the strong subalgebras paper of Dima.
Let $C$ be a congruence of a finite idempotent algebra $\fA$. Then 
\begin{enumerate} 
\item if $\fB \cabs \fA/_C$, then $\bigcup B \cabs \fA$. 
\item if $\fB \cabs \fA$, then $\fB/_C\cabs \fA/_C$. 
\end{enumerate} 
\end{lemma} 
\begin{proof}
For 1., we already know that $B' := \bigcup B \abs \fA$ (Lemma~\ref{lem:abs-hom}). 
Let $a \in A \setminus B'$. 
Suppose for contradiction that $(a,a) \in \langle (\{a\} \times B') \cup (B' \times \{a\}) \rangle_{\fA^2}$. Then there 
are a term $t(x_1,\dots,x_k,y_1,\dots,y_l)$ and $u_1,\dots,u_k,v_1,\dots,v_l \in B'$
such that 
$$(t^{\fA}(a,\dots,a,v_1,\dots,v_l),t^{\fA}(u_1,\dots,u_k,a,\dots,a)) = (a,a).$$
Note that $a/_C \notin B'/_C = B$
and that 
\begin{align*} (a/_C,u_1/_C),\dots,(a/_C,u_k/_C) & \in 
%\{a/_{\sim}\} \times B/_{\sim} = 
\{a/_C\} \times B \\
(v_1/_C,a/_C),\dots,(v_l/_C,a/_C) & \in 
%B/_{\sim} \times \{a/_{\sim}\} 
B \times \{a/_C\}
\end{align*} 
Thus, 
\begin{align*}
(a/_C,a/_C)
 = \; 
 %& 
%(f^{\fA/{\sim}}(a/{\sim},\dots,a/{\sim},v_1/{\sim},\dots,v_l/{\sim}),f^{\fA/{\sim}}(u_1/{\sim},\dots,u_k/{\sim},a/{\sim},\dots,a/{\sim})) \\
%= \; 
& (t^{\fA}(a,\dots,a,v_1,\dots,v_l),t^{\fA}(u_1,\dots,u_k,a,\dots,a))/_C \\
& 
 \in \langle (\{a/_C\} \times B) \cup (B  \times \{a/_C\})  \rangle_{(\fA/_C)^2}
\end{align*} 
which contradicts the assumption that $\fB \cabs \fA/_C$. 

\medskip 
For 2., it follows from Exercise~\ref{exe:abs-fact} that
$\fB/_C\abs\fA/_C$. Suppose, for a contradiction, that there is a
$C$-class $E$ disjoint from $B$ such that
\[
(E,E)\in
L' := \left\langle
(\{E\}\times B/_C)\cup(B/_C\times\{E\})
\right\rangle_{(\fA/_C)^2}.
\]
Since $\fA$ is idempotent, $E\leq\fA$. By
Lemma~\ref{lem:central-avoiding-square}, applied to $B$ and $E$,
there exists $e\in E$ such that
\[
L:=
\left\langle
(\{e\}\times B)\cup(B\times\{e\})
\right\rangle_{\fA^2}
\quad\text{satisfies}\quad
L\cap E^2=\emptyset.
\]
The quotient map sends $L$ onto $L'$. Since $(E,E) \in L'$, we have 
$L\cap E^2\neq\emptyset$, a contradiction. Therefore, 
$\fB/_C\cabs\fA/_C$.
\end{proof} 

%Note that the previous lemma also follows from the following powerful lemma. To see this, consider the structure $\bB := (A;\sim)$.????  

We also have the following powerful transfer theorem; for usual absorption rather than central absorption, this was Exercise~\ref{exe:abs-pp-trans}. 

\begin{lemma}\label{lem:pp-cabs}
Let $\bB$ and $\bB'$ be relational $\tau$-structures with the same domain.  
Suppose that $\Pol(\bB) = \Clo(\fA)$ and that $R^{\bB'} \cabs R^{\bB} \leq \fA^{\ar(R)}$ for every $R \in \tau$. 
Let $\phi(x_1,\dots,x_n)$ be a primitive positive $\tau$-formula. 
If $R \leq \fA^n$ is the relation defined by $\phi$ in $\bB$ and $R' \leq \fA^n$ is the relation defined by $\phi$ in $\bB'$, then $R' \cabs R$. 
\end{lemma}
\begin{proof} 
We need 
the following three claims, which hold for all algebras $\fA$. 

\medskip 
{\bf Claim 1.} If $\fC \cabs \fA$, then 
$\fC \times \fB \cabs \fA \times \fB$. 

Clearly, $\fC\times\fB \abs \fA\times\fB$
(Exercise~\ref{exe:abs-dummy}). Let
$(a,b)\in(A\times B)\setminus(C\times B)$, so that $a\notin C$, and
put
\[
L:=
\left\langle
\bigl(\{(a,b)\}\times(C\times B)\bigr)
\cup
\bigl((C\times B)\times\{(a,b)\}\bigr)
\right\rangle_{(\fA\times\fB)^2}.
\]
Consider the coordinate projection
\[
\pi_{1,3}\colon(\fA\times\fB)^2\to\fA^2,
\qquad
\pi_{1,3}\bigl((x,y),(x',y')\bigr):=(x,x').
\]
This is a homomorphism, and it maps the generating set of $L$ onto
$(\{a\}\times C)\cup(C\times\{a\})$. Therefore,
\[
\pi_{1,3}(L)
=
\left\langle
(\{a\}\times C)\cup(C\times\{a\})
\right\rangle_{\fA^2}.
\]
If $((a,b),(a,b))$ belonged to $L$, then its image $(a,a)$ would
belong to the subalgebra on the right, contradicting
$\fC\cabs\fA$. Hence $\fC\times\fB\cabs\fA\times\fB$.

\medskip 
{\bf Claim 2.} If $\fC_1 \cabs \fA$, 
$\fC_2 \cabs \fA$, then $C_1 \cap C_2 \cabs \fA$. By Corollary~\ref{cor:abs-intersect} we have $C_1 \cap C_2 \abs \fA$. 
Let $a \in A \setminus (C_1 \cap C_2)$. 
Then $a \notin C_i$ for some $i \in \{1,2\}$, 
Since $C_i \cabs \fA$, we have
$$(a,a) \notin \langle (\{a\} \times C_i) \cup (C_i \times \{a\}) \rangle_{\fA^2}.$$
The generating set obtained by replacing $C_i$ with $C_1 \cap C_2$ is contained in the one displayed above. By monotonicity of generated subalgebras,
$$\langle (\{a\} \times (C_1 \cap C_2)) \cup ((C_1 \cap C_2) \times \{a\}) \rangle_{\fA^2}
\subseteq \langle (\{a\} \times C_i) \cup (C_i \times \{a\}) \rangle_{\fA^2}.$$
Therefore $(a,a)$ is not in the subalgebra on the left, and $C_1 \cap C_2 \cabs \fA$. 

\medskip 
{\bf Claim 3.} If $\fC \cabs \fR \leq \fA^k$ and $I \subseteq [k]$, 
then $\pr_I(C) \cabs \pr_I 
(R) \leq \fA^I$. 
Let $\sim$ be the congruence defined on $\fA^k$ by $a_1 \sim a_2 \Leftrightarrow \pr_I(a_1) = \pr_I(a_2)$. Then the statement follows from the second item in Lemma~\ref{lem:central-abs-trans}.

\medskip 
By Claim 1, we may assume that all relation symbols that appear in $\phi$ have the same arity; we may also assume that each conjunct has no repeated entries. 
Claim 2 then implies that the quantifier-free part of $\phi$ defines in $\bB'$ a relation that centrally absorbs
the relation defined by this formula in $\bB$. 
Claim 3 then implies the statement. 
\end{proof} 

In particular, we obtain the following transfer principle for central absorption; the corresponding transfer result for usual absorption was lemma~\ref{lem:abs-trans-rel}. 

\begin{corollary}\label{cor:cabs}
Let $R \leq \fA_1 \times \cdots \times \fA_n$ and $B_i \cabs \fA_i$ for every $i \in [n]$.
Then $$R \cap (B_1 \times \cdots \times B_n) \cabs \fR.$$
\end{corollary}
\begin{proof} 
Immediate from Lemma~\ref{lem:pp-cabs} (see Proposition~\ref{prop:multisorted}). 
\end{proof} 

A more subtle source of centrally absorbing subalgebras 
is the following proposition. 

\begin{lemma}\label{lem:centrally-abs}
% Source? Cyclic terms paper?  
Let $\fA$ and $\fB$ be finite idempotent 
algebras such that $\fB$ is Taylor and has no proper 2-absorbing subuniverses.
% and $\Clo(\fB)$ does not have a minion homomorphism to the projections. 
Let $R \leq \fA \times \fB$ with left center $C$ 
be such that $\pi_1(R) = A$. 
%for every $a \in A$ there exists $b \in B$ such that $(a,b) \in R$. 
Then $C \cabs \fA$. 
\end{lemma}
% Source? 
\begin{proof}
By Corollary~\ref{cor:2-abs-centre} we have that $C \abs \fA$. If $C$ is not centrally absorbing, then for some $n,m \in {\mathbb N}$ there exists $a \in A \setminus C$ and 
a term operation $t$ of $\fA$ of arity $n+m$
and $c_1,\dots,c_m,d_1,\dots,d_n \in C$ 
 such that 
$$t(a,\dots,a,c_1,\dots,c_m) = a = t(d_1,\dots,d_n,a,\dots,a).$$
Note that $\{a\} + R$ (using the terminology from Exercise~\ref{exe:walk}) is a proper subalgebra of $\fB$: we have $\{a\} + R \neq B$ because $a \notin C$, and $\{a\} + R \neq \emptyset$ by assumption.  
Moreover, $\{a\}+R$ 
is 2-absorbing with respect to $f$ given by 
$$f(x,y) := t(\underbrace{x,\dots,x}_n,\underbrace{y,\dots,y}_m):$$ 
if $b \in \{a\}+R$ and $u \in B$, note that
$(c_1,u),\dots,(c_m,u) \in R$ and $(a,b) \in R$,
and hence %$f(a',u) \in a + R$, because 
\begin{align*}
f(b,u) & = \; t(b,\dots,b,u,\dots,u) \\
 & \in \{ t(a,\dots,a,c_1,\dots,c_m)\} + R = \{a\} + R.
\end{align*}
Similarly, $f(u,b) = t(u,\dots,u,b,\dots,b) \in \{t(d_1,\dots,d_n,a,\dots,a)\} + R = \{a\} + R$. 
This contradicts the assumption that $\fB$ has no proper 2-absorbing subuniverses. 
\end{proof}

%One application of centrally absorbing subalgebras is that they give 3-absorbing subalgebras. 

Interestingly, centrally absorbing subalgebras are 3-absorbing (Theorem~\ref{thm:centrally-3abs}). 
To prove this, we use essential relations. 
Recall that $m$-absorbing subalgebras $\fB$ imply the absence of $B$-essential relations of arity at most $m$ (Lemma~\ref{lem:essential-abs}). 
As we will see in Proposition~\ref{prop:central-abs-essential}, 
centrally absorbing subalgebras $\fB_1,\dots,\fB_k$ even imply the absence of $k$-ary relations, for $k \geq 3$, 
that are essential in the following generalised (multisorted) sense. 

\begin{definition}
Let $\fA_1,\dots,\fA_k$ be algebras with the same signature and for $i \in [k]$ let $B_i \leq \fA_i$.  
Then $R \leq \fA_1 \times \cdots \times \fA_k$ is called \emph{$(B_1,\dots,B_k)$-essential} if 
\begin{align*}
R \cap (B_1 \times \cdots \times B_{i-1} \times A_i \times B_{i+1} \times \cdots \times B_k) & \neq \emptyset & \text{ for all $i \in [k]$, but} \\
R \cap (B_1 \times \cdots \times B_k) & = \emptyset. 
\end{align*} 
\end{definition}

The following is immediate from the definition. 
\begin{lemma}\label{lem:essential-down-2}
If $R \leq \fA_1 \times \cdots \times \fA_k$ is $(B_1,\dots,B_k)$-essential, and $i \in [k]$,  then 
the $(k-1)$-ary relation 
$$R_{\backslash i} := \pr_{[k] \setminus \{i\}}(R \cap (A_1 \times \cdots \times A_{i-1} \times B_i \times A_{i+1} \times \cdots \times A_k))$$
is $(B_1,\dots,B_{i-1},B_{i+1},\dots,B_k)$-essential. 
\end{lemma}

The following lemma is from~\cite[Lemma 6.10]{Strong-Subalgebras-Published}. 
\begin{lemma}[Essential doubling, multisorted version]
\label{lem:doubling-dima}
Let $\fA_1,\dots,\fA_k$, for $k \geq 2$, be finite idempotent algebras with the same signature. 
Let $B_1 \cabs \fA_1,\dots,B_k \cabs \fA_k$, and let $R \leq \fA_1 \times \cdots \times \fA_k$ be $(B_1,\dots,B_k)$-essential. 
Then there exists a $(B_1,\dots,B_{k-1},B_1,\dots, B_{k-1})$-essential relation $R' \leq \fA_1 \times \cdots \times \fA_{k-1} \times \fA_1 \times \cdots \times \fA_{k-1}$. 
\end{lemma}
\begin{proof}
Define $E := \pr_k(R \cap (B_1 \times \cdots \times B_{k-1} \times A_k))$.
The relation $R\cap(B_1\times\cdots\times B_{k-1}\times A_k)$
is a subalgebra of
$\fA_1\times\cdots\times\fA_k$, and hence $E\leq\fA_k$.
Moreover, $E\neq\emptyset$, since $R$ is
$(B_1,\dots,B_k)$-essential.
We also have $E\cap B_k=\emptyset$. Indeed, if $e\in E\cap B_k$,
then the definition of $E$ yields
$(b_1,\dots,b_{k-1},e)\in R$
for some $b_i\in B_i$. This would imply
$R\cap(B_1\times\cdots\times B_k)\neq\emptyset$,
 contrary to the essentiality of $R$.

We may therefore apply
Lemma~\ref{lem:central-avoiding-square} to $\fA_k$, $B_k$, and $E$.
It yields an element $e\in E$ such that
\[
L:=
\left\langle
(\{e\}\times B_k)\cup(B_k\times\{e\})
\right\rangle_{\fA_k^2}
\]
satisfies $L\cap E^2=\emptyset$.
Let 
%$R' \leq \fA^{2k-2}$ 
$$R'\leq\fA_1\times\cdots\times\fA_{k-1}
\times\fA_1\times\cdots\times\fA_{k-1}$$
be given as the set of all tuples $(x_1,\dots,x_{k-1},y_1,\dots,y_{k-1})$
that satisfy 
\begin{align}
\exists x_k,y_k \big (R(x_1,\dots,x_k) \wedge L(x_k,y_k) \wedge R(y_1,\dots,y_k) \big).
\label{eq:doubling}
\end{align}
We verify that $R'$ is $(B_1,\dots,B_{k-1},B_1,\dots,B_{k-1})$-essential. 
First note that $L \cap E^2 = \emptyset$ implies that $R' \cap (B_1 \times \dots \times B_{k-1} \times B_1 \times \dots \times B_{k-1}) = \emptyset$. Next, for $i \in [k-1]$ we need to show that
\begin{align} R' \cap (B_1 \times \cdots \times B_{i-1} \times A_i \times B_{i+1} \times \cdots \times B_{k-1} \times B_1 \times \cdots \times B_{k-1}) \neq \emptyset.
\label{eq:first-doubling}
\end{align}
We may choose $(x_1,\dots,x_k) \in R \cap (B_1 \times \cdots \times B_{i-1} \times A_i \times B_{i+1} \times \cdots \times B_k)$, because $R$ is $(B_1,\dots,B_k)$-essential. 
We may also choose
$(y_1,\dots,y_k) \in R \cap (B_1 \times \cdots \times B_{k-1} \times \{e\})$
since $e \in E$. Note that $(x_k,y_k) = (x_k,e) \in L$, so $(x_1,\dots,x_k,y_1,\dots,y_k)$ satisfies the quantifier-free part of~\eqref{eq:doubling}, which proves~\eqref{eq:first-doubling}. 
If $i \in \{k,\dots,2k-2\}$, we set $j := i-k+1$ and show analogously that 
\begin{align*} R' \cap (B_1 \times \cdots \times B_{k-1} \times B_1 \times \cdots \times B_{j-1} \times A_j \times B_{j+1} \times \cdots \times B_{k-1}) \neq \emptyset.
\end{align*}
This proves that $R'$ is $(B_1,\dots,B_{k-1},B_1,\dots,B_{k-1})$-essential. 
\end{proof} 

\begin{corollary}[Essential doubling, one-sorted version]
% Source: Brady? Check, he says its from Marcin?
\label{cor:doubling}
Let $\fA$ be finite idempotent and let $C \cabs \fA$. 
Suppose that $R \leq \fA^k$, for $k \geq 3$, is $C$-essential.
Then there exists $R' \leq \fA^{2k-2}$ which is $C$-essential. 
\end{corollary}
\begin{proof} 
Immediate from Lemma~\ref{lem:doubling-dima} by taking all algebras and all centrally absorbing subalgebras equal.
\end{proof}

\begin{theorem}\label{thm:centrally-3abs}
Let $\fA$ be a finite idempotent algebra such that $C \cabs \fA$. 
Then $C$ 3-absorbs $\fA$. 
\end{theorem}
\begin{proof}
Suppose for contradiction that 
$\fC$ does not 3-absorb $\fA$. Then 
there exists a $C$-essential relation $R \leq \fA^3$ by Theorem~\ref{thm:essential-converse}. 
Applying Corollary~\ref{cor:doubling} sufficiently many times we may obtain $C$-essential relations of arbitrarily large arity. 
Then Theorem~\ref{thm:essential-converse} (combined with Remark~\ref{rem:abs-arity}) implies that
$\fC$ is not absorbing, contrary to our assumptions. 
\end{proof} 

\begin{proposition}\label{prop:central-abs-essential} 
Let $\fA_1,\dots,\fA_k$ be finite idempotent algebras and 
suppose that $\fB_1 \cabs \fA_1,\dots,\fB_k \cabs \fA_k$ and $k \geq 3$. 
Then there are no $(B_1,\dots,B_k)$-essential relations $R \leq \fA_1 \times \cdots \times \fA_k$. 
\end{proposition} 
\begin{proof}
Suppose for contradiction that $R \leq  \fA_1 \times \cdots \times \fA_k$ is $(B_1,\dots,B_k)$-essential. 
By repeatedly applying Lemma~\ref{lem:essential-down-2}
we find a ternary $(B_1,B_2,B_3)$-essential relation. 
By Lemma~\ref{lem:doubling-dima}, 
there is a $(B_1,B_2,B_1,B_2)$-essential relation, and by Lemma~\ref{lem:essential-down-2} applied with $i=2$ there is a 
$(B_1,B_1,B_2)$-essential relation. 
Finally, another application of Lemma~\ref{lem:doubling-dima} gives us a $(B_1,B_1,B_1,B_1)$-essential relation.
By applying Lemma~\ref{lem:doubling-dima} sufficiently many times, we obtain $B_1$-essential relations of arbitrarily large arity. This contradicts the assumption that $B_1 \abs \fA_1$ by Lemma~\ref{lem:essential-abs} combined with Remark~\ref{rem:abs-arity}. 
\end{proof}

We use this property to prove the so-called 
\emph{k-Helly property} for centrally absorbing subalgebras.  

\begin{corollary}\label{cor:helly}
Let $\fA$ be a finite idempotent algebra. For $k \geq 3$, let $\fB_1,\dots,\fB_k \cabs \fA$ 
be such that $\bigcap_{i \in [k] \setminus \{ j\}} B_i \neq \emptyset$ for every $j \in [k]$. 
Then $\bigcap_{i \in [k]} B_i \neq \emptyset$. 
\end{corollary} 
\begin{proof}
Suppose for contradiction that $B_1 \cap \cdots \cap B_k=\emptyset$. Let $R \leq \fA^k$ be the relation 
$\{(a,\dots,a) \mid a \in A\}$. 
Then 
\begin{align*}
R \cap (A \times B_2 \times \cdots \times B_k) 
& = \{(a,\dots,a) \mid a \in B_2 \cap \cdots \cap B_k \} \neq \emptyset, \\
\cdots \\
R \cap (B_1 \times \cdots \times B_{k-1} \times A) 
& = \{(a,\dots,a) \mid a \in B_1 \cap \cdots \cap B_{k-1}\}  \neq \emptyset, \\
%
%R \cap (B_1 \times A \times B_3) & \supseteq R \cap ((B_1 \cap B_3) \times A \times (B_1 \cap B_3)) = \{(a,a,a) \mid a \in B_1 \cap B_3) \neq \emptyset, \\
%R \cap (B_1 \times B_2 \times A) & \supseteq R \cap ((B_1 \cap B_2) \times (B_1 \cap B_2) \times A) = \{(a,a,a) \mid a \in B_1 \cap B_2) \neq \emptyset,  \\
\text{ and } R \cap (B_1 \times \cdots \times B_k) & = \{(a,\dots,a) \mid a \in B_1 \cap \cdots \cap B_k\} = \emptyset
\end{align*} 
and hence $R$ is $(B_1,\dots,B_k)$-essential, so by Proposition~\ref{prop:central-abs-essential} one of $B_1,\dots,B_k$ is not centrally absorbing. 
\end{proof} 

Combining these results with the proof from Section~\ref{sect:absorb}, we
obtain a strengthened form of the Absorption Theorem (Theorem~\ref{thm:absorb}).

\begin{theorem}
\label{thm:3-absorb}
Let $\fA$ and $\fB$ be finite idempotent algebras with the same signature such that 
$\fB$ is Taylor. 
%$\Clo(\fB)$ has no minion homomorphism to $\Proj$. 
Then 
%$\fA,\fB \in \HSP(\fC)$ and any
for every linked and subdirect $R \leq \fA \times \fB$ one of the following is true: 
\begin{enumerate}
\item $R = A \times B$; 
\item $\fA$ or $\fB$ has a proper binary or centrally absorbing subuniverse.
%\item $\fA$ has a proper centrally absorbing subuniverse. 
%\item $\fB$ has a proper $3$-absorbing subuniverse. 
\end{enumerate}
\end{theorem}
\begin{proof}
Suppose that 
$R \neq A \times B$, because otherwise item 1 of the theorem holds and we are done. 
Let $C$ be the left centre of $R$.
Note that $C \neq A$ because $R \neq A \times B$. 

Suppose that $\fB$ does not have a proper binary absorbing subuniverse. 
Then Corollary~\ref{cor:2-abs-centre} implies that  $C \abs \fA$.
If $C$ is non-empty,  then by Lemma~\ref{lem:centrally-abs} we have found a proper centrally absorbing subuniverse of $\fA$. 
%In this case, Proposition~\ref{prop:centrally-3abs} implies that $C$ 3-absorbs $\fA$ and item 2 of the theorem holds. 

Otherwise, suppose that $C=\emptyset$. Then the first alternative of
Proposition~\ref{prop:new-absorption} does not apply, and hence there
exists a subdirect relation $R'\leq\fB^2$ whose left centre $C'$ is a
proper subuniverse of $\fB$. Applying
Lemma~\ref{lem:centrally-abs} to $R'$, with both algebras equal to
$\fB$, shows that $C'\cabs\fB$.
\end{proof}

%\newpage
\begin{exercises}
\exercise[Blau]\label{exe:source-241} Show that Corollary~\ref{cor:helly} fails for $k=2$.
%\item %Show that Corollary~\ref{cor:helly} fails for ternary absorption instead of central absorption. Haeh? ist doch gar nicht zu Absorption.
%Find a finite idempotent algebra with three ternary absorbing subalgebras that do not satisfy the $2$-Helly property.
% Uebung oben ist staerker.
\exercise[Blau]\label{exe:source-242} Determine the centrally absorbing subalgebras of the algebra \\
$\fA_k := (\{0,1\};f_k)$ from Example~\ref{expl:bool-nu}, for all $k \geq 3$.
% Source: me
% Solution: we have \{0\},\{1\} \cabs A.
% For k = 3 both 0 and 1 are centrally
% absorbing since we have a majority operation.
% For k > 3, the Johnsson operation p witnesses
% that \{0\} is ternary absorbing:
% p(x,y,z) = x \wedge (y \vee z).
% Here 0 is absorbing:
% 0 \wedge ... = 0
% 1 \wedge (0 \vee 0) = 0
% it is even centrally absorbing:
% the relation x=1 or y=1 is preserved by
% f_k for all k \geq 0.
% But p is not showing that 1 is absorbing:
% 0 \wedge (1 \vee 1) = 0.
\exercise[Rot]\label{exe:source-243} Let $\fB$ be a finite algebra.
Prove that the following are equivalent.
\begin{itemize}
\item $\fB$ has a majority term;
\item every set of the form $\{b\}$, for $b \in B$, is a centrally absorbing subuniverse of $\fB$.
\end{itemize}
% forward: as in the example.
% backward: get absorption with respect to the same term for all singletons, which must then be a majority.
%
% not equivalent: every subalgebra of $\fB$ with more than one element has a proper centrally absorbing subuniverse.
% counterexample: semi-lattice on {0,1}.
\end{exercises}

\subsection{Paper, Scissors, Stone}
\label{sect:pss}
This section describes a fundamental example of a three-element algebra $\fA$ which shows some interesting behaviour and which provides important intuition for the abstract results in the following sections.
% (it already appeared in Exercise~\ref{exe:pss}). 
On the one hand, it is absorption-free, but on the other hand it is not hereditarily absorption-free. 

\begin{definition}[Paper-Scissors-Stone algebra]
\label{def:pss}
Let $\fA$ be the algebra \\
with the domain 
$A := \{0,1,2\}$ 
and let 
$\cdot \colon A^2 \to A$ 
be the binary operation \\
given by the multiplication  table on the right.  
\vspace{-2cm}
\begin{flushright}
\begin{tabular}{l|lll}
%[l|lll]
$\cdot$ & 0 & 1 & 2 \\
\hline
0 & 0 & 1 & 0 \\
1 & 1 & 1 & 2 \\
2 & 0 & 2 & 2  
\end{tabular}
\end{flushright}
%\end{center}
\end{definition}

\begin{wrapfigure}{r}{0.3\textwidth}
    \centering
    \includegraphics[width=0.25\textwidth]{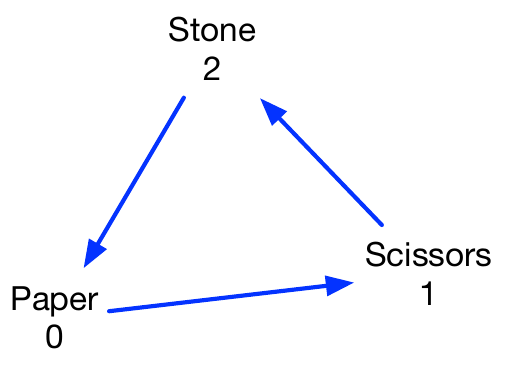}
\end{wrapfigure}

Note that $\fA$ has the automorphism
$$\rho \colon x \mapsto x+1 \mod 3.$$
Let $C_3 := \{(a,\rho(a)) \mid a \in A\}$ be the binary relation on $A$ which denotes the graph of $\rho$. All three 2-element subsets of $A$ are subuniverses of the algebra $\fA = (A;\cdot)$, and in each of the corresponding subalgebras the operation $\cdot$ denotes a semilattice operation; however, $\cdot$ itself is not a semilattice operation. 
We will see below (see Remark~\ref{rem:pss-absorption-free}) that none of the proper  subalgebras of $\fA$ is absorbing. 
However, $\{1\}$ is a proper absorbing subuniverse of the subalgebra of $\fA$ with domain $\{0,1\}$, so $\fA$ is not HAF, and in particular not Abelian by Theorem~\ref{thm:abelian-haf}. 
Note that for any $a,b \in A$
\begin{align}
(a \cdot \rho^{-1}(b)) \cdot b = b. \label{eq:pss-cancel}
\end{align}

%The relation $R$ 
%$C_3$
%is $B$-essential: 
%we have $R \cap (A \times B) = \{(2,0),(0,1)\}$
%and $R \cap (B \times A) = \{(0,1),(1,2)\}$, 
%but $R \cap B^2 = \emptyset$. 
%Hence, Lemma~\ref{lem:essential-abs} implies
%that $B$ is not $n$-absorbing for every $n \geq 2$.  Neither is it $1$-absorbing since $B$ is a poper subuniverse and $\fA$ is idempotent. 
%For the subuniverse $C = \{0\}$, the argument is similar, and every other proper  subuniverse is symmetric to $B$ or $C$ via $\rho$. 

The algebra $\fA$ is simple. Indeed, if $C$ is a congruence which contains $(0,1)$, then it must also contain $(0,1) \cdot (2,2) = (0,2)$,
and therefore also $(1,0)$ and $(2,0)$ by symmetry. By similar reasoning we conclude that $C = A^2$, which shows that $\fA$ has no proper congruences.

We first present an interesting relational description of $\Clo(\fA)$. 
%Let $\bA = (A;C_3)$ be the relational structure with the same domain $A$ as $\fA$ and a single binary relation 
Note that $\Inv(\fA)$ also contains the relation $$R_3^= := \big \{(x,y,z) \in A^3 \mid x \in \{0,1\} \wedge (x = 0 \Rightarrow y = z) \big \}.$$  
The relation $R_3^= \leq \fA^3$ is not subdirect, because the first argument cannot take value $2$.

\begin{lemma}\label{lem:subdirect-pss}
Let $R \leq {\fA}^n$, for $n \geq 1$, be subdirect. Then $R$ 
can be defined by a conjunction of atomic formulas over $(A;C_3)$. 
\end{lemma}
\begin{proof}
Our proof is by induction on $n$. 
For $n=1$ we have $R = A$ and hence $R$ can be defined by $x=x$. 
If $n = 2$, then $R$ is the graph 
of an automorphism of $\fA$ or linked, because $\fA$ is simple (Exercise~\ref{exe:simple-linked}). 
In the first case, either $C_3(x,y)$, $C_3(y,x)$, or $x=y$ defines $R$, and we are done, so let us assume that $R$ is linked. 

\medskip 
{\bf Claim.} $R = A^2$. Indeed, if $(u,v) \in A^2$,
 then the linkedness of $R$ implies that $(u,v) \in R$ or there exists a path 
$p_1,\dots,p_{2k} \in A$ for $k \geq 1$ such that
$(u,p_1),(p_2,p_1),(p_2,p_3),\dots,(p_{2k},v) \in R$. In the first case there is nothing to be shown. Otherwise, choose $k$ as small as possible. 
If $k=1$, we may assume that $p_2 \neq u$ and $p_1 \neq v$, because otherwise we are in the first case. 
Let $a,b \in A$ be such that $\{u,p_2,a\} = A = \{v,p_1,b\}$. Since $R$ is subdirect, there exist $a',b' \in A$ such that $(a,a'),(b',b) \in R$. 

Note that $(u,p_1) \cdot (p_2,v) \in \{(u,v),(u,p_1),(p_2,p_1),(p_2,v)\}$.
If $(u,p_1) \cdot (p_2,v) = (u,v)$ then $(u,v) \in R$, contrary to the minimal choice of $k$. 
If $(u,p_1) \cdot (p_2,v) = (p_2,v)$, we consider  the following subcases. 
\begin{enumerate}
\item $a' = v$. Then $(a,a') \cdot (u,p_1) = (u,a') = (u,v) \in R$, contradiction.  
\item $a' = p_1$. Then $(a,a') \cdot (p_2,v) = (a,v) \in R$, and we are in the first subcase. 
\item $a' = b$. Then $(a,a') \cdot (p_2,p_1) = (a,p_1) \in R$ and we are in the second subcase. 
\end{enumerate}
The case that $(u,p_1) \cdot (p_2,v) = (u,p_1)$ is similar. 
Finally, suppose that $(u,p_1) \cdot (p_2,v) = (p_2,p_1)$.  Again, we break into subcases. 
\begin{enumerate}
\item $a' = v$ and $b' = u$. Then 
$(a,a') \cdot (b',b) = (u,v) \in R$, a contradiction. 
\item $a' = p_1$ and $b' = p_2$. Then $(a,a') \cdot (b',b) = (a,b)$ and $(a,b) \cdot (p_2,v) = (a,v) \in R$. Moreover, $(u,p_1) \cdot (a,b) = (u,b) \in R$. 
Hence, $(a,v) \cdot (u,b) = (u,v) \in R$, 
and we are done.   
\item $a' = v$ and $b' = p_2$. Then $(p_2,p_1) \cdot (a,a') = (a,p_1) \in R$ and we are in subcase 2. 
\item $a' = p_1$ and $b' = u$. Then 
$(b',b) \cdot (p_2,p_1) = (p_2,b) \in R$ and we are again in subcase 2. 
\item $a' = b$. Then $(a,a') \cdot (p_2,v) = (a,v) \in R$ and $(a,a') \cdot (u,p_1) = (u,a') \in R$, and we are in subcase number one. 
\end{enumerate}
Finally, if $k \geq 2$, then we may assume that $\{u,p_2,p_4\} = A = \{v,p_1,p_3\}$. 
Then either $(u,p_1) \cdot (p_4,p_3)$
or $(u,p_1) \cdot (p_4,v)$ is from
$\{(u,v),(u,p_3),(p_4,p_1)\}$, and in each case we obtain a contradiction to the minimal choice of $k$. 
This concludes the proof of the claim. 

Now consider the case $n \geq 3$. If $R(x_1,\dots,x_n)$ implies $C_3(x_i,x_j)$ for some $\{i,j\} \in {[n] \choose 2}$, then $R$ has the definition $C_3(x_i,x_j) \wedge \psi$ in $\bA$, where
$\psi$ is the definition of $\pi_{[n] \setminus \{j\}}(R)$ in $\bA$, which exists by inductive assumption. Similarly we can treat the case that $x_i = x_j$ is implied instead of $C_3(x_i,x_j)$. 
Otherwise, we will show that $R = A^n$. Let $t \in A^n$. For any $a \in A$, the $(n-1)$-ary relation $R_a := \{\bar x \mid (\bar x,a) \in R \}$ 
is preserved by $\cdot$. 
%COPIED FROM THE PROOF OF ZHUK CASES. 
Moreover, we will prove that it is subdirect. Indeed, let $b \in A$. Note that the binary relation $\pi_{n-1,n}(R)$ equals $A^2$ by the case $n=2$, and hence in particular contains $(b,a)$. Therefore, there exists $c' \in A^{n-2}$ such that $(c',b,a) \in R$, and thus $(c',b) \in R_a$. Similar arguments apply to the other arguments of $R_a$, showing that $R_a \leq {\bf A}^{n-1}$ is subdirect. 

%The relations  
First consider the case $n=3$. Since $R_{t_3} \leq {\fA}^2$ is subdirect, by the case $n=2$ the
formula $R_{t_3}$ equals $A^2$, $=_A$, $C_3$, 
or $\{(y,x) \mid (x,y) \in C_3\}$.  
%\begin{align*}
%R' & := \{ (x_1,x_2) \mid (x_1,x_2,t_3,t_4,\dots,t_n) \in R\} \\
%R'' &  := \{ (x_1,x_3) \mid (x_1,t_2,x_3,t_4,\dots,t_n) \in R\}
%\end{align*}
In any case, $R_{t_3}$ contains $(t_1,t'_2)$ and $(t_1',t_2)$ for some $t_1',t_2' \in A$.
%Likewise, the relation $R' := \{(x,y) \mid (x,t_2,y) \in R\}$ 
%contains $(t_1'',t_3)$ and $(t_1,t_3'')$  for some $t_1'',t_3'' \in A$. 
If $t_1'=t_1$ or $t_2'=t_2$, 
%$t_1''=t_1$, or $t_3'' = t_3$, 
then $t \in R$ and we are done. 
If $t_1' = \rho^{-1}(t_1)$ and $t_2' = \rho^{-1}(t_2)$, then $(t'_1,t_2,t_3) \cdot (t_1,t'_2,t_3) = t \in R$ and we are again done. 
Hence, up to reordering the arguments of $R$ we may assume that $t_2'=\rho(t_2)$. Since $R_{t_3}$ is preserved by the automorphism $\rho^{-1}$,
we therefore get that
$(\rho^{-1}(t_1),t_2) \in R_{t_3}$. Therefore, $t' := (\rho^{-1}(t_1),t_2,t_3) \in R$. 
By similar reasoning with the relation $\{(x,y) \mid (t_1,x,y) \in R\}$ instead of $R_{t_3}$ we obtain
that $t'' := (t_1,\rho^{-1}(t_2),t_3) \in R$ or $t'' := (t_1,t_2,\rho^{-1}(t_3)) \in R$.  Applying $\cdot$ to $t',t'' \in R$, we again obtain $t \in R$. 

Finally, consider the case $n > 3$. Then for
$i,j \in {[n-1] \choose 2}$ we have that $\pi_{i,j,n}(R) = A^3$ by the case $n=3$. Hence, for any $a \in A$ we have $\pi_{i,j}(R_a) = A^2$, and it follows from the case $n-1$ that $R_a = A^{n-1}$. This means that $R = A^n$. 
%Note that $R' := \{(x_1,x_2,x_3,t_4,\dots,t_n) \mid x_1,x_2,x_3 \in A\}$ is preserved by $\cdot$ as well, and does not imply $C(x_i,x_j)$ and does not imply $x_i=x_j$ for every $\{i,j\} \in {\{1,2,3\} \choose 2}$. 
%Hence, it is full by the case $n=3$, and hence $t \in R$. 
\end{proof}

\begin{definition}
The class of \emph{pss-Horn clauses} is defined recursively as follows.
The formulas
\[
y\in\{c,d\},\qquad C_3(y,z),\qquad y=z,
\]
where $c,d\in A$, are pss-Horn clauses. Moreover, if $\chi$ is a
pss-Horn clause, $a\in A$, and $x$ is a variable, then
\[
x\in\{a,\rho(a)\}\wedge\bigl(x=a\Rightarrow\chi\bigr)
\]
is a pss-Horn clause. Variables occurring in a pss-Horn clause need
not be distinct.
\end{definition}

Thus, a pss-Horn clause has the form
\[
x_{i_1}\in\{a_1,\rho(a_1)\}\wedge
\left(x_{i_1}=a_1\Rightarrow
\left(
x_{i_2}\in\{a_2,\rho(a_2)\}\wedge
\left(x_{i_2}=a_2\Rightarrow
\left(\cdots\Rightarrow\psi\right)\right)
\right)\right),
\]
where $\psi$ is of the form $y\in\{c,d\}$, $C_3(y,z)$, or $y=z$.

\begin{lemma}[Separation by a pss-Horn clause]
\label{lem:pss-horn-separation}
Let $R\leq\fA^n$, where $n\geq1$, and let $t\in A^n\setminus R$.
Then there exists a pss-Horn clause which is satisfied by every tuple
in $R$ but not by $t$.
\end{lemma}

\begin{proof}
Put $R_0:=R$. Choose a sequence of distinct coordinates
$i_1,\dots,i_k$ of maximal length such that, recursively, for every $j\in[k]$
\begin{align*} 
\pi_{i_j}(R_{j-1})
 & \subseteq \{t_{i_j},\rho(t_{i_j})\} \\
\text{ 
and } \quad \quad 
R_j & = \{r\in R_{j-1}\mid r_{i_j}=t_{i_j}\}. 
\end{align*} 
Put $S:=R_k$ and
$J:=[n]\setminus\{i_1,\dots,i_k\}$. 
Since $\fA$ is idempotent, all the relations $R_j$, and hence $S$,
are subalgebras of $\fA^n$.

We first show that there is a formula $\psi$ of one of the forms
\[
y\in\{c,d\},\qquad C_3(y,z),\qquad y=z
\]
which is satisfied by every tuple in $S$ but not by $t$.

If $S=\emptyset$, we may take $\psi:=C_3(x_1,x_1)$. Thus, suppose
that $S\neq\emptyset$. If $t_i\notin\pi_i(S)$ for some $i\in J$,
then $\pi_i(S)$ is a nonempty proper subalgebra of $\fA$. Hence it
has the form $\{c,d\}$, where we permit $c=d$, and we may take
$\psi:=x_i\in\{c,d\}$. It remains to consider the case that $t_i\in\pi_i(S)$ for every $i\in J$.
Let
\begin{align*} 
M & :=\{i\in J\mid \pi_i(S)=A\} \\
\quad\text{and} \quad \quad 
N & :=J\setminus M.
\end{align*} 
For every $i\in N$, the maximality of the sequence
$i_1,\dots,i_k$ implies that
\[
\pi_i(S)\nsubseteq\{t_i,\rho(t_i)\}.
\]
Since $\pi_i(S)$ is a proper subalgebra containing $t_i$, it follows
that $\pi_i(S)=\{\rho^{-1}(t_i),t_i\}$.
On this two-element subalgebra, $t_i$ is absorbing for $\cdot$:
\[
t_i\cdot a=t_i\qquad\text{for every }a\in\pi_i(S).
\]

Suppose that $M\neq\emptyset$, and put $U:=\pi_M(S)$. The relation
$U$ is subdirect, so Lemma~\ref{lem:subdirect-pss} gives a definition
of $U$ by a conjunction of atomic formulas over $(A;C_3)$. If
$\pi_M(t)\notin U$, then one of these atomic formulas is satisfied by
every tuple in $U$ but not by $\pi_M(t)$. Taking this atomic formula
for $\psi$ proves the claim.

We finally show that the only remaining case is impossible. Thus,
assume that either $M=\emptyset$ or $\pi_M(t)\in U$. If $M\neq
\emptyset$, then $U$ is preserved by the coordinatewise application
of $\rho^{-1}$, because it is defined by atomic formulas over
$(A;C_3)$. Consequently, there are $p,q\in S$ such that
\[
p_i=t_i
\quad\text{and}\quad
q_i=\rho^{-1}(t_i)
\qquad\text{for every }i\in M.
\]
If $M=\emptyset$, choose $p,q\in S$ arbitrarily.

For every $i\in N$, choose $s^i\in S$ such that $s^i_i=t_i$.
If $N\neq\emptyset$, let $s\in S$ be any iterated product of the
tuples $s^i$, for $i\in N$. Since $t_i$ is absorbing in
$\pi_i(S)$, we have $s_i=t_i$ for every $i\in N$.
If $N=\emptyset$, choose $s\in S$ arbitrarily. Now put
\[
r:=(s\cdot q)\cdot p\in S.
\]
For $i\in M$, identity~\eqref{eq:pss-cancel} gives
\[
r_i=(s_i\cdot\rho^{-1}(t_i))\cdot t_i=t_i.
\]
For $i\in N$, the absorbing property of $t_i$ in $\pi_i(S)$ gives
$r_i=t_i$. Moreover, every tuple in $S$ has entry $t_{i_j}$ in
coordinate $i_j$, for every $j\in[k]$. Hence $r=t$, contradicting
$t\notin R$. This proves the existence of $\psi$. Starting with $\chi_k:=\psi$, define recursively, for
$j=k,k-1,\dots,1$,
\[
\chi_{j-1}:=
x_{i_j}\in\{t_{i_j},\rho(t_{i_j})\}
\wedge
\bigl(x_{i_j}=t_{i_j}\Rightarrow\chi_j\bigr).
\]
Then $\chi_0$ is a pss-Horn clause. The choice of
$i_1,\dots,i_k$ and the fact that $\psi$ holds on $S$ show that
every tuple in $R$ satisfies $\chi_0$. On the other hand, all the
implications in $\chi_0$ are activated by $t$, and $\psi$ is false
at $t$. Therefore, $t$ does not satisfy $\chi_0$.
\end{proof}

\begin{proposition}
\label{prop:subdirect-pss}
For every $R\subseteq A^n$, where $n\geq1$, the following are
equivalent.
\begin{enumerate}
\item $R$ is preserved by $\cdot$;
\item $R$ can be defined by a conjunction of pss-Horn clauses;
\item $R$ has a primitive positive definition in
$(\{0,1,2\};C_3,R_3^=)$.
\end{enumerate}
\end{proposition}

\begin{proof}
Suppose first that $R$ is preserved by $\cdot$. Then
$R\leq\fA^n$. For every $t\in A^n\setminus R$, choose, using
Lemma~\ref{lem:pss-horn-separation}, a pss-Horn clause $\chi_t$
which holds on $R$ but not at $t$. Then
\[
\bigwedge_{t\in A^n\setminus R}\chi_t
\]
defines $R$. This proves that 1.\ implies 2.

We next prove that 2.\ implies 3. First observe that all one- and
two-element unary relations on $A$ are primitively positively
definable in $(A;C_3,R_3^=)$. Indeed,
\begin{align*} 
U_{01}(x) & :=\exists y,z\;R_3^=(x,y,z)
 && \text{defines $\{0,1\}$} \\
U_{12}(x) & :=\exists u\,
\bigl(U_{01}(u)\wedge C_3(u,x)\bigr)
&& \text{defines $\{1,2\}$} \\
U_{20}(x) & :=\exists u\,
\bigl(U_{01}(u)\wedge C_3(x,u)\bigr)
&& \text{defines $\{2,0\}$.}
\end{align*}
Their pairwise intersections define the
singletons. For $a\in A$, let $\delta_a(x,u)$ be the following primitive
positive formula:
\[
\delta_a(x,u):=
\begin{cases}
U_{01}(u)\wedge x=u
  &\text{if }a=0,\\
U_{01}(u)\wedge C_3(u,x)
  &\text{if }a=1,\\
U_{01}(u)\wedge C_3(x,u)
  &\text{if }a=2.
\end{cases}
\]
Thus, $\delta_a(x,u)$ defines $\{(a,0),(\rho(a),1)\}$. 
We prove by induction over the recursive definition that every
pss-Horn clause is primitively positively definable. This is clear
for the basic clauses $y\in\{c,d\}$, $C_3(y,z)$, and $y=z$.
Suppose that $\chi(\bar y)$ is primitively positively definable and
consider
\[
\theta(x,\bar y):=
x\in\{a,\rho(a)\}\wedge
\bigl(x=a\Rightarrow\chi(\bar y)\bigr).
\]
Let $S$ be the relation defined by $\chi$. If $S=\emptyset$, then
$\theta$ is equivalent to $x=\rho(a)$, which is primitively
positively definable. Suppose that $S\neq\emptyset$, and let
$y_1,\dots,y_l$ be the distinct variables occurring freely in
$\chi$. Then $\theta$ is defined by
\[
\exists u,z_1,\dots,z_l\,
\left(
\delta_a(x,u)
\wedge
\chi(z_1,\dots,z_l)
\wedge
\bigwedge_{j=1}^l R_3^=(u,y_j,z_j)
\right).
\]
Indeed, if $x=a$, then $u=0$, and the displayed formula forces
$y_j=z_j$ for every $j\in[l]$. If $x=\rho(a)$, then $u=1$, and the
variables $z_1,\dots,z_l$ may be chosen as any tuple in the nonempty
relation $S$. This completes the induction. Since conjunctions of
primitive positive formulas are primitive positive, 2.\ implies 3.

Finally, we prove that 3.\ implies 1. The relation $C_3$ is preserved
by $\cdot$, because $\rho$ is an automorphism of $\fA$. To see that
$R_3^=$ is preserved by $\cdot$, let
\[
(x_1,y_1,z_1),(x_2,y_2,z_2)\in R_3^=
\]
and put
\[
(x,y,z):=(x_1\cdot x_2,y_1\cdot y_2,z_1\cdot z_2).
\]
Since $x_1,x_2\in\{0,1\}$, we have $x\in\{0,1\}$. If $x=1$, then
$(x,y,z)\in R_3^=$ immediately. If $x=0$, then
$x_1=x_2=0$, and hence $y_1=z_1$ and $y_2=z_2$. Therefore,
\[
y=y_1\cdot y_2=z_1\cdot z_2=z,
\]
so again $(x,y,z)\in R_3^=$. Thus, $\cdot$ preserves both basic
relations of $(A;C_3,R_3^=)$ and consequently every relation with a
primitive positive definition in this structure. Hence 3.\ implies
1.
\end{proof}

\begin{remark}\label{rem:pss-absorption-free}
The Paper-Scissors-Stone algebra $\fA$ is absorption free. 
First note that 
$B = \{0,1\}$ is not absorbing. 
%$1$-absorbing since $B$ is a proper subuniverse and $\fA$ is idempotent. 
Indeed, for every $n \geq 1$ the relation 
$$ R := \{(x_1,\dots,x_{n-1},y) \in A^n \mid x_1,\dots,x_{n-1} \in \{1,2\} \wedge x_1 = \cdots = x_{n-1} = 1 \Rightarrow y = 2\}$$
can be defined by a 
conjunction of 
pss-Horn clauses 
%, namely 
%\begin{align*}
%x_1,\dots,x_{n-1} \in \{1,2\} \wedge x_1 = \cdots = x_{n-1} = 1 \Rightarrow y  \in \{1,2\} 
%\wedge \; & 
%x_1,\dots,x_{n-1} \in \{1,2\} \wedge x_1 = \cdots = x_{n-1} = 1 \Rightarrow y  \in \{0,2\} 
%\end{align*}
and hence
is a subalgebra of $\fA^n$, 
%None of the proper  subalgebras of $\fA$ is absorbing: 
%$$ R := \{(x,y) \in A^2 \mid x \in \{0,1\} \wedge x=1 \Rightarrow y = 2\} \in \Inv(\fA)$$
%since it can be defined by a pss-Horn clause. 
%the relation $$ R := \{(x,y,z) \in A^3 \mid C_3(x,y) \wedge C_3(y,z) \}$$ 
and is $B$-essential: for every $i \in \{1,\dots,n-1\}$ 
we have 
\begin{align*}
R \cap (B^{i-1} \times A \times B^{n-i}) & = \{(1,\dots,1,2,1,\dots,1,0),(1,\dots,1,2,1,\dots,1,1)\} 
\\ R \cap (B^{n-1} \times A) & = \{(1,\dots,1,2)\}, 
\text{ but } \\
R \cap B^n & = \emptyset.
\end{align*} 
Hence, Lemma~\ref{lem:essential-abs} implies
that $B$ is not absorbing.  
The algebra $C = \{1\}$ is also not absorbing. 
For every \(n\geq1\), consider
\[
S_n:=\{1,2\}^n\setminus\{(1,\ldots,1)\}.
\]On \(\{1,2\}\), the operation \(\cdot\) is the semilattice operation with \(1\cdot1=1\) and all other products equal to \(2\). Hence, \(S_n\) is closed under coordinatewise application of \(\cdot\): if a tuple contains a \(2\), then its product with any tuple from \(\{1,2\}^n\) still contains a \(2\). Thus, \(S_n\leq\mathbf A^n\).
Moreover, \(S_n\cap C^n=\emptyset\), while for every \(i\in[n]\), the tuple having \(2\) in coordinate \(i\) and \(1\) everywhere else belongs to
\[
S_n\cap(C^{i-1}\times A\times C^{n-i}).
\]
Therefore, \(S_n\) is \(C\)-essential. Since such a relation exists for every \(n\geq1\), Lemma \(\ref{lem:essential-abs}\) shows that \(C\) is not \(n\)-absorbing for any \(n\). Consequently, \(C=\{1\}\) is not absorbing.

%For the subuniverse $C = \{0\}$, we consider the relation $C_3$ which is $C$-essential. 
%the argument is similar, and 
Every other proper  subuniverse is symmetric to $B$ or $C$ via $\rho$.
%However, $\fA$ is not hereditarily absorption free, because the subalgebra on $\{0,1\}$ has the proper absorbing subuniverse $\{1\}$. 
\end{remark}

%The path consist
Algorithms to solve $\Csp(\{0,1,2\};C_3,R_3^=)$ will be discussed in Section~\ref{sect:bounded-width}. 

\begin{exercises}
\exercise[Rot]\label{exe:source-245} Let $\fA = (\{0,1,2,3,4\}; \circ)$ be the idempotent algebra where $\circ$ is given by the rock, paper, scissors, lizard, Spock game: \begin{itemize}
\item Spock smashes scissors and vaporises rock,
\item scissors cuts paper and decapitates lizard,
\item paper disproves Spock and covers rock, \item rock crushes scissors and lizard, and
\item lizard eats paper and poisons Spock.
\end{itemize}
Determine the proper subalgebras and proper congruences of $\fA$. Which subalgebras are absorbing? Is $\fA$ Taylor, Abelian, absorption-free? Is $\fA$ subdirectly complete (Definition~\ref{def:sd-compl})?
% Solution sketch: The operation is conservative, so every subset is a subalgebra. The regular
% five-cycle tournament has no nontrivial compatible partition, hence the only congruences are
% equality and the universal relation. Strong connectivity rules out every proper absorbing subset.
% Commutativity and idempotence make $\circ$ a Taylor term; the two-element subalgebras are
% semilattices, so the algebra is not Abelian. It is therefore absorption-free and, by the
% simple-algebra four-cases argument (or a direct subdirect-product induction), subdirectly complete.
\exercise[Orange]\label{exe:source-244} Show that $\Csp(\{0,1,2\};C_3,R_3^=)$ can be solved by the $3$-consistency
procedure (see Exercise~\ref{exe:kcons-general}).
%\newpage
% Solution sketch: Run the canonical $3$-consistency tables. The binary relation $C_3$ fixes
% differences modulo three, while $R_3^=$ records exactly which triples have a repeated value. Every
% compatible partial map on at most three variables therefore has a unique consistent extension along
% a constraint. A nonempty fixed point can be extended variable by variable, so $3$-consistency
% decides the CSP.
\exercise[Weiss]\label{exe:source-246}
Is there a structure $\bA$ with a finite relational signature such that
a relation $R \subseteq A^n$ is preserved by the operation $\circ$ from Exercise~\ref{exe:source-245}
if and only if
$R$ has a
primitive positive definition in $\bA$?
% Solution sketch: Yes. The preceding algebra is a finite conservative quasi-primal algebra: its term
% clone consists of all conservative operations. Therefore take the structure on the same five-element
% domain whose basic relations are all unary subsets. Its polymorphisms are exactly the conservative
% operations, hence exactly the operations in $\Clo(\fA)$. There are only $2^5$ such relations, so the
% signature is finite; Pol--Inv gives the required equivalence.
\end{exercises}

%The relation $C_3$ is $\{0,1\}$-essential: 
%we have $C_3 \cap (\{0,1\} \times A) \neq \emptyset$ and $C_3 \cap \{0,1\} 
%A pss Horn clause 

\subsection{Subdirectly Complete Algebras} 
The property of the paper-scissors-stone algebra established in Lemma~\ref{lem:subdirect-pss} is of general importance when studying finite idempotent Taylor algebras.

% (i.e., $\Clo(\fA)$ does not have a minion homomorphism to $\Proj$, see Theorem~\ref{thm:taylor}). 

\begin{definition}\label{def:sd-compl}
Let $\fA$ be an algebra and let ${\mathfrak A}$ be
the relational structure with the same domain as $\fA$ 
whose relations are the graphs of the automorphisms of $\fA$. 
Then $\fA$ is called \emph{subdirectly complete} if every subdirect $R \leq {\fA}^n$, for every $n \in {\mathbb N}$,   
can be defined by a conjunction of atomic formulas over $\mathfrak A$. 
%2-decomposable, and every binary projection of $R$ is either 
\end{definition}

\begin{example}
The Paper-Scissors-Stone algebra (Definition~\ref{def:pss}) is subdirectly complete (Lemma~\ref{lem:subdirect-pss}). 
\end{example} 

%The next lemma will not be needed in the following, but is intended to illustrate the strength of  the property of being subdirectly complete. 
%\begin{lemma}
%Let $\fA$ be an algebra which is subdirectory complete. Then it is polynomially complete. 
%\end{lemma} 

\begin{lemma}\label{lem:sdcc}
Let $\fA$ be subdirectly complete and let $R \leq \fA^k$ be a relation that contains
$(a,\dots,a)$ for every $a \in A$. 
Then $R$ can be defined by a conjunction of formulas of the form $x_i = x_j$, for $i,j \in [k]$. 
%be an algebra with a congruence $\theta$. If $\fA/\theta$ is subdirectly complete 
\end{lemma} 
\begin{proof}
Clearly, $R$ is subdirect and hence has a definition by a conjunction of atomic formulas; each conjunct denotes the graph of an automorphism of $\fA$. 
Since the only automorphism that contains $(a,a)$ is the identity $\id_A$, the statement follows. 
\end{proof} 

\begin{remark}
Note that 
every subdirectly complete algebra is simple: this follows from Lemma~\ref{lem:sdcc} for $k=2$. 
\end{remark} 

\begin{exercises}
\exercise[Blau]\label{exe:source-247}
Show that every finite idempotent algebra which is subdirectly complete is polynomially complete (Exercise~\ref{exe:PC}).
% Source: Brady's notes, without proof.
% Proof sketch:
% Let $A$ be finite idempotent subdirectly complete algebra.
% Let R be an invariant relation.
% If R is binary and contains =_A, it is subdirect, hence conjunction of automorphism graphs, so either full or equal to A^2.
% If R is ternary such that binary projections are full, then it is subdirect, hence conjunction of automorphism graphs, so full.
\exercise[Blau]\label{exe:source-248} Find an example of a finite idempotent algebra which is polynomially complete but not subdirectly complete.
% Example: Pol([2];not x or not y).
% Solution sketch: Take the idempotent algebra whose term clone is $\Pol(\{0,1\};\neg x\vee\neg y)$.
% The binary relation makes every compatible subdirect product one of the permitted coordinate
% products, so the algebra is polynomially complete. Its proper compatible subdirect relation
% witnesses failure of subdirect completeness.
\exercise[Blau]\label{exe:source-249} Describe the clones of subdirectly complete Taylor algebras over the domain $\{0,1\}$.
% Source: Inspired by Cryptic Cor 4.7 in Brady's notes
% Solution: Pixley clone (Mino+Majo), the  clones with constants, and Pixley
% without the unary sets (i.e., with negation as an operation)
\end{exercises}

\subsection{The Four Cases for Taylor Algebras}
\label{sect:zhuk}
In this section we present a result which is essentially a consequence of the Absorption Theorem for finite \emph{simple} algebras. It is called `Zhuk's four cases' in the lecture notes of Zarathustra Brady~\cite{BradyNotes}, who cites~\cite{Strong-Subalgebras-Published}; 
and in~\cite{theoretics:11361}; 
a related predecessor result can be found in~\cite[Theorem 38]{absorption}. 

%but we have combindgled two cases with absorption into one, so we only show three cases here. The presentation is based on 

\begin{theorem}\label{thm:zhuk-simple}
Let $\fA$ be a simple finite idempotent Taylor algebra. 
%sucht that $\Clo(\fA)$ has no minion homomorphism to $\Proj$. 
Then at least one of the following four cases applies. 
\begin{enumerate}
\item $\fA$ has a proper binary absorbing subuniverse. 
\item $\fA$ has a proper centrally absorbing subuniverse. 
\item $\fA$ is affine. 
\item $\fA$ is subdirectly complete. 
\end{enumerate}
\end{theorem}
\begin{proof}
Suppose that $\fA$ has no proper binary or centrally absorbing subuniverse and is not affine. 
%Since $\fA$ is Taylor, Corollary~\ref{cor:trans} implies that $\fA$ has a transitive term operation. 
We prove by induction on $n \geq 1$ that every subdirect $R \leq {\fA}^n$ has a definition by a conjunction of atomic formulas in the structure ${\mathfrak A}$ whose relations are the graphs of the automorphisms of $\fA$. %If $R = A^n$
%then there is nothing to be shown; 
If $n=1$, then $R = A$ since $R$ is subdirect, and there is nothing to be shown. 
%Since $R$ is subdirect this covers the case that $n=1$. 
If $n=2$, 
%we consider the congruence 
%$C = \bigcup_{i \in {\mathbb N}} (R \circ R^{-1})^i$ (Exercise~\ref{exe:fin-congruence-generation}). 
%If $C = R \leq \fA^2$ 
then $R$ is linked or the graph of an automorphism of $\fA$, by the simplicity of 
$\fA$ (see Exercise~\ref{exe:simple-linked}).
In the latter case we are done, so suppose that
$R$ is linked. 
%, and we are done. Otherwise, by the assumption that $\fA$ is simple, 
%we must have that $C = A^2$,
%so $R$ is linked. 
%since otherwise $\fA$ has a proper congruence, contrary to the assumption that $\fA$ is simple. 
If $R = A^2$, then we are also done. 
Otherwise,  
 the Absorption Theorem in its strengthened version (Theorem~\ref{thm:3-absorb}) 
 applied to $R \leq \fA \times \fA$   
 implies 
that $\fA$ has a proper binary or centrally absorbing subuniverse. 
%a proper  3-absorbing subuniverse. TODO HERE. 
%, which is a contradiction to our assumptions. 

Now suppose that $n \geq 3$. 
We first consider the case that for some $\{i,j\} \in {[n] \choose{2}}$ the relation $R' := \pi_{i,j}(R)$ is the graph of an automorphism of $\fA$.
For the sake of notation, suppose that $j=n$. 
Then the relation  
$\pi_{[n-1]}(R)$ is subdirect and by the inductive assumption has a definition $\phi(x_1,\dots,x_{n-1})$ by a conjunction of atomic formulas over $\bA$. 
Then $\phi(x_1,\dots,x_{n-1}) \wedge R'(x_i,x_j)$ is a definition of $R$ over $\bA$ and we are done. 
Therefore, we may assume that for every $\{i,j\} \in {[n] \choose{2}}$ the relation $R' := \pi_{i,j}(R)$ is \emph{not} the graph of an automorphism, and hence $R' = A^2$ by the case $n=2$. 
We have to show that $R = A^n$. 

Note that for every $a \in A$ the $(n-1)$-ary relation 
$$ R_a := \{\bar x \mid (a,\bar x) \in R \}$$
is a subuniverse of $\fA^{n-1}$ because $\fA$ is idempotent. Moreover, $R_a$ 
is subdirect.
% i.e., for all $b,c \in A$ we have to prove
%that there exist $b',c' \in A$ such that $(b,c') \in R_a$ and $(b',c) \in R_a$.  
Indeed, let $b \in A$. 
Note that the binary relation 
$\pi_{1,2}(R)$ equals $A^2$ by the case $n=2$, 
and hence in particular it contains $(a,b)$. 
Hence, there exists $c' \in A^{n-2}$ such that $(a,b,c') \in R$, and thus $(b,c') \in R_a$. 
Similar arguments apply to the other arguments of $R_a$, showing that $R_a \leq \fA^{n-1}$ is subdirect. 
 
We first consider the case $n = 3$. 
Since $R_a \leq \fA^2$ is subdirect, by the case $n=2$ it is either the graph of an automorphism of $\fA$ or it equals $A^2$. First suppose that there exists an $a \in A$ such that $R_a = A^2$. 
Then 
$a$ is an element of the left center $C$ of $R$ if $R$ is considered as a subalgebra of $\fA \times \fA^2$. 
If $C = A$ then $R = A^3$ and we are done. Otherwise, $C$ is a proper subuniverse of $\fA$. 
%Note that $\Clo(\fA^2)$ does not have a minion homomorphism to $\Proj$; this follows from our assumptions, because $\Clo(\fA)$ has a minion homomorphism to $\Clo(\fA^2)$.
Clearly, $\fA^2$ is Taylor since $\fA$ is Taylor. 
Lemma~\ref{lem:abs-prod-2} implies that
$\fA^2$ does not have proper 2-absorbing subuniverses because $\fA$ has no proper 2-absorbing subuniverses. 
Since $R$ is subdirect, Lemma~\ref{lem:centrally-abs} implies that 
$C \cabs {\fA}$, 
contrary to our assumption that
$\mathbf{A}$ has no proper centrally absorbing subuniverse.
%Therefore, $C$ is a (proper) $3$-absorbing subuniverse of $\fA$ by Proposition~\ref{prop:centrally-3abs}.
%TODO HERE. 

%Corollary~\ref{cor:2-abs-centre} implies that 
%\begin{itemize}
%\item $\fA^2$ has a proper $2$-absorbing subuniverse, or 
%\item $C \abs \fA$. 
%\end{itemize} 
%In the first case, $\fA$ has a proper $2$-absorbing subuniverse as well (Exercise~\ref{exe:product-abs-trans-2}), contrary to our assumptions.  
%In the second case, we obtain that $C$ is a $3$-absorbing subuniverse of $\fA$ by Proposition~\ref{prop:centrally-3abs} (TODO: expand), which is also a contradiction to our assumptions. 

%and $\fA$ has a transitive term operation, we may apply  Proposition~\ref{prop:absorb-left-center} and 
%conclude that $\fA$ has a nonempty proper  
%3-absorbing subuniverse, a contradiction to our assumptions. 

Otherwise, for every $a \in A$ the relation $R_a$ is the graph of an automorphism of $\fA$. A similar argument applies to $R$ after permuting the arguments. So we may assume that for every $a \in A$ and every $i \in \{1,2,3\}$ the relation defined by $\exists x_i (R(x_1,x_2,x_3) \wedge x_i = a)$ is the graph of an automorphism of $\fA$. 
Then $\fA$ is abelian by 
Proposition~\ref{prop:prove-abelian}.
Since $\fA$ is Taylor, it therefore follows from Corollary~\ref{cor:abelian-affine} that $\fA$ is affine, contrary to our assumptions. 

Finally, suppose that $n>3$. Then for $i,j \in {\{2,\dots,n\} \choose 2}$ 
we have that 
$\pi_{\{1,i,j\}}(R) = A^3$ 
by the case $n=3$. Hence, for any $a \in A$ we have $\pi_{\{i,j\}}(R_a) = A^2$, and it follows from the case $n-1$ that $R_a = A^{n-1}$. This means that $R = A^n$. 
\end{proof}

%An algebra is \emph{non-trivial} if it has at least two elements. 

We can use Theorem~\ref{thm:zhuk-simple} to obtain a result about not necessarily simple finite idempotent Taylor algebras. This will not be needed in the proof of the cyclic terms theorem in Section~\ref{sect:cyclic}, but it will be used in the characterisation of finite-domain CSPs of bounded width in Section~\ref{sect:bounded-width}.

\begin{corollary}\label{cor:zhuk}
% This is essentially the important 
% Theorem 3.3 in Zhuk's strong subalgebras paper, but without splitting the absorption case into binary absorbing, PC, and central. 
% Corollary 3.12.12 in Brady's notes. 
Let $\fA$ be a finite idempotent Taylor algebra with at least two elements.  
Then at least one of the following is true. 
\begin{enumerate}
\item $\fA$ has a proper binary absorbing subuniverse. 
\item $\fA$ has a proper centrally absorbing subalgebra. 
\item $\fA$ has an affine quotient with at least two elements. 
\item $\fA$ has a subdirectly complete quotient with at least two elements. 
%Moreover, $\fA$ has a proper subalgebra $E \sdc \fA$ (Definition~\ref{def:sdc}). 
\end{enumerate}
\end{corollary}
\begin{proof}
If $\fA$ is simple, then the statement follows from Theorem~\ref{thm:zhuk-simple}. 
Otherwise, there exists a proper congruence $C$;
% be a maximal congruence on $\fA$, 
choose $C$ to be maximal with respect to inclusion, 
so that $\fA' := \fA/C$ is simple. Note that $\fA'$ has at least two elements because $C \neq A^2$. 
%$\fA$ has at least two elements. 
Hence, Theorem~\ref{thm:zhuk-simple} implies that $\fA'$ has a proper binary or centrally absorbing subalgebra $\fB$, is affine, or subdirectly complete.
If $\fA'$ has a proper binary absorbing subalgebra, then   
(the proof of) % commented out 4.12.25
% commented in since we need binary, 5.12.25.
Lemma~\ref{lem:abs-hom} shows that 
$\fA$ has a proper binary  absorbing subalgebra as well. 
If $\fA'$ has a proper centrally absorbing subalgebra, then Lemma~\ref{lem:central-abs-trans} shows that $\fA$ has a centrally absorbing subalgebra as well. 
Otherwise, $\fA$ has an affine quotient or a subdirectly complete quotient with at least two elements.
\end{proof} 

\begin{exercises}
\exercise[Blau]\label{exe:source-250} Use Theorem~\ref{thm:zhuk-simple} to give another proof of Lemma~\ref{lem:subdirect-pss}.
% Solution sketch: Apply Theorem~\ref{thm:zhuk-simple} to a minimal counterexample subdirect product.
% The polynomially complete and affine alternatives force the product to be full, while the absorbing
% alternative contradicts minimality. The remaining projective alternative yields a graph of an
% isomorphism, which is exactly the exceptional case in Lemma~\ref{lem:subdirect-pss}.
\exercise[Rot]\label{exe:source-254} Show that the finiteness assumption is needed in Theorem~\ref{thm:zhuk-simple}.
%$({\mathbb Q};\max)$:
% idempotent, simple since there are no non-trivial equivalence relations definable (even first-order), no non-trivial subalgebras, certainly not affine, certainly not PC.
\exercise[Orange]\label{exe:source-251} \label{exe:brady-one-1} Show that a subdirect relation $R \subseteq \{0,1,2\}^n$ is preserved by the Maltsev operation $m$ in Example~\ref{expl:m1} if and only if $R$ can be defined by a conjunction of graphs of permutations $\{(0,1),(1,2),(2,0)\}$ and $\{(0,0),(1,2),(2,1)\}$.

\par\noindent {\bf Hint.} Use a similar proof architecture as in the proof of Theorem~\ref{thm:zhuk-simple}. In the situation where we can apply  Proposition~\ref{prop:prove-abelian},
we obtain that $\fA$ is abelian and by Theorem~\ref{thm:abelian}, $\fA$ is affine with a central Maltsev operation $m'$. However,
$m' \neq m$, a contradiction (Exercise~\ref{exe:one-maltsev}).

% Own notes:
%
% Notes with Andrew:
% Claim: If R \leq A^n is subdirect, then
% R is pp-definable from graphs of automorphisms.
% By induction on n.
% n=2: The linking relation is a congruence.
% A is simple implies that subdirect R is the graph of automorphism of full.
% May suppose that there is no two element subset of coordinates such that the projection to the two arguments is the graph of an automorphism (because otherwiser we can reduce the arity of the relation).
% By inductive assumption: for all i \in [n], projection to all arguments except for i is the full relation.
% Consider binary relation R' between i-th projection and the rest. If (R') \circ (R')^{-1}
% is congruence by rectangularity. By simplicity, it is full or the identity. If it is full, then R = A^n.
% Otherwise, the projection of R to
% all but the i'th argument determines the i'th argument.
% Hence, it cannot be that for some a \in A R_a equals the full relation.
% So R_a equals a permutation for every a \in A. They must be disjoint.
% GENERAL CASE: decompose relation into subdirect part.
% Comment: Argument works more generally in conservative Maltsev if we forbid expl:m1 behavior.
\exercise[Orange]\label{exe:source-253} \label{exe:m2-2} Show that %there exists a finite set of relations such that
a relation $R \subseteq \{0,1,2\}^n$ is preserved by the Maltsev operation $m$ in Example~\ref{expl:m2}
if and only if it can be defined from the unary relations, $H$, and $L$.
% Solution sketch: Classify a preserved relation by its intersections with the unary fibres of $H$.
% Within the Boolean fibre, Maltsev closure makes every section an affine subspace and hence definable
% by the parity relations $L_n$; coordinates involving $2$ are controlled by $H$ and unary predicates.
% Induction on the number of coordinates combines these sections with $L$, proving definability. The
% converse is immediate because the listed relations are preserved by $m$.
\exercise[Schwarz]\label{exe:source-252} \label{exe:brady-one-2} Show that a subdirect relation $R \subseteq \{0,1,2\}^n$ is preserved by the Maltsev
operation $m$ in Example~\ref{expl:m1} if and only if $R$ can be defined primitively
positively
from the graphs of permutations
$\{(0,1),(1,2),(2,0)\}$ and
$\{(0,0),(1,2),(2,1)\}$
and from $\{(x,y,z) \in \{0,1\}^3 \mid x+y+z=0 \text{ mod } 2\}$.

\par\noindent {\bf Hints.} First prove the following substeps:
\begin{itemize}
\item Reduce the general case to the case that $|\pi_i(R)| \geq 2$ for every $i \in \{1,\dots,n\}$.
\item Reduce the general case to the case that $\pi_i(R) = \{0,1\}$ whenever
$|\pi_i(R)| = 2$ for some $i \in \{1,\dots,n\}$.
\item Reduce the general case to the case
where $R = R' \times \{0,1,2\}^m$ where
$R' \subseteq \{0,1\}^n$.
\item Show by induction on $m$
that a relation $R$ as in the previous item
can be defined by a conjunction of linear equations over $\{0,1\}$ (the most interesting step).
%Basis: Trivial if m=0.
%Inductive step: Suppose that the statement holds for m. We prove it for m+1.
%Assume R \leq {0,1}^n \times A^{m+1} and R projected onto A^m is the full relation.
%Set R' to be the binary relation obtained by considering R as a binary relation in the following way:
%R \leq ({0,1}^n \times A^m) \times A.

%Take any w \in A^m (this is the empty tuple if m=0). There exist extensions (r,w,0), (s,w,1), (t,w,2) in R. Because r,s,t \in {0,1}^n, m(r,s,t) = m(s,t,r). The same is clearly true for m(w,w,w). Therefore,
%((m(r,s,t),m(w,w,w)),m(0,1,2)) = ((m(r,s,t),m(w,w,w)),0) =: (u,0) \in R'
% ((m(s,t,r),m(w,w,w),m(1,2,0)) = (u,1)((m(r,s,t),m(w,w,w),1) \in R'.
% ((m(t,r,s),m(w,w,w),m(2,0,1)) = (u,2)((m(r,s,t),m(w,w,w),2) \in R'.
% We claim that for any tuple v in the projection of R onto {0,1}^n \times A^m, we have (v,0), (v,1), and (v,2) in R':
% Indeed, we know that there exists a \in {0,1,2} such that $(v,a) \in R'$.
% For the sake of notation, suppose that $a = 0$; other values of a can be treated analogously. Then
% m((u,1),(u,2),(v,0)) = (v,1)
% m((u,2),(u,1),(v,0)) = (v,2)
% By the inductive hypothesis, this projection is the set of all tuples that satisfy the formula \phi, so we are done.
\item Express linear equations over two-element subsets with primitive positive formulas over the given relations.
\end{itemize}
\end{exercises}
% END INLINED SOURCE: Zhuk.tex
% BEGIN INLINED SOURCE: cyclic.tex
% !TEX root = GH-UA.tex

%\section{Cyclic Polymorphisms}
%In this section, for several results that we present
%we do not give a proof. 

\section{Cyclic Terms}
\label{sect:cyclic}
An operation $c \colon A^n \to A$, for $n \geq 2$, is \emph{cyclic} if it satisfies for all $a_1,\dots,a_n \in A$ that
$c(a_1,\dots,a_n) = c(a_2,\dots,a_n,a_1)$. 
Cyclic operations are in particular Taylor operations.
Conversely, a result of Barto and Kozik (Theorem~\ref{thm:cyclic} below) implies that 
every Taylor operation on a finite set generates
a cyclic operation.

We start with some easy but useful observations about cyclic terms. 
The \emph{cyclic composition} $s \circlearrowleft t$ of a term $s$ of arity $k$ and a term $t$ 
of arity $l$ 
is the operation (or term) of arity $l$ defined by
$$(x_1,\dots,x_l) \mapsto 
s \big (t(x_1,\dots,x_l),t(x_2,\dots,x_l,x_1),\dots \big).$$

The following is easy to see. 
\begin{lemma}
Let $s \colon A^k \to A$ and
$t \colon A^l \to A$ be operations. 
\begin{itemize}
%\item
\item If $s$ is arbitrary and $t$ is cyclic, then $s \circlearrowleft t$ is cyclic. 
\item If $s$ is cyclic, $t$ is arbitrary, and $l$ divides $k$, then $s \circlearrowleft t$ is cyclic. 
\end{itemize}
\end{lemma}
\begin{proof}
For $i\geq1$, put
\[
u_i:=t(x_i,x_{i+1},\dots,x_{i+l-1}),
\]
where indices are read modulo $l$. Thus $u_{i+l}=u_i$ and
\[
(s\circlearrowleft t)(x_1,\dots,x_l)=s(u_1,\dots,u_k).
\]
Cyclically shifting $x_1,\dots,x_l$ replaces this expression by
$s(u_2,\dots,u_{k+1})$.

If $t$ is cyclic, then $u_1=\cdots=u_k$, so the two expressions are
equal. If $s$ is cyclic and $l$ divides $k$, then
$u_{k+1}=u_1$, and hence
\[
s(u_2,\dots,u_{k+1})=s(u_2,\dots,u_k,u_1)
=s(u_1,\dots,u_k).
\]
Thus $s\circlearrowleft t$ is cyclic in both cases.
\end{proof}

%\newpage

\begin{exercises}
\exercise[Rot]\label{exe:source-255} \label{exe:semilattice-cycl}
Show that if $\fA = (\{0,1\};\min)$
and $f \in \Clo(\fA)^{(k)}$ is cyclic, then
$$f(x_1,\dots,x_k) = \min(x_1,\dots,x_k).$$
%\bigvee_{i \in \{1,\dots,k\}} x_i.$$
% Solution sketch: Every term of the two-element semilattice is the minimum of a nonempty subset of
% its variables. If such a term is cyclic, rotating the variables leaves that subset invariant. The
% only nonempty subset of $[k]$ invariant under the $k$-cycle is all of $[k]$, giving the claimed
% operation.
\exercise[Rot]\label{exe:source-256}  If $s$ and $t$ are cyclic operations of arity $k$ and $l$, respectively,
then star composition $s*t$ (Definition~\ref{def:star})
is cyclic
after reordering the arguments, i.e.,
there is a permutation $\alpha$ of $[k l]$ such that $(s*t)_{\alpha}$ is cyclic.
% Solution sketch: Arrange the $kl$ variables in a $k\times l$ rectangle. A cyclic shift of the linear
% ordering can be chosen to move one step diagonally, wrapping around rows and columns. Cyclicity of
% $t$ handles the column rotation and cyclicity of $s$ the row rotation, so after the corresponding
% permutation of arguments $s*t$ is cyclic.
\exercise[Rot]\label{exe:source-257} Suppose that $\fA = (\{0,1\};\majority)$ and $f \in \Clo(\fA)^{(k)}$ is cyclic.
Show that
\begin{itemize}
\item $k \geq 3$;
\item if $r > k/2$ and $c \in A^k$ is such that $c_i = a$ for $i \leq r$
and $c_i = b$ otherwise, for distinct $a$ and $b$, then $f(c) = a$;
\item if $r,s,t$ are such that $r+s+t=k$, $r+s > t$, $s+t > r$, and $t+r > s$, then the operation
\begin{align}
(x,y,z) \mapsto f(\underbrace{x,\dots,x}_{r},\underbrace{y,\dots,y}_{s},\underbrace{z,\dots,z}_{t}) \label{eq:ternary}
\end{align}
is the ternary majority operation on $\{0,1\}$.
\label{exe:majority-cycl}
\end{itemize}
%{\bf Solution:} clearly, all operations
%in $\Clo(\fA)^{(2)}$ must be projections.
%For the second item, let $t(x,y,z)$ be the ternary operation
%$$(x,y,z) \mapsto f(\underbrace{x,\dots,x}_{k-r},y,\dots,y,\underbrace{z,\dots,z}_{k-r}).$$
%%%%defined in~(\ref{eq:ternary}).
%Then $t$ cannot be the minority operation, because the minority operation does not preserve the relation $\leq$, and is generated by the majority operation which does preserve $\leq$.  Moreover, we have $t(x,x,y)=t(y,x,x)$ by cyclically shifting the arguments of $f$, and hence $t$ ist either a majority or the second projection. In both cases, $f(c)=t(a,a,b)=a$, as required.

%For the third item, note that if
%$r+s>t$ then $$f(\underbrace{x,\dots,x}_{r},\underbrace{x,\dots,x}_{s},\underbrace{z,\dots,z}_{t})=x$$ by the second item.
%If $s+t>r$ or $t+r>s$ we use the same argument for a cyclic shift of the arguments of $f$.

\exercise[Orange]\label{exe:source-258} Suppose that $p$ is a prime and $\fA = (\{0,\dots,p-1\};m)$
where $m \colon A^3 \to A$ is given by
$m(x,y,z) = x-y+z \mod p$ and that $f \in \Clo(\fA)^{(k)}$ is cyclic. Show that if $r,s,t$ are such that
\begin{itemize}
\item $r+s+t = k$,
\item $r = t = k \mod p$
and
\item $s = -k \mod p$,
\end{itemize}
then the ternary operation
defined in (\ref{eq:ternary}) equals $x-y+z \mod p$.
\label{exe:affine-cycl}

%{\bf Solution:}
%Recall that all $k$-ary functions $f$ generated by $m$ have the form $\sum_{i=1}^k a_i x_i \mod p$.
%If $a_i \neq a_j$ for some $i,j$ then the operation is not cyclic, so suppose that $a_1 = a_2 = \cdots = a_k =: a$. Also note that $ka=1 \mod p$ since
%$f$ is idempotent. Then note that
%\begin{align*}
%f(\underbrace{x,\dots,x}_{r},\underbrace{y,\dots,y}_{s},\underbrace{z,\dots,z}_{t}) & = r a x + s a y + t a  z \\
%& = r k^{-1} x + s k^{-1} y + t k^{-1} z
%= x - y + z \mod p.
%\end{align*}
% ChatGPT:
%the inverse can be avoided entirely. From \(ka=1\) and
%\[
%r\equiv t\equiv k,\qquad s\equiv-k\pmod p,
%\]one directly obtains
%\[
%ra=ta=ka=1,\qquad sa=-ka=-1.
%\]That is cleaner and makes explicit that the necessary fact comes from \(ka=1\), not merely from the primality of \(p\).
\exercise[Blau]\label{exe:source-259} Does Exercise~\ref{exe:source-258} remain true if we drop the assumption that $p$ is prime?
% Hint: Z_2 times Z_2.
% there we can have TODO.
% Solution sketch: Yes, for the stated algebra on $\mathbb Z_p$ the proof does not use primality. A
% cyclic term has the form $f=a(x_1+\cdots+x_k)$ and idempotence gives $ka=1$ modulo $p$. From
% $r\equiv t\equiv k$ and $s\equiv-k$ one obtains $ra=ta=1$ and $sa=-1$, so the displayed ternary
% minor is $x-y+z$ for composite $p$ as well.
\end{exercises}

\subsection{Cyclic Relations}

When $a = (a_0,a_1,\dots,a_{k-1})$ is a $k$-tuple,
we write $\rho(a)$ for the $k$-tuple $(a_1,\dots,a_{k-1},a_0)$. 

\begin{definition}
A $k$-ary relation $R$ on a set $A$ is called \emph{cyclic} if for all $a \in A^k$
$$a \in R \Rightarrow \rho(a) \in R  \; .$$
\end{definition}

\begin{lemma}[from~\cite{cyclic}]\label{lem:cyclic}
A finite idempotent algebra  $\fA$ has a 
$k$-ary cyclic term if and only if 
every nonempty cyclic subalgebra of 
$\fA^k$ contains a constant tuple. 
\end{lemma}
\begin{proof}
Let $\tau$ be the signature of $\fA$. 
For the easy direction, suppose that $\fA$ has
a cyclic $\tau$-term $t(x_1,\dots,x_k)$. Let $a = (a_0,a_1,\dots,a_{k-1})$ be an arbitrary tuple in a cyclic subalgebra $\fR$ of $\fA^k$. As $R$ is cyclic, $\rho(a),\dots,\rho^{k-1}(a) \in R$, and since $\fR$ is a subalgebra
$$b := t^{\fA}(a,\rho(a),\dots,\rho^{k-1}(a)) \in R.$$ Since $t$ is cyclic, the $k$-tuple $b$ is constant. 

To prove the converse direction, we assume that
every nonempty cyclic subalgebra of $\fA^k$ contains a constant tuple. For a $\tau$-term $f(x_0,x_1,\dots,x_{k-1})$, let $S(f)$ be the set of all $a \in A^k$
such that $f^{\fA}(a) = f^{\fA}(\rho(a)) = \cdots = f^{\fA}(\rho^{k-1}(a))$. 
Choose $f$ such that $|S(f)|$ is maximal (here we use 
the assumption that $A$ is finite). If $|S(f)| = |A^k|$, then $f^{\fA}$ is cyclic and we are done.
Otherwise, arbitrarily pick  
$a = (a_0,a_1,\dots,a_{k-1}) \in A^k \setminus S(f)$.  
For $i \in \{0,\dots,k-1\}$, define $b_i := f(\rho^i(a))$ and $b := (b_0,\dots,b_{k-1})$, 
and let $B := \{b,\rho(b),\dots,\rho^{k-1}(b)\}$.  

We claim that the smallest subalgebra $\fC$ of $\fA^k$ that contains $B$ is cyclic. So let $c \in C$ be arbitrary. 
Since $\fC$ is generated by $B$, there exists
a $\tau$-term $s(x_0,x_1,\dots,x_{k-1})$ such that $c = s^{\fA}(b,\rho(b),\dots,\rho^{k-1}(b))$. Then $\rho(c) = s^{\fA}(\rho(b),\rho^2(b),\dots,\rho^{k-1}(b),b) \in C$, proving the claim. 

Since $C$ is cyclic, by our assumption it 
contains a constant tuple $d$. 
%Let $r \in \Clo^{(k)}(\fC)$ be such that $d = s(b,\rho(b),\dots,\rho^{k-1}(b))$. 
Then there exists a $\tau$-term $r(x_0,\dots,x_{k-1})$ 
such that $d = r^{\fC}(b,\rho(b),\dots,\rho^{k-1}(b))$. 
Note that 
$$r^{\fA}(b) = r^{\fA}(\rho(b)) = \cdots = r^{\fA}(\rho^{k-1}(b))$$ 
since $d$ is constant.
% (also see Figure~\ref{fig:cyclic}). 
It follows that $b \in S(r)$. 

\begin{figure}
\begin{center}
\includegraphics[scale=0.6]{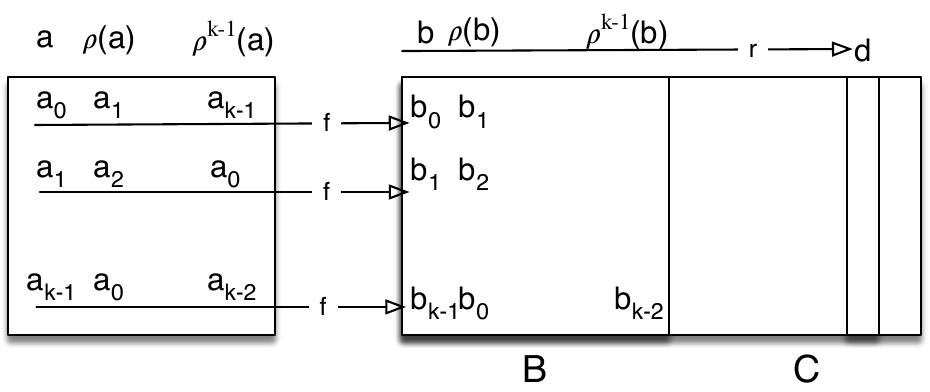} 
\end{center}
\caption{Diagram for the proof of Lemma~\ref{lem:cyclic}.}
\label{fig:cyclic}
\end{figure}

Now consider 
the $\tau$-term $t(x_0,x_1,\dots,x_{k-1})$ defined by
$$
t(x) := r \circlearrowleft f =  r(f(x),f(\rho(x)),\dots,f(\rho^{k-1}(x))) .
$$ 
where $x := (x_0,x_1,\dots,x_{k-1})$. 
We claim that $S(f) \subseteq S(t)$.
Let $e \in S(f)$.  
To show that $e \in S(t)$, note that for all $i \in \{0,\dots,k-1\}$ 
\begin{align*}
t^{\fA}(\rho^i(e)) = & \; r^{\fA}\big(f^{\fA}(\rho^i(e)),f^{\fA}(\rho^{i+1}(e)),\dots,f^{\fA}(\rho^{i-1}(e))\big)\\
= & \; r^{\fA}\big(f^{\fA}(e),f^{\fA}(\rho^1(e)),\dots,f^{\fA}(\rho^{k-1}(e)) \big) && \text{(since $e \in S(f)$)} \\
= & \; t^{\fA}(e).
\end{align*} 
Moreover, $a \in S(t)$, because 
\begin{align*}
t^{\fA}(\rho^i(a)) = & \; r^{\fA} \big (f^{\fA}(\rho^i(a)),f^{\fA}(\rho^{i+1}(a)),\dots,f^{\fA}(\rho^{i-1}(a)) \big) \\
= & \; r^{\fA}(b_i,b_{i+1},\dots,b_{i-1})  \\ %&& \text{} \\
= & \; r^{\fA}(\rho^i(b))
\end{align*}
is constant for all $i$ by the choice of $r$. We obtain a contradiction to the maximality of $|S(f)|$. 
\end{proof}

\begin{exercises}
\exercise[Blau]\label{exe:source-260} Show that the digraph $C_2^{++}$ from Exercise~\ref{exe:c2++} has a ternary cyclic polymorphism.
% Hint: cylic composition with large enough cyclic terms.
% on the bottom vertices, take majority, on all other input tuples return maximum.  ??
%\item Show that if $\fA$ has a cyclic term
%and $\fB$ has a cyclic term, then $\fA \times \fB$ has a cyclic term.
% Kommt spaeter nochmal.
\end{exercises}

\subsection{The Cyclic Terms Theorem}
\label{sect:cyclic-thm}
In this section we prove the following theorem of Barto and Kozik~\cite{cyclic}.

\begin{theorem}[of~\cite{cyclic}]\label{thm:cyclic}
Let $\fA$ be a finite algebra. Then the following are equivalent. 
\begin{enumerate}
\item $\fA$ has a Taylor term;
\item $\fA$ has a cyclic term;
\item for all prime numbers $p > |A|$, the algebra $\fA$ has a $p$-ary cyclic term.
\end{enumerate}
\end{theorem}
\begin{proof} 
The implication from 3.\ to 2.\ and the implication from 2.\ to 1.\ are
immediate. We prove the implication from 1.\ to 3. We first explain why
it suffices to prove this implication for idempotent algebras.

Suppose that the implication has been proved for all finite idempotent
algebras. Let $\fA$ be an arbitrary finite algebra with a Taylor term,
and put $\mathscr C:=\Clo(\fA)$. By
Proposition~\ref{lem:idempotent-reduct}, there exists an idempotent
operation clone $\mathscr D$ on a finite set $D$, with $|D|\leq |A|$,
and minion homomorphisms
$\xi\colon\mathscr C\to\mathscr D$ and 
$\eta\colon\mathscr D\to\mathscr C$. 
Let $\fD$ be the algebra whose basic operations are all the operations
in $\mathscr D$, so that $\Clo(\fD)=\mathscr D$. Since minion
homomorphisms preserve minors, they preserve height-one identities.
In particular, $\xi$ maps a Taylor operation of $\mathscr C$ to a
Taylor operation of $\mathscr D$. Hence, $\fD$ has a Taylor term.
Let $p>|A|$ be prime. Then $p>|D|$, and the idempotent case yields a
$p$-ary cyclic operation $c\in\mathscr D$. The cyclic identity is a
height-one identity, and therefore $\eta(c)$ is a $p$-ary cyclic
operation in $\mathscr C$. Thus, $\fA$ has a $p$-ary cyclic term.

It remains to prove the implication for finite idempotent algebras.
Let $\fA$ be such an algebra and let $p>|A|$ be prime. We proceed by
induction on $|A|$. 
For $|A|=1$ the statement is trivial.
For the induction step, we use Lemma~\ref{lem:cyclic}. Let $R \leq \fA^p$ be non-empty and cyclic. We have to show that $R$ contains a constant tuple. 

We may assume without loss of generality  that $R$ is subdirect: indeed, if $B := \pi_i(R)$ is a proper subuniverse of $\fA$, for some $i \in [p]$, then $\pi_j(R) = B$ for all $j \in [p]$ by the cyclicity of $R$, and hence 
$R \leq \fB^p$ contains a constant tuple by the inductive assumption. 

If there is $\{i,j\} \in {[p] \choose 2}$ such that $\pi_{i,j}(R)$  (Definition~\ref{def:proj}) is the graph of an automorphism $\alpha$ of $\fA$, then
$\pi_{j,2j-i}(R)$ is the graph of $\alpha$ as well, because $R$ is cyclic, and the same applies to $\pi_{2j-i,3j-2i}(R)$, etc. Moreover, $\alpha^p = \id_A$ since $R$ is cyclic and of arity $p$. 
 Since $p > |A|$ is prime, we must have $\alpha = \id_A$. This shows that $R$ 
 contains for every $a \in A$ the constant tuple $(a,\dots,a)$: by subdirectness, there exists a tuple $t = (t_1,\dots,t_p) \in R$ such that $t_i = a$; by what we have seen above, $t_j = a$, $t_{2j-i} = a$, etc, and since $p$ is prime we obtain that $t = (a,\dots,a)$.
 
 So we may suppose that for every $\{i,j\} \in {[p] \choose 2}$, the relation 
 $\pi_{i,j}(R)$ is not the graph of an automorphism of $\fA$. 

Suppose that $\fA$ has a proper congruence $C$.
Let $h$ be the homomorphism from $\fA$
to $\fA/C$. Since $C$ is proper, $|A/C|$ is strictly smaller than $|A|$. 
Let $h^*$ be the homomorphism from $\fA^p$ to $(\fA/C)^p$ obtained by applying $h$ componentwise. Then $h^*(R) \leq (\fA/C)^p$ (Lemma~\ref{lem:congruences}) is cyclic, so the inductive assumption implies that $h^*(R)$ contains a constant tuple $(a/C,\dots,a/C)$.
Note that $a/C$ is a nonempty proper subalgebra of $\fA$ since $\fA$ is idempotent and $C$ is proper.  
Then $R \cap (a/C)^p \leq \fA^p$ is non-empty and  cyclic, and hence contains a constant tuple by the inductive assumption. Thus, if $\fA$ is not simple we are done.

Now suppose that $\fA$ is simple. 
Note that in this case, $\pi_{i,j}(R)$ is linked for every $\{i,j\} \in {[p] \choose 2}$.  
Indeed, by assumption, $\pi_{i,j}(R)$ is not the graph of an automorphism of $\fA$. 
Since $R$ is subdirect, so is $\pi_{i,j}(R)$.
Hence, the simplicity of $\fA$ implies that 
$\pi_{i,j}(R)$ is linked
(Exercise~\ref{exe:simple-linked}).
Since $\fA$ is simple, one of the cases from Theorem~\ref{thm:zhuk-simple} applies.

First consider the case that $\fA$ is subdirectly complete. By our
assumption, $\pi_{i,j}(R)$ is not the graph of an automorphism of
$\fA$ for any distinct $i,j\in[p]$. Since every such projection is
subdirect, subdirect completeness implies that
$\pi_{i,j}(R)=A^2$.
By Definition~\ref{def:sd-compl}, the subdirect relation $R$ has a
definition by a conjunction of atomic formulas over the structure
whose relations are the graphs of the automorphisms of $\fA$. Consider
a conjunct $G_\alpha(x_i,x_j)$, where $G_\alpha$ is the graph of an
automorphism $\alpha$ of $\fA$. If $i\neq j$, then
$\pi_{i,j}(R)\subseteq G_\alpha$, contradicting
$\pi_{i,j}(R)=A^2$. If $i=j$, then the subdirectness of $R$ implies
that $\alpha(a)=a$ for every $a\in A$, so $\alpha=\id_A$ and the
conjunct is tautological. Thus the conjunction contains no
non-trivial conjuncts, and consequently $R=A^p$. In particular, $R$
contains a constant tuple.

 If $\fA$ is affine with underlying abelian group $(A;+,-,0)$, then for all $k \geq 1$ and $a_1,\dots,a_k \in {\mathbb Z}$ such that $a_1 + \cdots + a_k \equiv 1 \mod |A|$ the operation 
 $(x_1,\dots,x_k) \mapsto a_1 x_1 + \cdots + a_k x_k$ is a term operation of $\fA$ (combine Theorem~\ref{thm:abelian} and Exercise~\ref{exe:affine-maltsev-terms}). 
Since $p > |A|$ is prime, there
 exists $i$ such that $i p = 1 \mod |A|$.  
In particular,  
$(x_1,\dots,x_p) \mapsto i x_1 + \cdots + i x_p$ is a term operation of $\fA$, and clearly cyclic.  

The final case is that $\fA$ has a proper 2-absorbing or centrally absorbing subalgebra $U$. In both cases, $U$ is $3$-absorbing
 ($2$-absorbing trivially implies $3$-absorbing, and centrally absorbing implies $3$-absorbing by Theorem~\ref{thm:centrally-3abs}).
Define a directed graph $\bD$ whose vertices are pairs $(i,B)$ where $i \in [p]$ and $B \abs \fA$ is proper 3-absorbing, and whose edges are the pairs 
$((i,B),(j,B'))$ for distinct 
$i,j \in [p]$ with $B+\pi_{i,j}(R) \subseteq B'$ 
(the notation has been introduced in   Exercise~\ref{exe:walk}).\footnote{The author thanks Michael Pinsker for a suggestion how to simplify the argument in this case.} 

\medskip
We shall use the following restricted transitivity property of the edge
relation. Suppose that $\bD$ contains the edges
$((i,B),(j,B'))$ and $((j,B'),(k,B''))$, where $i\neq k$. Then
$\bD$ also contains the edge $((i,B),(k,B''))$. Indeed, if $t\in R$
and $t_i\in B$, then the first edge gives $t_j\in B'$, and the second
gives $t_k\in B''$. Hence,
\[
B+\pi_{i,k}(R)\subseteq B''.
\]
Also note that if $B \abs \fA$ is 3-absorbing, then
$B+\pi_{i,j}(R) \abs \fA$ is 3-absorbing as well
(Exercise~\ref{exe:walk}).

\medskip
{\bf Claim.} If $\bD$ contains the edge $((i,B),(j,B'))$, then
$((j,B'),(i,B))$ is not an edge. To prove this claim, suppose for
contradiction that $i,j\in[p]$ are distinct and
\[
B+\pi_{i,j}(R)\subseteq B'
\quad\text{and}\quad
B'+\pi_{j,i}(R)\subseteq B.
\]
For $X\subseteq A$, define
$X_0:=X\times\{0\}$ and $X_1:=X\times\{1\}$. Let $G$ be the bipartite
graph with vertex set $A_0\cup A_1$ and edge set
\[
\big\{\{(a,0),(a',1)\}\mid (a,a')\in\pi_{i,j}(R)\big\}.
\]
Then every neighbour of $B_0$ lies in $B'_1$, and every neighbour of
$B'_1$ lies in $B_0$. Hence, no edge of $G$ joins $B_0\cup B'_1$ to
its complement
\[
(A\setminus B)_0\cup(A\setminus B')_1.
\]
Both sets are non-empty because $B$ and $B'$ are non-empty proper
subuniverses of $\fA$. Moreover, $\pi_{i,j}(R)$ is subdirect, so $G$
has no isolated vertices. Thus, $G$ is disconnected, contradicting
that $\pi_{i,j}(R)$ is linked.

\medskip
We claim that $\bD$ is acyclic. Suppose otherwise, and choose a
directed cycle of minimum length,
\[
(i_0,B_0),(i_1,B_1),\dots,(i_{m-1},B_{m-1}),(i_0,B_0),
\]
where the subscripts are read modulo $m$. Since $\bD$ has no loops,
$m\geq2$, and the claim rules out $m=2$.
Suppose that $i_r\neq i_{r+2}$ for some $r$. The restricted
transitivity property gives an edge from $(i_r,B_r)$ to
$(i_{r+2},B_{r+2})$, producing a directed cycle of length $m-1$ and
contradicting the minimality of $m$. Thus, $i_r=i_{r+2}$ for every
$r$. Since consecutive indices are distinct, the indices alternate
between two values and $m$ is even.
Moreover, $B_r\subseteq B_{r+2}$ for every $r$. Indeed, for
$b\in B_r$, the subdirectness of $\pi_{i_r,i_{r+1}}(R)$ yields an
$a\in A$ such that
\[
(b,a)\in\pi_{i_r,i_{r+1}}(R).
\]
The first edge gives $a\in B_{r+1}$, and, since $i_{r+2}=i_r$, the
second edge gives $b\in B_{r+2}$. Going around the cycle, we obtain
\[
B_0\subseteq B_2\subseteq\cdots\subseteq B_{m-2}\subseteq B_0,
\]
and hence $B_0=B_2$. Therefore $(i_0,B_0)=(i_2,B_2)$, so the first two
edges of the cycle form a directed 2-cycle, contradicting the claim.
Thus, $\bD$ is acyclic.

Since $\bD$ is finite, non-empty, and acyclic, it must contain a sink $(i,B)$. Note that 
%$B$ is a proper 3-absorbing subuniverse of $\fA$ such that  % HIER NICHT GEBRAUCHT
$B + \pi_{i,j}(R) = A$ for all $j \in [p] \setminus \{i\}$. Since $R$ is cyclic, this implies that $\pi_I(R) \cap B^2 \neq \emptyset$ for all $I \in {[p] \choose 2}$. Hence, $R' := R \cap B^p$ is non-empty by 
Corollary~\ref{cor:essential}, 
%Proposition~\ref{prop:essential}, 
because $B$ is 3-absorbing. Since $|B| < |A|$, we obtain a constant tuple in $R' \subseteq R$ by the inductive assumption. 
\end{proof}

\begin{theorem}[Tractability Theorem, Version 5]
\label{thm:tractability-5}
Let $\bB$ be a relational structure with finite domain and finite signature. 
If $\bB$ has a cyclic polymorphism, then $\Csp(\bB)$ is in P. Otherwise, 
$\Csp(\bB)$ is NP-complete.
\end{theorem}

\begin{proof}
An immediate consequence of Theorem~\ref{thm:cyclic} and Theorem~\ref{thm:tractability-2}. 
%If $\bB$ does not have a Taylor polymorphism, then $K_3 \in \HI(\bB)$ by Corollary~\ref{cor:hpp},
%and $\Csp(\bB)$ is NP-hard by 
%Corollary~\ref{cor:pp-interpret-hard}. 
%Otherwise,  
%$\bB$ has a Taylor polymorphism by Corollary~\ref{cor:hpp}. 
\end{proof}

\begin{exercises}
\exercise[Blau]\label{exe:source-263} Give an immediate proof (without using results from the text) that $K_3$
does not have
weak near unanimity polymorphisms.
% Solution sketch: Every polymorphism of $K_3$ is essentially unary; after composing with an
% automorphism it is a projection. A projection of arity at least two gives different values on the
% tuples with the exceptional $y$ in different coordinates, so it cannot satisfy the weak
% near-unanimity identities.
\exercise[Blau]\label{exe:source-264} Find a cyclic term in the algebra
$({\mathbb Z}_p; m)$
where $m$ is given by $(x,y,z) \mapsto x-y+z$.
% Solution: look into the proof of the cyclic terms theorem!
\exercise[Blau]\label{exe:source-266} Let $\fA$ and $\fB$ be finite algebras with the same signature,
each with a Taylor term.
Show that $\fA \times \fB$ has a Taylor term.
% Solution: By this section, we know that
% there are terms e1 and e2 that are cyclic in \fA and in \fB, respectively, each of prime length larger than the domain size. Then use the
% Exercise~\ref{exe:source-265}.
\exercise[Rot]\label{exe:source-261} \label{exe:wnu} Use the results presented in the text to show that a finite idempotent algebra $\fA$ has a Taylor term if and only if
it has a \emph{weak near unanimity term}, i.e., a term $t$ of arity $n \geq 2$
such that $\fA$ satisfies
$$t(x,\dots,x,y) \approx t(x,\dots,x,y,x) \approx \cdots \approx t(y,x,\dots,x)$$
(this was first shown by Mar\'oti and McKenzie~\cite{MarotiMcKenzie}; see Section~\ref{sect:wnu} for
more facts about weak near unanimity operations).
% Solution sketch: A weak near-unanimity term is a Taylor term by its displayed nontrivial height-one
% identities. Conversely, the cyclic-term theorem gives a cyclic term of sufficiently large prime
% arity for every finite idempotent Taylor algebra; the standard cyclic-to-WNU construction in
% Section~\ref{sect:wnu} identifies suitable variables and yields a weak near-unanimity term.
\exercise[Rot]\label{exe:source-262} Show that a finite structure has a cyclic polymorphism if and only if it has
a weak near unanimity polymorphism (warning: the structure might have
non-idempotent operations!).
% Solution sketch: Pass first to the core of the finite structure. Every diagonal unary map of a
% polymorphism is then an automorphism, so composing with its inverse makes the operation idempotent.
% Apply the finite idempotent equivalence between cyclic and weak near-unanimity terms, and undo the
% automorphism composition. This works in both directions and does not assume the original
% polymorphisms idempotent.
\exercise[Orange]\label{exe:source-265} Let $\fA$ and $\fB$ be finite algebras with the same signature, each with a cyclic term.
Show that $\fA \times \fB$ has a cyclic term as well.
% SOLUTION:
%we know that ${\mathscr C}_i$ contains a
% cyclic operation $c_i \in {\mathscr C}^{(k)}$ for all arities $k$ that are larger than $|A_i|$. So we may assume $c_1$ and $c_2$ have the same arity $k$.
% Since ${\mathscr D}$ is subdirect, it contains operations $e_1 \coloneqq (c_1,d_1)$ and $e_2 \coloneqq (d_2,c_2)$. Let $c$ be the operation of ${\mathscr D}$ given by
%$$c(x_1,\dots,x_k) \coloneqq e_1 \big (e_2(x_1,\dots,x_k), e_2(x_2,\dots,x_k,x_1), \dots, e_2(x_k,x_1,\dots,x_{k-1}) \big ).$$
%We claim that $c$
%is cyclic.
%Let $(a_1,b_1),\dots,(a_k,b_k) \in A_1 \times A_2$.
%Then  \begin{align*} & c((a_1,b_1),\dots,(a_k,b_k))_1 \\
%= \;  & c_1 \big (d_2(a_1,a_2,\dots,a_k),d_2(a_2,\dots,a_k,a_1),\dots,d_2(a_k,a_1,\dots,a_{k-1}) \big )_1 \\
%= \;  &  c_1 \big (d_2(a_2,\dots,a_{k},a_1),
%d_2(a_3,\dots,a_k,a_1,a_2),
%\dots,d_2(a_1,a_2, \dots,a_{k}) \big)_1 \\
%= \;  & c \big ((a_2,b_2),\dots,(a_k,b_k),(a_1,b_1) )_1
%\end{align*}
%and
%\begin{align*}
%& c((a_1,b_1),\dots,(a_k,b_k))_2  \\
%= \;  & d_1 \big (c_2(a_1,a_2,\dots,a_k),c_2(a_2,\dots,a_k,a_1),\dots,c_2(a_k,a_1,\dots,a_{k-1}) \big )_2 \\
%= \;  & d_1 \big (c_2(a_2,\dots,a_{k},a_1),
%c_2(a_3,\dots,a_k,a_1,a_2),
%\dots,c_2(a_1,a_2,\dots,a_{k}) \big)_2 \\
%= \;  & c \big ((a_2,b_2),\dots,(a_k,b_k),(a_1,b_1) \big )_2
%\end{align*}
\exercise[Schwarz]\label{exe:source-267} Let $\fA$ and $\fB$ be finite algebras with the same signature,
each with a Maltsev term.
Is it true that $\fA \times \fB$ has a Maltsev term?
\ignore{
No. A product requires a single syntactic term that is Maltsev on both factors; the factors may have incompatible Maltsev terms.
Here is a finite counterexample. Let $V=\mathbb F_2^2$ and put
\[
P=\begin{pmatrix}1&0\\0&0\end{pmatrix},
\qquad
N=\begin{pmatrix}0&0\\1&0\end{pmatrix}.
\]Use the signature $(+,\oplus,g,0)$, with two binary operations and one quaternary operation. Define algebras $\mathbf A,\mathbf B$ on $V$ as follows:
In $\mathbf A$, $+$ is vector addition and $\oplus$ is constantly $0$.
In $\mathbf B$, $\oplus$ is vector addition and $+$ is constantly $0$.
Using ordinary vector addition on the right-hand sides, set
\[
g^{\mathbf A}(x_1,x_2,x_3,x_4)=Px_1+Nx_2,
\qquad
g^{\mathbf B}(x_1,x_2,x_3,x_4)=Px_3+Nx_4.
\]Then $x+y+z$ is a Maltsev term of $\mathbf A$, while $x\oplus y\oplus z$ is a Maltsev term of $\mathbf B$.
Nevertheless, $\mathbf A\times\mathbf B$ has no Maltsev term. Indeed, let $v=(0,1)^{\mathsf T}$. Since $Pv=Nv=0$, the set
\[
S=\{(v,0),(0,0),(0,v)\}
\]is a subalgebra of $\mathbf A\times\mathbf B$. All term operations of either factor are linear, and the Maltsev identities force any Maltsev operation on $V$ to be $m(x,y,z)=x+y+z$. Hence a Maltsev term $m$ on the product would satisfy
\[
m\bigl((v,0),(0,0),(0,v)\bigr)=(v,v)\notin S,
\]contradicting that every term operation preserves the subalgebra $S$.
Thus the statement becomes true only under the stronger assumption that $\mathbf A$ and $\mathbf B$ have a common Maltsev term. This counterexample is essentially Example 8.3 here: https://arxiv.org/abs/1311.2359.}
\end{exercises}

\subsection{Siggers Terms of Arity 4}
\label{sect:4ary}
Interestingly, whether 
a finite algebra has a Taylor term (equivalently: a cyclic term, or a weak near unanimity term from Exercise~\ref{exe:wnu}) can be tested by
searching for a single 4-ary term $s$ that satisfies
%$$s(x,y,z,y) \approx s(y,z,x,x) \, ,$$
%$$s(x,y,z,x) \approx s(y,z,x,z) $$ Fand ich schoen
% ist aber unpraktisch
$$s(a,r,e,a) 
\approx s(r,a,r,e) \, .$$
a so-called \emph{4-ary Siggers term}. 
%It follows that the question whether a given finite structure has a cyclic term is in the complexity class NP.
Note that this definition comes in numerous variants, because we may permute the arguments of $s$ and rename the variables of the identity and obtain equivalent conditions. For instance, if we cyclically
shift the first, second, and fourth argument, and fix the third, and rename $a$ to $x$, $r$ to $z$, and $e$ to $y$, we obtain the variant 
$$ s(x,x,y,z) \approx s(y,z,z,x)$$
which will be used below. 
Siggers originally found a 6-ary term (see Section~\ref{sect:6ary}), which has been improved later to the 4-ary term given above. 
The observation that this condition can be obtained by equating variables of a cyclic term of sufficiently high arity is from~\cite{Maltsev-Cond}; the proof below is based on a variant from~\cite{BradyNotes}
of their proof. 

% NEED: finiteness of free algebra!

\begin{theorem}\label{thm:4siggers}
%[due to~\cite{Siggers}; see~\cite{Maltsev-Cond}]
%Let $\fA$ be a finite 
%algebra. Then $\fA$ 
A finite algebra has a cyclic term if and only if 
it
%$\fA$ 
has a 4-ary Siggers term. 
\end{theorem}
\begin{proof}
Suppose that $\fA$ has a cyclic term. 
%Let $c(x_1,\dots,x_p)$ be a cyclic term of $\fA$ for some $p \geq 2$.
Choose numbers $a,b \in {\mathbb N}$ such that $2a+3b=p$ for some prime $p$ which is larger than $|A|$. 
By Theorem~\ref{thm:cyclic}, there exists a cyclic term $c$ of arity $p$. 
We  define $s(x,y,z,w)$ to be the term 
$$ s(x,y,z,w) := c(\underbrace{x,\dots,x}_b,\underbrace{y,\dots,y}_a,\underbrace{z,\dots,z}_b,\underbrace{w,\dots,w}_{a+b}) .$$
Then 
\begin{align*}
s(x,x,y,z) & = c(\underbrace{x,\dots,x}_b,\underbrace{x,\dots,x}_a,\underbrace{y,\dots,y}_b,\underbrace{z,\dots,z}_{a+b}) \\
& = c(\underbrace{y,\dots,y}_b,\underbrace{z,\dots,z}_a,\underbrace{z,\dots,z}_b,\underbrace{x,\dots,x}_{a+b}) 
 = s(y,z,z,x).
\end{align*}
%$$ c(x,x,\dots,x,y,\dots,y,w,z,\dots,z) \; ,$$
%where the variable $x$ occurs $k+1$ times, the variable $w$ occurs once, and the variables $y$ and $z$ occur $k$ times each. 
%Then 
%\begin{align*} s(x,y,z,y) \approx & \; c(x,x,\dots,x,y,y,\dots,y,y,z,\dots,z) && \text{($k+1$ times $x$, $k+1$ times $y$, $k$ times $z$)} \\
%= & c(x,x,\dots,x,y,y,\dots,y,w,z,\dots,z) , \\
%\approx & \; c(y,y,\dots,y,z,z,\dots,z,x,x,\dots,x) && \text{(by cyclicity of $c$)} \\
%\approx & \; s(y,z,x,x) \; .
%\end{align*}
Conversely, a Siggers term is a Taylor term, and therefore the other direction follows from Theorem~\ref{thm:cyclic}. 
\end{proof}

%While there is no single equation between ternary terms  
%system of equations of the form
%$\forall x,y. \; t(z_1,z_2,z_3) = t(z_1',z_2',z_3')$ 
%for a single ternary term $t$ 
%which would characterise the existence of Taylor terms,

The previous result is optimal in the sense that 
there is no equivalent characterisation 
using a single ternary Taylor term~\cite{Maltsev-Cond,KearnesMarkovicMcKenzie}. 
However, there is also a system of equations involving only ternary terms that characterises the existence of a Taylor term~\cite{JMMM}. Computationally, checking whether a given finite structure has polymorphisms satisfying these identities is easier than checking for a 4-ary Siggers polymorphism (for computer experiments, see~\cite{otrees}). 

\begin{proposition}[from~\cite{JMMM}]
\label{prop:pq}
Let $\fA$ be a finite idempotent algebra. Then $\fA$ has a Taylor term if and only if
$\fA$ has terms $p$ and $q$ satisfying the following identities (`$p$-$q$-terms'). 
\begin{align}
 q(y,x,x) \approx \; & q(x,x,y) \label{eq:eins} \\
 & q(x,x,y) \approx p(x,y,y) \label{eq:zwei} \\
 p(x,y,x) \approx \; & q(x,y,x)  \label{eq:drei}
\end{align}
\end{proposition}
\begin{proof}
First suppose that $\fA$ has a Taylor term and therefore a 4-ary Siggers term. 
Define
%$$ p(x,y,z) := s(x,x,y,z) \text{ and } q(x,y,z) := s(y,z,x,x) $$
$$ p(x,y,z) := s(y,z,x,x) \text{ and } q(x,y,z) := s(x,x,y,z) $$
and observe that the operations $p$ and $q$ satisfy the equations from the statement. 
\begin{align*}
q(y,x,x) & = s(y,y,x,x) \approx s(x,x,x,y) = q(x,x,y) \\
q(x,x,y) & = s(x,x,x,y) \approx s(y,y,x,x) = p(x,y,y) \\
p(x,y,x) & = s(y,x,x,x) \approx s(x,x,y,x) = q(x,y,x)   
\end{align*}

Conversely, let $\fA$ be an algebra that satisfies \eqref{eq:eins}, \eqref{eq:zwei} and~\eqref{eq:drei}. Then there is no $\xi \colon \Clo(\fA) \to \Proj$, because otherwise
\begin{align*}
\xi(q) & = \pi^3_2 && \text{ (because of \ref{eq:eins})} \\
\xi(p) & = \pi^3_1 && \text{ (because of \ref{eq:zwei})} \\
\xi(p) & = \pi^3_2 && \text{ (because of \ref{eq:drei})}
\end{align*}
which is a contradiction unless $|A|=1$. Hence, there is a Taylor term by Theorem~\ref{thm:taylor}. 
\end{proof}

\begin{exercises}
\exercise[Rot]\label{exe:source-268} Show that every algebra with a Maltsev term has a 4-ary Siggers term (directly, without using other results).
% Solution: define
% s(x_1,x_2,x_3,x_4) := m(x_2,x_4,x_3)
% Then s(x_1,x_2,x_3,x_2) = m(x_2,x_2,x_3) = m(x_3,x_1,x_1) = s(x2,x_3,x1,x1)
\end{exercises}

\subsection{Weak Near Unanimity Operations} 
\label{sect:wnu} 
A \emph{weak near unanimity operation} (defined in Exercise~\ref{exe:wnu})
is an operation that satisfies
$$w(x,\dots,x,y) \approx w(x,\dots,y,x) \approx \cdots \approx w(y,x,\dots,x) \, .
$$
We write $\WNU(k)$ for the
%the set of height-one identities that describe 
$k$-ary weak
near unanimity operations. 
Recall that 
a finite algebra has a Taylor term if and only if it has a weak near unanimity term (Exercise~\ref{exe:wnu}). 
Again, we warn the reader that some authors additionally assume that weak near unanimity operations are idempotent; we do not make this assumption since it gives us more flexibility of the terminology.

\begin{example}\label{expl:wnu}
The algebra $\fA_n := (\{0,\dots,n-1\};m)$ where $m(x,y,z) := x - y + z$ (see Exercise~\ref{exe:affine-maltsev-terms}) has a WNU$(k)$ term if and only if $\gcd(k,n)=1$: 
\begin{itemize}
\item if $\gcd(k,n) = 1$ then there is an $a \in \{0,\dots,n-1\}$ such that $ak \equiv 1 \mod n$. 
Hence, $\sum_i a x_i \in \WNU(k)$
and we have $\sum_i a = ka = 1$
so this operation is in $\Clo(\fA_n)$. 
\item Conversely, let $g \in \Clo(\fA_n)$ be a $\WNU(k)$ operation. In particular, we have 
\begin{align*}
a_1 = g(1,0,\dots,0) \equiv g(0,1,0,\dots,0) \equiv \cdots \equiv g(0,\dots,0,1) = a_k  \mod n
\end{align*}
and it follows that $a := a_1 \equiv \cdots \equiv a_{k}  \mod n$.  
But $1 = \sum_i a_i = k a \mod n$,
which implies that $n$ and $k$ are coprime. 
\end{itemize}
For example, $\Clo(\fA_6)$ has a WNU$(5)$ term, but not a WNU$(k)$ term for $k \leq 4$.  
\end{example}

\begin{exercises}
\exercise[Rot]\label{exe:source-269} % Source: Dima's strong subalgebras paper.
Let $(A;+,-,0)$ be a finite Abelian group.
Then every idempotent weak near-unanimity
operation that preserves the relation defined by $y_1+y_2 = y_3 + y_4$ is of the form
$(x_1,\dots,x_n) \mapsto t \cdot (x_1 + \cdots + x_n)$ for some $t \in {\mathbb N}_{\geq 1}$.
% Solution sketch: Preservation of $y_1+y_2=y_3+y_4$ makes the operation affine: $f(\bar x)=\sum_i a_i
% x_i+c$. Idempotence gives $c=0$ and $\sum_i a_i=1$. The WNU identities compare the tuples with one
% exceptional coordinate and force all coefficients $a_i$ to be equal to one endomorphism $t$. Hence
% $f=t(x_1+\cdots+x_n)$, with $nt=1$ on $A$.
\exercise[Rot]\label{exe:source-270} % Star product
Show that if a clone (not necessarily on a finite domain) contains $f \in \WNU(3)$
and $g \in \WNU(4)$, then it also contains an operation in $\WNU(12)$.
% Solution sketch: Form the star composition $f*g$ on a $3\times4$ array of variables. Reorder the
% twelve variables along a diagonal cycle. Moving the exceptional argument one step changes first only
% its position inside a $g$-block and then only the affected input of $f$; the WNU identities for $g$
% and $f$ show that all twelve exceptional positions have the same value. Idempotence is inherited, so
% the result lies in $\WNU(12)$.
\exercise[Rot]\label{exe:source-271} \label{exe:3-4-wnu}
% (Sebastian Meyer)
Show that if a clone (not necessarily on a finite domain) contains \emph{3-4 weak near unanimity operations}, i.e., operations $f \in \WNU(3)$ and $g \in \WNU(4)$ satisfying
$$\forall x,y. \;  f(y,x,x) = g(y,x,x,x),$$
then it also contains $h \in \WNU(13)$.
% Solution:
% take h(...) := g(f(123),f(456),f(789),g(10,11,12,13)).
% Then  h(x,y,...,y) = g(a,b,...,b)=g(b,...,b,a) = h(y,...,y,x)
\end{exercises}

\subsection{Summary of Equivalent Dichotomy Formulations}
In the following we list all the equivalent conditions on a finite structure $\bB$ with finite relational signature that imply that 
$\Csp(\bB)$ is in P. If $\bB$ does not satisfy these conditions, then $\Csp(\bB)$ is NP-complete. 

\begin{corollary}
\label{cor:dicho} 
Let $\bB$ be a finite structure with core $\bC$, let $\bD$ be the expansion of $\bC$ by all singleton unary relations, and let $\fD$ be a polymorphism algebra of $\bD$. 
 Then the following are equivalent. 
\begin{enumerate}
\item $K_3$ does not have a primitive positive interpretation in $\bD$; 
\item $\HSP(\fD)$ does not contain an at least 2-element algebra all of whose operations are projections;
\item $\HS(\fD)$ does not contain an at least 2-element algebra all of whose operations are projections (Corollary~\ref{cor:algebraic-dicho-summary});
\item there is no clone homomorphism from $\Pol(\bD)$ to $\Proj$ (Corollary~\ref{cor:algebraic-dicho-summary}); 
\item $\fD$ satisfies some non-trivial finite set of identities (Corollary~\ref{cor:algebraic-dicho-summary});
\item $\bB$ does not pp-construct $K_3$, i.e., $K_3 \notin \HI(\bB)$ (Corollary~\ref{cor:hi-hard}); 
\item $\bB$ does not pp-construct all finite structures (Corollary~\ref{cor:hi-hard}); 
\item there is no minion homomorphism
from $\Pol(\bB)$ to ${\bf Proj}$ (Corollary~\ref{cor:Taylor-minion}); 
\item $\Pol(\bB)$ contains a Taylor operation (Corollary~\ref{cor:Taylor-minion});
\item $\Pol(\bB)$ has a 6-ary Siggers operation (Theorem~\ref{thm:siggers}); 
\item $\Pol(\bB)$ contains a weak near-unanimity operation (Exercise~\ref{exe:source-262});
\item $\Pol(\bB)$ contains a cyclic operation (Theorem~\ref{thm:cyclic}); 
\item $\Pol(\bB)$ contains for all prime numbers $p> |B|$ a cyclic operation of arity $p$ (Theorem~\ref{thm:cyclic}); 
\item $\Pol(\bB)$ contains a 4-ary Siggers operation (Theorem~\ref{thm:4siggers}); 
\item $\Pol(\bB)$ has $p$-$q$-operations (Proposition~\ref{prop:pq}). 
\end{enumerate} 
\end{corollary}

\subsection{Undirected Graphs Revisited}
\label{sect:undir-revisited}
As another application of the cyclic term theorem, we obtain another proof (from~\cite{Pol}) of the classification of the complexity of $H$-colouring for finite undirected graphs $H$. 

%\begin{theorem}[second proof of the theorem of Hell and Ne\v{s}et\v{r}il taken from~\cite{cyclic}]
%Let $H$ be a finite undirected graph.
%If $H$ is bipartite then $\Csp(H)$ is in P. Otherwise, $\Csp(H)$ is NP-complete. 
%\end{theorem}
\begin{proof}[Second proof of Theorem~\ref{thm:HN}]
If the core $G$ of $H$ equals $K_2$ or has just one vertex, 
then $\Csp(H)$ can be solved in polynomial time, e.g.\ by the path-consistency procedure, Section~\ref{sect:PC}. 
Otherwise, $H$ is not bipartite
and there exists a cycle $a_0,a_1,\dots,a_{2k},a_0$ of odd length in $H$.
If $H$ has no Taylor polymorphism, 
then by Corollary~\ref{cor:Taylor-minion} 
%and  Corollary~\ref{cor:taylor-or-npc},
$\Csp(H)$ is NP-hard. 

Otherwise, if $H$ has a Taylor polymorphism, then Theorem~\ref{thm:cyclic} asserts that there
exists a $p$-ary cyclic polymorphism $c$ of $H$ where $p$ is a prime number greater than $\max\{2k,|V(H)|\}$. 
Since the edges in $H$ are undirected, we can also find a closed walk 
$a_0,a_1,\dots,a_{p-1},a_0$ in $H$. 
Then $c(a_0,a_1,\dots,a_{p-1}) = c(a_1,\dots,a_{p-1},a_0)$, which implies that
$H$ contains a loop, a contradiction to the assumption that the core of
$H$ has more than one element.  
\end{proof}

This proof naturally generalises to smooth digraphs that are strongly connected. In fact, the assumption that $H$ is strongly connected can be dropped.

%finite digraphs with no sources
%and no sinks; 
%such digraphs already appeared in Section~\ref{sect:alg-csp}. 
%\subsection{Digraphs without Sources and Sinks}
%In this section we present an important
%result about 
%Note that undirected graphs $(V,E)$, 
%viewed as directed graphs where for 
%every $\{u,v\} \in E$ we have $(u,v) \in E$
%and $(v,u) \in E$, are examples of such graphs. 

\begin{theorem}[Barto, Kozik, Niven~\cite{BartoKozikNiven}]\label{thm:without-sources-and-sinks}
Let $H$ be a smooth digraph. If $H$ has a Taylor polymorphism,
then $H$ is homomorphically equivalent 
to a disjoint union of cycles. 
\end{theorem}

We only present a proof for the special case where $H$ is strongly connected.  
In the proof we need the concept of \emph{algebraic length} of a graph.
A digraph containing a loop is defined to have \emph{algebraic length
one}. If a digraph is loopless and contains a cycle of positive net
length, then its \emph{algebraic length} is the minimum positive net
length of such a cycle.
% It is the minimum number $k \geq 1$ such that the graph contains a cycle of net length $k$. 

\begin{proof}[Proof of Theorem~\ref{thm:without-sources-and-sinks} if $H$ is strongly connected]
Suppose that $H$ is strongly connected. Let $d$ be the period of $H$,
that is, the greatest common divisor of the lengths of all closed
directed walks in $H$. We prove that $H$ is homomorphically equivalent
to $\vec C_d$.

Fix a vertex $r\in V(H)$. For every $v\in V(H)$, choose a directed path
$P_v$ from $r$ to $v$ and define
\[
 \lambda(v):=|P_v|\pmod d.
\]
This does not depend on the choice of $P_v$. Indeed, if $P$ and $P'$
are directed paths from $r$ to $v$, choose a directed path $Q$ from
$v$ to $r$. Then $PQ$ and $P'Q$ are closed directed walks, so their
lengths are divisible by $d$. Hence, $|P|\equiv |P'|\pmod d$.

If $(v,w)\in E(H)$, then appending this edge to a directed path from
$r$ to $v$ gives a directed path from $r$ to $w$. Consequently,
\[
 \lambda(w)=\lambda(v)+1\pmod d.
\]
Thus, $\lambda$ is a homomorphism from $H$ to $\vec C_d$.

It remains to construct a homomorphism from $\vec C_d$ to $H$. We first
recall a standard consequence of strong connectivity and the
definition of the period: for every sufficiently large integer $n$
divisible by $d$, there is a closed directed walk of length $n$ in
$H$. For completeness, fix $r\in V(H)$ and let
\[
 S:=\{\,|W| \mid W\text{ is a closed directed walk based at }r\,\}.
\]
The greatest common divisor of the elements of $S$ is $d$. Indeed, one
divisibility follows immediately from the definition of $d$. For the
other, let $W$ be a closed directed walk based at an arbitrary vertex
$v$, and choose directed paths $P$ from $r$ to $v$ and $Q$ from $v$ to
$r$. The two walks $PQ$ and $PWQ$ belong to $S$, and the difference of
their lengths is $|W|$. Hence, every common divisor of $S$ divides the
length of every closed directed walk in $H$. The assertion now follows
from the elementary fact that an additive submonoid of the natural
numbers with greatest common divisor $d$ contains every sufficiently
large multiple of $d$.

Choose a sufficiently large prime number $p>|V(H)|$ such that $H$ has
a closed directed walk
\[
 a_0,a_1,\dots,a_{pd}=a_0
\]
of length $pd$. By Theorem~\ref{thm:cyclic}, $H$ has a $p$-ary cyclic
polymorphism $c$. For $j\in\{0,\dots,d-1\}$, define
\[
 b_j:=
 c\bigl(a_j,a_{j+d},a_{j+2d},\dots,a_{j+(p-1)d}\bigr).
\]
For every $j<d-1$, the fact that $c$ is a polymorphism gives $(b_j,b_{j+1})\in E(H)$.
For the final edge, applying $c$ coordinatewise to the edges
\[
 (a_{d-1},a_d),\ (a_{2d-1},a_{2d}),\ \dots,\
 (a_{pd-1},a_{pd})
\]
gives an edge from $b_{d-1}$ to
$c\bigl(a_d,a_{2d},\dots,a_{(p-1)d},a_{pd}\bigr)
 =
 c\bigl(a_d,a_{2d},\dots,a_{(p-1)d},a_0\bigr)$.
Since $c$ is cyclic, the latter element equals
\[
 c\bigl(a_0,a_d,\dots,a_{(p-1)d}\bigr)=b_0.
\]
Therefore, $(b_{d-1},b_0)\in E(H)$. The map sending $j\in V(\vec C_d)$
to $b_j$ is consequently a homomorphism from $\vec C_d$ to $H$.
We have homomorphisms in both directions, so $H$ is homomorphically
equivalent to $\vec C_d$.
\end{proof}

\begin{corollary}[Loop Lemma]\label{cor:loop}
Let $\fA$ be a finite Taylor algebra and 
let $R \leq \fA^2$ be subdirect.
If the digraph $(A,R)$ has a connected component of algebraic length one, then $R$ has a loop. 
\end{corollary}
\begin{proof}
The digraph $(A,R)$ has no sources and sinks because $R$ is subdirect. Since it has the Taylor term operation of $\fA$ as a polymorphism, 
Theorem~\ref{thm:without-sources-and-sinks} implies that 
$(A,R)$ is homomorphically equivalent to a disjoint union of cycles. The only cycle that is homomorphically equivalent to a digraph of algebraic length one is the loop, so $(A,R)$ is homomorphically equivalent to a structure that contains a loop, so it must contain a loop. 
%and all cycles homomorphically map to the loop
\end{proof} 

If a graph $H$ is homomorphically
equivalent to a disjoint union of cycles, then
$\Csp(H)$ is in P (e.g., we can use the algorithm PC$_H$ to solve it; see Section~\ref{sect:PC}).  
On the other hand, a digraph without a Taylor polymorphism has an NP-hard CSP.
Therefore, 
Theorem~\ref{thm:without-sources-and-sinks} shows that the Feder-Vardi conjecture is true for digraphs without sources and sinks: their CSPs are in P or NP-complete. 

As another consequence we present a second proof that Taylor algebras have a 4-ary Siggers term, which is the original proof from~\cite{KearnesMarkovicMcKenzie}.

%Restricted to $\{x,y,z\}$, the relation $R$ looks as follows:
\begin{figure}
\begin{center}
\includegraphics[scale=.6]{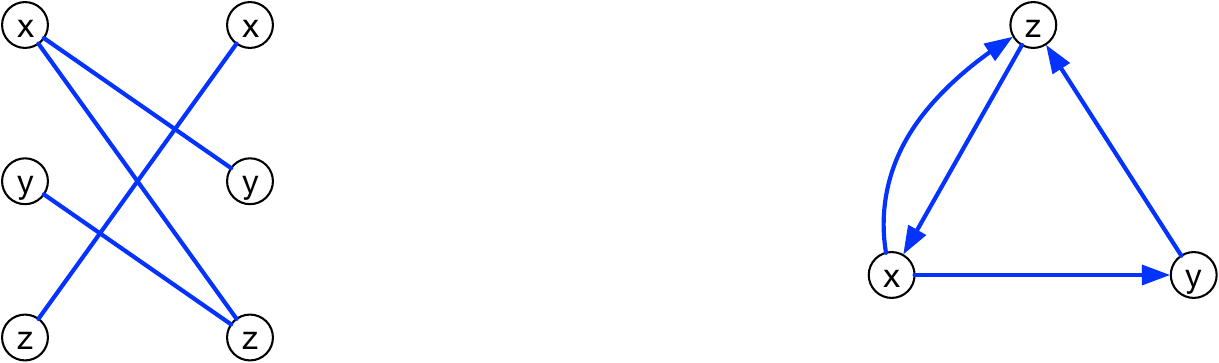}
\end{center}
\caption{The smooth digraph leading to the Siggers terms of arity four.} 
\label{fig:4siggers}
\end{figure}

\begin{proof}[Second proof of Theorem~\ref{thm:4siggers}]
Let $\fF$ be the free algebra with three generators $x,y,z$ in the variety generated by $\fA$ (see Section~\ref{sect:birkhoff}). Let $R \leq \fF^2$ be generated by 
$$\left \{ {x \choose y}, {y \choose z}, {z \choose x}, {x \choose z} \right \}.$$
Then $\fR$ is subdirect and $(F,R)$ is a smooth digraph of algebraic length 1 (see Figure~\ref{fig:4siggers} for the restriction of this digraph to $\{x,y,z\}$). Hence, the Loop Lemma (Corollary~\ref{cor:loop}) implies that $R$ contains a pair $(f,f)$, so there exists a term $t(x_1,x_2,x_3,x_4)$ such that 
$$t^{\fF^2}( {x \choose y}, {y \choose z}, {z \choose x}, {x \choose z} ) = {f \choose f}.$$
Thus, $t^{\fF}$ satisfies $t(x,y,z,x) = f = t(y,z,x,z)$. Permuting the arguments gives a term that satisfies the standard form of the 4-ary Siggers identity. 
\end{proof}
%Suppose that $\fA$ has a

\begin{exercises}
\exercise[Rot]\label{exe:source-272} Let $G$ and $H$ be finite smooth digraphs.
Show that if $\Csp(G \times H)$
can be
solved in polynomial time, then $\Csp(G)$ or $\Csp(H)$ can be solved
 in polynomial time as well (so we cannot use them to solve Exercise~\ref{exe:hard-factors-of-easy-product}).
% Solution sketch: If both smooth factors were hard, the cyclic-term dichotomy would give minion
% homomorphisms from each polymorphism minion to projections. For smooth digraphs these combine to
% such a homomorphism for the product, making $\Csp(G\times H)$ NP-hard. Contrapositively, a
% polynomial-time product has a tractable factor (subject to the standard P versus NP formulation used
% in the text).
\exercise[Schwarz]\label{exe:source-273} Let $G := (\{1,2,3,4\};E)$ be the digraph given by $$ E := \{(1,2),(1,3),(2,3),(3,2),(2,4),(3,4) \}.$$
Show that every finite structure has a primitive positive interpretation in $G$.
%\vspace{.2cm}
% Solution sketch: The two middle vertices pp-define a copy of the hard directed-edge gadget, while
% the common predecessor $1$ and common successor $4$ pp-define the required pins. Using two
% coordinates, the formula saying that exactly one coordinate traverses a middle edge pp-interprets
% $K_3$. Since $K_3$ pp-interprets every finite structure, transitivity gives the result.
\end{exercises}

% END INLINED SOURCE: cyclic.tex
% BEGIN INLINED SOURCE: bw.tex
% !TEX root = GH-UA.tex

\section{Bounded Width}
\label{sect:bounded-width}
Equipped with the universal-algebraic approach, 
we come back to one of the questions that occupied
us at the beginning of the course:
which $H$-colouring problems can be solved by the path-consistency procedure
(PC$_H$, introduced in Section~\ref{sect:PC})? We have seen in Section~\ref{ssect:majority} that if $H$ has a majority or a semilattice polymorphism, then PC$_H$ solves the $H$-colouring problem. But these were just sufficient, not necessary conditions. 

A necessary and sufficient polymorphism condition for solvability by PC$_H$ has been found by Barto and Kozik~\cite{BoundedWidth}. Their result is much stronger:
it characterises not just the strength of PC$_H$, but
more generally of $k$-consistency (introduced in Section~\ref{sect:PC}), and not just for $H$-colouring, but for
CSPs of finite relational structures in general (Section~\ref{sect:kcons}). 

\begin{theorem}\label{thm:bw-short}
Let $\bB$ be a finite structure with a finite relational signature. Then $\Csp(\bB)$ can be solved by the $k$-consistency procedure, for some $k \in {\mathbb N}$, if and only if 
$\bB$ has \emph{3-4 weak near unanimity} polymorphisms, i.e., operations $f \in \WNU(3)$ and $g \in \WNU(4)$ satisfying
$$\forall x,y. \;  f(y,x,x) = g(y,x,x,x).$$
\end{theorem} 

Note that for a given $\bB$, the condition given in Theorem~\ref{thm:bw-short} is obviously decidable. Many other equivalent 
characterisations of bounded width will be presented later in Theorem~\ref{thm:bounded-width}. 

It turns out that if $\Csp(\bB)$ can be solved by $k$-consistency, for some $k$, then it can already be solved by a particularly natural algorithm, namely by the singleton arc-consistency (SAC) procedure (Section~\ref{sect:SAC}).  
SAC is weaker than $3$-consistency in the sense that
if the SAC procedure detects that an instance is unsatisfiable, then so does the $3$-consistency procedure.\footnote{The converse of this statement in general does not hold: e.g., there are instances $\bA$ of $\Csp(K_3)$ where the $3$-consistency procedure for $\Csp(K_3)$ detects that $\bA$ is unsatisfiable, but SAC does not. In this sense, SAC is \emph{strictly} weaker than $3$-consistency.} However, SAC has the advantage that, unlike $3$-consistency, it only requires linear workspace and can be implemented to run in quadratic time. 
In fact, it can be solved by establishing an even weaker form of consistency, namely \emph{cycle-consistency}, which is still much stronger than arc-consistency (Theorem~\ref{thm:slac}). 
This strong result will be convenient when proving Theorem~\ref{thm:bw-short}. 

%Before we define the singleton arc consistency procedure in Section~\ref{sect:SAC}, we revisit in Section~\ref{sect:ACrevisited} the
%arc consistency procedure to generalise it to arbitrary relational structures and to characterise its expressive power using minion homomorphisms, because
%SAC is  based on AC. 

%Another important ingredient 

\subsection{$k$-Consistency}
\label{sect:kcons-revisited}
We have already briefly introduced the $k$-consistency procedure for digraph homomorphism problems in Section~\ref{sect:kcons}. 
%In this section, we treat the $k$-consistency procedure again, and in particular we discuss the following new aspects:
%\begin{itemize}
%\item the generalisation of the $k$-consistency procedure to general relational structures; 
%\item the $k$-consistency as a \emph{uniform} CSP algorithm. 
%\item two parameters for the $k$-consistency proce
%\end{itemize} 
% ...
%In the same way, the arc-consistency procedure can be viewed as a uniform CSP algorithm, in which case we simply write $\AC$ (rather than $\AC_{\bB}$). 
To discuss the idea of 
$k$-consistency for general 
relational structures more precisely, 
it is convenient to use a more flexible terminology. 
When generalising the 
3-consistency procedure for the $H$-colouring problem to
the $k$-consistency procedure for CSPs of arbitrary finite structures $\bB$,
there are two essential parameters:  
\begin{itemize}
\item The first is
the arity $l$ of the relations maintained for all
$l$-tuples of variables in the instance. 
For PC$_H$, for instance, we have $l=2$.
\item The second is the number of variables
considered at a time within the main loop of the algorithm.
For PC$_H$, for instance, we have $k=3$. 
\end{itemize}
Hence, for each pair $(l,k) \in {\mathbb N}^2$, we obtain
a different form of consistency procedure, called $(l,k)$-consistency procedure. 
When characterising the power of this procedure, the following definition will be useful. 

\begin{definition}\label{def:lk-strategy}
Let $\bA$ and $\bB$ be relational structures of the same signature,
and let $1\leq l\leq k$. If $U\subseteq A$, then a map
$s\colon U\to B$ is a \emph{partial homomorphism} if, for every
$r$-ary relation symbol $R$ and every
$(x_1,\dots,x_r)\in R^{\bA}$ with $x_1,\dots,x_r\in U$, we have
$(s(x_1),\dots,s(x_r))\in R^{\bB}$.

An \emph{$(l,k)$-strategy} from $\bA$ to $\bB$ is a family
\[
 \bigl(\mathcal H_U\bigr)_{U\subseteq A,\ |U|\leq l}
\]
of non-empty sets of partial homomorphisms from $U$ to $B$ with the
following extension property. For every $U\subseteq A$ with
$|U|\leq l$, every $s\in\mathcal H_U$, and every $V$ such that
$U\subseteq V\subseteq A$ and $|V|\leq k$, there exists a partial
homomorphism $t\colon V\to B$ extending $s$ such that $t|_W\in\mathcal H_W$
for every $W\subseteq V$ with $|W|\leq l$.

We say that \emph{the $(l,k)$-consistency procedure for $\Csp(\bB)$ accepts
$\bA$} if there exists an $(l,k)$-strategy from $\bA$ to $\bB$.
Equivalently, one starts, for every $U\subseteq A$ with $|U|\leq l$,
with the set of all partial homomorphisms from $U$ to $\bB$, and
repeatedly removes a map that does not have the extension required
above. The procedure rejects if one of these sets becomes empty.
We say that $\Csp(\bB)$ has \emph{width $(l,k)$} if every finite structure
$\bA$ that is accepted by the $(l,k)$-consistency procedure has a
homomorphism to $\bB$.
We say that $\Csp(\bB)$ 
has \emph{bounded width} if it 
has width $(l,k)$ 
for some $l,k \in \mathbb N$, and otherwise we say that it has \emph{unbounded width}. 
\end{definition}

\begin{remark} 
Note that it is easy to come up with finite 
structures $\bB$ whose CSP cannot
be solved by the $(l,k)$-consistency procedure if $\bB$ might contain relations of arity larger than $k$
(there is no possibility of the $(l,k)$-consistency
algorithm to take constraints into account that are imposed on more than $k$ variables). 
\end{remark}

\begin{remark} A CSP has bounded width if and
only if unsatisfiability of an instance of $\Csp(\bB)$ can be detected by a \emph{Datalog program} (see~\cite{FederVardi}). 
\end{remark}

The following lemma suggests that 
the universal-algebraic approach
can be used to study the question
for which structures $\bB$
the problem $\Csp(\bB)$ has bounded width. 

\begin{lemma}\label{lem:pp-datalog}
Let $\bB$ and $\bC$ be structures with finite relational signatures
such that $\bB \in\HI(\bC)$. If $\Csp(\bC)$ has bounded width, then so
does $\Csp(\bB)$.
\end{lemma}

\begin{proof}
We first consider the case that $\bB$ has a primitive positive
interpretation $I$ in $\bC$. Let $d$ be the dimension of $I$, let
$\delta_I(\bar y)$ be its domain formula, and let
$\iota\colon D\to B$ be its coordinate map. For every $r$-ary
relation symbol $R$ of $\bB$, let
$R_I(\bar y_1,\dots,\bar y_r)$ be the primitive positive formula
defining the preimage of $R^{\bB}$. Fix presentations of these
formulas as conjunctions of atomic formulas with existentially
quantified variables. We regard equality as a binary relation
interpreted by equality, so that equality atoms need not be eliminated
by identifying variables.

Let $\bA$ be an instance of $\Csp(\bB)$. We construct an instance
$T_I(\bA)$ of $\Csp(\bC)$ as follows. For every $a\in A$, introduce
a $d$-tuple $\bar y_a=(y_a^1,\dots,y_a^d)$ and a fresh copy of the
existential variables in $\delta_I(\bar y_a)$. For every
$(a_1,\dots,a_r)\in R^{\bA}$, introduce a fresh copy of the
existential variables in
$R_I(\bar y_{a_1},\dots,\bar y_{a_r})$. The constraints of
$T_I(\bA)$ are all the atomic formulas occurring in these copies of
the domain and relation formulas.

By the definition of a primitive positive interpretation, $\bA\to\bB$
if and only if $T_I(\bA)\to\bC$. Indeed, a homomorphism from $\bA$ to
$\bB$ can be lifted by choosing representatives in $D$ and witnesses
for the primitive positive formulas. Conversely, a homomorphism from
$T_I(\bA)$ to $\bC$ assigns to every $\bar y_a$ an element of $D$,
and composing with $\iota$ gives a homomorphism from $\bA$ to $\bB$.

Suppose that $\Csp(\bC)$ has width $(l,k)$. Put
$$q:=\max\bigl(\{1,d\}\cup\{\operatorname{ar}(R)\mid R\text{ is a
relation symbol of }\bB\}\bigr),$$ 
$L:=qk$, and $K:=q(k+1)$. We prove
that $\Csp(\bB)$ has width $(L,K)$.

Every variable $z$ of $T_I(\bA)$ has a support
$\operatorname{supp}(z)\subseteq A$, defined as follows. If
$z=y_a^i$, or if $z$ is a witness in the copy of the domain formula
belonging to $a$, then $\operatorname{supp}(z):=\{a\}$. If $z$ is a
witness in the copy of
$R_I(\bar y_{a_1},\dots,\bar y_{a_r})$ belonging to the constraint
$(a_1,\dots,a_r)\in R^{\bA}$, then
$\operatorname{supp}(z):=\{a_1,\dots,a_r\}$. Thus
$|\operatorname{supp}(z)|\leq q$. For a set $W$ of variables of
$T_I(\bA)$, put
$\operatorname{supp}(W):=\bigcup_{z\in W}\operatorname{supp}(z)$.
In particular, $|\operatorname{supp}(W)|\leq q|W|$.

Suppose that
$(\mathcal H_U)_{U\subseteq A,\ |U|\leq L}$ is an $(L,K)$-strategy
from $\bA$ to $\bB$. We construct an $(l,k)$-strategy
$(\mathcal G_W)_{W\subseteq T_I(A),\ |W|\leq l}$ from $T_I(\bA)$
to $\bC$.
Let $W\subseteq T_I(A)$ with $|W|\leq l$, and put
$U:=\operatorname{supp}(W)$. Then $|U|\leq ql\leq qk=L$. A map
$g\colon W\to C$ belongs to $\mathcal G_W$ if it can be obtained as
follows. Choose $s\in\mathcal H_U$. For every $a\in U$, choose a
representative $\bar c_a\in D$ such that
$\iota(\bar c_a)=s(a)$, and assign the entries of $\bar c_a$ to the
coordinate variables $y_a^1,\dots,y_a^d$. For every copy of a domain
or relation formula whose witness variables occur in $W$, choose a
full tuple of witnesses satisfying that formula. The map $g$ is the
restriction to $W$ of the resulting assignment. We retain the full
tuple of witnesses for every formula whose witnesses meet $W$, so
that these choices can be preserved when $W$ is enlarged.

The sets $\mathcal G_W$ are non-empty. To see this, choose
$s\in\mathcal H_U$ and representatives of its values. If a relation
gadget whose witnesses meet $W$ comes from
$(a_1,\dots,a_r)\in R^{\bA}$, then $a_1,\dots,a_r\in U$. Since $s$
is a partial homomorphism, $(s(a_1),\dots,s(a_r))\in R^{\bB}$.
Hence the chosen representatives satisfy
$R_I(\bar y_{a_1},\dots,\bar y_{a_r})$, and the required witnesses
exist.

The resulting map $g$ is a partial homomorphism. There is only one
minor point to verify. An atomic constraint of $T_I(\bA)$ whose
variables all belong to $W$ might come from a relation gadget whose
support is not contained in $U$. In this case the atom can only
involve coordinate variables, since every witness variable carries
the full support of its gadget. Let
$(a_1,\dots,a_r)\in R^{\bA}$ be the constraint producing this atom.
Since $|U\cup\{a_1,\dots,a_r\}|\leq L+q=K$, the extension property of
the $(L,K)$-strategy extends $s$ to a partial homomorphism on
$U\cup\{a_1,\dots,a_r\}$. Its values on
$a_1,\dots,a_r$ form a tuple in $R^{\bB}$. Consequently, the
corresponding interpretation formula holds for the chosen
representatives, and in particular the atomic constraint under
consideration is satisfied.

It remains to verify the extension property for the family
$(\mathcal G_W)$. Let $W_0\subseteq W_1\subseteq T_I(A)$ with
$|W_0|\leq l$ and $|W_1|\leq k$, and let
$g_0\in\mathcal G_{W_0}$. Put
$U_0:=\operatorname{supp}(W_0)$ and
$U_1:=\operatorname{supp}(W_1)$. Then
$|U_0|\leq ql\leq L$ and $|U_1|\leq qk=L$.

Choose a certificate for $g_0\in\mathcal G_{W_0}$, consisting of
$s_0\in\mathcal H_{U_0}$, representatives of its values, and the
chosen witnesses. By the extension property of the $(L,K)$-strategy,
there exists a partial homomorphism $s_1\colon U_1\to B$ extending
$s_0$ such that $s_1|_U\in\mathcal H_U$ for every $U\subseteq U_1$
with $|U|\leq L$. In particular, $s_1\in\mathcal H_{U_1}$.

Retain the representatives and witnesses already chosen for $g_0$.
Choose representatives for the new elements of $U_1$, and choose
witnesses for every new domain or relation gadget meeting $W_1$.
These witnesses exist because $s_1$ is a partial homomorphism. In this
way we obtain a partial homomorphism $g_1\colon W_1\to C$ extending
$g_0$.

Finally, let $W\subseteq W_1$ with $|W|\leq l$. Then
$|\operatorname{supp}(W)|\leq ql\leq L$, and hence
$s_1|_{\operatorname{supp}(W)}
\in\mathcal H_{\operatorname{supp}(W)}$. The representatives and
witnesses chosen above show that $g_1|_W\in\mathcal G_W$. Thus
$(\mathcal G_W)$ is an $(l,k)$-strategy from $T_I(\bA)$ to $\bC$.

Now suppose that the $(L,K)$-consistency procedure accepts $\bA$.
Then the $(l,k)$-consistency procedure accepts $T_I(\bA)$. Since
$\Csp(\bC)$ has width $(l,k)$, there is a homomorphism
$T_I(\bA)\to\bC$. It follows that $\bA\to\bB$. Hence
$\Csp(\bB)$ has width $(L,K)$.

Finally, suppose that $\bB$ is merely homomorphically equivalent to a
structure $\bD$ that has a primitive positive interpretation in
$\bC$. By what we have proved, $\Csp(\bD)$ has width $(L,K)$ for some
$L,K$. Let $e\colon\bB\to\bD$ and $f\colon\bD\to\bB$ be
homomorphisms. If $(\mathcal H_U)$ is an $(L,K)$-strategy from an
instance $\bA$ to $\bB$, then
$(\{e\circ s\mid s\in\mathcal H_U\})_{U\subseteq A,\ |U|\leq L}$ is
an $(L,K)$-strategy from $\bA$ to $\bD$. Therefore $\bA\to\bD$, and
composing with $f$ gives $\bA\to\bB$. Thus $\Csp(\bB)$ has bounded
width.
\end{proof}

\begin{remark} 
Recall from Corollary~\ref{cor:dual} that
for every structure $\bB$ with finite relational 
signature there exists a structure $\bC$ with maximal arity $2$ such that $\bB$ and $\bC$ are primitively positively bi-interpretable. It therefore follows from 
Lemma~\ref{lem:pp-datalog} that 
we may assume without loss of generality that $\bB$ has maximal arity $2$ when proving Theorem~\ref{thm:bw-short}. 
\end{remark}

\subsection{Unbounded Width}
In this section we prove that 
CSPs for solving 
linear equations 
over non-trivial abelian groups 
have unbounded width, i.e., 
for every $k \in {\mathbb N}$, 
there exists an unsatisfiable instance $\bA$ such that the $k$-consistency procedure 
does not derive `false' on $\bA$. 
Two proof sketches of this fact can be found 
in~\cite{FederVardi}. A stronger non-expressibility can be
found in~\cite{AtseriasBulatovDawar}
(the given CSP is not even expressible 
in least fixed point logic with counting quantifiers). 
The presentation in this section, however, is from~\cite{Book}. 

The \emph{girth}\index{girth (of a graph)} of an undirected graph $G$ is 
the length of a shortest cycle in $G$.
A graph is called \emph{$k$-regular}\index{regular graph} if every vertex has precisely $k$ neighbours. A 3-regular graph 
is also called \emph{cubic}\index{cubic graph}. 
A great deal is known about the existence of finite graphs of high girth. 

\begin{theorem}\label{thm:girth}
For every $k \in {\mathbb N}$ there exists
\begin{itemize}
\item a finite four-regular graph of girth at least $k$ (it is even known that there are such graphs of exponential size in $k$~\cite{JLR}); 
\item a finite cubic graph of girth at least $k$ with a Hamiltonian cycle (see 
the comments after the proof of Theorem 3.2 
in~\cite{BiggsCubic}). 
\end{itemize}
\end{theorem}

\begin{definition}[incidence graph of a structure]
\label{def:incidence-graph}
Let $\bA$ be a relational structure. 
Then the \emph{incidence graph}\index{incidence graph} $G(\bA)$ of $\bA$
is the bipartite graph whose set of vertices is 
the disjoint union of $A$ and the set of tuples in relations from $\bA$, 
and where an element from $A$ is connected to a tuple if and only if the element appears in that tuple. 
\end{definition}

Note that $G(\bA)$ is bipartite.  
We say that $\bA$ has \emph{girth $k$}\index{girth (of a structure)} if all tuples in relations from $\bA$ have pairwise distinct entries and the shortest
cycle of $G(\bA)$ has $2k$ edges. 
The following is a special case of Lemma 8.6.6.\ in~\cite{Book}. 

\begin{lemma}\label{lem:dom}
Let $\bB$ be a finite structure 
whose
signature $\tau$ consists of finitely many relation symbols
of arity at least 3
such that for every relation $R$ of $\bB$ of arity $r$, every $(r-1)$-ary projection of $R$ equals the full relation $B^{r-1}$. 
Suppose that 
there exist unsatisfiable instances of $\Csp(\bB)$ of arbitrarily large girth. 
Then $\Csp(\bB)$ has unbounded width. 
\end{lemma}

Let $G$ be an abelian group\index{abelian group} and $c \in G$. 
We define  
\begin{align*}
R^k_c & :=  \{(x_1,\dots,x_k) \in G^k \mid x_1+\cdots+x_k =  c\}. 
\end{align*}

\begin{theorem}[Feder and Vardi~\cite{FederVardi}]
\label{thm:ability-to-count}
Let $G$ be a finite abelian group with at least two elements, and let
$\bG$ be the structure with domain $G$ that contains the relation
$R^3_c$ for every $c\in G$. Then $\Csp(\bG)$ is not in Datalog.
\end{theorem}

\begin{proof}
Let $\bB$ be the structure with domain $G$ which contains, for every
$f\colon\{1,2,3\}\to\{-1,1\}$ and every $a\in G$, the relation
\[
R_{f,a}:=
\left\{
(x_1,x_2,x_3)\in G^3 \mid 
\sum_{i=1}^3 f(i)x_i=a
\right\}.
\]
These relations need not be primitively positively definable in
$\bG$. Instead, we show that $\bB$ has a two-dimensional primitive
positive interpretation in $\bG$.
Use $G^2$ as the domain of the interpretation and the coordinate map
\[
\iota\colon G^2\to G,\qquad \iota(x,y):=x-y.
\]
The kernel of $\iota$ is primitively positively definable in $\bG$:
\[
\iota(x,y)=\iota(x',y')
\quad\Longleftrightarrow\quad
\exists z\,
\bigl(
R^3_0(x,y',z)\wedge R^3_0(x',y,z)
\bigr).
\]
For $a\in G$, define the primitive positive formula
\begin{align*}
T_a(u_1,v_1,u_2,v_2,u_3,v_3)
\quad:\Longleftrightarrow\quad
\exists p,q,r\,
\bigl(&R^3_0(u_1,u_2,p)
\wedge R^3_0(v_1,v_2,q)\\
&{}\wedge R^3_0(q,u_3,r)
\wedge R^3_{-a}(p,v_3,r)\bigr).
\end{align*}
A direct calculation gives
\[
T_a(u_1,v_1,u_2,v_2,u_3,v_3)
\quad\Longleftrightarrow\quad
\sum_{i=1}^3(u_i-v_i)=a.
\]
For fixed $f$, put
\[
(u_i,v_i):=
\begin{cases}
(x_i,y_i) & \text{if }f(i)=1,\\
(y_i,x_i) & \text{if }f(i)=-1.
\end{cases}
\]
Then $T_a(u_1,v_1,u_2,v_2,u_3,v_3)$ defines the preimage of
$R_{f,a}$ under $\iota$. Hence $\bB$ has a primitive positive
interpretation in $\bG$. By Lemma~\ref{lem:pp-datalog}, it remains to
show that $\Csp(\bB)$ has unbounded width.

Every relation of $\bB$ satisfies the assumptions of
Lemma~\ref{lem:dom}. Fix $k\in{\mathbb N}$ and let $(V;E)$ be a cubic
graph of girth at least $k$, whose edges we orient arbitrarily. We
construct an instance $\bA$ of $\Csp(\bB)$ with domain $E$.

For every $v\in V$, choose an ordering $e_1,e_2,e_3$ of the edges
incident with $v$, and define
\[
f_v(i):=
\begin{cases}
-1 & \text{if $v$ is the initial vertex of $e_i$},\\
 1 & \text{if $v$ is the terminal vertex of $e_i$}.
\end{cases}
\]
Add the constraint $R_{f_v,0}(e_1,e_2,e_3)$. At one vertex, replace
this constraint by $R_{f_v,a}(e_1,e_2,e_3)$ for some
$a\in G\setminus\{0\}$.

This instance is unsatisfiable. Indeed, if $s\colon E\to G$ were a
solution, summing all its equations would give $0$ on the left,
because every edge occurs once with coefficient $1$ and once with
coefficient $-1$. The right-hand side is $a$, a contradiction.
Moreover, the incidence graph of $\bA$ is the subdivision of
$(V;E)$, so $\bA$ has girth at least $k$. Thus, Lemma~\ref{lem:dom}
shows that $\Csp(\bB)$ has unbounded width, and
Lemma~\ref{lem:pp-datalog} yields the same conclusion for
$\Csp(\bG)$.
\end{proof}
 
\begin{corollary}\label{cor:unbounded} 
Let $\bB$ be a structure with a finite domain and a finite relational signature. If
$\bB$ pp-constructs $({\mathbb Z}_p;+,1)$,
for some prime $p$, then $\Csp(\bB)$ has unbounded width. 
\end{corollary}
\begin{proof}
Having unbounded width is preserved by pp-interpretations (Lemma~\ref{lem:pp-datalog})
and homomorphic equivalence. 
For every $c \in {\mathbb Z}_p$, the relation $R^3_c$ is pp-definable in $({\mathbb Z}_p;+,1)$. 
Hence, the statement follows from Theorem~\ref{thm:ability-to-count}. 
\end{proof} 

\begin{corollary}
\label{cor:abelian-pp}
Let $\bB$ be a structure with a finite domain and a finite relational signature. 
Let $\fB$ be an idempotent algebra such that $\Clo(\fB) = \Pol(\bB)$. If $\HS(\fB)$ contains an abelian algebra with at least two elements, then $\bB$ pp-constructs
$({\mathbb Z}_p;+,1)$ for some prime $p$, and hence 
 $\Csp(\bB)$ has unbounded width. 
\end{corollary}
\begin{proof}
Since restricting to a subalgebra and taking a quotient can be implemented by primitive positive interpretations (Theorem~\ref{thm:pp-interpret}), we may assume that $\fB$ itself is abelian with at least two elements. If $\bB$ does not have a Taylor polymorphism, then it pp-constructs all finite structures (this follows from combining Corollary~\ref{cor:Taylor-minion} and Corollary~\ref{cor:hi-hard}; Exercise~\ref{exe:pp-construct-all-finite}).
Otherwise, by Corollary~\ref{cor:abelian-affine} we 
know that $\fB$ is affine, i.e., polynomially equivalent to a module $\fM$. Since $\fB$ is idempotent, the graph of addition $+$ in $\fM$ is preserved by $\Clo(\fB)$, 
so the structure $(M;+,(\{b\})_{b \in B})$ 
is pp-definable in $\bB$. Since $M$ is finite, it contains some element of order $p$, for $p$ prime. The set $\{x \in M \mid px=0\}$ is pp-definable in $\bB$; hence, we may assume that every non-zero element of $M$ has order exactly $p$. Thus, $(M;+)$ is isomorphic to 
%$({\mathbb Z}/p{\mathbb Z})^k$, 
$(({\mathbb Z}_p)^k;+)$, 
for some $k \in {\mathbb N}_{\geq 1}$. Let $c$ be any non-zero element of $M$. Then $(M;+,\{c\})$ is homomorphically equivalent to
$({\mathbb Z}_p;+,1)$. 
%$({\mathbb Z}/p{\mathbb Z};+,1)$. 
%primitively positively definable in $\bB$
\end{proof}

\subsection{Singleton AC}
\label{sect:SAC}
\emph{Singleton Arc-Consistency} (SAC) comes close to strategies that humans perform when solving Sudoku puzzles. 
It relies on the arc-consistency procedure from Section~\ref{sect:AC}, generalised to finite relational structures (Exercise~\ref{exe:gac}, Definition~\ref{def:ACB}). 

Let $\bB$ be a finite structure with a finite relational signature $\tau$. 
The algorithm $\SAC_{\bB}$
takes as input a $\tau$-structure $\bA$
and works as follows.  
First, we run $\AC_{\bB}$ on $\bA$; if it detects an inconsistency, we reject. Then we repeatedly perform the following steps. 
\begin{itemize} 
\item First create a copy $L'$ of every list $L$ computed by $\AC_{\bB}$.
\item Pick a variable $a$ and set $L'(a)$ to $\{b\}$ for some value $b \in L(a)$.
\item Again establish arc-consistency starting with the lists $L'$. 
\item If AC detects an inconsistency, we permanently remove the value $b$ from $L(a)$. 
\end{itemize}
This is done for all pairs $(a,b) \in A \times B$ until the lists $L$ do not change any more. 
If in this way one of the lists becomes empty, the procedure returns {\bf No}, otherwise {\bf Yes}. 
The pseudo-code is 
displayed in Figure~\ref{fig:sac}.

\begin{remark}\label{rem:sac-one-sided}
If the algorithm returns {\bf No}, then there is no homomorphism from $\bA$ to $\bB$. 
\end{remark}

\begin{figure}
\begin{center}
\fbox{
\begin{tabular}{l}
$\SAC_{\bB}(\bA)$ \\
Input: a finite structure $\bA$ with finite relational signature. \\
Data structure: a list $L(a) \subseteq B$ for each $a \in A$, initially set to $B$. \\
\\
Do \\
\hspace{.5cm}
Run $\AC_\bB(\bA)$ starting with the lists $L$; if it derives `false', return `false'. \\
\hspace{.5cm}
Create a copy $L'$ of the lists $L$ computed by AC. \\
\hspace{.5cm}
Pick some $a \in A$ and some value $b \in L(a)$ and set $L'(a) := \{b\}$. \\
\hspace{.5cm}
Run AC starting with the lists $L'$. \\
\hspace{.5cm}
If it derives false, remove $b$ from $L(a)$. \\
Loop until for no pair $(a,b) \in A \times B$ the value $b$ is removed from $L(a)$. \\
Return $\bf Accept$. 
\end{tabular}
}
\end{center}
\caption{The singleton arc-consistency procedure for $\Csp(\bB)$.}
\label{fig:sac}
\end{figure}

We say that $\SAC_{\bB}$ \emph{solves} 
$\Csp(\bB)$ if the converse is true as well, i.e., if the algorithm returns {\bf Yes}, then there is a homomorphism from $\bA$ to $\bB$.

%First, \emph{Linear Arc Consistency} (LAC) is the restriction of the arc consistency procedure for arbitrary relational signatures where, informally,  each inference uses at most one fact that has been derived previously. \emph{SLAC} is the extension of LAC which performs the following with an instance $I$ of $\Csp(\bB)$:

\begin{remark}\label{rem:SAC}
If $k$ is the maximal arity of $\bA$ and $\bB$, and $\SAC_{\bB}$ returns {\bf No}, then the 
$(2,\max(3,k))$-consistency procedure returns {\bf No} as well. Similarly, it can be seen that 
if $L(a)$ is the list computed by $\SAC_{\bB}$  for $a \in A$ at the final stage of the algorithm, then $L(a)$ is primitively positively definable in $\bB$. 
\end{remark}

\begin{definition}[Singleton arc-consistency]
\label{def:sac}
%If $\bA$ and $\bB$ are finite structures with finite relational signature $\tau$, 
An instance $\bA$ of $\Csp(\bB)$ together with sets $L(a) \subseteq B$ for every $a \in A$ is called \emph{singleton arc-consistent (with respect to $\bB$)}
if for every $a \in A$ and $b \in L(a)$,
there are lists 
$L'(a') \subseteq L(a')$ for every $a' \in A$ 
 with $L'(a) := \{b\}$ 
such that the instance $\bA$ together with the lists $L'$ 
is arc-consistent (Definition~\ref{def:ac}). 
\end{definition} 

\begin{remark}\label{rem:sac}
Note that $\bA$ together with the sets $L(a)$, for $a \in A$ computed by $\SAC_{\bB}$ at the final stage of the algorithm, is singleton arc-consistent. 
\end{remark}

\begin{example}\label{expl:2sat-sac}
It is easy to see that 
$\SAC_\bB$ solves $\Csp(\bB)$ for the structure $$\bB = (\{0,1\};R_{0,0},R_{0,1},R_{1,1})$$
with $R_{i,j} = \{0,1\}^2 \setminus \{(i,j)\}$ from Exercise~\ref{exe:2SAT}. 
%\{0,1\}^2 \setminus \{(0,0)\},\{0,1\}^2 \setminus \{(1,1)\},\{0,1\}^2 \setminus \{(0,1)\})$$ 
Note that $\bB$ is preserved by the majority operation on $\{0,1\}$ (see Lemma~\ref{lem:bij}; $\Csp(\bB)$ is essentially the 2SAT problem). 
We will show this from general principles in Example~\ref{expl:2sat-sac-2}. 
\end{example} 

In this section we present an algebraic characterisation of the power of $\SAC_{\bB}$, similar to our characterisation of the power of $\AC_{\bB}$ in Theorem~\ref{thm:ac-rev}. 
We start with an analog of the concept of the powerset graph $P(H)$ from Section~\ref{sect:PH} and the powerset structure $P(\bB)$ from Exercise~\ref{exe:gac}. 
%here we slightly deviate from the 
%it has first been described 
The following construction was found by  Chen, Dalmau, and Grussien~\cite{ACandFriends}. 
%our modification has the advantage that it will make it transparent why there is a algorithm that decides for given $\bB$ whether $\SAC_{\bB}$ solves $\Csp(\bB)$. 

%We write $1$ for the unique $\tau$-structure whose only element is $\emptyset$ and all of whose relations are non-empty. 
%Let $P_0({\bB})$ be the disjoint union of
%$P({\bB})$ and the one-element structure 
%whose relations 
%For $n \in {\mathbb N}$, will consider the $n$-th direct power of the powers structure $P(\bB)$

\begin{definition}
Let $I$ be a set. 
Then 
$P_{\SAC}^I(\bB)$
 is the substructure of $P({\bB})^I$ whose universe contains 
$s \colon I \to P({\bB})$ 
%$(S_1,\dots,S_d)$ 
if and only if 
%$$\emptyset \neq  \bigcup_{i \in [d]} S_i =  \bigcup_{i \in [d], |S_i| = 1} S_i.$$
$$\bigcup_{i \in I} s(i) =  \bigcup_{i \in I, |s(i)| = 1} s(i).$$
\end{definition}

The following is the analog of Lemma~\ref{lem:ac} for SAC instead of AC. 

\begin{lemma}\label{lem:sac}
Let $\bA$ and $\bB$ be structures with finite relational signature $\tau$ and finite domains. 
Then $\SAC_{\bB}$ accepts $\bA$ if and only if 
$\bA \to P^{A \times B}_{\SAC}(\bB)$. 
%$d := |\sum_{a \in A} |L(a)|$. 
% {\bF}_{{\bf M}_{\SAC}}(\bB)$. 
\end{lemma}
\begin{proof}
Suppose that $\SAC_{\bB}$ accepts an instance $\bA$. 
%Let $A = \{a_1,\dots,a_n\}$. 
%For $i \in [n]$, let $L(a_i)$ 
For $a \in A$, let $L(a)$ 
denote the (non-empty) set computed by the algorithm at the point of termination. 
%We use $A \times B$ as indices for the elements of $P^d_{\SAC}(\bB)$. 
%For $b \in B \setminus L(a)$, we define
%$h_{a,b}(a') := \emptyset$ for all $a' \in A$. 
%Now 
Let $b \in L(a)$. 
By the definition of the algorithm and Lemma~\ref{lem:ac}, there exists a homomorphism $h_{a,b}$ from 
$(\bA,\{a_1\},\{a_2\},\dots,\{a_n\},\{a\})$ 
to $P((\bB,L(a_1),\dots,L(a_n),\{b\}))$. 
For $b \in B \setminus L(a)$, we define
$h_{a,b}$ equal to $h_{a,b'}$ for some $b' \in L(a)$. 

Define $h(a')$ to be the function that maps
$(a,b)$ to $h_{a,b}(a') \subseteq B$. 
Since $h_{a,b}$ is in particular a homomorphism from $\bA$ to $P(\bB)$, 
we have that $h$ is a homomorphism from 
$\bA$ to $P(\bB)^{A \times B}$. In fact, it is even a homomorphism to $P^{A \times B}_{\SAC}(\bB)$, because for every $a' \in A$ we have 
\begin{align}
\bigcup_{a \in A,b \in B} h(a')(a,b) & = \bigcup_{a \in A, b \in L(a)} h_{a,b}(a') 
%\cup \underbrace{\bigcup_{a \in A, b \in B \setminus L(a)} h_{a,b}(a')}_{=\emptyset} 
\nonumber  \\
%\bigcup_{b \in B} h_{x,b}(x) = 
& = \bigcup_{a \in A, b \in L(a), |h_{a,b}(a')|=1} h_{a,b}(a') \label{eq:key} \\
&  = \bigcup_{a \in A, b \in B, |h(a')(a,b)| = 1} h(a')(a,b).  \nonumber 
\end{align}
%and this set is non-empty since $L(a)$ is non-empty. 
To see the inclusion 
$\subseteq$ in~\eqref{eq:key}, let $b' \in h_{a,b}(a')$. It follows that $b' \in L(a')$, and that 
$h_{a',b'}(a') = \{b'\}$, so $b'$ is also contained in the union shown in~\eqref{eq:key}. 

Conversely, suppose that $h$ is a homomorphism from $\bA$ to
$P^{A \times B}_{\SAC}(\bB)$. For every $x\in A$, put
\[
 U(x):=\bigcup_{(a,b)\in A\times B} h(x)(a,b).
\]
We prove, by induction over the execution of $\SAC_{\bB}$, that
\[
 U(x)\subseteq L(x)
\]
for every $x\in A$, where $L(x)$ denotes the current list of $x$.

The assertion holds after initialisation. Suppose that it holds before
an invocation of $\AC_{\bB}$ with the current lists $L$. Let $x\in A$
and $c\in U(x)$. Since
$h(x)\in P^{A\times B}_{\SAC}(\bB)$, the defining property of
$P^{A\times B}_{\SAC}(\bB)$ implies that there exists
$i\in A\times B$ such that
\[
 h(x)(i)=\{c\}.
\]
Define $h_i\colon A\to P(\bB)$ by
\[
 h_i(y):=h(y)(i).
\]
Since $h$ is a homomorphism into a substructure of
$P(\bB)^{A\times B}$, the map $h_i$ is a homomorphism from $\bA$ to
$P(\bB)$. Moreover, the induction hypothesis gives
\[
 h_i(y)\subseteq U(y)\subseteq L(y)
\]
for every $y\in A$. Lemma~\ref{lem:ac}, applied with the current lists,
therefore shows that no element of $h_i(y)$ is removed during this
execution of $\AC_{\bB}$. In particular, $c$ is not removed from
$L(x)$. Since $x$ and $c\in U(x)$ were arbitrary, the invariant still
holds after the execution of $\AC_{\bB}$.

It remains to consider the permanent deletions performed by
$\SAC_{\bB}$. Suppose that the algorithm tests a pair $(x,c)$ with
$c\in U(x)$. Again choose $i\in A\times B$ such that
$h(x)(i)=\{c\}$. When the algorithm copies the current lists to $L'$
and sets $L'(x):=\{c\}$, we have
\[
 h_i(y)\subseteq L'(y)
\]
for every $y\in A$. Indeed, this follows from the invariant when
$y\neq x$, and for $y=x$ it follows from $h_i(x)=\{c\}$. Hence,
Lemma~\ref{lem:ac} shows that the ensuing execution of
$\AC_{\bB}$ does not derive an inconsistency. Therefore, $c$ is not
permanently removed from $L(x)$. Thus, permanent deletions also
preserve the invariant.
Consequently, the final list $L(x)$ contains $U(x)$ for every $x\in A$.
The defining property of $P^{A\times B}_{\SAC}(\bB)$ implies that
$U(x)$ is non-empty. Hence, no final list is empty, and
$\SAC_{\bB}$ accepts $\bA$.
\end{proof}

\begin{proposition}\label{prop:sac}
Let $\bB$ be a structure with finite signature and finite domain. 
Then
% the following are equivalent:
%\begin{itemize}
%\item $\SAC_{\bB}$ solves $\Csp(\bB)$
%\item 
%\item 
%\end{itemize} 
$\SAC_{\bB}$ solves
$\Csp(\bB)$ if and only if 
$P^I_{\SAC}(\bB) \to \bB$ for every finite set $I$. 
\end{proposition}
\begin{proof}
Suppose that $\SAC_{\bB}$ solves
$\Csp(\bB)$, and let $I$ be a finite set. To show that $P^I_{\SAC}(\bB) \to \bB$, we view $P^I_{\SAC}(\bB)$ as an instance of $\Csp(\bB)$. 
Clearly, $P^I_{\SAC}(\bB) \to P^I_{\SAC}(\bB)$, and hence Lemma~\ref{lem:sac} implies that 
$\SAC_{\bB}$ accepts $P^I_{\SAC}(\bB)$. 
Thus, $P^I_{\SAC}(\bB) \to \bB$, because 
$\SAC_{\bB}$ solves
$\Csp(\bB)$. 

For the other direction, 
let $\bA$ be a finite structure with the same signature as $\bB$. 
If $\SAC_{\bB}$ rejects $\bA$, then clearly there is no homomorphism $\bA \to \bB$. 
Otherwise, by Lemma~\ref{lem:sac} there exists a homomorphism from $\bA$ to $P^{A \times B}_{\SAC}(\bB)$. 
By assumption, there exists a homomorphism $h \colon P^{A \times B}_{\SAC}(\bB) \to \bB$. 
%To show that $\SAC_{\bB}$ solves $\Cap(\bB)$, 
By composing with $h$ we obtain a homomorphism from $\bA$ to 
$\bB$. 
\end{proof} 

\subsection{Palette Totally Symmetric Polymorphisms} 
Similarly as the equivalence of the existence of a homomorphism from $P(\bB)$ to $\bB$ 
and the existence of totally symmetric polymorphisms of all arities of $\bB$ (Theorem~\ref{thm:totally-symmetric}),
we can characterise the existence of homomorphisms from $P^I_{\SAC}(\bB)$ 
to $\bB$ for all finite sets $I$ in terms of a polymorphism condition (this result is missing in~\cite{ACandFriends}). We use slightly simplified terminology of Zhuk~\cite{ZhukSAC}. 

\begin{definition}[Palette tuples]
Let $\ell,k \in {\mathbb N}_{\geq 1}$. 
%and $t \in B^{\ell \times k}$. 
%For $i \in [\ell]$, 
%the \emph{$i$-th block of $t$} is the tuple
%$(t_{(i-1)k+1},\dots,t_{ik}) \in B^k$. 
A matrix $t \in B^{\ell \times k}$ is called 
\emph{palette} if for every $i \in [\ell]$
and $j \in [k]$ there exists $i' \in [\ell]$ such that
the $i'$-th row
%column 
of $t$ equals $(t_{i,j},\dots,t_{i,j})$. 
\end{definition}

\begin{definition}\label{def:msac}
Let $\ell,k \in {\mathbb N}_{\geq 1}$. 
An operation $f \colon B^{\ell \times k} \to B$ is called \emph{$(\ell \times k)$-palette totally symmetric}
if $f(s)=f(t)$
 for all palette matrices $s,t \in B^{\ell \times k}$ such that 
% for every $i \in [\ell]$: too strong, this would imply via the second item that all entries are the same. 
there exists $i \in [\ell]$ with
 %and $i \in [\ell]$ such that
\begin{itemize}
\item the $i$-th %column 
row 
of $s$ and the $i$-th 
row
%column 
of $t$ have the same sets of entries, and 
\item in all other rows, 
%columns, 
$s$ and $t$ have the same entries. 
\end{itemize}
\end{definition}

\begin{example}\label{expl:2sat-palette}
For $\ell,k \in {\mathbb N}_{\geq 1}$, let $f \colon \{0,1\}^{\ell k} \to \{0,1\}$ be defined as follows
$$f(\underbrace{x_{1,1},\dots,x_{1,k}}_{\text{first row}},\dots,\underbrace{x_{\ell,1},\dots,x_{\ell,k}}_{\text{$\ell$-th row}}) := \begin{cases} x_{i,1} & \text{ if  $i$ is minimal such that } x_{i,1} = \cdots = x_{i,k} \\
x_{1,1} &  \text{ otherwise.} \end{cases} $$
This operation is $(\ell \times k)$-palette totally symmetric. Indeed, let $s,t \in \{0,1\}^{\ell \times k}$ be palette and $i \in [\ell]$ be such that 
\begin{itemize}
\item the $i$-th row
% column 
of $s$ and the $i$-th row
%column 
of $t$ have the same set of entries,
i.e., $\{s_{i,1},\dots,s_{i,k}\} = \{t_{i,1},\dots,t_{i,k} \}$,
\item $s_{i',j} = t_{i',j}$ for all $i' \in [\ell] \setminus \{i\}$ and $j \in [k]$. 
\end{itemize} 
Note that for every $i' \in [\ell]$ we have $s_{i',1} = \cdots = s_{i',k}$ if and only if
$t_{i',1} = \cdots = t_{i',k}$. 
Moreover, if $s_{i',1} = \cdots = s_{i',k}$
then $s_{i',1} = t_{i',1}$. 
Hence, 
$f(s) = f(t)$. 

Note that this operation preserves for every $p,q \in \{0,1\}$ the relation 
$R_{p,q} = \{0,1\}^2 \setminus \{(p,q)\}$ from
Exercise~\ref{exe:2SAT} and Example~\ref{expl:2sat-sac}. 
Indeed, suppose that $s,t \in \{0,1\}^{\ell \times k}$ are such that 
$(s_{i,j},t_{i,j}) \in R_{p,q}$ for every $i \in [\ell]$ and $j \in [k]$. 
Suppose for contradiction that 
$(f(s),f(t)) \notin R_{p,q}$, i.e., 
$f(s) = p$ and $f(t) = q$. 
If there exists $i$ such that $s_{i,1} = \cdots = s_{i,k}$, and $i$ is minimal with this property, then $p = f(s) = s_{i,1}$. 
Then $t_{i,j} \neq q$ for every $j \in [k]$, because $(s_{i,j},t_{i,j}) \in R_{p,q}$. 
This implies that $t_{i,1}=\cdots=t_{i,k}$.
The symmetric argument applies 
if there exists $i$ such that $t_{i,1} = \cdots = t_{i,k}$. Hence, exactly one of the following applies:
\begin{itemize}
\item the minimal $i$ such that $s_{i,1} = \cdots = s_{i,k}$ equals the minimal $i$ such that $t_{i,1} = \cdots = t_{i,k}$, 
\item for all $i \in [\ell]$ neither 
$s_{i,1} = \cdots = s_{i,k}$ nor $t_{i,1} = \cdots = t_{i,k}$. 
\end{itemize}
In the first case, we have $f(t) = t_{i,1} \neq q$, and hence $(f(s),f(t)) \in R_{p,q}$. 
In the second case, we have $f(s) = s_{1,1}$ and $f(t) = t_{1,1}$, and hence
$(f(s),f(t)) = (s_{1,1},t_{1,1}) \in R_{p,q}$. 
 \end{example} 

The following is similar to one of the implications in Theorem~\ref{thm:totally-symmetric}. 

\begin{proposition}\label{prop:palette} 
Let $\ell \in {\mathbb N}$. 
Let $\bB$ be a structure  
with 
domain $B = \{b_1,\dots,b_n\}$,
 maximal arity $m$, and 
an $(\ell \times mn)$-palette totally
symmetric polymorphism $f$. 
% for $k_1=\cdots=k_{\ell}=m |B|$. 
%for every $\ell,k_1,\dots,k_{\ell} \in {\mathbb N}$ a $(k_1,\dots,k_{\ell})$-palette totally symmetric polymorphism. 
Then 
$P^{[{\ell}]}_{\SAC}(\bB) \to \bB$.
% and $\SAC_{\bB}$ solves $\Csp(\bB)$.  
\end{proposition} 
\begin{proof}
%Let $B = \{b_1,\dots,b_n\}$ and
%let $f$ be 
% $(2n,\dots,2n)$-palette 
Let $s \colon [{\ell}] \to P(\bB)$ be an element of 
$P^{[{\ell}]}_{\SAC}(\bB)$.
For $i \in [{\ell}]$ let $k_i \in [n]$ be such that 
$s(i) = \{c^i_1,\dots,c^i_{k_i}\}$. 
%We know that $\bigcup_{i \in [{\ell}]} s(i) \neq \emptyset$; let $b_s \in [{\ell}]$ be minimal such that $b_s \in \bigcup_{i \in [{\ell}]} s(i)$. 
Define 
\begin{align}
h(s) := f(\underbrace{c^1_1,\dots,c^1_{k_1},\dots,c^1_{k_1}}_{\text{length } mn},\dots,\underbrace{c^{\ell}_1,\dots,c^{\ell}_{k_\ell},\dots,c^{\ell}_{k_\ell}}_{\text{length } mn})
\label{eq:h-s}
\end{align}
i.e., for every $i \in [{\ell}]$ the element $c^i_{k_i}$ is repeated so that the tuple $(c^i_1,\dots,c^i_{k_i},\dots,c^i_{k_i})$ has length $mn$. We view the tuple $c := (c^1_1,\dots,c^{\ell}_{mn})$ in~\eqref{eq:h-s} as a matrix from $B^{\ell \times mn}$. 
% If $\{c^i_1,\dots,c^i_{\ell_i}\} = \emptyset$ for some $i \in [{\ell}]$, then we replace $c^i_1,\dots,c^i_{\ell_i},\dots,c^i_{\ell_i}$ in the definition above by $b_s$. 

\medskip 
{\bf Claim 1.} $h$ is well defined. For this, we need to argue that for every $i \in [\ell]$, 
the result of $h(s)$ does not depend on the order and multiplicities of the arguments $c^i_1,\dots,c^i_{k_i}$.
%; this is clear if 
%$\{c^i_1,\dots,c^i_{\ell_i}\} = \emptyset$. 
%We have that 
%that $\bigcup_{b \in B} s(b) = \bigcup_{b \in B, |s(b)|=1 s(b)}$, 
%TODO
%Note that by assumption, for every $i \in [n]$ 
%Otherwise, 
Since $s$ is an element of $P^{[{\ell}]}_{\SAC}(\bB)$, 
for every 
$p \in [k_i]$ we have $c^i_p \in \bigcup_{i' \in [\ell]} s(i') = \bigcup_{i' \in [\ell], |s(i')|=1} s(i')$,
and hence there exists $i' \in [\ell]$ such that $s(i') = \{c^i_p\}$, i.e., 
%In other words, there exists $j \in [\ell]$ such that 
the $i'$-th row %column 
of $c$  
% := (c^1_1,\dots,c^1_{\ell_1},\dots,c^n_1,\dots,c^n_{\ell_n})$ 
equals
$(c^i_p,\dots,c^i_p)$. Hence, $c$ is palette. 
It follows that the definition of $h$ does not depend on the order and the multiplicities of the entries in each block. 

\medskip 
{\bf Claim 2.} $h$ is a 
homomorphism from
$P^{[\ell]}_{\SAC}(\bB)$ to $\bB$. The argument is similar as in the proof of the implication $3. \Rightarrow 1.$ of Theorem~\ref{thm:totally-symmetric}.
Let $R \in \tau$; we may assume without loss of generality that $R$ has arity $m$ (otherwise, we prove preservation of $R$ by applying the argument below to a padded relation of arity $m$). 
Let $(s_1,\dots,s_m) \in R^{P^{[\ell]}_{\SAC}(\bB)}$. 
For every $i\in[\ell]$ and $j\in[m]$, write
$s_j(i)=\{c^{i,j}_1,\dots,c^{i,j}_{k_{i,j}}\}$, where
$k_{i,j}\in[n]$, and put
$c^{i,j}_{k_{i,j}+1}=\cdots=c^{i,j}_n:=c^{i,j}_{k_{i,j}}$.
For 
%every $i\in[\ell]$, $j\in[m]$, and 
$u\in[n]$, the definition of
$R^{P(\bB)}$ allows us to choose a tuple
\[
 \bar c^i_{(j-1)n+u}
 =
 (c^i_{(j-1)n+u,1},\dots,c^i_{(j-1)n+u,m})
 \in R^{\bB}
\]
such that
$c^i_{(j-1)n+u,j}=c^{i,j}_u$ and
$c^i_{(j-1)n+u,p}\in s_p(i)$ for every $p\in[m]$.
It follows that for every $i\in[\ell]$ and $p\in[m]$ 
\[
 \{c^i_{v,p}\mid v\in[mn]\}=s_p(i).
\]
Applying $f$ to the tuples 
$$c^{1}_1,\dots,c^{1}_{mn},\dots,c^{\ell}_1,\dots,c^{\ell}_{mn}$$ 
yields a tuple $t$ in $R^{\bB}$, because $f$ is a polymorphism. 
However, $t = h(s_1,\dots,s_m)$, since $f$ is $(\ell \times mn)$-palette totally symmetric, 
so $h$ is indeed a homomorphism 
from
$P^{[\ell]}_{\SAC}(\bB)$ to $\bB$. 
\end{proof} 

\begin{example}\label{expl:2sat-sac-2}
We have shown in Example~\ref{expl:2sat-palette} that the structure
$\bB = (\{0,1\};R_{0,0},R_{0,1}$,
$R_{1,1})$ 
%$$\bB = (\{0,1\};\{0,1\}^2 \setminus \{(0,0)\},\{0,1\}^2 \setminus \{(1,1)\},\{0,1\}^2 \setminus \{(0,1)\})$$
from Example~\ref{expl:2sat-sac} 
has for every $\ell,k \in {\mathbb N}_{\geq 1}$ an $(\ell \times k)$-palette totally symmetric polymorphism.
Hence, $P^I_{\SAC}(\bB) \to \bB$ for every finite set $I$ by Proposition~\ref{prop:palette}. 
Then Proposition~\ref{prop:sac} implies 
that $\SAC_{\bB}$ solves $\Csp(\bB)$. 
\end{example} 

The converse implication in Proposition~\ref{prop:palette} also holds; this was originally shown by Zhuk~\cite{ZhukSAC}, using the Hales--Jewett theorem~\cite{Hales-Jewett}.

\begin{lemma}[Hales--Jewett theorem {\cite{Hales-Jewett}}]
\label{lem:hales-jewett}
Let $A$ and $C$ be non-empty finite sets. Then there exists
$L\in{\mathbb N}_{\geq1}$ such that, for every map
$\chi\colon A^L\to C$, there is a word
$w=(w_1,\dots,w_L)\in(A\cup\{\star\})^L$ in which $\star$ occurs at
least once and such that the value of $\chi(w(a))$ is independent of
$a\in A$. Here $w(a)\in A^L$ denotes the word obtained from $w$ by
replacing every occurrence of $\star$ by $a$.
\end{lemma}

We first introduce the minion that will be used in the proof.

\begin{definition}\label{def:minion-sac}
Let $X$ be a finite set. An element of
$({\bf M}_{\SAC})^{(X)}$ is a matrix
$M\colon X\times[r]\to\{0,1\}$, for some $r\geq1$, with the following
properties.
\begin{itemize}
\item Every column of $M$ contains at least one $1$.
\item If the $x$-th row of $M$ contains a $1$, then some column of
      $M$ is the characteristic vector of $\{x\}$.
\end{itemize}
The number $r$ is called the \emph{width} of $M$.
For a map $\alpha\colon X\to Y$, the minor $M_\alpha$ is the matrix
of the same width defined by
\[
 M_\alpha(y,j)=1
 \quad\text{if and only if}\quad
 M(x,j)=1\text{ for some }x\in\alpha^{-1}(y).
\]
\end{definition}

It is straightforward to verify that ${\bf M}_{\SAC}$ is a minion.

\begin{lemma}\label{lem:psac-minion}
Let $\bB$ be a finite structure. If
$P^{[r]}_{\SAC}(\bB)\to\bB$ for every $r \geq 1$, then there exists a
minion homomorphism
\[
 \xi\colon{\bf M}_{\SAC}\to\Pol(\bB).
\]
\end{lemma}

\begin{proof}
For every $r\geq1$, choose a homomorphism
$h_r\colon P^{[r]}_{\SAC}(\bB)\to\bB$.
For $M\in({\bf M}_{\SAC})^{(B)}$ of width $r$, define
$s_M(i):=\{b\in B\mid M(b,i)=1\}$ for $i\in[r]$.
Definition~\ref{def:minion-sac} ensures that
$s_M\in P^{[r]}_{\SAC}(\bB)$.
Put $h(M):=h_r(s_M)$.

We claim that $h\colon\bF_{{\bf M}_{\SAC}}(\bB)\to\bB$
is a homomorphism.
Let $(M_1,\dots,M_q)\in R^{\bF_{{\bf M}_{\SAC}}(\bB)}$,
witnessed by $N\in({\bf M}_{\SAC})^{(R^{\bB})}$ with
$N_{\pi_j}=M_j$ for every $j\in[q]$.
Since minors preserve width, these matrices have a common width $r$.
For each $i\in[r]$, the support of column $i$ of $N$ is a nonempty
subset of $R^{\bB}$ whose $j$-th projection is $s_{M_j}(i)$.
Hence $(s_{M_1},\dots,s_{M_q})\in R^{P^{[r]}_{\SAC}(\bB)}$,
so applying $h_r$ shows that $h$ preserves $R$.
Proposition~\ref{prop:free-struct} now yields the required
minion homomorphism.
\end{proof}

The next lemma contains the application of the Hales--Jewett theorem.

\begin{lemma}[Zhuk~\cite{ZhukSAC}]
\label{lem:palette-extraction}
Let $\bB$ be a finite structure, and suppose that there exists a
minion homomorphism
$\xi\colon{\bf M}_{\SAC}\to\Pol(\bB)$. Then $\bB$ has an
$(\ell\times k)$-palette totally symmetric polymorphism for all
$\ell,k\geq1$.
\end{lemma}

\begin{proof}
Fix $\ell,k\geq1$, put $q:=\ell k$, and partition $[q]$ into
consecutive sets $Q_1,\dots,Q_\ell$ of size $k$. 
%Let
%${\bf M}_0$ consist of all $0$-$1$ matrices with row set $[q]$ in
%which every column contains a $1$, but which need not satisfy the
%second condition from Definition~\ref{def:minion-sac}. 
We write
$M\circ N$ for the matrix obtained by concatenating the columns of
$M$ and $N$.

For $a\in[\ell]$, let $E_a$ be the matrix with one column, namely the
characteristic vector of $Q_a$. Let $\mathcal E$ be the set of all
matrices of the form
\[
 E_{a_1}\circ\cdots\circ E_{a_q},
 \qquad a_1,\dots,a_q\in[\ell],
\]
and let $I_q$ be the $q\times q$ identity matrix.

We view maps $\sigma\colon[q]\to[\ell]$ as matrices with $\ell$ rows
and $k$ columns. Let $\Omega$ be the set of all palette maps of this
form, and let $\omega$ be the set of all pairs
$(\sigma_1,\sigma_2)\in\Omega^2$ such that
$\sigma_1(Q_a)=\sigma_2(Q_a)$ for every $a\in[\ell]$. We make three observations.
\begin{enumerate}
\item If $E\in\mathcal E$ and
      $(\sigma_1,\sigma_2)\in\omega$, then
      $E_{\sigma_1}=E_{\sigma_2}$.
\item For every $\sigma\in\Omega$, there exists $E\in\mathcal E$
      such that $E_\sigma=(I_q)_\sigma$.
\item Every concatenation of matrices from
      $\mathcal E\cup\{I_q\}$ in which $I_q$ occurs at least once
      belongs to ${\bf M}_{\SAC}$.
\end{enumerate}
The first observation follows because every column of $E$ is the
characteristic vector of one of the sets $Q_a$. For the second, let
$b$ occur in the image of $\sigma$. Since $\sigma$ is palette, there
exists $u(b)\in[\ell]$ such that $\sigma$ is constantly $b$ on
$Q_{u(b)}$. For $v\in[q]$, take the $v$-th column of $E$ to be
$E_{u(\sigma(v))}$. Its $\sigma$-minor is the characteristic vector
of $\{\sigma(v)\}$, which is also the $v$-th column of
$(I_q)_\sigma$. The third observation follows because the copy of
$I_q$ supplies a unit-vector column for every row.

Let $\mathcal C$ be the finite set of maps from $\Omega$ to
$\Pol(\bB)^{(\ell)}\cup\{\star\}$. It is finite,  because $\bB$ is
finite. For $L\geq1$, define a colouring
$\chi_L\colon\mathcal E^L\to\mathcal C$ as follows. If
$E_1,\dots,E_L\in\mathcal E$ and $\sigma\in\Omega$, put
\[
 \chi_L(E_1,\dots,E_L)(\sigma)
 :=
 \begin{cases}
 \xi((E_1\circ\cdots\circ E_L)_\sigma),
 &\text{if }(E_1\circ\cdots\circ E_L)_\sigma
   \in{\bf M}_{\SAC},\\
 \star,
 &\text{otherwise.}
 \end{cases}
\]

By Lemma~\ref{lem:hales-jewett}, for some $L$ there is a word
$(T_1,\dots,T_L)\in(\mathcal E\cup\{X\})^L$ in which $X$ occurs and
such that the colour obtained by replacing every occurrence of $X$
by $E$ is independent of $E\in\mathcal E$. Let $M$ be obtained by
replacing every occurrence of $X$ by $I_q$ and concatenating the
resulting matrices. By observation 3, $M\in{\bf M}_{\SAC}$.

We claim that
\begin{equation}
 \xi(M)_{\sigma_1}=\xi(M)_{\sigma_2}
 \label{eq:palette-minors}
\end{equation}
for every $(\sigma_1,\sigma_2)\in\omega$. By observation 2, choose
$E_1,E_2\in\mathcal E$ such that
$(E_1)_{\sigma_1}=(I_q)_{\sigma_1}$ and
$(E_2)_{\sigma_2}=(I_q)_{\sigma_2}$. Let $M_1$ and $M_2$ be obtained
from $(T_1,\dots,T_L)$ by replacing $X$ by $E_1$ and $E_2$,
respectively. Then
\[
 M_{\sigma_1}
 =
 (M_1)_{\sigma_1}
 =
 (M_1)_{\sigma_2}.
\]
The second equality follows from observation 1. The choice of the
word $(T_1,\dots,T_L)$ gives
$\xi((M_1)_{\sigma_2})=\xi((M_2)_{\sigma_2})$, and the choice of
$E_2$ gives $(M_2)_{\sigma_2}=M_{\sigma_2}$. Since $\xi$ preserves
minors, this proves~\eqref{eq:palette-minors}.

Put $f:=\xi(M)$. Then $f$ is a polymorphism of $\bB$. Let
$s,t\in B^{\ell\times k}$ be palette matrices with the same set of
entries in every row. Every element that occurs in either matrix
occurs constantly in some row. Hence at most $\ell$ distinct elements
occur. Relabelling these elements by members of $[\ell]$, we can
write $s=\rho\circ\sigma_1$ and $t=\rho\circ\sigma_2$ for some
$\rho\colon[\ell]\to B$ and
$(\sigma_1,\sigma_2)\in\omega$. Equation~\eqref{eq:palette-minors}
implies
\[
 f(s)=f_{\sigma_1}(\rho)
 =f_{\sigma_2}(\rho)=f(t).
\]
Thus $f$ is $(\ell\times k)$-palette totally symmetric.
\end{proof}

\begin{theorem}\label{thm:palette} 
Let $\bB$ be a finite structure 
with 
domain $B = \{b_1,\dots,b_n\}$ 
and 
maximal arity $m$. 
Then the following are equivalent. 
\begin{enumerate}
\item $\bB$ has for all $\ell,k \in {\mathbb N}$ an $(\ell \times k)$-palette totally symmetric polymorphism. 
\item $\bB$ has for every $\ell \in {\mathbb N}$ an $(\ell \times nm)$-palette totally
symmetric polymorphism. 
\item $P^I_{\SAC}(\bB) \to \bB$ for every finite set $I$. 
\item There exists a minion homomorphism
${\bf M}_{\SAC}\to\Pol(\bB)$. 
\item $\SAC_{\bB}$ solves $\Csp(\bB)$.  
\end{enumerate}
\end{theorem}
\begin{proof} 
The
implication from 1.\ to 2.\ is immediate, and the implication from
2.\ to 3.\ is Proposition~\ref{prop:palette}. The equivalence of
3.\ and 5.\ is Proposition~\ref{prop:sac}. The implication from 3.\ to 4.\ is Lemma~\ref{lem:psac-minion}. 
The implication from 4.\ to 1.\ is 
Lemma~\ref{lem:palette-extraction}. 
\end{proof} 

\begin{exercises}
\exercise[Rot]\label{exe:source-274} Let $\ell,k \in {\mathbb N}_{\geq 1}$.
Show that every totally symmetric operation of arity $\ell k$ is
$(\ell \times k)$-palette totally symmetric.
% Source: me, 17.6.26
% Solution: Let $s,t \in B^{\ell \times k}$ be palette matrices such that for every $i \in [\ell]$, the $i$-th column of $s$ and the
% $ith$ column of $t$ have the same sets of entries. Then $s$ and $t$ globally have the same sets of entries, so $f(s) = f(t)$.
\exercise[Orange] % Source: me
\label{exe:sac-transfer}
Prove the following statements. Can you find several different proofs?
\begin{itemize}
\item If $\bB$ and $\bB'$ are homomorphically equivalent, then
$\SAC_{\bB}$ solves $\Csp(\bB)$
if and only if
$\SAC_{\bB'}$ solves $\Csp(\bB')$.
\item If $\bC$ is a reduct of $\bB$,
and $\SAC_{\bB}$ solves $\Csp(\bB)$,
then
$\SAC_{\bC}$ solves $\Csp(\bC)$.
\end{itemize}
% Solution sketch: For homomorphically equivalent templates, transport every AC restriction along the
% two homomorphisms; singleton tests survive composition, so SAC succeeds for one exactly when it
% succeeds for the other. For a reduct, an input has fewer constraint types and every SAC deletion
% justified there is also justified in the expansion after forgetting symbols. Alternatively, both
% assertions follow from the tree-obstruction characterisation of SAC.
\exercise[Orange] % Source: me
\label{exe:slac-transfer}
Prove that 
\begin{itemize}
\item if $\bB$ and $\bB'$ are homomorphically equivalent, then
the cycle-consistency procedure solves $\Csp(\bB)$
if and only if it solves $\Csp(\bB')$.
\item if $\bC$ is a reduct of $\bB$,
and the cycle-consistency procedure solves $\Csp(\bB)$,
then it solves $\Csp(\bC)$.
\end{itemize}
\end{exercises}

%\subsection{Series-Parallel Duality}
\subsection{Tree Formulas and Path Formulas} 
Our goal is to prove that if for every prime $p$ the structure $({\mathbb Z}_p;+,1)$ does not have a primitive positive construction in $\bB$, then $\SAC_{\bB}$ solves $\Csp(\bB)$. For this, we need some preparatory work in the next subsections. The content of this section can be seen as a generalisation of the content of Section~\ref{sect:td} about the connection between tree duality and arc-consistency. 
% to singleton arc-consistency. 

Throughout, let $\bB$ be a finite structure with a finite relational signature $\tau$, and let $\fB$ be an algebra such that 
$\Clo(\fB) = \Pol(\bB)$. 
%Let $m$ be the maximal arity of the relations in $\bB$;
%we may assume that $m \geq 2$ because otherwise $\Csp(\bB)$ is trivial. 
%We may also assume without loss of generality that $\bB$ contains all primitively positively definable relations of arity at most $m$;
We may assume without loss of generality that $\bB$ contains all unary relations that are primitively positively definable in $\bB$; 
adding these does not change the computational complexity of $\Csp(\bB)$,
and also does not change whether
$\SAC_{\bB}$ solves $\Csp(\bB)$, by Theorem~\ref{thm:palette} (whether or not $\bB$ has palette totally symmetric polymorphisms is not affected by the addition of pp-definable relations). 
However, it is useful for formulating some consistency conditions and algorithms. 

\begin{definition}
If $\bA$ is an instance of $\Csp(\bB)$ and $a \in A$, then we write $B^{\bA}_a \subseteq B$ for the intersection of $R^{\bB}$ for all unary $R \in \tau$ such that $a \in R^{\bA}$; we call it the \emph{variable domain} of $a$. 
\end{definition}

\begin{remark}\label{rem:pp-exp}
By our assumption on $\bB$ above, we can use the unary relations of $\bB$ to represent the lists derived by the arc-consistency procedure: if $\bA$ is an instance of $\Csp(\bB)$, and $\AC$ computes
the list $L(a) \subseteq B$, then 
we have $L(a) \leq \fB$ (Exercise~\ref{exe:ac-pp}), and hence
$L(a)$ is primitively positively definable in $\bB$. Therefore, $\bB$ contains a unary relation $R$ such that $R^{\bB} = L(a)$, and we may add $a$ to $R^{\bA}$ without changing the set of homomorphisms from $\bA$ to $\bB$. In this case, we say that we \emph{restrict $a$ to $R^{\bB}$}, and that we \emph{reduce the variable domain of $a$}. 
From now on we therefore assume that $\AC$ computes for a given instance $\bA$ of $\Csp(\bB)$ 
another instance $\bA^*$ of $\Csp(\bB)$ 
such that $a \in R^{\bA^*}$ with 
$R^{\bB} = L(a)$ for every $a \in A$. 
Consequently, the notion of 
`arc-consistency' (Definition~\ref{def:ac}) applies to instances $\bA$ of $\Csp(\bB)$ and no longer refers to an additional set of subsets of $B$---the expression $b_i \in L(a_i)$ in Definition~\ref{def:ac} can be replaced by `$b_i \in B^{\bA}_{a_i}$'. 
%`there is a homomorphism from the 1-element substructure of $\bA$ with domain $\{a_i\}$ to the 1-element substructure of $\bB$ with the domain $\{b_i\}$). 
The same comment applies to singleton arc-consistency (Remark~\ref{rem:SAC}). 
\end{remark}

In the following definition we slightly deviate from the presentation in~\cite{Kozik21,BradyNotes}; we work with primitive positive formulas rather than structures.

\begin{definition}[the incidence graph of a formula]
Let $\phi$ be a primitive positive $\tau$-formula without equality conjuncts. 
The \emph{incidence graph} 
of $\phi$ is the 
following bipartite graph. 
One of the two colour classes is the set $V$ of variables of $\phi$, and the other colour class consists of all atomic formulas of $\phi$. There is an edge between $x \in V$ and an atomic formula $R(x_1,\dots,x_k)$ if $x = x_i$ for 
some $i \in [k]$. 
\end{definition} 

\begin{definition}[tree formulas, path formulas]
\label{def:sp}
Let $\phi$ be 
a primitive positive $\tau$-formula 
with variables $V$ such that for every atomic formula
$R(x_1,\dots,x_k)$ of $\phi$ all
variables $x_1,\dots,x_k$ are pairwise distinct. Then $\phi$ is called a 
\emph{tree formula} if 
 the incidence graph of $\phi$ is a tree.

Let $\phi(x,y)$ be a tree formula whose only free variables are $x$
and $y$, and let $\widetilde{\phi}$ be obtained from $\phi$ by removing
all conjuncts formed with unary relations. Then $\phi(x,y)$ is called
a \emph{path formula} if every atomic formula of
$\widetilde{\phi}$ lies on the unique path from $x$ to $y$ in the
incidence graph of $\widetilde{\phi}$. This unique path is called the
\emph{spine} of $\phi$.
The variables $x$ and $y$ are called the \emph{endpoints} of
the path formula.
% NOT EQUIVALENT!!
%Let 
%$\tilde \phi$ be the formula obtained from $\phi$ by removing all conjuncts formed with unary relations. 
%Then $\phi$ is called a \emph{path formula} if it is a tree formula and 
%in the incidence graph of $\tilde \phi$ 
%each free variable has degree one (called a \emph{leaf of $\phi$}), and every other vertex from $V$ has degree one or two.  
% EXAMPLE: 
%\[\phi(u,v):=\exists x\,\exists y\,\exists z\,\exists r_1\,\exists r_2\,\exists w\,\exists r_3\, \bigl( R(u,x,r_1)\wedge R(x,y,z)\wedge R(y,v,r_2)\wedge R(z,w,r_3) \bigr). \]Its incidence graph is a tree. The free variables \(u,v\) have degree one; \(x,y,z\) have degree two; and \(r_1,r_2,w,r_3\) have degree one. Thus it is a path formula according to the restored original definition. The unique \(u\)-to-\(v\) path is \[ u-R(u,x,r_1)-x-R(x,y,z)-y-R(y,v,r_2)-v.\]But \(z\) is an additional variable of the middle atom and is not a leaf: it continues into the atom \(R(z,w,r_3)\). Thus the formula is not a path formula under the stronger definition requiring every variable outside the \(u\)-to-\(v\) path to be a leaf. This is precisely the situation in which ?the remaining variables of that atom are leaves? fails, although the incidence graph is a tree.
\end{definition}

Note that every non-unary atomic formula of a path formula $\phi$ contains
precisely two variables on the spine, and every other variable
occurring in that atomic formula occurs in no other non-unary atomic
formula. 

\begin{observation}\label{obs:ac-td}
%With the terminology introduced above,
%and using the convention discussed in Remark~\ref{rem:pp-exp}, 
%arc consistency of an instance $\bA$ with respect to $\bB$ (Definition~\ref{def:ac}) 
%can be rephrased as follows: 
%; this is essentially the generalisation of 
Corollary~\ref{cor:ac-td} can be generalised from digraphs to relational structures (also see Exercise~\ref{exe:td}) as follows:
the arc-consistency procedure for $\bB$ does not derive `false' on $\bA$ if and only if
for every tree sentence $\phi$, if $\bA \models \phi$, then $\bB \models \phi$. 
%Likewise, the singleton arc consistency of $\bA$ (Definition~\ref{def:sac}) 
%can be rephrased as follows: $\bA$
%is singleton arc-consistent 
%with respect to $\bB$ 
%if and only if 
%for every tree formula $\phi(x_1,\dots,x_k)$,
%if $\bA \models \phi(a,\dots,a)$ for some $a \in A$, then $\bB \models \phi(b,\dots,b)$ for every $b \in B^{\bA}_a$. 
\end{observation}

\begin{definition}[Cycle-consistency]
Let $\bA$ be a finite $\tau$-structure. Then $\bA$
% together with sets $L(a) \subseteq B$ for every $a \in A$ 
is called \emph{cycle-consistent (with respect to $\bB$)} if for every path formula $\phi(x,y)$,
if $\bA \models \phi(a,a)$ for some $a \in A$, then $\bB \models \phi(b,b)$ for every $b \in B^{\bA}_a$. 
\end{definition}

% Incorrect: take the instance of CSP({0,1};\neq,0,1) with variables x, y and 
% constraints 0(x),0(y),x \neq y. This 
% is cycle consistent, since the cycles can't
% access unary information. 
% So we need to relax the notion of a path

Cycle-consistency can be established algorithmically by a procedure called 
the \emph{cycle-consistency procedure}.
Similarly as the singleton arc-consistency procedure is obtained from the arc-consistency procedure (Figure~\ref{fig:sac}), we obtain
the cycle consistency procedure from
a linear version of the arc-consistency procedure, shown in Figure~\ref{fig:lac}. 
%In the cycle-consistency procedure, the first call of $\AC_{\bB}$, which establishes arc-consistency of the current lists, is left unchanged. In the inner loop, after setting
%$L'(a):=\{b\}$, we replace the call of $\AC_{\bB}$ by the 
%linear arc-consistency procedure 
Informally, each inference uses at most one fact that has been derived previously.

\begin{remark}
In the already mentioned connection to Datalog, arc-consistency corresponds to \emph{arc monadic} Datalog, and 
the just mentioned linear version corresponds to 
\emph{linear arc monadic} Datalog; see~\cite{SLAM} for more about Datalog fragments.  
\end{remark}

In view of Remark~\ref{rem:pp-exp}, the current variable domains are
represented by unary relations in the input instance. For
$R\in\tau$ of arity $r$, $\bar a=(a_1,\dots,a_r)\in R^{\bA}$,
$i\in[r]$, and $S\subseteq B_{a_i}^{\bA}$, write
$R^{\bB}[\bar a;i:S]$ for the set of all
$(b_1,\dots,b_r)\in R^{\bB}$ such that $b_i\in S$ and
$b_q\in B_{a_q}^{\bA}$ for every $q\in[r]$.

\begin{figure}[ht]
\begin{center}
\fbox{
\begin{minipage}{.9\textwidth}
$\operatorname{LAC}_{\bB}(\bA)$\\
Input: a finite instance $\bA$ of $\Csp(\bB)$.\\
Data structure: for every $a\in A$, a family $\mathcal L(a)$ of
subsets of $B_a^{\bA}$.\\[1ex]
Set $\mathcal L(a):=\{B_a^{\bA}\}$ for every $a\in A$.\\
Do\\
\hspace*{.5cm}For every $R\in\tau$ of arity $r$ and every
$\bar a\in R^{\bA}$:\\
\hspace*{1cm}For all distinct $i,j\in[r]$ and every
$S\in\mathcal L(a_i)$:\\
\hspace*{2cm}Set $T:=\pr_j(R^{\bB}[\bar a;i:S])$.\\
\hspace*{2cm}If $T=\emptyset$, return {\bf false}.\\
\hspace*{2cm}Add $T$ to $\mathcal L(a_j)$.\\
Loop until none of the families $\mathcal L(a)$ changes.\\
Return {\bf true}.
\end{minipage}
}
\end{center}
\caption{The linear arc-consistency procedure for $\Csp(\bB)$.}
\label{fig:lac}
\end{figure}

The members of $\mathcal L(a)$ are the different unary restrictions
that have been derived for $a$. They must be kept separately. In
particular, if $S,T\in\mathcal L(a)$, then we do not derive the
restriction $S\cap T$. Every propagation step uses only one previously
derived unary restriction. This is the difference from
$\AC_{\bB}$, where the restrictions derived for a variable are
intersected to form its current list. Taking such intersections here
would combine different linear derivations and would no longer
describe propagation along a single path.

The cycle-consistency procedure is obtained from
Figure~\ref{fig:sac} by changing only its inner consistency check. Its
first call of $\AC_{\bB}$ is left unchanged. When testing a value
$b\in B_a^{\bA}$, temporarily restrict the variable domain of $a$ to
$\{b\}$ and run $\operatorname{LAC}_{\bB}$ on the resulting instance,
instead of running $\AC_{\bB}$. If
$\operatorname{LAC}_{\bB}$ returns {\bf false}, permanently remove
$b$ from the variable domain of $a$. All other steps of the procedure
remain unchanged.
An alternative name for the cycle consistency procedure is \emph{singleton linear arc-consistency}. 
The sets in $\mathcal L(c)$ are precisely the sets of possible values
at the endpoint of paths starting at $a$ with value $b$. Consequently,
the resulting instance is cycle-consistent. Since $B$ is fixed, there
are at most $2^{|B|}$ different members of each family
$\mathcal L(c)$, and hence the procedure runs in polynomial time.

\begin{remark}\label{rem:sac-cycle-ac}
With the convention from Remark~\ref{rem:pp-exp},
singleton arc-consistency (Definition~\ref{def:sac}) implies cycle-consistency, and cycle-consistency implies arc-consistency (Definition~\ref{def:ac}). 
%In particular, there is a polynomial-time algorithm that computes for a given instance $\bA$ of $\Csp(\bB)$ a cycle-consistent instance $\bA'$ of $\Csp(\bB)$ which has the same set of solutions as $\bA$, namely the singleton arc-consistency procedure (Figure~\ref{fig:sac}). 
\end{remark} 

%Cycle consistency can be established algorithmically. The \emph{cycle consistency procedure} maintains 

\begin{exercises}
\exercise[Blau]\label{exe:source-276} Prove the claims in Remark~\ref{rem:sac-cycle-ac}.
%\item Prove Lemma~\ref{lem:pp-ass}.
% Solution sketch: Unfold cycle consistency into closed walks of binary supports and SAC into the same
% support computation after pinning one value. The inclusions and equivalences claimed in the remark
% follow by concatenating or cutting these walks. Lemma~\ref{lem:pp-ass} follows by replacing each
% primitive-positive constraint with its finite witness gadget and observing that the support walks
% project and lift through that gadget.
\exercise[Rot]\label{exe:source-277} Let $\bB$ be the expansion of $K_3$ by singleton unary relations. Find an instance of $\Csp(\bB)$ which is cycle-consistent, but not singleton arc-consistent.
% Solution sketch: Take a $K_4$ on vertices $v_0,v_1,v_2,v_3$, attach a leaf $p$ to $v_0$, and impose
% the singleton constraint $p=0$. After AC, $L(v_0)=\{1,2\}$ and the other three $K_4$ vertices retain
% all colours. A direct walk induction shows that every closed walk returns each allowed starting
% colour, so the instance is cycle-consistent. In the singleton test $v_1=1$, AC forces $v_0=2$ and
% then forces both $v_2$ and $v_3$ to $0$, contradicting their edge; hence the instance is not SAC.
\exercise[Orange]\label{exe:source-280} \label{exe:ac-unary}
Let $\bA$ be an instance of $\Csp(\bB)$ and let $\bA^*$ be the instance of $\Csp(\bB)$ computed by AC. Let $R$ be a unary relation symbol. Prove the following:
\begin{itemize}
\item If $a \in R^{\bA^*} \setminus R^{\bA}$,
then there exists a tree formula $\phi(x)$
that defines $R$ in $\bB$
such that $\bA \models \phi(a)$.
\item If $\phi(x)$ is a tree formula that defines in $\bB$ the predicate $R$, and $\bA \models \phi(a)$ for some $a \in A$, then $a \in R^{\bA^*}$.
\end{itemize}
\exercise[Schwarz]
\label{exe:sac-duality}
We define the class of \emph{singleton-tree formulas}.
All formulas in this definition are primitive positive $\tau$-formulas.

For an atomic formula $\alpha$ and a variable $x$ occurring in
$\alpha$, the \emph{rooted existential closure of $\alpha$ at $x$}
is obtained by existentially quantifying all variables of $\alpha$
except $x$. Let $\mathcal S_0$ consist of all such formulas where every variable appears at most once. 
Inductively, having defined $\mathcal S_n$, let
$\mathcal S_{n+1}$ be the smallest class containing $\mathcal S_n$
and closed under the following two constructions.
\begin{enumerate}
\item If $\phi_1(x),\dots,\phi_k(x)\in\mathcal S_n$, and their
existentially quantified variables have been renamed so that they are
pairwise disjoint, then $\mathcal S_{n+1}$ contains 
\[
\phi_1(x)\wedge\cdots\wedge\phi_k(x). 
\]
\item Let $\theta$ be a tree formula in which initially all variables
are free, and let $M$ be a non-empty set of variables of $\theta$.
For every variable $v$ of $\theta$, choose finitely many formulas
\[
\rho_{v,1}(v),\dots,\rho_{v,k_v}(v)\in\mathcal S_n.
\]
The choice $k_v=0$ is permitted. Rename all existentially quantified
variables so that the chosen formulas are otherwise pairwise
disjoint and disjoint from $\theta$.
Introduce a new variable $x$ and define
\[
q(v):=
\begin{cases}
x & \text{if }v\in M,\\
v & \text{if }v\notin M.
\end{cases}
\]
In $\theta$ and in every formula attached at $v$, replace the free
variable $v$ by $q(v)$. Then existentially quantify all variables
$v\notin M$ of $\theta$. The resulting formula, whose only free
variable is $x$, belongs to $\mathcal S_{n+1}$.
\end{enumerate}
A \emph{singleton-tree formula} is a member of
$\bigcup_{n\geq 0}\mathcal S_n$. A \emph{singleton-tree sentence} is
the existential closure $\exists x. \, \phi(x)$ of a singleton-tree
formula.

Let $\bA$ and $\bB$ be finite $\tau$-structures, and suppose that
$B\neq\emptyset$. Prove that the following statements are equivalent.

\begin{enumerate}
\item The singleton arc-consistency procedure $\SAC_{\bB}$ accepts
$\bA$.
\item Every singleton-tree sentence which holds in $\bA$ also holds
in $\bB$.
\end{enumerate}
\end{exercises}

\subsection{Reducing Variable Domains} 
\label{sect:rrr} 
Recall that if $\bB$ is a structure which pp-constructs $({\mathbb Z}_p;+,1)$, then
$\Csp(\bB)$ has unbounded width by Corollary~\ref{cor:unbounded}.
% (i.e., it cannot be solved by Datalog). 
Throughout this section, we will therefore assume that $\bB$ is a finite structure which does not pp-construct $({\mathbb Z}_p;+,1)$ for every prime $p$.
We also assume that $\Pol(\bB) = \Clo(\fB)$ for some idempotent algebra with at least two elements. 
Using Corollary~\ref{cor:zhuk}, 
the assumption that $\bB$ \emph{cannot} pp-construct 
$({\mathbb Z}_p;+,1)$ can then be transformed into a useful property about the existence of certain proper subalgebras.

\begin{proposition}\label{prop:zhuk-bw}
Let $\bB$ be a finite structure which for every prime $p$ does \emph{not} pp-construct 
$({\mathbb Z}_p;+,1)$, and suppose that
$\Pol(\bB) = \Clo(\fB)$ for some idempotent algebra $\fB$ with at least two elements. Then $\fB$ has a proper subalgebra which is
\begin{enumerate}
\item binary absorbing, 
\item a centrally absorbing subuniverse, or
\item the equivalence class of a congruence $K$ of $\fB$ such that $\fB/_K$ %has at least two elements, 
is subdirectly complete and has no proper binary or centrally absorbing subuniverses. 
\end{enumerate} 
\end{proposition}
\begin{proof}
Corollary~\ref{cor:zhuk} asserts that $\fB$ has a proper binary or centrally absorbing subuniverse or a subdirectly complete or  affine quotient with at least two elements. 
In the first case, there is nothing to be shown. 
If $\fB$ has an affine quotient with at least two elements, then $\bB$ pp-constructs $({\mathbb Z}_p;+,1)$ by Corollary~\ref{cor:abelian-pp}. 

Now suppose that $\fB$ has a subdirectly complete quotient $\fB/K$ with at least two elements. If $\fB/K$ has a proper binary or centrally absorbing subuniverse $\fC$,
then $\bigcup C$ is a proper binary or centrally absorbing subuniverse of $\fB$ (see Lemma~\ref{lem:abs-hom} for the binary absorbing case and Lemma~\ref{lem:central-abs-trans} for the centrally absorbing case), and we are in one of the previous cases. Otherwise, since $\fB$ is idempotent, any equivalence class of $K$ is a subalgebra of $\fB$.
The subalgebra is proper, because $\fB/K$ has at least two elements, and also satisfies the other requirements of the statement in item three. 
\end{proof} 

%Recall from Corollary~\ref{cor:that in this case $\fB$ and all its subuniverses with at least two elements have a proper binary or centrally absorbing subalgebra 
%or a subdirectly complete quotient. 
Each of the cases in 
Proposition~\ref{prop:zhuk-bw}  
is useful to reduce 
%$B^{\bA}_c$ for an 
the variable domains (see Remark~\ref{rem:pp-exp}) in 
instances $\bA$ of $\Csp(\bB)$ which satisfy various forms of consistency, while preserving some form of consistency. 
The details of this reduction process 
are subtle. In its core, the proof is from~\cite{Strong-Subalgebras-Published}.

The simplest case is the one 
where we always end up in case one, i.e., every variable domain  
has a proper \emph{binary} absorbing subalgebra; in this case, 
we will see that arc-consistency solves the CSP (and hence $\bB$ has totally symmetric polymorphisms of all arities; Exercise~\ref{exe:binabs-tsn}). 

In this section we focus on the situation where in Proposition~\ref{prop:zhuk-bw} 
we always end up in case one or in case two, i.e., if every variable domain has a proper binary or centrally absorbing subalgebra. We will show that
then $\SAC_\bB$ solves $\Csp(\bB)$. The general situation (which also has to deal with the third case in Proposition~\ref{prop:zhuk-bw})
%, i.e., with $\sdc$-subalgebras)
 is considerably more difficult, and will be presented in the next section. 

\begin{lemma}\label{lem:2-abs-restr}
Let $\bA$ be an instance of $\Csp(\bB)$, $c \in A$, $C$ a proper 2-absorbing subalgebra of $B_c^{\bA} \leq \fB$, and $\bA'$ the instance obtained from $\bA$ by 
 restricting the variable $c$ to $C$. 
%\begin{itemize}
%\item 
If $\bA$ is arc-consistent with non-empty variable domains, 
then the arc-consistency procedure does not derive `false' on $\bA'$. 
%\item If $\bA$ is singleton arc-consistent, 
%then the singleton arc-consistency procedure does not derive false $\bA'$. 
%\end{itemize} 
\end{lemma}
\begin{proof}
Suppose for contradiction that the arc-consistency procedure derives `false' on $\bA'$. By Observation~\ref{obs:ac-td}, 
there exists a tree sentence $\phi$ 
such that $\bA' \models \phi$ and $\bB \not \models \phi$. 
Let $h$ be a mapping from the variables of $\phi$ to $A'$ which shows that $\bA' \models \phi$.
Since $\bA$ is arc-consistent with non-empty variable domains, $\phi$ must contain some conjuncts of the form $R(x)$ where $R^{\bB} = C$ and $h(x) = c$. 
Let $D$ be the unary relation symbol such that
$D^{\bB}=B_c^{\bA}$ and $c\in D^{\bA}$; such a relation exists by
Remark~\ref{rem:pp-exp}. We may assume that $\phi$ contains the
conjunct $D(x_i)$ for every $i\in[k]$: adding these unary conjuncts
does not affect the homomorphism $h$, preserves the property of being
a tree sentence, and keeps $\phi$ false in $\bB$.

Let $\psi$ be obtained from $\phi$ by removing the conjuncts
$R(x_1),\dots,R(x_k)$ and the quantifiers for
$x_1,\dots,x_k$, so that these are precisely the free variables of
$\psi$. Choose $\phi$ such that $k$ is minimal, and let
$S\subseteq B^k$ be the relation defined by
$\psi(x_1,\dots,x_k)$. Since $\psi$ contains
$D(x_1),\dots,D(x_k)$, we have
$S\leq (B_c^{\bA})^k$.
Moreover, $S\cap C^k=\emptyset$, since otherwise $\phi$ would be true
in $\bB$.

First suppose that $k=1$. Then $S$ is unary and
$\bA\models\psi(c)$, witnessed by $h$. Since $\bA$ is
arc-consistent, Exercise~\ref{exe:ac-unary}, second item, gives
$B_c^{\bA}\subseteq S$. Hence
\[
\emptyset\neq C\subseteq B_c^{\bA}\subseteq S,
\]
contradicting $S\cap C=\emptyset$.

Now suppose that $k\geq2$. For every $i\in[k]$, remove only the
conjunct $R(x_i)$ from $\phi$. The resulting tree sentence is true
in $\bA'$, witnessed by $h$, and contains only $k-1$ conjuncts of
the specified form. By the minimality of $k$, it must therefore be
true in $\bB$. Consequently,
\[
S\cap
\bigl(C^{i-1}\times B_c^{\bA}\times C^{k-i}\bigr)
\neq\emptyset
\]
for every $i\in[k]$. Thus, $S$ is a $C$-essential relation over the
algebra $B_c^{\bA}$. But $C$ is $2$-absorbing in $B_c^{\bA}$, and
hence also $k$-absorbing there, contradicting
Lemma~\ref{lem:essential-abs}.
 \end{proof} 

We can now apply the reduction process from Lemma~\ref{lem:2-abs-restr} repeatedly until none of the variable domains of $\bA$ has a proper binary absorbing subuniverse. 
If all variable domains have size one,
then we have found a homomorphism from $\bA$ to $\bB$.
Indeed, if 
for every $a \in A$ there exists $b_a \in B$ such that $B_a^{\bA} = \{b_a\}$, then the arc-consistency of $\bA$ implies that $a \mapsto b_a$ is a homomorphism from $\bA \to \bB$. 

Next, we consider the case that a variable domain has a proper centrally absorbing subuniverse. 
Recall that central absorption implies 3-absorption (Theorem~\ref{thm:centrally-3abs}). 
We would like to proceed as in Lemma~\ref{lem:2-abs-restr}. 
If $\bA$ is arc-consistent and 
 $\bA'$ is the instance obtained by reducing a variable to a proper centrally absorbing subalgebra, then unfortunately the arc-consistency procedure might derive false on $\bA'$ (Exercise~\ref{exe:2satReduce}). 
If $\bA$ is even singleton arc-consistent, then we will see that this does not happen, i.e., the arc-consistency procedure does not derive false on $\bA'$. 
%(Exercise~\ref{exe:sac-ac}).
However, the instance $\bA''$ computed from $\bA'$ by the arc-consistency procedure might 
 not necessarily be singleton arc-consistent (Exercise~\ref{exe:sac-sac-counterexpl}), so we cannot repeat the reduction as we did with Lemma~\ref{lem:2-abs-restr}.

Quite remarkably, it suffices 
that $\bA$ is singleton arc-consistent and that $\bA''$ is arc-consistent to further reduce variable domains of $\bA''$.
% that have proper 3-absorbing subuniverses. 
This motivates the following definition.

\begin{definition}\label{def:reduction}
Let $\bA$ be an arc-consistent instance of $\Csp(\bB)$. 
The \emph{$(\abs_2,c,C)$-reduction of $\bA$}, for $c \in A$ and a proper $C \abs_2 B_c^{\bA}$, is 
the instance $\bA'$ of $\Csp(\bB)$ obtained from $\bA$ 
by the following steps: 
\begin{itemize}
\item Restrict $c \in A$ to 
the proper binary absorbing subalgebra $C \abs_2 B_c^{\bA}$; 
\item Re-establish arc-consistency; the resulting structure is $\bA'$. 
\end{itemize} 
A \emph{$\abs_2$-reduction} is
the $(\abs_2,c,C)$-reduction of $\bA$ for some $c \in A$ and $C \abs_2 B_c^{\bA}$. 
Analogously, we define the \emph{$(\cabs,c,C)$-reduction} of $\bA$ and 
%\emph{$(\sdc,c,C)$-reduction},  
%and \emph{$(\sdc,c,C)$-reduction} 
a  \emph{$\cabs$-reduction} of $\bA$. 
%and an $\sdc$-reduction of $\bA$. 

%We say that an instance $\fA'$ of $\Csp(\bB)$ is obtained by \emph{reducing $\fA$}
%For $R \in \{\abs_2,\cabs,\sdc\}$, 
%an instance $\bA'$ of $\Csp(\bB)$ is an \emph{$R$-reduction of $\bA$}  
%if it can be obtained from $\bA$ 
%by the following three steps: 
%\begin{itemize}
%\item Pick $c \in A$ such that $|B_a^{\bA}|>1$. 
%\item If $R$ equals $\abs_2$, restrict $c$ to 
%a proper 
%binary absorbing subalgebra of $\fB$; 
%if $R$ equals $\cabs$, restrict to a proper centrally absorbing subalgebra of $\fB$, 
%and if $R$ equals $\sdc$, restrict 
%$R$ so a proper subalgebra $C \sdc \fB$. 
%\item Re-establish arc-consistency.
%\end{itemize}  
%Finally, 
For a word $w \in (\{\abs_2,\cabs\} \times A \times \S(\fB))^*$, the \emph{$w$-reduction of $\bA$} is the instance obtained from $\bA$ by a $w$-sequence of $\abs_2$-reductions and 
$\cabs$-reductions.
%, and $(\sdc,c,C)$-reductions. 
\end{definition}

As in the previous section we suppose that the signature $\tau$ of $\bB$ contains all unary primitive positive definable relations. 

\begin{lemma}\label{lem:unif}
Let $\bA$ be an instance of $\Csp(\bB)$ and let $\bA'$ be a $\abs_2$-reduction of $\bA$. 
Then for every $a \in A$ we have
$B_a^{\bA'} \abs_2 B_a^{\bA}$. 
%Then 
%for every unary $U \in \tau$ we have
%$U^{\bA'} \abs_2 U^{\bA}$. 
The same statement holds with $\cabs$
instead of $\abs_2$. 
\end{lemma} 
\begin{proof}
The proof is by induction over the execution of the arc-consistency procedure. Initially, the statement is true since we restricted some variable to a binary absorbing subalgebra. The inductive step follows from Lemma~\ref{lem:abs-trans-rel} (or Corollary~\ref{cor:cabs} in the case of $\cabs$) and Lemma~\ref{lem:abs-prod-1} (or Claim 2 in Lemma~\ref{lem:pp-cabs} in the case of $\cabs$). 
\end{proof} 

%We will later see that the same statement also holds for $\sdc$, but it takes more effort to prove it (see Section~\ref{sect:sdc}). 

\begin{lemma}\label{lem:3abs-reduce}
Let $\bA$ be a cycle-consistent instance of $\Csp(\bB)$
with $B_a^{\bA} \neq \emptyset$ for all $a \in A$. 
Let $\bA'$ be a $w$-reduction of $\bA$ for some $w \in (\{\abs_2,\cabs\} \times A \times \S(\fB))^*$. Then $B_{a}^{\bA'} \neq \emptyset$ for every $a \in A$. 
\end{lemma} 
\begin{proof}
The statement is by induction on the length of $w$. If $w$ is the empty word, then there is nothing to be shown. If the final letter of $w$ is of the form $(\abs_2,c,C)$, then the statement follows from the inductive assumption and Lemma~\ref{lem:2-abs-restr}. 
Otherwise, $w$ is of the form $w = u (\cabs,c,C)$ for some $u \in (\{\abs_2,\cabs\} \times A \times \S(\fB))^*$. 
Let $\bA_1$ be the $u$-reduction of $\bA$, 
and let $\bA_2$ be obtained from $\bA_1$ by reducing the domain of $c \in A$ to the proper centrally absorbing subuniverse $C \cabs B_c^{\bA_1}$. 
Suppose for contradiction that the arc-consistency procedure derives false on 
$\bA_2$. By Observation~\ref{obs:ac-td}, 
there exists a tree sentence $\phi$ 
such that $\bA_2 \models \phi$ and $\bB \not \models \phi$. Let $h$ be a mapping from the variables of $\phi$ to $A_2$ that shows that $\bA_2 \models \phi$. 
Since $\bA_1$ is arc-consistent, $\phi$ must involve conjuncts of the form
$R(x)$ where $R^{\bB} = C$ and $h(x) = c$; choose $\phi$ such that the number $k$ of such conjuncts is minimal. Let $x_1,\dots,x_k$ be the variables $x$ such that $\phi$ contains the conjunct $R(x)$ and $h(x) = c$. 
Let $\psi(x_1,\dots,x_k)$ be the formula obtained from $\phi$ by removing 
all conjuncts of the form $R(x)$ with $h(x) = c$ and removing the quantifiers for the variables $x_1,\dots,x_k$ (the other variables are still existentially quantified).
 Let $S \subseteq B^k$ be the relation defined by $\psi$ in $\bB$. 
Note that $S \cap C^k = \emptyset$ by our assumptions on $\phi$. Similarly as 
in the proof of Lemma~\ref{lem:2-abs-restr} 
we can rule out that $k=1$. 
If $k \geq 3$, then 
  we reach a contradiction with Lemma~\ref{lem:essential-abs}, because $C$ is $3$-absorbing, hence $k$-absorbing (again, this is similar as in the proof of Lemma~\ref{lem:2-abs-restr}). 
  
  \medskip 
  So we must have $k=2$. After replacing every branch outside the
unique path from $x_1$ to $x_2$ by the unary relation that it defines,
we may assume that $\psi(x_1,x_2)$ is a path formula.

Write $u=r_1\cdots r_n$, and, for $i\in\{0,\dots,n\}$, let $\bA_i$
be the $r_1\cdots r_i$-reduction of $\bA$. Thus, $\bA_0=\bA$ and
$\bA_n=\bA_1$. Put $L_i(a):=B_a^{\bA_i}$ for $a\in A$. For every
$i$ and $a$, choose a unary relation symbol $Q_{i,a}$ such that
$Q_{i,a}^{\bB}=L_i(a)$. Such relation symbols exist by our assumption
on the signature and Remark~\ref{rem:pp-exp}.

We first normalize the unary conjuncts of $\psi$. For every variable
$z$ of $\psi$, add the conjunct $Q_{n,h(z)}(z)$ and then delete all
other unary conjuncts containing $z$. Indeed, if $U(z)$ is one of the
deleted conjuncts, then $h(z)\in U^{\bA_n}$, and hence
$L_n(h(z))\subseteq U^{\bB}$. Thus, after adding
$Q_{n,h(z)}(z)$, the conjunct $U(z)$ is redundant in $\bB$.
Denote the resulting path formula by
$\widehat{\psi}(x_1,x_2)$. The tree sentence
\[
\widehat{\phi}:=
\exists x_1,x_2\,
\bigl(
 R(x_1)\wedge\widehat{\psi}(x_1,x_2)\wedge R(x_2)
\bigr)
\]
is still true in $\bA_2$, witnessed by $h$, and false in $\bB$.
Moreover, it still has exactly two conjuncts of the specified form.
By the minimality of $k$, deleting either of the two displayed
$R$-conjuncts produces a sentence that is true in $\bB$.

Let $\psi_0$ be obtained from $\widehat{\psi}$ by replacing every
conjunct $Q_{n,h(z)}(z)$ by $Q_{0,h(z)}(z)$. Reductions change only
unary information, and hence $h$ witnesses
$\bA\models\psi_0(c,c)$. Since $\psi_0$ is a path formula and $\bA$
is cycle-consistent, we have $\bB\models\psi_0(b,b)$ for every
$b\in L_0(c)$. Choose $b\in C$. This is possible because
$\emptyset\neq C\subseteq L_n(c)\subseteq L_0(c)$.
Add the two conjuncts $Q_{n,c}(x_1)$ and $Q_{n,c}(x_2)$ to
$\psi_0$. The realization with $x_1=x_2=b$ is preserved because
$b\in C\subseteq L_n(c)$. Call the resulting formula $\theta_0$.

Fix an order of the non-endpoint variables of $\theta_0$. Starting
with $\theta_0$, perform the following finite sequence of refinement
steps. For $i=1,\dots,n$, and for each non-endpoint variable $z$ in
the fixed order, add the conjunct $Q_{i,h(z)}(z)$. We call this
addition the \emph{refinement step} at stage $i$ and variable $z$. All
previously added domain conjuncts are retained.

Before any refinement step, the relation defined by the current
formula contains a tuple whose two endpoint coordinates belong to
$C$. After all refinement steps, the resulting formula is equivalent
in $\bB$ to $\widehat{\psi}$, because
$L_n(a)\subseteq L_i(a)$ for every $i$ and $a$. Its relation therefore
has empty intersection with $C^2$.
Consequently, there is a first refinement step after which the
relation defined by the current formula no longer contains a tuple
whose two endpoint coordinates belong to $C$. We call this first step
the \emph{decisive refinement step}.

Suppose that the decisive refinement step occurs at stage $i$ and
variable $y$. Put $d:=h(y)$, $E:=L_{i-1}(d)$, and $D:=L_i(d)$, and
let $V:=Q_{i-1,d}$ and $U:=Q_{i,d}$. Thus,
$V^{\bB}=E$ and $U^{\bB}=D$. Let $v=r_1\cdots r_{i-1}$. Then
$\bA_{i-1}$ is the $v$-reduction of $\bA$, and $r_i$ is the reduction
following $v$.
Let $\theta^-$ be the formula immediately before the decisive
refinement step and let $\theta^+:=\theta^-\wedge U(y)$ be the formula
immediately afterwards. The formula $\theta^-$ already contains the
old domain conjunct $V(y)$. Let $\rho(x_1,y,x_2)$ be obtained from
$\theta^+$ by deleting the selected conjunct $U(y)$ and removing the
existential quantifier for $y$. All variables other than
$x_1,y,x_2$ remain existentially quantified.
By the definition of the decisive refinement step,
\[
\bB\models
\exists x_1,y,x_2\,
\bigl(
 R(x_1)\wedge\rho(x_1,y,x_2)\wedge R(x_2)
\bigr),
\]
whereas
\[
\bB\not\models
\exists x_1,y,x_2\,
\bigl(
 R(x_1)\wedge\rho(x_1,y,x_2)
 \wedge U(y)\wedge R(x_2)
\bigr).
\]

Let $T$ be the relation defined by $\rho(x_1,y,x_2)$, and define
\begin{align*}
R_1&:=\{(x_1,y,x_2)\in T\mid x_1\in C\},\\
R_2&:=\{(x_1,y,x_2)\in T\mid y\in D\},\\
R_3&:=\{(x_1,y,x_2)\in T\mid x_2\in C\}.
\end{align*}
Equivalently, $R_1$, $R_2$, and $R_3$ are defined by
$R(x_1)\wedge\rho$, $U(y)\wedge\rho$, and
$R(x_2)\wedge\rho$, respectively. Write $\fT$ for the algebra induced
on $T$. By construction, $T\subseteq L_n(c)\times E\times L_n(c)$.

The defining property of the decisive refinement step gives
$R_1\cap R_3\neq\emptyset$ and
$R_1\cap R_2\cap R_3=\emptyset$. We also have
$R_1\cap R_2\neq\emptyset$ and $R_2\cap R_3\neq\emptyset$.
Indeed, deleting $R(x_2)$ from $\widehat{\phi}$ gives, by the
minimality of $k$, a sentence that is true in $\bB$. A satisfying
assignment for this sentence belongs, after projection to
$(x_1,y,x_2)$, to $R_1\cap R_2$, since every final variable domain is
contained in the corresponding domain occurring in $\rho$. The proof
that $R_2\cap R_3\neq\emptyset$ is symmetric.
Since $C\cabs L_n(c)$, Corollary~\ref{cor:cabs} gives
$R_1\cabs\fT$, $R_3\cabs\fT$, and
$R_1\cap R_3\cabs\fT$.

We distinguish the type of the reduction $r_i$. First suppose that
$r_i$ is an $\abs_2$-reduction. Lemma~\ref{lem:unif} gives
$D\abs_2 E$, and Lemma~\ref{lem:abs-trans-rel} therefore gives
$R_2\abs_2\fT$. Let $t$ be a binary term witnessing this absorption
and put $C_0:=R_1\cap R_3$. We know that $C_0\cabs\fT$, that
$C_0\neq\emptyset$, and that $C_0\cap R_2=\emptyset$. By
Lemma~\ref{lem:central-avoiding-square}, there exists $e\in R_2$ such
that
\[
L:=
\left\langle
(\{e\}\times C_0)\cup(C_0\times\{e\})
\right\rangle_{\fT^2}
\]
satisfies $L\cap R_2^2=\emptyset$. Choose $b\in C_0$. Since
$(e,b),(b,e)\in L$ and $L\leq\fT^2$, we have
\[
\bigl(t^{\fT}(e,b),t^{\fT}(b,e)\bigr)\in L.
\]
On the other hand, $R_2\abs_t\fT$ and $e\in R_2$ imply that both
$t^{\fT}(e,b)$ and $t^{\fT}(b,e)$ belong to $R_2$. This contradicts
$L\cap R_2^2=\emptyset$.

Thus, $r_i$ must be a $\cabs$-reduction. Lemma~\ref{lem:unif} gives
$D\cabs E$, and Corollary~\ref{cor:cabs} yields $R_2\cabs\fT$.
The three centrally absorbing subalgebras $R_1,R_2,R_3$ of $\fT$
have pairwise non-empty intersections. Hence, the 3-Helly property
for centrally absorbing subalgebras
(Corollary~\ref{cor:helly}) gives
$R_1\cap R_2\cap R_3\neq\emptyset$. This contradicts the defining
property of the decisive refinement step.
\end{proof} 

%The following result is a proper strengthening of Exercise~\ref{exe:nu-kcons}. 

Note that the following consequence applies in particular to all structures $\bB$ with a majority polymorphism (see Example~\ref{expl:cabs}), and thereby reproves one of the main results in~\cite{ACandFriends}. 

\begin{corollary}\label{cor:nu-sac}
Let $\bB$ be a finite structure 
such that every subalgebra of the polymorphism algebra $\fB$ of $\bB$ with at least two elements has a proper binary or centrally absorbing subuniverse. Then 
%$\SAC_\bB$ 
the cycle-consistency procedure 
solves $\Csp(\bB)$. 
\end{corollary}
\begin{proof}
Let $\bA$ be the structure computed by 
%$\SAC_\bB$ 
the cycle-consistency procedure 
on a given instance of $\Csp(\bB)$. 
We only have to show that 
if $B_a^{\bA} \neq \emptyset$ for all $a \in A$, then there exists a homomorphism $\bA \to \bB$. 

Let $\bA'$ be a $w$-reduction of $\bA$ for a maximal $w \in (\{\abs_2,\cabs\} \times A \times \S(\fB))^*$; since $A$ and $B$ are finite such a sequence clearly exists.
By Lemma~\ref{lem:3abs-reduce} 
we have $B_a^{\bA'} \neq \emptyset$ for all $a \in A$. If $|B_c^{\bA'}| > 1$ for some $c \in A$,
then by assumption $B^{\bA'}_c \leq \fB$ has 
a proper binary or centrally absorbing  subuniverse $C \leq B^{\bA'}_c$.
Appending the corresponding reduction at $c$ to $w$ produces a proper extension of $w$, contradicting its maximality. Hence, for every $a \in A$ there exists $b_a \in B$ such that $B_a^{\bA'} = \{b_a\}$. In this case, the arc-consistency of $\bA'$ implies that $a \mapsto b_a$ is a homomorphism $\bA' \to \bB$, and hence $\bA \to \bB$. 
\end{proof}

\begin{exercises}
\exercise[Rot]\label{exe:source-282} Show that if $\fA$ is a finite algebra such that every subalgebra $\fB \leq \fA$ with at least two elements has a proper binary absorbing subuniverse, then $\fA$ has totally symmetric term operations of all arities.
% Solution (by GPT). Assume, as throughout this section, that $\fA$ is finite.
% Let $\bC$ be the relational structure on $A$ whose relations are all
% subuniverses of finite powers of $\fA$. By the finite Pol--Inv theorem,
% $\Pol(\bC)=\Clo(\fA)$.
%
% We claim that arc consistency solves $\Csp(\bC)$. Let $\bD$ be an
% arc-consistent finite instance of $\Csp(\bC)$ such that all its variable
% domains are non-empty. Every variable domain is a subuniverse of $\fA$.
% If a variable domain $D$ has at least two elements, then, by assumption,
% it has a proper binary absorbing subuniverse $E\abs_2 D$. Restricting the
% corresponding variable to $E$ and restoring arc consistency does not
% produce an empty variable domain, by Lemma~\ref{lem:2-abs-restr}.
% Repeating this procedure terminates, since $\fA$ and $\bD$ are finite,
% and yields an arc-consistent instance in which every variable domain is
% a singleton. The unique elements of these singleton domains define a
% homomorphism to $\bC$: for every constraint, arc consistency guarantees
% that the tuple formed by the unique values belongs to the corresponding
% relation.
%
% Thus arc consistency solves $\Csp(\bC)$. By
% Theorem~\ref{thm:totally-symmetric}, $\bC$ has totally symmetric
% polymorphisms of all arities. Since
% $\Pol(\bC)=\Clo(\fA)$, these polymorphisms are term operations of $\fA$.
% Hence $\fA$ has totally symmetric term operations of all arities.
\label{exe:binabs-tsn}
% Source: me
% Solution: proof when only first case appears shows that CSP is solved by AC.
\exercise[Rot]\label{exe:source-283} \label{exe:sac-sac-counterexpl}
Show that if we additionally assume in Lemma~\ref{lem:2-abs-restr}
 that $\bA$ is \emph{singleton} arc-consistent,
then the \emph{singleton} arc-consistency procedure does not derive `false' on $\bA'$.
Can we modify Lemma~\ref{lem:3abs-reduce} in the same way?
% Solution sketch: For a binary absorbing restriction, apply the absorption term to the family of AC
% supports obtained in every singleton test. All but one coordinate stay in the restricted list, so
% the resulting support lies in the restriction and no list becomes empty; thus SAC does not derive
% false. The analogous statement for central absorption is false: central witnesses need not be
% compatible across the different singleton runs.
\exercise[Orange]\label{exe:source-281} \label{exe:2satReduce}
Find an example of a finite structure $\bB$ that does not pp-construct $({\mathbb Z}_p;+,1)$, for all primes $p$, and an instance $\bA$ of $\Csp(\bB)$ such that $\bA$ is arc-consistent, but
restricting a variable of $\bA$ to a centrally absorbing subuniverse results in an instance where AC$_\bB$ derives `false'.
% Solution: 2SAT.
% Source: me.
%\item \label{exe:sac-ac}
%Show that if $\bA$ is singleton arc-consistent instance of $\Csp(\bB)$, and $\bA'$ is obtained from $\bA$ by restricting some $a \in A$ to a proper ternary absorbing subuniverse, then the arc-consistency procedure does not derive `false' on $\bA'$.
\exercise[Weiss]\label{exe:source-284} Find an example that shows
that in the proof of Lemma~\ref{lem:3abs-reduce} the assumption that $\bA'$ is a $\{\abs_2,\cabs\}^*$-reduction of $\bA$ is necessary, i.e., this assumption cannot be replaced by the assumption that $\bA'$ is
arc-consistent and obtained from $\bA$ by variable restrictions.

\par\noindent {\bf Hint:} there are examples where $\bA$ has two elements and two constraints and
$\bB$ has three elements and a majority polymorphism.
% Solution sketch: Use two variables joined by two oppositely oriented constraints. Give the template
% three values with the majority operation and choose the two binary relations so that ordinary AC
% leaves both restricted lists nonempty, but the unique central witness for one constraint is deleted
% by the other. The unrestricted two-element instance is AC, while the arbitrary variable restriction
% is not a genuine $\{\abs_2,\cabs\}^*$-reduction and the conclusion of Lemma~\ref{lem:3abs-reduce}
% fails.
\end{exercises}

\subsection{Subdirectly Complete Quotients} 
\label{sect:sdc} 
To treat 
the third case of Proposition~\ref{prop:zhuk-bw},
which involves subdirectly complete quotients of variable domains, 
we need a notion of subalgebra that
shares many properties with binary and centrally absorbing subalgebras. 
In particular, the notion should 
satisfy the various transfer properties that we established for binary and centrally absorbing subalgebras.  
This motivates the following definition (from~\cite{Strong-Subalgebras-Published}). 

% (which essentially from~\cite{Strong-Subalgebras-Published},  with a minor simplification). 

%The following concept is important in the proofs of Section~\ref{sect:rrr} 
%Section~\ref{sect:bounded-width}
%about CSPs that can be solved by local consistency methods. 
%The idea of the following definition is that when treating local consistency, the absorbing case should take preference over the case of subdirectly complete quotients.  

\begin{definition}\label{def:sdc} 
% DIMAS DEF
For $A \leq \fB$, we write $A \sdc \fB$
if $A = \emptyset$, or if 
there exists a congruence $K$ of $\fB$ 
such that $A$ is a congruence class of $K$, and $\fB/_{K} \simeq \fB_1 \times \cdots \times \fB_k$ for some $k \in {\mathbb N}$ such that for each $i \in [k]$ 
\begin{enumerate}
\item $\fB_i$ is subdirectly complete, and 
\item $\fB_i$ does not contain a proper binary or centrally absorbing subuniverse. 
\end{enumerate} 
\end{definition}

% A good example in this context is:
% Z/6Z, with congurence mod 2 and mod 3.
% We have the 4-ary subdirect 
% R \leq (Z/2Z)^2 times (Z/3Z)^2.
% e.g. defined by x=y+1 mod 2
% and x=y_1 mod 3.

% MY DEF: 
%\begin{definition}\label{def:sdc} 
%For $E \leq \fB$, we write $E \sdc \fB$
%if $E = \emptyset$, or if there exists a congruence $K$ of $\fB$ 
%such that 
%\begin{enumerate}
%\item $E$ is a congruence class of $K$, 
%\item $\fB/_K$ is subdirectly complete, and 
%\item $\fB/_K$ does not contain a proper binary or centrally absorbing subuniverse. 
%\end{enumerate} 
% \end{definition} 
%If we want to extend the results from the previous section to the general case, we need to treat variable restrictions to subalgebras $C \sdc \fB$ (see Definition~\ref{def:sdc}). Such subalgebras share many properties with centrally absorbing subalgebras. 

Lemma~\ref{lem:multi-sdc} below, which is from~\cite[Lemma 6.19]{Strong-Subalgebras-Published}, 
can be viewed as a multisorted version of subdirect completeness (see Proposition~\ref{prop:multisorted}), and is the key for all the transfer properties for $\sdc$ that we prove later. 
Throughout this section, 
$\fB_1,\dots,\fB_k$ denote finite idempotent Taylor algebras with the same signature. 

% SIMPLE VERSION: 
%\begin{lemma}\label{lem:multi-sdc}
%Let $R \leq \fA_1 \times \cdots \times \fA_k$ be subdirect where 
%$\fA_1,\dots,\fA_k$ are subdirectly complete algebras without proper binary or centrally absorbing subuniverses. 
%Let $\bA(k)$ be the relational structure 
%with domain $A(k) := A_1 \cup \cdots \cup  A_k$ whose
%relations are the graphs of isomorphisms $\fA_i \to \fA_j$, for $i,j \in [k] \setminus \{1\}$. 
%Then $R$ can be defined by a conjunction of atomic formulas in $\bA(k)$. 
%\end{lemma} 

%In the definition of $\sdc$(Definition~\ref{def:sdc})
%we require 
%, i.e., not having binary or centrally absorbing subuniverses,  is already useful in the proof of Lemma~\ref{lem:multi-sdc},
%which can be viewed as a multisorted version of subdirect completeness (see Proposition~\ref{prop:multisorted}). 

\begin{lemma}\label{lem:multi-sdc}
Let $R \leq \fB_1 \times \cdots \times \fB_k$ be subdirect such that
\begin{itemize} 
\item $\fB_2,\dots,\fB_k$ are subdirectly complete,
\item $\fB_1$ has no proper centrally absorbing subuniverse, and 
\item $\fB_2,\dots,\fB_k$ have no proper binary or centrally absorbing subuniverses. 
\end{itemize} 
Let $\bB(k)$ be the relational structure 
whose domain $B(k)$ is the disjoint union $B_1 \uplus \cdots \uplus  B_k$ and whose
relations are the graphs of isomorphisms $\fB_i \to \fB_j$, for $i,j \in \{2,\dots,k\}$, 
and surjective homomorphisms 
$\fB_1 \to \fB_i$ for $i \in [k]$. 
Then $R$ can be defined by a conjunction of atomic formulas in $\bB(k)$. 
\end{lemma} 

In the proof of Lemma~\ref{lem:multi-sdc}
we need the following concept. 

\begin{definition}
If $R \subseteq B_1 \times \cdots \times B_k$ and $i \in [k]$, we say that \emph{the $i$-th coordinate of $R$ is uniquely determined} if for all $b_1 \in B_1, \dots, b_k \in B_k$ 
there exists at most one $b_i' \in B_i$
such that $(b_1,\dots,b_{i-1},b_i', b_{i+1},\dots,b_k) \in R$. 
\end{definition}
 
\begin{observation}
If $R \leq \fB_1 \times \fB_2$ is subdirect and the second coordinate of $R$ is uniquely determined, then $R$ is the graph of a surjective homomorphism $\fB_1 \to \fB_2$. If both coordinates are uniquely determined, then 
$R$ is the graph of an isomorphism $\fB_1 \to \fB_2$. 
\end{observation}

\begin{proof}[Proof of Lemma~\ref{lem:multi-sdc}] 
We prove the statement by induction on the arity $k$ of $R$. 
If $k=1$, then $R = B_1 = B(1)$ by the assumption that $R$ is subdirect,
and hence can be defined by the empty conjunction. 

%If $k=2$, then $S := R \circ R^{-1} \leq (\fA_1)^2$ (recall the definition of $R^{-1}$ from Exercise~\ref{exe:converse})  
%is subdirect, because $R$ is subdirect.
%Since $\fA_1$ is subdirectly complete, 
%$S$ is either the graph of an automorphism or $(A_1)^2$. 
%If it is $(A_1)^2$, then 
%It follows that $R$ is the graph of an injection from $A_1$ to $A_2$.
%The same argument with $R^{-1} \circ R$ shows that $R^{-1}$ is the graph of an injection as well. Hence, $R$ is the graph of a bijection. 

Suppose now that $k \geq 2$. 
If $R = B_1 \times \cdots \times B_k$, then 
it can be defined by $\bigwedge_{i \in [k]} \id_{B_i}(x_i,x_i)$, so suppose that this is not the case. 

Consider the case that $R' := \pr_{[k-1]}(R) \neq B_1 \times \cdots \times B_{k-1}$. 
Then $R' \leq \fB_1 \times \cdots \times \fB_{k-1}$ is subdirect, and 
by the inductive assumption it has 
a definition by a conjunction of atomic formulas $\phi(x_1,\dots,x_{k-1})$ in $\bB(k-1)$. 
%Since $R' \neq B_1 \times \cdots \times B_{k-1}$, the formula $\phi$ has at least  one such conjunct $\alpha_1(x_i,x_j)$, for distinct $i,j \in [k-1]$, $i>j$, and the respective relation $R_h$ denotes the graph of  a surjective homomorphism  $h \colon \fB_i \to \fB_j$. Let $R'' := \pr_{[k] \setminus \{j\}}(R)$. By the inductive assumption, $R''$ has a definition $\alpha_2(x_1,\dots,x_{j-1},x_{j+1},\dots,x_k)$ by conjunctions of atomic formulas  in $\bB(k)$. 
%Then $\alpha_1 \wedge \alpha_2$ is a conjunction of atomic formulas over $\bB(k)$ that defines $R$. 
Since $R'\neq B_1\times\cdots\times B_{k-1}$, the formula $\phi$
contains a conjunct $\alpha_1(x_p,x_\ell):=R_h(x_p,x_\ell)$,
where $p\neq\ell$, $\ell\in\{2,\dots,k-1\}$, and $R_h$ is the graph
of a surjective homomorphism $h\colon\fB_p\to\fB_\ell$. 
Let $R'':=\pr_{[k]\setminus\{\ell\}}(R)$. 
By the inductive assumption, $R''$ has a definition 
$\alpha_2(x_1,\dots,x_{\ell-1},x_{\ell+1},\dots,x_k)$ 
by a conjunction of atomic formulas in $\bB(k)$. We claim
that $\alpha_1\wedge\alpha_2$ defines $R$.
Indeed, every tuple in $R$ satisfies both formulas. Conversely,
suppose that $\bar a$ satisfies $\alpha_1\wedge\alpha_2$. Since
$\alpha_2$ defines $R''$, there exists $b_\ell\in B_\ell$ such that
the tuple $\bar b$ obtained from $\bar a$ by replacing $a_\ell$ with
$b_\ell$ belongs to $R$. In particular, $\bar b$ satisfies
$\alpha_1$, and hence
\[
b_\ell=h(b_p)=h(a_p)=a_\ell.
\]
Thus $\bar a=\bar b\in R$, proving the claim.

Similarly, we can treat the case that
$\pr_I(R) \neq \prod_{i \in I} B_i$ for any $I \subsetneq [k]$. 
Hence, we may assume that for every $I \subsetneq [k]$ we have 
\begin{align} 
\pr_I(R) = \prod_{i \in I} B_i. 
\label{eq:full-proj}
\end{align} 
%We claim that 
%this case is impossible. Is possible for k = 2, so present it differently. 

%\item $\fA_i$ has a non-trivial congruence. 
%\item there exists $l \in [k]$ and a central relation $R \leq \fA_l \times \fA_i$. 
%\end{enumerate}
%By the symmetric nature of the statement, it suffices to prove the claim for $i = 1$. 
For $i \in [k]$ and $s \in [|B_i|]$, 
let $\chi_{i,s}(y_1,\dots,y_s)$ be the formula 
$$\exists x_1,\dots,x_{i-1},x_{i+1},\dots,x_k \bigwedge_{j \in [s]}  R(x_1,\dots,x_{i-1},y_j,x_{i+1},\dots,x_k).$$

{\bf Claim 1.} 
For every $i \in \{2,\dots,k\}$, the formula 
$\chi_{i,|B_i|}$ does not define $(B_i)^{|B_i|}$. 

For the sake of the notation, we assume that $i = k$. If the claim is not true, then for every $j \in [k-1]$ there is  
$b_j \in B_j$ such that for every $a \in B_k$ we have
$(b_1,\dots,b_{k-1},a) \in R$. 
Let $l \geq 2$ be the minimal number such that 
$$T := \{(a_1,\dots,a_l) \mid (b_1,\dots,b_{k-l},a_1,\dots,a_l) \in R\} \neq B_{k-l+1} \times \cdots \times B_k.$$
%$$T := \{(a_1,\dots,a_l) \mid (a_1,\dots,a_l,b_{l+1},\dots,b_k) \in R\} \neq A_1 \times \cdots \times A_l.$$
Such an $l$ exists, because for $l = k$ we get that $T = R \neq (B_1 \times \cdots \times B_k)$. 
Choose $(a_1,\dots,a_l) \notin T$. 
Then the relation $$T' := \{(x,y) \mid (x,a_2,\dots,a_{l-1},y) \in T\} \leq \fB_{k-l+1} \times \fB_k$$ contains
$\{b_{k-l+1}\} \times B_k$ by the minimality of $\ell$, and hence has a non-empty left center $C$. 
By assumption, $\fB_k$ has no proper binary absorbing subuniverses. 
Moreover, \eqref{eq:full-proj} implies that $\pr_1(T') = B_{k-l+1}$. 
%\begin{align}
%\pi_{\{1,\dots,l-1,l+1,\dots,k\}}(R) = A_1 \times \cdots A_{l-1} \times A_{l+1} \times \cdots \times A_k.
%\label{eq:full-proj}
%\end{align} 
Hence, 
by Lemma~\ref{lem:centrally-abs}   
we have $C \cabs \fB_{k-l+1}$. 
Note that $(a_1,a_l) \notin T'$, 
%, we have $T' \neq A_{k-l+1} \times A_k$, 
thus $a_1 \notin C \neq B_{k-l+1}$. 
Therefore, $C$ is a proper centrally absorbing subuniverse of $\fB_{k-l+1}$, 
a contradiction to our assumptions.

\medskip 
{\bf Claim 2.} 
For every $i \in \{2,\dots,k\}$, 
%one of the following cases applies. 
%\begin{enumerate} 
%\item 
the $i$-th coordinate of $R$ is uniquely determined. 

Let $S_i \leq \fB_i^2$ be the relation %$$\pi_{[k] \setminus \{i,j\}}
defined by $\chi_{i,2}(y_1,y_2)$. 
Clearly, $S_i$ is reflexive and in particular subdirect. 
Since $\fB_i$ is subdirectly complete,
we must have $S_i = \id_{B_i}$
or $S_i = (B_i)^2$. 
In the first case, the $i$-th coordinate of $R$ is uniquely determined. Suppose for contradiction that the second case applies.  Let $l$ be minimal such that the relation $T$ defined by $\chi_{i,l}(x_1,\dots,x_l)$ is not equal to 
$(B_i)^l$; such an $l$ exists because of Claim~1. 
Note that $T$ contains all tuples with repeated entries, by the minimal choice of $l$. 
Pick $(a_1,\dots,a_l) \in (B_i)^l \setminus T$. 
Then the relation 
$$T' := \{(x,y) \mid (a_1,\dots,a_{l-2},x,y) \in T\} \leq (\fB_i)^2$$ 
contains $\{a_1,\dots,a_{l-2}\} \times B_i$, and hence has a non-empty left center $C$. 
By assumption, $\fB_i$ has no proper binary absorbing subuniverses. Clearly, 
$\pi_1(T') = B_i$. Hence, $C \cabs \fB_i$ by Lemma~\ref{lem:centrally-abs}. 
Since $T \neq (B_i)^l$, we have $T' \neq (B_i)^2$, and thus $C \neq B_i$. Therefore, $C$ is a proper centrally absorbing subuniverse of $\fB_i$, 
a contradiction to our assumptions. 
This concludes the proof of Claim 2. 

\medskip 
Claim 2 implies that if $k=2$, then $R$ is the graph of a surjective homomorphism from $\fB_1$ to $\fB_2$ and we are done. So suppose that
$k \geq 3$.
If $|B_k|=1$, then 
$R=B_1\times\cdots\times B_k$ by~\eqref{eq:full-proj}, which has the definition $\bigwedge_{i\in[k]}\id_{B_i}(x_i,x_i)$ of the required form. 
So suppose that $|B_k| > 1$. 
We then consider $L \leq (\fB_k)^4$ defined by the formula $\eta(z_1,z_2,z_3,z_4)$ given by 
\begin{align*}
\exists x_1,x_2,\dots,x_{k-1},x_1',x_2' 
\big (&R(x_1,x_2,x_3,\dots,x_{k-1},z_1) \wedge 
R(x_1,x_2',x_3,\dots,x_{k-1},z_2) \\
\wedge \; & 
R(x_1',x_2,x_3,\dots,x_{k-1},z_3)
\wedge 
R(x_1',x_2',x_3,\dots,x_{k-1},z_4) \big )
\end{align*} 
Clearly, $L$ contains 
$(a_k,a_k,a_k,a_k)$ 
for every $a_k \in B_k$: 
to see this,
we can find $a_1,\dots,a_{k-1}$ such that $(a_1,\dots,a_{k-1},a_k) \in R$ since $R$ is subdirect, 
and we then choose $x_1=x'_1:=a_1$, $x_2=x'_2 :=a_2$,
$x_3 := a_3,\dots,x_{k-1} := a_{k-1}$ to satisfy all four conjuncts of $\eta(a_k,a_k,a_k,a_k)$. 

We claim that every three-coordinate projection of $L$ is full.
Let $a,b,c\in B_k$, and choose
$x_1\in B_1,x_3\in B_3,\dots,x_{k-1}\in B_{k-1}$ arbitrarily.
By~\eqref{eq:full-proj}, we may successively choose
$x_2,x_2'\in B_2$, $x_1'\in B_1$, and $d\in B_k$ such that
\begin{align*}
&(x_1,x_2,x_3,\dots,x_{k-1},a)\in R,\\
&(x_1,x_2',x_3,\dots,x_{k-1},b)\in R,\\
&(x_1',x_2,x_3,\dots,x_{k-1},c)\in R,\\
&(x_1',x_2',x_3,\dots,x_{k-1},d)\in R.
\end{align*}
Thus, $(a,b,c,d)\in L$, and hence
$\pr_{\{1,2,3\}}(L)=(B_k)^3$. Exchanging $x_1$ with $x_1'$ and/or
$x_2$ with $x_2'$ gives the same conclusion for every
three-coordinate projection of $L$. In particular, every binary
projection of $L$ is full.

We have already seen that $L$ contains $(a,a,a,a)$ for every
$a\in B_k$. Since $\fB_k$ is subdirectly complete,
Lemma~\ref{lem:sdcc} implies that $L$ can be defined by a conjunction
of equalities between its coordinates. Since $|B_k|>1$ and every
binary projection of $L$ is full, no equality between distinct
coordinates can occur. Therefore, $L=(B_k)^4$.
%Since $\fB_k$ is subdirectly complete and by~\eqref{eq:full-proj} we can conclude that  $L=(B_k)^4$. 

In particular, $(a,a,a,b) \in L$ for distinct $a,b \in B_k$ (recall that $|B_k|>1$). 
Let $x_i \in B_i$, for $i \in [k-1]$, and let $x'_1 \in B_1$ and $x'_2 \in B_2$ be witnesses for the existentially quantified variables in $\eta(a,a,a,b)$. Since the second coordinate of $R$ is uniquely determined and $z_1=a=z_2$, the first two conjuncts in $\eta(a,a,a,b)$ imply that $x_2 = x'_2$.
Since the last coordinate of $R$ is uniquely determined, the final two conjuncts in $\eta(a,a,a,b)$ imply that $a=z_3=z_4=b$, 
a contradiction to $a \neq b$. 
\end{proof} 

\begin{corollary}\label{cor:multi-sdc-product}
Suppose that $R\leq\fB_1\times\cdots\times\fB_k$ satisfies the
assumptions of Lemma~\ref{lem:multi-sdc}. Then there exists a set
$I\subseteq[k]$ such that the projection
\[
\pr_I\colon\fR\to\prod_{i\in I}\fB_i
\]
is an isomorphism. In particular, $\fR$ is isomorphic to a product
of some of the algebras $\fB_1,\dots,\fB_k$.
\end{corollary}

\begin{proof}
By Lemma~\ref{lem:multi-sdc}, the relation $R$ has a definition
$\phi(x_1,\dots,x_k)$ by a conjunction of atomic formulas whose
binary relations are graphs of isomorphisms between
$\fB_i$ and $\fB_j$, for $i,j\in\{2,\dots,k\}$, or graphs of
surjective homomorphisms from $\fB_1$ to $\fB_i$.

Consider the graph on $[k]$ in which distinct $i,j\in[k]$ are
adjacent if $x_i$ and $x_j$ occur together in a conjunct of $\phi$.
Choose one index $i_C$ from every connected component $C$, choosing
$i_C=1$ for the component containing $1$, and put
$I:=\{i_C\mid C\text{ is a connected component}\}$.

We claim that $\pr_I$ is an isomorphism. It is injective because, in
a component not containing $1$, the entry at coordinate $i_C$ of a
tuple in $R$ uniquely determines all the other entries in that
component via isomorphisms. In the component containing $1$, all
entries are uniquely determined by the first entry via homomorphisms
from $\fB_1$ and isomorphisms.

To prove surjectivity, let $(a_i)_{i\in I}\in\prod_{i\in I}B_i$.
For every connected component $C$, the subdirectness of $R$ yields
a tuple $r^C\in R$ whose $i_C$-th entry is $a_{i_C}$. Define a tuple
$r\in B_1\times\cdots\times B_k$ by taking, for every component $C$,
the entries of $r^C$ at the coordinates belonging to $C$. Every
conjunct of $\phi$ contains variables from a single connected
component. Hence, $r$ satisfies $\phi$, and therefore $r\in R$.
Moreover, $\pr_I(r)=(a_i)_{i\in I}$, proving that $\pr_I$ is
surjective.

Finally, $\pr_I$ is a homomorphism because it is a coordinate
projection. Its inverse is also a homomorphism: in every connected
component, each non-chosen coordinate is obtained from the chosen
coordinate by a composition of the homomorphisms and isomorphisms
whose graphs occur in $\phi$. Thus $\pr_I$ is an isomorphism.
\end{proof}

Similarly as for absorption and central absorption, there are a number of transfer principles for $\sdc$.

\begin{corollary}\label{cor:intersect-sdc}
% Source: 6.19.2 in Dima.
% Potential Problem: Definition is slightly different. 
Let $\fB$ be an algebra and 
let $A_1,A_2 \sdc \fB$. 
Then $A_1 \cap A_2 \sdc \fB$. 
\end{corollary} 

\begin{proof}
If $A_1=\emptyset$, $A_2=\emptyset$, or $A_1\cap A_2=\emptyset$,
then the statement follows immediately from Definition~\ref{def:sdc}.
Hence, suppose that $A_1,A_2$, and $A_1\cap A_2$ are non-empty.

For $i\in\{1,2\}$, let $K_i$ be a congruence of $\fB$ such that
$A_i$ is a congruence class of $K_i$. By Definition~\ref{def:sdc},
there are subdirectly complete algebras
$\fC_1,\dots,\fC_m,\fD_1,\dots,\fD_n$, without proper binary or
centrally absorbing subuniverses, such that
\[
\fB/K_1\simeq\fC_1\times\cdots\times\fC_m
\quad\text{and}\quad
\fB/K_2\simeq\fD_1\times\cdots\times\fD_n.
\]
Compose the map $b\mapsto(b/_{K_1},b/_{K_2})$ with these
isomorphisms, and denote the resulting homomorphism by
\[
h\colon B\to
C_1\times\cdots\times C_m\times D_1\times\cdots\times D_n.
\]
Its image $R:=h(B)$ is a subdirect subalgebra of
\[
\fC_1\times\cdots\times\fC_m\times
\fD_1\times\cdots\times\fD_n,
\]
and the kernel of $h$ is $K_1\cap K_2$. Hence,
$\fB/(K_1\cap K_2)\simeq\fR$.
By Corollary~\ref{cor:multi-sdc-product}, $\fR$ is isomorphic to a
product of some of the algebras
$\fC_1,\dots,\fC_m,\fD_1,\dots,\fD_n$. Thus, 
$\fB/(K_1\cap K_2)$ is isomorphic to a product of subdirectly
complete algebras without proper binary or centrally absorbing
subuniverses. Since $A_1\cap A_2$ is a congruence class of
$K_1\cap K_2$, Definition~\ref{def:sdc} yields
$A_1\cap A_2\sdc\fB$.
\end{proof}

\begin{lemma}\label{lem:sdc1}
Let $\fR \leq \fB_1 \times \cdots \times \fB_k$ be subdirect and $A_i \sdc \fB_i$ for every $i \in [k]$. Then 
$R \cap (A_1 \times \cdots \times A_k) \sdc \fR$. 
\end{lemma} 
\begin{proof}
If $A_i=\emptyset$ for some $i\in[k]$, then the relation in the conclusion is empty, and the statement follows from Definition~\ref{def:sdc}.
Otherwise, 
each $A_i$ is a congruence class of
a congruence $K_i$ of $\fB_i$ such that $\fB_i/K_i$ is isomorphic to a product of subdirectly complete algebras without proper binary or centrally absorbing subalgebras.  
Let $K_i'$ be the equivalence relation on $R$ defined as
$$K_i' := \{(r,s) \in R^2 \mid (r_i,s_i) \in K_i\}.$$ 
Let $E_i$ be the class  of $K_i'$ given by 
$E_i := \{r \in R \mid r_i \in A_i\}$.  
Since $R \leq \fB_1 \times \cdots \times \fB_k$ is subdirect, $\fR / K_i'$ is isomorphic to $\fB_i / K_i$. 
It follows that 
%$\fR/K_i'$ is subdirectly complete and does not contain proper binary or centrally absorbing subalgebras. 
%Therefore, 
$E_i \sdc \fR$. 
Note that $R \cap (A_1 \times \cdots \times A_k) = E_1 \cap \cdots \cap E_k$. 
Hence, the statement follows by repeated application of
Corollary~\ref{cor:intersect-sdc}.
\end{proof} 

\begin{lemma}\label{lem:sdc2}
Let $\fR \leq \fB_1 \times \fB_2$ be subdirect such that $\fB_1$ has no proper centrally absorbing subuniverses. 
Let $A_2 \sdc \fB_2$. Then 
$A_1 := \pr_{1}(R \cap (B_1 \times A_2)) \sdc \fB_1$. 
\end{lemma} 
\begin{proof}
If $A_2 = \emptyset$, then $A_1 = \emptyset$ and there is nothing to be shown. Otherwise, 
let $K$ be a congruence of $\fB_2$
with congruence class $A_2$ 
such that there is an isomorphism
$\iota$ 
between 
$\fB_2/_K$ and a product $\fC_1 \times \cdots \times \fC_k$ of subdirectly complete algebras without proper binary or centrally absorbing subuniverses. 
Then the relation 
$$S := \{(b_1,\iota(b_2/_K)) \mid (b_1,b_2) \in R\} \leq \fB_1 \times \fC_1 \times \cdots \times \fC_k$$
satisfies the assumptions of Lemma~\ref{lem:multi-sdc},
and $S$ can be defined by a conjunction of atomic formulas that 
denote isomorphisms 
$\fC_i \to \fC_j$ for $i,j \in [k]$
and surjective homomorphisms
$h_{j} \colon \fB_1 \to \fC_{i_j}$ for $j \in [\ell]$ and $i_1,\dots,i_{\ell} \in [k]$.
For $i \in [k]$, let $c_i \in C_i$ be such that $\iota(A_2) = (c_1,\dots,c_k)$. 

Note that 
$D_j := h_{j}^{-1}(c_{i_j}) \sdc \fB_1$ for each $j \in [\ell]$.  
Clearly, $D_j$ is a class of the kernel of $h_{j}$, which is a congruence $K_j$ of $\fB_1$. 
Moreover, $\fB_1/_{K_j}$ is isomorphic to $\fC_{i_j}$ and hence 
subdirectly complete and without proper binary or centrally absorbing subalgebras. Then  
\begin{align*}
A_1 = \pi_1(R \cap (B_1 \times A_2)) 
& = \pi_1(S \cap (B_1 \times \{c_1\} \times \cdots \times \{c_k\})) \\
& = \bigcap_{j \in [\ell]} h_{j}^{-1}(c_{i_j}). 
\end{align*} 
Hence, $A_1 \sdc \fB_1$ by Corollary~\ref{cor:intersect-sdc}. 
\end{proof}

\begin{lemma}\label{lem:sdc3}
Let $\fR \leq \fB_1 \times \cdots \times \fB_k$ be subdirect such that $\fB_1$ has no proper centrally absorbing subuniverses. 
Let $A_i \sdc \fB_i$ for every $i \in [k]$. Then 
$$A := \pr_{1}(R \cap (A_1 \times  \cdots \times A_k)) \sdc \fB_1.$$
\end{lemma} 
\begin{proof}
Let $D:= \pr_{\{2,\dots,k\}}(R)$.  
Then $R$ can be viewed as a subdirect relation $R \leq \fB_1 \times \fD$. Let $E := D \cap (A_2 \times \cdots \times A_k)$. 
By Lemma~\ref{lem:sdc1} we have
$E \sdc \fD$. Then $\pr_1(R \cap (B_1 \times E)) \sdc \fB_1$ by Lemma~\ref{lem:sdc2},
and $A = A_1 \cap  \pr_1(R \cap (B_1 \times E)) \sdc \fB_1$ by Corollary~\ref{cor:intersect-sdc}. 
\end{proof} 

Subalgebras $B \sdc \fB$ satisfy a strong form of the $k$-Helly property (compare with Corollary~\ref{cor:helly}). 

\begin{lemma}\label{lem:sdc-helly}
Let $A_1, \dots, A_k \sdc \fB$, for $k \geq 2$, be such that 
$A_i \cap A_j \neq \emptyset$ for all distinct $i,j \in \{1,\dots,k\}$. Then 
$\bigcap_{i \in [k]} A_i \neq \emptyset$. 
\end{lemma} 
\begin{proof}
For every $i\in[k]$, let $K_i$ be a congruence of $\fB$ such that
$A_i$ is a congruence class of $K_i$, and choose an isomorphism
\[
\iota_i\colon \fB/K_i\to
\fC_{i,1}\times\cdots\times\fC_{i,m_i},
\]
where every $\fC_{i,r}$ is subdirectly complete and has no proper
binary or centrally absorbing subuniverse. Write
\[
a_i=(a_{i,1},\dots,a_{i,m_i}):=\iota_i(A_i).
\]
%Here, $A_i$ is viewed as an element of the quotient $B/K_i$.
For $i\in[k]$ and $r\in[m_i]$, let
$h_{i,r}\colon B\to C_{i,r}$
be the composition of the quotient map $B\to B/K_i$, the isomorphism
$\iota_i$, and the projection onto the $r$-th coordinate. Put
\[
J:=\{(i,r)\mid i\in[k],\ r\in[m_i]\}
\]
and consider the joint image
\[
S:=\bigl\{(h_{i,r}(b))_{(i,r)\in J}\mid b\in B\bigr\}
 \leq \prod_{(i,r)\in J}\fC_{i,r}.
\]
The relation $S$ is subdirect. Hence, Lemma~\ref{lem:multi-sdc}
gives a definition of $S$ by a conjunction $\phi$ of atomic formulas
whose relations are graphs of homomorphisms between the factors
$\fC_{i,r}$. We claim that the tuple
\[
a:=(a_{i,r})_{(i,r)\in J}
\]
satisfies $\phi$. Consider a conjunct of $\phi$ involving coordinates
$(i,r)$ and $(j,s)$. If $i=j$, choose $b\in A_i$. If $i\neq j$,
choose $b\in A_i\cap A_j$, which is possible by assumption. In both
cases, the tuple $(h_{p,q}(b))_{(p,q)\in J}$ belongs to $S$ and its
coordinates $(i,r)$ and $(j,s)$ are $a_{i,r}$ and $a_{j,s}$,
respectively. Hence, the tuple $a$ satisfies the conjunct. Since this
holds for every conjunct of $\phi$, we have $a\in S$. Consequently, there exists $b\in B$ such that
$h_{i,r}(b)=a_{i,r}$ for all $(i,r)\in J$. Therefore, for every
$i\in[k]$,
\[
\iota_i(b/_{K_i})=(a_{i,1},\dots,a_{i,m_i})
=\iota_i(A_i).
\]
Since $\iota_i$ is injective, $b/_{K_i}=A_i$, and hence $b\in A_i$.
Thus $b\in\bigcap_{i\in[k]}A_i$, as required.
\end{proof}

To verify that $A \sdc \fB$, the following lemma is useful.

\begin{lemma}\label{lem:abs-to-factor}
Let $R \leq \fB_1 \times \cdots \times \fB_k$ be a proper binary or centrally absorbing subuniverse. Then there exists $i \in [k]$ such that $\fB_i$ has a proper binary or centrally absorbing subuniverse.
\end{lemma} 
\begin{proof}
The proof is by induction on $k$. For $k = 1$, the statement is trivial. 
For $k > 1$, first consider the case that $\pi_i(R) \neq B_i$ for some $i \leq k$. Then 
the statement follows from Lemma~\ref{lem:abs-prod-1} (in the case of binary absorption)
and Lemma~\ref{lem:pp-cabs} (in the case of central absorption). 

Otherwise, $\pi_i(R) = B_i$ for all $i \leq k$, i.e., $R$ is subdirect. 
Since $R$ is a proper subuniverse, 
there exists $(b_1,\dots,b_k) \in (B_1 \times \cdots \times B_k) \setminus R$. Then 
by the idempotence of $\fB_k$ 
$$R' := \{(b'_1,\dots,b'_{k-1}) \mid (b'_1,\dots,b_{k-1}',b_k) \in R\} \leq \fB_1 \times \cdots \times \fB_{k-1}.$$
Then $R'$ is non-trivial: it does not contain $(b_1,\dots,b_{k-1})$, and it is non-empty because $\pi_k(R) = B_k$.
%Moreover,  by Lemma~\ref{lem:abs-prod-1} 
%and Lemma~\ref{lem:pp-cabs}
%it is binary or centrally absorbing, and the statement follows from the inductive assumption. 
Moreover, $R'$ is binary or centrally absorbing in
$\fB_1\times\cdots\times\fB_{k-1}$, respectively. Indeed, suppose
first that $R\abs_t\fB_1\times\cdots\times\fB_k$. Given tuples in
$R'$ in all but at most one argument of $t$, extend each of them by
the coordinate $b_k$, and extend the remaining tuple in the same
way. Applying $t$ gives a tuple in $R$ whose last coordinate is
$t(b_k,\dots,b_k)=b_k$, by idempotence. Its projection onto the first
$k-1$ coordinates therefore belongs to $R'$, proving that
\[
R'\abs_t\fB_1\times\cdots\times\fB_{k-1}.
\]

Now suppose that $R\cabs\fB_1\times\cdots\times\fB_k$. The preceding
argument shows that $R'$ is absorbing. Let
$a\notin R'$, where
$a\in B_1\times\cdots\times B_{k-1}$, and put
$\widehat a:=(a,b_k)$. Then $\widehat a\notin R$. If $(a,a)$ belonged
to
\[
\left\langle
(\{a\}\times R')\cup(R'\times\{a\})
\right\rangle,
\]
we could extend every tuple occurring in the generators by the
coordinate $b_k$. Idempotence would then give
\[
(\widehat a,\widehat a)\in
\left\langle
(\{\widehat a\}\times R)\cup(R\times\{\widehat a\})
\right\rangle,
\]
contradicting $R\cabs\fB_1\times\cdots\times\fB_k$. Hence $R'$ is
centrally absorbing as well, and the statement follows from the
inductive assumption.
\end{proof}

\begin{lemma}\label{lem:sdc-cabs-disjoint} 
Let $A_1,A_2 \leq \fB$ be non-empty.
Suppose that $\fA_1 \sdc \fB$ and that $\fA_2 \abs_2 \fB$ or $\fA_2 \cabs \fB$. 
Then $A_1 \cap A_2 \neq \emptyset$. 
%\begin{itemize}
%\item If $\fB_2 \abs_2 \fA$, then $\fB_1 \abs_2 \fA$.
%\item If $\fB_2 \cabs \fA$, then $\fB_1 \cabs \fA$. 
%\end{itemize} 
\end{lemma} 
\begin{proof}
Suppose for contradiction that 
$A_1 \cap A_2 = \emptyset$. 
By assumption, there exists 
a congruence $K$ with congruence class $A_1$ such that $\fB/_K$ is isomorphic to a product of  subdirectly complete algebras without proper binary or centrally absorbing subalgebras. 
Then 
%$K' \cap (B_2)^2$ is a congruence of $\fB_2$, and $\fB_2/K'$ 
$\fA_2/_K$ 
is a subalgebra of $\fB/_K$ (see Remark~\ref{rem:congnot} and Lemma~\ref{lem:congruences}). 
If $A_2 \abs_2 \fB$,
then $\fA_2/_K \abs_2 \fB/_K$ (Exercise~\ref{exe:abs-fact}). Likewise, if $A_2 \cabs \fB$, then
$\fA_2/K \cabs \fB/_K$ (Lemma~\ref{lem:central-abs-trans}).
Then $\fA_2/_K$ is a proper subalgebra, 
because $A_1 \cap A_2 = \emptyset$.  
By Lemma~\ref{lem:abs-to-factor}, 
we obtain a contradiction to the assumption that the factors of 
$\fB/_K$ do not have proper binary or centrally absorbing subuniverses. 
\end{proof} 

%In the case that the variable domain $B^{\bA}_c \leq \fB$, for some $c \in A$, has a subdirectly complete quotient $B^{\bA}_c/K$, 
%we consider two cases:
%\begin{itemize}
%\item $B^{\bA}_c/K$ has a proper binary or centrally absorbing subuniverse $\fC$. In this case, $\bigcup C$ is a proper binary or centrally absorbing subuniverse of $B^{\bA}_c$ (see Lemma~\ref{lem:abs-hom} for the binary absorbing case and Lemma~\ref{lem:central-abs-trans} for the centrally absorbing case), and we proceed by restricting the variable domain of $c$ to $\bigcup C$ as discussed before. 
%\item Otherwise, let $E$ be an arbitrary congruence class $E$ of $K$; we then have $E \sdc B^{\bA}_c$ (Definition~\ref{def:sdc}). In this case, 
%we restrict $c$ to $E$. 
%\end{itemize} 
%We claim that we can then re-establish arc-consistency without deriving false. 

%Recall our notation 
%Note that $E$ is a subalgebra of $B^{\bA}_c$.
%Similarly as before, we only need that the original instance is singleton arc-consistent,
%and that an instance obtained from the original instance by variable restriction to absorbing subuniverses is arc-consistent, to further reduce the variable domain of $c$, and to

\begin{lemma}\label{lem:sdc-coordinate-slice}
Let $\fR\leq \fB_1\times\cdots\times\fB_m$
be subdirect, let $j\in[m]$, and let $D\sdc\fB_j$. Then
\[
\{r\in R\mid r_j\in D\}\sdc\fR.
\]
\end{lemma}

\begin{proof}
For $i\neq j$, put $D_i:=B_i$, and put $D_j:=D$. For every
$i\neq j$ we have $D_i\sdc\fB_i$: this is witnessed by the universal
congruence, whose quotient is the one-element algebra. Hence,
Lemma~\ref{lem:sdc1} gives
\[
R\cap(D_1\times\cdots\times D_m)\sdc\fR,
\]
and this intersection is precisely $\{r\in R\mid r_j\in D\}$.
\end{proof}

\begin{lemma}\label{lem:sdc-relative-intersection}
Let $A\sdc\fB$, and let $\emptyset\neq C\leq\fB$. Suppose that
\[
C\abs_2\fB,\qquad C\cabs\fB,\qquad\text{or}\qquad C\sdc\fB.
\]
Then
\[
A\cap C\sdc\fC.
\]
\end{lemma}

\begin{proof}
If $A\cap C=\emptyset$, then the statement follows from
Definition~\ref{def:sdc}. Hence, suppose that $A\cap C\neq\emptyset$,
and let $K$ be a congruence of $\fB$ witnessing $A\sdc\fB$.

First suppose that $C\abs_2\fB$ or $C\cabs\fB$. Every class of $K$
is an $\sdc$-subalgebra of $\fB$, since the same quotient $\fB/K$
occurs in Definition~\ref{def:sdc}. Lemma~\ref{lem:sdc-cabs-disjoint}
therefore implies that $C$ meets every class of $K$. Consequently,
the homomorphism
\[
\fC\longrightarrow\fB/K,\qquad c\longmapsto c/_{K},
\]
is surjective and has kernel $K\cap C^2$. Thus,
\[
\fC/(K\cap C^2)\simeq\fB/K.
\]
Since $A\cap C$ is a class of $K\cap C^2$, Definition~\ref{def:sdc}
gives $A\cap C\sdc\fC$.

It remains to consider the case $C\sdc\fB$. Let $L$ be a congruence
witnessing this inclusion, and choose isomorphisms
\[
\fB/K\simeq\mathbf P_1\times\cdots\times\mathbf P_p,
\qquad
\fB/L\simeq\mathbf Q_1\times\cdots\times\mathbf Q_q,
\]
where all factors are subdirectly complete and have no proper binary
or centrally absorbing subuniverses. Let
\[
h\colon\fB\longrightarrow
\mathbf P_1\times\cdots\times\mathbf P_p
\times
\mathbf Q_1\times\cdots\times\mathbf Q_q
\]
be the product of the two quotient maps, followed by these
isomorphisms, and put $\fR:=h(\fB)$.

Apply the connected-component description from the proof of
Corollary~\ref{cor:multi-sdc-product} to $\fR$. The factors occurring
above are subdirectly complete and therefore simple. Hence every
surjective homomorphism between two non-one-element factors that
occurs in that description is an isomorphism. Fixing the
$\mathbf Q_1,\dots,\mathbf Q_q$ coordinates to the tuple
corresponding to the $L$-class $C$ therefore leaves a subalgebra
isomorphic to a product of some of the factors
$\mathbf P_1,\dots,\mathbf P_p$, possibly together with one-element
factors.

This fixed-coordinate subalgebra is $h(C)$. Projection onto the
$\mathbf P_1,\dots,\mathbf P_p$ coordinates identifies it with the
image of $C$ in $\fB/K$. Hence,
\[
\fC/(K\cap C^2)
\]
is isomorphic to a product of subdirectly complete algebras without
proper binary or centrally absorbing subuniverses. Since $A\cap C$
is a class of $K\cap C^2$, Definition~\ref{def:sdc} again gives
$A\cap C\sdc\fC$.
\end{proof}

We also have a version of 
Lemma~\ref{lem:unif} for $\sdc$.
If $\fA$ is an instance of $\Csp(\bB)$, then  
the concept of the \emph{$(\sdc,c,C)$-reduction and a $\sdc$-reduction of $\bA$}
is defined analogously as for $(\cabs,c,C)$-reductions of $\fA$ (Definition~\ref{def:reduction}).

\begin{lemma}\label{lem:unif-sdc}
Let $\bA$ be an instance of $\Csp(\bB)$ 
such that no variable domain has a proper centrally absorbing subuniverse. 
Let $\bA'$ be a $\sdc$-reduction of $\bA$. 
Then for every $a \in A$ we have
$B_a^{\bA'} \sdc B_a^{\bA}$. 
\end{lemma} 
\begin{proof}
The proof is by induction over the execution of the arc-consistency procedure. Initially, the statement is true since we restricted some variable $c \in A$ to $C \sdc B^{\bA}_c$. The inductive step follows from %Lemma~\ref{lem:sdc1} and 
Lemma~\ref{lem:sdc3}. 
\end{proof}

We can now state and prove the missing lemma for the general case of reducing variable domains (the proof of Theorem 5.7 in~\cite{Strong-Subalgebras-Published}). 

\begin{lemma}\label{lem:subdirect-complete-reduce}
Let $\bA$ be a 
cycle-consistent 
instance of $\Csp(\bB)$ whose variable domains are non-empty. 
Let $\bA_1$ be a $w$-reduction of $\bA$ 
for some $w \in (\{\abs_2,\cabs,\sdc\} \times A \times \S(\fB))^*$
such that every 
$\sdc$-reduction is only applied if all variable domains do not have proper centrally absorbing subuniverses. 
Then $B_a^{\bA_1} \neq \emptyset$ for every $a \in A$. 
\end{lemma} 
\begin{proof}
We argue by induction on the length of $w$. The case where $w$ is
empty is immediate. If the final letter of $w$ is of the form
$(\abs_2,c,C)$, the statement follows from the inductive assumption
and Lemma~\ref{lem:2-abs-restr}.

It remains to treat the case where the final letter has type
$\cabs$ or $\sdc$. Write
\[
w=u(\diamond,c,C),
\qquad
\diamond\in\{\cabs,\sdc\},
\]
and write $u=r_1\cdots r_n$. For $i\in\{0,\dots,n\}$, let $\bA_i$
be the $r_1\cdots r_i$-reduction of $\bA$. Thus,
$\bA_0=\bA$ and $\bA_n=\bA_1$. Let $\bA^+$ be obtained from
$\bA_n$ by restricting the domain of $c$ to $C$, before restoring
arc consistency.

Suppose for contradiction that the arc-consistency procedure derives
false on $\bA^+$. By Observation~\ref{obs:ac-td}, there is a tree
sentence $\phi$ such that
\[
\bA^+\models\phi
\qquad\text{and}\qquad
\bB\not\models\phi.
\]
Let $h$ witness $\bA^+\models\phi$. Let $R$ be the unary relation
symbol used to impose the restriction at $c$, so that
$R^{\bB}=C$. Choose $\phi$ so that the number $\ell$ of conjuncts
$R(x)$ with $h(x)=c$ is minimal, and denote their variables by
$x_1,\dots,x_\ell$.
Put $F:=B_c^{\bA_n}$. 
Add to $\phi$ the old-domain conjunct at each $x_j$. Remove the
selected conjuncts $R(x_1),\dots,R(x_\ell)$ and the quantifiers for
$x_1,\dots,x_\ell$, and denote the resulting formula by
$\psi(x_1,\dots,x_\ell)$. Let $H$ be the relation defined by $\psi$
in $\bB$. Then $H\leq F^\ell$, and the minimality of $\ell$ gives
\[
H\cap C^\ell=\emptyset
\]
and
\[
H\cap
\bigl(C^{j-1}\times F\times C^{\ell-j}\bigr)
\neq\emptyset
\qquad
\text{for every }j\in[\ell].
\]
Moreover, $H$ is subdirect in $F^\ell$. Indeed, for every $j$, the
unary tree formula obtained by quantifying all variables except
$x_j$ is true at $c$ in the arc-consistent instance $\bA_n$.
Exercise~\ref{exe:ac-unary} therefore shows that its interpretation
contains $F$, while the retained old-domain conjunct gives the
reverse inclusion.

The case $\ell=1$ is impossible by the same argument. If
$\ell\geq3$ and $C\cabs\fF$, then $C$ is $\ell$-absorbing in $\fF$,
contradicting Lemma~\ref{lem:essential-abs}. If $\ell\geq3$ and
$C\sdc\fF$, put
\[
H_j:=\{a\in H\mid a_j\in C\}
\qquad
(j\in[\ell]).
\]
Lemma~\ref{lem:sdc-coordinate-slice} gives $H_j\sdc\fH$ for every
$j$. The displayed essentiality conditions imply that the $H_j$
have pairwise non-empty intersections, whereas
\[
\bigcap_{j\in[\ell]}H_j=H\cap C^\ell=\emptyset.
\]
This contradicts Lemma~\ref{lem:sdc-helly}. Consequently, $\ell=2$.

As in the proof of Lemma~\ref{lem:3abs-reduce}, replace every branch
outside the path from $x_1$ to $x_2$ by the unary relation it defines.
For every variable $z$ of the resulting path formula, add a unary
conjunct whose interpretation is $B_{h(z)}^{\bA_0}$ and delete all
other unary conjuncts containing $z$. Denote the resulting path
formula by $\psi_0(x_1,x_2)$, let
\[
Z=\{x_1,x_2,y_1,\dots,y_m\}
\]
be its set of variables, and let $S\leq B^Z$ be the relation defined
by its quantifier-free matrix.

For $i\in\{0,\dots,n\}$ and $z\in Z$, put
$L_i(z):=B_{h(z)}^{\bA_i}$ 
and
\[
S^{(i)}:=S\cap\prod_{z\in Z}L_i(z).
\]
The relation $S^{(i)}$ is subdirect in
$\prod_{z\in Z}L_i(z)$. To see this, fix $z\in Z$, adjoin to
$\psi_0$ all unary domain conjuncts with interpretations $L_i(z')$,
for $z'\in Z$, and quantify every variable except $z$. This is a
unary tree formula true at $h(z)$ in $\bA_i$. Since $\bA_i$ is
arc-consistent, Exercise~\ref{exe:ac-unary} gives $\pr_z(S^{(i)})=L_i(z)$.
Put $L_{n+1}(x_1)=L_{n+1}(x_2):=C$ 
and, for $0\leq p,q\leq n+1$ and $0\leq r\leq n$, define
\[
S_{p,q,r}:=
S\cap
\left(
 L_p(x_1)\times L_q(x_2)
 \times\prod_{j=1}^m L_r(y_j)
\right).
\]
The failure of the final restriction gives
\[
S_{n+1,n+1,n}=\emptyset.
\]
The minimality of $\ell=2$ gives
\[
S_{n+1,n,n}\neq\emptyset,
\qquad
S_{n,n+1,n}\neq\emptyset.
\]
On the other hand, $\bA\models\psi_0(c,c)$. Choose any $b\in C$.
Since $C\subseteq L_n(x_1)\subseteq L_0(x_1)$, cycle consistency
gives $\bB\models\psi_0(b,b)$, and hence
$S_{n+1,n+1,0}\neq\emptyset$.
It follows that $n\geq1$. Choose $k<n$ minimal such that
$S_{n+1,n+1,k+1}=\emptyset$.
Thus, $S_{n+1,n+1,k}\neq\emptyset$.

For $i\in[n]$, let $\mathcal T_i$ be the type of the reduction
$r_i$. Lemmas~\ref{lem:unif} and~\ref{lem:unif-sdc} show that, for
every $z\in Z$, the inclusion
$L_i(z)\leq L_{i-1}(z)$ 
has type $\mathcal T_i$. In the binary absorbing case, the proof of
Lemma~\ref{lem:unif} shows more precisely that the same binary term
witnesses all these inclusions for the fixed reduction $r_i$.

We claim that, for every $i\in\{k,\dots,n\}$,
$S_{i,i,k+1}$ is a subalgebra of $S_{i,i,k}$ of type
$\mathcal T_{k+1}$. For $i=k$, this follows from the subdirectness of
$S^{(k)}=S_{k,k,k}$ and the appropriate transfer result:
Lemma~\ref{lem:abs-trans-rel}, Corollary~\ref{cor:cabs}, or
Lemma~\ref{lem:sdc1}.

Suppose that the claim holds for $i<n$. Since $S^{(i)}\subseteq S_{i,i,k}$,
the two endpoint projections of $S_{i,i,k}$ are
$L_i(x_1)$ and $L_i(x_2)$. Restricting these two coordinates to
$L_{i+1}(x_1)$ and $L_{i+1}(x_2)$ therefore shows that
$S_{i+1,i+1,k}$ is a subalgebra of $S_{i,i,k}$ of type
$\mathcal T_{i+1}$. Here one uses
Lemma~\ref{lem:abs-trans-rel}, Corollary~\ref{cor:cabs}, or
Lemma~\ref{lem:sdc1}, according to the type. Now
\[
S_{i+1,i+1,k+1}
=
S_{i,i,k+1}\cap S_{i+1,i+1,k}.
\]
If $\mathcal T_{k+1}$ is $\abs_2$ or $\cabs$, its defining property
restricts to every subalgebra, and hence the displayed intersection
has type $\mathcal T_{k+1}$ in $S_{i+1,i+1,k}$. If
$\mathcal T_{k+1}$ is $\sdc$, the same conclusion follows from
Lemma~\ref{lem:sdc-relative-intersection}. This proves the claim.

Put
\[
X:=S_{n,n,k},
\qquad
X_1:=S_{n+1,n,k},
\qquad
X_2:=S_{n,n+1,k},
\qquad
X_3:=S_{n,n,k+1}.
\]
The endpoint projections of $X$ are both equal to $L_n(c)$, because
$S^{(n)}\subseteq X$. Hence $X_1$ and $X_2$ are subalgebras of
$\fX$ of the same type as $C\leq L_n(c)$: this follows from
Corollary~\ref{cor:cabs} in the centrally absorbing case and from
Lemma~\ref{lem:sdc-coordinate-slice} in the $\sdc$ case. The claim
shows that $X_3$ has type $\mathcal T_{k+1}$ in $\fX$.

The three subalgebras have pairwise non-empty intersections. Indeed,
\[
X_1\cap X_2=S_{n+1,n+1,k}\neq\emptyset
\]
by the minimality of $k$, while
\[
S_{n+1,n,n}\subseteq X_1\cap X_3,
\qquad
S_{n,n+1,n}\subseteq X_2\cap X_3.
\]
Nevertheless,
\[
X_1\cap X_2\cap X_3
=
S_{n+1,n+1,k+1}
=
\emptyset.
\]

Suppose first that $C\cabs L_n(c)$. Then $X_1,X_2\cabs\fX$, and
their non-empty intersection
\[
X_0:=X_1\cap X_2
\]
centrally absorbs $\fX$. If $X_3\cabs\fX$, Corollary~\ref{cor:helly}
contradicts the empty triple intersection. If $X_3\sdc\fX$, the same
contradiction follows from Lemma~\ref{lem:sdc-cabs-disjoint}.

Finally, suppose that $X_3\abs_2\fX$, witnessed by a binary term
$t$. Since $X_0\cap X_3=\emptyset$, Lemma~\ref{lem:central-avoiding-square}
gives $e\in X_3$ such that
\[
M:=
\left\langle
(\{e\}\times X_0)\cup(X_0\times\{e\})
\right\rangle_{\fX^2}
\]
satisfies $M\cap X_3^2=\emptyset$. Choose $b\in X_0$. Since
$(e,b),(b,e)\in M$, closure under $t$ gives
\[
\bigl(t^{\fX}(e,b),t^{\fX}(b,e)\bigr)\in M.
\]
Both coordinates belong to $X_3$, because $X_3\abs_t\fX$. This
contradicts $M\cap X_3^2=\emptyset$.

It remains to treat the case that $C\sdc L_n(c)$. Then
$X_1,X_2\sdc\fX$, and
\[
X_0:=X_1\cap X_2\sdc\fX
\]
by Corollary~\ref{cor:intersect-sdc}. If $X_3\abs_2\fX$ or
$X_3\cabs\fX$, Lemma~\ref{lem:sdc-cabs-disjoint} contradicts
$X_0\cap X_3=\emptyset$. If $X_3\sdc\fX$, the pairwise
non-emptiness of $X_1,X_2,X_3$ contradicts
Lemma~\ref{lem:sdc-helly}.

All possible types of the final letter and of $r_{k+1}$ lead to a
contradiction. Therefore, the arc-consistency procedure cannot derive
false after the final restriction, completing the induction.
\end{proof}

The lemmata in this section lead to the following theorem, which was first proved by Kozik~\cite{Kozik16}, but with a different proof. Since it is the main result of the previous sections, we give a self-contained statement without implicit global assumptions on $\bB$. 

\begin{theorem}\label{thm:slac} 
Let $\bB$ be a finite structure which does not pp-construct $({\mathbb Z}_p;+,1)$ for every prime $p$ and such that
$\Pol(\bB)$ is idempotent. 
If $\bA$ is a cycle-consistent instance of $\Csp(\bB)$ such that $B_a^{\bA} \neq \emptyset$ for every $a \in A$, then there is a homomorphism from $\bA$ to $\bB$. In particular, the cycle-consistency procedure solves $\Csp(\bB)$. 
\end{theorem} 
\begin{proof} 
If $|B|=1$, the statement is trivial, so we may assume that $|B|>1$, and can work with the global assumptions made in the previous sections. 

Starting with the empty word, we construct a word $w$ of
$\{\abs_2,\cabs,\sdc\}$-reductions. For every word $w$ constructed in
this way, let $\bA_w$ denote the $w$-reduction of $\bA$.
Suppose that some current variable domain $B_c^{\bA_w}$ has more than
one element. If there exist $c\in A$ and a proper subuniverse $C\abs_2 B_c^{\bA_w}$,
then we replace $w$ by $w(\abs_2,c,C)$. If no such binary absorbing
subuniverse exists, but there exist $c\in A$ and a proper subuniverse $C\cabs B_c^{\bA_w}$,
then we replace $w$ by $w(\cabs,c,C)$.
Finally, suppose that no current variable domain has a proper binary
or centrally absorbing subuniverse. Choose $c\in A$ such that
$|B_c^{\bA_w}|>1$. Proposition~\ref{prop:zhuk-bw}, applied to
$B_c^{\bA_w}$, then yields a proper subuniverse $C\sdc B_c^{\bA_w}$, 
and we replace $w$ by $w(\sdc,c,C)$.
Thus, every $\sdc$-reduction is performed only when none of the
current variable domains has a proper centrally absorbing
subuniverse. Lemma~\ref{lem:subdirect-complete-reduce} therefore
implies that every instance $\bA_w$ constructed above has non-empty
variable domains. Moreover, every step strictly decreases
$\sum_{a\in A}|B_a^{\bA_w}|$, 
so the construction terminates. Let $\bA':=\bA_w$ be the final
instance. By termination and non-emptiness, for every $a\in A$ there
is an element $b_a\in B$ such that
$B_a^{\bA'}=\{b_a\}$. Since $\bA'$ is arc-consistent, the map
$a\mapsto b_a$ is a homomorphism from $\bA'$ to $\bB$, and hence
$\bA\to\bB$.
The final statement follows from Remark~\ref{rem:sac-one-sided}
and Remark~\ref{rem:sac}. 
\end{proof}

\begin{exercises}
\exercise[Weiss]\label{exe:source-285} Show that the definition of $\sdc$ cannot be simplified in the following natural way.
Specifically, show that for the following definition
of $\leq'_{\text{sdc}}$, we do not obtain closure under intersections (i.e., Corollary~\ref{cor:intersect-sdc} fails):
$A \leq'_{\text{sdc}} \fB$ holds if for $A \leq \fB$,
and $A = \emptyset$ or there exists a congruence $K$ of $\fB$
such that
\begin{enumerate}
\item $A$ is a congruence class of $K$,
\item $\fB/_K$ is subdirectly complete, and
\item $\fB/_K$ does not contain a proper binary or centrally absorbing subuniverse.
\end{enumerate}
% Solution sketch: Take an algebra with two incomparable congruences whose selected classes meet
% nontrivially, while the quotient by either congruence is subdirectly complete and absorption-free.
% Each selected class then satisfies $\leq'_{\rm sdc}$, but their intersection is not a class of any
% congruence with the required quotient. The concrete four-element square algebra from the preceding
% subdirect-completeness examples supplies these two coordinate congruences.
\end{exercises}

\subsection{The Bounded Width Theorem} 
The following theorem strengthens Theorem~\ref{thm:bw-short}. 
The implication $\ref{eq:width-bounded}. \Rightarrow \ref{eq:width-lineq}.$ is Corollary~\ref{cor:unbounded}. 
The converse implication from $\ref{eq:width-lineq}. \Rightarrow \ref{eq:width-bounded}.$ was open for many years and has been conjectured by Larose and Zadori~\cite{LaroseZadori} (and in equivalent form, by Feder and Vardi~\cite{FederVardi}; also see~\cite{AbilityToCount}),
and was proven by Barto and Kozik~\cite{BoundedWidth}.

\begin{theorem}
%[Barto and Kozik~\cite{BoundedWidth}]
\label{thm:bounded-width}
Let $\bB$ be a finite structure with finite relational signature. 
Then the following are equivalent. 
\begin{enumerate}
\item \label{eq:width-bounded} $\Csp(\bB)$ has width $(l,k)$ for some $l,k \in {\mathbb N}$.
%The variety generated by $\fA$ contains neither a trivial algebra nor a vector space. 
\item \label{eq:width-lineq}
For every prime number $p$, the structure $({\mathbb Z}_p;+,1)$ does not have a primitive positive construction in $\bB$. 
%$\HH(\PP(\bB))$ does not 
%contain $({\mathbb Z}_p;R_+,\{1\})$. 
% L_0,\dots,L_{p-1})$ where $p$ is a prime and $$L_i := \{(x,y,z) \mid x+y+z=i \mod p\}.$$ 
%\item \label{eq:bw-affine}
%$\bB$ does not pp-construct a structure $\bC$ with at least two elements such
%that there exists an idempotent affine algebra $\fA$ with 
%$\Clo(\fA) = \Pol(\bC)$. 
\item \label{eq:width-slac}
$\Csp(\bB)$ can be solved
by the cycle-consistency procedure. 
\item \label{eq:width-sac}
$\Csp(\bB)$ can be solved
by the singleton arc-consistency procedure (SAC). 
\item \label{eq:width-2k} $\Csp(\bB)$ 
has width $(2,k)$, where $k:=\max(3,r)$ and $r$ is the maximal arity of $\bB$. 
%\item \label{eq:width-wnu} 
%There exists an $n \in \mathbb N$ 
%such that
 \item \label{eq:width-3/4}
$\Pol(\bB)$ contains \emph{3-4 weak near unanimity operations}, i.e., operations $f \in \WNU(3)$ and $g \in \WNU(4)$ satisfying
$$f(y,x,x) \approx g(y,x,x,x) \, .$$
\end{enumerate}
\end{theorem}
%We do not give a complete proof of the important Theorem~\ref{thm:bounded-width}, but only show some of the easy implications, and we explain how to deduce the remaining implications from
%statements that can be found
%explicitly in the literature. 

\begin{proof}
%Note that each of the items is preserved 
%by homomorphic equivalence of $\bB$, so we may assume that $\bB$ is a core. 

The implication $\ref{eq:width-bounded}. \Rightarrow \ref{eq:width-lineq}.$ is Corollary~\ref{cor:unbounded}. 
 
 The implication
$\ref{eq:width-lineq}.\Rightarrow\ref{eq:width-slac}.$ is the most
substantial, and essentially Theorem~\ref{thm:slac}. Let $\bC$ be the core of $\bB$, and let $\bD$ be the
expansion of $\bC$ by all singleton unary relations. By
Theorem~\ref{thm:wonderland}, $\bD$ has a primitive positive
construction in $\bB$. Hence, $\bD$ does not pp-construct
$({\mathbb Z}_p;+,1)$ for any prime $p$: otherwise, the composition
of the two primitive positive constructions would give such a
construction in $\bB$.
Theorem~\ref{thm:slac}, applied to
$\bD$, shows that the cycle-consistency procedure 
solves $\Csp(\bD)$. In particular, it solves $\Csp(\bC)$, and thus also $\Csp(\bB)$ (see Exercise~\ref{exe:slac-transfer}). 
 
The implication
$\ref{eq:width-slac}.\Rightarrow\ref{eq:width-sac}.$ is immediate (see Remark~\ref{rem:sac-cycle-ac}). 
 
The implication $\ref{eq:width-sac}.\Rightarrow \ref{eq:width-2k}.$ is easy: a derivation of `false' by $\SAC_{\bB}$ can be simulated 
by a derivation of `false' by the $(2,k)$-consistency procedure where $k:=\max(3,r)$ and $r$ is the maximal arity of the relations in $\bB$.

The implication $\ref{eq:width-2k}. \Rightarrow \ref{eq:width-bounded}.$ is trivial.

We are left with the task to prove that \ref{eq:width-3/4}.\ is equivalent to the other items. 
For the implication
$\ref{eq:width-3/4}.\Rightarrow\ref{eq:width-lineq}.$, recall that
the identities in~\ref{eq:width-3/4} are preserved by primitive
positive constructions. Suppose for contradiction that
$({\mathbb Z}_p;+,1)$ has ternary and quaternary weak near unanimity
polymorphisms $f$ and $g$ satisfying
$f(y,x,x)\approx g(y,x,x,x)$. Every $n$-ary polymorphism $h$ of
$({\mathbb Z}_p;+,1)$ is of the form
$h(x_1,\dots,x_n)=\lambda_1x_1+\cdots+\lambda_nx_n$, where
$\lambda_1+\cdots+\lambda_n=1$; see Example~\ref{expl:affine}. The weak near unanimity identities
therefore imply that
$f(x,y,z)=\alpha(x+y+z)$ and
$g(x,y,z,u)=\beta(x+y+z+u)$, where $3\alpha=1$ and $4\beta=1$.
Comparing the coefficients of $y$ in
$f(y,x,x)\approx g(y,x,x,x)$ gives $\alpha=\beta$. Hence
$3\alpha=4\alpha=1$, which is impossible in ${\mathbb Z}_p$.
Thus $({\mathbb Z}_p;+,1)$ does not have a primitive positive
construction in $\bB$.

The proof of the implication
$\ref{eq:width-slac}.\Rightarrow\ref{eq:width-3/4}.$ is adapted from
a beautiful proof of a related statement in~\cite{Maltsev-Cond}.
The implications proved above show that
$\ref{eq:width-slac}.\Rightarrow\ref{eq:width-lineq}.$ Hence, $\bB$
does not pp-construct $({\mathbb Z}_p;+,1)$ for any prime $p$.

Let $\fB$ be a polymorphism algebra of $\bB$, and let $\fF$ be the
free algebra for $\HSP(\fB)$ over $\{x,y\}$
(Definition~\ref{def:free}). Let $R$ be a ternary and $S$ a
quaternary relation symbol, and let $\bF$ be the $\{R,S\}$-structure
with domain $F$ where
\begin{itemize}
\item $R^{\bF}$ is the subalgebra of $\fF^3$ generated by
$\{(y,x,x),(x,y,x),(x,x,y)\}$, and
\item $S^{\bF}$ is the subalgebra of $\fF^4$ generated by
$\{(y,x,x,x),(x,y,x,x),(x,x,y,x),(x,x,x,y)\}$.
\end{itemize}
Both relations are preserved by $\Clo(\fF)$ and are symmetric under
permutations of their coordinate positions.
The algebra $\fF$ can be realised as the algebra of binary term
operations of $\fB$, and hence as a subalgebra of the finite power
$\fB^{B^2}$. Since
$\Clo(\fF)\subseteq\Pol(\bF)$, Theorem~\ref{thm:pp-interpret} shows
that $\bF$ has a primitive positive interpretation in $\bB$.

Let $\bE$ be a core of $\bF$, viewed as a substructure of $\bF$.
Choose a homomorphism $q\colon\bF\to\bE$. The restriction of $q$ to
$E$ is an endomorphism of the finite core $\bE$, and hence an
automorphism. Replacing $q$ by
$\bigl(\left.q\right|_E\bigr)^{-1}\circ q$, we may therefore assume
that $q(e)=e$ for every $e\in E$.
Let $\bG$ be the expansion of $\bE$ by all singleton unary relations.
%Choose a core $\bE$ of $\bF$ which is a retract of $\bF$, and let $r\colon\bF\to\bE$ be a retraction. Let $\bG$ be the expansion of $\bE$ by all singleton unary relations. 
By
Theorem~\ref{thm:wonderland}, $\bG$ has a primitive positive
construction in $\bF$, and hence in $\bB$. It follows that $\bG$ does
not pp-construct $({\mathbb Z}_p;+,1)$ for any prime $p$. Moreover,
$\Pol(\bG)$ is idempotent.

Let $\bA$ have domain $A=\{1,\dots,3|F|+1\}$ and let
\begin{align*}
R^{\bA}&:=\{(a,b,c):|\{a,b,c\}|=3\},\\
S^{\bA}&:=\{(a,b,c,d):|\{a,b,c,d\}|=4\}.
\end{align*}
We view $\bA$ as an instance of $\Csp(\bG)$ by interpreting every
singleton unary relation of $\bG$ as the empty relation in $\bA$.
Thus, every variable domain of $\bA$ equals $E$. We claim that $\bA$ is cycle-consistent with respect to $\bG$. Put
\[
Q:=\left\langle
\{(y,x),(x,y),(x,x)\}
\right\rangle_{\fF^2}.
\]
For every two distinct coordinate positions $i,j$, the projection
$\pr_{i,j}(R^{\bF})$ equals $Q$, and the same holds for
$\pr_{i,j}(S^{\bF})$. Indeed, projection is a homomorphism, and the
projections of the respective generating tuples are precisely
$(y,x)$, $(x,y)$, and $(x,x)$.

Let $e\in E$. Since $\fF$ is generated by $x$ and $y$, there exists a
binary term $p$ such that $e=p(x,y)$. Put $c:=p(x,x)$ and
$e':=q(c)$. Then
\begin{align*}
(e,c)&=p\bigl((x,x),(y,x)\bigr)\in Q,\\
(c,e)&=p\bigl((x,x),(x,y)\bigr)\in Q,\\
(c,c)&=p\bigl((x,x),(x,x)\bigr)\in Q.
\end{align*}
Since $q$ is a homomorphism from $\bF$ to $\bE$ and fixes $e$, 
%Since $r$ is a retraction and $e\in E$, 
it follows that every
relevant binary projection of both $R^{\bE}$ and $S^{\bE}$ contains
$(e,e')$, $(e',e)$, and $(e',e')$.

Let $\phi(u,v)$ be a path formula such that
$\bA\models\phi(a,a)$ for some $a\in A$. The formula $\phi$ cannot
contain an atom formed with one of the singleton unary relations,
since all these relations are empty in $\bA$. Consider the sequence
of relational atoms on the path from $u$ to $v$, and let $m$ be its
length. The case $m=0$ is immediate. Moreover, $m\neq1$: otherwise,
the homomorphism witnessing $\bA\models\phi(a,a)$ would map two
distinct coordinates of a tuple in $R^{\bA}$ or $S^{\bA}$ to $a$,
whereas all entries of every tuple in these relations are pairwise
distinct. Hence, $m\geq2$.

Assign $e$ to the two endpoints of the path and $e'$ to all its
internal variables. The consecutive pairs on the path are
\[
(e,e'),(e',e'),\dots,(e',e'),(e',e).
\]
Each pair can be extended to a tuple in the relation of $\bE$
corresponding to the respective atom. The values of the remaining
variables of that atom can be chosen independently, since these
variables are leaves of the path formula. This gives an assignment
witnessing $\bG\models\phi(e,e)$. Since $e\in E$ was arbitrary,
$\bA$ is cycle-consistent with respect to $\bG$.

Theorem~\ref{thm:slac}, applied to $\bG$, now yields a homomorphism
$h\colon\bA\to\bG$. After forgetting the singleton unary relations,
we may view $h$ as a homomorphism from $\bA$ to $\bF$. Since
$|A|>3|F|$, there exist distinct $a,b,c,d\in A$ such that
$t:=h(a)=h(b)=h(c)=h(d)$. Since
$(a,b,c,d)\in S^{\bA}$, we have $(t,t,t,t)\in S^{\bF}$. By the
definition of $S^{\bF}$, there exists a term
$g(x_1,x_2,x_3,x_4)$ such that
\[
g(y,x,x,x)=g(x,y,x,x)=g(x,x,y,x)=g(x,x,x,y)=t.
\]
Similarly, $(a,b,c)\in R^{\bA}$, so $(t,t,t)\in R^{\bF}$. Hence,
there exists a term $f(x_1,x_2,x_3)$ such that
\[
f(y,x,x)=f(x,y,x)=f(x,x,y)=t.
\]
The statement follows from Lemma~\ref{lem:easy-free}.
\end{proof}

\begin{remark} 
Another condition that is equivalent to the conditions listed in Theorem~\ref{thm:bounded-width} is that $\bB$ has a binary polymorphism $f_2$ and polymorphisms $f_n \in \WNU(n)$ for every $n \geq 3$ 
and $$f_n(x,y,\dots,y) \approx f_2(x,y) \, .$$
This follows from~\cite[Proposition 4.1]{JMMM}. %\item \label{eq:width-3}
\end{remark}

\begin{remark} 
Yet another condition which is equivalent to the conditions listed in Theorem~\ref{thm:bounded-width} is that $\bB$ has ternary polymorphisms $p,q$ such that $p \in \WNU(3)$ and 
\begin{align*}
p(x,x,y) \approx q(x,y,x) \quad \text{ and } \quad q(x,x,y) \approx q(x,y,y) .
\end{align*}
This condition from~\cite[Corollary 3.3]{JMMM} appears to be the most useful one for checking whether $\Csp(\bB)$ has bounded width for a given finite structure $\bB$. 
\end{remark} 

We close this section with some immediate consequences of Theorem~\ref{thm:bounded-width}. 

\begin{corollary}
Let $H$ be a finite digraph.
Then the path-consistency procedure solves
$\Csp(H)$ if and only if $H$ 
has weak near unanimity polymorphisms $f$ and $g$
satisfying 
$$f(y,x,x) \approx g(y,x,x,x).$$
\end{corollary}

Another remarkable consequence is that
for the $H$-colouring problem, 
$(2,3)$-consistency is as powerful as
$(2,k)$-consistency for all $k \geq 3$ (the \emph{`collapse of the bounded width hierarchy'}; we already stated this in Theorem~\ref{thm:width-collaps}). 
%One technical step of the proof of Theorem~\ref{thm:bounded-width} is to reduce the
%argument to an argument about the strength of
%$(2,3)$-consistency via 
%Corollary~\ref{cor:dual}. 

Quite remarkably, the results in this section imply strong results in universal algebra that were only discovered in the study of local consistency (but that can be stated without mentioning local consistency at all). 

\begin{corollary}\label{cor:bw-alg}
% Thm 4.15 in Dima's Strong Subalgebras paper. 
Let $\fA$ be a finite idempotent algebra. Then the following are equivalent.
\begin{itemize}
\item $\fA$ has 3-4 weak near unanimity terms. 
%\item $\fA$ has for every $k \geq 3$ a weak near unanimity terms of arity $k$. 
\item $\HS(\fA)$ does not contain an abelian algebra with at least two elements. 
\end{itemize} 
\end{corollary}
\begin{proof}
By Corollary~\ref{cor:finite-signature-arity} applied for $k=4$, we may 
choose a finite-signature structure $\bB$ on $A$ containing all singleton unary relations such that $\Clo(\fA)\subseteq\Pol(\bB)$, with equality in arities $3$ and $4$.
If $\HS(\fA)$ has no nontrivial abelian member, neither does $\HS(\fB)$ for a polymorphism algebra $\fB$ of $\bB$ expanding $\fA$.
The Taylor characterisation, Corollary~\ref{cor:zhuk}, and the reduction argument of Theorem~\ref{thm:slac} therefore give bounded width; Theorem~\ref{thm:bounded-width} supplies the required terms.
Conversely, an abelian $\fC\in\HS(\fA)$ inherits these terms and is affine by Corollary~\ref{cor:abelian-affine}.
Their identities force $3\alpha=4\alpha=\operatorname{id}_C$ for an endomorphism $\alpha$ of the underlying abelian group, whence $\operatorname{id}_C=0$ and $|C|=1$.
\end{proof}

%\subsection{Cycle Consistency Solves Paper-Scissors-Stone}
%\subsection{-,0,+}
% 3.6.1 in Brady

\begin{exercises}
\exercise[Blau]\label{exe:source-287} Prove Corollary~\ref{cor:bw-alg}.
% Solution sketch: Apply Theorem~\ref{thm:bounded-width}: the stated algebraic condition is equivalent
% to bounded width, and the fixed width-consistency procedure decides every instance. Since the
% template and the width are fixed, the procedure runs in polynomial time, proving the corollary.
\exercise[Orange]\label{exe:source-288} Show that $\Csp(\bB)$ has bounded width if and only if $\Pol(\bB)$ contains for every $n \geq 3$ an operation from $\WNU(n)$.
% Solution sketch: Bounded width gives the SD$(\wedge)$ condition and hence WNU terms of every arity
% $n\geq3$ by the term theorem in the text. Conversely, WNU operations in all those arities supply the
% finite family of consistency identities required by the bounded-width theorem; a fixed sufficiently
% large pair of arities already yields a bounded-width algorithm.
\exercise[Weiss]\label{exe:source-286} Show that our proof of the implication
$\ref{eq:width-slac}. \Rightarrow \ref{eq:width-3/4}.$ in the proof of Theorem~\ref{thm:bounded-width} fails if we replace the cycle-consistency procedure by the singleton arc-consistency procedure. Specifically, show that there are finite structures $\bB$ such that the instance $\bA$ of $\Csp(\bG)$, for $\bA$ and $\bG$ constructed in this proof, is (cycle-consistent, but) not singleton arc-consistent.
\end{exercises}

\section{Uniform Algorithms}
\label{sect:uniform}
Previously, we have fixed a structure $\bB$ and studied the computational complexity of $\Csp(\bB)$. However, some of the algorithms that we have encountered can easily be adapted to the situation where $\bB$ is a finite structure that is part of the input: this is for instance the case for the arc-consistency procedure (Section~\ref{sect:AC}), the 
path-consistency procedure (Section~\ref{sect:PC}), the $k$-consistency procedure (Section~\ref{sect:kcons}), and singleton arc-consistency (Section~\ref{sect:SAC}).
However, this was not the case for the algorithm for $\Csp(\bB)$ when $\bB$ has a Maltsev polymorphism (Section~\ref{sect:buldalalg}). 

 We suppose that both structures $\bA$ and 
$\bB$ are given \emph{explicitly}, i.e., represented by listing all tuples for each of the relations. 

\begin{definition}[uniform CSP algorithm]
\label{def:uniform}
A \emph{(one-sided correct) uniform CSP algorithm} is an algorithm that takes as input a pair $(\bA,\bB)$ of explicitly represented finite structures, and returns `Yes' or `No' such that if the algorithm returns `No', then there is no homomorphism from $\bA$ to $\bB$. 
If ${\mathcal C}$ is a class of relational structures, each with a finite domain and finitely many relations, then we say that a uniform algorithm \emph{solves ${\mathcal C}$} if the algorithm returns `Yes' on input $(\bA,\bB)$ with $\bB \in {\mathcal C}$ if and only if $\bA \to \bB$. 
\end{definition}

For example, the arc-consistency procedure  is a uniform (and polynomial-time, see Remark~\ref{rem:AC-uniform}) CSP algorithm, and it solves the class of all finite structures $\bB$ with tree duality (Theorem~\ref{thm:td}). 
In contrast, it is not known whether there exists a polynomial-time uniform algorithm that solves the class of all finite structures $\bB$ with a Maltsev polymorphism. Finding polynomial-time uniform algorithms for larger and larger classes $\mathcal C$ is an active line of research in constraint satisfaction. 
%Uniform algorithms are particularly useful 

In this section we present another important  uniform polynomial-time CSP algorithm, 
the affine integer programming relaxation (AIP). 
%For each of these algorithms $A$, 
We also present a characterisation of the largest class $\mathcal C$ that is solved by this algorithm. This characterisation is inspired by 
the characterisation for the arc-consistency procedure (Section~\ref{sect:ACrevisited}).
This class contains many important CSPs of unbounded width. It is a current research theme in constraint satisfaction research to combine it with bounded width algorithms in order to obtain a polynomial-time uniform algorithm that solves the class of all finite structures $\bB$ with a Taylor polymorphism. 
%It will be particularly elegant to  present them in the terminology of minions from Section~\ref{sect:minions};
%this is why we revisit the arc consistency procedure in Section~\ref{sect:ACrevisited}. 

%Let ${\mathcal C}$ be a class of finite relational structures. We say that an algorithm is
%Definition of uniform algorithm. 

\subsection{The Affine Integer Programming Relaxation}
%In this section we define the basic 
%and characterise the power of the resulting polynomial-time algorithm for CSPs. 
The idea of the affine integer programming 
relaxation (AIP)  is to formulate the question 
whether there is a homomorphism from 
$\bA$ to $\bB$ as a 0-1 integer program in a canonical way, and to then relax the domains of the variables in the program to ${\mathbb Z}$ (instead of $\{0,1\}$).
The motivation is that
\begin{itemize}
\item systems of linear equations over the integers can be solved in polynomial time (see, e.g.,~\cite{Schrijver}),
and 
\item this is a flexible way of dealing with equation systems over finite rings that can be expressed in $\bB$. 
\end{itemize}

\begin{definition}[$\AIP(\bA,\bB)$]
\label{def:AIP}
Let $\bA$ and $\bB$ be structures with finite domains and the same finite relational signature $\tau$. Then $\AIP(\bA,\bB)$ is the following system of linear equations.

For every $x \in A$ and $b \in B$ there is a variable $\mu(x,b)$, and for every $R \in \tau$ of arity $k$, $y \in R^{\bA}$, and $t \in R^{\bB}$ there is another variable $\mu_{R}(y,t)$. 
The linear equations are:
\begin{align}
 \sum_{b \in B} \mu(x,b) & = 1 && \text{ for every } x \in A \label{eq:vars} \\
 \sum_{t \in R^{\bB}} \mu_R(y,t) & = 1 && \text{ for every } R \in \tau, y \in R^{\bA} 
 \label{eq:rel-vars}\\
 \mu(y_i,b) & = \sum_{t \in R^{\bB}, t_i = b} \mu_{R}(y,t) && \text{ for every } R \in \tau, y \in R^{\bA}, i \in [\ar(R)] , b \in B. 
 \label{eq:constraints}
\end{align}
%\begin{align*}
% \sum_{b \in B} \mu(x,b) & = 1 && \text{ for every } x \in A \\
% \mu_R(y_i,b) & = \sum_{t \in R^{\bB}, t_i = b} \mu_{R}(y,t) && \text{ for every } R \in \tau, y \in R^{\bA}, i \in [\ar(R)] , b \in B. 
%\end{align*}
\end{definition}

\begin{remark}
\eqref{eq:rel-vars} is redundant: it follows from~\eqref{eq:vars} together with~\eqref{eq:constraints}. 
\end{remark}

\begin{remark}\label{rem:sol}
Note that if there is a homomorphism $h$ from $\bA$ to $\bB$, then 
the system~\eqref{eq:vars} together with~\eqref{eq:constraints} has the following solution: 
\begin{itemize}
\item set $\mu(x,b)$ to $1$ if $h(x) = b$, and to $0$ otherwise, 
and 
\item set $\mu_R(y,t)$ to $1$ if 
$(h(y_1),\dots,h(y_{\ar(R)})) = t$, and to $0$ otherwise. 
\end{itemize} 
Conversely, every solution to ~\eqref{eq:vars} and~\eqref{eq:constraints}
which only takes values in $\{0,1\}$ corresponds to a homomorphism from $\bA$ to $\bB$. 
\end{remark}

Remark~\ref{rem:sol} implies that if $\AIP(\bA,\bB)$ does not have a solution where the variables take values in ${\mathbb Z}$, then there is no homomorphism from $\bA$ to $\bB$. 

\begin{definition}
We say that \emph{AIP solves $\Csp(\bB)$} if
for every finite structure $\bA$, there is 
a homomorphism from $\bA$ to $\bB$
if and only if $\AIP(\bA,\bB)$ has a solution where the variables take values in ${\mathbb Z}$. 
\end{definition} 

\begin{remark}
If AIP solves $\Csp(\bB)$, then 
$\Csp(\bB)$ is in P, because 
$\AIP(\bA,\bB)$ can be solved in polynomial time over ${\mathbb Z}$, as we have mentioned earlier~\cite{Schrijver}. 
Since the size of $\AIP(\bA,\bB)$ is polynomial in the representation size of $\bA$ and $\bB$,
  the resulting algorithm is uniform (for the class of all $\bB$ such that AIP solves $\Csp(\bB)$, in the sense of Definition~\ref{def:uniform}).  
\end{remark} 

\begin{remark} 
In the solution in Remark~\ref{rem:sol}, all the variables take values in $\{0,1\}$, but the constraints are still interpreted over ${\mathbb Z}$, rather than ${\mathbb Z}_2$. 
Over ${\mathbb Z}_2$, we could find such solutions in polynomial time. In contrast, the relation  defined by $x+y+z = 1$ over ${\mathbb Z}$ and restricted to $\{0,1\}$ has an NP-hard CSP  (Exercise~\ref{exe:oit}). 
\end{remark}

\begin{remark} 
Note that $\AIP(\bA,\bB)$ can be seen
as an instance of the CSP for the structure
with the infinite domain ${\mathbb Z}$ and the countably infinite signature which contains relations for every
linear equation with integer coefficients.
Since we do not need this perspective, we do not formalise this here (we focus on \emph{finite domain}  CSPs in this text, and furthermore we have defined the CSP only for \emph{finite relational signatures}). 
\end{remark} 

\subsection{Alternating Polymorphisms}
Our goal in this section is to characterise the power of AIP 
algebraically. We first present the relevant sets of height-one identities. 

\begin{definition}\label{def:altern}
An operation $f$ of arity $2n+1$, for $n \geq 1$, is called 
\emph{alternating} if it satisfies 
$$f(x_1,\dots,x_{2n+1}) \approx f(x_{\pi(1)},\dots,x_{\pi(2n+1)})$$
for all permutations $\pi$ that \emph{preserve parity}, i.e., $\pi(i) \equiv i \pmod 2$ for all $i \in [2n+1]$, and 
$$f(x_1,\dots,x_{2n-1},y,y) \approx f(x_1,\dots,x_{2n-1},z,z).$$
\end{definition} 

\begin{example}
Examples of alternating operations come from 
\emph{alternating sums}:
for example, the operation $f \colon {\mathbb Z}^{2n+1} \to {\mathbb Z}$, for $n \geq 1$, given by $$(x_1,\dots,x_{2n+1}) \mapsto x_1 - x_2 + x_3 - \cdots + x_{2n+1}$$
is alternating. 
This example can be generalised from ${\mathbb Z}$ to any abelian group. 
\end{example} 

The following example is from~\cite{BrakensiekG21}. 

\begin{example}
Let $a(x_1,\dots,x_{2n+1}):=x_1-x_2+x_3-\cdots+x_{2n+1}$ on ${\mathbb Z}$. The \emph{alternating threshold function of arity $2n+1$} is the Boolean function $t \colon {\mathbb Z}_2^{2n+1} \to {\mathbb Z}_2$ defined by  
$$t(x_1,\dots,x_{2n+1}) := \begin{cases}
1 & \text{ if } x_1-x_2+x_3-\cdots + x_{2n+1} > 0 \\
0 & \text{otherwise.} \end{cases}$$
This example is related to the previous one.
Note that the algebra $({\mathbb Z}_2;t)$ is the reflection of $({\mathbb Z};a)$ with respect to $(h,g)$ where $h \colon {\mathbb Z} \to {\mathbb Z}_2$ maps $x$ to $1$ if $x \geq 1$, and to $0$ otherwise, 
and where $g \colon {\mathbb Z}_2 \to {\mathbb Z}$ is the inclusion map (see Definition~\ref{def:refl}). 
\end{example}

Theorem~\ref{thm:AIP} below implies 
that AIP solves $\Csp(\bB)$ if and only if $\bB$ has an alternating polymorphism of arity $2n+1$ for every $n \geq 1$; 
it is a special case of a more general theorem about the power of AIP for promise CSPs~\cite[Theorem 7.19]{BartoBKO21}. 
For the proof, it will be convenient to follow the strategy that we used to prove the various characterisations of the power of the arc-consistency procedure; it is obtained from specialising the proof of Theorem 7.19 in~\cite{BartoBKO21}. In particular, we define a concept that plays the role of the powerset structure (Exercise~\ref{exe:gac}).

\begin{definition}[$\IP(\bB)$]
\label{def:IP}
Let $\bB$ be a finite structure with finite relational signature $\tau$. Then $\IP(\bB)$ is the $\tau$-structure whose domain consists of all functions $\phi \colon B \to {\mathbb Z}$ such that 
\begin{align}
\sum_{b \in B} \phi(b) = 1.
\label{eq:ipvars1}
\end{align}
 If $R \in \tau$ has arity $k$, 
then the relation $R^{\IP(\bB)}$ consists of the set of all $k$-tuples $(\phi_1,\dots,\phi_k)$ such that there exists a function $\gamma \colon R^{\bB} \to {\mathbb Z}$ such that 
\begin{align}
\sum_{t \in R^{\bB}} \gamma(t) & = 1 \label{eq:ipvars} \\
\text{ and } \quad \phi_i(b) & = \sum_{t \in R^{\bB}, t_i = b} \gamma(t) && \text{for all $b \in B$ and $i \in [k] $.} \label{eq:ipconstraints}
\end{align}
\end{definition} 

The following lemma is analogous to Lemma~\ref{lem:ac} for the arc-consistency procedure. 

\begin{lemma}\label{lem:AIP} 
Let $\bA$ and $\bB$ be finite structures with finite relational signature $\tau$. 
Then the following are equivalent. 
\begin{itemize}
\item $\AIP(\bA,\bB)$ has a solution where the variables take values in ${\mathbb Z}$;
\item $\bA$ has a homomorphism to $\IP(\bB)$. 
\end{itemize} 
\end{lemma} 
\begin{proof}
Suppose first that $h\colon\bA\to\IP(\bB)$ is a homomorphism. Set
$\mu(x,b):=h(x)(b)$. For every $R\in\tau$ of arity $k$ and
$y\in R^{\bA}$, choose a function
$\gamma_{R,y}\colon R^{\bB}\to{\mathbb Z}$ witnessing that
$(h(y_1),\ldots,h(y_k))\in R^{\IP(\bB)}$, and set
$\mu_R(y,t):=\gamma_{R,y}(t)$. The defining equations of
$\IP(\bB)$ are precisely the equations of $\AIP(\bA,\bB)$, so
these values form an integer solution.

Conversely, given an integer solution $\mu$ of $\AIP(\bA,\bB)$,
define $h(x)(b):=\mu(x,b)$. Equation~\eqref{eq:vars} shows that
$h(x)\in\IP(\bB)$. Moreover, for $R\in\tau$ and $y\in R^{\bA}$,
the function $t\mapsto\mu_R(y,t)$ witnesses that
$(h(y_1),\ldots,h(y_k))\in R^{\IP(\bB)}$, by the remaining
equations of $\AIP(\bA,\bB)$. Hence $h$ is a homomorphism from
$\bA$ to $\IP(\bB)$.
\end{proof}

\begin{theorem}\label{thm:AIP}
% SOURCE? 
Let $\bB$ be a finite non-empty structure with finite relational signature $\tau$. Then the following are equivalent.
\begin{enumerate}
\item AIP solves $\Csp(\bB)$; 
%$\AIP(\bA,\bB)$ has a solution over ${\mathbb Z}$. 
%\item $\bB$ has a pp-construction in 
\item $\IP(\bB)$ homomorphically maps to $\bB$. 
\item $\bB$ has a pp-construction in $({\mathbb Z};\{(a,b,c) \mid a+b=c\},\{1\})$;
\item There is a minion homomorphism from
$\Clo({\mathbb Z};(x,y,z) \mapsto x-y+z)$ to $\Pol(\bB)$; 
%where 
%$m$ is given by $(x,y,z) \mapsto x-y+z$; 
\item $\bB$ has an alternating polymorphism of arity $2n+1$ for every $n \geq 1$. 
\end{enumerate} 
\end{theorem} 
\begin{proof}
%[Proof of Theorem~\ref{thm:AIP}]
$1.\Rightarrow 2.$ 
%Suppose that AIP solves $\Csp(\bB)$. 
By compactness\footnote{See, e.g.,~\cite{BodMathLogic}.}, it suffices to verify that every finite substructure $\bA$ of $\IP(\bB)$ homomorphically maps to $\bB$. 
Since $\IP(\bB) \to \IP(\bB)$, we have
$\bA \to \IP(\bB)$, and by Lemma~\ref{lem:AIP} we get that 
 $\AIP(\bA,\bB)$ has a solution. 
Since AIP solves $\Csp(\bB)$, this implies that there exists a homomorphism from $\bA$ to $\bB$. 

\medskip 
$2.\Rightarrow 1.$ Suppose that $\IP(\bB)$ has a homomorphism $h$ to $\bB$. 
Let $\bA$ be a finite $\tau$-structure. 
If $\AIP(\bA,\bB)$ has no solution, then $\bA \not \to \bB$. If $\AIP(\bA,\bB)$ does have a solution, then $\bA \to \IP(\bB)$ by Lemma~\ref{lem:AIP}. 
Since $\IP(\bB)$ has a homomorphism to $\bB$, we compose the homomorphisms and obtain a homomorphism from $\bA$ to $\bB$, as desired. 

\medskip 
$2.\Rightarrow 3.$ 
We show that $\bB$ is homomorphically equivalent to a $|B|$-th pp-power $\bP$ of $({\mathbb Z};\{(a,b,c) \mid a+b = c\};\{1\})$. We view the domain elements of $\bP$ as functions from $B$ to ${\mathbb Z}$. 
For $R \in \tau$ of arity $k$, 
the relation $R^{\bP}$ contains all tuples 
$(\phi_1,\dots,\phi_k)$ such that 
\begin{itemize}
\item $\phi_i$ satisfies~\eqref{eq:ipvars1}, for every $i \in \{1,\dots,k\}$, and 
\item there exists 
$\gamma \colon R^{\bB} \to {\mathbb Z}$ such that~\eqref{eq:ipvars} and~\eqref{eq:ipconstraints} are satisfied. 
\end{itemize} 
Note that this relation is pp-definable as a $|B| \cdot k$-ary relation over $({\mathbb Z};\{(a,b,c) \mid a+b = c\};\{1\})$. 

We construct homomorphisms 
$h \colon \bP \to \bB$ and $g \colon \bB \to \bP$. 
For $b \in B$, let $\chi_b$ be given by $\chi_b(b) = 1$ and $\chi_b(b') = 0$ for $b' \in B \setminus \{b\}$. 
Then $g(b) := \chi_b$ is a homomorphism from $\bB$ to $\bP$.

Conversely, let $f\colon\IP(\bB)\to\bB$ be the homomorphism which
exists by assumption, and let $h\colon P\to B$ be any extension of
$f$. We claim that $h$ is a homomorphism. Indeed, let $R\in\tau$ be
$k$-ary and let $(\phi_1,\dots,\phi_k)\in R^{\bP}$. By the definition
of $R^{\bP}$, every $\phi_i$ satisfies
$\sum_{b\in B}\phi_i(b)=1$.
Thus, every $\phi_i$ belongs to the domain of $\IP(\bB)$. Moreover, the
function $\gamma$ witnessing that
$(\phi_1,\dots,\phi_k)\in R^{\bP}$ also witnesses that
$(\phi_1,\dots,\phi_k)\in R^{\IP(\bB)}$. 
Since $h$ agrees with $f$ on the domain of $\IP(\bB)$ and $f$ is a
homomorphism, it follows that
\[
\bigl(h(\phi_1),\dots,h(\phi_k)\bigr)
=
\bigl(f(\phi_1),\dots,f(\phi_k)\bigr)
\in R^{\bB}.
\]
Hence, $h$ is a homomorphism. Notice that the values chosen for $h$ on
$P\setminus\IP(\bB)$ are irrelevant, since such elements do not occur
in any tuple of any relation of $\bP$.

\medskip 
$3.\Rightarrow 4.$ follows from general principles (see Remark~\ref{rem:forward-lin-birk}) and the observation that $(x,y,z) \mapsto x-y+z$ is idempotent and preserves both the addition graph $\{(a,b,c) \mid a+b=c\}$ and the singleton $\{1\}$. 
%$\Pol({\mathbb Z};\{(a,b,c) \mid a+b+c = 0\};\{1\}) = \Clo({\mathbb Z};$. 

\medskip 
$4.\Rightarrow 5.$ is clear, because $\Clo({\mathbb Z};(x,y,z) \mapsto x-y+z)$ contains the alternating sum operation of arity $2n+1$, for every $n \geq 1$, which is alternating, and minion homomorphisms preserve the existence of such operations. 

\medskip 
$5.\Rightarrow 2.:$ 
For $\ell \geq 1$, define 
$\IP_{\ell}(\bB)$ to be the $\tau$-structure 
whose domain is the set of all functions 
$\phi \colon B \to {\mathbb Z}$ such that $\sum_{b \in B} | \phi(b) | \leq 2 \ell +1$ and
$\sum_{b \in B} \phi(b) = 1$. 
The relations of $\IP_{\ell}(\bB)$
are defined similarly as in $\IP(\bB)$
with the difference that for $R \in \tau$
the witnessing function $\gamma \colon R^{\bB} \to {\mathbb Z}$ additionally satisfies
$\sum_{t \in R^{\bB}} |\gamma(t)| \leq 2 \ell + 1$. Note that every finite substructure of
$\IP(\bB)$ is a substructure of $\IP_{\ell}(\bB)$ for a sufficiently large $\ell$, so by compactness
%\footnote{See, e.g.,~\cite{BodMathLogic}.} 
it suffices to find a homomorphism $h_\ell \colon \IP_{\ell}(\bB) \to \bB$ for every $\ell \geq 1$. Fix a $(2 \ell +1)$-ary alternating operation $a_{2 \ell + 1} \in \Pol(\bB)$. For an element $\phi$ of 
$\IP_{\ell}(\bB)$, 
define $h_{\ell}(\phi)$ as follows.
We claim that we can find 
$c_1,\dots,c_{2 \ell +1} \in B$ such that for every $b \in B$ the difference of the number of times $b$ appears among the $c_i$'s with odd and with even indices is exactly $\phi(b)$: e.g., if $B = \{b_1,\dots,b_n\}$,
recall that $\phi(b_1) + \cdots + \phi(b_n) = 1$. 
For $b\in B$, put
\[
p_b:=\max(\phi(b),0)
\qquad\text{and}\qquad
q_b:=\max(-\phi(b),0),
\]
and let
\[
P:=\sum_{b\in B}p_b
\qquad\text{and}\qquad
N:=\sum_{b\in B}q_b.
\]
Since $\sum_{b\in B}\phi(b)=1$, we have $P-N=1$. Moreover,
\[
P+N=\sum_{b\in B}|\phi(b)|\leq 2\ell+1.
\]
It follows that $N\leq\ell$ and $P=N+1\leq\ell+1$.

We now construct $c_1,\dots,c_{2\ell+1}$. For every $b\in B$, place
$p_b$ copies of $b$ in odd positions and $q_b$ copies of $b$ in even
positions. This uses $P$ of the $\ell+1$ odd positions and $N$ of the
$\ell$ even positions. Since $P=N+1$, the same number
\[
\ell-N=\ell+1-P
\]
of odd and even positions remains. Fix an arbitrary $b_0\in B$ and
fill the remaining positions in odd-even pairs with $b_0$; more
explicitly, after permuting positions of the same parity if necessary,
put
\[
c_{2j}=c_{2j+1}:=b_0
\qquad\text{for }j=N+1,\dots,\ell.
\]
Then, for every $b\in B$,
\[
\bigl|\{i\text{ odd}\mid c_i=b\}\bigr|
-
\bigl|\{i\text{ even}\mid c_i=b\}\bigr|
=p_b-q_b=\phi(b).
\]
Define
\[
h_\ell(\phi):=
a_{2\ell+1}(c_1,\dots,c_{2\ell+1}).
\]
This value depends only on the signed multiplicities $\phi(b)$:
parity-preserving permutations allow us to reorder the odd and even
positions independently, and the second alternating identity allows
us to replace any equal odd-even pair by any other equal odd-even
pair.

It remains to prove that $h_\ell$ is a homomorphism. Let $R\in\tau$
be $k$-ary and let
\[
(\phi_1,\dots,\phi_k)\in R^{\IP_\ell(\bB)}.
\]
Choose a witnessing function $\gamma\colon R^{\bB}\to{\mathbb Z}$.
Thus
\[
\sum_{t\in R^{\bB}}\gamma(t)=1,
\qquad
\sum_{t\in R^{\bB}}|\gamma(t)|\leq 2\ell+1,
\]
and
\[
\phi_i(b)=
\sum_{\substack{t\in R^{\bB}\\t_i=b}}\gamma(t)
\qquad
\text{for every $i\in[k]$ and $b\in B$.}
\]

Applying the signed-multiset construction above to $\gamma$, we obtain
tuples
\[
t_1,\dots,t_{2\ell+1}\in R^{\bB}
\]
such that, for every $t\in R^{\bB}$,
\[
\bigl|\{j\text{ odd}\mid t_j=t\}\bigr|
-
\bigl|\{j\text{ even}\mid t_j=t\}\bigr|
=\gamma(t).
\]
Here the unused positions are filled with equal odd-even pairs of an
arbitrary tuple from $R^{\bB}$; such a tuple exists because
$\sum_{t\in R^{\bB}}\gamma(t)=1$.

For every $i\in[k]$ and $b\in B$, it follows that
\begin{align*}
&\bigl|\{j\text{ odd}\mid t_j(i)=b\}\bigr|
-
\bigl|\{j\text{ even}\mid t_j(i)=b\}\bigr|\\
&\qquad =
\sum_{\substack{t\in R^{\bB}\\t_i=b}}\gamma(t)
=\phi_i(b).
\end{align*}
Consequently, by the well-definedness of $h_\ell$,
\[
h_\ell(\phi_i)
=
a_{2\ell+1}\bigl(t_1(i),\dots,t_{2\ell+1}(i)\bigr)
\qquad
\text{for every $i\in[k]$.}
\]
Since $a_{2\ell+1}$ is a polymorphism of $\bB$ and
$t_1,\dots,t_{2\ell+1}\in R^{\bB}$, we obtain
\[
\bigl(
a_{2\ell+1}(t_1(1),\dots,t_{2\ell+1}(1)),
\dots,
a_{2\ell+1}(t_1(k),\dots,t_{2\ell+1}(k))
\bigr)
\in R^{\bB}.
\]
This tuple is precisely
\[
\bigl(h_\ell(\phi_1),\dots,h_\ell(\phi_k)\bigr).
\]
Thus $h_\ell\colon\IP_\ell(\bB)\to\bB$ is a homomorphism. Since every
finite substructure of $\IP(\bB)$ is contained in
$\IP_\ell(\bB)$ for some $\ell$, compactness yields a homomorphism
$\IP(\bB)\to\bB$.
\end{proof} 

\begin{corollary}
There is a uniform polynomial-time algorithm for the class of all finite structures 
with finite relational signature and 
alternating polymorphisms of arity $2n+1$ for all $n \geq 1$. 
\end{corollary}

%Note that the result does not follow straightforwardly from Gaussian elimination, because 

In particular, we obtain an alternative proof of the following special case that does not rely on the Bulatov-Dalmau algorithm from Section~\ref{sect:buldalalg} (also see Exercise~\ref{exe:GoldmannRussell}). 
The result can also be shown by  linear algebra, generalising Gaussian elimination (see~\cite{GoldmannRussell}), but this is non-trivial and will not be covered in this course. 

\begin{corollary}
Let $G$ be a finite abelian group with elements $g_1,\dots,g_n$. 
%Then $\Csp({\mathbb Z}_n;\{(x,y,z) \mid x+y+z=1\},\{1\})$ 
Then $$\Csp(G;\{(x,y,z) \mid x+y+z=0\},\{g_1\},\dots,\{g_n\})$$ 
can be solved in polynomial time. 
\end{corollary} 

% TODO: 
% every one-sided correct algo can be turned into a singleton algorithm
% Singleton AIP+AC
% D_8.
% Besserer Algorithmus: 
% Speichere fuer jede Teilmenge von B
% eine ganze Zahl. 
% Viele Gleichungen "=1" fuer verschiedene Partitionen von B. 
% Gleichungen fuer Constraints bewirken
% mindestens AC. 
% Liefert leider in dieser Form keinen uniformen Algorithmus mehr. 
% Wie laeuft das auf D_8? 

% END INLINED SOURCE: uniform.tex
%\input few.tex
%\input conservative.tex
%\input minimal-taylor.tex

\section{Open Problems}
The Feder and Vardi dichotomy conjecture~\cite{FederVardi} has been the outstanding open problem in the field; it was solved in 2017 by Bulatov~\cite{BulatovFVConjecture} and, independently, by
Zhuk~\cite{ZhukFVConjecture}. The border between polynomial and NP-hard cases has numerous equivalent logical and algebraic characterisations, for example characterisations based on primitive positive constructability and 
characterisations based on identities that are satisfied by the polymorphism clone (see Corollary~\ref{cor:dicho}). 

There are many interesting problems in the field that are still left open. 
We start with open research problems where all the relevant concepts have already been introduced in the course.  

\begin{enumerate}
\item Is there a uniform algorithm for finite-domain CSPs (see Section~\ref{sect:uniform}), i.e., is there a polynomial-time algorithm that takes as input a pair $(\bA,\bB)$ of finite structures with the same finite relational signature, where $\bB$ is promised to have a Taylor polymorphism (for equivalent formulations of this promise, see Corollary~\ref{cor:dicho}), that decides whether there is a homomorphism from $\bA$ to $\bB$? This question is also open if we replace `Taylor polymorphism' by `Maltsev polymorphism'. 
\item Is there a polynomial-time algorithm to determine whether a given core structure $\bB$ has a Siggers polymorphism? Is this true for the special case where $\bB$ is a digraph or an orientation of a tree? This problem is known to be NP-complete if $\bB$ is not required to be a core structure~\cite{MetaChenLarose}. 
This problem is known to be in P if $\bB$ is a core and there is a uniform algorithm for finite-domain CSPs (as in the previous question; see~\cite{MetaChenLarose}). 
\item Is the class of all finite structures (with potentially infinite signature!), ordered by pp-constructability and factored by the respective equivalence relation~\cite{PPPoset,Collapse,meyerstarke2025}, countably infinite or uncountably infinite? Is it a lattice? Are there infinite ascending chains? Can we answer these questions when restricting to finite directed graphs~\cite{smooth-digraphs,maximal-digraphs,StarkeDiss}? 
\item What is the computational complexity of  determining whether 
a given finite core  structure $H$ has tree duality? Is this problem in P? Is it in P if $H$ is a digraph or even an orientation of a tree? 
\item (Bul\'in~\cite{Bulin18}) Is it true that the CSP of an orientation of a tree is in P if and only if it can be solved by $k$-consistency, for some $k$? 
%\item Is it true that if $H$ is an orientation of a tree, and $\Csp(H)$ has a Siggers polymorphism, 
%then $\Csp(H)$ can be solved by 
%the path-consistency procedure? 
\item Is it true that \emph{most} orientations of finite trees are hard, i.e., is it true that 
the probability that an orientation of a tree drawn uniformly at random from the set of all such trees with vertex set $\{1,\dots,n\}$ is NP-hard tends to 1 as $n$ tends to infinity~\cite{otrees}? The answer is yes if we ask the question for random labelled digraphs instead of random labelled trees~\cite{LuczakNesetril}. 
\item 
 Determine the smallest trees whose CSP is P-hard (under logspace reductions, and assuming that $\operatorname{NL}\neq\operatorname{P}$). 
 % TODO: add these concepts the the complexity primer at the end. 
 It is known that they must have at least 16 vertices, since all smaller trees have a majority polymorphism and thus are in NL~\cite{otrees}. 
 \item Is the CSP of most digraphs with a Taylor polymorphism P-hard? 
\item Is the CSP of most digraphs with a Taylor polymorphism not solvable by $k$-consistency, for all $k$? 
\end{enumerate}
The following open problems require knowledge of concepts that have not been covered in this course; however, references are provided where these concepts are defined formally. 
\begin{enumerate}
\item Prove that a finite-domain CSP is in P if and only if it can be expressed in \emph{Choiceless Polynomial Time}~\cite{blass1999choiceless}. 
%\item Prove that a finite-domain CSP is in P if and only if it can be expressed in \emph{fixed point logic} expanded by a \emph{rank operator} (with respect to some finite abelian group~\cite{gradel2019rank,dawar2009logics}).
\item (Dalmau~\cite{LinearDatalog}) Is it true that if $\Csp(H)$ is in NL, then $\Csp(H)$ is in linear Datalog? 
Is this at least true for digraphs $H$? The same question would already be interesting for orientations of trees. 
\item (Egri-Larose-Tesson~\cite{EgriLaroseTessonLogspace})
Is it true that if $\Csp(H)$ is in L, then $\Csp(H)$ is in symmetric Datalog? Is this at least true for digraphs $H$? It would already be interesting for orientations of trees. 
\item (Larose-Tesson~\cite{LaroseTesson})
Is it true that if the polymorphism algebra of $H$ generates a congruence join-semidistributive variety, then $\Csp(H)$ is in linear Datalog? Is this at least true for digraphs $H$? It would already be interesting for orientations of trees. 
\item Is it true that if $\Csp(H)$ is not P-hard under logspace reductions, then it is in NC?
It is known that NC is closed under logspace reductions, and it is believed that P is different from NC. Moreover, the CSP for the structure $(\{0,1\}; \{0,1\}^3 \setminus \{(1,1,0)\}, \{0\}, \{1\})$ is P-hard (see Exercise~\ref{exe:horn}). 
Is it true that if $\Csp(H)$ does not pp-construct this structure then $\Csp(H)$ is in NC? 
\end{enumerate}
\noindent 
%\newpage
Finally we pose some curious open questions for concrete finite digraphs.
% where we do not know the answer. 
%finite combinatorial puzzles that 
\begin{enumerate}
\item Is the CSP of the orientation of a tree displayed on the right \\
in NL~\cite{otrees}? 
\vspace{-.9cm}
\begin{flushright}
\begin{tikzpicture}[scale=.7]
% Level 0
\node[bullet] (12) at (1.5,0) {};
% Level 1
\node[bullet] (1) at (0.5,1) {};
\node[bullet] (10) at (3,1) {};
\node[bullet] (11) at (1.5,1) {};
% Level 2
\node[bullet] (2) at (0,2) {};
\node[bullet] (0) at (1,2) {};
\node[bullet] (7) at (3,2) {};
\node[bullet] (13) at (2,2) {};
% Level 3
\node[bullet] (3) at (0,3) {};
\node[bullet] (6) at (1,3) {};
\node[bullet] (8) at (3,3) {};
\node[bullet] (14) at (2,3) {};
% Level 4
\node[bullet] (4) at (0,4) {};
\node[bullet] (9) at (3,4) {};
\node[bullet] (15) at (2,4) {};
% Level 5
\node[bullet] (5) at (0,5) {};
\path[->,>=stealth']
(1) edge (2)
(1) edge (0)
(2) edge (3)
(3) edge (4)
(4) edge (5)
(0) edge (6)
(7) edge (6)
(7) edge (8)
(10) edge (7)
(8) edge (9)
(11) edge (0)
(11) edge (13)
(12) edge (11)
(13) edge (14)
(14) edge (15)
;
\end{tikzpicture}
\end{flushright}
\vspace{-3.7cm}
\item What are the smallest digraphs with a Taylor polymorphism that \\
cannot be solved by Datalog? 
\end{enumerate}

\medskip 
For CSPs over infinite domains, there are numerous open problems, \\
and I invite 
the reader to have a look at~\cite{Book}. 

\bibliographystyle{abbrv}
\bibliography{local}

\appendix
% BEGIN INLINED SOURCE: O-Notation.tex
%!TEX root = Graph-Homomorphisms.tex

\section{O-notation}
The letters $o$ and $O$ stand for the \emph{order} of growth of a function. The \emph{big-O notation} expresses asymptotic upper bounds, while the \emph{little-o notation} expresses strictly smaller order of growth. We mention that there exists related
notation to describe other kinds of bounds on asymptotic growth, e.g., $\Theta$, $\Omega$, $\omega$, of which we only need $\Theta$ in this text, so we skip the definitions of the others. 

Let $g \colon {\mathbb R} \to {\mathbb R}$ be eventually nonzero
(we use ${\mathbb R}$ for convenience; similar definitions exist for other domains such as ${\mathbb N}$ and ${\mathbb Q}$).
Then $O(g)$ is the set of all functions $f \colon {\mathbb R} \to {\mathbb R}$ such that there exist $c>0$ and $x_0 \in {\mathbb R}$ with $|f(x)| \leq c|g(x)|$ for all $x \geq x_0$. 
Note that 
$$f \in O(g) \Leftrightarrow \limsup_{x \rightarrow \infty} \left | \frac{f(x)}{g(x)} \right| < \infty.$$
In typical usage, the formal definition of $O(g)$ is not used directly; rather, we first use the following simplification rules:
\begin{itemize}
\item if $g(x)$ is a sum of several terms, if there is one with largest growth rate, then we drop all other terms; 
\item if $g(x) = c \cdot f(x)$ and $c$ is a constant that does not depend on $x$, then $c$ can be omitted.
%\item if $g(x) = f_1(x) f_2(x)$ then $O(g) = O(f_1) O(f_2) 
\end{itemize}
When we write $O(g)$, we typically choose 
$g$ to be as simple as possible. 
$O$-notation can also be used within arithmetic terms. For example, $h + O(g)$ denotes the set of functions of the form $h + f$ for 
$f \in O(g)$. In other words, $k \in h + O(g)$ 
is equivalent
to $k - h \in O(g)$. 

We write $o(g)$ for the set of all functions $f \colon {\mathbb R} \to {\mathbb R}$
such that for every $\epsilon \in {\mathbb R}_{>0}$ there exists $x_0 \in {\mathbb R}$ such that $|f(x)| \leq \epsilon |g(x)|$ for all $x \geq x_0$. 
Informally, $f \in o(g)$ means that $g$ grows much faster than $f$. For example, $x \mapsto 2x$ is in $o(x \mapsto x^2)$, and $x \mapsto 1/x$ is in $o(1)$. Note that $o(g) \subseteq O(g)$, and that 
$$f \in o(g) \Leftrightarrow \lim_{x \to \infty} \frac{f(x)}{g(x)} = 0.$$
Similarly as in the case of the $O$-notation we may use the $o$-notation in arithmetic expressions. Note that if $f \in o(g)$ and 
$c$ is a constant, then $cf \in o(g)$. 
Frequent notation is to write
$f \ll g$ (or $g \gg f$) if $f \in o(g)$. 

We write $\Theta(g)$ for the set of
all functions $f$ such that there are constants $c,C>0$ and $x_0 \in {\mathbb R}$ such that $c|g(x)| \leq |f(x)| \leq C|g(x)|$ for every $x \geq x_0$. 
In other words, $f \in \Theta(g)$ if $f \in O(g)$ and $g \in O(f)$. 

Finally, we write $f(x) \sim g(x)$ if 
$$ \lim_{x \to \infty} \frac{f(x)}{g(x)} = 1$$
and we say that $f$ and $g$ are \emph{asymptotically equivalent} (for $x \to \infty$). 
% END INLINED SOURCE: O-Notation.tex
% BEGIN INLINED SOURCE: Complexity-Expanded.tex
%!TEX root = GH-UA.tex

\section{Basics of Complexity Theory}
\label{sect:complexity}

This appendix recalls the complexity-theoretic notions that are used in the
course.  It is not intended as a replacement for an introductory course in
complexity theory; for a more systematic treatment we refer
to~\cite{Papa,GareyJohnson}.

\medskip
{\bf Words, languages, and decision problems.}
For a finite set $A$, we write $A^*$ for the set of all finite words over the
alphabet $A$.  A word of length $n$ over $A$ can be viewed as a function from
$\{1,\dots,n\}$ to $A$.  We write $\epsilon$ for the empty word and $|w|$ for
the length of a word $w$.

A \emph{decision problem} is a question with a yes-or-no answer.  After choosing
an encoding of its instances by words over $\{0,1\}$, it can be identified with
the set $L \subseteq \{0,1\}^*$ of encodings of yes-instances.  Such a set is
also called a \emph{formal language}.  Equivalently, we may use its
characteristic function $f_L \colon \{0,1\}^* \to \{0,1\}$, where $f_L(w)=1$
if and only if $w \in L$.  Complexity classes such as P and NP are usually
defined as classes of languages.  In informal statements we also say that the
corresponding decision problems are in P or in NP.

The instances in this text are normally finite graphs, digraphs, or relational
structures rather than words.  This causes no difficulty, because finite
structures admit natural finite encodings.

\medskip
{\bf Coding finite structures.}
Let $\tau$ be a fixed finite relational signature.  A finite $\tau$-structure
$\bA$ with domain $\{1,\dots,n\}$ can be encoded by writing $n$ in unary and,
for every relation symbol $R \in \tau$, listing the tuples in $R^{\bA}$.  We
use separators such as $\#$ to make the encoding unambiguous.  Alternatively,
we may use the characteristic table of each relation.  Since the signature and
all arities are fixed, these standard encodings can be translated into one
another in polynomial time.

For example, a digraph $G$ with $n$ vertices can be encoded by its $n \times n$
adjacency matrix.  The resulting word has length of order $n^2$.  An adjacency
list can be shorter for sparse digraphs, but the two representations are still
polynomially related.  Therefore the assertion that a problem is solvable in
polynomial time does not depend on which of these standard explicit encodings
we choose.  It can depend on unreasonable encodings: writing $n$ in unary or
in binary, for example, may change the input length exponentially.  Whenever
we speak about the size of a graph or structure, a standard explicit encoding
is understood.

\medskip
{\bf Turing machines and running time.}
There are several mathematically precise models of computation.  The standard
one is the Turing machine.  A deterministic one-tape Turing machine consists of
a finite set $Q$ of states, a finite tape alphabet $\Gamma$ containing the input
symbols $0,1$ and a blank symbol, a read-write head, and a finite transition
function.  A transition depends only on the current state and the symbol under
the head; it changes the state, writes a symbol, and moves the head one cell to
the left or to the right.  There are distinguished start, accept, and reject
states.  The input word is initially written on the tape and all other cells
are blank.

A machine \emph{decides} a language if it halts on every input and accepts
precisely the words in the language.  Its running time on inputs of length $n$
is the largest number of transitions performed on any such input.  We say that
it runs in \emph{polynomial time} if this number is at most $O(n^k)$ for some
$k \in {\mathbb N}$.  Multi-tape Turing machines, random-access machines, and
the usual programming-language models give the same notion of polynomial-time
computability, up to a polynomial change in running time.  This robustness is
one reason why polynomial time is used as the basic mathematical model of
efficient computation.

\begin{definition}
The class $\operatorname{P}$ consists of all languages that are decided by a
deterministic Turing machine in polynomial time.
\end{definition}

The exponent in the running-time bound is allowed to depend on the algorithm,
but not on the input.  Thus an algorithm with running time $n^{100}$ is a
polynomial-time algorithm, while one with running time $2^n$ is not.  This
definition is asymptotic and deliberately ignores constant factors; see the
preceding appendix on $O$-notation.

\medskip
{\bf Fixed-template constraint satisfaction problems.}
The distinction between fixed and variable parts of the input is essential in
this course.  In $\Csp(\bB)$ the finite template $\bB$ is fixed once and for all;
only the finite instance $\bA$ is part of the input.  Consequently $|B|$ is a
constant in the complexity analysis of $\Csp(\bB)$.  An algorithm may depend on
$\bB$, and even the exponent in its polynomial running-time bound may depend on
$\bB$.

This does not make the problem trivial.  If $|A|=n$ and $|B|=q$, then trying all
maps from $A$ to $B$ takes $q^n$ attempts.  This is exponential in $n$ whenever
$q \geq 2$, even though $q$ is constant.  The algorithms studied in the course,
such as arc-consistency, path-consistency, and the Bulatov-Dalmau algorithm,
avoid this exhaustive search for suitable templates.

The \emph{uniform CSP} has both $\bA$ and $\bB$ as input.  A uniform
polynomial-time algorithm must have a running-time bound that is polynomial in
the combined encoding length, with one exponent valid for all templates.  The
question in Section~\ref{sect:uniform} asks for such an algorithm under an
algebraic promise on $\bB$.  This is a \emph{promise problem}: the algorithm is
required to answer correctly when the promised condition holds, and no
requirement is imposed on inputs that violate the promise.

A \emph{meta-problem} asks about the complexity or the algorithms of
$\Csp(\bB)$ when the template $\bB$ itself is the input.  For example, the
question whether the arc-consistency procedure solves $\Csp(\bB)$ is a meta-problem.  The
power structure $P(\bB)$ used in Section~\ref{sect:AC} has exponentially many
vertices in $|B|$; this is why constructing it explicitly can lead to an
exponential-time meta-algorithm even though every individual execution of 
the arc-consistency procedure is polynomial in the size of its instance.

\medskip
{\bf Certificates and NP.}
Many problems have the following form: a yes-answer can be justified by a
short object whose correctness is easy to check.  For graph colouring, the
object is a colouring; for $\Csp(\bB)$, it is a homomorphism to $\bB$.

\begin{definition}
The class $\operatorname{NP}$ consists of all languages $L$ for which there are
a deterministic polynomial-time Turing machine $M$ and a number
$d \in {\mathbb N}$ such that, for every word $w$,
\begin{multline*}
w \in L \quad\text{if and only if there exists } a \in \{0,1\}^*
\text{ such that}\\
|a| \leq |w|^d+d \quad\text{and}\quad M(w\#a) \text{ accepts.}
\end{multline*}
The word $a$ is called a \emph{certificate} or \emph{witness}.
\end{definition}

The letters NP stand for \emph{nondeterministic polynomial time}.  An
equivalent definition uses nondeterministic Turing machines: such a machine may
choose among several possible transitions, and accepts if at least one
computation path accepts within polynomial time.  Guessing the certificate and
then running the verifier gives the equivalence with the definition above.

\begin{proposition}
For every finite relational structure $\bB$ with finite signature,
$\Csp(\bB)$ is in NP.
\end{proposition}
\begin{proof}
On input $\bA$, a certificate specifies a map $h \colon A \to B$.  Since $B$ is
fixed, this takes $O(|A|)$ symbols.  The verifier checks, for every relation
symbol $R$ and every tuple in $R^{\bA}$, whether its componentwise image under
$h$ belongs to $R^{\bB}$.  This takes polynomial time in the explicit encoding
size of $\bA$.
\end{proof}

The same proof applies to the uniform CSP: the certificate has length
$O(|A|\log |B|)$, and relation membership can be checked from the encoding of
$\bB$.  Notice that NP only requires short certificates for yes-instances.  It
does not provide certificates for no-instances.

Clearly $\operatorname{P} \subseteq \operatorname{NP}$: a verifier may ignore
the certificate and run the polynomial-time decision algorithm.  The question
whether $\operatorname{P}=\operatorname{NP}$ is one of the central open
problems in mathematics and theoretical computer science.

\medskip
{\bf Polynomial-time reductions.}
A reduction translates instances of one problem into instances of another
while preserving yes- and no-answers.

\begin{definition}
Let $L_1,L_2 \subseteq \{0,1\}^*$.  A \emph{polynomial-time many-one
reduction} from $L_1$ to $L_2$ is a polynomial-time computable function
$g \colon \{0,1\}^* \to \{0,1\}^*$ such that
$$
w \in L_1 \quad\text{if and only if}\quad g(w) \in L_2.
$$
We then write $L_1 \leq_m^{\operatorname{P}} L_2$.
\end{definition}

The expression ``$L_1$ reduces to $L_2$'' indicates that $L_2$ is at least as
hard as $L_1$: an algorithm for $L_2$, preceded by the translation $g$, gives
an algorithm for $L_1$.  The direction of this implication is worth keeping in
mind.

\begin{proposition}
Polynomial-time many-one reductions have the following properties.
\begin{enumerate}
\item They are transitive.
\item If $L_1 \leq_m^{\operatorname{P}} L_2$ and $L_2$ is in P, then $L_1$ is
in P.
\item If $L_1$ is NP-hard and $L_1 \leq_m^{\operatorname{P}} L_2$, then $L_2$
is NP-hard.
\end{enumerate}
\end{proposition}
\begin{proof}
For the first statement, compose the two reduction functions.  A composition
of polynomial-time computable functions is polynomial-time computable; the
length of the intermediate word is polynomially bounded.  The second statement
follows by composing the reduction with the decision algorithm for $L_2$, and
the third follows from the first.
\end{proof}

Many reductions in this course have substantially better running time.  For
example, Lemma~\ref{lem:pp-reduce} replaces every constraint by a fixed
primitive positive definition and yields a linear-time reduction.  Primitive
positive interpretations and constructions similarly produce efficient
reductions between CSPs.

\medskip
{\bf Hardness and completeness.}
Let $\mathcal C$ be a complexity class and fix a notion of reduction.  A
language $L$ is \emph{$\mathcal C$-hard} if every language in $\mathcal C$
reduces to $L$.  It is \emph{$\mathcal C$-complete} if it is $\mathcal C$-hard
and also belongs to $\mathcal C$.

In particular, a problem is \emph{NP-hard} if every problem in NP has a
polynomial-time many-one reduction to it, and \emph{NP-complete} if it is both
NP-hard and in NP.  The $3$-colouring problem, equivalently $\Csp(K_3)$, is
NP-complete.  Hence, to prove that a problem $L$ is NP-hard, it is enough to
give a polynomial-time reduction from $3$-colouring to $L$; transitivity then
supplies reductions from all problems in NP.

Since every fixed finite-domain CSP is in NP, an NP-hardness proof for such a
CSP automatically proves NP-completeness.  This explains why the main text
occasionally states only the stronger part that is relevant for the argument,
namely NP-hardness.

\begin{proposition}
If an NP-hard problem belongs to P, then $\operatorname{P}=\operatorname{NP}$.
\end{proposition}
\begin{proof}
Every problem in NP reduces to the NP-hard problem and can therefore be solved
in polynomial time.  Hence $\operatorname{NP} \subseteq \operatorname{P}$; the
reverse inclusion always holds.
\end{proof}

Thus, assuming $\operatorname{P}\neq\operatorname{NP}$, an NP-hard problem has
no polynomial-time algorithm.  NP-hardness alone does not imply membership in
NP: optimisation problems and even undecidable problems can be NP-hard.

The dichotomy theorem for finite-domain CSPs says much more than that every
such CSP is either in P or not in P.  Assuming $\operatorname{P}\neq
\operatorname{NP}$, Ladner's theorem~\cite{Ladner} gives problems in NP that
are neither in P nor NP-complete.  The dichotomy theorem asserts that none of
these intermediate degrees occurs among finite-domain CSPs.

\medskip
{\bf Complement classes.}
For a language $L$, let
$\overline L:=\{0,1\}^*\setminus L$.  The class \emph{coNP} consists of the
languages whose complements are in NP.  Equivalently, a language is in coNP if
its no-instances have polynomial-size certificates verifiable in polynomial
time.  We have
$$
\operatorname{P} \subseteq \operatorname{NP}\cap\operatorname{coNP}.
$$
It is not known whether NP equals coNP.  If $\operatorname{P}=\operatorname{NP}$,
then P, NP, and coNP are all equal.  The prefix ``co'' refers to complement and
not to a complement of the class as a set: coNP contains P, rather than being
disjoint from it.

\medskip
{\bf Space complexity, L, and NL.}
Besides time, one can measure the amount of memory used by a computation.  To
make the measure meaningful for small space bounds, the input is placed on a
read-only tape and only cells on separate work tapes are counted.  A machine
uses logarithmic space if on inputs of length $n$ it visits only $O(\log n)$
work-tape cells.

\begin{definition}
The class $\operatorname{L}$ consists of the languages decidable by a
deterministic Turing machine in logarithmic space.  The class
$\operatorname{NL}$ is defined in the same way using a nondeterministic Turing
machine.
\end{definition}

A logspace machine has only polynomially many configurations, and the
reachability of an accepting configuration can be tested in polynomial time.
Consequently,
$$
\operatorname{L}\subseteq\operatorname{NL}\subseteq\operatorname{P}.
$$
It is known that $\operatorname{NL}=\operatorname{coNL}$.  A central complete
problem for NL is directed reachability: given a digraph $G$ and two vertices
$s,t$, decide whether $G$ contains a directed path from $s$ to $t$.  For
undirected graphs the analogous problem is in L, by Reingold's
theorem~\cite{Reingold}.

A \emph{logspace reduction} is a many-one reduction whose output is produced
by a deterministic logarithmic-space transducer.  Such reductions are also
polynomial-time reductions.  They are used when polynomial-time reductions
would be too coarse.  For example, saying that a problem is P-hard under
logspace reductions records that its difficulty is already present even when
the translation is allowed very little memory.

The appearance of NL in the main text is motivated by algorithms and
expressibility results for special CSPs.  For example, the open problems ask
whether certain orientation-of-tree CSPs lie in NL and whether every CSP in NL
can be expressed in linear Datalog.

\medskip
{\bf Parallel computation and NC.}
Boolean circuits provide a model of parallel computation.  A circuit of
polynomial size has polynomially many gates.  Its \emph{depth} is the length of
the longest path from an input gate to the output gate and measures parallel
time when gates on the same level are evaluated simultaneously.

For $k \geq 1$, the class $\operatorname{NC}^k$ consists, informally, of the
problems solvable by suitably uniform families of polynomial-size circuits of
depth $O((\log n)^k)$ and bounded fan-in.  We put
$$
\operatorname{NC}:=\bigcup_{k\geq 1}\operatorname{NC}^k.
$$
Thus NC is a standard model for problems that admit algorithms with
polylogarithmic parallel time and polynomially many processors.  We have
$\operatorname{NC}\subseteq\operatorname{P}$, and it is widely believed that
the inclusion is strict.

A problem is \emph{P-hard} in the usual fine-grained sense used in this course
if every problem in P reduces to it by a logspace reduction.  It is
\emph{P-complete} if it is P-hard and belongs to P.  Since NC is closed under
logspace reductions, a P-hard problem can belong to NC only if
$\operatorname{P}=\operatorname{NC}$.  This is the significance of the
P-hardness questions in the section on open problems: they ask not only for a
polynomial-time algorithm, but whether the problem is likely to be inherently
sequential.

The choice of reduction matters.  ``P-hard'' in this paragraph always means
hard under logspace reductions, whereas ``NP-hard'' in the main text may safely
be read as hardness under polynomial-time many-one reductions unless a
different reduction is explicitly mentioned.

\medskip
{\bf Exponential time and meta-problems.}
For completeness, define
$$
\operatorname{EXPTIME}
  := \bigcup_{k\geq 1}\operatorname{DTIME}(2^{n^k})
$$
and
$$
\operatorname{NEXPTIME}
  := \bigcup_{k\geq 1}\operatorname{NTIME}(2^{n^k}).
$$
The first class uses deterministic machines and the second nondeterministic
machines.  Clearly
$$
\operatorname{P}\subseteq\operatorname{EXPTIME}
\quad\text{and}\quad
\operatorname{NP}\subseteq\operatorname{NEXPTIME}.
$$
The inclusions are strict in the deterministic case by the time hierarchy
theorem, while the relations between several other classes in this appendix
are open.

These exponential classes arise naturally for CSP meta-problems.  If a
construction from a template $\bB$ has $2^{|B|}$ elements, then explicitly
constructing it and guessing a homomorphism may be possible in NEXPTIME, even
though no polynomial-time meta-algorithm is known.  Proposition~\ref{prop:power-set-exptime}
is an example where additional structure improves such a nondeterministic
exponential bound to a deterministic exponential bound.

\medskip
{\bf Logic and complexity.}
There is a useful logical description of NP.  Existential second-order logic
extends first-order logic by existential quantification over relations.

\begin{theorem}[Fagin~\cite{Fagin}]
Let $\tau$ be a finite relational signature.  A class $\mathcal C$ of finite
$\tau$-structures is in NP if and only if there exists an existential
second-order $\tau$-sentence $\Phi$ such that, for every finite
$\tau$-structure $\bA$,
$$
\bA \in \mathcal C \quad\text{if and only if}\quad \bA \models \Phi.
$$
\end{theorem}

For a fixed finite template $\bB$, the existentially quantified relations may
encode a homomorphism from the input structure to $\bB$.  If
$B=\{b_1,\dots,b_q\}$, introduce a unary relation $X_i$ for each $b_i$.
First-order clauses require that every input element lies in exactly one
$X_i$, and further clauses require that every input relation tuple receives a
tuple of colours that belongs to the corresponding relation of $\bB$.  This
gives another proof that $\Csp(\bB)$ is in NP.  The quantified relations are
unary, and the input relations occur monotonically; this special form is the
starting point for the connection between finite-domain CSPs and MMSNP
mentioned in the introduction.  More on descriptive complexity can be found
in~\cite{Libkin}.

\medskip
{\bf A small map of the classes used in the course.}
The unconditional inclusions discussed above can be summarised as follows:
$$
\operatorname{L}
\subseteq \operatorname{NL}
\subseteq \operatorname{P}
\subseteq \operatorname{NP}
\subseteq \operatorname{NEXPTIME},
$$
and independently $\operatorname{NC}\subseteq\operatorname{P}$ and
$\operatorname{P}\subseteq\operatorname{EXPTIME}$.  Most equalities between
neighbouring classes in this display are not known.  The equality
$\operatorname{NL}=\operatorname{coNL}$ is known; the famous unresolved
equality is $\operatorname{P}=\operatorname{NP}$.

\begin{exercises}
\exercise[Blau]\label{exe:source-289} Show that polynomial-time many-one reducibility is reflexive and
transitive.
% Solution sketch: The identity map is a polynomial-time reduction, proving reflexivity. If $f$
% reduces $L_1$ to $L_2$ and $g$ reduces $L_2$ to $L_3$, then $g\circ f$ is computable in polynomial
% time because the length of $f(x)$ is polynomially bounded. This proves transitivity.
\exercise[Blau]\label{exe:source-290} Let $\bB$ be a fixed finite relational structure.  Give an explicit
certificate for a yes-instance of $\Csp(\bB)$ and estimate its encoding length.
% Solution sketch: A certificate is the table of a map $h:A\to B$. It uses $|A|\lceil\log_2|B|\rceil$
% bits. For every input relation tuple, verify that its coordinatewise image lies in the fixed
% corresponding relation of $\bB$; this takes polynomial time.
\exercise[Blau]\label{exe:source-292} Explain why the brute-force algorithm for $\Csp(\bB)$ runs in
exponential time when $|B|\geq 2$, although $\bB$ is fixed.
% Solution sketch: There are $|B|^{|A|}$ candidate maps $A\to B$. Although $|B|$ is constant, this is
% $2^{|A|\log_2|B|}$ and is exponential in the input size when $|B|\geq2$. Checking a single candidate
% is only polynomial.
\exercise[Rot]\label{exe:source-293} Show that a deterministic logspace reduction is a polynomial-time
reduction.  You may assume that a halting logspace transducer does not repeat a
configuration.
% Solution sketch: A logspace transducer has only polynomially many configurations: its state,
% input-head position, work tape, and output-control data use $O(\log n)$ bits. If it halts without
% repeating a configuration, it runs for polynomially many steps. Hence its output and running time
% are polynomial, so it is a polynomial-time reduction.
\exercise[Rot]\label{exe:source-294} Suppose that a P-hard problem under logspace reductions is in NC.  Prove
that $\operatorname{P}=\operatorname{NC}$.
% Solution sketch: Let $L$ be P-hard under logspace reductions and suppose $L\in\mathrm{NC}$. Every
% language in P has a logspace reduction to $L$; logspace reductions are computable in NC and NC is
% closed under composition. Thus every P language lies in NC. Since $\mathrm{NC}\subseteq\mathrm P$,
% equality follows.
\exercise[Orange]\label{exe:source-291} Suppose that $L_1$ and $L_2$ are polynomial-time equivalent.  Show that
$L_1$ is in P if and only if $L_2$ is in P, and that $L_1$ is NP-hard if and
only if $L_2$ is NP-hard.
% Solution sketch: Compose a polynomial-time decision procedure with the reduction in either direction
% to transfer membership in P. For NP-hardness, compose every reduction to one language with the
% reduction from that language to the other. Polynomial reductions are transitive, so hardness
% transfers both ways.
\end{exercises}
% END INLINED SOURCE: Complexity-Expanded.tex
\end{document}

%% file: local-defs.tex
\usepackage{a4wide,amsmath,amsxtra,amsthm,mathrsfs,amssymb,amsfonts,stmaryrd}
 \usepackage{makeidx}  
  \makeindex

\DeclareMathOperator{\sdc}{\leq_{\text{sdc}}}

\DeclareMathOperator{\majority}{majority}
\DeclareMathOperator{\minority}{minority}
\DeclareMathOperator{\median}{median}
\DeclareMathOperator{\pr}{\pi}
\DeclareMathOperator{\Proj}{\bf Proj}

\DeclareMathOperator{\Ker}{Ker}
\DeclareMathOperator{\PC}{PC}
\DeclareMathOperator{\OIT}{1IN3}
\DeclareMathOperator{\NAE}{NAE}
\DeclareMathOperator{\WNU}{WNU}
\DeclareMathOperator{\inv}{Inv}
\DeclareMathOperator{\Con}{Con}

\renewcommand{\P}{\operatorname{P}}
\renewcommand{\S}{\operatorname{S}}
\DeclareMathOperator{\Pfin}{P^{\fin}}
\newcommand{\R}{\operatorname{Refl}}
\DeclareMathOperator{\RP}{\R P}

\DeclareMathOperator{\RPfin}{\R P^{\fin}}

\newcommand{\cl}[1]{\langle #1 \rangle}

\DeclareMathOperator{\CD}{{\mathfrak S}} % canonical Database
\DeclareMathOperator{\CQ}{\phi} % canonical Query

\DeclareMathOperator{\comp}{comp} % canonical Query

\newcommand{\abs}{\vartriangleleft}
\newcommand{\cabs}{\vartriangleleft_{\text{cen}}}

\DeclareMathOperator{\Minion}{Minion}

\newcommand{\bA}{\mathfrak{A}}
\newcommand{\bB}{\mathfrak{B}}
\newcommand{\bC}{\mathfrak{C}}
\newcommand{\bD}{\mathfrak{D}}
\newcommand{\bE}{\mathfrak{E}}

\newcommand{\bF}{\mathfrak{F}}
\newcommand{\bG}{\mathfrak{G}}

\newcommand{\bH}{\mathfrak{H}}
\newcommand{\bL}{\mathfrak{L}}
\newcommand{\bM}{\mathfrak{M}}
\newcommand{\bS}{\mathfrak{S}}

\newcommand{\bP}{\mathfrak{P}}

\newcommand{\mN}{{\mathbb N}}

\newcommand{\cB}{{\mathscr B}}
\newcommand{\cE}{{\mathscr E}}
\newcommand{\cF}{{\mathscr F}}
\newcommand{\cC}{{\mathscr C}}
\newcommand{\cD}{{\mathscr D}}

\newcommand{\cO}{{\mathscr O}}
\newcommand{\fA}{{\bf A}}
\newcommand{\fX}{{\bf X}}
\newcommand{\fQ}{{\bf Q}}
\newcommand{\fB}{{\bf B}}
\newcommand{\fC}{{\bf C}}
\newcommand{\fD}{{\bf D}}
\newcommand{\fG}{{\bf G}}
\newcommand{\fH}{{\bf H}}
\newcommand{\fL}{{\bf L}}
\newcommand{\fS}{{\bf S}}
\newcommand{\fE}{{\bf E}}
\newcommand{\fR}{{\bf R}}
\newcommand{\fF}{{\bf F}}
\newcommand{\fM}{{\bf M}}
\newcommand{\fN}{{\bf N}}
\newcommand{\fT}{{\bf T}}
\DeclareMathOperator{\Cg}{Cg} 
\DeclareMathOperator{\pp}{pp} 
\DeclareMathOperator{\Clo}{Clo} 
 
\DeclareMathOperator{\fin}{fin}
\DeclareMathOperator{\HI}{HI}
\DeclareMathOperator{\HS}{HS}
\DeclareMathOperator{\SP}{SP}
\DeclareMathOperator{\SH}{SH}
\DeclareMathOperator{\PH}{PH}
\DeclareMathOperator{\HP}{HP}
\DeclareMathOperator{\PS}{PS}
\DeclareMathOperator{\HSP}{HSP}
\DeclareMathOperator{\HHH}{H}
\DeclareMathOperator{\PPP}{P}
\DeclareMathOperator{\PPPfin}{P^{\fin}}
\DeclareMathOperator{\SSS}{S}
\DeclareMathOperator{\HSPfin}{HSP^{\fin}}

\DeclareMathOperator{\id}{id}
\DeclareMathOperator{\PP}{I}
\DeclareMathOperator{\HH}{H}
\DeclareMathOperator{\CC}{C}

\newcommand{\I}{\operatorname{I}}
\usepackage{epsf}
\usepackage{epsfig}
\usepackage{graphicx}
\usepackage{url}
\usepackage{cite}
\usepackage{amsxtra, amssymb}
\usepackage{eucal, url}
\usepackage{xspace, varioref}

\newtheorem{theorem}{Theorem}[section]
\newtheorem{proposition}[theorem]{Proposition}
\newtheorem{conjecture}{Conjecture}
\newtheorem{corollary}[theorem]{Corollary}
\newtheorem{lemma}[theorem]{Lemma}

\newtheorem{question}{Question}

\theoremstyle{definition}
\newtheorem{definition}[theorem]{Definition}
\newtheorem{example}[section]{Example}
\newtheorem{remark}[theorem]{Remark}
\newtheorem{observation}[theorem]{Observation}

\DeclareMathOperator{\NExpTime}{NExpTime}
\DeclareMathOperator{\AC}{AC}
\DeclareMathOperator{\SAC}{SAC}

\DeclareMathOperator{\Aut}{Aut}
\DeclareMathOperator{\Sym}{Sym}
\DeclareMathOperator{\Pol}{Pol}
\DeclareMathOperator{\End}{End}
\DeclareMathOperator{\Inv}{Inv}

\DeclareMathOperator{\Csp}{CSP}

\DeclareMathOperator{\Hom}{H}
\DeclareMathOperator{\HomE}{H}
\DeclareMathOperator{\Var}{Var}

\newcommand{\true}{{\it true}}
\newcommand{\ignore}[1]{}

\newtotcounter{exercise}

\newlength{\exerciselabelwidth}
\newskip\exercisegap
\newif\iffirstexerciseinseries

\newenvironment{exercises}{%
  \par\medskip
  \firstexerciseinseriestrue
  \exercisegap=1.05\baselineskip
    plus 0.15\baselineskip
    minus 0.10\baselineskip
  \setlength{\belowdisplayskip}
    {0.3\baselineskip plus 0.1\baselineskip
                         minus 0.1\baselineskip}%
  \setlength{\belowdisplayshortskip}
    {0.15\baselineskip plus 0.1\baselineskip}%
  \begin{list}{}{%
    \setlength{\labelwidth}{\exerciselabelwidth}%
    \setlength{\labelsep}{0.6em}%
    \setlength{\leftmargin}{\dimexpr\labelwidth+\labelsep\relax}%
    \setlength{\rightmargin}{0pt}%
    \setlength{\itemsep}{0pt}%
    \setlength{\parsep}{0.15\baselineskip}%
    \setlength{\topsep}{0.3\baselineskip}%
    \setlength{\partopsep}{0pt}%
  }%
}{%
  \end{list}%
}

\newcommand{\exercise}[1][]{%
  \iffirstexerciseinseries
    \firstexerciseinseriesfalse
  \else
    \par\addvspace{\exercisegap}%
  \fi
  \refstepcounter{exercise}%
  \item[\textbf{Exercise \theexercise.}]%
  \if\relax\detokenize{#1}\relax
  \else
    \leavevmode
    \makebox[0pt][l]{%
      \hspace{\dimexpr\linewidth+0.75em\relax}%
      \smash{\raisebox{-0.45\height}{%
        \includegraphics[
          height=10mm,
          width=\marginparwidth,
          keepaspectratio
        ]{#1.jpg}%
      }}%
    }%
  \fi
  \ignorespaces
}